\documentclass[groupeaddress,aps,prb,article,nofootinbib,10pt]{revtex4-2}
\usepackage{array}
\usepackage{titlesec}
\usepackage{amsbsy}
\usepackage{tikz}
\usetikzlibrary{calc}
\usetikzlibrary{positioning}

\titleformat{\paragraph}
{\normalfont\normalsize\bfseries}{\theparagraph}{1em}{}
\titlespacing*{\paragraph}{0pt}{3.25ex plus 1ex minus .2ex}{1.5ex plus .2ex}
\renewcommand*{\theparagraph}{\roman{paragraph})}

\usepackage{amssymb, amsmath,amsthm,bm,bbm}    
\usepackage{graphicx}   
\usepackage{verbatim}   
\usepackage{color}      
\usepackage[utf8]{inputenc}
\usepackage{newunicodechar}
\newunicodechar{⁢}{\,}
\usepackage[labelformat=simple]{subcaption}
\usepackage{mathrsfs} 
\usepackage{dsfont}
\usepackage[linktocpage = true]{hyperref}  
\usepackage{cancel}
\usepackage{graphicx}
\usepackage{caption} 
\usepackage{tcolorbox}
\usepackage{yfonts}
\usepackage{multirow}
\definecolor{green}{RGB}{35,142,35}
\usepackage{float}
\usepackage{upgreek}

\usepackage{empheq} 
 
\usepackage{ulem}

\usepackage[framemethod=tikz]{mdframed}

 \usepackage{physics}
 
 \usepackage{bigints}

 \usepackage{braket}
 \usepackage{xparse}

\usepackage[utf8]{inputenc}
\DeclareUnicodeCharacter{2215}{/}

\hypersetup{
    colorlinks   = true,
    linkcolor = blue,
    citecolor    = red
}  
\makeatletter
\def\p@subsection{}
\makeatother

\usepackage{etoolbox}
\usepackage{cleveref}
\makeatletter
\appto{\appendix}{%
\@ifstar{\def\theequation@prefix{A. }}%
  {}%
}
\makeatother

\usepackage{multirow}

\allowdisplaybreaks

\newcommand{\eq}[1]{Eq. \eqref{#1}}
\newcommand{\fig}[1]{Fig. \ref{#1}}
\newcommand{\tab}[1]{Table \ref{#1}}

\newcommand{\dslash}{\not{\hbox{\kern-2pt $\partial$}}}
\newcommand{\bqa}{\begin{eqnarray}} 
\newcommand{\eqa}{\end{eqnarray}}
\newcommand{\nn}{\nonumber \\}

\newcommand{\beq}{\begin{equation}}
\newcommand{\eeq}{\end{equation}}

\def\be{\begin{eqnarray}}
\def\ee{\end{eqnarray}}

\newcommand\HD{{\bf H}_d}

\newcommand\thetasq{\Theta (\theta,\vec q)}
\newcommand\deltaq{\Delta (\delta,\theta,\vec q)}

\newcommand\thetasqp{\Theta (\theta',\vec q)}
\newcommand\deltaqp{\Delta (\delta',\theta',\vec q)}

\newcommand\edim{{\pmb e}}
\newcommand\lambdadim{{\pmb \lambda}}

\newcommand{\mtheta}{ \theta_{\overline{12}} }
\newcommand\sigp{\mathfrak{s}^t}

\newcommand\tcold{{\theta_{cs}}}

\newcommand\bth{{\bar \theta}}
\newcommand\sqmu{{\sqrt{\mu}}}
\newcommand\cN{{\mathbb{N}}}

\newcommand{\Lc}{{}}

\newcommand{\snut}{\mathfrak{s}^{(\nu,t)}}
\newcommand\KFthetadim{{\bf K}_{F,\theta}}

\newcommand\KFtheta{K_{F,\theta}}
\newcommand\KFAVdim{{\bf k}_{F}}
\newcommand\KFAV{k_{F}}

\newcommand\Dtwon{\Delta^{(2n)}{}}
\NewDocumentCommand{\av}{e{_^}}{%
  \ensuremath{%
    {\mathfrak V}%
    \IfValueT{#1}{_{#1}}
    \IfValueTF{#2}{^{\prime\,#2}}{^{\prime}}
  }%
}

\NewDocumentCommand{\as}{e{_^}}{%
  \ensuremath{%
    {\mathbb S}%
    \IfValueT{#1}{_{#1}}
    \IfValueTF{#2}{^{\prime\,#2}}{^{\prime}}
  }%
}

\NewDocumentCommand{\aV}{e{_^}}{%
  \ensuremath{%
    V%
    \IfValueT{#1}{_{#1}}
    \IfValueTF{#2}{^{\prime\,#2}}{^{\prime}}
  }%
}
\NewDocumentCommand{\aS}{e{_^}}{%
  \ensuremath{%
    S%
    \IfValueT{#1}{_{#1}}
    \IfValueTF{#2}{^{\prime\,#2}}{^{\prime}}
  }%
}
\NewDocumentCommand{\aVM}{e{_^}}{%
  \ensuremath{%
    V%
    \IfValueT{#1}{_{#1}}
    \IfValueTF{#2}{^{\prime M\,#2}}{^{\prime M}}
  }%
}
\NewDocumentCommand{\aVS}{e{_^}}{%
  \ensuremath{%
    V%
    \IfValueT{#1}{_{#1}}
    \IfValueTF{#2}{^{\prime S\,#2}}{^{\prime S}}
  }%
}

\newcommand\wII{w}

\newcommand\aVP[2]{\aV^{(#1,#2)}}

\newcommand\disc{\eta_{P,y}}
\newcommand\discinf{\eta_{P,-\infty}}
\newcommand\etaPy{\eta_{P,y}}
\newcommand\etaPI{\eta_{P,\infty}}
\newcommand\etaPIII{\eta_{P,-\infty}}
\newcommand\etaPII{\eta_{P,w}}

\newcommand\etag{\eta}

\newcommand\hII{h_w}
\newcommand\hIII{h_{-\infty}}
\newcommand\deltaaVUV{\delta \aV}

\newcommand\gtt{\mathsf{g}_{\theta,\theta}}

\DeclareRobustCommand{\halfominus}{
  \begin{tikzpicture}[baseline=(base)]
    \node[inner sep=0pt] (base) at (0,0) {};
    \draw[thick] (0,0) circle (0.06);
    \fill[black] (0.06,0) arc (0:180:0.06) -- cycle; 
    \draw[thick] (-0.06,0) -- (0.06,0); 
  \end{tikzpicture}
}
\newcommand\dist{\mathscr{G}_\mu}
\newcommand\renf{\tilde{\boldsymbol{K}}}
\newcommand\renm{\tilde{\boldsymbol{k}}}
\newcommand\fk{\mathscr{K}\left((\mathbf{L}+\renf_2),(\mathbf{L}+\renf_3)\right)}
\newcommand\renfc{\tilde{\boldsymbol{T}}}
\newcommand\fkc{\mathscr{K}}
\newcommand\renfcs{\boldsymbol{T}}
\newcommand\renfca{\boldsymbol{t}}

\newcolumntype{P}[1]{>{\centering\arraybackslash}p{#1}}

\begin{document}

\title{
\textbf{
Classification of non-Fermi liquids and universal superconducting fluctuations
}}

\author{Shubham Kukreja$^\dagger$}
\author{Dawson M. Willerton$^\dagger$}
\author{Sung-Sik Lee}
\affiliation{$^{1}$Department of Physics \& Astronomy, McMaster University, Hamilton ON L8S 4M1, Canada}
\affiliation{$^{2}$Perimeter Institute for Theoretical Physics, Waterloo ON N2L 2Y5, Canada}
\affiliation{$^{\dagger}$ 
These are the co-first authors.
}

\date{\today}

\begin{abstract}
In quantum critical metals, a plethora of different non-Fermi liquids arises depending on the nature of critical fluctuations coupled to Fermi surfaces. 
In this paper, we classify non-Fermi liquids 
that arise from long-wavelength ($q=0$) critical fluctuations
and characterize their 
universal superconducting fluctuations.
The essential tool is the projective fixed points, 
which generalizes the notion of fixed points to fixed trajectories that take into account the incessant running of the Fermi momentum under the renormalization group flow. 
Based on the topology of bundles of projective fixed points,
non-Fermi liquids are first grouped into seven superuniversality classes. 
Each superuniversality class includes multiple universality classes, which are further classified by the universal pairing interactions and emergent symmetries.
Despite the pairing interaction generated by critical fluctuations, some non-Fermi liquids remain stable down to zero temperature due to the incoherence of excitations and the lack of scale invariance caused by Fermi momentum.
Depending on the strength and span of the universal pairing interaction in momentum space, the emergent symmetry of non-Fermi liquids may or may not be lower than that of Fermi liquids. 
In non-Fermi liquids that become superconductors at low temperatures, the universal data of the parent metal determine the lower bound for the superconducting transition temperature and the associated pairing symmetry.
In superuniversality classes that contain non-Fermi liquids prone to non-s-wave superconducting instabilities, 
the critical angular momentum above which pairing instability becomes inevitable is sensitive to the Fermi momentum,
and the associated superconducting transition  temperature oscillates as a function of the density.
We use physical examples, as well as a toy model, to elucidate the universal low-energy physics of all superuniversality classes.
\end{abstract}

\maketitle
\newpage
\tableofcontents
\newpage


\newpage

\section*{Notation and Glossary}

\begin{itemize}

\item $\Lambda'$: the high-energy cutoff below which the low-energy effective theory is valid

\item $\Lambda$: a UV energy cutoff, which is generally lower than $\Lambda'$, below which a non-Fermi liquid physics sets in

\item $T_c$ : superconducting transition temperature

\item $\mu$: floating energy scale at which the scale-dependent coupling functions are matched with the vertex function

\item $l= \log \Lambda/\mu$:  logarithmic length scale

\item $l_{SC}$:  logarithmic length scale at which superconductivity occurs

\item $\KFthetadim$ : the magnitude of Fermi momentum at angle $\theta$

\item 
$\KFAVdim = \frac{1}{2\pi} \int d \theta \KFthetadim$ : 
the average Fermi momentum

\item 
$\bar \theta$ : proper angular coordinate,
where $d \bar \theta / \sqmu$
represents the number of scatterings a fermion needs to undergo with critical bosons at energy scale $\mu$ to traverse between $\theta$  and $\theta + d \theta$ near the Fermi surface

\item 
$\cN(l)$ :
the number of scatterings that a fermion needs to undergo with critical bosons to traverse the entire Fermi surface once at scale $l$

\item
$y^{(m)}(l) = 
 \log \left[ 
         \frac{m}{\cN(l)} \right]$ :
the logarithm of the angular momentum $m$ normalized by $\cN(l)$

\item 
$\aV_{y}$
: the pairing interaction at the logarithmic angular momentum $y$

\item Projective fixed point (PFP): a one-dimensional trajectory of the renormalization group flow along which 
$\KFAVdim/\mu$ grows incessantly 

\item $\pm\infty$-asymptotic fixed points: 
the small/large $\KFAVdim/\mu$ limit of a projective fixed point 

\item 
$\aV^\bullet_{\pm \infty}/
\aV^\circ_{\pm \infty}/
\aV^{\halfominus}_{\pm \infty}$:
stable/unstable/marginal $\pm \infty$-asymptotic fixed point

\item 
${\cal B}^{IR (UV)}_{x}$ : the basin of attraction of the asymptotic fixed point $x$ under the renormalization group flow with decreasing (increasing) energy scale



\item Metallic PFP:
the PFP that emanates from 
$\aV^\bullet_{\infty}$ or
$\aV^\halfominus_{\infty}$

\item Separatrix PFP:
the PFP that forms the boundary of
${\cal B}^{IR}_{\aV^\bullet_{-\infty}}$
or
${\cal B}^{IR}_{\aV^\halfominus_{-\infty}}$

\item Universality class (also called individual universality class): a class of microscopic theories that are in one phase

\item Superuniversality class: a group of universality classes that share a common topology of bundles of projective fixed points

\item Critical superuniversality class: a superuniversality class that arises at the boundary between two topologically stable superuniversality classes

\item Stable non-Fermi liquid: an individual non-Fermi liquid universality class that does not require fine tuning of the bare fermion-fermion coupling

\item Critical non-Fermi liquid: an individual non-Fermi liquid universality class that requires some fine tuning of the bare fermion-fermion coupling

\item Quasi-universal non-Fermi liquid: an individual non-Fermi liquid universality class that is realized between $\Lambda$ and $T_c$ in the presence of hierarchy $T_c/\Lambda \ll 1$

\end{itemize}

\newpage

\section{Introduction}
\label{sec:intro}

Fermi liquids\cite{LANDAU,LANDAU2} are fragile states of matter that become unstable against superconductivity at low temperatures in the presence of arbitrarily weak attractive interactions\cite{PhysRevB.42.9967,POLCHINSKI1,SHANKAR}.
At the same time, superconducting instabilities of Fermi liquids are weak and non-universal: an attractive pairing interaction perturbatively added to Fermi liquids grows only logarithmically with decreasing temperature\cite{BCS1,BCS2},
and the superconducting transition temperature and the pairing symmetry are sensitive to the bare interaction introduced at high-energy scales.
These reflect the indecisive nature of Fermi liquids, which are neither stable as metals nor strongly superconducting.
It is, then, curious to know whether and how these delicate features of Fermi liquids are altered in various non-Fermi liquids that arise in the presence of quantum critical fluctuations\cite{
HERTZ,
MILLIS,
VARMALI,
POLCHINSKI2,
PLEE1,
ALTSHULER,
YBKIM,
ABANOV2,
SSLEE,
MAX0,
MAX2,
NAYAK2,
SCHOFIELD,
DENNIS,
SENTHIL,
MROSS,
FITZPATRICK,
SHOUVIK2,
SCHLIEF,
PhysRevX.11.021005,
PhysRevB.107.165152,
annurev:/content/journals/10.1146/annurev-conmatphys-031218-013339,
Chang_2025}.
More specifically, the question can be refined into the following ones:
\begin{itemize}
    \item 
{\it 
Can non-Fermi liquids remain stable down to zero temperature in spite of the strong pairing interaction generated from critical fluctuations? 
} 
    \item 
{\it 
In non-Fermi liquids that become superconductors at low temperatures, 
how do the universal data of the parent metals constrain the pairing symmetry and the transition temperature of the superconducting states?
}
    \item 
{\it
How are the distinct non-Fermi liquids classified?
}
\end{itemize}
The goal of this paper is to address these questions for non-Fermi liquids that support hot Fermi surfaces of incoherent fermions in the clean limit\footnote{
For clean non-Fermi liquids with hot spots, see Refs. \cite{ABANOV3, MAX2, SCHLIEF, BORGES2023169221}.
For recent progress in understanding the interplay between quantum critical fluctuations and superconductivity in the presence of disorder, see Refs. \cite{ doi:10.1126/science.abq6011,
PhysRevLett.132.236501, 
2025arXiv250611952E,
10.1143/PTP.32.37,
PhysRevLett.133.186502,
sachdev2025foot}.}
Understanding superconducting fluctuations is important not only for determining the ultimate fate of metals in the low temperature limit\cite{
STEWART,
taillefer2010scattering,
2023arXiv231007539S,
jin2011link,
sarkar2019correlation,
analytis2014transport,
yuan2022scaling,
Jiang:2023aa,
Nguyen:2021aa,
Zhang:2025aa,
Keimer2015179,
phillips2022stranger,
Cai:2023aa},
but also for classifying the universal low-energy behaviors of metals, as the superconducting fluctuations encode universal properties of non-Fermi liquids, 
such as emergent symmetry\cite{2005cond.mat..5529H,PhysRevX.11.021005}.

\begin{figure}[th]
\centering
\includegraphics[width=0.28\linewidth]{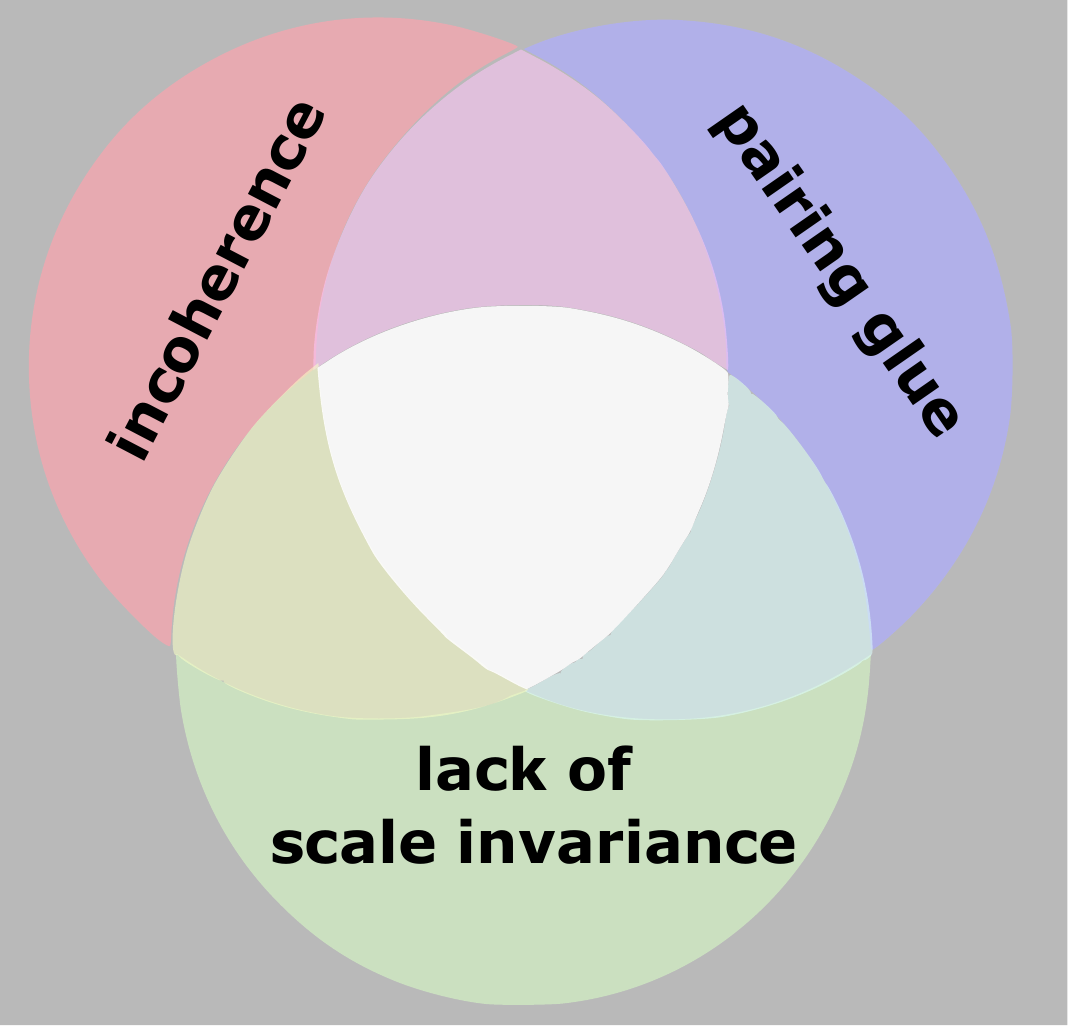}
\caption{
Three ingredients that determine universal superconducting fluctuations in non-Fermi liquids.
}
\label{fig:triangle}
\end{figure}

The possibility of realizing non-Fermi liquids that are stable against superconductivity arises due to the pair-breaking effect of incoherence\cite{
Balatsky01101993,
PhysRevLett.74.2575,
doi:10.1142/S0217979296000349,
PhysRevB.100.115132,
BORGES2023169221}.
Fermions strongly dressed by quantum fluctuations 
acquire some resistance to superconducting instabilities because Cooper pairs inherit a large uncertainty in energy from the incoherent constituents and are less prone to Bose-Einstein condensation.
However, the pair-breaking effect of incoherence is only one of multiple factors that determine the fate of non-Fermi liquids.
The critical fluctuations that cause incoherence also provide a glue for the formation of Cooper pairs that promotes superconducting instabilities\cite{
PhysRevD.59.094019,
PhysRevB.91.115111}. 
Generally, the more incoherent the excitations are, the stronger the universal pairing interactions become.
Precisely weighing these two opposing forces is further complicated by the lack of scale invariance, which is caused by the Fermi momentum\cite{
BORGES2023169221,
PhysRevB.110.155142}.
Therefore, superconducting fluctuations and instabilities of non-Fermi liquids are determined by the {\it scale-dependent competition between pair-breaking incoherence and pair-forming glue}, as illustrated in  \fig{fig:triangle}.

\begin{figure}[th]
\centering
\includegraphics[width=0.75\linewidth]{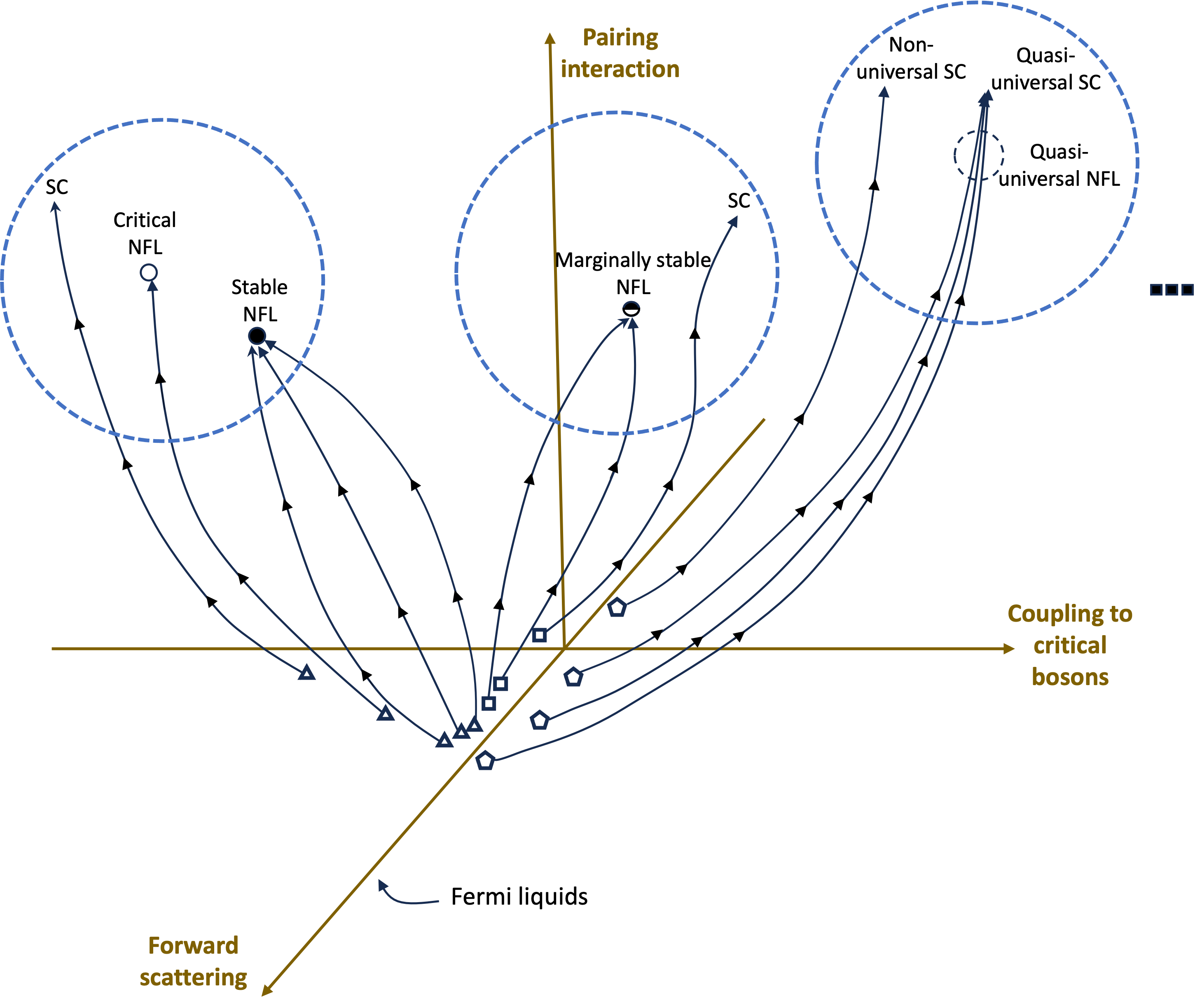}
\caption{
A schematic renormalization group (RG) flow from Fermi liquids to non-Fermi liquids and their descendant superconductors.
The RG flows from Fermi liquids to non-Fermi liquids are triggered by coupling Fermi surfaces with various types of critical fluctuations and tuning the four-fermion coupling.
Different types of polygons represent microscopic theories that exhibit qualitatively different behaviors at low energies.
For example, the triangles, that represent a Fermi surface coupled with U(1) gauge field, can flow to a stable non-Fermi liquid (NFL), a superconductor (SC) or a critical NFL,
depending on the bare fermion-fermion coupling.
The pentagons represent an Ising-nematic quantum critical metal, which becomes unstable against pairing irrespective of the bare coupling.
A subset of such unstable non-Fermi liquids may still exhibit quasi-universal physics at intermediate energy scales 
due to an incomplete but significant focus of the RG flow toward a universal profile. 
The squares represent a critical case that is between the former two classes of non-Fermi liquids.
}
\label{fig:fromFLtoNFLs}
\end{figure}

There are many different types of non-Fermi liquids, and their behaviors vary among them.
A schematic renormalization group (RG) flow that illustrates various possibilities is shown in  \fig{fig:fromFLtoNFLs}.
If the incoherence dominates over the universal pairing interaction, 
the resulting non-Fermi liquid can be stable against pairing down to zero temperature as long as the bare pairing interaction is not too attractive.
If the pairing glue generated from critical fluctuations prevails over incoherence, however, the ground states will be superconductors, irrespective of the bare coupling. 
In the latter case, the resulting superconducting instabilities are expected to be more universal and stronger.
They will be less sensitive to the microscopic details than in Fermi liquids because a non-Fermi liquid generally has fewer tunable (marginal) parameters in each universality class.\footnote{
Although there can be many distinct non-Fermi liquid universality classes, as will be discussed in this paper.
}
For example,  superconducting instabilities can be driven by the universal pairing interaction that is largely insensitive to the bare coupling defined at the lattice scale\cite{
PhysRevD.59.094019,PhysRevB.91.115111}.
Generic non-Fermi liquids are also expected to host stronger superconductivity because the critical fluctuations that generate the pairing glue are strongly coupled with fermions.
When superconductivity dominates the incoherence, one generically expects it to win by a finite `margin' 
rather than `marginally'; although, a priori, it is hard to know which one wins.

\begin{figure}[th]
\centering
\includegraphics[width=0.25\linewidth]{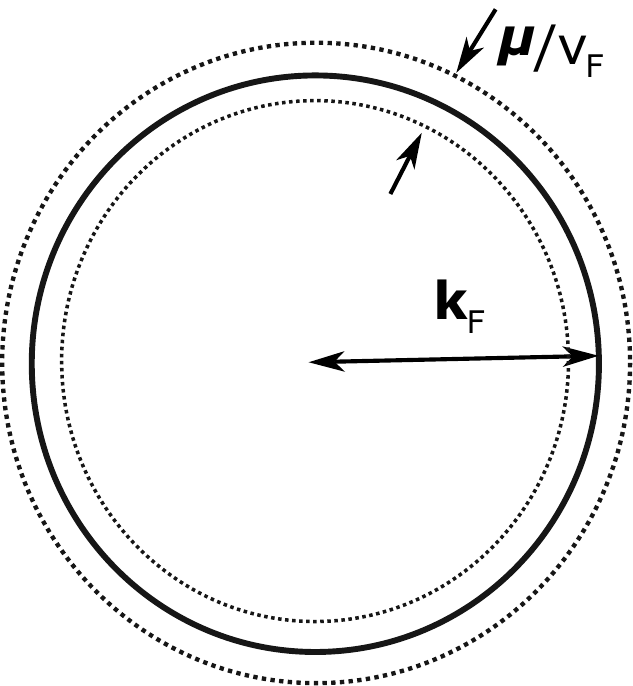}
\caption{
The Fermi momentum is relevant:
the dimensionless ratio between Fermi momentum $\KFAVdim$ and the thickness of the shell of energy $\mu$ near the Fermi surface continues to increase as $\mu$ is lowered.
$v_F$ is Fermi velocity.
}
\label{fig:kFisRelevant}
\end{figure}

How can one systematically chart the landscape of non-Fermi liquids that exhibit different behaviors at low energies?
The incoherence of excitations and their universal pairing interaction are intrinsic features that define non-Fermi liquids as low-temperature states of quantum matter.
Therefore, one should be able to classify non-Fermi liquids and characterize their superconducting fluctuations through low-energy effective theories.
Traditionally, it was thought that metallic universality classes, like other phases of matter, can be identified with scale-invariant fixed points of the renormalization group (RG) flow, and their instabilities can be understood as runaway flows from (real or complex) fixed points\cite{PhysRevB.42.9967,POLCHINSKI1,SHANKAR}.
This idea is at the heart of the patch theory, which selects a few patches of the Fermi surface and approximates them with non-compact manifolds\cite{POLCHINSKI2}.
One can hope to define a fixed point for the patch theory because a non-compact Fermi surface has no finite scale, and a scale transformation leaves it invariant.

However, replacing the Fermi surface with a few non-compact patches is justified only if inter-patch couplings are weak.
When there are strong inter-patch couplings, one must consider the entire Fermi surface as it is.
Closed Fermi surfaces cannot be scale-invariant due to the finite scales associated with their sizes.
The Fermi momentum $\KFAVdim$ is a strictly relevant parameter.
{\it Relative} to the floating momentum scale lowered under a coarse-graining procedure, 
such as the thickness of the low-energy shell near the Fermi surface,
the {\it dimensionless} Fermi momentum increases without bound\cite{BORGES2023169221,PhysRevB.110.155142} 
(see \fig{fig:kFisRelevant}).


\begin{figure}[th]
\centering
\includegraphics[width=1.0\linewidth]{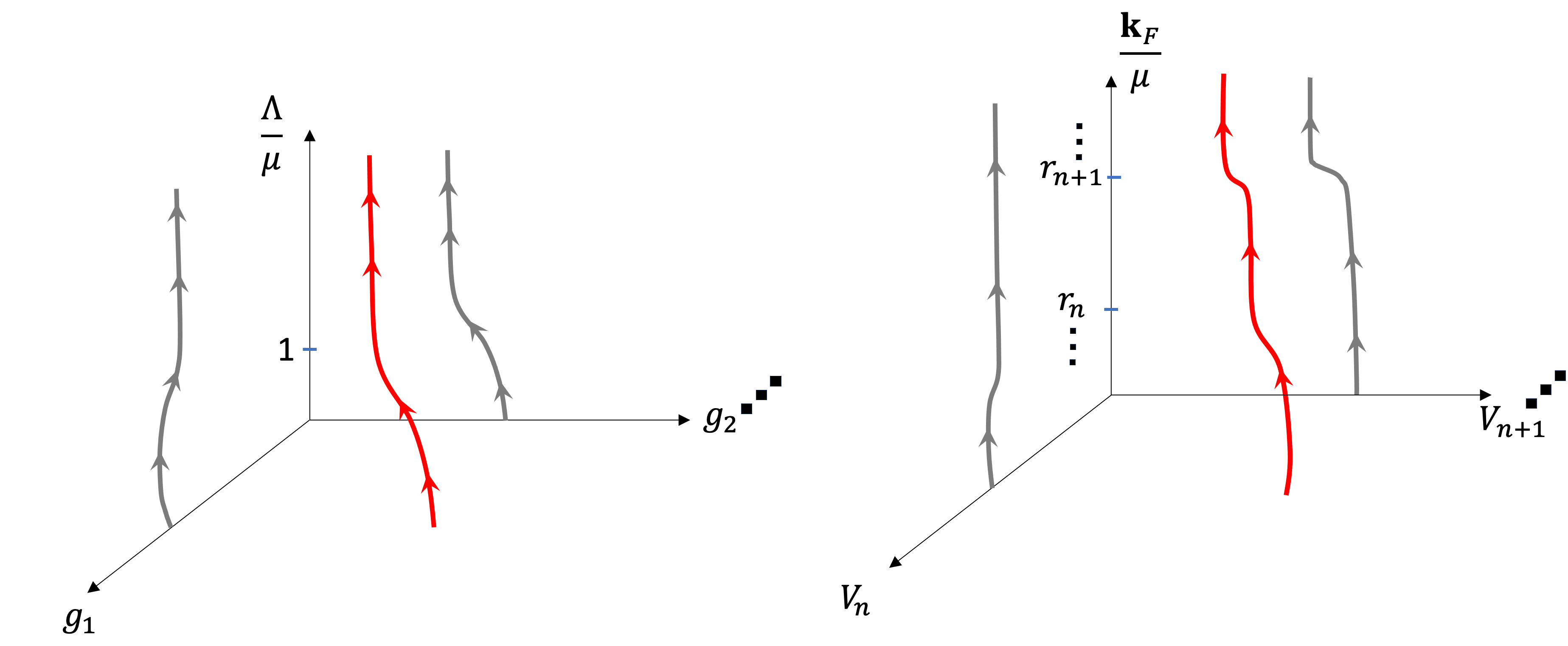}
\caption{
The trajectory of RG flow and its projections to the subspace of various couplings.
(a) Even in non-metallic critical systems,
the dimensionless ratio between a UV scale $\Lambda$ associated with irrelevant couplings and the RG energy scale $\mu$ increases as $\mu$ decreases. 
Thus, the RG flow of marginal/relevant couplings are generally scale-dependent at high energies.
However, there exists a crossover scale $\mu \sim \Lambda$ below which {\it all} low-energy observables are determined by the $\Lambda/\mu \rightarrow \infty$ limit.
(b) In metals, $\KFAVdim/\mu$ formally plays a similar role as $\Lambda/\mu$.
However, the situation is different because there are infinitely many couplings and infinitely many crossover energy scales that become vanishingly small.
The incessant cascade of crossovers is a part of the universal low-energy physics.
}
\label{fig:crossovers}
\end{figure}

As a matter of fact, all effective theories come with some scales.
Associated with an irrelevant coupling $g_i$, one can define a momentum scale $\Lambda_i$, below which the coupling is suppressed. 
One can study the 
physics of crossover
by tracking the RG flow of the irrelevant coupling and its feedback to other couplings as the energy scale is lowered across $\Lambda_i$.
Along the RG trajectory, 
$\Lambda_i/\mu$ continues to grow, and the flows of the other couplings are generally scale-dependent.
In non-metallic systems, however, there exists a non-zero energy scale $\Lambda \equiv \min_i \Lambda_i$ below which the flow of relevant and marginal couplings is captured by the $\Lambda_i/\mu \rightarrow \infty$ limit modulo corrections that vanish in negative powers of $\Lambda_i/\mu$.
Since the $\Lambda_i/\mu \rightarrow \infty$ limit is well defined, one can set  $\Lambda_i/\mu = \infty$ upfront to study the low-energy physics using {\it renormalizable theories} that include only the relevant and marginal couplings.
This is illustrated in \fig{fig:crossovers} (a).
One may hope to treat $\KFAVdim/\mu$ in the same way as $\Lambda_i/\mu$ for irrelevant couplings.
However, this cannot be done\cite{BORGES2023169221}.
Since there are infinitely many couplings in metals, the crossover energy scales can be arbitrarily small.
We elaborate this point below.

\begin{figure}[th]
\centering
\begin{subfigure}{.27\textwidth}
  \centering
  \includegraphics[width=1.0\linewidth]{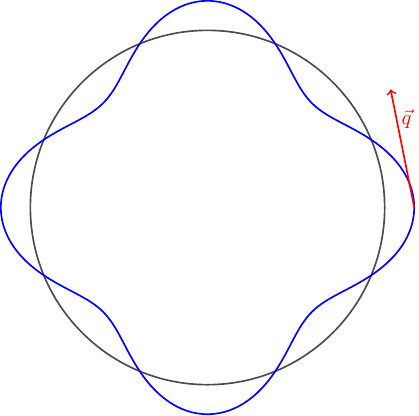}
  \caption{}
  \label{fig:Circle_Wave1}
\end{subfigure}%
~~~~~
\begin{subfigure}{.27\textwidth}
  \centering
  \includegraphics[width=1.0\linewidth]{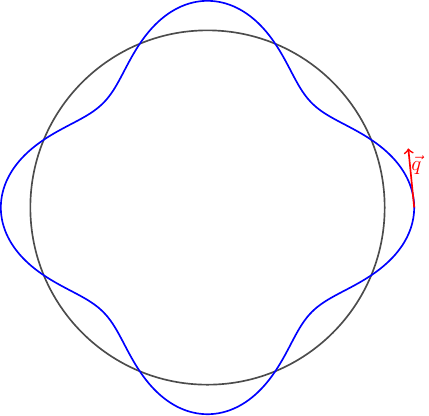}
  \caption{ 
}
  \label{fig:Circle_Wave2}
\end{subfigure}%
~~~~~
\begin{subfigure}{.27\textwidth}
  \centering
  \includegraphics[width=1.0\linewidth]{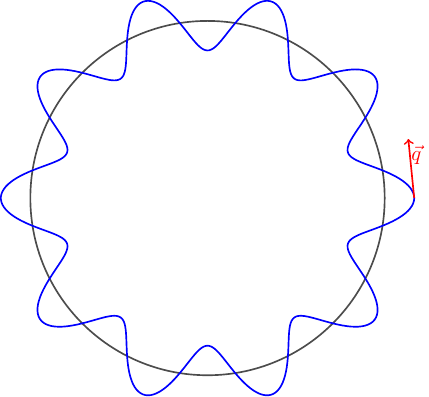}
  \caption{ 
}
  \label{fig:Circle_Wave3}
\end{subfigure}
\caption{
The origin of the angular momentum dependent crossover in the universal pairing interaction generated by critical fluctuations.
Suppose that the critical fluctuations induce scattering with a positive-definite scattering amplitude.
The typical momentum $q$ carried by critical fluctuations depend on the energy scale.
(a) At a high energy, $q$ connects peaks and troughs 
of the Cooper pair wavefunction of angular momentum $n$, mediating an attractive interaction in that channel.
(b) 
At low energies, $q$ becomes smaller than the pitch of that wavefunction $2\pi \KFAVdim/n$. 
Critical fluctuations create scatterings primarily within branches of the same sign, mediating a repulsive interaction.
(c) At a higher angular momentum channel $n'$, the same $q$ in (b) connects peaks and troughs of the pairing wavefunction, still mediating the attractive interaciton.
In the angular momentum channel $n'$, the pairing interaction crossovers to a repulsive one at a lower energy scale.
}
\label{fig:attractive_to_repulsive}
\end{figure}

The difference between $\KFAVdim$ and other high-energy cutoffs manifests in how the Fermi momentum controls the universal pairing interaction in quantum critical metals.
As the energy scale is lowered, the momenta carried by bosonic critical fluctuations become smaller, and the fermion-fermion interaction generated from them becomes increasingly singular. 
In non-metallic critical states, only the interaction at a specific momentum, e.g., zero momentum, matters for low-energy physics because gapless modes exist only at discrete points in momentum space.
In metals, however, the entire profile of the momentum-dependent interaction remains important at low energies because there are infinitely many gapless fermions with continuously varying momentum.
For example, the coupling function in the pairing channel  $V_{\vec k, \vec k'}$  scatters a low-energy Cooper pair from momenta ($\vec k$, $-\vec k$) to ($\vec k'$, $-\vec k'$).
Being an interaction vertex for gapless modes on the Fermi surface, 
$V_{\vec k',\vec k}$ for all $\vec k'$ and $\vec k$ on the Fermi surface is low-energy data.
While $|\vec k'- \vec k|$
ranges from $0$ to $2\KFAVdim$, the contribution of the critical fluctuation to  $V_{\vec k',\vec k}$ is peaked at $|\vec k'- \vec k|\sim q_\mu$ at energy scale $\mu$, 
where $q_\mu$ is a $\mu$-dependent momentum that approaches zero with a decreasing energy scale $\mu$. 
Therefore, the RG flow of the coupling function $V_{\vec k',\vec k}$ depends on $\KFAVdim/\mu$ that keeps growing under the RG flow.

This scale-dependence effectively makes the angular momentum of the Cooper pair run under the RG flow. 
Let $V_n$ be the pairing interaction at angular momentum $n$.
The way $V_n$ is renormalized by critical fluctuations is sensitive to whether $q_\mu$ is greater than or less than the crossover scale 
$2\pi \KFAVdim/n$, 
as illustrated in \fig{fig:attractive_to_repulsive}\footnote{
Without full rotational symmetry, the original angular momentum is not a good quantum number. 
Even in such cases, one can still use the same integer to label Cooper pair wavefunctions preserved under interaction.
}.
As $q_\mu$ is lowered across the crossover scale, the contribution of the critical fluctuations to $V_n$ typically change sign from a repulsive interaction to an attractive one (or vice versa).
Since the renormalization of $V_n$ is controlled by the dimensionless ratio $n q_\mu/\KFAVdim$, under the RG flow, $V_n$ runs with lowering $\mu$ as if $n$ itself is gradually lowered. 
Namely, $V_{n'}$ of large $n'$ is renormalized at a low-energy scale in the same way that $V_n$ of small $n$ is renormalized at a high energy scale.

The pairing interactions in low angular momentum channels eventually become scale invariant at sufficiently low energies. 
One may conclude that full scale invariance should emerge once $V_n$ for all $n$ saturate to their low-energy limits at sufficiently small $\mu$. 
However, this never happens at any non-zero energy scale because $n$ is unbounded, and the crossover energy scale approaches zero in the large $n$ limit.
At any non-zero $\mu$, no matter how small it is, $V_n$ at sufficiently large $n$ has yet to reach its `low-energy' limit.
This is illustrated in \fig{fig:crossovers} (b).

\begin{figure}[t]
\centering
\includegraphics[width=0.6\linewidth]{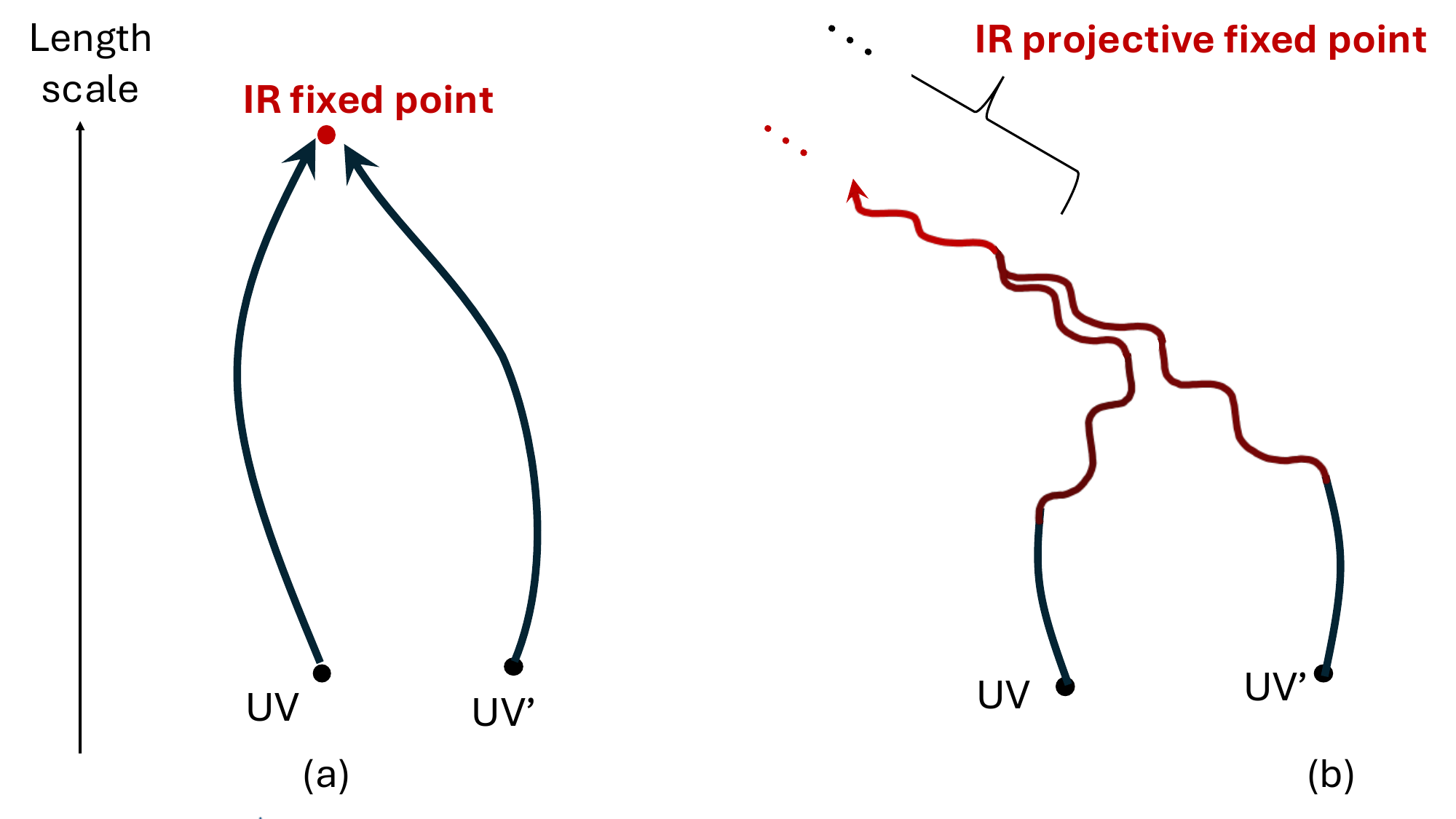}
\caption{
(a) A universality class of non-metallic phase is identified with an infrared fixed point of the RG flow toward which microscopic theories within a basin of attraction are attracted.
(b) In metals, there is no scale invariant fixed point because incessant crossovers persist down to zero energy limit in the infinite dimensional space of coupling functions. 
Nonetheless, a sense of universality arises as the RG trajectories of microscopic theories within a basin of attraction converge at low energies.
Consequently, a metallic universality class is identified with a one-dimensional RG trajectory that emerges in the low-energy limit.
}
\label{fig:FPvsPFP}
\end{figure}

The incessant crossovers that continue at arbitrarily low energies are among the universal low-energy behaviors of metals.
Therefore, each metallic universality class should be identified with the RG {\it trajectory} indefinitely extended from a UV cutoff scale to the zero energy limit, not just a fixed point (see \fig{fig:FPvsPFP}).
The one-dimensional RG trajectory, toward which microscopic theories within a common basin of attraction converge, is referred to as a {\it projective fixed point}\cite{BORGES2023169221,PhysRevB.110.155142}.
They are the closest objects that can be regarded as `fixed points' in metals, but without strict scale invariance.
Due to the flow of the dimensionless Fermi momentum, a sense of scale invariance can be defined only modulo a rescaling of the Fermi momentum.

\begin{figure}[t]
\centering
\begin{subfigure}{.5\textwidth} 
    \centering
\includegraphics[width=0.8\linewidth]{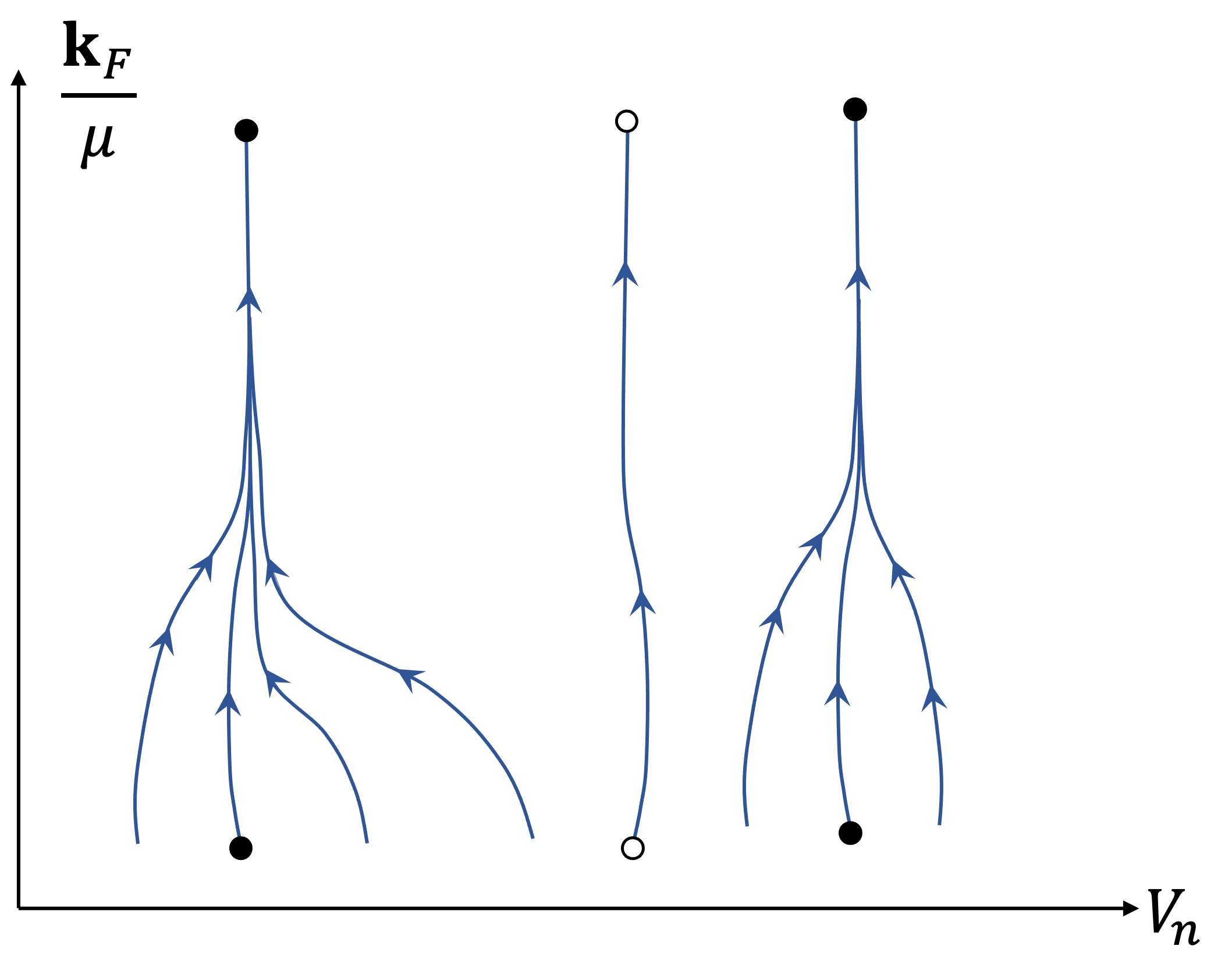}
\caption{}
 \label{}
 \end{subfigure}%
 \begin{subfigure}{.5\textwidth} 
    \centering
\includegraphics[width=0.8\linewidth]{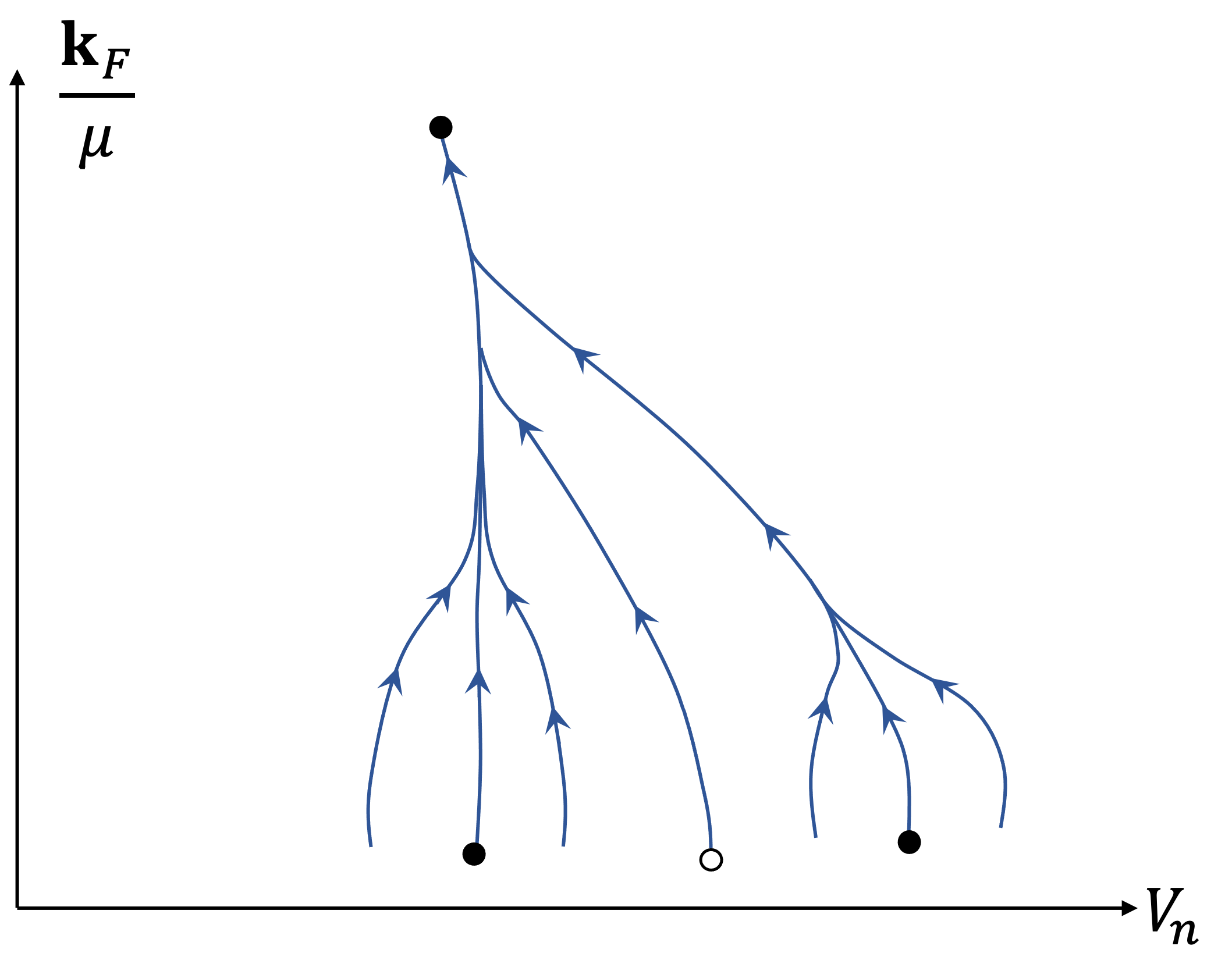}
\caption{}
\label{}
\end{subfigure}
\caption{
Projective fixed points 
drawn in the plane of $\KFAVdim/\mu$ and one coupling, such as the pairing interaction $V_n$ in the angular momentum channel $n$.
The bundle consists of RG trajectories with various bare couplings.
Within this subspace, there can be asymptotic fixed points in the long-distance and short-distance limits; 
the filled (open) circles denote stable (unstable) asymptotic fixed points.
Since the crossover from a short-distance asymptotic fixed point to a long-distance asymptotic fixed point occurs at an arbitrarily low energy in the large $n$ limit, the crossovers themselves are parts of the universal low-energy data of a 
metallic universality class.
The trajectories can be first classified by the topology of the bundle that are robust against smooth deformations.
The topological data consists of the number and type of the asymptotic fixed points, and
the connectivity formed between the asymptotic fixed points in the  small $\KFAVdim/\mu$ region and those in the large $\KFAVdim/\mu$ region.
The bundles in (a) and (b) are topologically distinct, giving rise to distinct superuniversality classes.
The topological phase transition from (a) to (b),
which can be induced by tuning kinematic data or marginal parameters of the theory, occurs when a pair of long-distance asymptotic fixed points are annihilated.
}
\label{fig:bundletopology}
\end{figure}

In this paper, we classify universality classes of non-Fermi liquids through their projective fixed points and characterize their superconducting fluctuations and emergent symmetries. 
There is a large body of literature on superconductivity in non-Fermi liquids, 
which is primarily based on 
the Eliashberg formalism, 
the RG flow of the pairing interaction for low angular momentum channels, 
and numerical simulations\cite{
PhysRevD.59.094019,
PhysRevB.91.115111,
PhysRevB.102.024524,
PhysRevB.102.024525,
PhysRevB.102.094516,
PhysRevB.103.024522,
PhysRevB.103.184508,
PhysRevB.104.144509,
PhysRevB.95.165137,
PhysRevB.92.205104,
PhysRevB.94.115138,
CHUBUKOVSC,
PhysRevB.91.115111,
ABANOV3,
PhysRevB.95.174520,
LEDERERREV,
LEDERER,
PhysRev.135.A550,
PhysRevB.69.020505,
PhysRevB.95.165137,
HAUCK2020168120,
PhysRevB.107.L041103}.
In this work, we use the field-theoretic functional renormalization group formalism\cite{BORGES2023169221} for capturing all universal low-energy crossovers that are present in non-Fermi liquids.
The field-theoretic functional RG is similar to the traditional functional RG\cite{
POLCHINSKIFRG,
WETTERICH,
MORRIS,
REUTER,
ROSA,
HOFLING,
HONERKAMP,
GIES,
GIES2,
BRAUN,
METZNER,
doi:10.1080/00018732.2013.862020,
PhysRevB.61.13609,
PhysRevLett.102.047005,
PhysRevB.61.7364,
SCHERER,
JANSSEN2,
MESTERHAZY,
PhysRevB.87.045104,
EBERLEINMETZNER,
PLATT,
WANGEBERLEIN,
JANSSEN,
MAIEREBERLEIN,
EBERLEIN2,
EBERLEIN3,
JAKUBCZYK,
MAIER,
TORRES} 
in that it keeps track of the RG flow in the space of coupling functions, but it has a different goal.
The goal of the field-theoretic functional RG is to isolate the minimal set of coupling functions, in terms of which all low-energy observables are determined within errors that vanish as positive powers of the energy scale.
While its validity is limited to the low-energy realm, it is easier to control where it is valid. 
Therefore, it uses renormalizable field theories, which include all gapless degrees of freedom, but with the minimal set of coupling functions needed for characterizing universal low-energy physics.
However, some complications arise in metallic renormalizable theories, compared with relativistic field theories.
Low-energy modes can, in principle, exchange large momenta and bring the Fermi momentum scale into the low-energy scaling, altering the nature of an operator from what the dimensional analysis naively suggests\cite{ BORGES2023169221, 
PhysRevLett.128.106402,f3lc-rq2n,IPSITA}.
In local theories, scattering processes with general momentum transfer are required to be present.\footnote{
For recent progress made in solvable non-local theories and their superconducting instabilities, see Refs.  \cite{doi:10.1143/JPSJ.61.2056,Phillips:2020aa,Huang:2022aa,Mai:2026aa}.
}
One should include all couplings that give rise to infrared singularities, even though they are seemingly `irrelevant' by power counting.

From the universal low-energy sector of the functional RG trajectories, we consider a bundle of projective fixed points that captures the functional RG flow associated with a collection of bare coupling functions.
Here, it is convenient to consider bundles composed of varying non-marginal couplings only because marginal couplings act as fixed parameters at low energies. 
Those bundles evolve as the marginal parameters or the kinematic data, such as the spatial dimension and the field content of the theory, are tuned.
However, the topological features of the bundle, such as the connectivity of the short-distance part of the bundle with the long-distance part, are robust against smooth deformation.
This is illustrated in \fig{fig:bundletopology}.
Based on the topology of the bundle of projective fixed points, universality classes of non-Fermi liquids can be grouped into {\it superuniversality classes},
each of which generally contains multiple universality classes that share the same topology.
Distinct non-Fermi liquids within each superuniversality class are further classified by their dynamical properties, such as emergent symmetries and critical exponents.

\begin{figure}[t]
\centering
\includegraphics[width=0.3\linewidth]{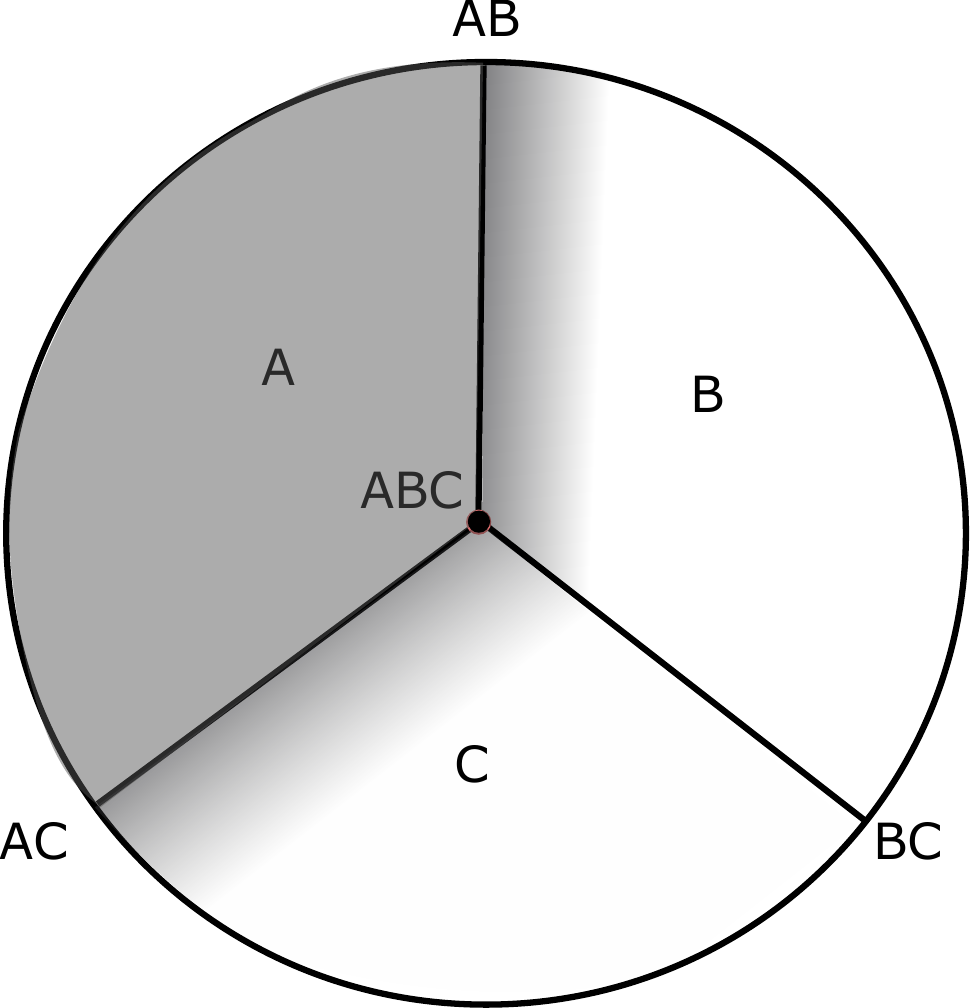}
\caption{
The schematic phase diagram for superuniversality classes.
There are three topologically stable superuniversality classes (A,B,C), divided by critical superuniversality classes (AB, AC, BC).
Class ABC at the center is a double-critical superuniversality class.
Classes A, AB, AC and ABC host stable non-Fermi liquids, while all non-Fermi liquids in other classes are bound to become superconductors at sufficiently low temperatures.
In classes B, C and BC, individual universality classes proximate to class A (in the shaded region) exhibit quasi-universal behaviors, governed by the non-Fermi liquids realized in intermediate energy scales.
}
\label{fig:super_phasediagram}
\end{figure}

We show that there are seven distinct superuniversality classes of non-Fermi liquids. 
Three of them are topologically stable.
The remaining ones describe critical superuniversality classes associated with changes in the topology of the bundles of projective fixed points.
The schematic phase diagram for superuniversality classes is shown in \fig{fig:super_phasediagram}.
The three generic superuniversality classes are as follows:
\begin{itemize}
\item 
Stable non-Fermi liquid superuniversality class (class A)

This superuniversality class contains the most stable type of non-Fermi liquids.
A necessary condition to realize stable non-Fermi liquids is that the universal pairing interaction generated from critical fluctuations is repulsive (or not too attractive) in the s-wave channel.
However, this is not a sufficient condition.
Even if the universal pairing interaction is repulsive in the s-wave channel, there always exist higher angular momentum channels in which the pairing interactions are attractive.
The more repulsive the interaction is in the s-wave channel, the stronger the attractive interaction becomes at high angular momentum channels.
Based on this observation, it may be argued that most, if not all, non-Fermi liquids are unstable against superconductivity in one angular momentum channel or another.
Nonetheless, non-Fermi liquids have two defense mechanisms against superconductivity.
The first, which applies to all angular momentum channels, is the pair-breaking effect of incoherence\cite{
Balatsky01101993,
PhysRevLett.74.2575,
doi:10.1142/S0217979296000349,
PhysRevB.100.115132,
BORGES2023169221}.
Incoherent fermions are intrinsically less susceptible to pairing instability.
Thanks to this pair-breaking effect,
incoherent fermions can tolerate attractive interactions to some extent,
and the Kohn-Luttinger instability\cite{PhysRevLett.15.524} can be avoided. 
The second mechanism is effective in deterring instabilities in non-zero angular momentum channels when the universal pairing interaction is strongly attractive in non-zero angular momentum channels and repulsive in the s-wave channel. 
In such cases, the attractive interactions generated in the non-zero angular momentum channels are bound to be transient in RG `time'.
Namely, the attractive interaction turns into a repulsive one at sufficiently low energies due to the effective running of the angular momentum.
If superconductivity does not arise before the sign change occurs, superconducting instability is averted by the crossover of the universal pairing interaction generated by critical fluctuations.
In such cases, the pairing interaction exhibits a non-monotonic behavior as a function of energy scale.

\item
s-wave superconducting superuniversality class (class C)

This superuniversality class contains non-Fermi liquids in which the universal pairing interaction generated by critical fluctuations 
is strong enough that even the incoherence of fermions does not prevent an instability
in the s-wave channel.
The s-wave channel is special in that it acts as the `fixed-point' for the running angular momentum, and the second defense mechanism discussed above does not apply.
If the attractive interaction is stronger than a critical strength in the s-wave channel, 
it is permanent in the RG time, and a superconducting instability is inevitable.

\item
Non-s-wave superconducting superuniversality class (class B)

This superuniversality class includes non-Fermi liquids that are prone to instabilities in non-s-wave pairing channels.
There exists a non-zero critical angular momentum above which pairing instabilities are unavoidable irrespective of the bare couplings.
Unconventional pairing arises in non-Fermi liquids if the universal pairing interaction is repulsive in the s-wave channel but strongly attractive in non-zero angular momentum channels. 
Such non-Fermi liquids are rarer than the previous two cases because the universal attractive interaction needs to be strong enough to overcome both the pair-breaking effect of incoherence and its transient nature.
Because superconducting instabilities in this class are controlled by the running of the angular momentum, which is caused by the Fermi momentum, the superconducting transition temperature and the pairing symmetry are strongly dependent on the Fermi momentum.

\end{itemize}

Each of the three topologically stable superuniversality classes occupies a finite region in the space of the marginal couplings and kinematic data.
The boundaries between generic superuniversality classes are spaces of codimension one in which critical superuniversality classes are realized.
There are three distinct critical superuniversality classes associated with three different boundaries (AB, AC and BC). 
The critical superuniversality classes describe topological phase transitions of the bundle of projective fixed points.
For example, superuniversality class A transitions to class C as the universal pairing interaction in the s-wave channel becomes strong enough to overcome the pair-breaking effect of incoherent fermions.
At the boundary, a critical superuniversality class (class AC) is realized, characterized by the strongest possible universal s-wave pairing interaction that incoherent fermions can withstand without instability.
The superuniversality class AB is realized when the universal attractive interaction at non-zero angular momenta reaches the critical strength and is on the verge of becoming the non-s-wave superconducting superuniversality class (class B).
Similarly, the superuniversality class BC arises when the universal pairing interaction in the s-wave channel reaches critical strength, while the interaction in non-zero angular momentum channels is strong enough to cause instability in non-s-wave pairing channels.
Finally, the multi-critical superuniversality class (class ABC) is realized in the codimension-two space where all three generic classes meet.
The phase diagram of the seven superuniversality classes takes the form of \fig{fig:super_phasediagram}.

 \begin{figure}[htbp]
    \centering
\begin{subfigure}{.60\textwidth} 
    \centering
\includegraphics[width=1.0\linewidth]{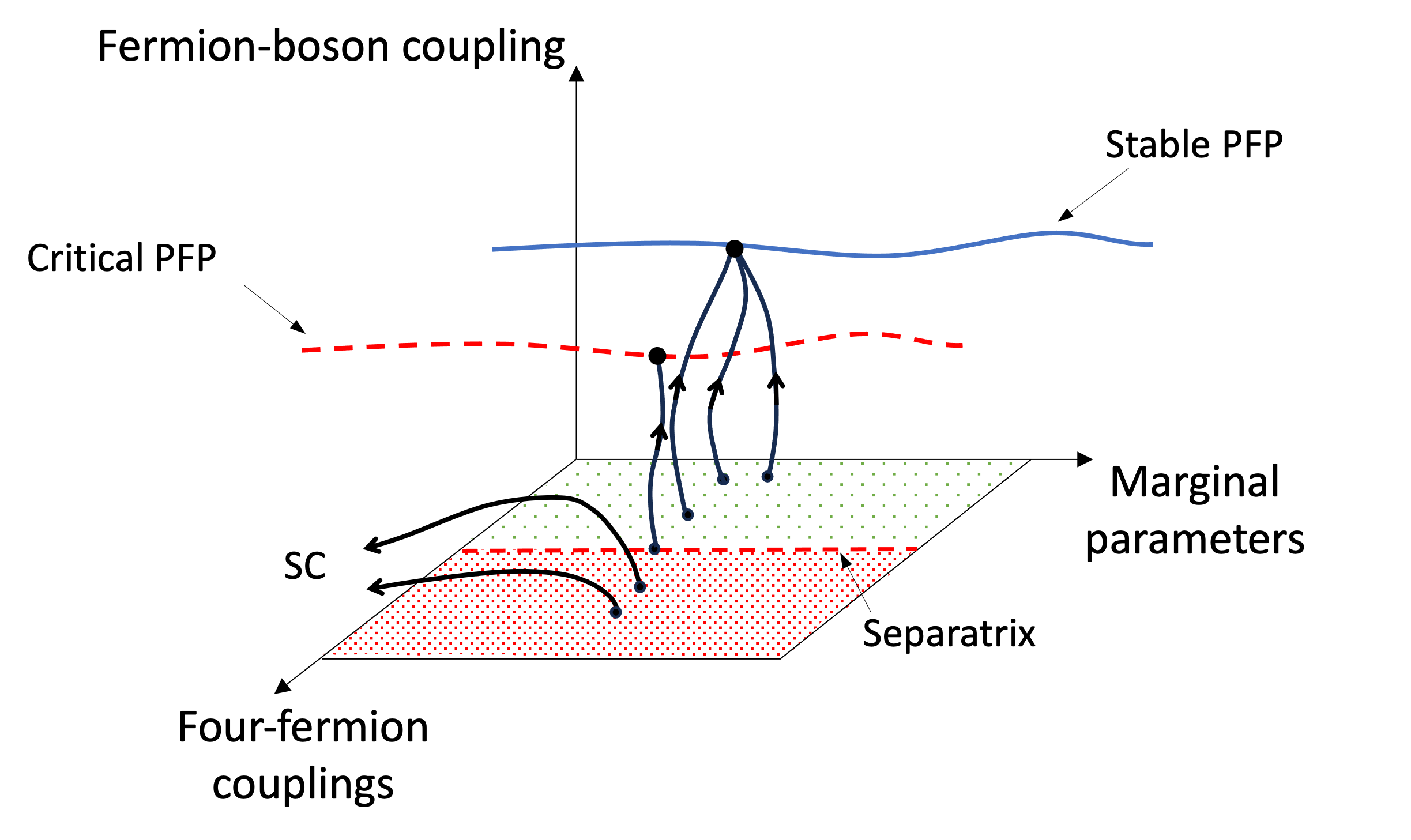}
\caption{}
 \label{fig:NFLtoSCA}
 \end{subfigure}%
 \begin{subfigure}{.4\textwidth} 
    \centering
\includegraphics[width=1.0\linewidth]{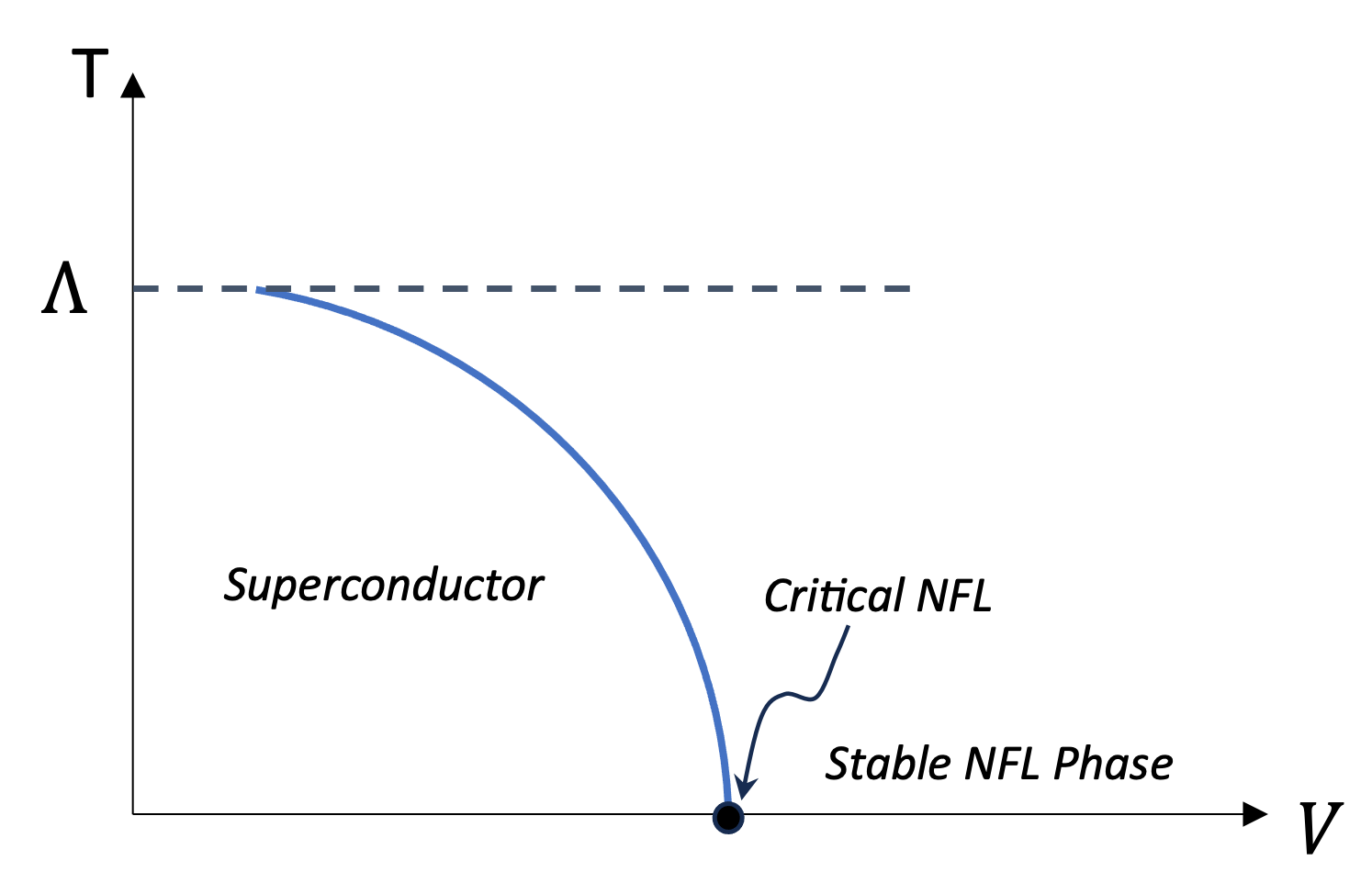}
\caption{}
\label{fig:StableNFL}
\end{subfigure}
        \caption{
A schematic RG flow and the associated phase diagram for superuniversality classes A, AB, AC and ABC.
Here, PFP denotes projective fixed point.
The running Fermi momentum is not shown.
The marginal parameters are the Fermi velocity and the shape of the Fermi surface.
In these classes, the stable non-Fermi liquid state is realized over an extended region of the fermion-fermion coupling.
As an attractive bare $4$-fermion (or $4n$-fermion) coupling is cranked up, a quantum phase transition occurs from the stable non-Fermi liquid to a (charge-$2n$) superconducting state.
The critical point is described by a critical non-Fermi liquid.
}
\label{fig:NFLtoSC}
\end{figure}

Each superuniversality class generally includes multiple universality classes.
Two non-Fermi liquids that arise from distinct superuniversality classes are necessarily distinct.
Nonetheless, it is convenient to broadly categorize individual universality classes into 
stable non-Fermi liquids,
critical non-Fermi liquids,
and superconductors.

\begin{enumerate}

\item  Stable non-Fermi liquids

Each of the superuniversality classes A, AB, AC and ABC supports one stable non-Fermi liquid universality class that can be realized without fine-tuning of the bare four-fermion coupling.
They are characterized by unique universal superconducting fluctuations.
Due to the unboundedness of angular momentum and the running of the Fermi momentum, there always exist channels with large enough angular momentum whose universal couplings undergo crossovers with lowering energy. 

\item Critical non-Fermi liquids

The stable non-Fermi liquids serve as platforms for generating infinitely many critical non-Fermi liquids.
The absence of an immediate charge $2$-superconducting instability opens the possibility of realizing a broad range of critical non-Fermi liquids with a judicious choice of the microscopic interaction. 
A critical non-Fermi liquid refers to an individual non-Fermi liquid universality class, which is not to be confused with a critical superuniversality class.
It is realized by tuning an irrelevant four-fermion or higher-order coupling to a critical strength, which drives a quantum phase transition from a stable non-Fermi liquid to a state with broken symmetry. 
Of particular interest is the critical states that arise at continuous phase transitions between the stable non-Fermi liquid and a charge $2n$-superconductor, driven by a $4n$-fermion coupling for positive integer $n$.
This is illustrated in \fig{fig:NFLtoSCA}.

 \begin{figure}[htbp]
    \centering
\begin{subfigure}{.60\textwidth} 
    \centering
\includegraphics[width=1.0\linewidth]{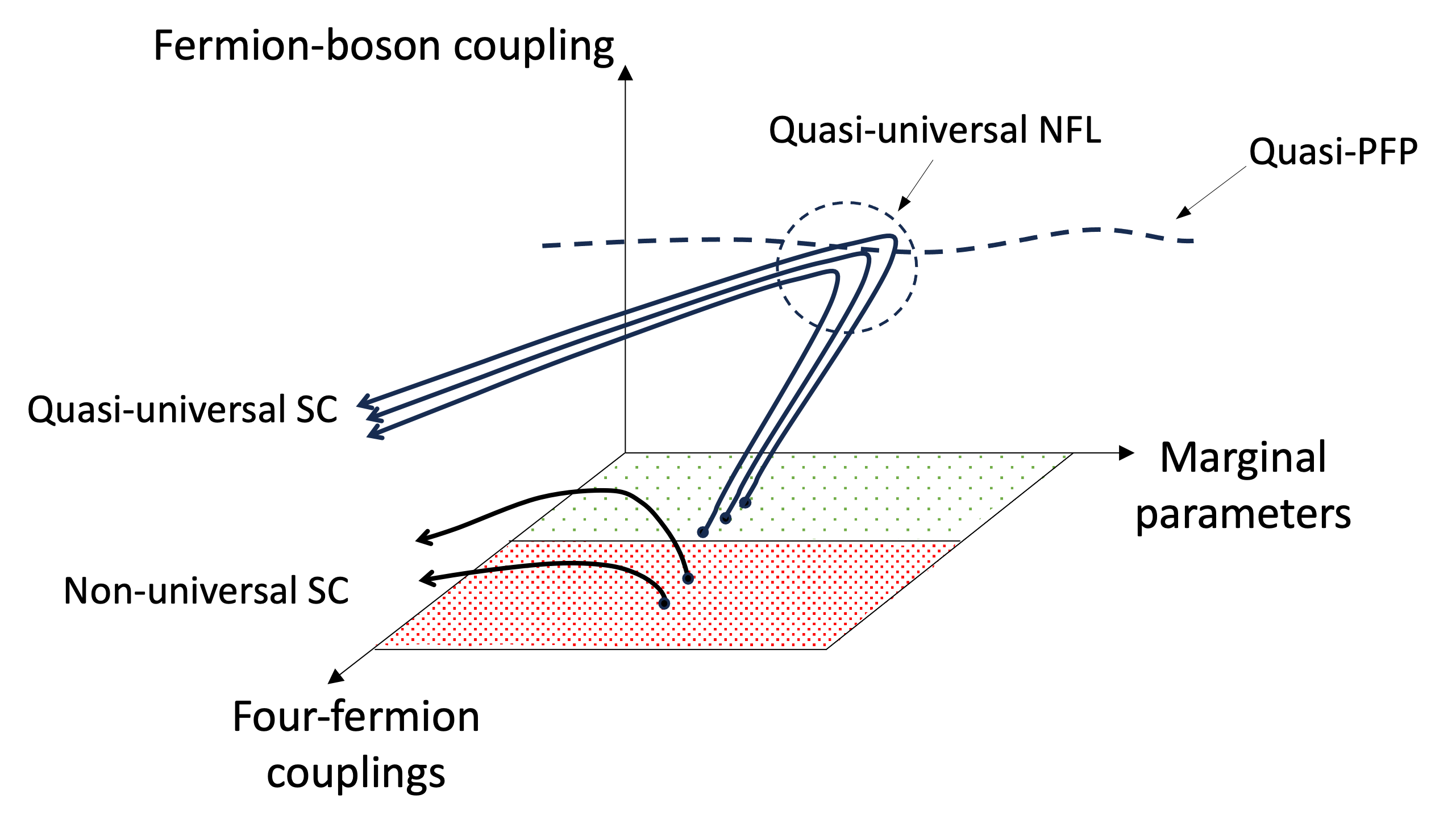}
\caption{}
 \label{fig:Quasi_Universal_SC_RG_Flow}
 \end{subfigure}%
 \begin{subfigure}{.40\textwidth} 
    \centering
\includegraphics[width=1.0\linewidth]{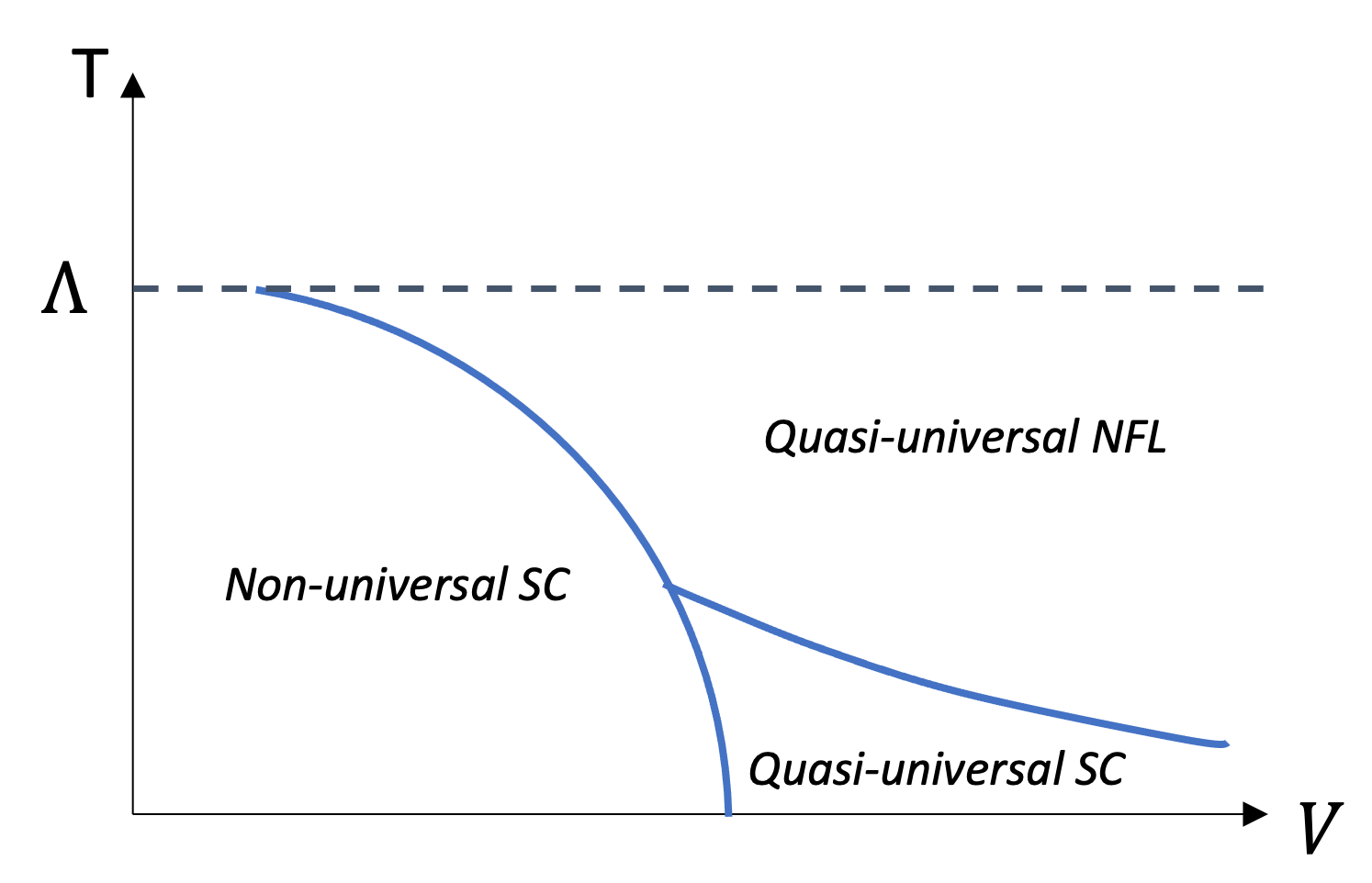}
\caption{}
\label{fig:NFLtoSCBC_b}
\end{subfigure}
        \caption{
A schematic RG flow and the associated phase diagram for superuniversality classes B, C and BC in proximity with class A.
If the bare fermion-fermion couplings are not too attractive, the theory is first attracted toward 
a quasi-projective fixed point, exhibiting a quasi-universal non-Fermi liquid behavior within an intermediate energy scale before it becomes a superconductor at low energies.
In that case, the pairing symmetry and the transition temperature are largely determined from the universal data of the non-Fermi liquid. 
At a sufficiently strong attractive bare fermion-fermion coupling, the ground state switches to a `non-universal' superconductor picked by the bare coupling (likely through a first-order transition).
}
\label{fig:NFLtoSCBC}
\end{figure}

\item Superconductors

Superconductors that arise from non-Fermi liquids can be loosely divided into two types, depending on the ratio $T_c/\Lambda$, where
$T_c$ is the superconducting transition temperature
and $\Lambda$ is the energy scale below which the non-Fermi liquid physics sets in.

\begin{itemize}

\item Non-universal superconductors

If $T_c/\Lambda \sim 1$,
the superconductivity emerges out of a normal state that hasn't yet settled into a universal non-Fermi liquid state.
In these cases, superconducting instabilities are driven by `high-energy' physics.
Such non-universal superconductors can arise in any superuniversality classes if the bare four-fermion couplings are strongly attractive in any angular momentum channel 
(see  \fig{fig:Quasi_Universal_SC_RG_Flow}).
Non-universal superconductors can also arise without a strong bare pairing interaction for non-Fermi liquids in superuniversality classes B, C and BC 
if the universal pairing interaction is strong enough 
that $T_c$ is intrinsically comparable with $\Lambda$ \cite{PhysRevB.91.115111}.

\begin{figure}[th]
\centering
\begin{subfigure}{.5\textwidth} 
    \centering
\includegraphics[width=0.5\linewidth]{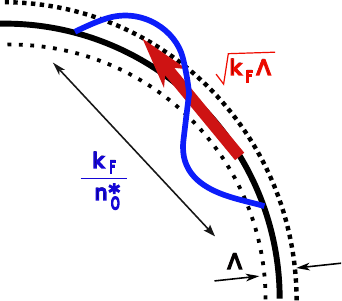}
\caption{}
 \label{fig}
 \end{subfigure}%
 \begin{subfigure}{.5\textwidth} 
    \centering
\includegraphics[width=1.0\linewidth]{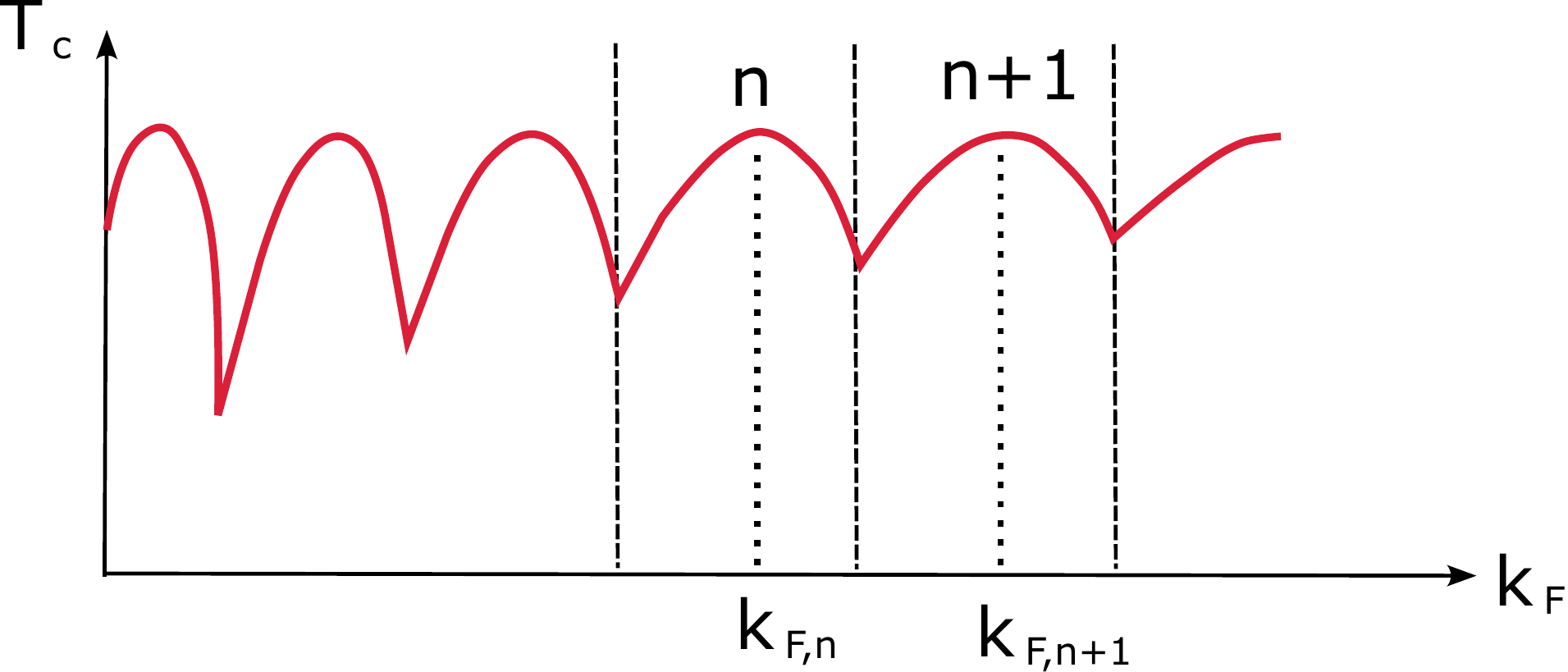}
\caption{}
 \label{fig}
 \end{subfigure}
\caption{
(a) Suppose a metal settles into a non-Fermi liquid in superuniversality class B at energy scale $\Lambda$.
The critical fluctuations that scatter fermions within the shell of energy $\Lambda$ near the Fermi surface carry momentum that is order of $q \sim \sqrt{\KFAVdim \Lambda}$.
Typically, in class B, the critical fluctuations generate a positive interaction vertex for Cooper pairs that suppresses the pairing in the s-wave channel.
However, an oscillating phase of the Cooper pair wavefunction can turn the positive vertex to an attractive pairing interactions at non-zero angular momentum channels\cite{SCALAPINO}.
The strongest attractive interaction is generated for an optimal angular momentum $n_O^* \sim \sqrt{\frac{\KFAVdim}{\Lambda}}$ 
for which $q$ connect the peak of the pair wavefunction with the trough.
(b) At a generic $\KFAVdim$, $n_O^*$ is not an integer and the superconducting instability arises in the angular momentum channel $n$ that is closest to $n_O^*$.
$n_O^*$ becomes integers at a discrete set of `resonant' Fermi momenta. 
As $\KFAVdim$ is continuously increased, $n$ increases in a stepwise manner, exhibiting
a local maximum in $T_c$ at each resonant Fermi momentum. 
}
\label{fig:Tc_oscillation_schematic}
\end{figure}

\item Quasi-universal superconductors

Non-Fermi liquids in superuniversality classes B, C and  BC are bound to become superconductors at low temperatures irrespective of the bare four-fermion coupling.
However, some non-Fermi liquids in these superuniversality classes are proximate to the superuniversality class A,
indicated by the shaded region in \fig{fig:super_phasediagram}.
In those non-Fermi liquids, the universal pairing interaction is relatively weak that a hierarchy of $T_c/\Lambda \ll 1$ is present.
This creates a large window of energy scale controlled by a non-Fermi liquid state. 
Accordingly, the nature of the superconducting state that emerges at low energies is largely determined by the universal data of the non-Fermi liquid\footnote{
The two-dimensional antiferromagnetic quantum critical metal at a small nesting angle belongs to this type\cite{BORGES2023169221}.}.
This situation is depicted in \fig{fig:NFLtoSCBC}.
A candidate for the quasi-universal superconductor is the Fermi surface coupled with multiple critical bosons that mediate the pairing interaction with opposite signs, thus weakening the net universal pairing interaction without suppressing $\Lambda$.
For classes B, C and BC, the quasi-universal superconductors will be the focus of our discussion.

\begin{itemize}

\item Quasi-universal non-Fermi liquids

For those non-Fermi liquids that arise within the window of intermediate energy scales between $\Lambda$ and $T_c$, the physics still depends on the microscopic information to a certain degree because there is only a finite RG time before the RG flow is cut off by a superconducting instability.
However, 
for theories with $T_c/\Lambda \ll 1$,
a quasi-universal behavior emerges in the intermediate energy scale between $\Lambda$ and $T_c$ with a weak dependence on the microscopic details.
These non-Fermi liquid states are characterized by the quasi-universal pairing interactions that mediate strong inter-patch couplings.
In these cases, one can first understand the physics of non-Fermi liquids realized in the intermediate energy scales and then determine the universal properties of the low-temperature superconducting states in terms of the universal data of the normal state.\\

\item S-wave superconductors

Non-Fermi liquids in class C have a non-zero minimum $T_c$\cite{
PhysRevD.59.094019,PhysRevB.91.115111} in the s-wave channel. 
In other words, the pairing instability in the s-wave channel is inevitable irrespective of how the bare coupling is chosen in that channel with the lower bound fixed by the universal data of the normal metal.
Superconducting instability in the s-wave channel is largely insensitive to the running of the Fermi momentum because the s-wave channel is the fixed point of the running angular momentum.
Consequently, the lower bound of $T_c$ is independent of the Fermi momentum.
\\

\item Non-s-wave superconductors

For all non-Fermi liquids in superuniversality classes B and BC, non-s-wave superconducting instabilities are unavoidable. 
There exists a critical angular momentum above which superconductivity is inevitable, irrespective of their bare couplings.
For the angular momentum channels above the critical angular momentum, there exists a universal lower bound for $T_c$.
Contrary to the s-wave superconductors, the Fermi momentum plays prominent roles in determining the properties of the non-s-wave superconductors.
For example, the superconducting transition temperature $T_c$ exhibits an oscillatory behavior as a function of the Fermi momentum (or doping),
as is illustrated in \fig{fig:Tc_oscillation_schematic}.
Such relations between $T_c$ and $\KFAVdim$ arise because the superconducting instability is triggered by a `resonance' between the momentum of critical fluctuations and the pitch of the pairing wavefunction $2\pi \KFAVdim/n$.
%

\end{itemize}

\end{itemize}

\end{enumerate}

The stable, critical, and quasi-universal non-Fermi liquids represent distinct metallic states.
The former two persist down to zero temperature, whereas the last one is realized only in intermediate energy scales.
The strength and range of the inter-patch coupling created by the universal pairing interaction are the key determining factors of the emergent symmetry.

\begin{figure}[t]
\centering
\includegraphics[width=0.3\linewidth]{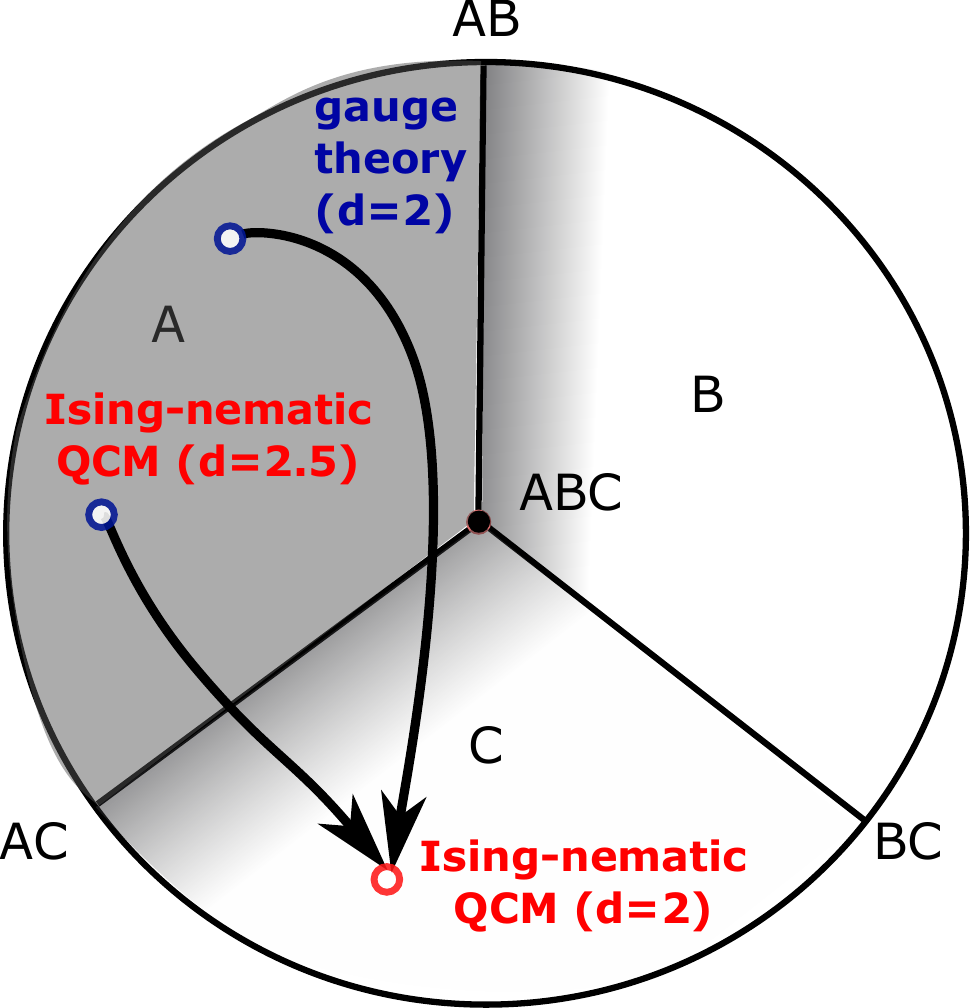}
\caption{
Two one parameter families of theories considered in this paper as physical examples.
The first describes the Ising-nematic quantum critical metal with varying space dimension $2 \leq d \leq 5/2$.
The second describes the Fermi surface coupled to both the U(1) gauge field and the Ising-nematic critical boson.
As the ratio between the two Yukawa couplings, which is marginal to the leading order in $\epsilon$, is tuned, the theory interpolates the pure U(1) gauge theory with decoupled Ising-nematic boson to the pure Ising-nematic critical metal with decoupled gauge boson.
}
\label{fig:super_phasediagram_path}
\end{figure}

Here is the outline of the paper.
Sec. \ref{sec:theory} introduces the low-energy effective theories that describe Fermi surfaces coupled with critical bosons of various types.
Although we are ultimately interested in the non-Fermi liquids that arise in the space dimension $d=2$, we use the dimensional regularization scheme that tunes the codimension of the Fermi surface.
It provides controlled access to the low-energy physics of the models near the upper critical dimension $d_c=5/2$. 
It is possible that non-perturbative effects, not taken into account near the upper critical dimension, render a two-dimensional non-Fermi liquid belong to a universality class different from what is predicted near the upper critical dimension.
However, the general classification itself, discussed in Sec. \ref{sec:classification}, does not rely on the perturbative control of a particular theory.
In Sec. \ref{sec:beta}, we discuss the beta functionals that describe the functional renormalization group flow of the Fermi momentum, Fermi velocity, Yukawa coupling functions, and the pairing interaction. 
We then introduce the central object used in our classification: projective fixed points (PFPs) that geometrically capture the simultaneous flow of the Fermi momentum and the coupling functions.
In particular, our focus is on the projective fixed points realized in the space of the pairing interaction.
Sec. \ref{sec:classification} begins with the classification of the topology of the bundles of projective fixed points.
Each topologically distinct class, referred to as superuniversality class, contains multiple non-Fermi liquid universality classes. 
We then provide a general discussion on the individual non-Fermi liquid universality classes that arise from each superuniversality class with emphasis on their universal properties and the emergent symmetries.
Before we delve into the real examples, in Sec. \ref{sec:toymodel}, we introduce a toy model whose projective fixed points can be obtained exactly. 
The `phase diagram' of the superuniversality classes constructed thereby reveals the physical nature of parameters that are to be tuned to traverse across different superuniversality classes. 
Then, we proceed to the physical examples.
To facilitate the determination of the superuniversality classes
in Sec. 
\ref{sec:diagnostics}, we introduce some practical diagnostics.
Secs. \ref{sec:ex1}-
\ref{sec:ex3} discuss concrete examples of physical theories that realize three of the seven superuniversality classes.
In Sec. \ref{sec:ex1}, we consider the U(1) gauge theory coupled with the Fermi surface, which will be shown to be in class A in all dimensions to the leading order in the $\epsilon=5/2-d$ expansion.
Sec. \ref{sec:ex2} considers an Ising-nematic critical metal.
We show that it evolves from class A to class C through critical class AC as $d$ is lowered from $5/2$ toward $2$.
Sec. \ref{sec:ex3} is dedicated to a theory with multiple critical bosons, where the Fermi surface, which is already coupled with the U(1) gauge field is undergoing an Ising-nematic phase transition.
In this example, there is an additional marginal parameter, to the leading order in $\epsilon$, which can be tuned to interpolate the pure gauge theory and the pure Ising-nematic theory.
Various physical theories considered in these three sections are summarized in \fig{fig:super_phasediagram_path}.
We discuss the physical properties of individual non-Fermi liquid universality classes that arise from each of those superuniversality classes in detail.
In Sec. \ref{sec:ex4}, we return to the toy model to discuss the universal low-energy physics of the non-Fermi liquids contained in the remaining four superuniversality classes not realized by the theories considered in this paper.
This section also serves as a summary of superuniversal properties that only depend on superuniversality classes.
In Sec. \ref{sec:conclusion}, we conclude with a summary and discussion.

From this point on, the paper can be read in multiple ways.
Secs. 
\ref{sec:theory}-\ref{sec:classification} 
are recommended to all readers as they motivate the following discussions and introduce the basic terminology.
The reader who is mainly interested in the general classification of superuniversality classes can continue to read Secs. \ref{sec:toymodel} and
\ref{sec:ex4}.
The reader who are interested in the physical examples can choose to read one or more of Secs. \ref{sec:ex1} - \ref{sec:ex3} after reading Sec. \ref{sec:diagnostics}.

\section{Low-energy effective field theory}
\label{sec:theory}

We start with a general theory that describes a Fermi surface coupled with a collection of critical bosons in (2+1)-dimensions,
\begin{equation}
\begin{aligned}
S & =    
\int d_f^3 {\bf k} ~
\psi_{j}^\dagger (k_0,\delta,\theta )
\Bigl[ 
i k_0   +v_{F,\theta  }\delta   \Bigr] \psi_{j}(k_0,\delta,\theta ) 
+ \frac{1}{2} 
\sum_t \int d_b^3 {\bf q}~
|{\bf q}|^2 
\phi_t^*(\mathbf{q})
\phi_t({\bf q}) \\ &
+ \frac{1}{\sqrt{N}}
\sum_t
\int d_f^3 {\bf k} ~ d_b^3 {\bf q}~
\edim^t_{\thetasq,\theta}(\varphi)
\phi_t(\mathbf{q})
\psi^\dagger_{j}\left(k_0+q_0,
 \deltaq,
  \thetasq 
\right)\psi_{j}(k_0,\delta,\theta ) 
+ \int d_f^3 {\bf k} ~ d_f^3 {\bf k}' ~ d_b^3 {\bf q}~
\lambdadim_{\left(\begin{smallmatrix}
    \thetasq  & \theta^{\prime} 
    \\ \theta      & \thetasqp   
    \end{smallmatrix}\right)}
~  
\times \\ & \hspace{1cm}
\psi^{\dagger}_{j_1}\left(k_0+q_0,
 \deltaq,
  \thetasq 
\right)
\psi_{j_1}(k_0,\delta,\theta)
\psi^\dagger_{j_2}
(k^{\prime}_0,\delta^{\prime},\theta^{\prime} )
\psi_{j_2}
\left(k^{\prime}_0+q_0, \deltaqp, \thetasqp  \right).
\label{eq:action_2d}
\end{aligned}
\end{equation}
Here, $\psi_j(k_0,\delta,\theta)$ is the fermionic field with frequency $k_0$ and Fermi-polar coordinates $(\delta,\theta)$:
the spatial momentum of the fermion  is written as 
$\vec{k} = (\KFthetadim +\delta)(\cos(\theta),\sin(\theta))$, where  $\KFthetadim$ is the angle-dependent Fermi momentum,
and $\delta$ is the radial displacement of the fermion away from the Fermi surface.
We consider the cases in which the Fermi surface is globally convex and $\KFthetadim$ is unique for each $\theta$. 
$j$ is the flavour index that runs from 1 to $N$.
Repeated indices are understood to be summed over. 
$\phi_t(\mathbf{q})$ represents a real boson of type $t$ with $\mathbf{q} = (q_0,\vec{q})$,
where $q_0$ is the frequency and the spatial momentum is $\vec{q} = q(\cos(\varphi),\sin(\varphi)) $, with ($q$,$\varphi$) being the usual polar coordinates.
$\int d_f^3 {\bf k} \equiv   \int\frac{dk_0}{2\pi} \int  \frac{d\delta}{2\pi}   \int_{-\pi}^{\pi} \frac{d\theta }{2\pi}   \KFthetadim$
and
$\int d_b^3 {\bf q}  \equiv \int \frac{dq_0}{2\pi}   \int \frac{dqq}{2\pi}   \int_{-\pi}^{\pi} \frac{d\varphi }{2\pi}$ represent the Cartesian measure of fermion and boson, respectively, expressed in terms of their respective coordinate systems. 
$v_{F,\theta}$ is the angle-dependent Fermi velocity.
$\edim^t_{\thetasq,\theta  }(\varphi)$ is the Yukawa coupling function that describes the interaction in which a low-energy fermion at angle $\theta$ absorbs 
momentum  $\vec{q} = q(\cos(\varphi),\sin(\varphi))$ from the boson of type $t$.
$\left(\thetasq, \deltaq \right)$ denotes the Fermi-polar coordinate of the momentum $\vec k + \vec q$.
For $q,\delta\ll \KFthetadim$, 
$\thetasq$ and $\deltaq$ can be written as
\bqa
\thetasq 
= \theta+\mathscr{A}_{\varphi,\theta}\frac{q}{\KFthetadim}
\label{angle_scattered}, ~~~~~
\deltaq = \delta+\mathscr{F}_{\varphi,\theta}q+\mathscr{G}_{\varphi,\theta}\frac{q^2}{\KFthetadim}+\frac{q\delta}{\KFthetadim}\mathscr{I}_{\varphi,\theta},
\label{eq:ThetaDelta}
\eqa
where
\begin{equation}
    \begin{aligned}
        &\mathscr{A}_{\varphi,\theta} = 
       \sin(\varphi-\theta),
       ~~\mathscr{F}_{\varphi,\theta} =  \left(
        \cos(\varphi -\theta )-
        \sin(\varphi -\theta )
        \frac{{\bf K}^{\prime}_{F,\theta  }}{\KFthetadim}\right),\\
      &\mathscr{G}_{\varphi,\theta} =  
      \frac{1}{2}\left( 
        \sin^2(\varphi -\theta )\left[1-\frac{{\bf K}^{\prime\prime}_{F,\theta}}{\KFthetadim}\right]+\sin\left(2\left(\varphi-\theta\right)\right)\frac{{\bf K}^{\prime}_{F,\theta}}{\KFthetadim}\right),~~
        \mathscr{I}_{\varphi,\theta} = \frac{\KFthetadim^\prime}{\KFthetadim}\sin\left(\varphi-\theta\right).
\label{eq:scattering_functions}
    \end{aligned}
\end{equation}
Here, $\KFthetadim^{\prime}$ and $\KFthetadim^{\prime\prime}$
represent the first and second derivatives, respectively, of the Fermi momentum with respect to $\theta$.
At low energies, $\vec q$ is small, which allows us to expand 
$
\edim^t_{\thetasq,\theta  }(\varphi)
=
\edim^t_{\theta,\theta  }(\varphi) + ..
$.
We denote the leading order term as
$\edim^t_{\theta}(\varphi)= 
\edim^t_{\theta,\theta}(\varphi)$.
$\lambdadim_{\left(\begin{smallmatrix}
    \thetasq  & \theta^{\prime} 
    \\ \theta      & \thetasqp   
    \end{smallmatrix}\right)}$
is the four-fermion coupling function for the short-range interaction in which fermions at angles $\theta$ and $\thetasqp$ scatter to angles $\thetasq$ and $\theta^{\prime}$, respectively. 
For the four-fermion coupling,
$\vec{q}$ represents the momentum transfer.
One can, in principle, include the quartic boson couplings, but they are not important at low energies.

\begin{figure}[th]
\centering
\begin{subfigure}{.22\textwidth} \centering \includegraphics[width=1.0\linewidth]{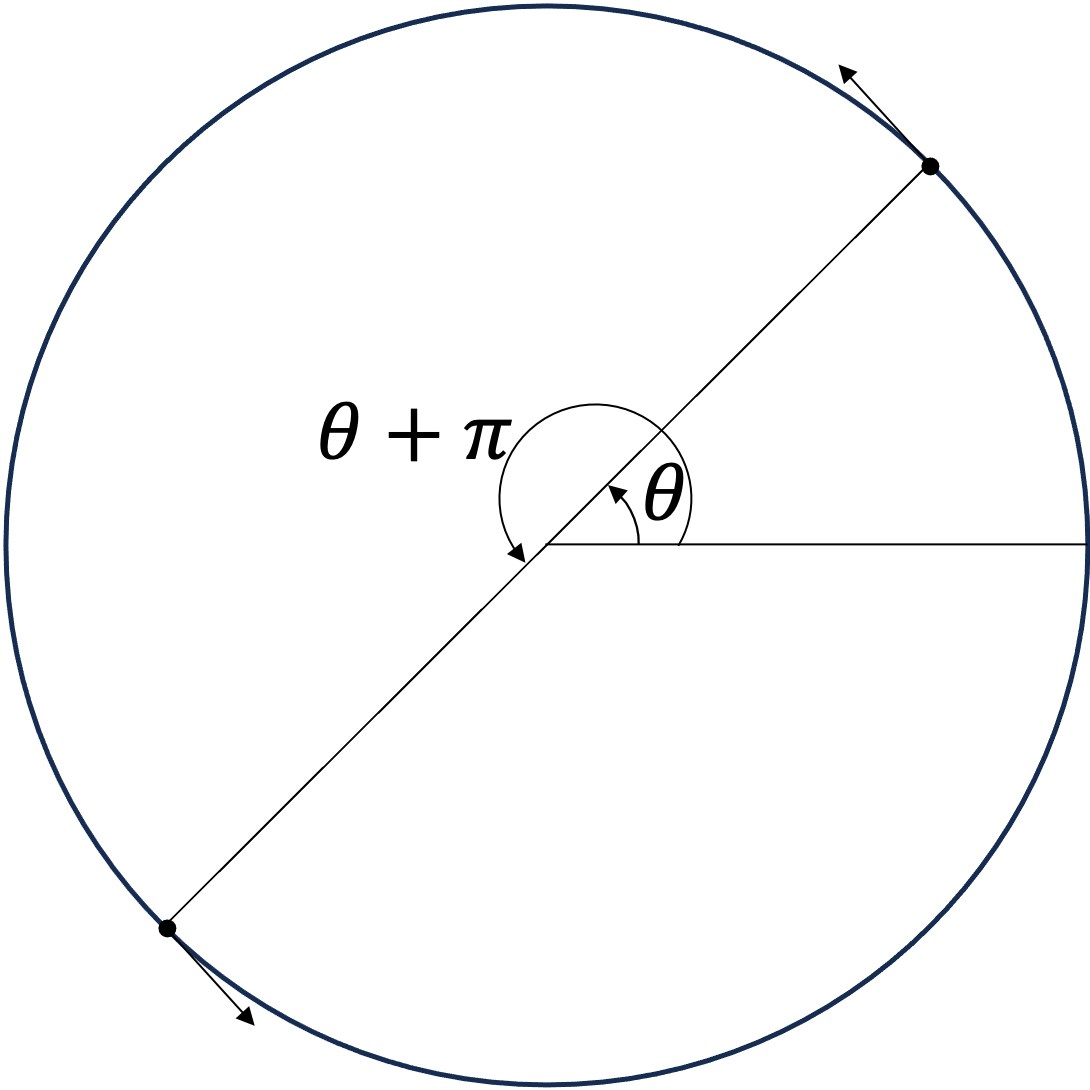} 
\caption{} 
\label{fig:Pairing_Interaction1} 
\end{subfigure}%
\hspace{0.1\textwidth}
\begin{subfigure}{.3\textwidth}
\centering
\includegraphics[width=1.0\linewidth]{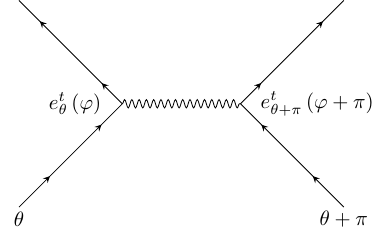}
\caption{  }
\label{fig:two_body_4f}
\end{subfigure}%
\caption{
Critical bosons generate the pairing interaction that scatters Cooper pairs tangentially along the Fermi surface.
The sign of the resulting four-fermion vertex is given by \eq{eq:sttheta}.
}
\label{fig:pairing_interaction}
\end{figure}

The number and type of critical bosons, as well as the angle-dependence of the coupling functions, are the kinematic data that determine the fate of the theory in the low-energy limit.
For the Ising-nematic quantum critical metal
that describes a phase transition that spontaneously breaks $C_{2n}$-symmetry to
$C_n$-symmetry, 
$\edim_{\theta}(\varphi)  \sim \cos( n\theta)$\footnote{
For general $n$, the symmetry can be broken down to various different subgroups\cite{Sayyad_2023}.
}.
For the Fermi surface coupled with U(1) gauge field $\vec a$, $\phi$ represents the transverse gauge field 
 in the Coulomb gauge through 
$a_i = \frac{1}{\sqrt{-\partial_i^2}}\epsilon_{ij} \partial_j \phi$.
The minimal coupling between the gauge field and current, $\vec a \cdot \vec j$, gives
$\edim_{\theta}(\varphi) 
\sim i \sin(\varphi-\theta)$.
While there are infinitely many different ways for a critical boson to couple with the Fermi surface, 
they can be divided into two groups based on the sign of the four-fermion interaction generated by the boson.
Integrating out the critical boson of type $t$ generates the four-fermion interaction in the pairing channel,
\bqa
\frac{ s^t_{\theta}(\varphi) }{N |{\bf q}|^2}
\psi^\dagger_{j}\left(k_0+q_0, \deltaq, \thetasq  \right)
\psi_{j}(k_0,\delta,\theta ) 
\psi^\dagger_{j'} \left(k_0'-q_0,  \Delta(\delta',\theta+\pi,-\vec q), \Theta(\theta+\pi,-\vec q) \right)
\psi_{j'} (k_0',\delta^{\prime},\theta+\pi ),
\label{eq:VatTreeLevel}
\eqa
where 
\bqa
s^t_\theta(\varphi) \equiv -
\edim^t_{\theta}\left(
\varphi \right)
\edim^t_{\theta+\pi}\left( 
\varphi+\pi \right).
\label{eq:sttheta}
\eqa
$s^t_\theta(\varphi)$ determines the nature of the interaction mediated by the critical boson for Cooper pairs, as shown in \fig{fig:pairing_interaction}.
For the Ising-nematic theories with even $n$, $s^t_\theta(\varphi) < 0$ except for the discrete cold spots at which
the coupling vanishes.
In this case, the critical fluctuations mediate an attractive pairing interaction in the s-wave channel.
For the Ising-nematic theories with odd $n$
and the gauge theory, $s^t_\theta(\varphi)>0$, which gives rise to a repulsive interaction in the s-wave channel.
In general, we refer to the theories  with positive (negative)  $s^t_\theta(\varphi)$ as s-wave repulsive (s-wave attractive)  theories.
More generally, a Fermi surface can be coupled with multiple bosons of different types.
For example, the critical metal that describes a nematic quantum phase transition in the presence of the U(1) gauge field has two critical bosons, $\phi_1$ and $\phi_2$ 
with 
$\edim^1_{\theta}(\varphi)  \sim \cos( n\theta)$
and
$\edim^2_{\theta}(\varphi) 
\sim i \sin(\varphi-\theta)$.
Theories with multiple bosons of mixed signs of
$s^t_{\theta}(\varphi)$ are called hybrid theories.

Since the theory becomes strongly coupled at low energies, we introduce a small parameter to perturbatively access the interacting non-Fermi liquids by increasing codimension of the Fermi surface.
However, our general classification is largely independent of the perturbative scheme. 
The generalized $(d+1)$-dimensional theory describes the one-dimensional Fermi surface embedded in $d$-dimensional momentum space coupled with the critical bosons,
\begin{equation}\begin{aligned}
   S & =   \int^{'}   d_{f}^{d+1} {\bf k}~
 \tilde \Psi_j(\mathbf{K},\delta,\theta )
\Bigl[ 
i {\bf \tilde \Gamma} \cdot{\bf K} + i v_{F,\theta  } \delta \Bigr]
\Psi_{j}(\mathbf{K},\delta,\theta ) 
+ 
\frac{1}{2} \sum_t \int d_{b}^{d+1} {\bf q}~ 
 q^2
|\phi_{t}(\mathbf{q})|^2 \\
&+ 
\frac{\mu^\frac{3-d}{2}}{\sqrt{N}} 
\sum_t \int^{'} 
d_{f}^{d+1} {\bf k}~
\int
d_{b}^{d+1} {\bf q}~
\edim^t_{\thetasq,\theta  }(\varphi)
\phi_{t}(\mathbf{q})   
\tilde{\Psi}_{j}\left(\mathbf{K}+\mathbf{Q},\deltaq,
\thetasq 
\right)\mathscr{T}_{t}\Psi_{j}(\mathbf{K},\delta,\theta )\\
&+ \mu^{1-d}
\sum_{\nu}
\sum_{s=d,e}
\int^{'}
d_{f}^{d+1} {\bf k}~
d_{f}^{d+1} {\bf k}'~
\int
d_{b}^{d+1} {\bf q}~~
\lambda^{(\nu,s)}_{ \theta',\theta}(\vec q)
~
{\bf O}^{(\nu,s)}_{\Lc} ( \mathbf{K}^{\prime},  \delta^\prime, \theta'; \mathbf{K}, \delta, \theta; \mathbf{Q}, \vec q).
\label{action_low_energy}
\end{aligned}\end{equation}
Here, 
$
\Psi_j({\bf K},\delta,\theta ) = \left( 
\begin{array}{c}
	\psi_{j}({\bf K},\delta ,\theta  ) \\
\psi_{j}^\dagger(-{\bf K},\delta ,\theta  +\pi) 
\end{array}
\right)$
is a two-component spinor that combines fermion field at angle $\theta$ and $\theta+\pi$.
$(d+1)$-dimensional momentum is composed of 
the original two-dimensional vector $\vec k$ and
the $(d-2)$-dimensional vector 
$(k_0,k_1,..,k_{d-2})$ which includes the frequency $k_0$ and $(d-2)$ additional components for the added codimensions.  
$(\delta,\theta)$ and $(\delta',\theta')$ are the fermionic polar coordinates
for $\vec k$ and $\vec k'$, respectively.
The bosonic measure is 
$\int d_{b}^{d+1} {\bf q} \equiv   \int\frac{d^{d-1}{\bf Q}}{2\pi} \int  \frac{dq  q}{2\pi}   \int_{-\pi}^{\pi} \frac{d\varphi }{2\pi}$ with $\mathbf{q}=(\mathbf{Q},\vec{q})$ being the $(d+1)$-dimensional boson frequency-momentum vector.  
However, the fermionic measure is written as 
$\int^{'} d_{f}^{d+1} {\bf k} \equiv   \int\frac{d^{d-1}{\bf K}}{(2\pi)^{d-1}} \int  \frac{d\delta}{2\pi}   \int_{-\pi/2+\theta_0}^{\pi/2+\theta_0} \frac{d\theta }{2\pi}   \KFthetadim$,
where the angle is integrated over the fundamental domain $[-\pi/2+\theta_0, \pi/2+\theta_0]$ of width $\pi$ because the spinors at angle $\theta$ and $\theta+\pi$ satisfy the constraint,
\bqa
\Psi_j({\bf K},\delta,\theta+\pi )  
= \sigma_x
\Psi_j^*(-{\bf K},\delta,\theta ) .
\label{eq:Psiconstraint}
\eqa
The center of the fundamental domain is denoted as $\theta_0$.
Because of \eq{eq:Psiconstraint},
the action is independent of $\theta_0$ which is a gauge freedom. 
For now, $\theta_0$ is left unfixed, but it will be fixed to a specific value later when it is needed to simplify calculations. 
Although the spinors are independent only within the fundamental domain,
$\Psi_j({\bf K},\delta,\theta)$ is defined everywhere in the extended domain, $\theta \in [-\pi, \pi)$.
This allows us to seamlessly write the Umklapp-like scatterings where a fermion is scattered across the boundaries of the fundamental domain.
For example, the boson-fermion or fermion-fermion interaction in \eq{action_low_energy} can scatter a fermion from an angle $\theta$ within the fundamental domain to $\thetasq$, which can be anywhere in the extended domain.
We use
$\tilde \Psi = 
-i \Psi^\dagger  \sigma_z$
as the conjugate of $\Psi$.\footnote{
In the earlier works\cite{DENNIS}, 
$\bar\Psi = \Psi^\dagger \sigma_y$ was used as the conjugate field.
Whether we use $\bar \Psi$ or $\tilde \Psi$  in the path integration is a mere convention.
Here, we use $\tilde \Psi$ as it simplifies the discussion of the symmetry of various types of theories.
}
$\tilde {\mathbf{\Gamma}} = (\tilde \gamma_0,\tilde \gamma_1,...,\tilde \gamma_{d-2})$
together with
$\mathbf{K}=
(k_0,k_1,..,k_{d-2})$ 
transform as $(d-1)$-dimensional vectors under the $SO(d-1)$ rotation.
The remaining gamma matrices 
$\{ 
\tilde \gamma_{d-1},..,
\tilde \gamma_2
\}$
form a vector under the $SO(4-d)$.
In $d=2$, the choice of 
$\tilde \gamma_0 = i \sigma_z$,
$\tilde \gamma_{1}= -i \sigma_x$
and 
$\tilde \gamma_2 = -i \sigma_y$ 
reproduces the original action.
The gamma matrices are normalized as $\{ \tilde \gamma_a, \tilde \gamma_b \} = -2 \delta_{a,b}$.
In $d=2$, the cubic interaction takes the form of
$\phi    \tilde{\Psi}_{j} \mathscr{T} \Psi_{j} $,
where
$\mathscr{T}
= i \mathbbm{1}$ for the s-wave attractive theories
and 
$\mathscr{T}=
\tilde \gamma_{1} \tilde \gamma_{2}
$
for the s-wave repulsive theories.
In $d=2$,
both
$ i \tilde{\Psi}_{j} \mathbbm{1} \Psi_{j} $
and 
$ \tilde{\Psi}_{j} \tilde \gamma_{1} \tilde \gamma_{2} \Psi_{j} $
are scalars under $SO(d-1) \times SO(4-d)$.
In general $d$, the former remains a scalar, but the latter forms the
\(  \frac{(4 - d)(3 - d)}{2} \)-dimensional representation spanned by 
$\{ \tilde \gamma_{\alpha} \tilde \gamma_{\beta}
| d-1 \leq \alpha<\beta \leq 2 \}$.
Accordingly, the boson that couples to the fermion bilinear becomes a scalar and
a rank-two antisymmetric tensor for the s-wave attractive and repulsive theories, respectively.
Each non-Fermi liquid has a fixed set of $t$ that determines the number and type of critical bosons in the system.
For example, 
\( t \in \{1\} \) with \( \mathscr{T}_1 = i\mathbbm{1} \) 
for the Ising-nematic critical metal with an even $n$,
and
\( t = \{ (\alpha, \beta) |  d-1 \leq \alpha< \beta \leq 2 \} \) 
with 
\(  \mathscr{T}_{\alpha\beta} = \tilde{\gamma}_\alpha \tilde{\gamma}_\beta \) 
for the U(1) gauge theory.
For the hybrid theory with both types of bosons,
\( t \in \{1\} \cup
 \{ (\alpha, \beta) |  d-1 \leq \alpha, \beta \leq 2 \} \).
%
$\lambda^{(\nu,s)}_{ \theta',\theta}(\vec q)$ is the channel-dependent four-fermion coupling associated with the four-fermion operator,
\bqa
&& 
{\bf O}^{(\nu,s)} 
(\mathbf{K}^{\prime},  \delta^\prime, \theta'; \mathbf{K}, \delta, \theta; \mathbf{Q}, \vec q) 
\equiv 
T^{(\nu,s)}_{\left(\begin{smallmatrix}     j_1  & j_2      \\ j_4      & j_3       \end{smallmatrix}\right)}
o_{m;j_1,j_4}^{(\nu)}  ( \mathbf{k}', \mathbf{k}'+\mathbf{q})
o_{m;j_2,j_3}^{(\nu)}  ( \mathbf{k}+\mathbf{q}, \mathbf{k}).
\label{eq:Onus}
\eqa
Here,
\bqa
o_{m;j_1,j_2}^{(\nu)}  ( \mathbf{k+q}, \mathbf{k})
  \equiv 
\tilde{\Psi}_{j_1}\left(\mathbf{K}+\mathbf{Q},
 \deltaq, \thetasq 
\right)
\tilde{I}^{(\nu)}_m
\Psi_{j_2}(\mathbf{K},\delta,\theta),
\label{eq:bilinear}
\eqa
represents a particle-hole fermion bi-linear with momentum ${\bf q}$ for $\nu=F_\pm$ and a particle-particle fermion bi-linear with the center of mass momentum ${\bf q}$ for $\nu=P$ created at angle $\theta$ and $\theta+\pi$. 
$\nu = F_+(F_-)$  represents the forward-scattering in even (odd) angular momentum channel, while $\nu = P$ denotes the pairing channel. 
For each $\nu=F_+,F_-,P$, 
$m$ is summed over 
$1$, $(d-1)$ and $(4-d)$ matrices, respectively,
where their elements are given by
\begin{equation}\begin{aligned}
   \tilde{I}^{(F_+)} = i\mathbbm{1}, ~~~
   \tilde{I}^{(F_-)} = 
   (\tilde \gamma_0,..,\tilde \gamma_{d-2}), ~~~
\tilde I^{P} =  i(\tilde{\gamma}_{d-1},\tilde \gamma_d,...,\tilde \gamma_2).
   \label{4f_channels}
\end{aligned}\end{equation}
$T^{(\nu,s)}_{\left(\begin{smallmatrix}     j_1  & j_2      \\ j_4      & j_3       \end{smallmatrix}\right)}$ 
for $s=d,e$ contracts the flavour indices in the direct and exchange channels,
respectively,
\bqa
T^{(\nu,s)}_{\left(\begin{smallmatrix}     j_1  & j_2      \\ j_4      & j_3       \end{smallmatrix}\right)} =
\left\{
\begin{array}{cl}
\delta_{j_1,j_4} \delta_{j_2,j_3}, 
&   \nu=F_\pm,~  s=d     \\
\delta_{j_1,j_3} \delta_{j_2,j_4},
&  \nu=F_\pm, ~  s=e 
~~~\mbox{or}~~~ \nu=P, ~ s=d
\\
\delta_{j_1,j_2} \delta_{j_3,j_4},
&  \nu=P, ~ s=e     
\end{array}
\label{Flavour_Tensor}
\right..
\eqa%
In addition to $Z_2 : \psi \rightarrow -\psi$ under which fermion bi-linears are invariant,
the theory is also invariant under $SO(d-1) \times SO(4-d)$ in general dimensions.
In $d=2$, the symmetry group is reduced to
$Z_2 \times SO(2)$,
where $SO(2)$ is identified as $U(1)/Z_2$ of the original U(1) charge conservation.
Under the $SO(2)$ group,
$o^{(F\pm)}_{m;j_1,j_2}$ is a singlet
and
$o^{(P)}_{m;j_1,j_2}$ 
transforms as a vector.
In $d=3$, the original U(1) symmetry is lowered to $Z_2$ 
by the Higgs mass term that 
gaps out fermions 
away from the line nodes in the direction of the added dimensions.
However, the new $SO(2)$ symmetry, which rotates $k_0$ and $k_1$, arises.
The symmetry of the theory in general $d$ is
$Z_2 \times SO(d-1) \times SO(4-d)$.

It is instructive to write the pairing vertex for Cooper pairs with zero center of mass momentum explicitly in $d=2$,
\bqa
{\bf O}^{(P,s)}_{\theta', \theta}
 &=
2
T^{(P,s)}_{\left(\begin{smallmatrix}     j_1  & j_2      \\ j_4      & j_3       \end{smallmatrix}\right)}\left[\psi^{\dagger}_{j_1}\left(\theta^{\prime}\right)\psi^{\dagger}_{j_4}\left(\theta'+\pi \right)\psi_{j_2}\left(\theta+\pi\right)\psi_{j_3}\left(\theta\right)+\psi_{j_2}^{\dagger}\left(\theta\right)\psi^{\dagger}_{j_3}\left( \theta+\pi\right)\psi_{j_1}\left( \theta^{\prime}+\pi\right)\psi_{j_4}\left(\theta^{\prime}\right)
        \right].
\eqa
This is Hermitian and symmetric under the exchange of $\theta$ and $\theta'$.
Accordingly, the coupling function  
$\lambda^{(P,s)}_{ \theta',\theta}(\vec q)$ must be real and symmetric.
Besides this, there is the anti-symmetric Hermitian operator,
\bqa
{\bf O}^{'(P,s)}_{\theta', \theta}
 &=
2 i
T^{(P,s)}_{\left(\begin{smallmatrix}     j_1  & j_2      \\ j_4      & j_3       \end{smallmatrix}\right)}\left[\psi^{\dagger}_{j_1}\left(\theta^{\prime}\right)\psi^{\dagger}_{j_4}\left(\theta'+\pi \right)\psi_{j_2}\left(\theta+\pi\right)\psi_{j_3}\left(\theta\right)
-\psi_{j_2}^{\dagger}\left(\theta\right)\psi^{\dagger}_{j_3}\left( \theta+\pi\right)\psi_{j_1}\left( \theta^{\prime}+\pi\right)\psi_{j_4}\left(\theta^{\prime}\right)
        \right]
\label{eq:antisymO}
\eqa
and the anti-symmetric real coupling function,
$\lambda^{'(P,s)}_{ \theta',\theta}(\vec q)$. 
\eq{eq:antisymO} can be alternatively written as
\bqa
\sum_{m_1,m_2=1}^2
\epsilon^{m_1,m_2}
T^{(P,s)}_{\left(\begin{smallmatrix}     
j_1  & j_2     \\ 
i_1  & i_2       \end{smallmatrix}\right)}
o_{m_{1};j_1,i_1}^{(P)}( \mathbf{k},
\mathbf{k})
o_{m_{2};j_{2},i_{2}}^{(P)}  ( \mathbf{k}', \mathbf{k}').
\eqa
In general $d$, it is generalized to a $2(4-d)$-fermion vertex that is invariant under $Z_2 \times SO(4-d)$,
\bqa
&& 
{\bf O}^{'(P,s)}_{\theta_1,..,\theta_{4-d}}  
\equiv
\sum_{m_1,m_2,..,m_{4-d}=d-1}^2
\epsilon^{m_1,m_2,..,m_{4-d}}
T^{(P,s)}_{\left(\begin{smallmatrix}     
j_1  & j_2 & .. & j_{4-d}    \\ 
i_1  & i_2 & .. & i_{4-d}      \end{smallmatrix}\right)}
o_{m_{1};j_1,i_1}^{(P)}  ( \mathbf{k}_1
,
\mathbf{k}_1)
..
o_{m_{4-d};j_{4-d},i_{4-d}}^{(P)}  ( \mathbf{k}_{4-d},
\mathbf{k}_{4-d}),
\label{eq:Onusasym}
\eqa
where
$\epsilon^{m_1,m_2,..,m_{4-d}}$ is the $(4-d)$-dimensional Levi-Civita tensor,
and its coupling function is a totally anti-symmetric tensor of $(4-d)$ angles,
$\lambda^{'(P,s)}_{ \theta_1,..,\theta_{4-d}}$.
In the most general case, one has to consider both
$\lambda^{(P,s)}_{ \theta_1,\theta_{2}}$
and
$\lambda^{'(P,s)}_{ \theta_1,..,\theta_{4-d}}$.
However, the latter breaks the time-reversal symmetry.
In this paper, we focus on the theories with time-reversal symmetry, and set  
$\lambda^{'(P,s)}_{ \theta_1,..,\theta_{4-d}}=0$.

%

\section{Beta functionals}
\label{sec:beta}

In this section, we discuss the beta functionals for the coupling functions.
To the leading order in the $\epsilon$-expansion, the feedback of the four-fermion coupling to the rest of the coupling functions can be ignored.
Therefore, we first discuss the beta functionals and their solutions for the Fermi momentum, Fermi velocity and the Yukawa coupling functions.
Then, we discuss the beta functional for the four-fermion coupling function.
Since the forward scattering is irrelevant in all $d$, at least to the leading order in $\epsilon$, the Landau function is completely fixed by other coupling functions\cite{PhysRevB.110.155142}.
Therefore, we will focus on the four-fermion coupling function in the pairing channel.

\subsection{Fermi momentum, Fermi velocity and fermion-boson coupling functions}

All coupling functions are defined to be dimensionless angle-dependent vertex functions defined at the energy scale $\mu$
(see Appendix \ref{app:RGscheme}).
In particular, the dimensionless Fermi momentum reads
\begin{equation}
    \begin{aligned}
        \KFtheta = \frac{\KFthetadim}{\mu}.
    \end{aligned}
    \label{eq:KFdimless}
\end{equation}
$\mu$ is the energy scale that is lowered to generate the functional RG flow of the coupling functions.
The angle-dependent Fermi momentum can be further decomposed into the average Fermi momentum 
and the normalized function as
\bqa
\KFAVdim=\frac{1}{2\pi}\int d\theta \KFthetadim, ~~~~~
\kappa_{F,\theta}= \KFthetadim/\KFAVdim.
\label{eq:kappa_kfav}
\eqa
$\KFAVdim$ set the scale for the overall size of the Fermi surface 
and 
$\kappa_{F,\theta}$
specifies the shape of the Fermi surface.
All low-energy observables at and below the energy scale $\mu$
can be expressed in terms of
$\{v_{F,\theta},\KFtheta, e^t_{\theta}(\varphi), \lambda_{\theta_1,\theta_2}\left(\vec{q}\right)\}$
up to errors that vanish in powers of $\mu$ in the low-energy limit\cite{PhysRevB.110.155142}.
The expansion is organized by the effective Yukawa coupling,
$g^t_{\theta} = \frac{1}{N}
\frac{\left|e^{t}_{\theta}\right|^{4/3} |X_{\theta}| |\chi_{\vartheta^{-1}(\theta)}|^{1/3}}{v_{F,\theta}^{1/3} K_{F,\theta}^{1/3}}$
given by the ratio of the Yukawa coupling of boson $t$ to the product of the Fermi velocity and Fermi momentum at angle $\theta$ with $X_{\theta} = \sin\left(\vartheta^{-1}(\theta) - \theta\right)$ and $\chi_{\varphi} = \sin(\varphi - \vartheta(\varphi))\left[
1
-\frac{\mathbf{K}''_{F,\vartheta(\varphi)}}{\mathbf{K}_{F,\vartheta(\varphi)}}
+\left(\frac{\mathbf{K}'_{F,\vartheta(\varphi)}}{\mathbf{K}_{F,\vartheta(\varphi)}}\right)^2
\right] \quad + \cos(\varphi - \vartheta(\varphi))\frac{\mathbf{K}'_{F,\vartheta(\varphi)}}{\mathbf{K}_{F,\vartheta(\varphi)}}$,
where
$\vartheta(\varphi) = \varphi - \arctan\left(\frac{\mathbf{K}_{F,\vartheta(\varphi)}}{\mathbf{K}^\prime_{F,\vartheta(\varphi)}}\right)$ is the angle at which 
$\vec{q} = q(\cos(\varphi),\sin(\varphi))$ is tangential to the Fermi surface.
Furtheremore, the dynamical critical exponent and anomalous dimensions of the fields are determined by the net effective Yukawa coupling, 
\bqa 
\bar g_{\theta} = \sum_tg^t_{\theta} 
\label{eq:sumofg}
\eqa
given by the sum of $g^t_\theta$  over all critical bosons.
In particular, each boson contributes to the anomalous dimension.
To the leading order, 
the dynamical critical exponent $z$, 
the boson anomalous dimension $\eta_\phi$,
and the angle-dependent fermion anomalous dimension 
$\eta_{\psi,\theta}$ are given by
\bqa
z = 1 + u_1(d)\bar g_0, ~~~
\eta_\phi =  -\frac{d-1}{2}u_1(d)\bar g_0, ~~~
\eta_{\psi,\theta} = \frac{u_1(d)}{2} \left( \bar g_{\theta} - d\, \bar g_0 \right),
\label{eq:zandetas}
\eqa
where
$u_1(d) = \frac{1}{3\sqrt{3}} \frac{\Omega_{d-1}}{(2\pi)^{d-1}} 
\frac{\Gamma\left(\frac{d}{2}\right)\Gamma\left(\frac{d-1}{3}\right)\Gamma\left(\frac{d-1}{2}\right)\Gamma\left(\frac{11-2d}{6}\right)}
{\Gamma\left(\frac{1}{2}\right)\Gamma\left(\frac{d-1}{6}\right)\Gamma\left(\frac{5d-2}{6}\right)}$.
Each boson contributes additively to the dynamical critical exponent and the anomalous dimensions of fields, irrespective of the sign of the pairing interaction it generates.
If the Fermi surface is coupled to both attractive and repulsive bosons, the net pairing interaction is determined by the difference of $g^t_\theta$, whereas the incoherence is given by their sum.
This makes it possible to tune the strength of the universal pairing interaction relative to the incoherence in hybrid theories that include multiple critical bosons.

The beta functionals of the Fermi momentum, Fermi velocity, and the effective Yukawa coupling function read 
\begin{align}
&
\frac{d K_{F,\theta}}{dl'} = K_{F,\theta}, 
~~~~~ 
\frac{d v_{F,\theta}}{dl'} = u_1(d) \left( \bar g_0 - \bar g_{\theta} \right) v_{F,\theta}, \label{eq:betavf} \\
& \frac{d \bar g_{\theta}}{dl'} 
= \bar g_{\theta} \left[ \frac{5 - 2d}{3} + \frac{5 - 2d}{3} u_1(d)\, \bar g_0 - u_1(d) \bar g_{\theta}  \right], 
~~~~~
\frac{d 
}{dl'} 
\left( \frac{\bar g^t_{\theta}}{ \bar g_{\theta} }
\right)
= 0.
\label{eq:betae}
\end{align}
Here, 
$l' \equiv \log \Lambda'/\mu$ is the logarithmic length scale associated with $\mu$ relative to a high-energy cutoff scale 
$\Lambda'$ below which our low-energy effective theory is valid.
With increasing $l'$, the coupling functions evolve as\cite{PhysRevB.110.155142}
and
\bqa 
& K_{F,\theta}(l') =  K_{F,\theta}^{UV} e^{l'},
~~~~~
 v_{F,\theta}(l') = \frac{\bar g_0^{UV}}{\bar g_\theta^{UV}}\frac{\bar g_{\theta}(l')}{\bar g_{0}(l')} v_{F,\theta}^{UV}, \nn
& \bar g_\theta(l') = \frac{
        \bar g^{UV}_\theta
        \bar g^{UV}_0e^{\frac{5-2d}{3}l}\left(\frac{2(d-1)}{5-2d}
        u_1(d)\bar g_0^{UV}\left(e^{\frac{5-2d}{3}l'}-1\right)+1\right)^{\frac{5-2d}{2(d-1)}}}{
        \bar g^{UV}_{\theta}\left(\frac{2(d-1)}{5-2d}
        u_1(d)\bar g_0^{UV}\left(e^{\frac{5-2d}{3}l'}-1\right)+1\right)^{\frac{3}{2(d-1)}}+
        \left (\bar g^{UV}_0-\bar g^{UV}_\theta\right)
        }.
\label{eq:gbartheta}
\eqa
In the low-energy limit, the net effective Yukawa coupling flows to an angle-independent value
\begin{align}
\bar g^{*} = \frac{5 - 2d}{2(d - 1)} \frac{1}{u_1(d)},
\label{eq:gstar}
\end{align}
except for the cold spots
at which  $\bar g_\theta^{UV}=0$,
if cold spots exist.
The angle-dependent Fermi velocity becomes 
$v_{F,\theta}^*  = 
 \frac{\bar g_0^{UV}}{\bar g_\theta^{UV}}
 v_{F,\theta}^{UV}$.
While the Fermi velocity is subject to renormalization, it saturates to a function that depends on the bare Fermi velocity and the bare coupling.
Therefore, the Fermi velocity is a marginal parameter whose renormalized value can be changed by tuning the bare parameters.
At the one-loop order, the relative couplings are also marginal as $g^t_\theta/\bar g_\theta$ is not renormalized.
\footnote{
However, the relative magnitude may be fixed at higher orders\cite{Ipsita_gauge}. 
For instance, the fermion self-energy can, in principle, be renormalized by the pairing interaction, which couples fermions on antipodal patches. 
In those quantum corrections, different bosons generally contribute differently to the self-energy, and the RG flow depends on the relative ratio between the Yukawa couplings.}
The marginal parameters of the theory at the one-loop order consist of 
$\left\{
\kappa_{F,\theta},
v_{F,\theta},
\frac{\bar g^t_{\theta}}{ \bar g_{\theta} }
\right\}$.
It is likely that only
$\kappa_{F,\theta}$
and $v_{F,\theta}$ remain marginal once higher-order corrections are included.
In the low-energy limit, the dynamical critical exponent and anomalous dimensions become
\begin{align}
z^{*} = \frac{3}{2(d - 1)},~~~\eta^{*}_{\phi} = \frac{2d - 5}{4}, ~~~ \eta^{*}_{\psi,\theta} = \frac{2d - 5}{4}.
\end{align}
The fermion anomalous dimension becomes angle-independent, except for the measure zero cold spots if present.

\subsection{Four-fermion coupling in pairing channel }
\label{sec:4fermion}

We now turn to the beta functional for the four-fermion coupling functions in the pairing channel.
Our primary goal is to examine the RG flow of the pairing interaction governed by the universal physics of the non-Fermi liquid at low energies.
Let us first discuss the range of energy scales in which the following discussion is valid.
$\Lambda'$ is the `true' high-energy cutoff scale, such as the bandwidth, below which the low-energy effective theory is valid.
We now consider $\Lambda$, which is lower than $\Lambda'$, below which the effective Yukawa coupling $\bar g_\theta$ in \eq{eq:gbartheta} is saturated to \eq{eq:gstar} almost everywhere on the Fermi surface.
To be precise, let's suppose that there exists a cold spot $\tcold$ at which the bare Yukawa coupling vanishes ($\bar g_{\tcold}(0)=0$).
Such cold spots arise in the Ising-nematic quantum critical metal, for example. 
At the cold spots, $\bar g_{\tcold}(l)$ never saturates to \eq{eq:gstar}.
For the angle-dependent bare Yukawa coupling $\edim^{t}_\theta$ that vanishes linearly in $\theta-\tcold$ near the cold spot, the effective coupling behaves as $\bar g_{\theta} \sim |\theta-\tcold|^{4/3}$ at the energy scale $\Lambda'$.
At energy scale $\Lambda$, $\bar g_\theta$ reaches \eq{eq:gstar} within an error that vanishes as
$\left( \frac{\Lambda}{\Lambda'} \right)^{
\frac{5-2d}{3}}$ 
except for the `lukewarm' region of angular span
$\Delta \theta \sim 
\left( \frac{\Lambda}{\Lambda'} \right)^{\frac{3(5-2d)}{8(d-1)}}$ near the cold spot.
\footnote{
According to \eq{eq:gbartheta}, the effective Yukawa coupling near the cold spot saturates to the fixed-point value only at the logarithmic length scale $l_\theta \sim  \frac{8(d-1)}{3(5-2d)} \log \frac{1}{|\Delta \theta|}$.
}
Here, we choose $\Lambda$ so that 
$\Delta \theta \ll 1$.
Below that energy scale $\Lambda$, the superconducting fluctuations are dominated by hot electrons away from the cold spots. 
To the leading order in $\Lambda/\Lambda'$, we now set $\bar g_\theta = \bar g^*$ at {\it all} $\theta$.
In the rest of the paper, we will use $\Lambda$ as our UV cutoff scale and 
$l \equiv \log \Lambda/\mu$ 
as the logarithmic length scale instead of 
$l' \equiv \log \Lambda'/\mu$.

The beta functional for the angle-dependent pairing interaction at zero center of mass momentum is
(see Appendix \ref{app:qc} for derivations)
\begin{equation}
    \begin{aligned}
      \frac{\mathrm{d} \tilde{V}^{\pm}}{dl} 
         =
        \tilde{W}_d \tilde{V}^{\pm}
         -\tilde{\mathcal{U}}\cdot \tilde{V}^{\pm}
        -\tilde{V}^{\pm}\cdot \tilde{\mathcal{U}} 
        -R_d
        \left(
        \tilde{V}^{\pm}
        \cdot
        \tilde{V}^{\pm}
        +
\tilde{V}^{\pm}
\cdot
\frac{\tilde{h}^{\pm}}{4}
 + 
\frac{\tilde{h}^{\pm}}{4}
\cdot
\tilde{V}^{\pm}
\right) 
+\tilde S^{\pm},
\label{main:betalambda_convolution0}  
    \end{aligned}
\end{equation}
where 
$\tilde V^{\pm}_{\theta_1,\theta_2} = 
\sqrt{\frac{K_{F,\theta_1}K_{F,\theta_2}}{v_{F,\theta_1}v_{F,\theta_2}}}
\left[
\lambda^{(P,d)}_{\theta_1,\theta_2}(\vec q=0)\pm \lambda^{(P,e)}_{\theta_1,\theta_2}(\vec q=0)
\right]$ represents the effective pairing interaction that scatters a Cooper pair at angle $\theta_2$ and $\theta_2+\pi$ to 
$\theta_1$ and
$\theta_1 + \pi$.
The superscript $\pm$ denotes the symmetric and anti-symmetric representations of the Cooper pair under the flavour group.
For the SU(2) spin, $+$ and $-$ denote the triplet and singlet representations, respectively.
The fermionic statistics forces
$\tilde V^{+}_{\theta_1,\theta_2}$ 
($\tilde V^{-}_{\theta_1,\theta_2}$)
to be odd (even) under the $\pi$ rotation of the angles around the Fermi surface :
$\tilde V^{\pm}_{\theta_1+\pi,\theta_2} = \tilde V^{\pm}_{\theta_1,\theta_2+\pi} = \mp  \tilde V^{\pm}_{\theta_1,\theta_2}$. 
$\tilde V^{\pm}_{\theta_1,\theta_2}$ is real and symmetric under the exchange of $\theta_1$ and $\theta_2$ in the presence of time reversal symmetry.
The original four-fermion interaction $\lambda$ has been normalized by the Fermi velocity because this ratio controls the physical observables, such as the anomalous dimensions.
In addition, the Fermi momentum multiplied to $\lambda$ accounts for the phase space available to Cooper pairs at angle $\theta$; 
the larger the Fermi momentum, the greater the phase space per unit angle.
With this normalization, the measure of the angular integration for virtual Cooper pairs in the loop takes the simple form:
$  \left(A\cdot B\right)_{\theta_1,\theta_2} = \int_{-\frac{\pi}{2}}^{\frac{\pi}{2}}\frac{d\theta}{2\pi} A_{\theta_1,\theta}B_{\theta,\theta_2}$.
%
$\tilde{W}_{d} = \left[ 2-d +(3-3d)(z-1)
\right]$ 
is the tree-level scaling dimension of $\tilde V$,
which is bigger than that of $\lambda$ by $1$ due to the contribution of Fermi momentum,
corrected by the dynamical critical exponent. 
$\tilde{\mathcal{U}}_{\theta_1,\theta_2} = 4\pi\eta_{\psi,\theta_1} 
\delta\left(\theta_1-\theta_2\right)$  
is a diagonal matrix that represents the shift of the scaling dimension due to the anomalous dimension of the fermion field.
$\tilde{V}^{\pm} \cdot \tilde{V}^{\pm}$
represents the renormalization of the quartic fermion coupling by itself,
which is responsible for the BCS instability in Fermi liquids.
$\tilde{V}^{\pm}
\cdot
\frac{\tilde{h}^{\pm}}{4}$
and 
$
\frac{\tilde{h}^{\pm}}{4}
\cdot
\tilde{V}^{\pm}
$
describe the mixing of the quartic paring interaction mediated by the critical boson.
$\tilde{h}^{\pm}_{\theta_1,\theta_2}$
mixes
$\tilde{V}^{\pm}_{\theta_1',\theta_1}$
with
$\tilde{V}^{\pm}_{\theta_1',\theta_2}$,
and
$\tilde{V}^{\pm}_{\theta_2,\theta_2'}$
with
$\tilde{V}^{\pm}_{\theta_1,\theta_2'}$\cite{PhysRevB.110.155142}.
The mixing matrix is given by 
(see Appendix \ref{app:lambda1} for details)
\begin{equation}
    \begin{aligned}
\tilde{h}^{\pm}_{\theta_1,\theta_2}
&= 
\sum_t
\mathfrak{s}^{(t)}
\left\{\frac{K_{F,\mtheta}(e^t_{\mtheta})^2\mu^2}{N v_{F,\mtheta}}
        \frac{1}{q(\theta_1,\theta_2)^{2}+ \alpha_d\beta_d\left(
\frac{Ng^t_{\mtheta}\mathbf{K}_{F,\mtheta}}{\left|X_{\mtheta}\right|\left|\chi_{\vartheta^{-1}\left(\mtheta\right)}\right|v_{F,\mtheta}}
\right)^{\frac{3}{2}}
        \frac{\mu
        ^{\frac{3}{2}}}{\sqrt{q(\theta_1,\theta_2)^{2}+\mu^2}}}\mp(\theta_1\rightarrow\theta_1+\pi)\right\}.
\label{eq:htilde_explicit}
    \end{aligned}
\end{equation}
Here, the contributions of all critical bosons are summed over with 
$\mathfrak{s}^{(t)}=\frac{d}{4-d}$ ($-1$)
for the boson that mediates a repulsive (attractive) pairing interaction in the s-wave channel.
\bqa
R_d
 =
  \frac{
 2^{d-1}
 \Omega_{d-1}
 \Gamma\left(\frac{d+1}{2}\right)\Gamma\left(\frac{4-d}{2}\right)\Gamma^2\left(\frac{d}{2}\right)}{
 \Gamma\left(\frac{3}{2}\right)(2\pi)^{d-1}\Gamma(d)}, 
~~~~
\alpha_d =\left(\frac{R_d(2\pi)^{d-1}}{2\Omega_{d-1}}\right)^{\frac{d-1}{d-2}}, 
~~~~
\beta_d = \frac{\Gamma^{2}(\frac{d}{2})}{2^{d-1}\pi^{\frac{d-1}{2}}\Gamma(d)\Gamma(\frac{d-1}{2})|cos(\frac{\pi d}{2})|}
\label{eq:alphad_betad}
\eqa
are $O(1)$ constants.
$q\left(\theta_1,\theta_2\right)$\footnote{
$q\left(\theta_1,\theta_2\right) 
= \sqrt{\mathbf{K}^2_{F,\theta_1}+\mathbf{K}^2_{F,\theta_2}-2\mathbf{K}_{F,\theta_1}\mathbf{K}_{F,\theta_2}\cos\left(\theta_1-\theta_2\right)}$}
is the momentum that connects 
angles $\theta_1$ and $\theta_2$ on the Fermi surface.
The $1/q\left(\theta_1,\theta_2\right)^2$ decay of the mixing matrix at large 
$q\left(\theta_1,\theta_2\right)$ 
reflects the fact that the mixing induced by the critical boson decays according to the  boson propagator at large momenta.
The anomalous dimension of the boson, if present, would modify the exponent in $q$\cite{BORGES2023169221}. 
However, it can be ignored to the leading order in $\epsilon$
because the Yukawa vertex already suppresses the magnitude of the mixing matrix. 
Finally, 
$\tilde S^{\pm}_{\theta_1,\theta_2}$, which is generated from \fig{ladder1},
represents the pairing interaction generated by the critical boson. 
It is closely related to the mixing matrix $\tilde{h}^{\pm}$ because the same critical bosons generate both the mixing and the pairing interaction.
$\tilde S^{\pm}_{\theta_1,\theta_2}$
can be viewed as the vertex obtained by mixing the four-fermion vertex generated by the critical boson in \fig{fig:two_body_4f} through another critical boson.
In particular, 
$\tilde S^{\pm}_{\theta_1,\theta_2}$ can be written as the convolution of two mixing matrices for $|\theta_1-\theta_2| \gg \sqrt{\mu/\KFAVdim}$.
This allows us to write
$
\tilde S^{\pm}_{\theta_1,\theta_2}
=
-R_d \frac{
\tilde{h}^{\pm}
\cdot
\tilde{h}^{\pm}
}{16}
+
\delta \tilde S^{\pm}_{\theta_1,\theta_2} $,
where
$\delta \tilde S^{\pm}_{\theta_1,\theta_2}$
is the difference between
$\tilde S^{\pm}_{\theta_1,\theta_2}$
and the convolution of two mixing matrices, which vanishes at large 
$|\theta_1-\theta_2|$.
Being quartic in the fermion-boson coupling,
$\delta \tilde S^{\pm}_{\theta_1,\theta_2}(\mu)$ is negligible to the leading order in the $\epsilon$ expansion.
Its explicit expression can be
found in \eq{appendix:source}, but we won't need it.

Notably, the beta functional depends on the RG energy scale $\mu$ explicitly through the energy-dependent width of the mixing matrix,
$\Delta \theta \sim 
( \mu/ \KFthetadim )^{1/2}$.
The width decreases with decreasing energy scale because momenta carried by critical bosons decrease with decreasing $\mu$.
Since the variation of the coupling functions is negligible within this range,
the mixing matrix that connects $\theta_1$ and $\theta_2$ can be expressed in terms of the coupling functions evaluated at the middle point,
$\mtheta = \frac{\theta_1+\theta_2}{2}$ 
at low energies.
While the momentum range of critical bosons becomes small at low energies, the four-fermion coupling itself can induce large-angle scatterings, depending on the angular profile of 
$\tilde V^\pm_{\theta_1,\theta_2}$.
The symmetrization/anti-symmetrization ensures that
$\tilde{h}^{\pm}_{\theta_1,\theta_2}$
satisfies the same boundary condition as
$\tilde V^{\pm}_{\theta_1,\theta_2}$.
It is noted that $\tilde{h}_{\theta_1,\theta_2}^{\pm}$ can multiply $\tilde V^{\pm}$ either from the left or right in the beta functional because the pairing vertex can mix through either incoming or outgoing Cooper pairs.

\begin{figure}[th]
\centering
\begin{subfigure}{.35\textwidth} \centering \includegraphics[width=0.8\linewidth]{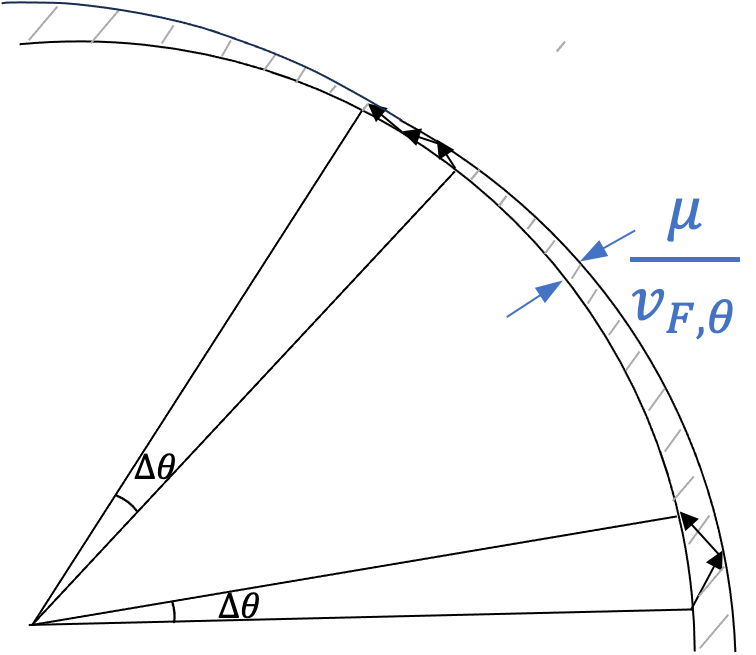} \caption{} 
\label{fig:bartheta} \end{subfigure}%
\begin{subfigure}{.40\textwidth}
\centering
\includegraphics[width=0.8\linewidth]{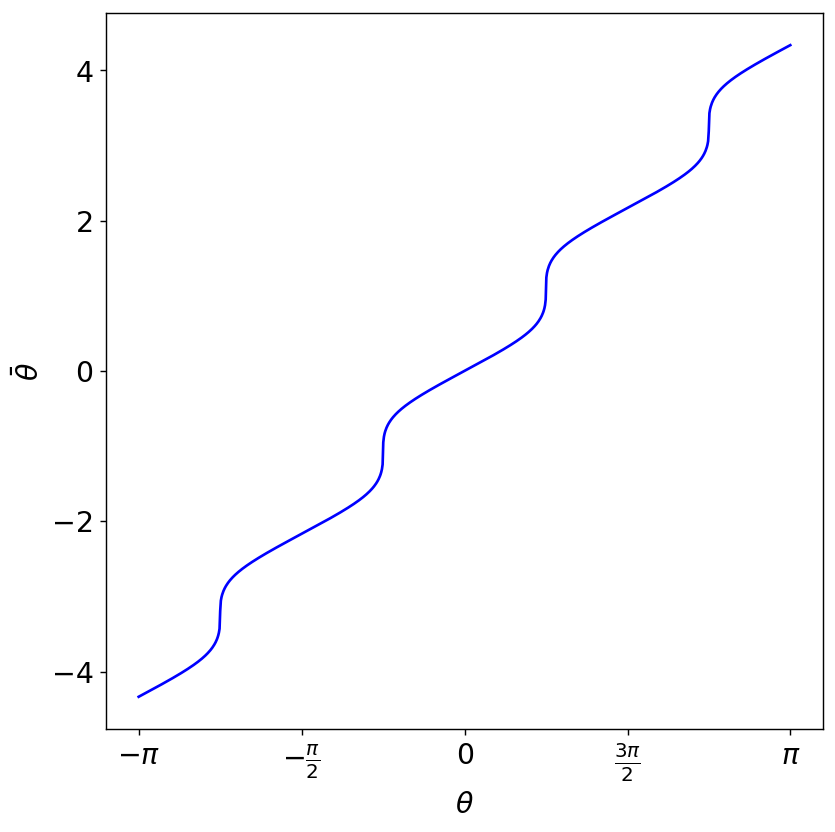}
 \caption{ 
}
\label{fig:theta_bar_vs_theta}
\end{subfigure}%
\caption{
(a) For the $C_4$ symmetric Ising-critical metal with the bare Yukawa coupling $e_\theta \sim \cos(2 \theta)$ and the angle-independent bare Fermi velocity, the renormalized Fermi velocity becomes bigger near the cold spots near $\theta=\pm \pi/4, \pm 3\pi/4$\cite{PhysRevB.110.155142}.
The thickness of the momentum shell of low-energy fermions with energy less than $\mu$, denoted by the shaded region, scales as $\mu/v_{F,\theta}$.
Near the cold spot, the large Fermi velocity pinches the shell, forcing fermions to go through more tangential scattering with smaller momentum transfers to traverse the same angular displacement $\Delta \theta$, compared to the region far away from the cold spot.
(b) Consequently, $\bar \theta$, which is a measure of the proper distance felt by those low-energy fermions, increases more steeply near the cold spots.
For the plot, we use bare parameters, $\KFthetadim = 1+0.025 \cos(4\theta)$,  $e_\theta = 0.1\cos(2\theta)$, $v_{F,\theta} = 1$ and  $N$ = 1 in $d = 2$.
}
\label{Fig: C4FS_Rescaled_Single}
\end{figure}

In the low-energy limit, the anomalous dimension of the fermion $\eta_{\psi,\theta}$ becomes angle independent.
Furthermore, the weight of the mixing matrix given by
\begin{equation}
    \begin{aligned}
\int_{-\frac{\pi}{2}}^{\frac{\pi}{2}}\frac{d\theta'}{2\pi}
        \tilde{h}^\pm_{d;\theta,\theta'}\left(\mu\right)
= \frac{2\eta_d}{\bar g^{*} R_d}\mathfrak{s}^{(t)}g^{t*}
    \end{aligned}
    \label{eq:weight_htilde}
\end{equation} 
with 
\begin{equation}
    \begin{aligned}
        \eta_d = \frac{1}{3\sqrt{3}}\frac{R_d\bar g^*}{\left(\alpha_d\beta_d\right)^{\frac{1}{3}}}
        \label{eq:etad}
    \end{aligned}
\end{equation}
also become independent of $\theta$ in the low-energy limit.
This suggests that there is an emergent rotational symmetry in the low-energy limit although $v_{F,\theta}$ and $K_{F,\theta}$ are generally angle-dependent.
The emergent rotational symmetry can be made explicit in a new angular coordinate $\bar \theta$,
called proper angle,
related to the original angle through the
non-linear coordinate transformation\cite{PhysRevB.110.155142},
$\theta = a(\bar \theta)$ 
with
\bqa
a^{\prime}(\bar{\theta}) \equiv\frac{d\theta}{d\bar{\theta}} =  \beta_d^{\frac{1}{3}}\left(\frac{N
\bar g^{*}
\left|X_{\theta}\right|}{
|\chi_{\vartheta^{-1}(\theta)}|
\KFthetadim
v^{*}_{F,\theta}
}
\right)^{\frac{1}{2}}.
\label{eq:theta_bartheta}
\eqa
Here,  $\bar g^* = \sum_t g^{t*}$.
The proper angle $\bar \theta$ has dimension $1/2$ because of $1/\sqrt{\KFthetadim}$ in $d\theta/d\bar\theta$. 
The non-linear coordinate transformation in \eq{eq:theta_bartheta} maps a fixed angular span of $d \theta$ to a larger $d \bar \theta$ in the region where the $\KFthetadim v_{F,\theta}^*$ is greater.
This can be understood in the following way.
At energy scale $\mu$, the thickness of the low-energy shell scales as 
$\mu/v_{F,\theta}$ at angle $\theta$, 
and the magnitude of the maximum momentum that a fermion can absorb in the tangential direction while staying within the shell of energy $\mu$ is 
\bqa
q_\theta(\mu) = \sqrt{\frac{\mu \KFthetadim}{v_{F,\theta}}}.
\label{eq:qmutheta}
\eqa
This is also reflected in the width of $\tilde{h}^{\pm}_{\theta_1,\theta_2}$ in \eq{eq:htilde_explicit}. 
Therefore, the number of scatterings a fermion must go through to traverse the angular displacement of $\Delta \theta $ while staying within the low-energy shell goes as $
\frac{\KFthetadim \Delta \theta}{q_\theta(\mu)}
\sim \sqrt{\frac{v_{F,\theta}
\KFthetadim}{\mu}} 
\Delta \theta
$.
The coordinate transformation in
\eq{eq:theta_bartheta} is chosen so that $\Delta \bar \theta$ is proportional to this number of scatterings.
To be quantitative,
we introduce an angle-dependent metric 
\bqa
\gtt =
\frac{1}{
4 \pi^2
\beta_d^{2/3}}
\frac{
|\chi_{\vartheta^{-1}(\theta)}|
\KFthetadim
v^{*}_{F,\theta}
}{N 
\mu
\bar g^{*} \left|X_{\theta}\right|},
\label{eq:gthetatheta}
\eqa
such that the proper distance between $\theta$ and 
$\theta + \Delta \theta$ is given by
$\sqrt{\gtt} \Delta \theta$.
In the region with larger  $\gtt$,  it takes more scatterings for fermions to traverse a fixed angular span $\Delta \theta$, 
as illustrated in   \fig{fig:theta_bar_vs_theta}.
The new angular coordinate 
$\bar \theta$ corresponds to
this proper distance,
$\bar \theta(\theta) = 
2\pi \sqrt{\mu}
\int_0^\theta d\theta' \sqrt{\gtt(\theta')}$ 
with $\bar\theta(0) = 0$,
where 
the factor of $\sqmu$ is added for convenience so that the overall size of $\bar \theta$ flows to a finite number in the low-energy limit.
The folded angular space is spanned by 
$-\bar \theta_{max} \leq \bar \theta <  \bar \theta_{max}$ 
with
\bqa
\bar \theta_{max} 
=\bar \theta( \pi/2 ) =
\gamma
\sqrt{\KFAVdim},
\label{eq:thetamax}
\eqa
where 
$\gamma  = 
2\pi \sqrt{\frac{\mu}{\KFAVdim}}
\int_{0}^{\frac{\pi}{2}} d \theta
\sqrt{ \gtt (\theta)}
=
\frac{\pi}{2} 
\sqrt{\frac{\mu}{\KFAVdim}}
\cN(l) 
$
and 
\bqa
\cN(l) = 
\int_{\pi}^{-\pi} 
d \theta
\sqrt{ \gtt (\theta)}
\label{eq:cN}
\eqa
is the total number of scatterings a fermion needs to undergo to traverse around the Fermi surface once at energy scale $\mu$.
$\gamma$ is a dimensionless parameter fixed by the marginal coupling functions.

To keep the uniformity of the angular measure, we further rescale the pairing interaction as 
\begin{equation}
\begin{aligned}
\bar{V}^{\pm}_{\bar{\theta}_1,\bar{\theta}_2}
    \equiv
\sqrt{
\mu
a^\prime(\bar{\theta}_1) a^\prime(\bar{\theta}_2)}
    \tilde V^{\pm}_{\theta_1,  \theta_2}.
\end{aligned}
\label{eq:lambdabar}
\end{equation}
For this rescaled pairing interaction, \eq{main:betalambda_convolution0} becomes
\begin{equation}
    \begin{aligned}
        \frac{\mathrm{d} \bar{V}^{\pm}}{dl} 
         &=
        -\HD\bar{ V}^{\pm}
         -
         R_d
        \left(\bar{V}^{\pm}+\frac{\bar{h}^{\pm}}{4}\right) \cdot  \left(\bar{V}^{\pm}+\frac{\bar{h}^{\pm}}{4}\right)
        + \delta \bar{S}^{\pm}_{\bar{\theta}_1,\bar{\theta}_2},
\label{eq:beta_function_nonlinear}
    \end{aligned}
\end{equation}
where
$        (A \cdot B)_{\bar{\theta}_1,\bar{\theta}_2} = \int_{-\bar{\theta}_{\mathrm{max}}}^{\bar{\theta}_{\mathrm{max}}} \frac{d\bar{\theta}}{2\pi\sqrt{\mu}} A_{\bar{\theta}_1,\bar{\theta}} B_{\bar{\theta},\bar{\theta}_2}$
and
$\delta \bar{S}^{\pm}_{\bar{\theta}_1,\bar{\theta}_2}
    \equiv
\sqrt{
\mu
a^\prime(\bar{\theta}_1) a^\prime(\bar{\theta}_2)}
    \delta \tilde S^{\pm}_{\theta_1,  \theta_2}$.
$\HD =d-\frac{3}{2}+3(d-1)(z-1)+4\eta_{\psi}$.
$\bar{h}^{+}_{\bar{\theta}_1,\bar{\theta}_2}$ 
($\bar{h}^{-}_{\bar{\theta}_1,\bar{\theta}_2}$)
satisfies the anti-periodic (periodic) boundary condition when one of the angle is shifted by $2 \bar \theta_{max}$.
At low energies, it is sharply peaked at $\bar{\theta}_1 \approx  \bar{\theta}_2$.
For $|\bar{\theta}_1 - \bar{\theta}_2| \ll \bar \theta_{max}$, it becomes
\begin{equation}
 \begin{aligned}
\bar{h}^{\pm}_{\bar{\theta}_1,\bar{\theta}_2} = 
        \sum_t \sigp
        \frac{
        \left(g^{t*}\right)^\frac{3}{2}
        }{\beta_d^{\frac{1}{3}} 
        \bar g^{*\frac{1}{2}}
        }\frac{ \left|\bar{\theta}_1-\bar{\theta}_2\right|\mu}{
       \left|\bar{\theta}_1-\bar{\theta}_2\right|^3 
      +
\alpha_d \left( \frac{g^{t*}}{ \bar g^{*}}  \mu \right)^{\frac{3}{2}} 
}.
\label{eq:hbar}
\end{aligned}
\end{equation}
In the proper angular coordinate $\bar \theta$, the width of the mixing matrix for boson of type $t$ is given by 
$\Delta \bar \theta_t  \sim  \alpha_d^{1/3}  \left( \frac{g^{t*}}{ \bar g^{*}}  \mu \right)^{\frac{1}{2}}  $.
Therefore, the sum of its squares is universally given by
\bqa
\sum_t (\Delta \bar \theta_t)^2  \sim  \alpha_d^{2/3} \mu,
\label{eq:thetabarscale}
\eqa
where $\alpha_d$ defined in \eq{eq:alphad_betad}
is O(1) constant.

The beta functional in
\eq{eq:beta_function_nonlinear} 
is composed of the three main ingredients illustrated in \fig{fig:triangle}.
Firstly, $\HD$ contains the effects of fermion incoherence through the dynamical critical exponent and the anomalous dimension.
The more incoherent the fermion is, the larger $\HD$ becomes, making the pairing interaction less relevant.
This encodes the pair-breaking effect of fermion incoherence.
Secondly, $\bar{h}^{\pm}$ describes the universal pairing interaction generated by the critical fluctuations.
In addition to the conventional BCS scattering process, denoted by
$\bar{V}^{\pm} \cdot   \bar{V}^{\pm}$,
the critical bosons mix the four-fermion coupling defined at different angles through
$\bar{V}^{\pm} \cdot \frac{\bar{h}^{\pm}}{4}$
and
$\frac{\bar{h}^{\pm}}{4} \cdot \bar{V}^{\pm} $.
Furthermore, the critical bosons
themselves generate a pairing interaction through
$\frac{\bar{h}^{\pm}}{4} \cdot \frac{\bar{h}^{\pm}}{4}$.
If $\bar{h}^{\pm}$, as a matrix defined in the angular space, supports an eigenvector with a sufficiently negative eigenvalue, 
the four-fermion coupling becomes relevant, leading to a superconducting instability.
Because $\bar{h}^{\pm}$ is traceless, it always supports both positive and negative eigenvalues, except for the fine-tuned case in which  $\bar{h}^{\pm}$ is identically zero.
Thirdly, the RG flow equation, which describes the renormalization of the pairing interaction from $\mu$ to $\mu e^{-dl}$, does depend on the energy scale $\mu$ explicitly.
In particular, the width of 
$\bar{h}^{\pm}_{
\bar{\theta}_1,\bar{\theta}_2}$
as a function of  $\bar{\theta}_1-\bar{\theta}_2$ scales as $\sqrt{\mu}$.
The explicit appearance of the RG scale in the beta function is a unique phenomenon that arises in metals, in contrast to relativistic field theories. 
It is caused by the running of the dimensionless Fermi momentum, $\KFthetadim/\mu$.
Therefore, superconducting fluctuations of non-Fermi liquids are determined by the scale-dependent competition between pair-breaking incoherence and pairing-forming glue.

\begin{figure}[ht]
\centering
\includegraphics[width=0.45\linewidth]{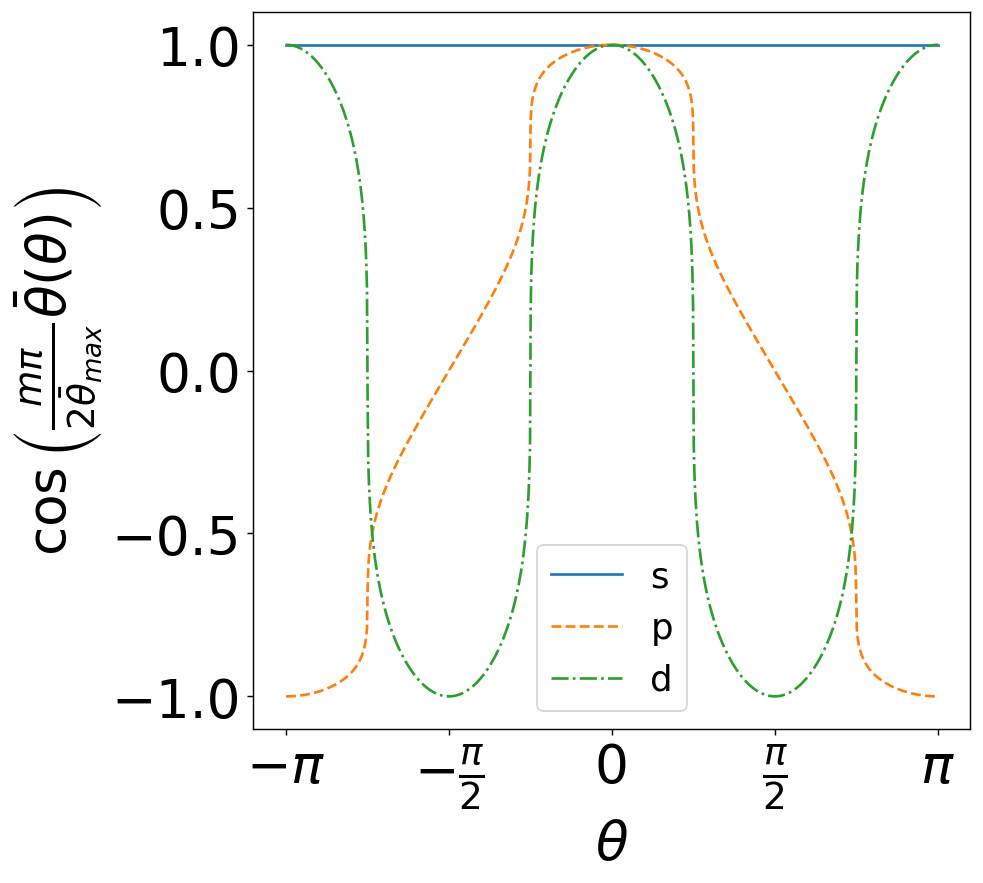}
\caption{
The eigen functions of the renormalized pairing interaction $\aV_{\theta_1,\theta_2}$
that emerge in the low-energy limit.
While the eigen functions are labeled by `angular momentum' $m \in \mathbb{Z}$ that counts the number of nodes, they don't have the perfect sinusoidal form in $\theta$ due to the angle-dependent metric created by critical fluctuations.
Here, we consider the $C_4$ symmetric Fermi surface at the Ising-nematic quantum critical point with bare parameters, 
$\KFthetadim = 1+0.025 \cos(4\theta)$, 
$e_\theta = 0.1 \cos(2\theta)$,
$v_{F,\theta}= 1$
and
$N=1$ 
in $d=2$.
In the low-energy limit, the effective Yukawa coupling
$g^*_\theta$ becomes independent of angle, but $v^*_{F,\theta}$ becomes larger near the cold spots 
at $\theta = \pm \pi/4, \pm 3\pi/4$
(see \eq{eq:gbartheta})\cite{PhysRevB.110.155142}.
Consequently, the metric $\gtt$ in \eq{eq:gthetatheta} becomes larger near the cold spots,
causing the phase of the pairing wavefunction to change more rapidly as a function of $\theta$. 
}
\label{fig:pairing_wavef_original}
\end{figure}

In \eq{eq:beta_function_nonlinear},
the quantum corrections depend on 
the short-range four-ferminon interaction ($\bar V$) and the interaction mediated by the critical bosons ($\bar h_d$)  through the net two-body interaction
$ \aV^{\pm}_{\bar{\theta}_1,\bar{\theta}_2} = \bar{V}^{\pm}_{\bar{\theta}_1,\bar{\theta}_2} + \frac{\bar{h}^{\pm}_{\bar{\theta}_1,\bar{\theta}_2}}{4} $. 
From now on, we consider the RG flow of that net two-body interaction. 
It is also convenient to consider the pairing interaction in the angular momentum basis. 
Due to the emergent rotational symmetry,
the angular momentum conjugate to $\bar \theta$ is conserved,
and there is no mixing between different angular momentum channels in the low-energy limit.
The coupling in the angular momentum channel $m \in \mathbb{Z}$ is 
\bqa
\av_m = \int_{-\bar{\theta}_{\text{max}}}^{\bar{\theta}_{\text{max}}}
\frac{d\bar{\theta}}{2\pi\sqrt{\mu}}\, 
\aV^{\pm}_{\bar{\theta}_1,\bar{\theta}_1+\bar{\theta}} 
\, e^{i\frac{m \pi}{2\bar\theta_{\text{max}}}\bar{\theta}},
\label{eq:FourierV}
\eqa
where $m$ are 
even (odd) integers for
$\aV^{-}$
($\aV^{+}$).
$\av_m$ is real because 
$\aV^{\pm}_{\bar{\theta}_1,\bar{\theta}_2}$ as a matrix of two continuous indices $\bar \theta_1$ and $\bar \theta_2$ is Hermitian.
Since $\av_m^+$ and $\av_m^-$ have different conjugate momenta, we drop $\pm$ in $\av_m$.
In the original angular space, the pairing wavefunction associated with angular momentum $m$ is deformed from pure sinusoidal functions due to the non-linear relation between $\theta$ and $\bar \theta$.
Nonetheless, $m$ is still referred to as angular momentum since it is quantized and counts the number of nodes that the Cooper pair wavefunction has around the Fermi surface.
See Fig. \ref{fig:pairing_wavef_original} for the pairing wavefunction $\cos\left( \frac{m \pi}{2 \bar \theta_{max}} \bar \theta(\theta) \right)$ plotted in the original angular space for a few selected angular momenta $m$.

The coupling in the angular momentum basis obeys the beta function,
\begin{equation}
\frac{d \av_m}{dl} = - R_d \av^2_{m} + \Big(\tfrac{1}{2}-\HD\Big)\av_m + \as_{m}(l),
\label{eq:RG_xspace0}
\end{equation}
where
$\as_{m}(l) = \int_{-\bar\theta_{\text{max}}}^{\bar\theta_{\text{max}}}
~
\frac{d\bar{\theta}}{2\pi\sqrt{\mu}} 
\aS^{\pm}_{\bar{\theta}_1,\bar{\theta}_1+\bar{\theta}}
\, e^{i\frac{m \pi}{2\bar\theta_{\text{max}}}\bar{\theta}}$
with
$\aS^{\pm}_{\bar{\theta}_1,\bar{\theta}_1+\bar{\theta}}
=
\left(\HD-\frac{d}{d\ln\mu} \right)\frac{\bar h^\pm_{\theta_1,\theta_1+\bar \theta}}{4}
+ 
\delta S^{\pm}_{\bar{\theta}_1,\bar{\theta}_1+\bar{\theta}}
$
being the source term generated from the critical boson.
To the leading order in $\epsilon$,
$\delta S^{\pm}_{\bar{\theta}_1,\bar{\theta}_1+\bar{\theta}}$ can be ignored and will be dropped henceforth.
It is noted that $\as_m(l)$ depends not only on $m$ but also on the RG scale because the width of  
$\aS^{\pm}_{\bar{\theta}_1,\bar{\theta}_1+\bar{\theta}}$ 
is $\sqrt{\mu}$.
Furthermore, the tracelessness of \eq{eq:hbar} guarantees that
\bqa
\sum_{m=-\infty}^\infty \as_{m}(l) = 0.
\label{eq:Smtraceless}
\eqa
This implies that the universal pairing interaction generated from critical fluctuations cannot be entirely repulsive or attractive across all angular momentum channels. 

\begin{figure}[t]
\begin{subfigure}{.4\textwidth} \centering 
\includegraphics[width=0.8\linewidth]{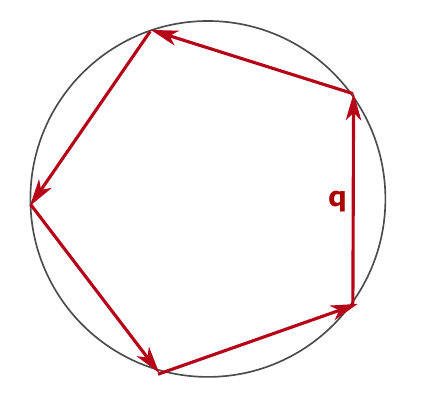}
\caption{  }
\label{fig:meaning_of_ym_a} 
\end{subfigure}%
\begin{subfigure}{.40\textwidth}
\centering
\includegraphics[width=0.8\linewidth]{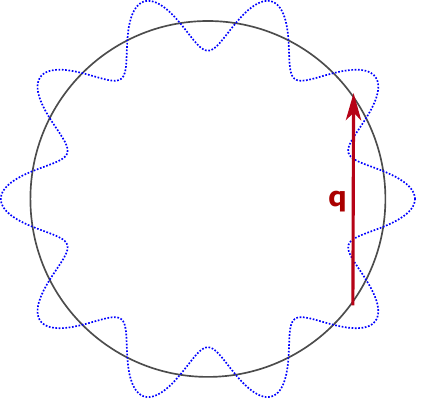} 
\caption{  }
\label{fig:cN}
\end{subfigure}%
\caption{
At energy scale $\mu=\Lambda e^{-l}$, the typical momentum carried by critical boson (represented by the arrows) is given by $q \sim \sqrt{\mu \KFthetadim} \sim e^{-l/2}$.
(a) $\cN(l)$ denotes the number of steps needed to traverse the entire Fermi surface with each step of size $q$ at scale $l$.
(b) The universal pairing interaction generated in angular momentum channel $m$ at scale $l$ is a function of $x=m/\cN(l)$, 
which represents the number of oscillations that the pairing wavefunction undergoes with momentum transfer $q$.
In this figure, $m=10$ and $\cN(l)=5$, which gives $x=2$.
As $l$ increases,  
$x$ decreases as $e^{-l/2}$ since a fermion scattered by a boson experiences fewer oscillations.
}
\label{fig:meaning_of_ym}
\end{figure}

In principle, one must solve the scale-dependent beta function for each angular momentum channel separately.
However, this task can be greatly simplified because the coupling of a large angular momentum channel is renormalized at low energy in the same manner as the coupling of a small angular momentum is renormalized at high energy.
This scaling relation arises because the pairing source for the angular momentum channel m is determined by 
$\frac{1}{m}\sqrt{\frac{\KFAVdim}{\mu}}$ which is the ratio
between the pitch of the pair wavefunction along the Fermi surface $\KFAVdim/m$
and the typical momentum $q \sim \sqrt{\mu \KFAVdim}$ that critical fluctuations carry at energy $\mu$.
To check this explicitly, we introduce a rescaled angular variable $\hat \theta = \bar \theta e^{l/2}$ and single out the scale dependence in the source term as
$\as_m(l) = \int_{-\bar \theta_{\text{max}} e^{l/2}}^{\bar \theta_{\text{max}} e^{l/2}}
\frac{d\hat{\theta}}{2\pi\sqrt{\Lambda}} 
\hat S^{\pm}_{\hat{\theta}_1,\hat{\theta}_1+\hat{\theta}}
\, e^{i\frac{ \pi m \sqmu }{2\bar\theta_{\text{max}}}
\frac{\hat{\theta}}{\sqrt{\Lambda}}
}$.
Here, 
$\hat S^{\pm}_{\hat{\theta}_1,\hat{\theta}_2}
=
\aS^{\pm}_{e^{-l/2}\hat{\theta}_1,e^{-l/2}\hat{\theta}_2}$ is independent of $l$ for fixed $\hat \theta_i$.
At sufficiently large $l$, we can set $\bar \theta_{max} e^{l/2}$ to $\infty$ provided that 
$\hat S^{\pm}_{\hat{\theta}_1,\hat{\theta}_2}$ decays sufficiently fast in $\hat \theta$.
In this low-energy limit, we note that
$\as_m(l) = 
\aS_{m e^{l'/2}}(l+l')$.
This implies the followings.
Firstly, the scale-dependence of the beta function is entirely encoded in the rescaled angular momentum $\tilde m=me^{-l/2}$, which runs under the RG flow.
Secondly, the beta function at a fixed rescaled angular momentum $\tilde m$ takes the same form irrespective of what the actual angular momentum $m$ is.
In particular, the source term is a function of one variable $x = \frac{ \pi m \sqmu }{2\bar\theta_{\text{max}}}$.
Its physical meaning becomes clear once it is written as $x = m/\cN(l)$, where
$\cN(l)$ defined in \eq{eq:cN} denotes the number of tangential scatterings a fermion need to go through with bosons at scale $l$ to go around the Fermi surface once.
Hence, $x$ corresponds to the number of oscillations that the pairing wavefunction has over the range of the typical momentum a critical boson carries at scale $l$.
See \fig{fig:meaning_of_ym} for illustration.

This allows us to consider a single flow equation in the plane of the rescaled angular momentum and the coupling, instead of solving an infinite set of scale-dependent beta functions. 
In formalizing this, it is more convenient to consider the logarithm of $x$ that runs linearly in $l$.
Since the rescaling does not change the sign of the rescaled angular momentum, it is sufficient to consider $m \geq 0$ at a time.
In the presence of time reversal symmetry, $\av_{-m}=\av_m$.
If not, we can consider the flow of the time-reversal counterpart ($\av_m$ with $m \leq 0$) separately.
So, we label angular momentum in terms of the logarithm of $\log \frac{m}{\cN(l)}$,
\begin{equation}
    \begin{aligned}
    y^{(m)}(l) = 
         \log \left[ \frac{m}{\cN(l)} \right]
         =
         \log \left[ m \frac{ \sqrt{\Lambda} \pi } {2 \bar{\theta}_{\max}}  \right]
    -\frac{l}{2}.
    \end{aligned}
    \label{eq:y}
\end{equation} 
We then introduce a two-dimensional plane $(y,\aV)$ to represent the coupling in each angular momentum channel. 
At scale $l$, the point $\left( y^{(m)}(l), \av_m(l) \right)$ in the plane represents the running logarithmic angular momentum and the value of the coupling in angular momentum channel $m$.
The RG flow of the coupling function in the pairing channel can be now understood as the flow of a collection of points 
in the $(y, \aV)$ plane.
At the UV cutoff scale ($l=0$), the set 
$\left\{ (y^{(m)}(0), \av_m(0)) | m \in \mathbb{Z}+ \right\}$ 
specifies the bare coupling in all angular momentum channels.
We note that the UV couplings for the pairing interaction can be tuned in two ways.
One can tune the bare four-fermion coupling $\aV_{y^{(m)}}$ at fixed $y^{(m)}$, which affects the vertical location in the plane.
One can also tune the Fermi momentum (or the fermion density), which shifts the horizontal location through the $\KFAVdim$ dependence in $y^{(m)}$.
With increasing $l$, the coupling in each angular momentum channel flows, obeying the beta functional.
At the same time, the rescaled angular momentum is dilated by $e^{-l/2}$, 
which causes $y$ to shift as $y \rightarrow y - l/2$. 
Therefore, each point defined in the space of $(y,\aV)$ flows,
obeying a two-dimensional flow equation:
\bqa
\frac{\mathrm{d}}{dl}
\left( y, \aV \right)
=
\left( -\frac{1}{2}, 
\beta_{\aV}(y) 
\right).
\label{eq:flowofyaV}
\eqa
Here, $-\frac{1}{2}$ is the horizontal component of the flow velocity,
and
\begin{equation}
\begin{aligned}
\beta_{\aV}(y) = -
R_d\aV^{{}^2{}}_y+\left(\frac{1}{2}-\HD\right)\aV^{{}}_y
        +
        \aS_y
\label{eq:beta_vertical}
\end{aligned}
\end{equation}
is the vertical flow velocity.
$\aS_{y} = 
\as_{\frac{2\bar \theta_{max}e^y}{\pi\sqrt{\mu}}}$
is the source for the net two-body interaction generated from the critical bosons at the logarithmic angular momentum $y$.
In the $\KFAVdim \gg \mu$ limit, $\aS_y$ becomes
\begin{equation}
    \begin{aligned}
\aS_{y} &= 
        \sum_t \frac{\sigp g^{t*}}{8\sqrt{3} \pi ^{3/2}\alpha_d^{\frac{1}{3}}\beta_d^{\frac{1}{3}}} 
        \left\{2 (\HD-1) G_{1,7}^{5,1}\left(\frac{\left(g^{t*}\right)^3 e^{6y} \alpha_d ^2}{\bar{g}^{*3} 46656}|
\begin{array}{c}
 -\frac{1}{3} \\
 0,\frac{1}{6},\frac{1}{3},\frac{2}{3},\frac{2}{3},\frac{1}{2},\frac{5}{6} \\
\end{array}
\right)
\right. \\
&
\hspace{3.2cm}
\left.
+(2 \HD+1) G_{1,7}^{5,1}\left(\frac{\left(g^{t*}\right)^3 e^{6y}  \alpha_d ^2 }{\bar{g}^{*3} 46656}|
\begin{array}{c}
 \frac{2}{3} \\
 0,\frac{1}{3},\frac{2}{3},\frac{2}{3},\frac{7}{6},\frac{1}{2},\frac{5}{6} \\
\end{array}
\right) \right\}.
\label{eq:Sx}
    \end{aligned}
\end{equation}

\begin{figure}[htbp]
    \centering
\includegraphics[height=7cm,width=10cm]{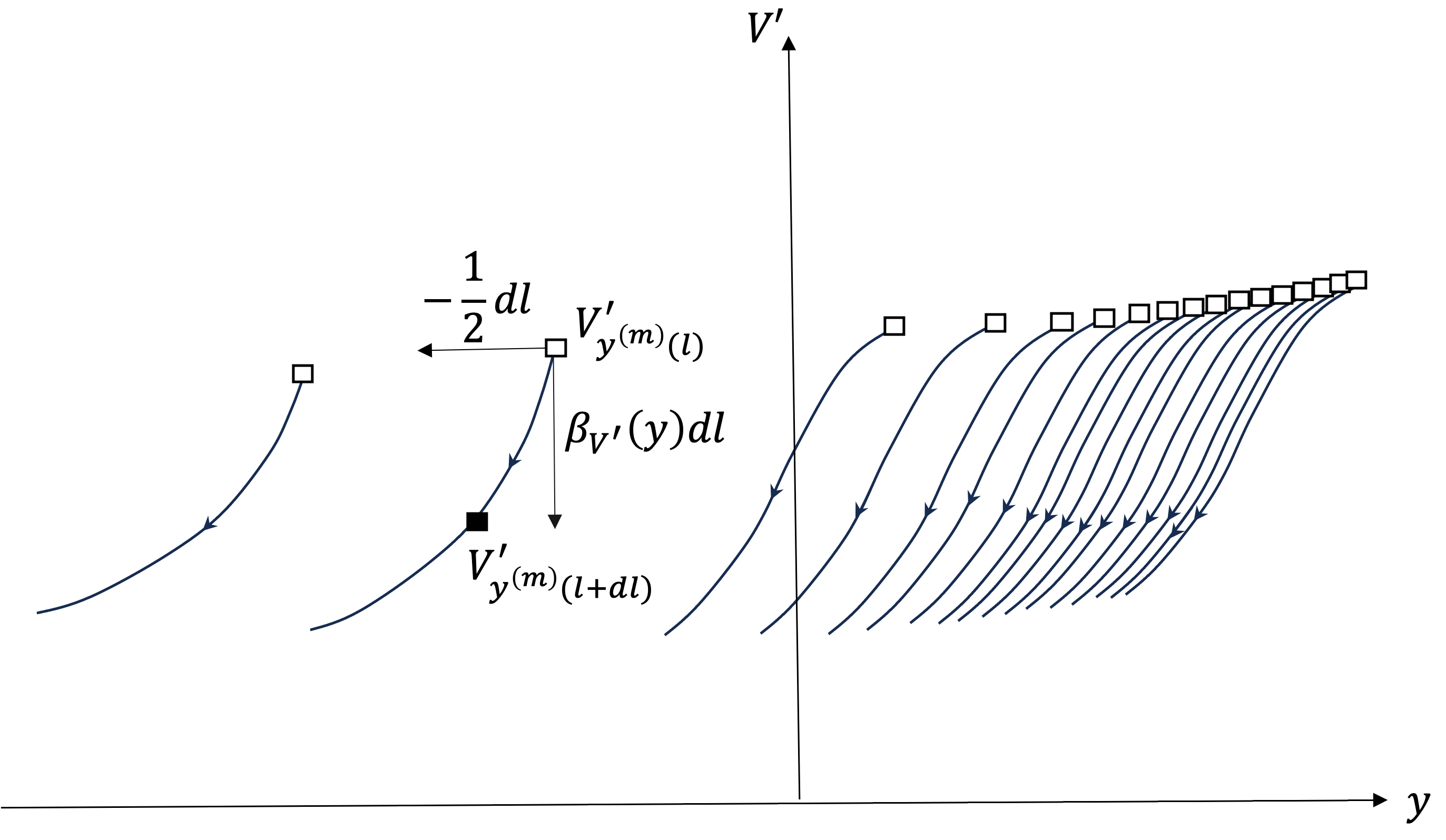}
\label{Fig:Infinitesimal_RG}
\caption{
At a scale $l$, the pairing interaction is specified by an infinite set of points, $\left( y^{(m)}(l), \aV_{y^{(m)}(l)}(l) \right)$ in the $y-\aV$ plane.
With increasing $l$, each point drifts with the flow velocity given in \eq{eq:flowofyaV}.
The horizontal velocity is fixed to be $-1/2$, but the vertical velocity $\beta_{\aV}(y)$ depends on both $y$ and $\aV$.
Each coupling takes a trajectory in the plane, which is a particular solution of \eq{eq:fp_eq_log}. 
}
\label{Fig:RG_Flow_Dynamics}
\end{figure}

Under the infinitesimal RG step,
the $y$ coordinate changes by $-\frac{1}{2} dl$
and
the $\aV$ coordinate by
$\beta_{\aV}(y) dl$,
as is shown in  
\fig{Fig:RG_Flow_Dynamics}
Namely, each point that represents the pairing interaction in one angular momentum channel flows with the velocity 
$\vec U = \left( -\frac{1}{2},
      \beta_{\aV}
      \right)
$ in the two-dimensional plane.
The vertical velocity 
$\beta_{\aV}(y)$ captures the flow caused by angular-momentum-dependent quantum corrections.
The horizontal velocity $-1/2$
reflects the fact that the rescaled angular momentum has the scaling dimension $-1/2$ and runs.
At general $l>0$, the renormalized coupling is captured by the set of $(y_{m}(l), \aV_{y^{(m)}(l)})$.
The advantage of  \eq{eq:flowofyaV} is two-fold.
Firstly, it is independent of the RG scale, unlike  \eq{eq:beta_function_nonlinear}. 
The scale-dependence of the beta function in 
\eq{eq:beta_function_nonlinear} has been traded for the flow of $y$, which effectively makes
$\beta_{\aV}(y)$ scale-dependent.
Namely, the pairing interaction at each angular momentum channel is effectively subject to the scale-dependent quantum corrections due to the horizontal flow of $y$.
Secondly, this single two-dimensional flow equation describes the flow of the couplings in all angular momentum channels.
Since the flow velocity depends only on $y$ and $\aV$, the RG flow of all angular momentum channels can be viewed as the collection of trajectories followed by the infinite set of `particles' scattered in the plane, where each particle obeys the same flow equation.
The collective behavior of those particles determines the fate of the theory in the low-energy limit.
Therefore, the RG flow of the coupling functions is entirely encoded in the solutions of
\begin{equation}
    \begin{aligned}
-\frac{1}{2} \partial_y\aV_y
= -
R_d\aV^{{}^2{}}_y+\left(\frac{1}{2}-\HD\right)\aV^{{}}_y
        +
        \aS_y.
\label{eq:fp_eq_log}
    \end{aligned}
\end{equation}
\eq{eq:fp_eq_log}, which is referred to as the {\it projective fixed point (PFP) equation}, is the central object that will be used repeatedly in the rest of the paper.
Henceforth, we use $\aV_y$ without argument $l$ to denote a solution of \eq{eq:fp_eq_log}, which represents a geometric trajectory of the RG flow in the plane.


\section{Classification of metallic universality classes}
\label{sec:classification}

Distinct universality classes of matter are usually identified as `fixed points' of the RG flow. 
In metals, the RG flow equations include the incessant growth of the dimensionless Fermi momentum in \eq{eq:betae}.
Unlike other high-energy scales, the flow of the Fermi momentum introduces explicit scale dependence in the beta functional of the pairing interaction 
in \eq{eq:fp_eq_log}.
What sets metals apart from other forms of matter is the presence of infinitely many couplings and arbitrarily small crossover energy scales.
In this section, we classify distinct non-Fermi liquid universality classes through the solutions of 
\eq{eq:fp_eq_log}.
Obviously, its solutions generally depend on the four-fermion couplings defined at the UV scale.
Therefore, we divide our discussion into two parts.
We first consider the general solutions of \eq{eq:fp_eq_log}, and classify them based on the topology of the `phase space' portraits of the differential equation\cite{Arnold1988}.
Each topologically distinct general solution defines a {\it superuniversality class}.
Each superuniversality class generally contains multiple non-Fermi liquid universality classes that share the same topological properties.
We then identify individual non-Fermi universality classes in terms of their universal superconducting fluctuations.

\subsection{Projective fixed points}

Let's begin our discussion by asking whether there can be fixed points at which couplings cease to flow under the RG flow.
Generally, the answer is no because of the 
$y$-dependence in the beta function and the flow of $y$, which
prevents couplings from settling into a fixed value.
However, the beta function approaches $y$-independent forms asymptotically in the large $|y|$ limit.
Therefore, we first consider these asymptotic limits.
In the large angular momentum limit ($y \rightarrow \infty$), $\aS_{\infty}=0$.
This is guaranteed by the fact that the quantum corrections generated at a non-zero RG energy scale are smooth in the momentum space due to locality,
and their Fourier transformations  vanish in the large angular momentum limit. 
In the small angular momentum limit ($y \rightarrow -\infty$), the source saturates to a non-zero value, $\aS_{-\infty}$ whose sign is set by the universal pairing interaction generated in the s-wave channel through  \eq{eq:sttheta}.
Generally, the source behaves as
\bqa
\aS_y \approx 
\begin{cases}
0 & \text{for $y \gg w_L$} \\
\aS_{-\infty} & \text{for $y \ll w_S$} 
\end{cases},
\label{eq:wLwS}
\eqa
where $w_L$ and $w_S$ are crossover angular momentum scales.
The $\infty$ and $-\infty$ asymptotic behaviors hold in $y \gg w_L$
and $y \ll w_S$, respectively.
The crossover scales $w_L$ and $w_S$ depend on the theory.
However, $w_L$ and $w_S$ can be shifted by changing the overall scale of the proper angle $\bar \theta$ in  \eq{eq:theta_bartheta}.
Here, we already made a specific choice of $\bar \theta$ such that  \eq{eq:thetabarscale} is satisfied.
In this choice, the universal pairing interaction is peaked around $\Delta \bar \theta / \sqmu \sim 1$.
Since $e^y$ is conjugate to $\bar \theta/\sqmu$,
the crossover between the two opposite asymptotic limits occurs around $y \sim 0$.
Therefore, we should view only a combination $w_L-w_S$ as an independent parameter of the theory.

In the large $|y|$ limit, the beta function becomes independent of $y$ and one can consider {\it asymptotic fixed points}
at which 
 $\beta_{\aV}(y)$ vanishes.
These asymptotic fixed points play an important role in the classification of metallic universality classes.
For sufficiently large and negative $\aV_y$, it always flows to $-\infty$ at low energies.
Therefore, $\aV_y = -\infty$ always acts as an asymptotic fixed point.
The number and type of additional asymptotic fixed points is determined by the discriminant 
\bqa
    \etaPy = (2\HD - 1)^2 + 16 R_d \aS_y 
\label{eq:disc_y}
\eqa
in the $y \rightarrow \pm \infty$ limit.
The discriminant encapsulates the competition between the incoherence of fermions encoded in the scaling dimension $\HD$ 
and the universal pairing interaction $\aS_y$ generated from critical fluctuations.
In the large $\HD$ limit with a fixed $\aS_y$, 
$\etaPy > 0$.
In this case, fermions are highly incoherent, and they are resistant to pairing instabilities.
In the opposite limit with a large and negative $\aS_{y}$, $\etaPy < 0$ and fermions become susceptible to pairing instabilities.
Consequently, the number of additional asymptotic fixed points decreases with decreasing $\etaPy$.
There are $2$, $1$ and $0$ additional asymptotic fixed points for 
$\etaPy>0$,
$\etaPy=0$
and
$\etaPy<0$, respectively, in the large $|y|$ limit.

Let us first consider the $y \rightarrow \infty$ limit.
Here, we assume $(\HD-1/2) > 0$, which is satisfied by all non-Fermi liquids considered here. 
The equality is saturated for the non-interacting theory in $d=2$.
Since $\etaPI > 0$,
three asymptotic fixed points arise at $y=\infty$, 
\bqa
\aV^\bullet_{\infty}=0, ~~~~
\aV^\circ_{\infty} =- \frac{1}{R_d} \left(\HD-\frac{1}{2}\right), ~~~~ 
\aV^{SC}_{\infty}=-\infty.
\label{eq:plusinfasymfp}
\eqa
$\aV^\bullet_{\infty}$ and $\aV^\circ_{\infty}$ describe fixed points with zero and finite pairing interactions in the infinite angular momentum limit.
For $(\HD-1/2) > 0$, 
which is satisfied for all physical examples considered here,
the former is stable  and the latter is unstable against linear perturbations.\footnote{
For $(\HD-1/2)<0$, 
the following discussion holds once 
$\aV^\bullet_\infty$
and
$\aV^\circ_\infty$ are switched.
}
$\aV^{SC}_{\infty}$ 
represents the stable fixed point of infinitely negative interaction, corresponding to a superconducting (SC) asymptotic fixed point.
For a large negative  $\aV^{SC}_{y}$, 
the fermion bi-linear
$\tilde{\Psi}_{j}\left(\mathbf{K}+\mathbf{Q},
 \deltaq, \thetasq 
\right)
\tilde{I}^{(\nu)}_m
\Psi_{j'}(\mathbf{K},\delta,\theta)
$ in \eq{eq:Onus} is expected to acquire a non-zero expectation value at $({\bf Q},\vec q)=0$,
picking one of the $4-d$ direction for $m=d-1,..,2$ in the pairing channel.
It breaks 
$Z_2\times SO(d-1)\times SO(4-d)$
to
$Z_2\times SO(d-1)$.
In $d=2$, this is nothing but the spontaneous breaking of the 
charge 
$U(1)$ symmetry to its $Z_2$ subgroup.
Even in general $d$, this marks a non-trivial symmetry-breaking phase transition.
$\aV^{SC}_{\infty}$ is stable because $\aV_y$ with a sufficiently negative initial value always flows to $-\infty$.
We refer to 
$\aV^\bullet_{\infty}$/
$\aV^\circ_{\infty}$/
$\aV^{SC}_{\infty}$
as 
{\it 
stable/
unstable/
superconducting 
$\infty$-asymptotic 
fixed points}, respectively.

One can also look for asymptotic fixed points in the $y\rightarrow -\infty$ limit where the beta function also becomes independent of $y$.
At $y=-\infty$, $\aV^{SC}_{-\infty}=-\infty$ remains to be a fixed point.
For $\discinf > 0$,  there exist one stable  
($\aV^\bullet_{-\infty}$) and one unstable  ($\aV^\circ_{-\infty}$) additional fixed points at 
\bqa
\aV^\bullet_{-\infty} = \frac{-(2\HD-1) + \sqrt{\discinf}}{4R_d}, 
&&
\aV^\circ_{-\infty} = \frac{-(2\HD-1) - \sqrt{\discinf}}{4R_d}.
\label{eq:minusinfaf}
\eqa
They are referred to as {\it  stable and unstable  $-\infty$-asymptotic  fixed points}, respectively.
If $\discinf < 0$, there is no additional asymptotic fixed point for real $\aV_{-\infty}$ besides the superconducting one at $y=-\infty$. 
In this case, $V_{y}$, which is real for the Hermitian theories, is bound to flow to $\aV^{SC}_{-\infty}$ at $y=-\infty$.
Namely, the strong attractive interaction generated by the critical fluctuations makes normal states unstable against superconductivity.
If $\discinf = 0$, there is a 
{\it marginal $-\infty$-asymptotic fixed point}, which is denoted as $\aV^{\halfominus}_{-\infty}$.
As $\discinf$ changes sign from positive to negative, the asymptotic fixed points collide and become complex\cite{BORGES2023169221}.

Since $y$ decreases with increasing length scale, the $\infty$-asymptotic fixed points correspond to the `short'-distance limit of a projective fixed point (but still above the length scales associated with high-energy scales).
It can also be viewed as the small $\KFAVdim$ limit where the pairing interactions in most angular momentum channels are renormalized in the same way that the coupling in the infinite angular momentum channel is renormalized
(see \fig{fig:attractive_to_repulsive}).
On the other hand, the $-\infty$-asymptotic fixed points correspond to the long-distance (large $\KFAVdim$) limit of a projective fixed point.
There is a sense in which $\infty$ and $-\infty$ asymptotic fixed points act as UV and IR fixed points, respectively, as the pairing interaction of a large but fixed angular momentum channel flows from a stable $\infty$ asymptotic fixed point to a stable $-\infty$ asymptotic fixed point under the RG flow.
However, one key difference from the conventional RG flow lies in the fact that there are infinitely many couplings.
In the angular momentum channel $n$, the coupling crossovers from the $\infty$-asymptotic fixed point to the $-\infty$-asymptotic fixed point at a scale $\mu_n \sim \frac{\KFAVdim}{n^2}$;
the crossover occurs when the typical momentum of critical boson in \eq{eq:qmutheta} becomes comparable to the pitch of the pairing wavefunction $\KFAVdim/n$ around the Fermi surface.
Since $\mu_n$ goes to zero in the large $n$ limit, there are infinitely many angular momentum channels whose couplings are near the stable $\infty$-asymptotic fixed point at any non-zero energy scale.
Therefore, not only the $-\infty$-asymptotic fixed points but also the $\infty$-asymptotic fixed points and the crossovers between them play important roles in determining the infrared physics of the theory. 
The presence of the infinite set of couplings and the incessant running of $y$ prevent us from using the usual notion of fixed points in metals.
In particular, it is not possible for all couplings to be close to either of the asymptotic fixed points within errors that vanish with powers of $e^{-l}$ at any finite $l$.
{\it
Since the entire set of couplings constitutes the low-energy observables, metals cannot exhibit scale invariance as a whole.
}%

The lack of scale invariance in metal is a direct consequence of the running of the Fermi momentum under the RG flow.
While there is no conventional fixed point in metals,
universality classes are characterized by projective fixed points (PFPs),
which represent one-dimensional manifolds defined in the space of coupling functions that include Fermi momentum\cite{PhysRevB.110.155142}.
These manifolds are non-compact as $\KFAV$ increases toward $\infty$.
Being a subset of the full coupling functions, the pairing interaction also exhibits PFPs, which correspond to $y$-dependent profiles of  $\aV_y$ that satisfy  \eq{eq:fp_eq_log}.
A PFP of the pairing interaction can be understood as a one-dimensional manifold that is invariant under the two-dimensional flow equation in \eq{eq:flowofyaV}.
\eq{eq:fp_eq_log} is a first-order differential equation,  and the general solution is a one-parameter family of solutions.
We refer to this set as {\it general PFP}.
A {\it particular PFP} is the solution that satisfies an `initial' condition imposed at one $y$.
%
Since couplings at different angular momentum channels can be tuned independently, they generally traverse different particular PFPs under the RG flow.

The first step for classifying universality classes of non-Fermi liquids is to classify general PFPs, which are independent of the bare four-fermion couplings.
General PFPs are determined from the kinematic data (such as the spatial dimensions and the number/type of critical bosons) and the marginal coupling functions (Fermi velocity and the dimensionless shape of the Fermi surface).
A superuniversality class, which corresponds to a topologically distinct general PFP, contains multiple non-Fermi liquid universality classes that have distinct universal four-fermion coupling functions.
Once the superuniversality classes are identified through general PFPs, we identify individual non-Fermi liquid universality classes contained within each superuniversality class.
Distinct non-Fermi liquid universality classes within one superuniversality class can be accessed by tuning the bare four-fermion couplings.
Among the infinitely many particular PFPs within one general PFP, the PFPs connected to the asymptotic fixed points constrain the behavior of all other PFPs to the extent that classifying general PFPs amounts to identifying all possible ways in which the asymptotic fixed points are connected through particular PFPs. 
Before we delve into a systematic classification,
we first illustrate the basic idea using some examples.

\begin{figure}[th]
\centering
\begin{subfigure}{.4\textwidth}
\centering
\includegraphics[width=1.0\linewidth]{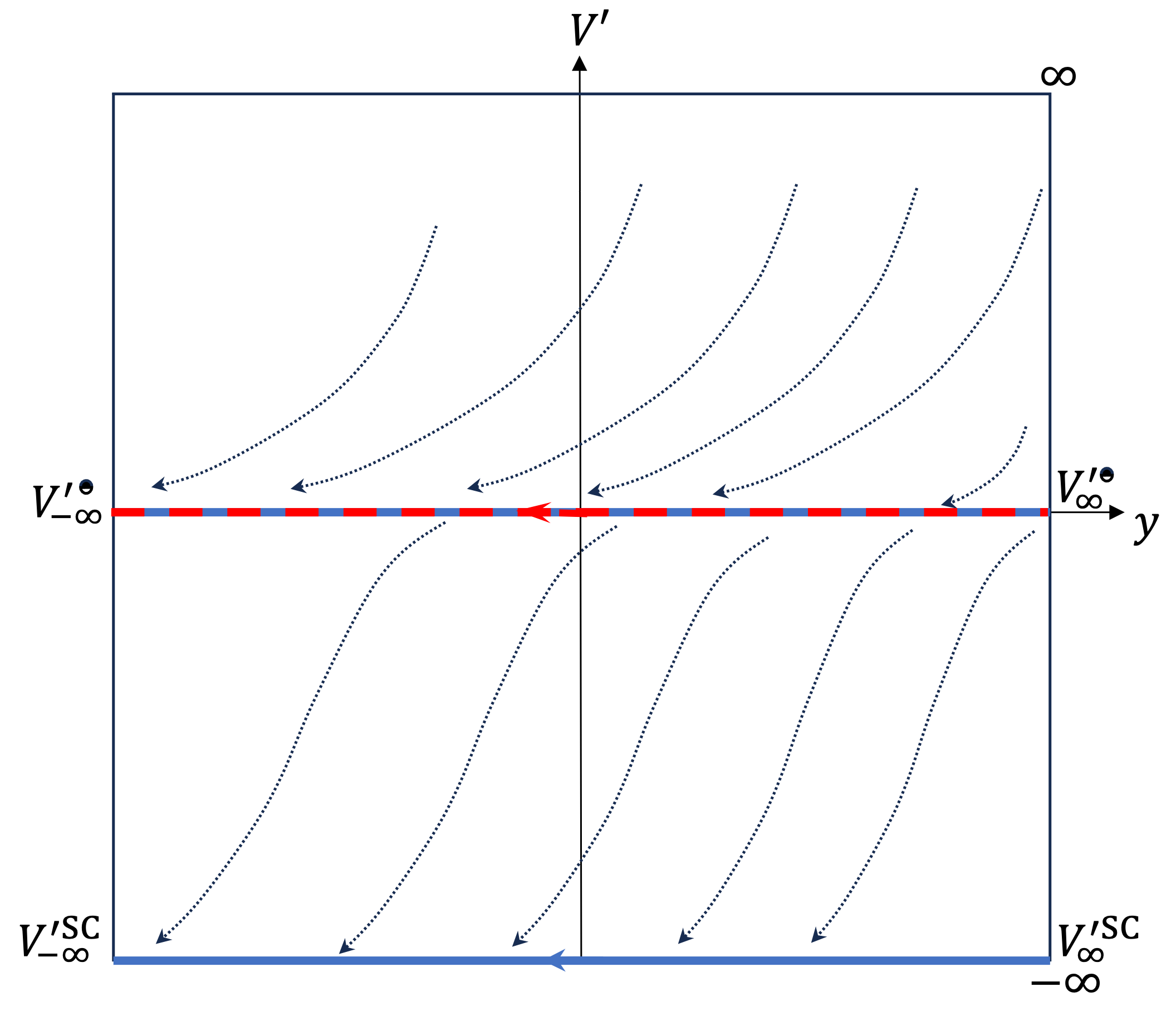}
  \caption{}
  \label{fig:PFP_fl}
\end{subfigure}%
\begin{subfigure}{.4\textwidth}
\centering
\includegraphics[width=1.0\linewidth]{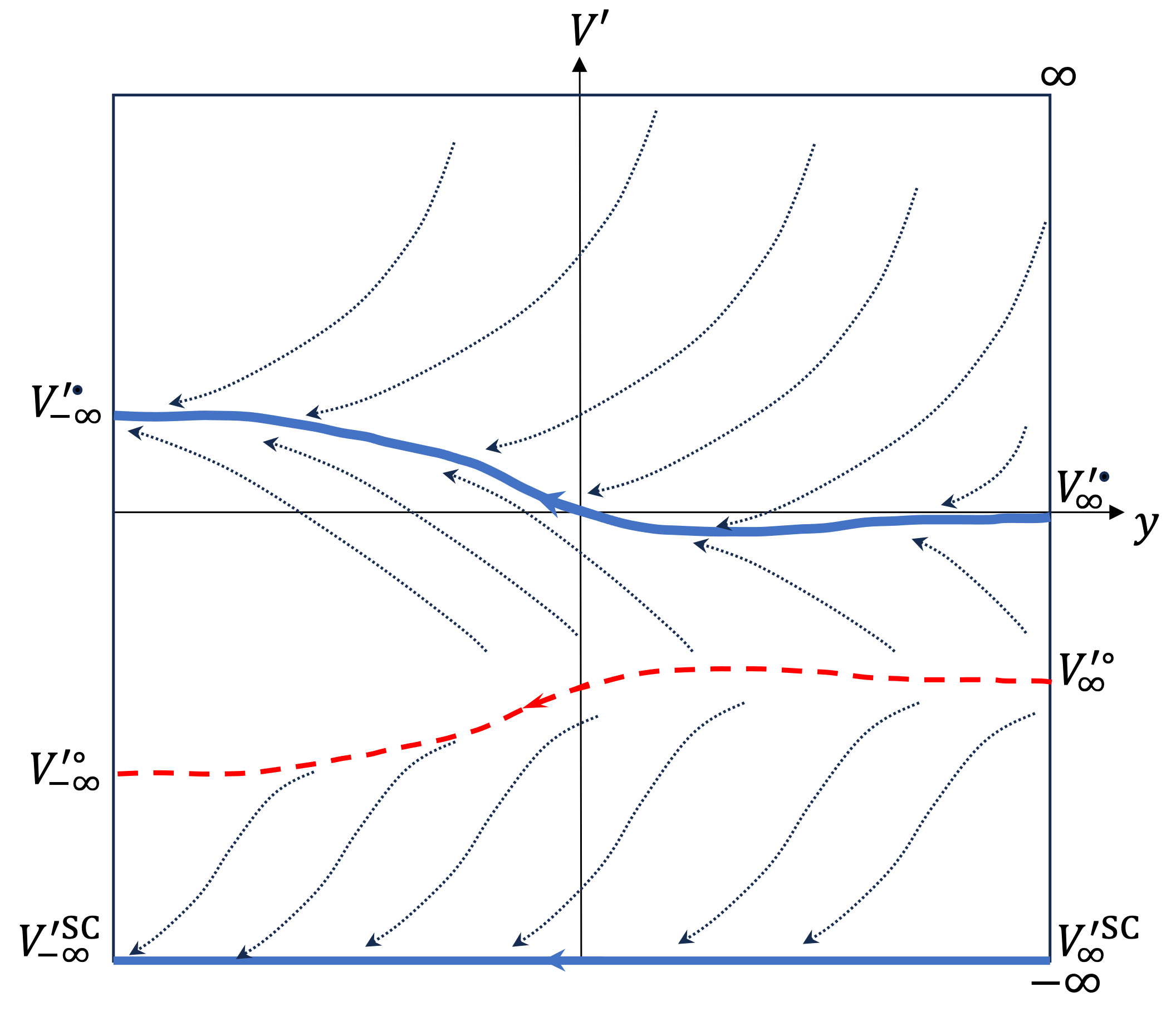}
  \caption{}
  \label{fig:PFP_nfl1}
\end{subfigure}
\caption{
The portraits of the general PFPs (the collection of all particular PFPs)
that arise from 
(a) the Fermi liquids in $d=2$,
and
(b) the U(1) gauge theory at a finite chemical potential.
The metallic and separatrix PFPs are denoted by the blue solid curve and the red dashed curve, respectively.
In Fermi liquids, the metallic PFP which emanates from  $\aV^{\halfominus}_{\infty}$ and the separatrix that forms the boundary of the basin of attraction of $\aV^{\halfominus}_{-\infty}$ are the same.
In the U(1) gauge theory, the metallic PFP emanates from  $\aV^\bullet_{\infty}$ and the separatrix is the boundary of the basin of attraction of $\aV^\bullet_{-\infty}$, which are distinct.
}
\label{fig:PFP_example}
\end{figure}

The first example is Fermi liquids in $d=2$.
For Fermi liquids,  $z=1$, $\eta_\psi=0$ and $\aS_y=0$.
This leads to only marginal asymptotic fixed points, $\aV^{\halfominus}_{\pm \infty}=0$ at $\pm \infty$.
The PFP equation 
in \eq{eq:fp_eq_log} is reduced to
\begin{equation}
    \begin{aligned}
\frac{1}{2} \partial_y
\aV_y
- R_d 
\aV^2_y
=0.
\label{eq:fp_eq_log_FL}
    \end{aligned}
\end{equation}
Besides the superconducting PFP 
($\aV^{SC}_y=-\infty$), there exists another particular PFP of importance: 
$\aV^{\halfominus}_y=0$ 
that connects the marginal asymptotic fixed points at $\pm \infty$.
PFPs that are above (below)  $\aV^{\halfominus}_y$ 
are attracted to 
$\aV^{\halfominus}_y$ 
($\aV^{SC}_y$)
in the small $y$ limit.
Therefore,
$\aV^{\halfominus}_y$ acts as the attractor from the repulsive side, and as the separatrix that divides the basins of attraction of
$\aV^{\halfominus}_y$ 
and
$\aV^{SC}_y$.
The corresponding diagram of general PFP is shown in \fig{fig:PFP_fl}.

As the second example, let us consider the general PFP shown in \fig{fig:PFP_nfl1}. 
Here, each line represents a particular PFP with the arrows pointing toward the direction of increasing length scale.
This example is drawn from the U(1) gauge theory coupled with the Fermi surface, but we do not have to worry about the details yet.
For now, we are mainly concerned with the topology formed by the collection of PFPs.
In addition to the superconducting asymptotic fixed points,
there are stable and unstable asymptotic fixed points at both $y=\infty$ and $-\infty$. 
The existence of the stable $-\infty$-asymptotic fixed point is attributed to the fact that the U(1) gauge field mediates a repulsive pairing interaction in the s-wave channel.
Naturally, the superconducting asymptotic fixed points at $y=\pm \infty$ are interpolated by the PFP, $\aV_y=-\infty$.
For understanding the behaviors of other PFPs, we start by defining the basin of attraction of the stable $-\infty$-asymptotic fixed points,
\bqa
{\cal B}^{IR}_{\aV^\bullet_{-\infty}} 
= \{ (y,\aV) | \mbox{the PFP that goes through $(y,\aV)$ converges to  $\aV^\bullet_{-\infty}$ in the $y \rightarrow -\infty$ limit} \}.
\label{eq:basinIR}
\eqa
${\cal B}^{IR}_{\aV^\bullet_{-\infty}}$ is an open set in the usual topology of the plane.
Its closure is denoted as
$\bar {\cal B}^{IR}_{\aV^\bullet_{-\infty}}$, and
$\partial {\cal B}^{IR}_{\aV^\bullet_{-\infty}}
\equiv
\bar {\cal B}^{IR}_{\aV^\bullet_{-\infty}} -{\cal B}^{IR}_{\aV^\bullet_{-\infty}}$ represents its boundary.
One can also define the basins of attraction for 
$\aV^{SC}_{-\infty}$
and
$\aV^\circ_{-\infty}$, respectively.
It is noted that  $\partial {\cal B}^{IR}_{\aV^\bullet_{-\infty}} = 
\partial {\cal B}^{IR}_{\aV^{SC}_{-\infty}}
=
{\cal B}^{IR}_{\aV^\circ_{-\infty}}
$.
One crucial factor for deciding the general PFP is which basin of attraction 
$\aV^\bullet_\infty$ belongs to.
$\aV^\bullet_\infty$
can be 
either in 
${\cal B}^{IR}_{\aV^\bullet_{-\infty}}$,
${\cal B}^{IR}_{\aV^{SC}_{-\infty}}$,
or
${\cal B}^{IR}_{\aV^\circ_{-\infty}}$.
In the next section, we will go over all possibilities.
Here, we focus on the current example in which
$\aV^\bullet_\infty
\in 
{\cal B}^{IR}_{\aV^\bullet_{-\infty}}$.
Namely, the PFP that emanates from  $\aV^\bullet_\infty$ converges to $\aV^\bullet_{-\infty}$ in the small $y$ limit.
It is denoted as the thick blue line in \fig{fig:PFP_nfl1}.
Naturally, $\partial {\cal B}^{IR}_{\aV^\bullet_{-\infty}}$ is a particular PFP that acts as the separatrix.
It turns out that this separatrix PFP connects $\aV^\circ_\infty$ and $\aV^\circ_{-\infty}$.
To see that there must exist a PFP that connects them, it is useful to consider the UV beta function, which is the negative of
\eq{eq:flowofyaV},
$\frac{\mathrm{d} 
}{d \log \mu}  
(y, \aV_{y})
=
\left( \frac{1}{2},
-\beta_{\aV}(y) 
\right)$.
This UV beta function describes the evolution of couplings with increasing {\it energy} scale $\mu$.
The trajectories of PFPs and the asymptotic fixed points of the UV beta functions are identical to those of the IR beta functions.
However, the direction of the flow is reversed, and the stability of the asymptotic fixed points is inverted. 
Namely, the stable (unstable) asymptotic fixed points of the IR beta function become unstable (stable) under the reversed RG flow. 
Under the reversed RG flow,
$\aV^\circ_{\infty}$ is an attractor, and it has an extended basin of attraction,
\bqa
{\cal B}^{UV}_{\aV^\circ_{\infty}} 
= \{ (y,\aV) | \mbox{the PFP that goes through $(y,\aV)$ converges to  $\aV^\circ_{\infty}$ in the $y \rightarrow \infty$ limit} \}.
\label{eq:basinUV}
\eqa
The closure and boundary are defined similarly.
Under the reversed RG flow,
$\aV_{\infty}=\infty$ is another stable $\infty$-asymptotic fixed point with its own basin of attraction.
The PFP that interpolates $\aV^\bullet_{\infty}$ and $\aV^\bullet_{-\infty}$ is 
the separatrix PFP under the reversed RG flow; a PFP that is above (below) that PFP heads to $+\infty$ ($\aV^\circ_{\infty}$) under the reversed flow.
This guarantees 
$\aV^\circ_{-\infty}
\in
{\cal B}^{UV}_{\aV^\circ_{\infty}}$,
and
the PFP that emanates from
$\aV^\circ_{-\infty}$
converges to
$\aV^\circ_{\infty}$
under the reversed RG flow.
This implies that there exists a PFP that connects  $\aV^\circ_{\infty}$ and $\aV^\circ_{-\infty}$,
which is denoted by the dashed red line in \fig{fig:PFP_nfl1}.
This PFP is the separatrix PFP under the original IR RG flow and is identical to 
$\partial {\cal B}^{IR}_{\aV^\bullet_{-\infty}}$.

The particular PFPs that emanate from  
$\aV^\bullet_\infty$ and $\aV^{SC}_\infty$ are denoted as 
$\aV^\bullet_y$ and  $\aV^{SC}_y$.
The separatrix PFP is denoted as  $\aV^\circ_y$. 
The PFP that connects the two stable asymptotic fixed points acts as an attractor under the original PFP equation
(not the reversed one) because PFPs that deviate slightly away from it at a finite $y$ are attracted to it as $y$ decreases.
Therefore, the pairing interactions that emerge in the low-energy limit are controlled by this PFP.
The topology of the general PFP is completely fixed by 
$\aV^\bullet_y$, 
$\aV^{SC}_y$ and
$\aV^\circ_y$: 
the region above (below) 
$\aV^\circ_y$ 
corresponds to the basin of attraction of 
$\aV^\bullet_y$
($\aV^{SC}_y$) under the IR beta function,
and
the points on $\aV^\circ_y$ stay on it.
In this example, the number and type of asymptotic fixed points and their connectivity are robust against small deformations, such as a change in the spatial dimension and a deformation of the marginal coupling functions (Fermi velocity and shape of the Fermi surface).
Therefore, this general PFP is topologically stable.

The above examples illustrate that the PFP that emanates from the stable or marginal $\infty$-asymptotic fixed point attracts nearby PFPs at least at large $y$. 
Since that PFP governs the universal pairing interaction in the normal state, it is referred to as {\it metallic PFP} and denoted as $\aVM_y$.
On the other hand,
 $\partial {\cal B}^{IR}_{\aV^\bullet_{-\infty}}$
 or
 $\partial {\cal B}^{IR}_{\aV^\halfominus_{-\infty}}$
(the boundary of the basin of attraction of  $\aV^\bullet_{-\infty}$ or $\aV^\halfominus_{-\infty}$)
is the {\it separatrix PFP}, denoted as $\aVS_y$.
In the first example,  $\aVM_y= \aVS_y= \aV^{\halfominus}_y$. 
In the second example, 
$\aVM_y=\aV^\bullet_y$
and 
$\aVS_y=\aV^\circ_y$, which are distinct.

\begin{figure}[th]
\centering
\includegraphics[width=0.4\linewidth]{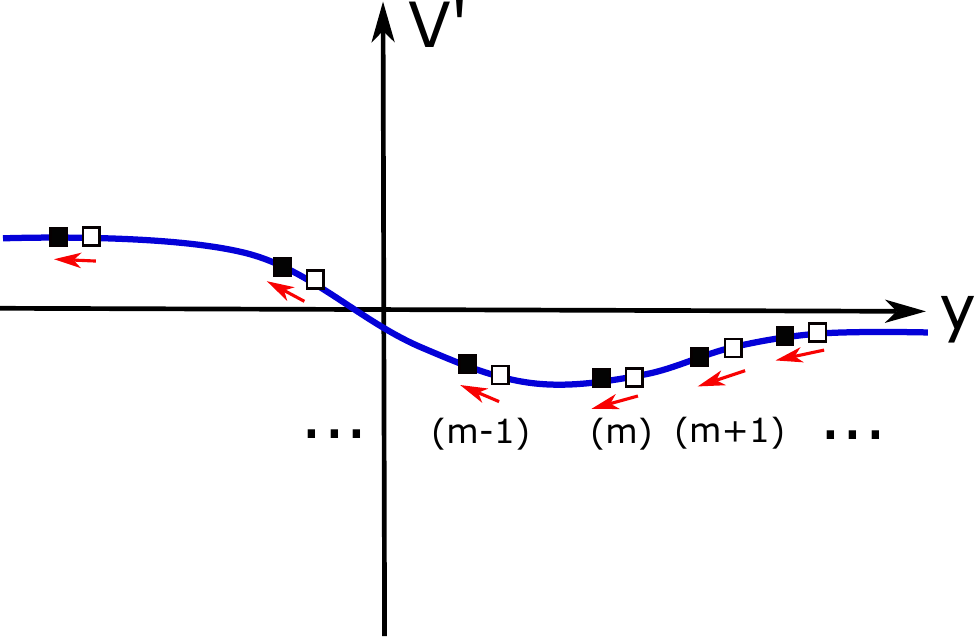}
\caption{
In a metal that supports a general PFP of the form shown in \fig{fig:PFP_nfl1}, 
a stable non-Fermi liquid arises
if the bare pairing interactions lie above the separatrix PFP.
At low energies, the pairing interactions in all angular-momentum channels flow into the universal profile set by the metallic PFP.
For generic non-Fermi liquids, the metallic PFP has a non-trivial $y$-dependence, which reflects the scale-dependent pairing interaction generated by critical fluctuations.
Due to the drift of $y$ under the RG flow, the universal couplings associated with the infinite chain of angular momenta undergo the incessant crossovers within the metallic PFP. 
}
\label{fig:lack_of_scale_invariance}
\end{figure}

In Fermi liquids, the metallic PFP is simply $\aV^M_y=0$.
Therefore, there are no intrinsic superconducting fluctuations in the low-energy limit.
In the second example, the universal pairing interactions that emerge at low energies is controlled by the metallic PFP which has a non-trivial $y$-dependence.
Because the angular momentum is unbounded and $y$ shifts incessantly under the RG flow, there always exist some angular momentum channels that are in the region of crossover from the $\infty$ asymptotic region to the $-\infty$ asymptotic region
(see \fig{fig:lack_of_scale_invariance}).
This incessant crossover prevents the non-Fermi liquid from exhibiting a scale invariant universal pairing interaction, even if there is no instability.

\subsection{Linear stability of PFP}

Since the topology of general PFPs is determined by the metallic and separatrix PFPs, it is useful to understand their linear stability.
The stability of a PFP can be understood in terms of the linearized flow of a small perturbation added to the PFP.
Let 
$\aV_y$ 
be a PFP.
We consider a nearby PFP,
$
\aV^{{}}_y    =
 \aV_y +
\delta\aV^{{}}_y$.
The deviation $\delta \aV_y$  as a function of $y$ obeys
\begin{equation}
 \begin{aligned}
 \frac{1}{2} \partial_y   \delta\aV_y 
+ \chi_y         \delta\aV_y = 0,
\label{eq:linear_beta}
\end{aligned}
\end{equation}
to the linear order in $\delta \aV_y$,
where
\bqa
\chi_y        =
 -\frac{1}{2}\left( 4 R_d\aV_y+2\HD-1\right).
 \label{chiy}
 \eqa
This equation describes how the deviation evolves with decreasing $y$ through
$\delta\aV_{y'} = A_{y';y} \delta\aV_y$,
where
\bqa
A_{y';y}=
\exp{2 \int_{y'}^y dy''  \chi_{y''}
 }.
 \label{eq:Ayl}
 \eqa
Here, the sign of $\chi_y$ determines the stability of a PFP locally at $y$.
For $\chi_y<0$ ($\chi_y>0$),
the nearby PFP approaches (moves away from) 
$\aV_y$ with decreasing $y$.
We say that a PFP is locally stable, unstable, and marginal at $y$ 
if  
$\chi_y < 0$.
$\chi_y > 0$,
and
$\chi_y = 0$,
respectively.
From \eq{chiy}, we note that a PFP is locally 
stable (unstable) 
at $y$ if 
$\aV_y > -\frac{ 2\HD-1}{ 4 R_d }$
($\aV_y < -\frac{ 2\HD-1}{ 4 R_d }$).

\begin{figure}[th]
	\begin{center}
\includegraphics[width=0.5\textwidth]{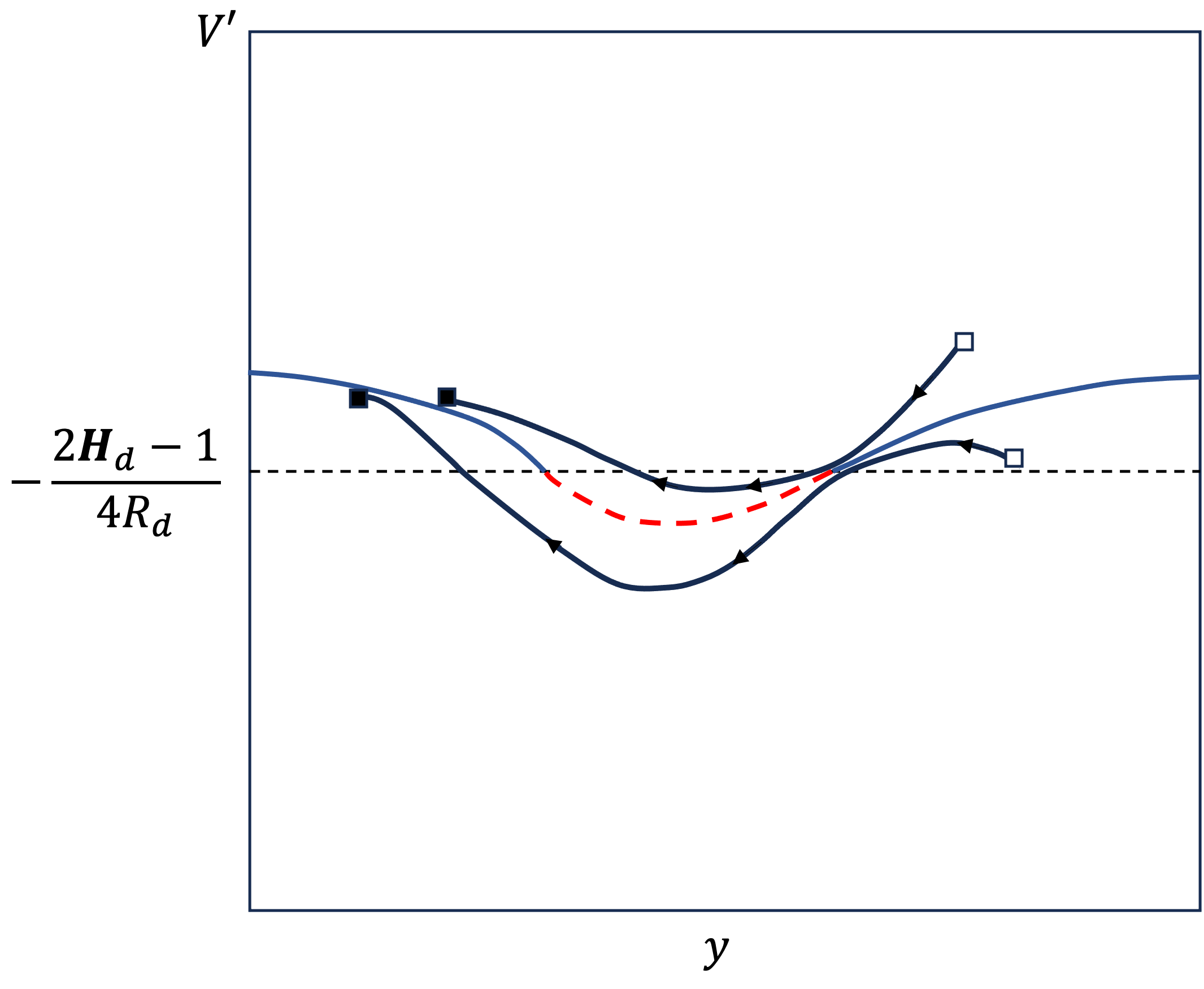}
	\end{center} 
\caption{ 
PFPs are locally stable (unstable) in the region with 
$\aV_y > -\frac{ 2\HD-1}{ 4 R_d }$
($\aV_y < -\frac{ 2\HD-1}{ 4 R_d }$).
A PFP which is locally unstable within finite segments of $y$ is globally stable as long as nearby PFPs are attracted to the PFP outside the finite locally unstable region.
}
\label{fig:global_local_stability}
\end{figure}

Another notion of stability, which turns out to be the more useful one, is global stability.
\begin{itemize}
    \item 
A PFP is globally stable 
if 
for all $y$ and $\epsilon'>0$, there exists $y''$ such that $A_{y';y} < \epsilon'$ for all $y'<y''$.
In this case, PFPs that are close to the PFP at any point in $y$ become arbitrarily close to it at sufficiently small $y$.

\item 
A PFP is marginal if $A_{y';y}$ 
approaches a non-zero finite value in the small $y'$ limit for some $y$. 

\item 
A PFP is unstable if there exists a $y$ for which $A_{y';y}$ becomes arbitrarily large for some $y' < y$.
In this case, a perturbation added to the PFP at $y$ grows without a bound as $y$ approaches $y'$.
\end{itemize}
A PFP that is locally stable at all $y$ is globally stable.
However, the converse is not necessarily true.
$\aV_y$,
which is unstable locally in some regions of $y$,
is globally stable if it satisfies the following conditions:
(a) it is regular, that is, $\aV_y$ is finite for all $y$,
(b) $\chi_y > 0$ only within a finite support  
$X=\cup_i \{ y | y^-_i < y < y^+_i \}$, where $y^\pm_i$ are finite,
(c) there exist positive constants $\delta''$ and $\epsilon''$ such that $\chi_y < -\epsilon''$ in the supplement of $X_{\delta''}
=\cup_i \{ y | y^-_i-\delta'' < y < y^+_i+\delta'' \}$.
$\aV_y$ that satisfies these conditions is globally stable
because a PFP that is near 
$\aV_y$ at a $y$
can diverge away from 
$\aV_y$ with decreasing $y$ 
only in $X$ and
the diverging rate $\chi_y$ is bounded.
In the small $y$ limit, however, the neighboring PFP is destined to be attracted toward $\aV_y$ with $A_{y';y}$ approaching $0$ due to the attraction it experiences over an arbitrarily large range of $y$ outside $X_{\delta''}$. 
This is illustrated in \fig{fig:global_local_stability}.
The metallic PFP 
in \fig{fig:PFP_nfl1}
is globally stable.
This follows from the fact that the presence of asymptotic fixed points in both $y \rightarrow \pm \infty$ limits guarantees that the region $X$ of negative $\chi_y$ is finite, if present.
The separatrix PFP in \fig{fig:PFP_nfl1} is globally unstable as a small perturbation diverges with decreasing $y$.
The separatrix PFP in \ref{fig:PFP_fl}, which is also the metallic PFP,
is marginal.
For marginal PFP, one has to include non-linear effects to understand stability.
In Fermi liquids, the separatrix PFP is marginally stable (unstable) for perturbations with 
$\delta \aV_y>0$
($\delta \aV_y<0$).

\subsection{Classification of superuniversality classes}

\begin{figure}[th]
	\begin{center}
\includegraphics[width=1.0\textwidth]{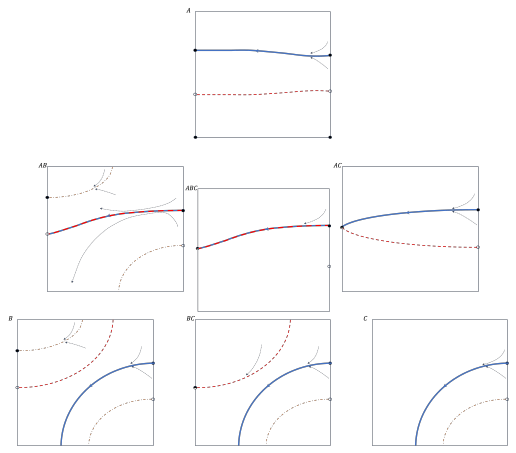}
	\end{center} 
\caption{ 
Portraits of the seven topologically distinct general PFPs, each of which corresponds to a superuniversality class. 
Three at the `corners' (A,B,C) are the topologically stable classes that are robust against small changes of the marginal couplings and kinematic data.
Three at the `edges' (AB, BC, AC) are critical classes that require one fine tuning.
They describe superuniversality phase transition between two stable classes. 
The one at the center (ABC) is the double critical class.
The thick blue line represents the metallic PFP that emanates from  $\aV^\bullet_\infty$.
The thick red dashed line represents the separatrix PFP that forms the boundary of the basin of attraction for the stable or marginal $-\infty$ asymptotic fixed point.
In classes AB and ABC, the metallic PFP and the separatrix PFP coincide.
}
\label{fig:PFP}
\end{figure}

In this section,  we classify all topologically distinct general PFPs, each of which defines one superuniversality class.
Distinct general PFPs are characterized by 
(i) the number and local stability of asymptotic fixed points and 
(ii) their connectivity formed by particular PFPs.
These are {\it topological} data that are insensitive to the geometric shape of PFPs.
We first classify topologically stable classes of PFPs.

\begin{description}
    \item[Class A]
    {\bf Stable non-Fermi liquid superuniversality class} 
    
At $y=\infty$, there are always one stable 
($\aV^\bullet_\infty$)
and one unstable 
($\aV^\circ_\infty$)
asymptotic fixed points  shown in  \eq{eq:plusinfasymfp}.
\footnote{ 
Because $\aS_\infty=0$,
$\eta_{P,\infty}>0$.} 
At $y=-\infty$, there can be $2$, $1$ or $0$ asymptotic fixed points on the real axis of $\aV_{-\infty}$.
Let us first consider the cases with 
two $-\infty$-asymptotic fixed points.
From the quadratic nature of the beta function,
one ($\aV^\bullet_{-\infty}$)
is stable 
and the other ($\aV^\circ_{-\infty}$)
is unstable.
Since $\beta_{\aV}(y)$ in \eq{eq:beta_vertical} is negative for large $|\aV|$,
$\aV^\bullet_{-\infty} > \aV^\circ_{-\infty}$.
The first type of general PFPs is the one in which 
$\aV^\bullet_{\infty} \in 
{\cal B}^{IR}_{\aV^\bullet_{-\infty}}$.
This implies that the metallic PFP ($\aVM_y$), which emanates from  $\aV^\bullet_{\infty}$,
converges to 
$\aV^\bullet_{-\infty}$ in the small $y$ limit.
In this case,  $\aV^\circ_{\infty}$
and 
$\aV^\circ_{-\infty}$
must be connected by a PFP as is discussed below  \eq{eq:basinIR}.
That PFP is the separatrix PFP, $\aVS_y$ which forms 
$\partial {\cal B}^{IR}_{\aV^\bullet_{-\infty}}$.
The metallic and separatrix PFPs give rise to the topology of 
\fig{fig:PFP} (A) for the general PFP,
which is identical to the second example we considered in \fig{fig:PFP_example}.
We refer to the superuniversality class of this general PFP as
{\it stable non-Fermi liquid superuniversality class} (or class A, in short).
As the name suggests, this superuniversality class contains non-Fermi liquids that survive down to zero temperature without fine tuning the bare pairing interaction.

\item[Class B]
{\bf Non-s-wave superconducting superuniversality class} 

Let us proceed with the next class where there are still two $-\infty$-asymptotic fixed points but with a distinct connectivity to the $+\infty$-asymptotic fixed points.
If $\aV^\bullet_{\infty} \notin
{\cal B}^{IR}_{\aV^\bullet_{-\infty}}$,
$\aV^\bullet_{\infty}$ must be {\it generically} in 
${\cal B}^{IR}_{\aV^{SC}_{-\infty}}$ because
$\aV^{SC}_{-\infty}$ is
the only other stable $-\infty$-asymptotic fixed point. 
Therefore, the metallic PFP must reach $\aV=-\infty$.
Alternatively, the metallic PFP can flow into  $\aV^\circ_{-\infty}$, but it requires fine-tuning. 
Such critical classes will be discussed separately.
The particular PFP that emanates from $\aV^\circ_{\infty}$ must also reach $\aV=-\infty$ because it has to stay below the metallic PFP.\footnote{
Distinct PFPs cannot cross due to the uniqueness of the solution of the first-order differential equation.
}
The fates of PFPs that are connected to 
the $-\infty$-asymptotic fixed points 
can be readily understood through the reverse RG flow. 
Since  $\aV^\circ_{-\infty}$ is not in 
$\bar {\cal B}^{UV}_{\aV^\circ_{\infty}}$ under the reversed RG flow\footnote{
If it were, there would be a PFP that connects $\aV^\circ_{\pm\infty}$.
But this is not the case.
},
the PFP that is connected to the former must reach $\aV=\infty$, the only other stable $\infty$-asymptotic fixed point for the reversed RG flow.
The PFP connected to 
$\aV^\bullet_{-\infty}$
must reach  $\aV=\infty$ as well because it should stay above the PFP connected to 
$\aV^\circ_{-\infty}$.
This fixes the topology of this class of general PFP to be that of \fig{fig:PFP} (B).
This class is referred to as 
{\it non-s-wave superconducting superuniversality class} (or class B) because a non-s-wave superconducting instability is unavoidable in this class.

\item[Class C]
{\bf S-wave superconducting superuniversality class}

The final superuniversality class that can be realized without fine-tuning is associated with general PFPs with no $-\infty$-asymptotic fixed points.
In this case, $\discinf < 0$, 
and the vertical beta function is negative definite in the $y \rightarrow -\infty$ limit. 
Therefore, all PFPs that emanate from $y=\infty$ reach $\aV=-\infty$ at some $y$.
The topology of the general PFP in this class is depicted in  
\fig{fig:PFP} (C).
We call this {\it s-wave superconducting superuniversality class} (or class C)
because the s-wave superconducting instability is present in all non-Fermi liquids in this superuniversality class.
\end{description}

The above three cases exhaust all generic superuniversality classes that can be realized without fine tuning of marginal couplings or the kinematic data.
Now, we consider the classes that require fine tuning.
Those critical superuniversality classes can be viewed as transitions between superuniversality classes (not between individual phases).
These superuniversality phase transitions are associated with changes in the topology of the portraits of general PFPs.

\begin{description}

\item[Class AB]
{\bf NFL to non-s-wave SC critical superuniversality class}

The critical superuniversality class that is at the border between classes A and B has two asymptotic fixed points, both at $y=\infty$ and $-\infty$, but they are connected differently from classes A and B.
Namely, $\aV^\bullet_\infty$ is neither in ${\cal B}^{IR}_{\aV^\bullet_{-\infty}}$ 
nor in ${\cal B}^{IR}_{\aV^{SC}_{-\infty}}$.
Instead, $\aV^\bullet_\infty$ is in ${\cal B}^{IR}_{\aV^\circ_{-\infty}}$, as shown in \fig{fig:PFP} (AB).
Because $\aV^\circ_{-\infty}$ is locally unstable, this requires tuning one parameter; hence, it corresponds to a critical superuniversality class. 
In this class, the metallic PFP coincides with the separatrix PFP because the PFP that emanates from $\aV^\bullet_{\infty}$ forms 
$\partial {\cal B}^{IR}_{\aV^\bullet_{-\infty}}$
This class is referred to as  {\it NFL to non-s-wave SC critical superuniversality class} (or class AB).

    \item[Class AC] 
{\bf NFL to s-wave SC critical superuniversality class}

In the critical superuniversality class that arises at the transition from classes A to C,
the $-\infty$-asymptotic fixed points are on the verge of moving off the real axis as they merge into a marginal fixed point at 
$\aV^{\halfominus}_{-\infty}$ while
$\aV^{\bullet}_{\infty} \in {\cal B}^{IR}_{\aV^\halfominus_{-\infty}}$
but 
$\aV^{\bullet}_{\infty} \notin \partial {\cal B}^{IR}_{\aV^\halfominus_{-\infty}}$.
Namely, the metallic PFP converges to the marginal $-\infty$-asymptotic fixed point within the interior of
${\cal B}^{IR}_{\aV^\halfominus_{-\infty}}$.
The separatrix PFP that forms 
$\partial {\cal B}^{IR}_{\aV^\halfominus_{-\infty}}$
connects
$\aV^{\circ}_{\infty}$ 
and
$\aV^{\halfominus}_{-\infty}$.
Points above (below) the separatrix PFP flow toward the metallic (superconducting) PFP in the small $y$ limit.
This class is called {\it NFL to s-wave SC critical superuniversality class} (or class AC), and its general PFP is illustrated in \fig{fig:PFP} (AC).

\item[Class BC]
{\bf S-wave to non-s-wave SC critical superuniversality class}

This critical superuniversality class describes the transition from classes B to C.
In this case, the $-\infty$-asymptotic fixed points collide to form a marginal fixed point,
while the metallic PFP diverges to $-\infty$ at a finite $y$, as is the case in class B.
This case, illustrated in \fig{fig:PFP} (BC), is called  {\it s-wave to non-s-wave SC critical superuniversality class} (or class BC).
\end{description}

Finally, there is a multi-critical superuniversality class where all three generic superuniversality classes converge.

\begin{description}
    \item[Class ABC] {\bf Double-critical superuniversality class}
    
This multi-critical superuniversality class can be reached from one of the critical superuniversality class with one additional fine-tuning.
For example, one can start from class AC 
(NFL to s-wave SC critical superuniversality class)
and tune a second parameter such that the metallic PFP and the separatrix PFP merge 
while maintaining the marginality of 
$\aV^{\halfominus}_{-\infty}$,
as is shown in \fig{fig:PFP} (ABC). 
This is referred to as {\it
double-critical superuniversality class}.
Once the metallic PFP leaves the basin of attraction of  $\aV^{\halfominus}_{-\infty}$, the metallic PFP becomes divergent,
and the general PFP becomes that of class BC (s-wave to non-s-wave critical superuniversality class).
Similarly, this double-critical superuniversality class can be understood as the critical superuniversality class realized in the transition 
from classes AC to AB,
or 
from  classes AB to BC.
The general PFP for the double-critical class is shown in \fig{fig:PFP} (ABC).

\end{description}

The primary goal of the rest of this paper is to understand the universal and superuniversal properties of individual non-Fermi liquids that arise in each class.
First, we do this by focusing on the general topological constraints imposed on non-Fermi liquids in each class.
In Secs. \ref{sec:ex1}- \ref{sec:ex4}, we revisit them more quantitatively through examples.

\subsection{Classification of individual universality classes}
\label{sec:individual_classes}

\begin{figure}[th]
\centering
\begin{subfigure}{.33\textwidth}
\centering
\includegraphics[width=1.0\linewidth]{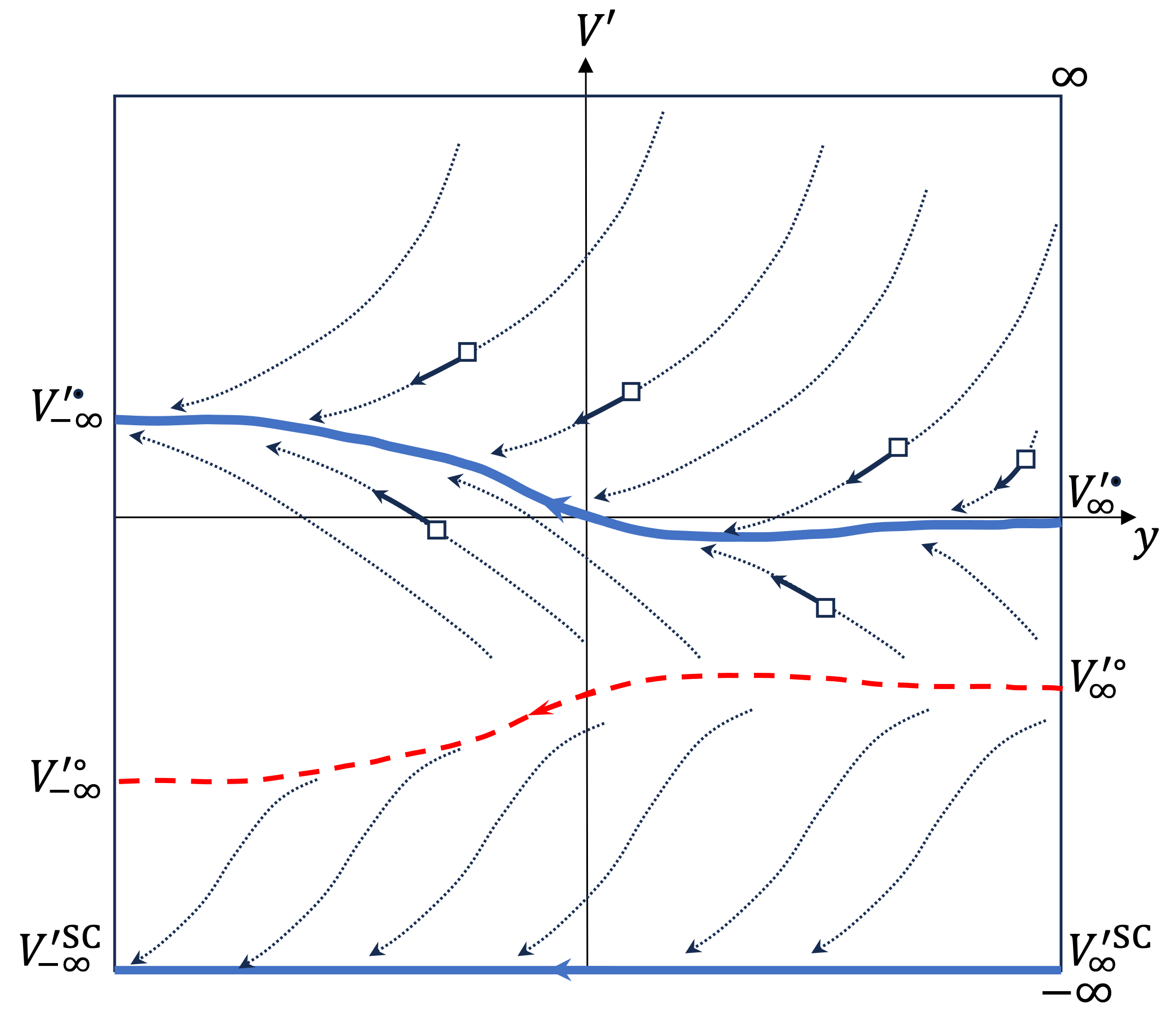}
  \caption{}
  \label{fig:PFP_example_stableNFL}
\end{subfigure}%
\begin{subfigure}{.33\textwidth}
\centering
\includegraphics[width=1.0\linewidth]{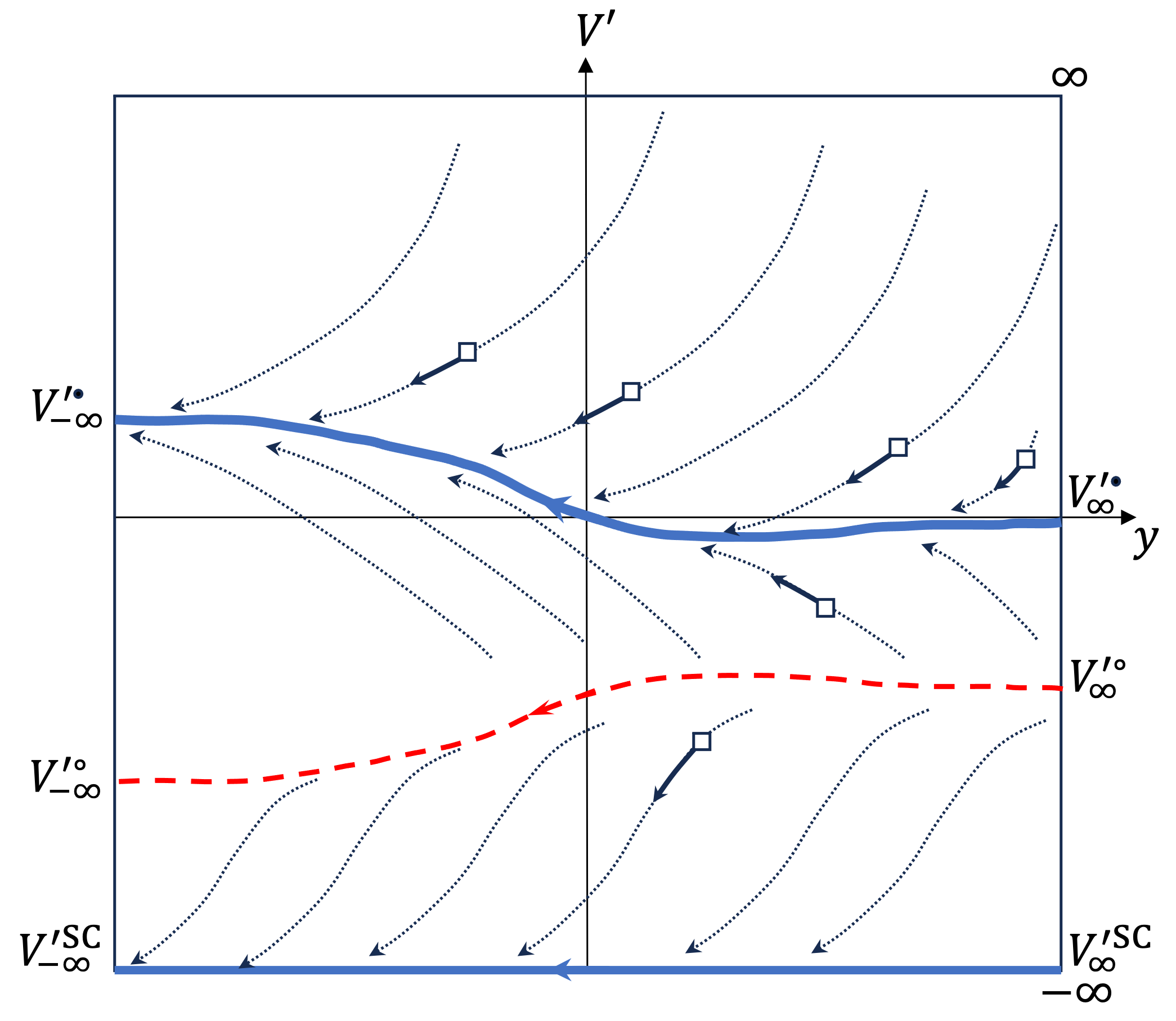}
  \caption{}
  \label{fig:PFP_example_SC}
\end{subfigure}%
\begin{subfigure}{.33\textwidth}
\centering
\includegraphics[width=1.0\linewidth]{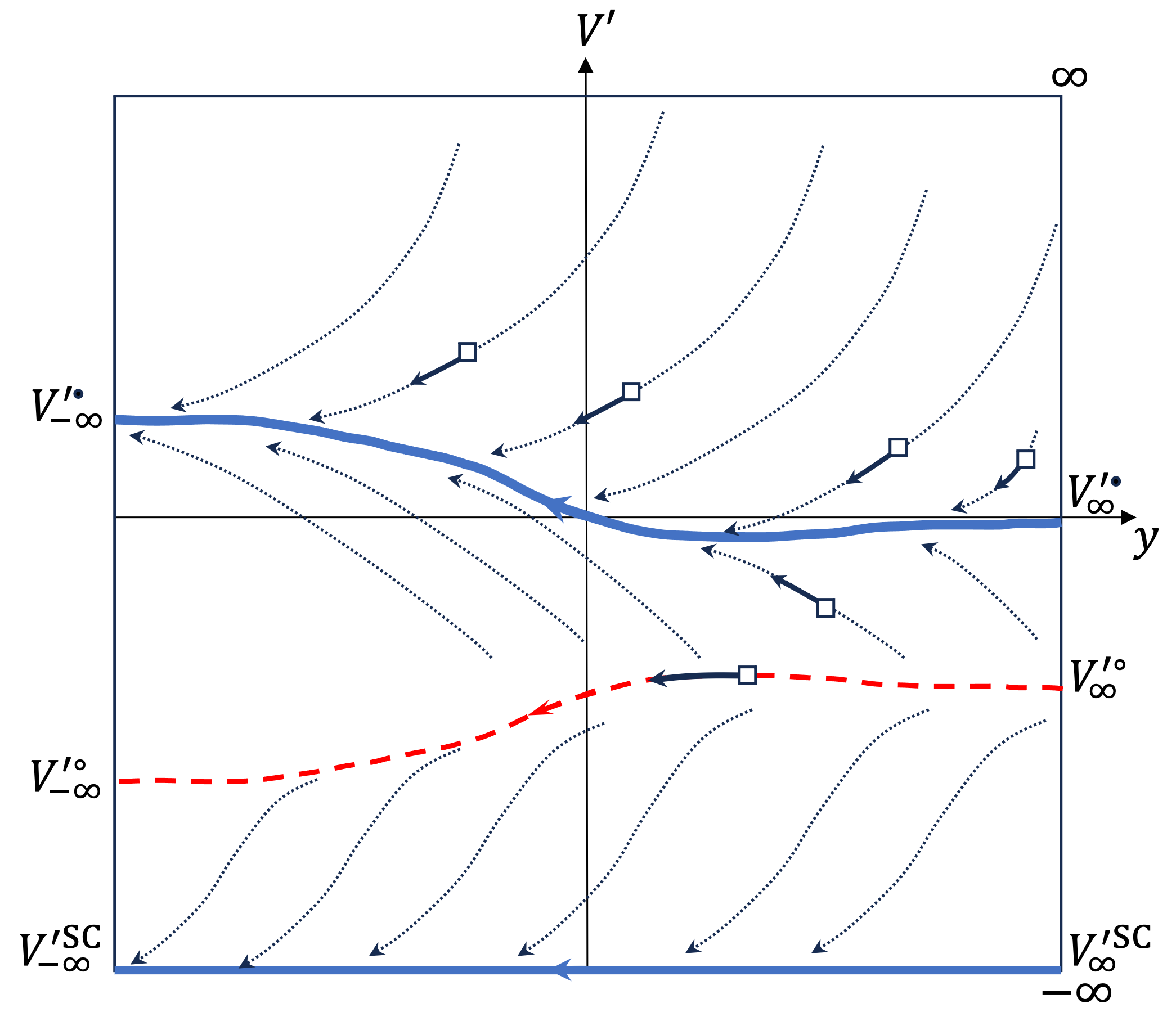}
\caption{}
\label{fig:PFP_example_criticalNFL}
\end{subfigure}
\caption{
Non-Fermi liquid to superconductor quantum phase transition in superuniversality class A.
The similar picture holds for superuniversality classes AB, AC and ABC.
(a) If all bare couplings denoted by open squares are above the separatrix PFP (dashed line), they are attracted to the metallic PFP (solid line above the separatrix PFP) at low energies, realizing the stable non-Fermi liquid.
(b) If one of the couplings is outside the basin of attraction of the metallic PFP, the ground state is the superconducting state.
(c) At the critical point, one of the couplings is on the separatrix PFP, realizing a critical non-Fermi liquid differentiated from the parent non-Fermi liquid by the critical pairing interaction.
}
\label{fig:PFP_example_NFL}
\end{figure}

Within each superuniversality class, there are multiple universality classes that are differentiated by their dynamical properties, such as superconducting fluctuations/instabilities and emergent symmetries.
Here, we discuss individual universality classes that arise from each superuniversality class and explain their key properties in general terms.

\begin{itemize}

\item Stable non-Fermi liquids

The superuniversality classes A, AB, AC and ABC support regular metallic PFPs:
they are non-divergent and extended from $y=\infty$ to $-\infty$. 
Furthermore, there exists a separatrix PFP.
In classes A and AC, the separatrix PFP is distinct from the metallic PFP, but in classes AB and ABC, the two coincide.
If the bare couplings in all angular momentum channels are above the separatrix PFP, the metallic states survive to zero temperature.
These non-Fermi liquids are stable against small perturbations of the bare four-fermion couplings because the region above the separatrix is extended in the $(y,\aV)$ plane.
The stable non-Fermi liquids that arise in different superuniversality classes are differentiated from each other in terms of their universal pairing interactions that emerge in the low-energy limit. 
In classes A, AC and ABC, the metallic PFP is the attractor for the couplings above the separatrix PFP (for example, see \fig{fig:PFP_example_stableNFL} for class A).
Consequently, the universal pairing interactions are entirely fixed by the metallic PFP in these classes.

\begin{figure}[th]
    \centering
\includegraphics[width=0.5\linewidth]{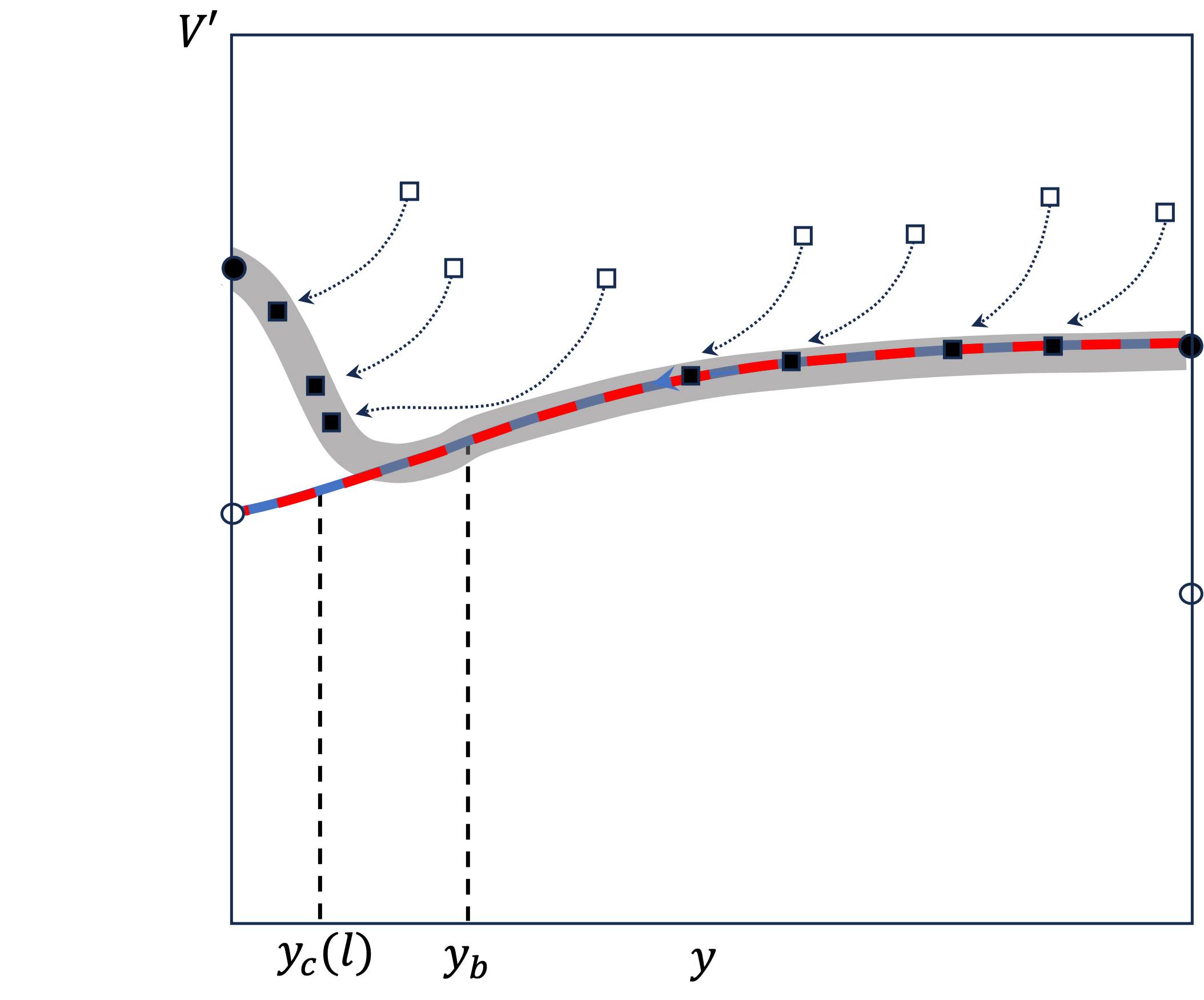}
\caption{
In class AB, the metallic PFP, which is the same as the separatrix PFP, 
is locally stable (unstable) for $y>y_b$ $(y<y_b)$.
At a large but finite $l$,
the couplings that are initially above the separatrix PFP end up being close to the metallic PFP
($\aV^\bullet_{-\infty}$) at large (small) angular momentum channels. 
Consequently, the quasi-universal pairing interaction that emerges at large $l$, which is denoted as the thick grey line, interpolates the metallic PFP to $\aV^\bullet_{-\infty}$  across an $l$-dependent crossover angular momentum scale $y_c(l)$.
}
\label{fig:universalVstableNFLclassB}
\end{figure}

The stable non-Fermi liquid realized in superuniversality class AB is unique in that the metallic PFP is connected to $\aV^\circ_{-\infty}$ whose basin of attraction is only the separatrix itself.
While couplings at large angular momentum channels are attracted to the metallic PFP at intermediate energy scales, they eventually crossover to 
$\aV^\bullet_{-\infty}$ in the low-energy limit.
Consequently, the pairing interaction that emerges at low energies from a generic choice of bare couplings interpolates the metallic PFP at large $y$ to $\aV^\bullet_{-\infty}$ at small $y$, as shown in \fig{fig:universalVstableNFLclassB}.
Due to this crossover, which becomes sharper with lowering energy, the universal pairing interaction that arises in class AB exhibits a stronger $y$-dependence.
This will be discussed in greater detail in Sec. \ref{sec:ex4}.

\item Critical non-Fermi liquids 

In each superuniversality class (A, AB, AC, ABC) 
that supports the stable non-Fermi liquid, there are infinitely many critical non-Fermi liquid universality classes that can be obtained by tuning irrelevant couplings such as the four-fermion or higher-order couplings. 
Let us first consider the four-fermion coupling.
If the coupling in one angular momentum channel is in the basin of attraction of the superconducting PFP, a superconducting instability occurs in that channel 
(\fig{fig:PFP_example_SC}). 
A `critical' non-Fermi liquid that describes a phase transition between the stable non-Fermi liquid and the superconducting state has the coupling in one angular momentum channel on the separatrix PFP with all other couplings above the separatrix PFP.
These are illustrated in
\fig{fig:PFP_example_criticalNFL} for class A.
This critical non-Fermi liquid has one relevant direction in the space of four-fermion coupling.
The non-Fermi liquid to superconductor phase transition can be induced either by tuning the strength of the coupling or by tuning the Fermi momentum.
The former and latter change the vertical and horizontal location of the bare coupling in the $(y,\aV)$ plane, respectively.
Critical non-Fermi liquids that arise from distinct superuniversality classes exhibit different critical exponents and represent distinct universality classes.
There also exist multi-critical non-Fermi liquids where the pairing interactions in multiple angular momentum channels are placed on the separatrix through fine tuning.
More generally, one can crank up a higher-order fermion coupling, such as the eight-fermion coupling.
They can drive a phase transition from the stable non-Fermi liquid to a superconductor with charge $4$ or higher.
Generally, those critical points represent yet another distinct non-Fermi liquids.

\item{Superconductors}

The superuniversality classes B, C and BC do not have regular metallic PFP. 
This implies that there exist angular momentum channels for which the pairing interaction flows to $-\infty$ at low energies, irrespective of the choice of the bare coupling.
Therefore, all non-Fermi liquids in these superuniversality classes become superconductors in the low-energy limit.
The nature of the superconducting states realized at low temperatures is largely determined by the superuniversality class if there is a large separation of scale between $\Lambda$ and the superconducting transition temperature.
Below, we discuss the universal features of the superconducting state in such cases.

\begin{itemize}

\item Superconductors in class C

In this class, the universal pairing interaction is strong enough that incoherence of fermions can not prevent the pairing instability in the s-wave channel 
even if instabilities in non-s-wave channels can be avoided for some choices of bare four-fermion couplings.
While the superconducting transition temperature $T_c$ depends on the bare four-fermion coupling, 
the minimum s-wave superconducting transition temperature is fixed by the universal data of the superuniversality class.
\\

\item Superconductors in class B

\begin{figure}[th]
\centering
\begin{subfigure}{.45\textwidth}
\centering
\includegraphics[width=1.0\linewidth]{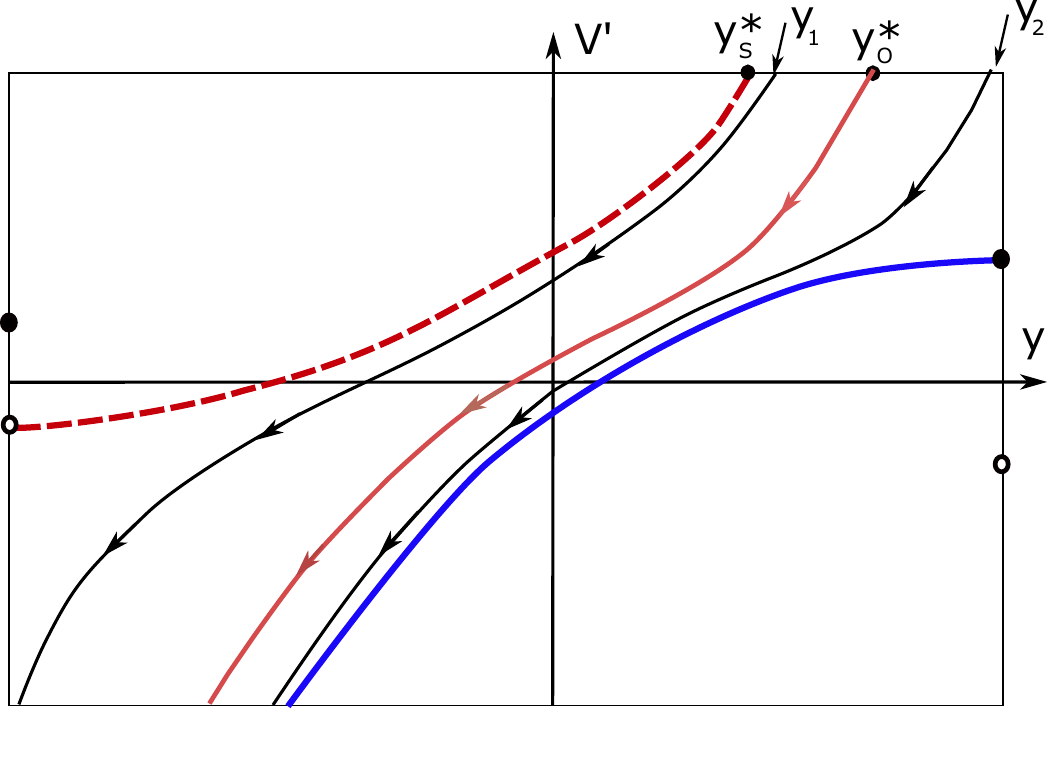}
  \caption{}
  \label{fig:classBdetails}
\end{subfigure}%
\hspace{1cm}
\begin{subfigure}{.45\textwidth}
\centering
\includegraphics[width=1.0\linewidth]{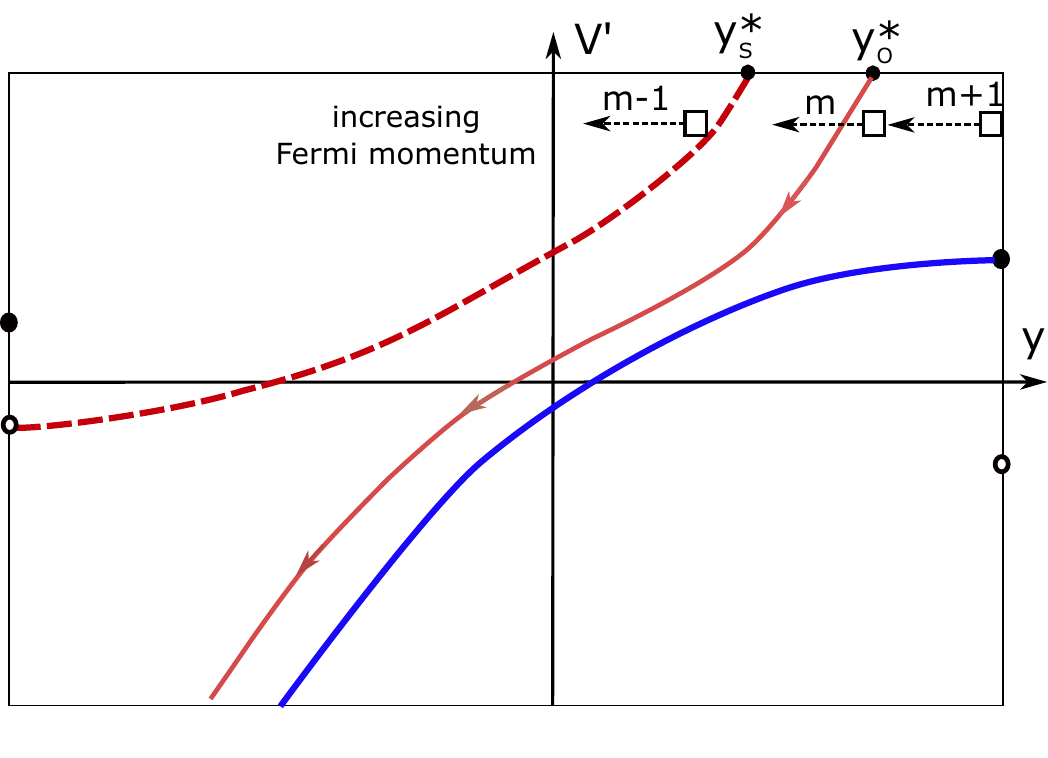}
\caption{}
\label{fig:whyTc_oscillate_classB}
\end{subfigure}
\caption{
The metallic PFP (thick blue solid line) 
and the separatrix PFP (red dashed line) in superuniversality class B.
The separatrix (metallic) PFP diverges to $\infty$ ($-\infty$) at $y^*_S$ ($y^*_M$).
(a) 
Consider the case where all bare couplings are repulsive and large.
In this case, there is no superconducting instability for the angular momentum channels with $y^{(m)}(0) \leq y^*_S$ because they flow to 
$\aV^\bullet_{-\infty}$ 
or
$\aV^\circ_{-\infty}$ 
in the low-energy limit.
Among the angular momentum channels with 
$y^{(n)}(0) > y^*_S$, the superconductivity arises in the channel whose 
$y^{(n)}(0)$ is closest to the optimal $y$, denoted as $y_O^*$.
The existence of $y_O^*$ can be understood as follows.
If $y^{(n)}(0)$ is too close to $y_S^*$ ($y_1$ in the figure), it tracks the separatrix PFP for a large RG time before it diverges.
If $y^{(n)}(0)$ is too large ($y_2$ in the figure), it is first attracted to the metallic PFP and diverges near $y^*_M$, which also makes $l_{SC} \sim 2(y_2-y^*_M)$ large.
Because $l_{SC}$ diverges in both limits, there should exist $y^*_O$ between $y^*_S$ and $\infty$ at which $l_{SC}$ is minimized.
(b) 
At a given $\KFAVdim$, the channel whose $y^{(n)}(0)$ is closest to $y^*_O$ 
(denoted as $m$ in the figure) 
determines $T_c$.
As $\KFAVdim$ is tuned with the bare coupling fixed, the relative position of $y^{(m)}(0)$ to $y^*_O$ changes, creating a local maximum of $T_c$ at a specific choice of $\KFAVdim$ that satisfies $y^{(m)}(0)=y^*_O$.
If $\KFAVdim$ is increased, the bare couplings gradually shifts to left in the $(y,\aV)$ plane until 
$y^{(m+1)}(0)$ becomes closer to $y_O^*$
than $y^{(m)}(0)$ beyond a certain critical value of $\KFAVdim$.
This causes the jump of the angular momentum channel of superconducting instability from $m$ to $m+1$ and an oscillation of $T_c$ as a function of $\KFAVdim$,
as is shown in \fig{fig:Tc_oscillation_schematic}.
}
\label{fig:classB}
\end{figure}

In class B, a superconducting instability is inevitable above a non-zero critical angular momentum $y^*_S$, while the instability is absent below the critical angular momentum for sufficiently repulsive bare couplings.
In $y^{(m)}(0)  < y^*_S$, which includes the s-wave channel, there is an extended basin of attraction that flows to $\aV^\bullet_{-\infty}$ under the RG flow.
If the bare couplings lie in 
${\cal B}^{IR}_{V^\bullet_{-\infty}}$, 
they flows to $\aV^\bullet_{-\infty}$ at low energies without causing an instability. 
The boundary of ${\cal B}^{IR}_{V^\bullet_{-\infty}}$ is the separatrix PFP connected to $\aV^\circ_{-\infty}$.
The separatrix PFP diverges to $\infty$ at $y_S^*$ under the reverse RG flow.\footnote{If it does not diverge, it should be connected to 
$\aV^\bullet_\infty$
or
$\aV^\circ_\infty$, 
which results in either class AB or A.
}
Since all points in the $(y,\aV)$ plane with $y > y_S^*$ lie outside ${\cal B}^{IR}_{\aV^\bullet_{-\infty}}$, angular momentum channels with $y^{(m)}(0) > y_S^*$ cannot escape superconducting instabilities, no matter how large and repulsive the bare couplings are.
In the limit that the bare couplings are large and repulsive, the angular momentum channel that exhibits the smallest $l_{SC}$, the length scale for superconducting instability, is given by the channel whose $y^{(m)}(0)$ is closest to an optimal $y$, denoted as $y^*_O$.
The optimal angular momentum exists between $y^*_S$ and $\infty$
because $l_{SC}$ diverges as $y^{(m)}(0)$ approaches either $y^*_S$ or $\infty$
(see \fig{fig:classBdetails}).
If $\KFAVdim$ is increased, 
$\{ y^{(n)}(0) \}$ shifts to the left, and $m$ for which $y^{(m)}(0)$ is closest to $y^*_O$ jumps from one angular momentum to the next at discrete values of $\KFAVdim$.
This results in a stepwise increase of $m$ for superconducting instability and an oscillation of $T_c$ as a function of $\KFAVdim$, as is shown in  \fig{fig:Tc_oscillation_schematic}.
See \fig{fig:whyTc_oscillate_classB} for a more detailed illustration.

\begin{figure}[th]
\centering
\begin{subfigure}{.45\textwidth}
\centering
\includegraphics[width=1.0\linewidth]{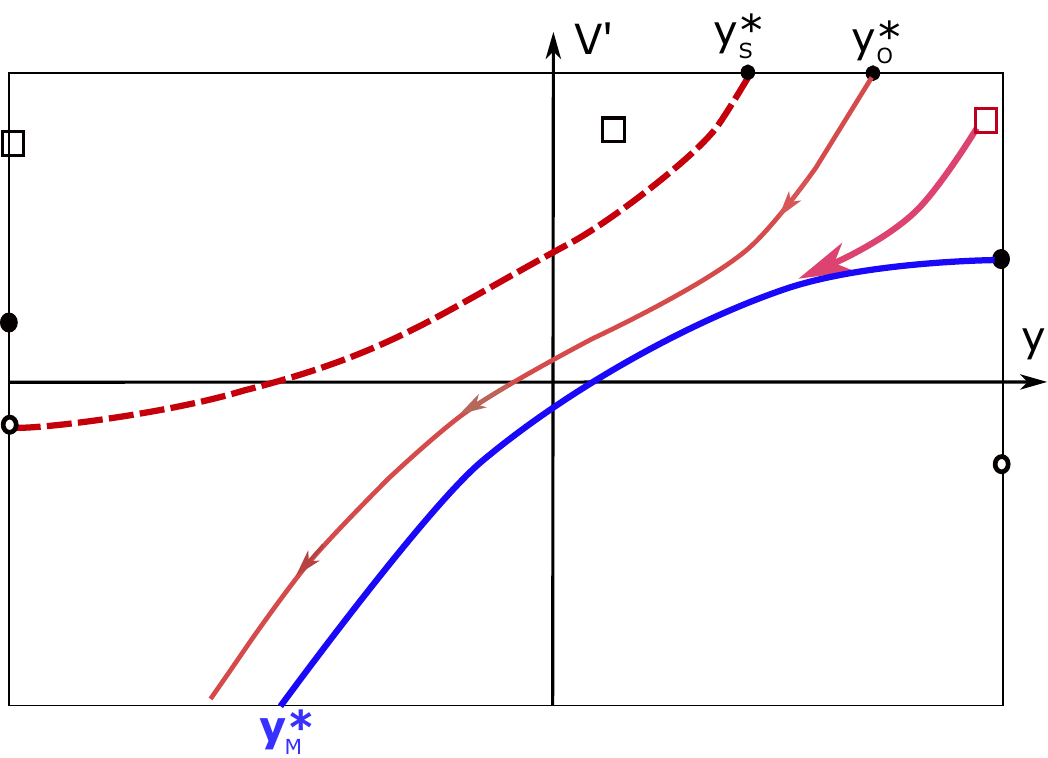}
\caption{}
\label{fig:smallkF}
\end{subfigure}
\hspace{1cm}
\begin{subfigure}{.45\textwidth}
\centering
\includegraphics[width=1.0\linewidth]{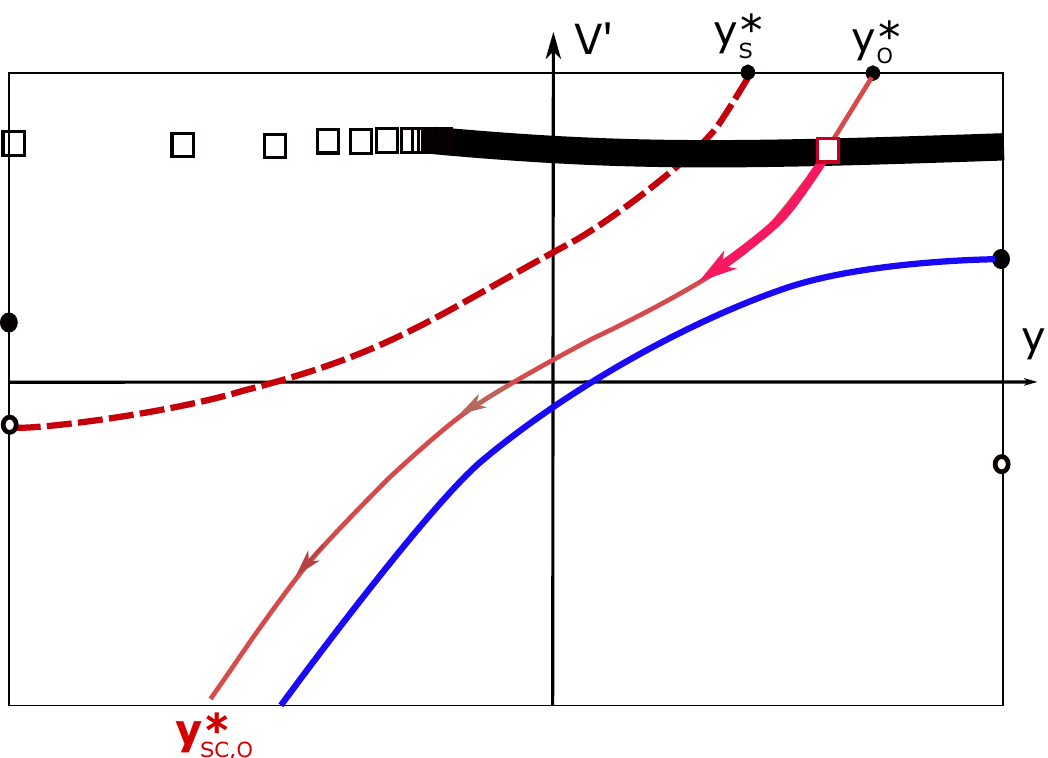}
  \caption{}
  \label{fig:largekF}
\end{subfigure}%
\caption{
(a) For a small $n_O$ in \eq{eq:nO}, 
the set of  $\left( y^{(m)}(0), 
\aV_{y^{(m)}(0)}(0) 
\right) $ 
associated with the bare couplings is spare in the $y$ axis around $y^*_O$.
Consequently, the angular momentum channel closest to $y^*_O$ is generically well separated from $y^*_O$. 
If the bare coupling is not far from the metallic PFP, it is first attracted to the metallic PFP before it diverges near $y^*_M$.
(b) In the large $n_O$ limit,
the bare couplings become dense in the $y$ axis near $y^*_O$.
If the bare coupling is large and weakly dependent on the angular momentum, the superconducting instability occurs in the channel with $y^{(m)}(0) \approx y^*_O$.
}
\label{fig:largekF_smallkF}
\end{figure}

The superconducting transition temperature is not universal because it depends on the bare coupling.
However, $T_c/\KFAVdim^z$,
where $z$ is the dynamical critical exponent, 
becomes insensitive to the bare coupling in certain limits.
Such universality arises when the set of discrete bare couplings composed of 
$\left(
y^{(m)}(0),
\aV_{y^{(m)}(0)}(0) \right)$ 
becomes dense or sparse near $y_S^*$.
The angular momentum that corresponds to $y_O^*$ is given by 
\bqa
n_O = \frac{2 \gamma \sqrt{\KFAVdim}}{\pi \sqrt{\Lambda}} e^{y_O^*}
\label{eq:nO}
\eqa
from \eq{eq:y}.
Near this angular momentum, the spacing between bare couplings in the y-axis is
$y^{(n_O+1)}(0) -y^{(n_O)}(0) = \log (n_O+1)- \log n_O$.
For a small $n_O$, 
$y^{(n)}(0)$ becomes sparse near $y^*_O$, and the distance between 
$y^{(n)}(0)$ with $y^{(n)}(0) > y^*_S$
and $y^*_O$ is generically not small, as illustrated in \fig{fig:smallkF}.
As a result, there is some RG time during which the coupling is attracted to the metallic PFP before it diverges at the superconducting scale.
In such cases, $y^{(n)}(0) - l_{SC}/2$ is close to $y_M^*$
when the bare coupling is not too far from the metallic PFP.
Since $y^{(n)}(0) \sim
\log \frac{n\sqrt{\Lambda}}{\sqrt{\KFAVdim}}$ 
and 
$T_c \sim e^{-z l_{SC}}$,
the ratio between 
$T_{c,n}$ and $\KFAVdim^z$ becomes
\bqa
\frac{\Lambda^{z-1} T_{c,n}}{\KFAVdim^z} \approx \bigg(\frac{2\gamma}{n\pi} \bigg)^{2z} 
e^{2z y_M^*}.
\label{eq:TckFyO}
\eqa

In the large $n_O$ limit, on the other hand,
$y^{(n_O+1)}(0) -y^{(n_O)}(0) \sim 1/n_O$,
which causes $y^{(n)}(0)$ to be densely populated near $y^*_O$, as shown in \fig{fig:largekF}. 
If the bare coupling is strongly repulsive and weakly depends on $y$, 
there exists an angular momentum channel whose bare coupling is very close to the PFP that emanates from 
$(y_O^*, \infty)$.
In that case,  superconducting instability occurs in channel $n$ with 
$y^{(n)}(0) \approx y_O^*$
and the superconducting scale $l_{SC}$ satisfies
$y^{(n)}(0) - l_{SC}/2 \approx y_{SC,O}^*$, where 
$y_{SC,O}^*$ is the location at which the PFP from
$(y_O^*, \infty)$ diverges to $-\infty$.
In this case, $T_c/\KFAVdim^z$ is given by 
\eq{eq:TckFyO}
with $y_M^*$ replaced with
$y_{SC,O}^*$.
In Sec. \ref{sec:ex4}, we will discuss this in detail, including the corrections that depend on the bare coupling.
\\

\item Superconductors in class BC

The superconducting state in this class is similar to that of class B in that the non-s-wave superconducting instability is inevitable.
The universal $T_c/\KFAVdim^z$ relation and the oscillatory behavior of $T_c$ vs $\KFAVdim$ hold similarly in this class.
The main difference is that the pairing interactions at low angular momenta, including the s-wave channel, approach their universal values more `slowly' with lowering energy scale due to the marginality of the $-\infty$-asymptotic fixed point to which the couplings are attracted from above in the low-angular momentum channels.

\end{itemize}

\item Quasi-universal non-Fermi liquids

In superuniversality classes B, C and BC, the normal state is not stable at zero temperature.
Because the RG flow is cut off by the superconductivity at low energies, one can't have a true universal behavior in the non-Fermi liquid states realized above $T_c$.
Nonetheless, the non-Fermi liquids exhibit quasi-universal behavior if $\Lambda/T_c \gg 1$, where $\Lambda$ is the energy scale below which the non-Fermi liquid physics sets in.
This quasi-universal behavior is governed by a PFP, which is referred to as the regularized metallic PFP, toward which couplings are attracted at intermediate energy scales between $\Lambda$ and $T_c$. 
The regularized metallic PFP is identical to the metallic PFP in the range of $y$ in which the metallic PFP is locally stable and attractive.
In the region of $y$ where the metallic PFP is locally unstable or absent, the coupling function is attracted to the profile for which the RG flow is stalled over a finite but large window of length scale.

\end{itemize}

As discussed above, there are three large families of non-Fermi liquids.
The first is the family of stable non-Fermi liquids realized in classes A, AB, AC and ABC.
The second is composed of critical non-Fermi liquids obtained by tuning some irrelevant couplings to critical values in the stable non-Fermi liquids.
These two families are stable down to zero temperature.
The third includes the quasi-universal non-Fermi liquids that arise in the intermediate energy scales above non-zero superconducting transition temperatures in classes B, BC and C.
In the stable non-Fermi liquids, the scattering of Cooper pairs decays fast enough in the angle around the Fermi surface that the inter-patch coupling is negligible. 
Due to the suppression of the large-angle scatterings, these non-Fermi liquids possess the same emergent symmetry as Fermi liquids.
On the other hand, the latter two families invariably exhibit a slow decay of large-angle scatterings of Cooper pairs. 
Due to the strong inter-patch couplings, the emergent symmetry is lower than that of Fermi liquids.
To understand this, we need more dynamical information, which will be discussed in detail through examples in Sec. \ref{sec:ex1}-\ref{sec:ex4}.
In the next section, we discuss some subtleties that arise in determining the emergent symmetry of metals.

\subsection{Emergent symmetries of metals}
\label{sec:symmetry}

In Fermi liquids, the number of particles is independently conserved at `every' angle in the low-energy limit due to the lack of non-forward scatterings\cite{2005cond.mat..5529H}.
Consequently, the U(1) group for global number conservation is promoted to the loop U(1) (LU(1)) group, under which the fermion field at angle $\theta$ is transformed as
\bqa
T_\zeta : 
\psi(\theta) \rightarrow e^{i \zeta(\theta)} \psi(\theta), 
\eqa
where $\zeta(\theta)$ is a general smooth function of angle.
For simplicity, we focus on $d=2$ in this section.
The emergent symmetry of non-Fermi liquids is less clear because of the presence of universal non-forward interactions, such as the pairing interaction.
The emergent symmetry depends on the type of non-Fermi liquids: 
some non-Fermi liquids have the full LU(1) group\cite{PhysRevX.11.021005},  but others only retain a proper subgroup of the LU(1) group.
In this section, we discuss a subtlety that arises from the projective nature of metallic fixed points when determining the emergent symmetry.

\newcommand{\SK}{SingK^{(n)}}

While microscopic symmetry is a property of the Hamiltonian, the emergent symmetry is a property of low-energy observables.
Roughly speaking, a symmetry emerges if low-energy observables are invariant under the corresponding transformation, even if it is not a symmetry of the microscopic Hamiltonian.
As an example of low-energy observables, let us consider the $2n$-point fermion vertex function,
\bqa
\Gamma &=& 
\sum_n
\int 
\left[ \prod_{i=1}^{2n}
d\omega_i d\vec k_i \right]
\Gamma^{(2n)}\left( \{\omega_i\},\{\vec k_i\} \right)
\delta\left( \sum_{i=1}^n \omega_i  - \sum_{i=n+1}^{2n} \omega_i \right) 
\delta\left( \sum_{i=1}^n \vec k_i  - \sum_{i=n+1}^{2n} \vec k_i \right)  \times \nn
&& 
\psi^\dagger(\omega_1,\vec k_1)
\psi^\dagger(\omega_2,\vec k_2)
..
\psi^\dagger(\omega_n,\vec k_n)
\psi(\omega_{n+1},\vec k_{n+1})
\psi(\omega_{n+2},\vec k_{n+2})
..
\psi(\omega_{2n},\vec k_{2n})
\eqa
measured at low frequencies.
In general, one can choose any momenta $\{ \vec k_i \}$, but
for low-energy observables, 
we focus on the set of momenta at which the vertex function is non-analytic as a function of frequencies in the $\omega_i \rightarrow 0$ limit,
$\SK=
\left\{ \{\vec k_i \} 
\Bigl|
\Gamma^{(n)}\left( 
\{\omega_i\},\{ \vec k_i\} \right) 
\mbox{ is non-analytic
at} ~~ \omega_i=0
\Bigr.
\right\}$.
Momenta outside $\SK$ only contribute to the short-time (high-energy) dynamics and can be ignored in the long-time (low-energy) limit.
In metal, $\SK$ corresponds to the set of Fermi momenta, $\{ \KFthetadim | 0 \leq \theta < 2\pi \}$.
A necessary condition for
$T_\zeta$ to be an emergent symmetry is that
the vertex function is invariant under $T_\zeta$ 
in the low-frequency limit 
at all external momenta on the Fermi surface.
In Fermi liquids, only the forward scatterings survive in the low-energy limit, and the LU(1) indeed emerges as a low-energy symmetry.

In non-Fermi liquids, in addition to the forward scattering, there exist non-zero universal pairing interactions generated from the critical fluctuations. 
Because the forward scattering does not change the symmetry, even in non-Fermi liquids, we focus here on the pairing channel.
The universal pairing interaction is encoded in the coupling function via the renormalization condition,
\bqa
\lambda^{(P)}_{\theta_1,\theta_2}(\mu) 
\sim
\mu^{d-1}
\Gamma^{(4)}\left( 
\{\omega_i\},
\{
{\bf K}_{F,\theta_2},
{\bf K}_{F,\theta_2+\pi},
{\bf K}_{F,\theta_1+\pi},
{\bf K}_{F,\theta_1}
\} 
\right)_{\omega_i = \alpha_i \mu},
\eqa
where $\alpha_i$ are RG scheme-dependent numbers that are $O(1)$.
The precise relation between the coupling function and the vertex function depends on the RG scheme encoded in the choice of $\alpha_i$.
See Appendix \ref{app:RGscheme}
for the scheme used in this paper.
The factor of $\mu^{d-1}$ is introduced to make the left-hand side a dimensionless quantity.
The one-particle irreducible vertex function 
$\lambda^{(P)}_{\theta_1,\theta_2}$ 
is related to the net two-body interaction 
$\aV_{\bar \theta_1, \bar \theta_2}$ through 
\bqa
\aV^\pm_{\bar \theta_1, \bar \theta_2}
=
\sqrt{ \mu a^\prime(\bar{\theta}_1) a^\prime(\bar{\theta}_2)}
\sqrt{\frac{K_{F,\theta_1}K_{F,\theta_2}}{v_{F,\theta_1}v_{F,\theta_2}}}
\left[
\lambda^{(P,d)}_{\theta_1,\theta_2}(\vec q=0)\pm \lambda^{(P,e)}_{\theta_1,\theta_2}(\vec q=0)
\right]
+    
\frac{\bar{h}^{\pm}_{d;\bar{\theta}_1,\bar{\theta}_2}}{4},
\eqa
where the net two-body interaction is written as a function of the proper angle $\bar \theta$,
which is in one-to-one correspondence with the original angle $\theta$ (see \eq{eq:theta_bartheta}).
Since physical observables such as the anomalous dimension generated from the four-fermion coupling and the critical boson is determined from $\aV^\pm$, 
it is convenient to use $\aV^\pm_{\bar \theta_1, \bar \theta_2}$
to determine the emergent symmetry.
Under $T_\zeta$, 
$\aV^\pm_{\bar \theta_1, \bar \theta_2}(\mu)$
is transformed into
\bqa
[T_\zeta\aV^\pm]_{\bar \theta_1,\bar \theta_2}(\mu) =
\exp\left\{-i\left[\zeta(\theta_1) + \zeta(\theta_1+\pi) - \zeta(\theta_2) - \zeta(\theta_2+\pi)\right]\right\} 
\aV^\pm_{\bar \theta_1, \bar \theta_2}(\mu).
\eqa
To be concrete, let us consider an example of the form,
\bqa
\aV^\pm_{\bar \theta + \Delta \bar \theta, \bar \theta}(\mu) \sim
\left\{
\begin{array}{cc}
1, 
& \mbox{for} ~~~
|\Delta \bar \theta| \ll 
\sqrt{\mu}, 
\\
\left|
\frac{\sqrt{\mu}}{\Delta  \bar \theta}
\right|^\Delta 
& \mbox{for} ~~~
\sqrt{\mu} \ll |\Delta \bar \theta| \ll 1
\end{array} 
\right.
\label{eq:lambdaa}
\eqa
with $\Delta>0$.
In Secs. \ref{sec:ex1} - \ref{sec:ex4}, we will see that this type of universal pairing interaction arises in physical examples, 
where $\Delta$ is the universal exponent specific to each universality class.
In \eq{eq:lambdaa}, the pairing interaction vanishes in the small $\mu$ limit for any fixed $\theta_1 \neq \theta_2$.
As a result,
\bqa
\aV^\pm_{\bar \theta_1, \bar \theta_2}(\mu)
-
[T_\zeta\aV^\pm]_{\bar \theta_1,\bar \theta_2}(\mu)
\label{eq:difflambda}
\eqa
goes to zero for any smooth $\zeta(\theta)$ 
in the small $\mu$ limit.
From this, one may conclude that the full LU(1) symmetry emerges in the low-energy limit as long as $\Delta>0$. 
However, this is not necessarily true
because there can be other low-energy observables that are not invariant under the general LU(1) transformation for some $\Delta>0$.
This subtlety arises because metals have infinitely many low-energy observables, and observables that probe the entire Fermi surface are more sensitive than others to large-angle scatterings.

\begin{figure}[th]
\centering
\begin{subfigure}{.4\textwidth}
  \centering
\includegraphics[width=0.5\linewidth]{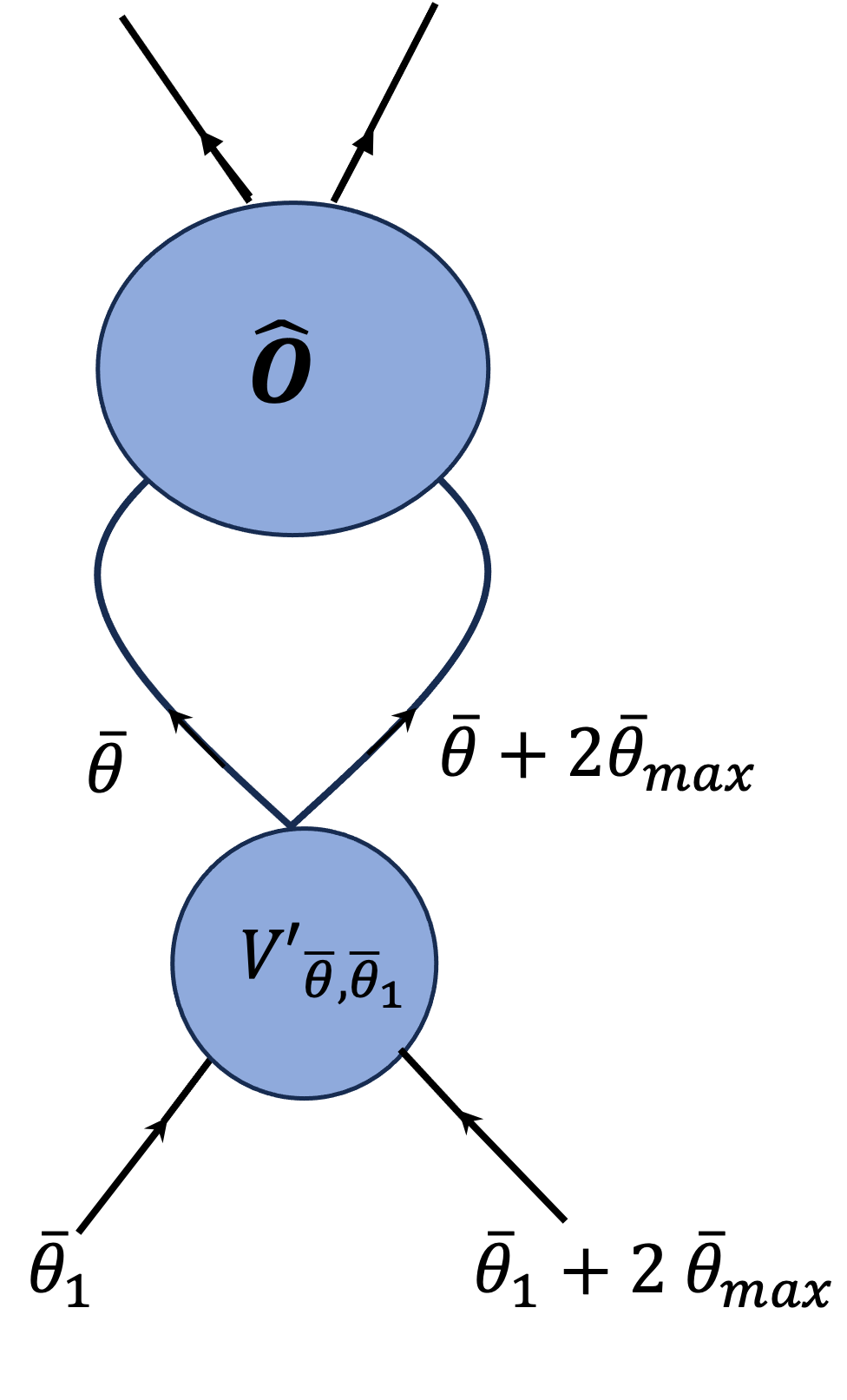}
  \caption{
}
 \label{fig:}
\end{subfigure}%
\begin{subfigure}{.55\textwidth}

  \centering
\includegraphics[width=0.55\linewidth]{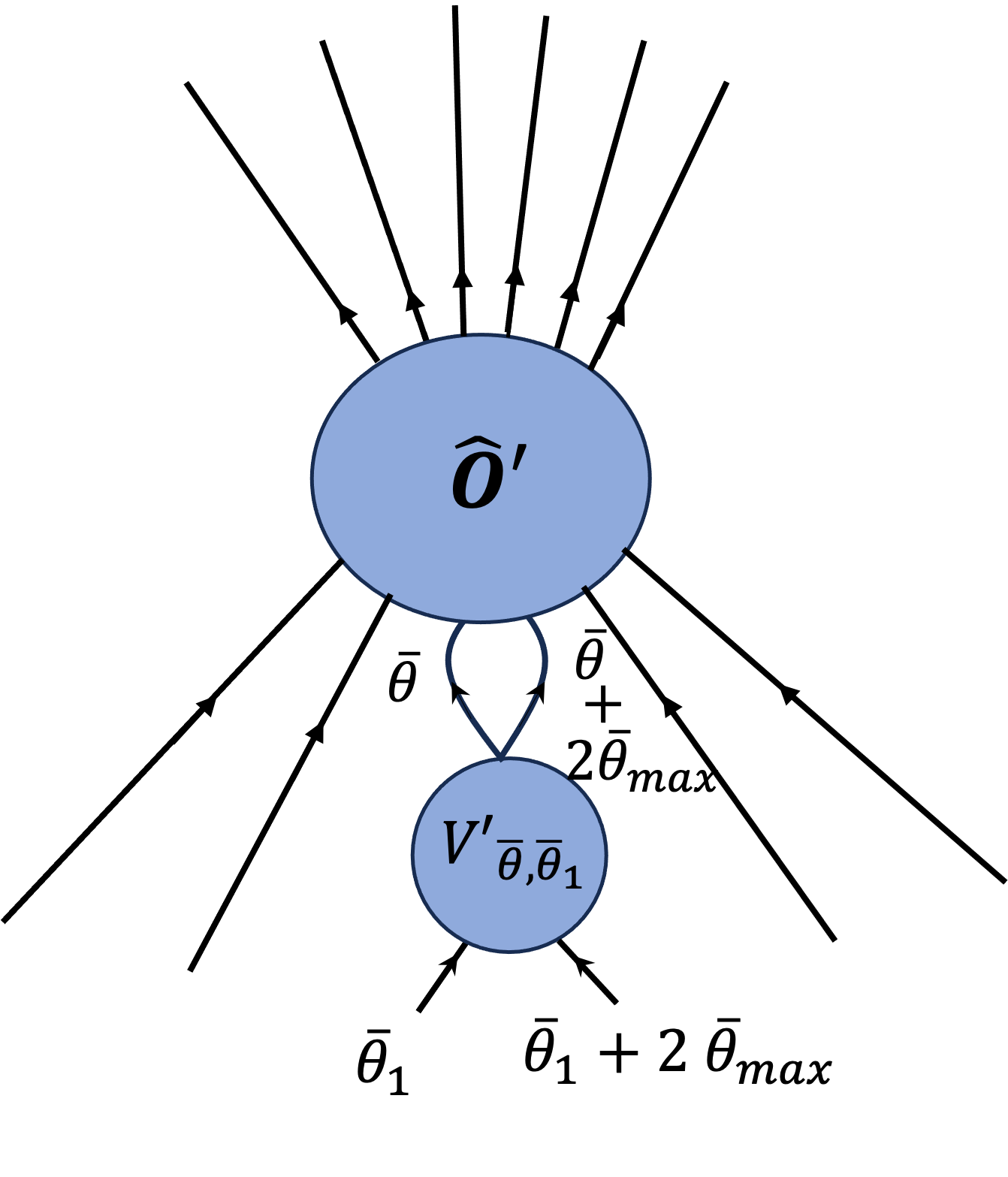}
  \caption{
}
\label{fig:}
\end{subfigure}%
\caption{
Either the four-fermion operator in the pairing channel (a) or a higher-order composite operator that includes a Cooper pair in the external legs (b) acquires an anomalous dimension through virtual Cooper pairs that can be created across the Fermi surface.
The external legs dressed by $\aV$ are in the Cooper channel; $\theta(\bar \theta_1 + 2 \bar \theta_{max}) =
\theta(\bar \theta_1)+\pi$. 
}    \label{Fig:UV/IR}
\end{figure}

One such observable is the anomalous dimension of composite operators of fermions.
The vertex correction shown in \fig{Fig:UV/IR} contributes
\bqa
\eta[\aV^\pm] =
\int 
\frac{d \bar \theta}{2\pi \sqrt{\mu}} \aV^\pm_{\bar \theta_1,\bar \theta_1+\bar \theta}
\label{eq:etaV}
\eqa
to the anomalous dimension of the composite operator whose external momenta for the pair of fermions are in the s-wave channel.
In the beta function in \eq{eq:RG_xspace0},
the term that is quadratic in 
$\av$ (which is the same as $\aV$) is nothing but the contribution of the anomalous dimension of the two-body interaction generated from the two-body interaction itself.
If $\Delta \leq 1$,
the contribution from 
$\bar \theta \gg 
\sqrt{\mu}$
is large although $\aV^\pm_{\bar \theta_1, \bar \theta_1+\bar \theta}$ is negligible at individual $\bar \theta$.
What is crucial is the measure 
$\frac{d\bar \theta}{\sqrt{\mu}}$ 
that increases with decreasing $\mu$ in the expression of the anomalous dimension.
It captures the fact that the critical boson carries a typical momentum $q_\theta(\mu)\sim \sqrt{\mu \KFAVdim}$ at the energy scale $\mu$ as shown in \eq{eq:qmutheta}, and 
two fermions separated by an angle larger than $q_\mu/\KFthetadim$ contribute to the anomalous dimension independently\cite{PhysRevB.78.085129}.
$\frac{\Delta \bar \theta}{\sqrt{\mu}}$
denotes the number of such patches within $\Delta \bar \theta$ that additively contribute to the anomalous dimension.
As a result, the net anomalous dimension is proportional to the size of the Fermi surface measured in the proper angle.
The angular metric 
in \eq{eq:gthetatheta},
which increases with decreasing $\mu$, accounts for the fact that the number of such patches increases with decreasing energy.
For \(\Delta \leq 1\), the integral becomes `UV'-divergent in the \(K_F \to \infty\) limit, as the contribution from large-angle scatterings accumulates.
Therefore,
\bqa
\lim_{\mu \rightarrow 0}
\eta[T_\zeta \aV^\pm] -  \eta[\aV^\pm] 
\neq 0
\eqa
for general smooth function of $\zeta(\theta)$
if $\Delta \leq 1$. 
In this case, the anomalous dimension is invariant 
only under the proper subgroup $\text{OLU}(1)$ \cite{PhysRevB.110.155142} for which $\zeta(\theta)$ is constrained to satisfy
\begin{equation}
    \begin{aligned}
        \zeta(\theta + \pi) = -\zeta(\theta).
        \label{eq:olu1}
    \end{aligned}
\end{equation}

Another example that we will encounter in physical examples has a universal pairing interaction of the form,
\bqa
\aV^\pm_{\bar \theta + \Delta \bar \theta, \bar \theta}(\mu) \sim
\frac{\sqrt{\mu}}{\bar \theta_{max}}.
\label{eq:olu2} 
\eqa
At each angle, it has a vanishingly small overall amplitude in the low-energy limit. 
However, its amplitude does not decay at large angles. 
The large-angle scatterings add up to give rise to a non-trivial anomalous dimension in \eq{eq:etaV} in the low-energy limit.
Therefore, the universal pairing interaction of the form in \eq{eq:olu2} also breaks the LU(1) to OLU(1).

\section{Toy model}
\label{sec:toymodel}

As the first example, we consider a toy model for which the beta functional of the pairing interaction is exactly solvable.
While the toy model is not realistic, its role is to offer an economical and analytical route to realizing distinct superuniversality classes, given that topology of general PFPs is insensitive to the details of the theory. 
Furthermore, it provides useful insight into how various features of general PFPs are affected by the tunable parameters of the model.

\begin{figure}[th]
\centering
\begin{subfigure}{.4\textwidth}
\centering
\includegraphics[width=0.9\linewidth]{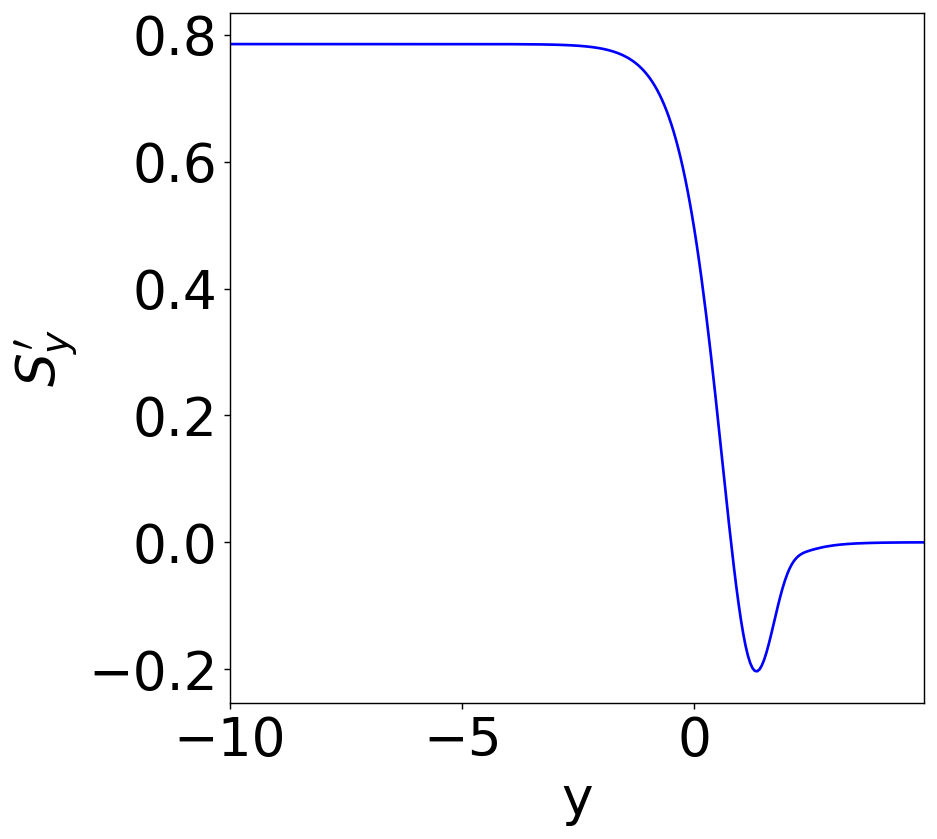}
\caption{}
\label{fig:Source_d2_U1}
\end{subfigure}
\begin{subfigure}{0.45\textwidth}
\centering
\includegraphics[width=0.9\linewidth]{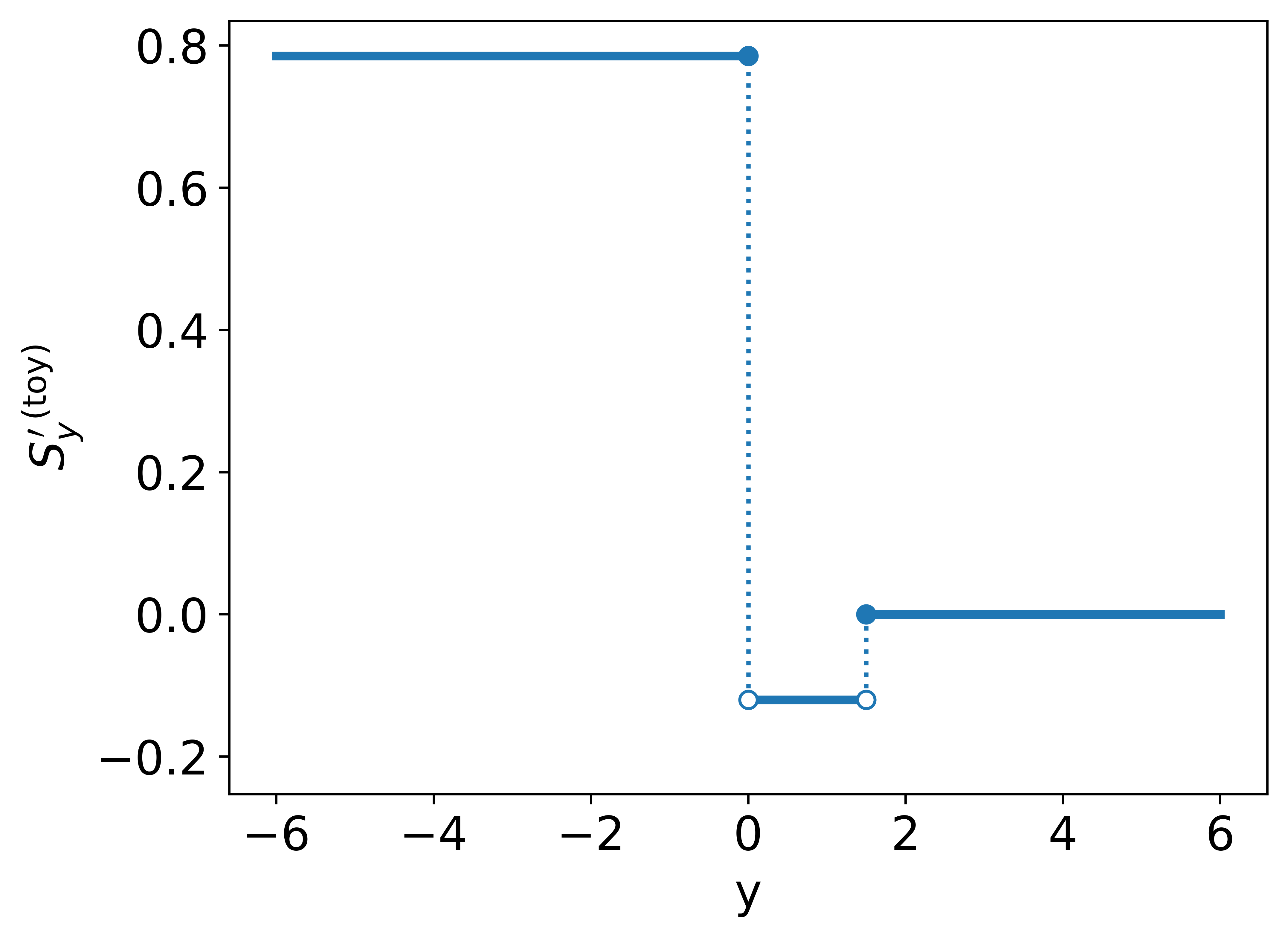}
\caption{}
\label{fig:toy_mimic_d2_U1_source}
\end{subfigure}
\caption{(a) The source $\aS_y$ that describes the universal pairing interaction generated from the U(1) gauge field as a function of the logarithmic angular momentum $y$ in $d=2$. (b) A toy source $\aS^{(\text{toy})}$ that has been tailored to mimic the source of the gauge theory.
The example shown uses $\hIII = S'_{-\infty}$ and $\hII = -0.12$, $\wII = 1.5$ for the intermediate region. }
\end{figure}

We begin by writing the differential equation for PFPs as
\bqa
\frac{1}{2}\partial_y \aV_y  = R_d \bigg(\aV_y + \frac{2\HD-1}{4R_d} \bigg)^2 - \frac{1}{16R_d} \etaPy,
\eqa
where $\etaPy = (2\HD-1)^2 + 16R_d\aS_y$ is the $y$-dependent discriminant.
This differential equation has qualitatively distinct solutions, depending on 
the scaling dimension $\HD$
and the angular momentum-dependent source $\aS_y$.
However, it is difficult to solve it exactly because of the $y$-dependence of the source.
For example, the source generated from the U(1) gauge field is shown in 
\fig{fig:Source_d2_U1}.
The source has three qualitatively important features.
Firstly, it vanishes in the large $y$ limit, as is required by locality.
The pairing interaction generated from critical fluctuations at a non-zero energy scale must be an analytic function of angle, and its Fourier component must vanish in the large angular momentum limit.
Secondly, the source in the $y \rightarrow -\infty$ limit is dictated by the s-wave pairing interaction generated by critical fluctuations.
Its sign is given by \eq{eq:sttheta}.
For the U(1) gauge field, the Amperian law implies that a pair of fermions with opposite momenta repel each other, resulting in a repulsive source.
On the other hand, the $C_4$ symmetric Ising nematic critical fluctuations generate an attractive interaction in the s-wave channel, 
and its $\aS_y$ is opposite to that of the U(1) gauge theory.
$\aS_y$ is well approximated by $\aS_{\infty}$ and 
$\aS_{-\infty}$
in 
$y \gg w_L$
and 
$y \ll w_S$, respectively,
where $w_L$ and $w_L$ denote the crossover scales.
Finally, there is a transient region of finite $y$ between $w_S$ and $w_L$ in which $|\aS_y|$ is significant, and the sign of the source is opposite to that of the s-wave limit.
The existence of the transient region is guaranteed by the traceless condition of the source 
in \eq{eq:Smtraceless}: 
if the source is repulsive (attractive) in the s-wave limit, there must be a range of angular momentum in which it is attractive (repulsive). 
In the U(1) gauge theory, the interaction vertex is positive definite in angular space, giving rise to the repulsive pairing interaction in the s-wave channel. 
Nonetheless, it gives rise to an attractive interaction in non-zero angular momentum channels in which critical bosons predominantly scatter Cooper pairs between parts of the Fermi surface that have opposite phases in the pairing wavefunction.

For our toy model, we consider a piece-wise constant $\aS_y$,
which captures these three features of the sources.
The source of the toy model reads
\bqa
\aS^{(\text{toy})}_y = \begin{cases}
0 ~~~~~~~~ \text{for} \quad y \ge w_L 
~&~ \text{(region I)}\\
\hII ~~~~~~ \text{for} \quad w_S < y < w_L 
~&~ \text{(region II)}\\
\hIII ~~~ \text{for} \quad y \le w_S 
~&~ \text{(region III)}
\end{cases}.
\eqa
As discussed below \eq{eq:wLwS},
only $w_L-w_S$ is the independent parameter.
Accordingly, the superuniversality class only depends on the width of the intermediate region, $w \equiv w_L-w_S$.
From now on, we set $w_S=0$ and $w_L=w$.
The region I represents the large $y$ limit where the source of the four-fermion coupling is negligible.
The region II with width $\wII$ represents a crossover region where the source takes a `transient' value $\hII$.
The region III is the asymptotic small-angular momentum region where the source takes a constant value of $\hIII$, which represents the source in the s-wave limit.
The tunable parameters of the toy model are $\wII$, $\hII$ and $\hIII$ along with the scaling dimension $\HD$.
We will see that the toy model with these parameters is general enough to provide access to all superuniversality classes discussed in the previous section.

To begin our analysis of the toy model, let us first consider the general PFP solution. 
This is given by the piecewise function
$\aV_{y}$ valid in each region, where
\bqa
\aV_{y} = \frac{1}{4R_d}\Big[-  \sqrt{\etag}  \tanh \Big(\frac{1}{2}  \sqrt{\etag}y + \Phi\Big) - 2\HD+1\Big],  \label{eq:generalsol0}
\eqa
where
\bqa
\etag = \begin{cases}
   \etaPI, & \text{$y \geq \wII$}, \\
   \etaPII, & \text{for $0<y<\wII$}, \\
   \etaPIII, & \text{for $y \leq 0$} 
\end{cases}
\eqa
represents the discriminant in each region. 
This expression with real $\Phi$ makes the reality of $\aV_y$ manifest in the region with a positive discriminant 
$\etag$.
For $\etag<0$, we analytically continue 
$\sqrt{\etag} 
\rightarrow
i\sqrt{-\etag}$
and 
$\Phi \rightarrow i\tilde{\Phi}$ 
to write the general solution as
\bqa
\aV_{y} = \frac{1}{4R_d} \Big[ \sqrt{-\etag} \tan \Big(\frac{1}{2}  \sqrt{-\etag}y +\tilde{\Phi}\Big)-2\HD+1 \Big].
\label{eq:generalsol2}
\eqa
Because $\aS_y=0$ in region I, $\etaPI\ge 0$.
While $\etaPII$ and $\etaPIII$ can take either sign, 
we can focus on the cases with $\etaPII<0$ without loss of generality.
This is because an intermediate region of $\etaPII\ge0$ can be essentially merged into region I and does not generate topologically new classes.
The constants
$\Phi$ (alternatively $\tilde{\Phi}$) are 
fixed by a boundary condition that a particular PFP satisfies.
The most important particular solutions for identifying the topological class of a general PFP are the metallic and separatrix PFPs. 
In the toy model, the metallic PFP is the one that satisfies $\lim_{y\rightarrow \infty} 
\aV_y =
\aV^\bullet_{\infty}$,
\bqa
\aV^M_y = \frac{1}{4R_d}\begin{cases}
\sqrt{\etaPI} - 2\HD + 1 &\text{for} \quad y \ge \wII\\
\sqrt{-\etaPII} \tan \Big(\frac{1}{2} \sqrt{-\etaPII }y 
+\tilde{\Phi}_{II}^M
\Big) -2\HD + 1&\text{for} \quad  0 < y < \wII\\
-\sqrt{\etaPIII}\tanh \Big(\frac{1}{2} \sqrt{\etaPIII}y + \Phi_{III}^{M}\Big) -2\HD + 1~~~ &\text{for} \quad y \le 0
\end{cases}.
\label{eq:generalsol}
\eqa
It is understood that 
$\sqrt{\etaPIII}$
and
$\Phi_{III}^{M}$ are 
imaginary if
$\etaPIII<0$.
For the metallic PFP,
the phases in regions II and III are 
\bqa
\tilde{\Phi}_{II}^M 
&=& 
-\frac{1}{2}\sqrt{-\etaPII}\wII + \arctan \Big(\frac{\sqrt{\etaPI}}{\sqrt{-\etaPII}}\Big), \nn
\Phi_{III}^M &=& -\text{arctanh} \left\{\frac{\sqrt{-\etaPII}}{\sqrt{\etaPIII}}\tan\Big[-\frac{1}{2}\sqrt{-\etaPII}\wII + \arctan \Big(\frac{\sqrt{\etaPI}}{\sqrt{-\etaPII}}\Big)\Big]\right\}.
\label{eq:toy_gn_mtl}
\eqa
It is noted that in the first line of \eq{eq:generalsol}, the combination $\sqrt{\etaPI}-2\HD + 1$ is equal to zero because $\aS_\infty=0$.
However, we will continue writing expressions as has been done above to keep the separate roles of $\sqrt{\etaPI}$ and $-2\HD +1$ more manifest.
For $\etaPIII \geq 0$,
there exists the separatrix PFP that satisfies 
$\lim_{y \rightarrow -\infty}
\aV^S_y = 
\frac{1}{4R_d}\left(-\sqrt{\etaPIII} -2\HD + 1\right)$,
\bqa
\aV^S_y =\frac{1}{4R_d} \begin{cases}
-\sqrt{\etaPI}\tanh \Big(\frac{1}{2} \sqrt{\etaPI}y + \Phi_{I}^S\Big)-2\HD + 1  &\text{for} \quad y \ge \wII
\\
\sqrt{-\etaPII} \tan \Big(\frac{1}{2} \sqrt{-\etaPII }y 
+
\tilde{\Phi}^S_{II}
\Big) -2\HD +1 &\text{for} \quad  0 < y < \wII \\
-\sqrt{\etaPIII} -2\HD + 1 ~~~ &\text{for} \quad y \le 0
\end{cases}
\label{eq:toy_gn_spx}
\eqa
with 
\bqa
\Phi_{I}^S &=& -\frac{1}{2}\sqrt{\etaPI}\wII - \text{arctanh}\left\{ \frac{\sqrt{-\etaPII}}{\sqrt{\etaPI}} \tan \Big[\frac{1}{2}\sqrt{-\etaPII}\wII - \arctan \Big(\frac{\sqrt{\etaPIII}}{\sqrt{-\etaPII}}\Big)  \Big]\right\}, \nn
\tilde{\Phi}^{S}_{II}
&=& -\arctan \Big(\frac{\sqrt{\etaPIII}}{\sqrt{-\etaPII}}\Big). \label{eq:pb_spx} 
\eqa

\begin{figure}[H]
    \centering
    \includegraphics[width=0.5\linewidth]{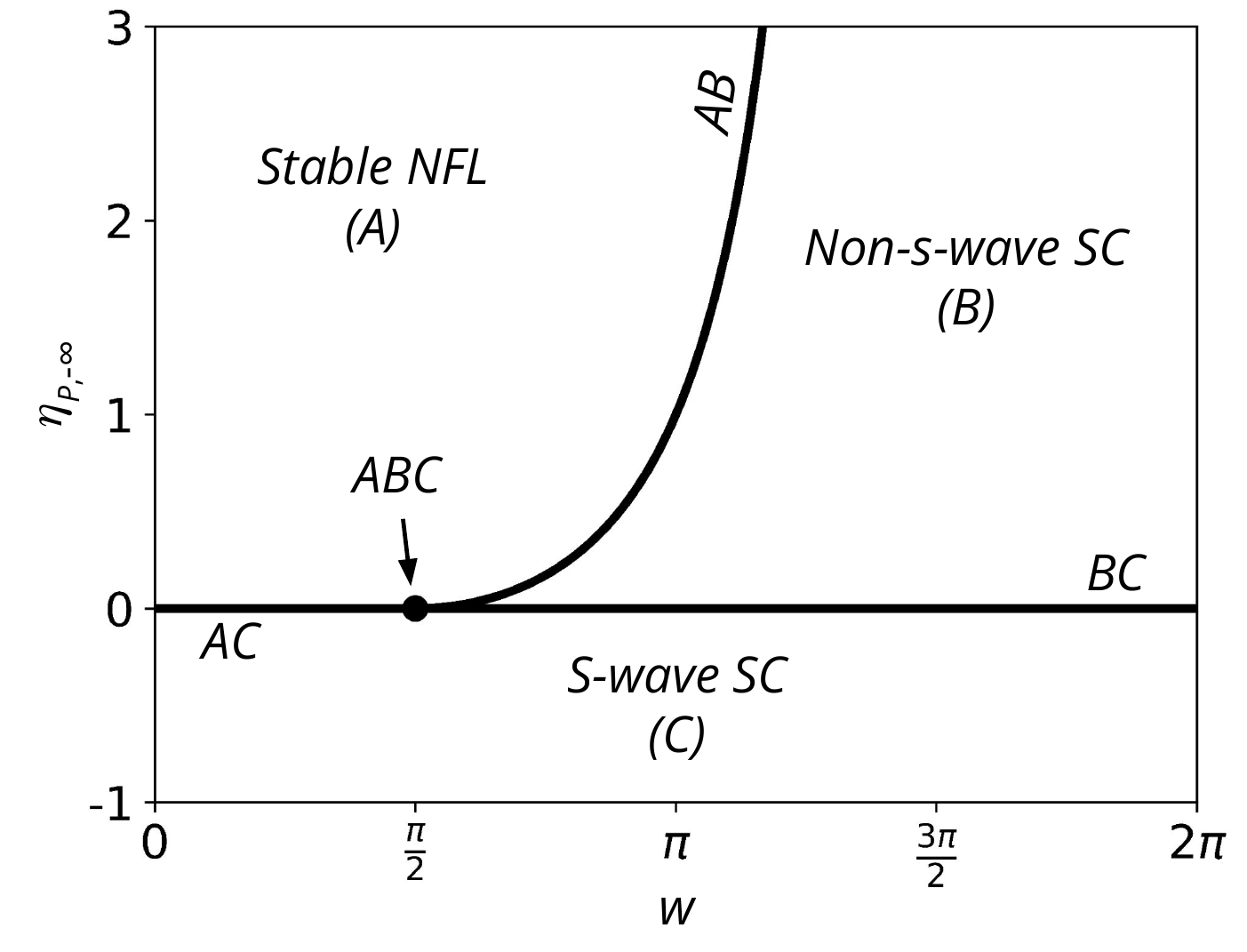}
    \caption{
The superuniversality phase diagram of the toy model with
$\HD=1$ and  $\etaPII=-1$.
The boundary between classes A and B represents the line of critical widths given by
$\etaPIII(w) = \tan^2\big[ \frac{1}{2}(w - \frac{\pi}{2} )\big]$.
}
    \label{fig:toy_phasediagram}
\end{figure}

The topology of the general PFPs is determined by the behaviors of the metallic and separatrix PFPs.
While all parameters of the toy model can be tuned, we do not need to do so because all seven superuniversality classes can be realized for fixed $\HD$ and  $\etaPII<0$.
Therefore, we only tune $\etaPIII$ and $\wII$
(note that $\etaPI$ is fixed by $\HD$).
$\etaPIII$ controls the strength of the universal pairing interaction relative to the incoherence of fermions in the s-wave channel.
The stronger the attractive s-wave pairing interaction is, and the more coherent the fermions are, the smaller $\etaPIII$ becomes.
When $\etaPIII$ is negative, the metallic PFP behaves as a tangent function in region III and is guaranteed to diverge.
On the other hand, 
$\wII$ is the range of the non-zero angular momentum on the logarithmic scale in which the attractive interaction is strong enough to overcome the pair-breaking effect of incoherent fermions and make $\etaPII$ negative. 
For a wide `well', the coupling of a large angular momentum channel is renormalized with the attractive pairing interaction over a long RG time before the drift of $y$ brings the coupling to region III, increasing the propensity for non-s-wave superconductivity.
If the width is sufficiently large, the metallic PFP diverges to $-\infty$, which can occur either in region II or III.
Since the topology of the general PFP does not depend on where the metallic PFP diverges, for the topological classification, we focus on the cases in which the metallic PFP diverges in region III.
This is more convenient for studying the phase transition to class A because the metallic PFP is bound to diverge in region III near the boundary with class A.
In this case, both metallic and separatrix PFPs remain regular in region II, and 
the width is bounded by $\wII < \wII_{max} \equiv \frac{1}{\sqrt{-\etaPII}}\Big[\pi + 2 \min \Big\{\arctan{\Big(\frac{\sqrt{\etaPI}}{\sqrt{-\etaPII}}\Big)}, \arctan{\Big(\frac{\sqrt{\etaPIII}}{\sqrt{-\etaPII}}\Big)} \Big\}\Big]$.

Fig. \ref{fig:toy_phasediagram} is the `phase diagram' of superuniversality classes realized in the plane of $\etaPIII$ and $\wII$. 
For $\etaPIII < 0$, the attractive pairing interaction in the s-wave channel is strong enough to overcome the incoherence.
In this case, the s-wave superconducting instability is unavoidable, and class C is realized.
For $\etaPIII > 0$, there are two asymptotic fixed-points in the small $y$ regime -- one stable and one unstable.
In such situations, the s-wave superconducting instability can be avoided with a repulsive bare coupling. 
Then, either class A, AB, or B is realized depending on the choice of $\wII$. 
If $\wII$ is smaller than the critical width,
\bqa
\wII_c \equiv  \frac{2}{\sqrt{-\etaPII}} \Big[ \arctan\Big(\frac{\sqrt{\etaPIII}}{\sqrt{-\etaPII}}\Big)  + \arctan\Big(\frac{\sqrt{\etaPI}}{\sqrt{-\etaPII}}\Big)\Big],
\eqa
the range of non-zero angular momenta with strong attractive interaction is not wide enough to cause a singularity in the metallic PFP.
In this case, the real part of $\Phi^M_{III}$ is negative, and the metallic PFP, which takes the form of either a hyperbolic tangent or hyperbolic cotangent in region III, remains regular for $y<0$.
Therefore, the general PFP belongs to the stable NFL class (class A). 
For large enough widths
($\wII > \wII_c$)
\footnote{But still smaller than $\wII_{max}$ so that the profile stays regular in region II.}, 
the metallic PFP lies outside ${\cal B}^{IR}_{\aV^\bullet_{-\infty}}$, and
enters region III below $\aV^\circ_{-\infty}$.
In this case, the non-s-wave SC class (class B) is realized. 
If $\wII = \wII_c$, the metallic PFP flows to $\aV^\circ_{-\infty}$ in the small $y$ limit.
This corresponds to class AB. 
$\wII_c$ marks the critical point for the topological phase transition between classes A and B.
%

The final three superuniversality classes are realized when $\etaPIII = 0$. 
In this case, there exists a marginal $-\infty$ asymptotic fixed point. 
This requires fine-tuning of the universal pairing interaction in the s-wave channel relative to the anomalous dimension.
On the critical line of $\etaPIII=0$, the sign of $ \wII - \wII_c$ determines which of the classes AC, ABC, or BC is realized. 
For this, it is useful to have the profile of the metallic PFP in the $\etaPIII\to 0^+$ limit,
\bqa
\aV^M_{y,III} =\frac{1}{4R_d}\Bigg\{- \frac{\sqrt{-\etaPII} \tan \Big[\frac{1}{2}\sqrt{-\etaPII}(\wII - \wII_c)\Big]}{1 + \frac{1}{2} \sqrt{-\etaPII} \tan \Big[\frac{1}{2}\sqrt{-\etaPII}(\wII - \wII_c)\Big]  \,y}-2\HD + 1 \Bigg\}    ~~~~\text{for } \sqrt{\etaPIII} = 0,
\eqa
where the argument of the tangent function has been written in terms of the deviation $\delta w = \wII-\wII_c$ from critical width as defined above. 
When 
$\wII < \wII_c$, the metallic PFP is {\it within} 
${\cal B}^{IR}_{\aV^\halfominus_{-\infty}}$,
and it converges to $\aV_{-\infty}^{\halfominus}$ in the small $y$ limit.
This corresponds to class AC.
When $\wII > \wII_c$, the prefactor on $y$ in the denominator is positive, and the metallic PFP diverges at a finite $y_M^*$ in region III.
In this case, the metallic PFP is {\it outside} 
${\cal B}^{IR}_{\aV^\halfominus_{-\infty}}$,
and the critical class BC is realized.
When $\wII$ is tuned towards the critical value from above, $y_M^* \to -\infty$ as the tangent function approaches zero. 
At $\wII = \wII_c$, the metallic PFP is in
$\partial {\cal B}^{IR}_{\aV^\halfominus_{-\infty}}$, and the multi-critical universality class ABC is realized.

\section{Diagnostics}
\label{sec:diagnostics}

Later in this paper, we will consider physical examples.
The first challenge that we face in discussing physical examples is to 
determine which superuniversality class a given theory belongs to.
Ideally, one can do this through the general solution of the PFP equation.
However, it is generally difficult to solve the equation exactly.
Fortunately, it is often possible to infer the superuniversality class through less direct but more practical routes that rely on our ability to place it within a suitable bound. 
We begin this section by developing practical diagnostics that will be useful later in the following sections.

\begin{enumerate}

\item 
{\bf A necessary condition for regular metallic PFP (classes A, AB, AC and ABC)}

In classes A, AB, AC and ABC, the metallic PFP is non-divergent.
In those classes, \eq{eq:fp_eq_log} has a regular solution that is extended from $y=-\infty$ to $\infty$.
In the large $l$ limit, the projective fixed point equation for a regular PFP can be written in the rescaled angular space $\hat \theta = \bar \theta e^{l/2}$, which is extended from $-\infty$ to $\infty$.
For $\hat{V}^{\pm}_{\hat{\theta}_1,\hat{\theta}_2} =    V^{' \pm}_{\hat{\theta}_1e^{-l/2}, \hat{\theta}_2 e^{-l/2}}$,
\eq{eq:fp_eq_log} becomes\cite{BORGES2023169221}
\begin{equation}
    \begin{aligned}
-\HD\hat{V}^{\pm}-\frac{1}{2}\left(L\cdot\hat{V}^{\pm}+\hat{V}^{\pm}\cdot L^{\dagger}\right)-R_d
 \hat{V}^{\pm}\cdot\hat{V}^{\pm}
 + \hat{S}^{\pm}_{\hat{\theta}_1,\hat{\theta}_2} 
      =0,
      \label{eq:dilated_RG_main}
    \end{aligned}
\end{equation}
where 
$\hat{S}^{\pm}_{\hat{\theta}_1,\hat{\theta}_2} =    S^{' \pm}_{\hat{\theta}_1e^{-l/2}, \hat{\theta}_2 e^{-l/2}}$
and
$L$ is an operator that dilates the rescaled angular variable,
$L_{\hat{\theta}_1,\hat{\theta}_2} = 2\pi\sqrt{\Lambda}\hat{\theta}_1\partial_{\hat{\theta}_1}\delta\left(\hat{\theta}_1-\hat{\theta}_2\right)$.
Thanks to the rescaling of the angular variable, 
\eq{eq:dilated_RG_main} is independent of scale $l$.
This projective fixed point equation can be cast into a simple form,
\begin{equation}\begin{aligned}
    U_{1}U_{2} = D,
\end{aligned}\end{equation}
where 
\begin{equation}\begin{aligned}
    U_{1} =  \left(\hat{V}^{\pm}+\frac{\HD I+L}{2R_d}\right),~~~
    U_2 =  \left(\hat{V}^{\pm}+\frac{\HD I+L^\dagger}{2R_d}\right),~~~
    D = \frac{1}{4R_d^2}\left(\HD I+L\right)\cdot\left(\HD I+L^{\dagger}\right)
    +
    \frac{\hat{S}^{\pm}}{R_d}
    \label{eq:u1_u2_d_main}
\end{aligned}\end{equation}
with
$I_{\hat{\theta}_1,\hat{\theta}_2} =2\pi\sqrt{\Lambda}\delta\left(\hat{\theta}_1-\hat{\theta}_2\right)$. 
For a Hermitian $\hat V^{\pm}$,
$U_{2}$ = $U_{1}^{\dagger}$ 
and \eq{eq:u1_u2_d_main} admits a solution only if $D$ is non-negative.
This leads to a necessary condition for a regular metallic PFP:
\bqa
\mbox{
A regular metallic PFP exists only if all eigenvalues of $D$ are non-negative.
}
\label{eq:ineq_2_5}
\eqa

If $D$ has negative eigenvalues,
there is no regular solution that is Hermitian.
In this case, the system must be in class B, C or BC, and Hermitian PFPs are bound to be singular.
However, there still exist regular non-Hermitian solutions (complex PFPs).
Although they are not directly physical, the RG flow in the space of Hermitian theories can be constrained by complex PFPs.
An example will be discussed in Sec.
\ref{sec:ex2}.

This diagnostic can be used beyond the metals with hot Fermi surface. 
Even if there is no emergent rotational symmetry,  \eq{eq:dilated_RG_main} is valid.
It was used to show that the antiferromagnetic quantum critical metal must be unstable against pairing instability in two dimensions in the limit that the nesting angle is small\cite{BORGES2023169221}.

\item
{\bf A sufficient condition for class A (stable NFL superuniversality class)
}

To understand whether there is a superconducting instability, it is helpful to consider the discriminant in \eq{eq:disc_y}.
If \eq{eq:disc_y} is non-negative at all $y$, 
$\beta_{\aV}(y)=0$ at 
$\aV^\bullet_y =  \frac{ - \left(2 \HD-1 \right) + \sqrt{ \etaPy } }{4R_d}$ 
and
$\aV^\circ_y =
\frac{ - \left(2 \HD-1 \right) - \sqrt{ \etaPy } }{4R_d}$.
In this case, the metallic PFP  
($\aVM_y$) must stay above the horizontal line of 
$\aV^{H}=-\frac{\left(2 \HD-1 \right)}{4R_d}$:
because $\beta_{\aV}(y) \geq 0$ at $\aV^H$,
the metallic PFP, which starts at $\aV^\bullet_\infty > \aV^H$ in the large $y$ limit, does not dip below that line as $y$ decreases.
This excludes a singular metallic PFP in this case.
Therefore, we conclude that
\bqa 
\mbox{
A general PFP is in class A if 
}
\etaPy  \ge 0, \quad \forall  y.
\label{eq:ineq_1}
\eqa 

\item 
{\bf A sufficient condition for class C (s-wave SC superuniversality class)}

If it were not for the lack of scale invariance, 
\eq{eq:ineq_1} would also be a necessary condition for the stable NFL superuniversality class. 
With the incessant horizontal flow, however, metallic states acquire some tolerance against pairing instability, even in the presence of a strong attractive interaction within a finite range of $y$.
If $\eta_{P;y} >0$ at small $y$, the coupling in each angular momentum channel spends only a finite RG `time' in the region of a negative discriminant, and the metallic PFP may remain regular. 
For instance, suppose there is only a small region in the $y$ space over which $\etaPy < 0$. 
The metallic PFP decreases with decreasing $y$ in that region but may not diverge to $-\infty$ before the discriminant becomes positive, pushing the metallic PFP toward $\aV^\bullet_{-\infty}$ at small $y$.
Therefore, it takes a sufficiently large region of negative discriminant for the metallic PFP to diverge.
In general, it is difficult to determine whether the metallic PFP is regular or not.
An exception is the situation in which the discriminant is negative in the $y\rightarrow -\infty$ limit. 
Here, the RG time that each channel is exposed to a strong attractive interaction is infinite, and the metallic PFP is bound to diverge at small $y$.
The s-wave channel ($y=-\infty$) is special in that its beta function does not change with decreasing energy.
This leads to the following general statement:
\bqa
\mbox{A general PFP is in class C if  }
\discinf < 0.
\label{eq:ineq_2}
\eqa

\item 
{\bf A sufficient condition for regular metallic PFP (classes A, AB, AC and ABC)}

If \eq{eq:ineq_2} is not satisfied,
the s-wave SC superuniversality class is excluded.
However, it may still belong to the non-s-wave SC superuniversality class, 
depending on whether the minimum value of $\aV$ on the metallic PFP (denoted as $\aVM_{min}$) is finite or not. 
The condition we discuss here is useful for excluding such non-s-wave SC superuniversality class when there is a finite range of $y$ in which $\eta_{P;y}<0$.
Suppose there exist $y_c$ below which the discriminant is positive definite
and $\aV_c$ such that 
\bqa
\beta_{\aV_c}(y) \ge 0, \quad \forall y \le y_c.
\eqa
In this case, the stable NFL superuniversality class is realized as far as 
$\aVM_{y_c} \ge \aV_c$ because 
$\aVM_{y}$ for $y<y_c$ stays above $\aV_c$.
Although it is hard to compute 
$\aVM_{y_c}$
exactly,
we can place a lower bound on it.
For this, we rewrite the PFP equation as
\begin{equation}
  -  \partial_y \aV_y= -2 R_d \aV_y\bigg(\aV_y+ \frac{2\HD-1}{2R_d} \bigg) + 2\aS_y.
\end{equation}
In this form, we note that so long as 
$\aVM_y \geq -\frac{2\HD-1}{2R_d}$ for $y>y_c$, $\aVM_{y_c}$ is bounded as 
\footnote{This statement use the fact that $\aV^\bullet_\infty = 0$.},
\begin{equation}
\aVM_{y_c} \geq
 2 \int^\infty_{y_c} dy \, \aS_y \Theta(-\aS_y).
\label{eq:cond_finite_fp}
\end{equation}
Therefore, we conclude that 
\bqa
\mbox{
A general PFP is in class
A, AB, AC or ABC if 
}
 2 \int^\infty_{y_c} dy  \aS_y \Theta(-\aS_y) \geq
\mbox{max} \left( \aV_{c},~
-\frac{2\HD-1}{2R_d} \right).
\label{eq:ineq_3}
\eqa

\end{enumerate}

In the next three sections, we discuss three physical examples of non-Fermi liquids.
In each example, we identify its superuniversality class and discuss the universal properties of non-Fermi liquids that arise from it.
While we are ultimately interested in the theories at $d=2$, we consider them in general $d$ between the physical dimension $2$ and the upper critical dimension $5/2$ 
primarily because
we do not yet have non-perturbative tools for studying general strongly coupled theories at $d=2$ with a few exceptions\cite{SHOUVIK2,SCHLIEF}.
The perturbative approach may place a specific two-dimensional theory into the wrong superuniversality class.
Indeed, we will discuss an example in which the superuniversality class changes as the spatial dimension is lowered from $5/2$ toward $2$.
However, the perturbative solution allows us to understand how the behavior of the theory evolves as the dimension approaches the physical dimension.
Here, the space dimension also serves as a theoretical knob that can be tuned to explore different non-Fermi liquid universality classes that are inaccessible within the currently known examples. 
Unfortunately, the physical examples that we discuss here do not span all seven superuniversality classes.
In the last section, we return to the toy model to discuss some universal properties of non-Fermi liquids that arise from superuniversality classes not realized through the physical examples.


\section{Example 1:
U(1) gauge field coupled to Fermi surface}
\label{sec:ex1}

The first physical example is the Fermi surface coupled to the dynamical U(1) gauge field.
This theory describes, for example, the U(1) spin liquid state that supports the spinon Fermi surface\cite{PhysRevB.72.045105,PhysRevLett.95.036403}.
As the theory becomes strongly coupled in $d=2$\cite{SSLEE}, we consider the theory in general $2 \leq d \leq 5/2$, where $d_c=5/2$ is the upper critical dimension\cite{DENNIS}.
Our regularization scheme, 
introduced in Sec. \ref{sec:theory},
is symmetric under
$Z_2\times 
SO(d-1)\times 
SO(4-d)$.
In $d=2$, it is reduced to $Z_2 \times SO(2)$ with $SO(2) = U(1)/Z_2$ for the U(1) charge conservation.
A further superconducting instability, which can occur in the presence of strong attractive interaction, would further break $SO(4-d)$.
In $d=2$, for example, a charge $2$ superconductivity completely breaks $SO(2)$, leaving only $Z_2$.
The transverse component of the U(1) gauge field in $d=2$ is viewed as the rank $2$ anti-symmetric tensor of $SO(2)$.
In general $d$, the rank $2$ anti-symmetric tensor of $SO(4-d)$ has $\frac{(4-d)(3-d)}{2}$ components.

\subsection{The metallic PFP}

\begin{figure}[htbp]
    \centering
        \includegraphics[height=7
        cm,width=7
        cm]{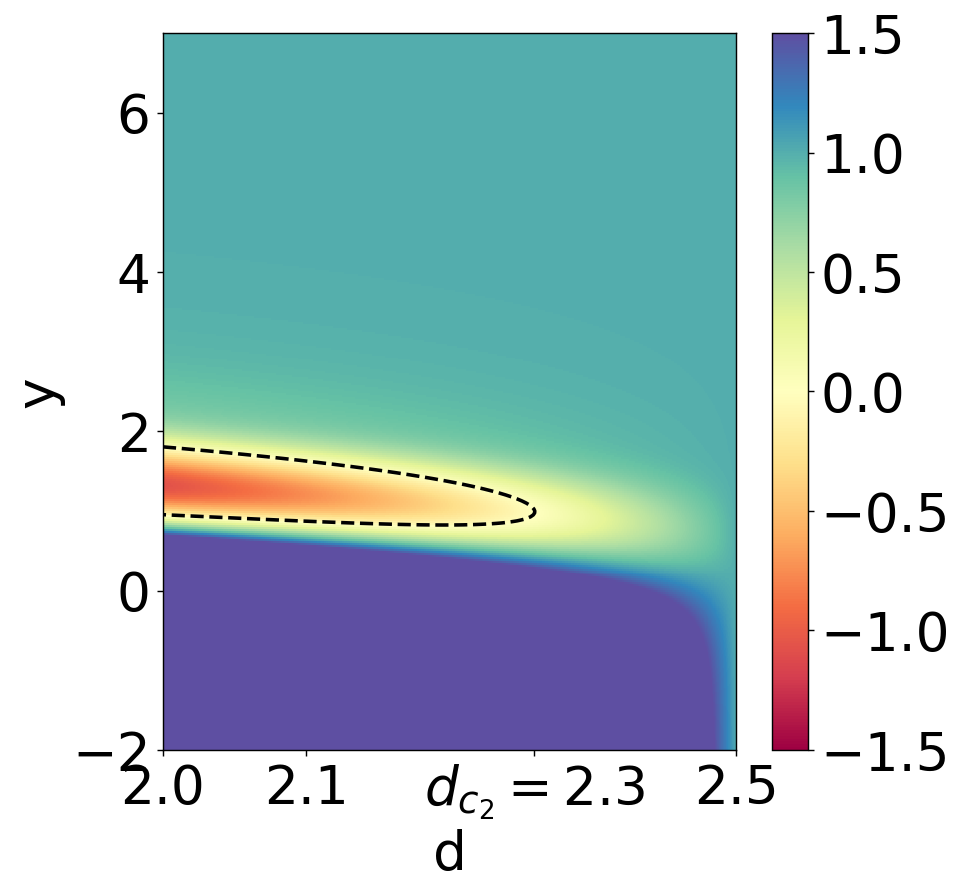}
        \caption{
The $y$-dependent discriminant $\eta_{P,y}$ of the U(1) gauge theory in space dimensions between $d=2$ and $5/2$.
In $d<d_{c_2}$, there exists a range of angular momentum, bounded by the dashed line, in which the attractive interaction generated by the gauge field is strong enough to dominate over the anomalous dimension to make $\eta_{P,y}<0$.
 }
\label{fig:eta_py_U(1)}
\end{figure}

We begin the discussion on the U(1) gauge theory 
by displaying the discriminant as a function of $y$ and $d$ in \fig{fig:eta_py_U(1)}.
In all $d$, the discriminant is positive in the $y \rightarrow \infty$ limit because the pairing source generated from the gauge field vanishes in the large angular momentum limit.
In the $y \rightarrow -\infty$ limit, the discriminant is even more positive because the gauge field mediates a repulsive interaction in the $s$-wave pairing channel. 
This rules out the s-wave SC superuniversality class (class C).
In an intermediate range of $y$, $\aS_y$ is negative.
Whether the attractive interaction at those non-zero angular momenta is strong enough to overcome the incoherence and make the discriminant negative depends on the dimension.
In dimensions above $d_{c_2} = 2.3$, $\eta_{P,y}$ remains positive for all $y$.
In those dimensions, the diagnostic in \eq{eq:ineq_1} immediately implies that the non-Fermi liquid belongs to class A (the stable NFL superuniversality class).
In dimensions below $d_{c_2}$, $\eta_{P,y}$ is negative within a window of $y$, $X=\{y|y^-<y<y^+\}$
with $y^-, y^+ \sim O(1)$.

This is in accordance with the fact that at lower dimensions, the gauge coupling becomes stronger, generating a stronger pairing interaction.
At scale $l$, the angular momentum channels with $y^{(m)}(l) \in X$ are renormalized by the attractive pairing interaction that is strong enough to overcome the pair-breaking effect of incoherence.
This occurs as the typical momentum of the critical boson in \eq{eq:qmutheta} connects a crest and a nearby trough of the pairing wavefunction, satisfying 
$q_\theta(\mu) \sim \frac{\KFthetadim}{m}$.
Namely, the oscillating phase of the pairing wavefunction turns the matrix element of the Cooper pair from positive to negative within a range of angular momenta
(see \fig{fig:Circle_Wave1}).
If there were no drift of $y$ under the RG flow, the attractive interaction strong enough to make the discriminant negative would create superconducting instabilities in $d<d_{c_2}$ at a non-zero angular momentum channel.
However, we have to take into account the effective running of angular momentum to determine the fate of the theory.
The question is whether that attractive interaction is strong enough to make the metallic PFP diverge to $-\infty$ at a finite $y$, realizing the non-s-wave superuniversality class (class B).
Alternatively, the metallic PFP can remain regular, keeping the system in the stable NFL superuniversality class (class A).

\begin{figure}[h]
    \centering
        \includegraphics[height=6cm,width=7cm]{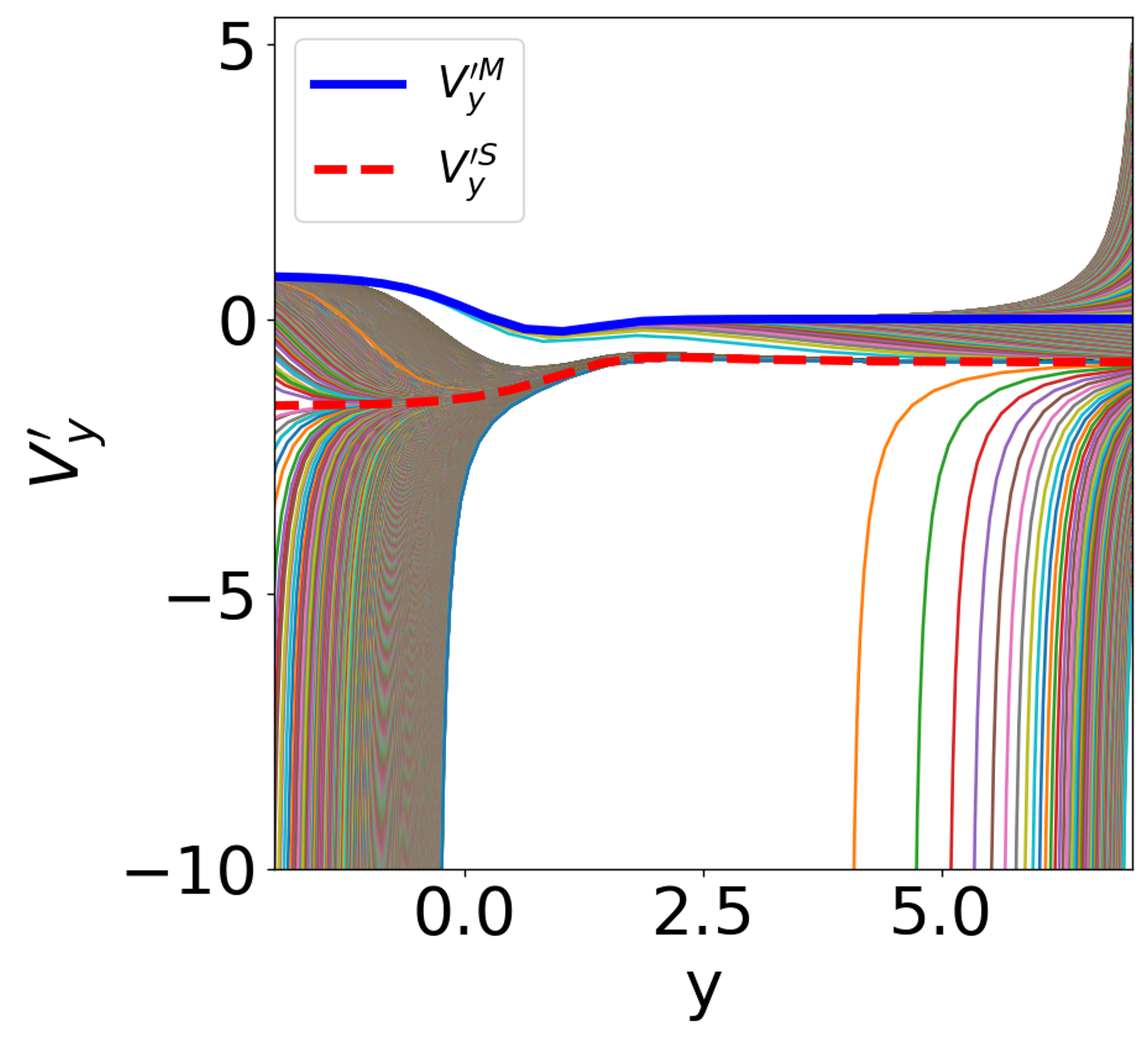}
        \caption{
The portrait of the general PFP for the U(1) gauge theory in $d=2$, obtained to the leading order in the $\epsilon$-expansion.
        }
        \label{Fig:U1(1)_phase}
\end{figure}

To determine the fate of the theory in $d<d_{c_2}$, we resort to the diagnostics in \eq{eq:ineq_3}.
For this, we choose $y_c$ to be the point at which the source changes sign ($\aS_{y_c}=0$) and $\aV_c$ to be
\( \aV^\circ_{\infty}  =
-\frac{2\HD-1}{2R_d} \). 
This choice is convenient because 
\( \beta_{\aV_c}(y) = \aS_y >0\) for all \( y < y_c \).
Numerically, we find that
$ 2 \int^\infty_{y_c} dy  \aS_y \Theta(-\aS_y) >
\aV_{c}$.
\footnote{
In particular,
$\int_{y_c}^\infty dy \, \aS_y > -0.2$  and $\aV_{c} = -\frac{2\HD-1}{2R_d}< -0.78$
everywhere in $2 \leq d \leq 5/2$.
}
This shows that the metallic PFP is regular and that the theory belongs to the stable NFL superuniversality class at all $2 \leq d \leq 5/2$ to the leading order in the $\epsilon$-expansion.
The regularity of the metallic PFP  implies that there exists a separatrix PFP that divides the basin of attraction for the metallic PFP from that for the superconducting PFP.
The metallic PFP and the separatrix PFP then completely determine the topology of the general PFP, as shown in  \fig{Fig:U1(1)_phase}.

In principle, higher-order corrections in the $\epsilon$-expansion can make the metallic PFP singular in $d=2$.
In that case, the U(1) gauge theory would be in the non-s-wave SC superuniversality class (B).
Despite this uncertainty in the fate of the theory in $d=2$, the PFP equation illustrates the mechanism by which a non-Fermi liquid, subject to strong attractive interactions in non-zero angular momentum channels, can avert superconducting instabilities.
The first is the incoherence of electrons.
The large anomalous dimension of incoherent electrons makes the scaling dimension of the pairing interaction $\HD$ large, which makes $\disc$ less negative.
The second is the constant drift of $y$ that arises from the effective flow of angular momentum.
Although fermions are subject to strong attractive interactions mediated by the gauge field in non-zero angular momentum channels, that interaction is only `transient' in RG time because the nature of the pairing interaction changes from attractive to repulsive in the low-energy limit.
This is because, as energy decreases, the critical boson transfers increasingly small momenta to Cooper pairs, and the interaction generated by the gauge field becomes increasingly more s-wave-like, which is repulsive.

\begin{figure}[th]
  \centering
  \includegraphics[width=0.4\linewidth]{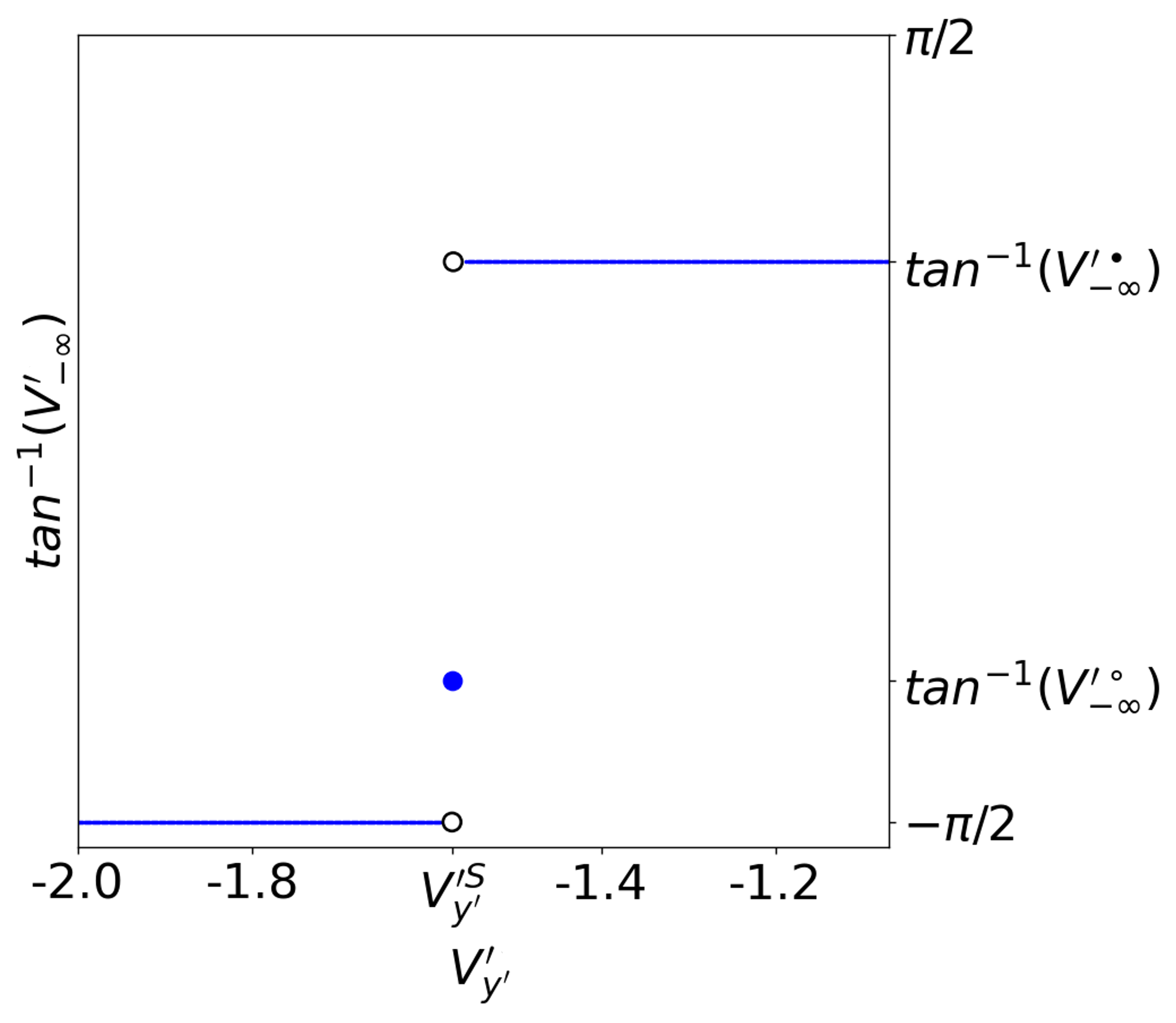}
  \caption{
The asymptotic value of 
$\tan^{-1} \aV_{y}$ in \eq{eq:U1_smally}
in the $y \rightarrow -\infty$ limit plotted as a function of 
$\aV_{y^\prime}$ at $y^\prime \approx -7$
in $d=2$.
}
  \label{fig:V_IR_U1_d2}
  
\end{figure}

  
%

Although it is difficult to obtain a closed-form expression of PFPs, the profile of PFPs can be analytically computed in the small $y$ limit.
For this, we expand the pairing source at small $y$ as
$S^\prime_y = \frac{d}{4-d} \frac{\eta_d}{4R_d} \left(1 - \frac{\alpha_d^{2/3}}{(3-d)(4-d)} e^{2y}\right)+S^\prime_2 ye^{4y}
+ ..$,
where \( \eta_d 
\), 
defined in Eq. (\ref{eq:etad}), 
denotes the magnitude of the anomalous dimension generated by the critical fluctuations for the s-wave pairing interaction
and
$S^\prime_2 = -\frac{3\sqrt{3}}{16\pi}\frac{d\eta_d \alpha_d^{\frac{4}{3}}}{R_d(4-d)^3(3-d)^2}$. 
In the small $y$ limit, the general PFP is written as
\begin{equation}
    \begin{aligned}
\aV_{y} =
         \begin{cases}
             -
          \frac{1}{4R_d}
         \left[1+\mathscr{C}_{y}\left(\frac{c_1\left[J_{-1-\frac{1}{2} \sqrt{\discinf} }\left(\mathscr{C}_{y}\right)-J_{1-\frac{1}{2} \sqrt{\discinf} }\left(\mathscr{C}_{y}\right)\right]}{c_1J_{-\frac{1}{2} \sqrt{\discinf} }\left(\mathscr{C}_{y}\right)+\Gamma\left(1+\frac{\sqrt{\discinf} }{2}\right)J_{\frac{1}{2} \sqrt{\discinf} }\left(\mathscr{C}_{y}\right)}
         \right.\right. \ 
         \\ \hspace{2.5cm}
         \left.\left.+\frac{\Gamma\left(1+\frac{\sqrt{\discinf} }{2}\right)\left[J_{-1+\frac{1}{2} \sqrt{\discinf} }\left(\mathscr{C}_{y}\right)-J_{1+\frac{1}{2} \sqrt{\discinf} }\left(\mathscr{C}_{y}\right)\right]}{c_1J_{-\frac{1}{2} \sqrt{\discinf} }\left(\mathscr{C}_{y}\right)+\Gamma\left(1+\frac{\sqrt{\discinf} }{2}\right)J_{\frac{1}{2} \sqrt{\discinf} }\left(\mathscr{C}_{y}\right)}\right)\right]
         ~~&d\neq d_{c_1},\\
          
         -
          \frac{1}{4R_d}
          \left[1+\mathscr{C}_{y}\left(\frac{
          \left[J_{0}\left(\mathscr{C}_{y}\right)-J_{2}\left(\mathscr{C}_{y}\right)\right]
          - c^{\prime}_1
          2\left[Y_{0}\left(\mathscr{C}_{y}\right)-Y_{2}\left(\mathscr{C}_{y}\right)\right]}{
          J_{1 }\left(\mathscr{C}_{y}\right)
          -c^{\prime}_1
          2Y_{1 }\left(\mathscr{C}_{y}\right)}\right)\right]
        ,~~&d=
        d_{c_1}
         \end{cases}.
         \label{eq:U1_smally}
    \end{aligned}
\end{equation}
Here, $J_\alpha(x)$ and $Y_\alpha(x)$ are the Bessel functions of the first and second kinds, respectively.  
$\mathscr{C}_{y} = \sqrt{\frac{d\eta_d}{(4-d)^2(3-d)}}\alpha_d^{\frac{1}{3}} e^y$. 
When $\discinf$ becomes $4$, which happens at $d=d_{c_1} \approx 2.37$,
the expression in general $d$ exhibits an apparent singularity.
However, the divergence is actually canceled only to give rise to a logarithmic correction in $e^y$.
$c_1$ ($c_1'$) is a constant of integration that specifies a particular PFP.
We can express $c_1$($c_1^\prime$) in terms of the coupling at a fixed $y'$,
\begin{equation}
    \begin{aligned}
        c_{1} &= -
        \Gamma \left(1+\frac{\sqrt{\discinf} ^{ }}{2}\right)
        \frac{\mathscr{C}_{y'}^{-1}\left(1+4R_d \aV^{*}_{y^\prime}\right) J_{\frac{\sqrt{\discinf} ^{ }}{2}}\left(\mathscr{C}_{y'}\right)+ \ J_{-1+\frac{\sqrt{\discinf} ^{ }}{2}}\left(\mathscr{C}_{y'}\right)- J_{1+\frac{\sqrt{\discinf} ^{ }}{2}}\left(\mathscr{C}_{y'}\right)}{\mathscr{C}_{y'}^{-1}\left(1+4R_d\aV^{*}_{y^\prime}\right)  J_{-\frac{\sqrt{\discinf} ^{ }}{2}}\left(\mathscr{C}_{y'}\right)+  J_{-1-\frac{\sqrt{\discinf} ^{ }}{2}}\left(\mathscr{C}_{y'}\right)- J_{1-\frac{\sqrt{\discinf} ^{ }}{2}}\left(\mathscr{C}_{y'}\right)}\\
        c_1^\prime &=  
\frac{J_1\left(\mathscr{C}_{y'}\right)\left(1+4R_d\aV^{*}_{y'}\right)+\mathscr{C}_{y'}\left(J_0\left(\mathscr{C}_{y'}\right)-J_2\left(\mathscr{C}_{y'}\right)\right)}{
2Y_1\left(\mathscr{C}_{y'}\right)(1+4R_d\aV^{*}_{y'})+2\mathscr{C}_{y'}\left(Y_0\left(\mathscr{C}_{y'}\right)-Y_2\left(\mathscr{C}_{y'}\right)\right)}.
\label{eq:c1_c1prime}
    \end{aligned}
\end{equation}
As $\aV_{y'}$ is decreased from $\infty$ to $-\infty$,
$\lim_{y \rightarrow -\infty}\aV_{y}$
exhibits a jump from 
$\aV^\bullet_{-\infty}$
to
$-\infty$
through 
$\aV^\circ_{-\infty}$,
as shown in Fig. \ref{fig:V_IR_U1_d2}.
This is expected because 
$\aV^\bullet_{-\infty}$ 
and
$\aV^{SC}_{-\infty}$ 
are stable asymptotic fixed points with extensive
basins of attraction,
while $\aV^\circ_{-\infty}$ is an unstable asymptotic fixed point whose basin of attraction is only the separatrix PFP.
The separatrix PFP is reached at $c_1=0$.
In the small $y$ limit, the separatrix PFP takes the form of
\begin{equation}
    \aVS_{y} = \aV^\circ_{-\infty} + 
    b^{S} e^{2y}
    + 
b^{S'} y e^{4y} + ..
\label{eq:VySPXu1}
\end{equation}
in the $y \rightarrow -\infty$ limit, where 
$    b^{S} = \frac{d\eta_d \alpha_d^{2/3}}{2(4-d)^2(3-d) R_d (2 + \sqrt{\discinf})}$
and 
$b^{S'}= \frac{3\sqrt{3}}{8\pi}\frac{d\eta_d\alpha_d^{\frac{4}{3}}}{
(4+\sqrt{\discinf})
R_d(4-d)^3(3-d)^2}$.

\begin{figure}[th]
  \centering
  \includegraphics[width=0.4\linewidth]{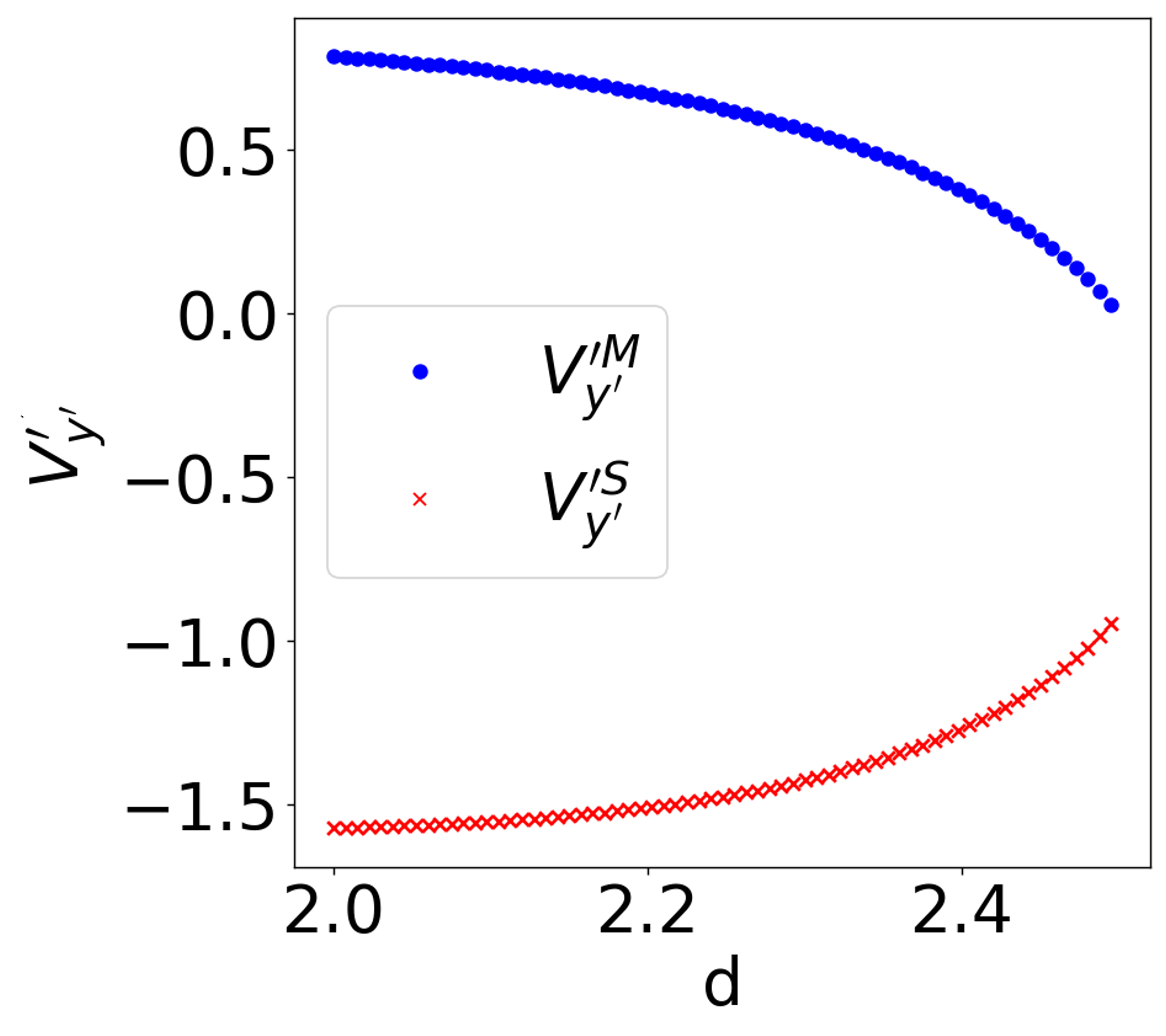}
  \caption{
$\aVM_{y^\prime}$ and $\aVS_{y^\prime}$ evaluated at $y^\prime \approx -7$ as functions of $d$.
Since two distinct PFP cannot intersect, the metallic PFP is strictly above the separatrix PFP at all $y$ in the U(1) gauge theory. 
}
  \label{fig:VMy_VSy_U1}
  
\end{figure}
The metallic PFP is strictly above the separatrix PFP in all $d$, as shown in Fig. \ref{fig:VMy_VSy_U1}.  
This shows that the U(1) gauge theory is in class A in all $d$ to the leading order in the $\epsilon$ expansion.
Despite the fact that the gauge field mediates an attractive interaction that is strong enough to overcome the incoherence of fermions at non-zero angular momentum channels near $d=2$, the running of the Fermi momentum prevents superconducting instability by limiting the window of length scales in which each angular momentum channel is exposed to the strong attractive renormalization.
Had we not taken into account the projective nature of the fixed point associated with the running Fermi momentum,
we would have reached a different conclusion.

In the small $y$ limit, the metallic PFP becomes
\begin{equation}
    \begin{aligned}
          \aVM_y = 
              \aV^\bullet_{-\infty}+b^M
              e^{\sqrt{\discinf} y}+b^{M\prime} e^{2 y},
          \label{stable_sol}
    \end{aligned}
\end{equation}
where
$b^M = -\frac{\sqrt{\discinf}\Gamma\left(1-\frac{\sqrt{\discinf}}{2}\right) }{
2 R_d c_1}
\left(
\frac{\mathscr{C}_0}{2}
\right)^{\sqrt{\discinf} }$
and $b^{M\prime} =  \frac{d\eta_d \alpha_d^{2/3}}{2(4-d)^2(3-d) R_d (2 - \sqrt{\discinf})}$. 
In $d>d_{c_1}$ ($d<d_{c_1}$), 
$\discinf < 4$ ($\discinf > 4$)
and the second (third) term in Eq. (\ref{stable_sol}) is dominant in the small $y$ limit. 
In the $d\rightarrow d_{c_1}$ limit, 
$\discinf \rightarrow 4$ and $c_1\rightarrow 1$.
At $d_{c_1}$, the singularities in $b^M$ and $b^{M'}$ are canceled, giving rise to a logarithmic correction,
$\aVM_y = 
\aV^\bullet_{-\infty}+\frac{3\alpha_d^{\frac{2}{3}}}{8(3-d)(4-d)R_d} ye^{2 y}$.

\subsection{Stable non-Fermi liquid and critical non-Fermi liquids}

Now, we consider individual non-Fermi liquid universality classes that arise in this stable NFL superuniversality class. 

\subsubsection{Stable non-Fermi liquid}

\begin{figure}[h]
    \centering
    \begin{subfigure}{.5\textwidth}
  \centering
\includegraphics[width=0.8\linewidth]{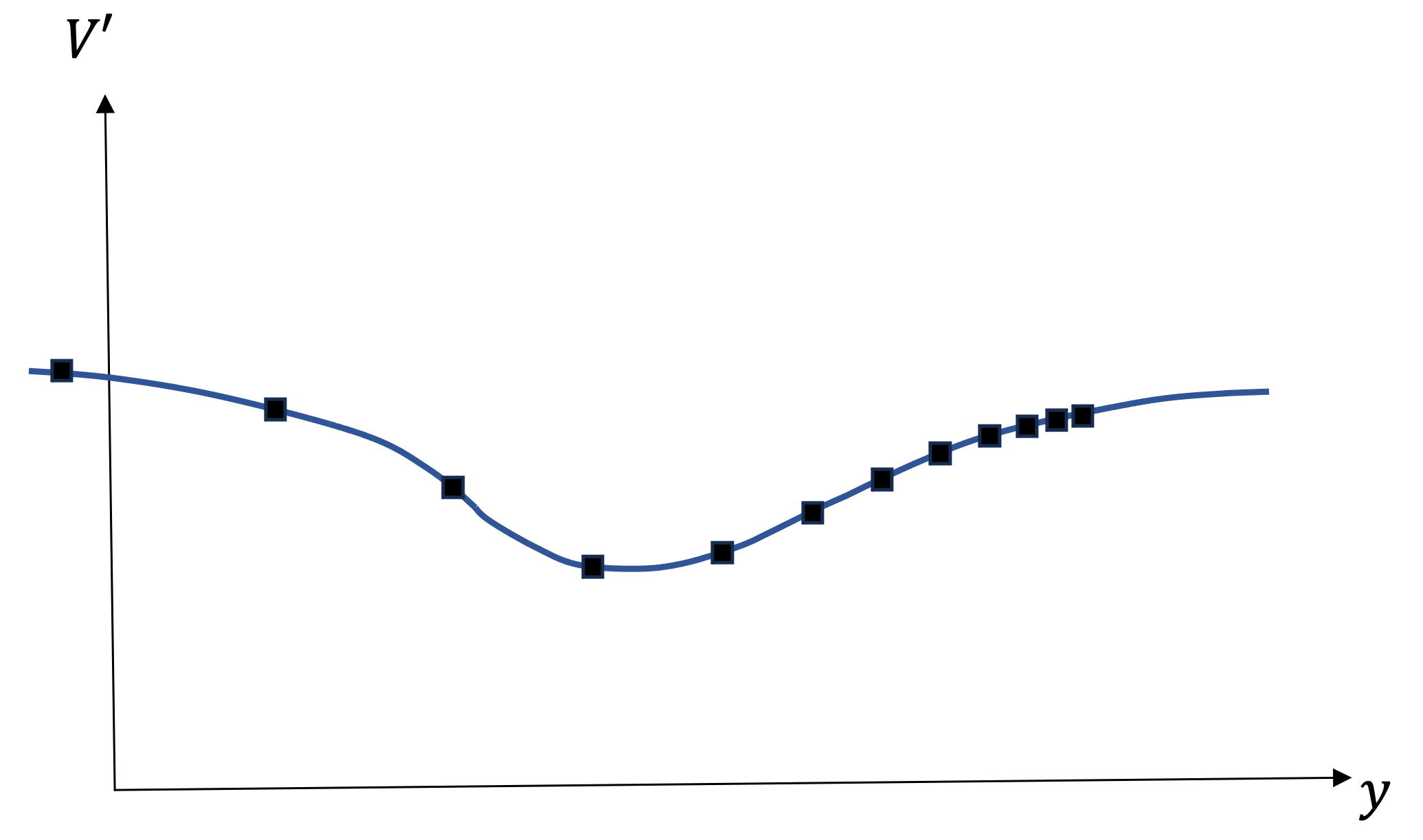}
  \caption{
}
 \label{}
\end{subfigure}%
\begin{subfigure}{.5\textwidth}

  \centering
\includegraphics[width=0.8\linewidth]{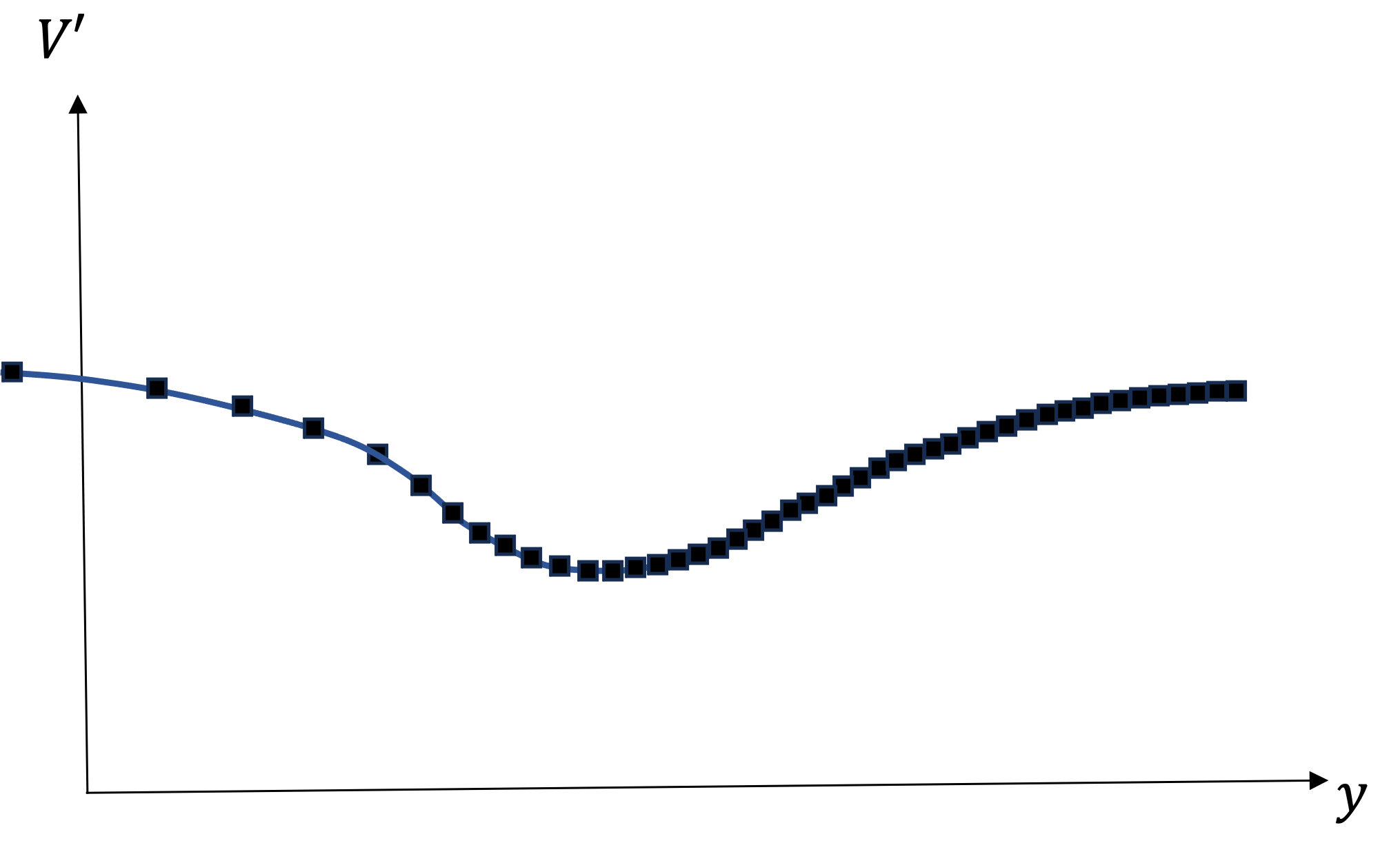}
  \caption{
}
\label{}
\end{subfigure}
\caption{
(a) At a finite $l$, the region of crossover, where $\aV^M_y$ changes significantly as a function of $y$, is occupied with a sparse set of angular momenta. 
The discreteness of $m$ in \eq{eq:InverseFourierV} prevents a scale invariance of the universal pairing interaction 
even if all couplings are already on the metallic PFP.
(b) In the large $l$ limit, the region of crossover is densely populated. 
The region of $y \sim \log \Lambda/\KFAVdim - l/2$ is still occupied with points with large 
spacings, $y^{(m+1)}(l) - y^{(m)}(l) \sim O(1)$, but $\aV^M_y$ at small $y$ reaches $\aV^\bullet_{-\infty}$ up to a correction suppressed by a positive power of $e^y$.
Therefore, \eq{eq:InverseFourierV} can be turned into \eq{eq:InverseFourierV2} up to a correction that vanishes as $e^{-l/2}$.
}
\label{Fig:emergent_scale_invariance}
\end{figure}  

If all bare four-fermion couplings lie above the separatrix PFP, they flow to the metallic PFP at low energies.
The resulting low-energy state is a stable non-Fermi liquid that can be reached without fine-tuning of the four-fermion coupling.
In this non-Fermi liquid phase, two marginal coupling functions $\kappa_{F,\theta}$ and $v_{F,\theta}$ completely determine the universal low-energy physics.
In particular, the superconducting fluctuation in each angular momentum channel is dictated by the metallic PFP in the low-energy limit.
In the angular space, the coupling function at scale $l$ is given by the inverse Fourier transform of the metallic PFP,
\bqa
\aV^{\pm}_{\bar{\theta}_1,\bar{\theta}_1+\bar{\theta}}(l)
=
\frac{\pi \sqrt{\mu} }{\bar{\theta}_{\text{max}} }
\sum_m
\aV^M_{y^{(m)}(l)}  e^{-i\frac{m \pi}{2\bar\theta_{\text{max}}} \bar{\theta}}.
\label{eq:InverseFourierV}
\eqa
Here, $m$ is summed over even (odd) integers for 
$\aV^{-}_{\bar{\theta}_1,\bar{\theta}_1+\bar{\theta}}$
($\aV^{+}_{\bar{\theta}_1,\bar{\theta}_1+\bar{\theta}}$).
The measure of the inverse Fourier transformation  is determined by \eq{eq:FourierV}.
If the renormalized couplings haven't yet flowed to the metallic PFP completely, \eq{eq:InverseFourierV} would have corrections. 
Such corrections are exponentially small in $l$ because the metallic PFP is globally stable in class A and any deviation from it decays exponentially in $l$.
If all couplings are precisely on the metallic PFP, there are no such corrections.
Even in that case, however, 
\eq{eq:InverseFourierV} does not exhibit a scale invariance at a finite $l$ because the coupling still flows within the metallic PFP.
Furthermore, the discreteness of $m$ prevents one from relating the coupling in one angular momentum channel with another under a continuous scale transformation.
This is illustrated in  \fig{fig:lack_of_scale_invariance}.
This is a key difference between metallic projective fixed points and the usual fixed points of relativisitic systems.
In the latter cases, physical observables exhibit the scale invariance even at finite length scales provided that all couplings are chosen to be the fixed point values - this is hard to achieve, in practice, but can be done, in principle.
In metals, however, one cannot have a scale invariance at any finite length scale even if all couplings are tuned to be on the projective fixed point because of the incessant running of the coupling within the projective fixed point.

Even when the couplings are on the metallic PFP,
only in the large $l$ limit does a sense of scale invariance arise in metals.
In the large $l$ limit, the spacing between $y^{(m)}(l)$ approaches zero at any finite $y$. 
This allows one to treat $y$ as a continuous variable 
(see \fig{Fig:emergent_scale_invariance}) 
and write the renormalized coupling function as
\bqa
\aV^{\pm}_{\bar{\theta}_1,\bar{\theta}_1+\bar{\theta}}(l)
=
2\int_{-\infty}^\infty dy ~e^y~
\aV^M_{y} 
\cos\left(e^y 
\frac{\bar{\theta}}{\sqrt{\mu}}\right)
+ O(e^{-l/2}).
\label{eq:InverseFourierV2}
\eqa
The factor of $2$ is the contribution of negative $m$ in the presence of time reversal symmetry.
Without time reversal symmetry, one needs to add the negative $m$ contribution separately.
The difference between the Riemann sum in 
\eq{eq:InverseFourierV}
and the integration in 
\eq{eq:InverseFourierV2},
scales as $e^{-l/2}$. 
Normally, one would immediately ignore such corrections at large $l$.
In metals, however, we need to be more careful.
Since some observables, such as the anomalous dimension of operators shown in \fig{Fig:UV/IR}, are given by the integration over the entire Fermi surface with the measure $\frac{d \bar \theta}{\sqrt{\mu}}$,
the corrections that are of the order of $e^{-l/2}$ can make an order of $O(1)$ difference if the support of the correction is extensive in the space of $\bar \theta$.
In the present non-Fermi liquid,
$\aV^{\pm}_{\bar{\theta}_1,\bar{\theta}_1+\bar{\theta}}(l)$ itself goes to zero in the $\bar \theta/\sqrt{\mu} \rightarrow \infty$ limit, and
the support of the correction in the space of $\bar \theta$  also vanishes in the low-energy limit.
Therefore, in this case, the $O(e^{-l/2})$ correction can be safely ignored in \eq{eq:InverseFourierV2}.

\begin{figure}[ht]
\centering
\includegraphics[width=0.4\linewidth]{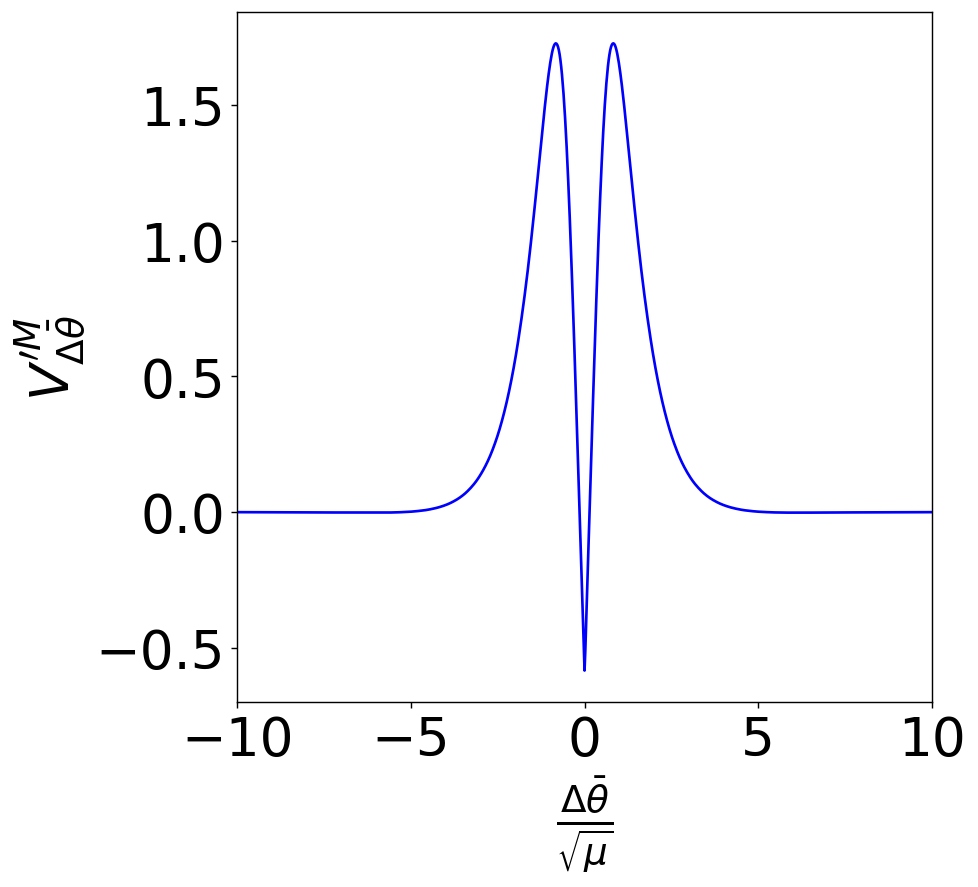}
\caption{Profile of metallic PFP in the rescaled angular space for U(1) gauge theory in $d =2$.
}
\label{Fig:U1_MFP_d2} 
\end{figure}

The universal coupling function becomes scale invariant if the low-energy limit is taken with fixed 
$\bar \theta /\sqrt{\mu}$.
The profile of the universal pairing interaction 
is shown in \fig{Fig:U1_MFP_d2}.
It consists of an attractive core at small angular separation and a repulsive interaction for $| \bar \theta_1 - \bar \theta_2| > \sqrt{\mu}$.
For 
$ |\bar{\theta}_1 - \bar{\theta}_2| \gg \sqrt{\mu}$ 
but with
$ |\bar{\theta}_1 - \bar{\theta}_2| \ll \sqrt{\KFAVdim}$,
the universal pairing interaction is governed by the small $y$ limit of \eq{stable_sol},
\begin{equation}
    \begin{aligned}
\aVM_{\bar{\theta}_1,\bar{\theta}_2} \sim
\left|\frac{\sqrt{\mu}}{\bar{\theta}_1-\bar{\theta}_2}\right|^{1+\sqrt{\discinf} }.
 \label{eq:stable_large_mom}
\end{aligned}
\end{equation}
It decays in $\left| \sqrt{\mu}/\Delta \bar \theta \right|^\Delta$ with the exponent $\Delta$ that monotonically increases with decreasing $d$ from $\Delta=2$ at $d=5/2$.
At the leading order in the $\epsilon$-expansion,  $\Delta \approx 4$ at $d=2$.
Consequently, the net two-body interaction at large angular separation decays faster than
$\sqrt{\mu}/|\bar \theta_1 - \bar \theta_2|$.
The particle number in each patch remains conserved in the low-energy limit, and the full $LU(1)$ symmetry emerges.
This implies that inter-patch couplings are unimportant at low energies, and the patch theory is a legitimate description.
Although the separatrix is not directly related to physical observables\footnote{The separatrix can be realized only if the pairing interactions at all angular momentum channels are tuned to be on the separatrix.}, its profile is given by 
\begin{equation}
\aVS_{\bar{\theta}_1,\bar{\theta}_2} \sim \left| \frac{\sqrt{\mu}}{\bar{\theta}_1 - \bar{\theta}_2} \right|^5.
\label{eq:VthetaSPXu1}
\end{equation}

In this stable non-Fermi liquid, the four-fermion coupling and higher-order couplings are fixed by the marginal parameters $(\kappa_{F,\theta},v_{F,\theta})$.
Consequently, small perturbations added to that universal coupling flow to zero at low energies, and they are irrelevant.
If one of those irrelevant couplings is dialed up beyond a critical strength, however, a superconducting instability arises.
In the next two sections, we discuss new types of non-Fermi liquids that emerge at such phase transitions.
\\

\subsubsection{
Critical non-Fermi liquids at charge-$2$ superconducting critical points}

If the bare 4-fermion coupling in one angular momentum channel is within ${\cal B}^{IR}_{\aV^{SC}_{-\infty}}$,
the coupling in that angular momentum channel diverges to $-\infty$ and the system becomes superconducting at low energies.
At the critical point that divides the superconducting phase and the stable non-Fermi liquid phase, a new type of non-Fermi liquid arises (see \fig{fig:NFLtoSC}).
The resulting non-Fermi liquid is described by the pairing interaction in one angular momentum channel on the separatrix PFP and all others on the metallic PFP.
This critical non-Fermi liquid, which is distinct from the stable one, has one relevant perturbation because the pairing interaction on the separatrix PFP requires fine-tuning. 
There also exist multi-critical non-Fermi liquids with couplings in $x>1$ different angular momentum channels on the separatrix PFP.
We refer to them as the \( x^{\text{th}} \) \textit{critical} non-Fermi liquid  ($\text{NFL}_\text{x}$). 

Let us first  consider $\text{NFL}_1$
and
extract the critical exponent $z \nu$ in 
\bqa
T_c \sim | \deltaaVUV |^{z \nu},
\label{eq:TcdeltaV}
\eqa
where $T_c$ is the superconducting transition temperature and $\deltaaVUV$
is the deviation of the bare coupling in angular momentum channel $n$ away from the critical strength. 
In the low-energy limit, the universal pairing interaction at general angular momentum $m$ becomes
\begin{equation}
   \av_{m}(l) = 
    \begin{cases}
        \aVM_{y^{(m)}(l)}, & m \neq n \\
        \aV^{S}_{y^{(m)}(l)}, & m = n,
    \end{cases}
\end{equation}
where
$y^{(m)}(l)$ is defined in \eq{eq:y}.
In the angular space, the universal pairing interaction at energy scale $\mu$ becomes
\begin{equation}
    \begin{aligned}
\aV^*_{\bar{\theta}_1,\bar{\theta}_2} = \aVM_{\bar{\theta}_1,\bar{\theta}_2}
+\frac{2 \pi\sqrt{\mu}}{\bar\theta_{\text{max}}}
\left(
\aV^\circ_{-\infty}
-\aV^\bullet_{-\infty}
\right)
\cos\left(\frac{n \pi (\bar\theta_1-\bar\theta_2)}{
2 \bar\theta_{\text{max}}
}\right),
\label{eq:cNFL1}
    \end{aligned}
\end{equation}
where $\aVM_{\bar{\theta}_1,\bar{\theta}_2}$ is the universal interaction of the stable non-Fermi liquid in \eq{eq:stable_large_mom}.
It is assumed that $l = \log \Lambda/\mu$ is large enough that 
$ \aVM_{y^{(n)}-l/2} \approx \aV^{\bullet}_{-\infty} $
and
$ \aV^S_{y^{(n)}-l/2} \approx \aV^{\circ}_{-\infty} $.
Similarly, the universal pairing interaction for $\text{NFL}_x$ takes the form of
\begin{equation}
    \begin{aligned}
\aV^*_{\bar{\theta}_1,\bar{\theta}_2} = \aVM_{\bar{\theta}_1,\bar{\theta}_2}
+\frac{2\pi\sqrt{\Lambda}e^{-l/2}}{\bar\theta_{\text{max}}}
\left(
\aV^\circ_{-\infty}
-\aV^\bullet_{-\infty}
\right)
\sum_{i=1}^x
\cos\left(\frac{n_i \pi (\bar\theta_1-\bar\theta_2)}{
2 \bar\theta_{\text{max}}
}\right),
\label{eq:cNFLx}
    \end{aligned}
\end{equation}
where $\{ n_i | 1 \leq i \leq x \}$ is the set of angular momentum channels whose pairing interactions are tuned to be on the separatrix.

In the small $\mu$ limit,
$\aV^*_{\bar\theta_1,\bar\theta_2}$ in Eqs. 
\eqref{eq:cNFL1} 
and
\eqref{eq:cNFLx}
approach $\aVM_{\bar\theta_1,\bar\theta_2}$ 
at any fixed $(\bar \theta_1, \bar \theta_2)$.
One may conclude that the difference between 
$\aV^*_{\bar\theta_1,\bar\theta_2}$ 
and
$\aV^M_{\bar\theta_1,\bar\theta_2}$ 
is negligible and that the critical non-Fermi liquid also respects the LU(1) symmetry, 
as is the case for the parent stable non-Fermi liquid.
However, this is not the case because of the $O(e^{-l/2})$ correction.
Unlike \eq{eq:InverseFourierV},
the support of 
the $O(e^{-l/2})$ correction
is extensive;
its magnitude does not decay at large $\bar \theta_1 - \bar \theta_2$.
Consequently, the anomalous dimension generated by this universal coupling function for composite operators, whose expression is similar to Eq. \eq{eq:etaV} but in angular momentum channels that include $n_i$, is not invariant under the general LU(1) transformation due to large-angle scatterings.
This implies that the critical non-Fermi liquid only has the $\text{OLU}(1)$ symmetry (see \eq{eq:olu1}), and the patch description is not valid.
Furthermore, the universal pairing interaction does not exhibit scale invariance, either in the space of angular momentum or in angular space.

Suppose the bare coupling in the angular momentum channel $n$ is tuned away from the separatrix PFP by $\deltaaVUV$. 
For $\deltaaVUV<0$, the ground state becomes a superconductor. 
The relation between $T_c$ and  $\deltaaVUV$ can be obtained through the PFP that passes through $\left(  y_0,  \aV^S_{y_0}+  \deltaaVUV \right)$,
where $y_0 \equiv y^{(n)}(0)= \log \left[ \frac{ \sqrt{\Lambda}} {2 \bar{\theta}_{\max}}  n \pi  \right]$.
If that PFP diverges to $-\infty$ at $y_{SC}$, the superconducting transition temperature is expressed as 
$T_c = \Lambda e^{-2z \left(
 y_0-y_{SC}
 \right)
 }
$, where $z$ is the dynamical critical exponent
and the factor of $2$ in the exponent arises due to the fact that $y$ runs as $-l/2$ with increasing logarithmic length scale $l$.
Since the $T_c$ vs. $\deltaaVUV$ relation takes different forms depending on how $\deltaaVUV$ and $n$ are tuned, we consider three different cases separately.

\begin{itemize}
\item 
Small $\deltaaVUV$ limit for a small fixed $n$

Let us first consider the small  $\deltaaVUV$  limit with a fixed $n$ with $y_0 < w_L$, where $w_L$ is the crossover logarithmic angular momentum above which the $\infty$-asymptotic behavior holds (see \eq{eq:wLwS}).
In this case, the coupling does not spend any significant RG time in the $\infty$-asymptotic region.
For small $\deltaaVUV$, the coupling diverges to $-\infty$ surely in the $-\infty$-asymptotic region.
The evolution of the PFP can be understood in three steps.
We first choose a small but fixed scale $y'$ such that 
$y' \ll w_S$.
For $y< y'$, the PFP is well described by the $y \rightarrow -\infty$ limit of the PFP equation.
In the first stage, the PFP is evolved from  $y_0$ to  $y'$.
In the small $\deltaaVUV$ limit, $\delta \aV$ at $y'$ should remain small and proportional to $\deltaaVUV$.
For $y<y'$, the growth of the deformation is well approximated by 
$\delta \aV_y = \delta \aV_{y'}e^{\sqrt{\discinf} 
(y' - y)}$ through the PFP equation linearized near the separatrix.
$\delta \aV_y$ becomes $O(1)$ at $y_1 \sim  y' - \frac{1}{\sqrt{\discinf}} \log \frac{1}{|\delta \aV_{y'}|}$. 
In $y<y_1$, we can no longer use the linearized PFP equation; however, the PFP is expected to diverge at $y_{SC} = y_1 - O(1)$.
Therefore, the superconducting transition temperature, which can be written as 
$T_c = \Lambda e^{- 2z ( y_0-y_1 + O(1) )}$,
scales as
\bqa
T_c \sim 
 \Lambda 
 \left|
 \deltaaVUV \right|^{\frac{2z}{\sqrt{\discinf}}}
 \label{eq:TcdelVA}
\eqa
in $\deltaaVUV$.
The critical exponent is determined from the discriminant in the $-\infty$ asymptotic region because the most growth of the coupling occurs in the small $y$ limit.

\item
Small $\deltaaVUV$ limit for a large fixed $n$

\begin{figure}[ht]
    \centering
   \includegraphics[width=0.45 \textwidth]{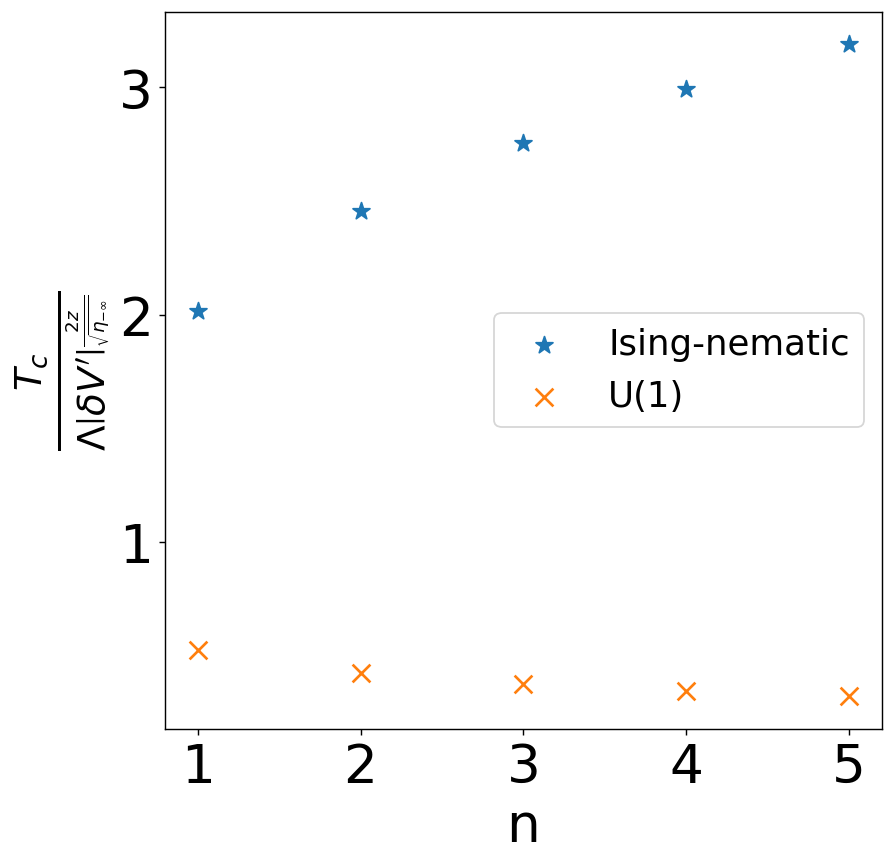}
    \caption{
$\frac{T_c}{\Lambda\left|\delta V^\prime \right|^{\frac{2z}{\sqrt{\eta_{-\infty}}}}}$ 
for the critical non-Fermi liquid realized in the U(1) gauge theory 
and the Ising-nematic quantum critical metal in 
$d\approx 2.485$. 
Both theories belong to class A in that dimension.
In the U(1) gauge theory (Ising-nematic critical metal), the pairing interaction is repulsive (attractive) in the s-wave channel and  $\etaPI/\etaPIII < 1$ ($\etaPI/\etaPIII > 1$).
This results in the opposite trends for the superconducting transition temperature as a function of the angular momentum when the stable non-Fermi liquids are deformed by a four-fermion coupling beyond the critical strength.
For the plot, we use 
$|\delta V'| = 0.1$.
}
\label{fig:Tcvsn}
\end{figure}

If $y_0 \gg w_L$, the coupling spends a significant RG time in the $\infty$-asymptotic region before reaching $y'$.
Nonetheless, it diverges to $-\infty$ in the $-\infty$-asymptotic region in the small $\deltaaVUV$ limit.
The only additional factor we need to consider compared to the previous case is the growth of the coupling that occurs between
$y_0$ and $y'$: 
$\delta \aV_{y'} \sim 
\deltaaVUV
e^{\sqrt{\etaPI} y_0}$.
Keeping all $y_0$ dependences in
$T_c \sim \Lambda e^{- 2z ( y_0-y_1 )}$, we obtain
\bqa
T_c \sim 
\Lambda 
n^{
2z \left( 
\sqrt{\frac{\etaPI}{\etaPIII}}
-1  
\right)
}
\left|
 \deltaaVUV \right|^{\frac{2z}{\sqrt{\discinf}}}.
 \label{eq:TcdelVA2}
\eqa
The dependence on $\deltaaVUV$ is the same as \eq{eq:TcdelVA}.
What is new here is the dependence on $n$.
As $n$ increases, $T_c$ increases (decreases) for 
$\etaPI > \etaPIII$
($\etaPI < \etaPIII$).
As $n$ increases,
the coupling spends longer RG time 
in the $\infty$ asymptotic region
and less RG time 
in the $-\infty$ asymptotic region.
If the growth is faster (slower) in the $\infty$ asymptotic region than in the $-\infty$ asymptotic region, $T_c$ is enhanced (suppressed) with increasing $n$.
It is noted that 
$\etaPI/\etaPIII > 1$ 
($\etaPI/\etaPIII < 1$)
for theories in which the universal pairing interaction is attractive (repulsive) 
in the s-wave channel.
In the U(1) gauge theory, $T_c$ is suppressed with increasing $n$ due to the repulsive nature of the universal pairing interaction in the s-wave channel.
This feature is determined entirely by the discriminant in the asymptotic regions and does not depend on the details of the crossovers.
Hence, the way $T_c$ depends on $n$ in critical non-Fermi liquids can be used to infer the nature of the critical fluctuations.
In the next section, we will discuss the Ising-nematic critical metal.
It is still in class A above a critical dimension, but the attractive interaction in the s-wave channel gives rise to the opposite trend as is shown in 
\fig{fig:Tcvsn}.

\item
Large $n$ limit for a small but fixed $\deltaaVUV$ 

In this limit, the PFP that goes through 
$\left( 
y_0,  \aV^S_{y_0}+  \deltaaVUV \right)$ 
diverges in the region with 
$y \gg w_L$.
Since the growth of the coupling is controlled by the discriminant in the $\infty$ asymptotic region,
the critical exponent is given by that of the $\infty$ asymptotic fixed point,
\bqa 
T_c \sim  \Lambda  \left| \deltaaVUV  \right|^{ \frac{2z}{\sqrt{\etaPI}} }. 
 \label{eq:TcdelVA3}
\eqa
\end{itemize}

\subsubsection{
Critical non-Fermi liquids
at charge-$2n$ superconducting critical points  
}
\begin{figure}[th]
\centering
\begin{subfigure}{.45\textwidth}
  \centering
\includegraphics[width=0.7\linewidth]{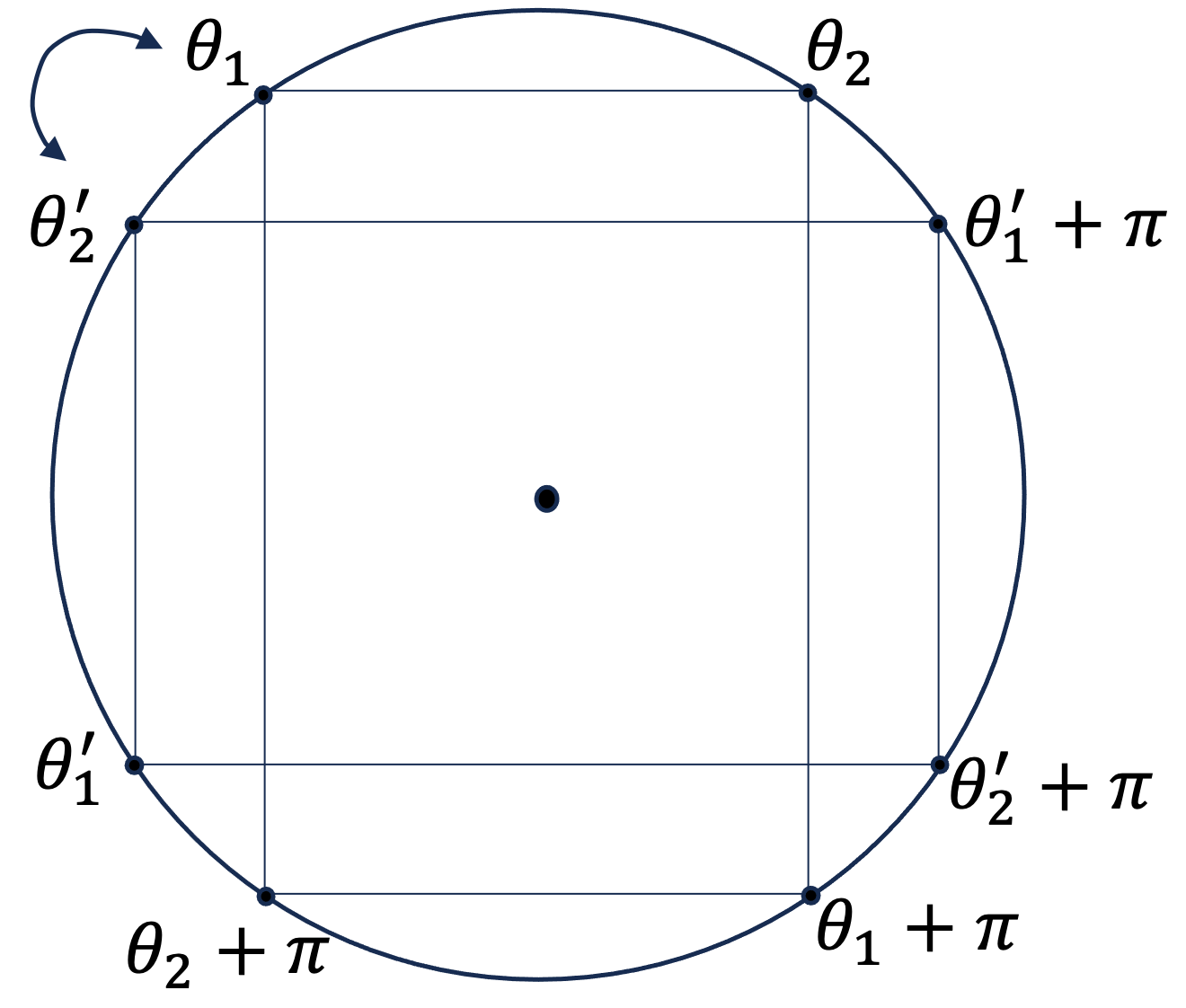}
  \caption{
}
 \label{fig:4fermion}
\end{subfigure}%
\begin{subfigure}{.45\textwidth}

  \centering
\includegraphics[width=0.65 \linewidth]{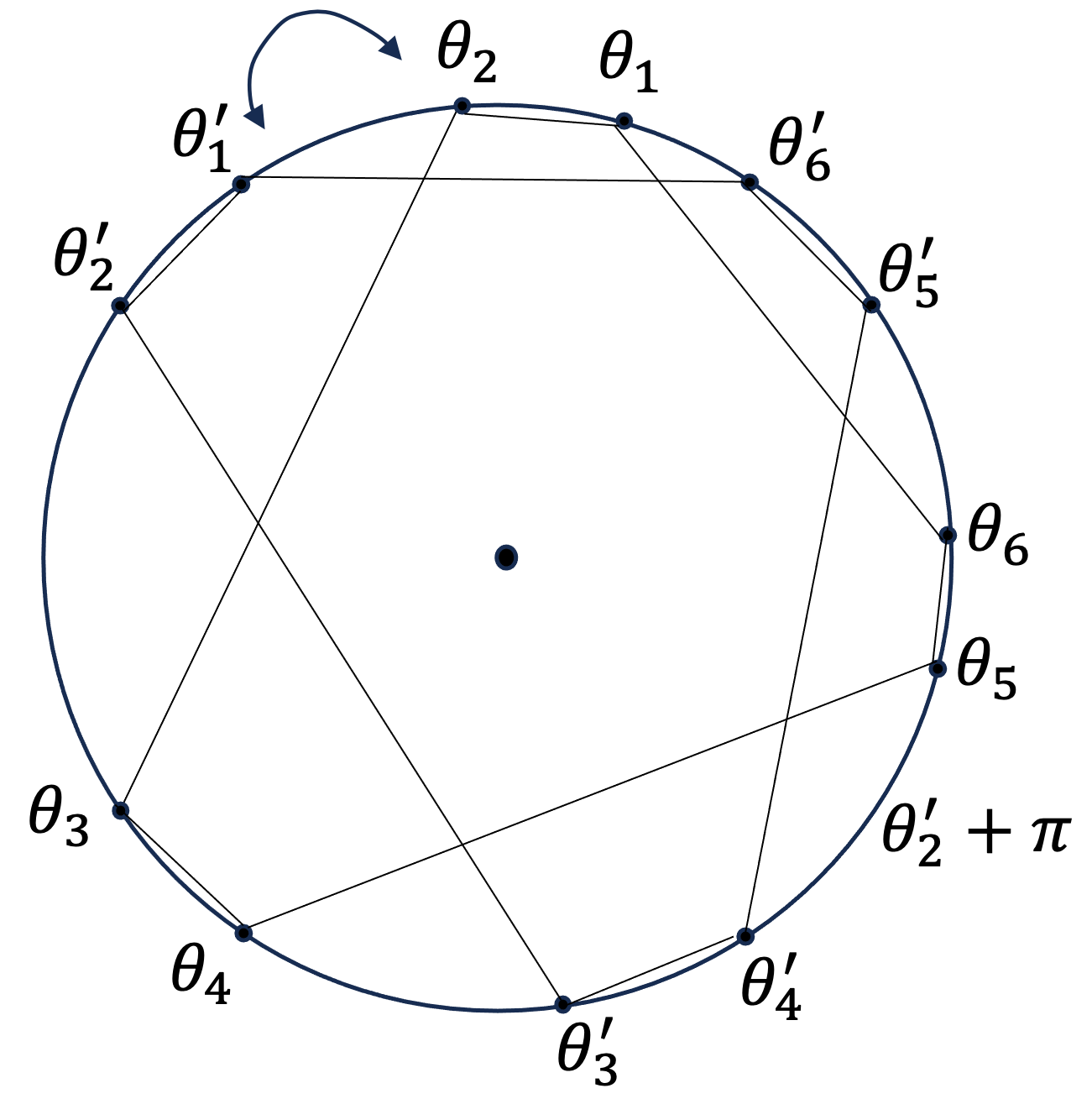}
  \caption{
}
\label{fig:6fermion}
\end{subfigure}%
    \caption{
The $4n$-fermion interaction in the pairing channel describes scatterings of $2n$ fermions placed near the Fermi surface with zero center of mass momentum 
to a different low-energy configuration of $2n$ fermions.
(a) For $n=2$, a low-energy $4$-fermion state must be composed of pairs of fermions located at antipodal points on the Fermi surface.
(b) For $n>2$, however, there are enough internal degrees of freedom so that the $2n$ low-energy fermions do not have to be made of pairs of fermions with opposite momenta.
}  \label{fig:4and6fermions}
\end{figure}

Now let us consider critical non-Fermi liquids that arise at the phase transitions from the stable non-Fermi liquid to a charge $2n$ superconductor with $ n \geq 2$.
In stable non-Fermi liquids, there is no immediate charge-$2$ superconducting instability that preempts other instabilities, and it is, in principle, possible to realize a charge $2n$-superconductivity with $n\geq 2$.
Such phase transitions can be driven by tuning the strength of the $4n$-fermion interaction.
For example, let us consider an angle-independent $4n$-fermion interaction,
\bqa
S_{4n}  &=&
 \mu^{1-(2n-1)d}
\int^{'}
\prod_{i=1}^{2n}\left[
d_{f}^{d+1} {\bf k}_i
d_{f}^{d+1} {\bf k}_i' \right] 
~
\delta^{d+1} \left( \sum_{a=1}^{2n} 
({\bf k}_a + {\bf k}'_a)
\right) ~
\lambda^{(4n)}_{ (\theta_1',..,\theta_{2n-2}'); (\theta_1,..,\theta_{2n-2}) }
\times \nn &&
~~~~~~~~~~~~~~~~~~~~
T_{R,R'}^{A,A'} 
o^{(2n) R}_{A}( {\bf k}_1,{\bf k}_2,.., {\bf k}_{2n})
o^{(2n) R'}_{A'}( {\bf k}'_1,{\bf k}'_2,.., {\bf k}'_{2n})
\nn
\eqa
where
\bqa
&& 
o^{(2n) R}_{A}( {\bf k}_1,{\bf k}_2,.., {\bf k}_{2n})
=
t_{A}^{i_1,j_1,..,i_n,j_n}
s^{R}_{m_1,..,m_n} 
~
o_{m_1;i_1,j_1}^{(P)}  ( \mathbf{k}_1, -\mathbf{k}_2)
o_{m_2;i_2,j_2}^{(P)}  ( \mathbf{k}_3, -\mathbf{k}_4)
..
o_{m_n;i_n,j_n}^{(P)}  ( \mathbf{k}_{2n-1}, -\mathbf{k}_{2n}) 
\label{eq:2nfermionint}
\eqa
represent $2n$-fermion operators that carry charge $2n$ or $-2n$ under the global U(1) in $d=2$:
$o^{(2n)R'}_{A'} \sim \psi^{2n}$
and
$o^{(2n)R}_{A} \sim (\psi^{\dagger })^{2n}$. 
In general $d$, the $2n$-fermion operator in the $2n$-particle channel is selected by the choice of $t^{i_1,j_1,..,i_n,j_n}_A$  
that specifies the quantum numbers of the operator under the $SO(4-d)$ spinor symmetry.
Similarly, $s^{R}_{m_1,..,m_n}$ determines the quantum numbers of the operator for the flavour symmetry.
$T^{A,A'}_{R,R'}$ is an invariant tensor under the symmetry.
%
$\theta_a$ and $\theta_a'$ are the polar angles associated with  ${\bf k}_a$ and ${\bf k}'_a$, respectively.
The interaction vertex describes the scattering of $2n$ low-energy fermions with the center of momentum close to zero, to another set of $2n$ low-energy fermions.
The associated coupling function $\lambda^{(4n)}$
generally depends on $2\times(2n-2)$ angles because $(2n-2)$ of the $2n$ incoming (outgoing) fermions can be placed anywhere around the Fermi surface, modulo some global constraints, while keeping all fermions close to the Fermi surface with zero total momentum.
The examples of $n=2,3$ are illustrated in \fig{fig:4and6fermions}.
At a large and negative $4n$-fermion coupling, the $2n$-fermion operator will condense, breaking the $U(1)$ to $Z_{2n}$ in $d=2$.
This is a charge $2n$-superconductor.

If the transition is continuous, a metallic state with critical $2n$-pairing fluctuations arises at the critical point.
In addition to the original critical bosons, the effective theory includes a new gapless boson $\Dtwon^R_A$ 
coupled with the $2n$-linear fermionic operator through,
\bqa
S_{\Delta}  &=&
\int^{'}
d_{b}^{d+1} {\bf q}
\left( |{\bf Q}|^2 + c_\Delta^2 |\vec q|^2 \right)
T_{R,R'}^{A,A'} 
\Dtwon^R_A({\bf q})
\Dtwon^{R'}_{A'}(-{\bf q}), \nn
S_{\Delta-\psi}  &=&
 \mu^{\frac{3-(2n-1)d}{2}}
\int^{'}
\prod_{i=1}^{2n}\left[
d_{f}^{d+1} {\bf k}_i
\right] 
d_{b}^{d+1} {\bf q}
~
\delta^{d+1} \left( \sum_{a=1}^{n} 
{\bf k}_a  + {\bf q}
\right) ~
e^{(2n)}_{(\theta_1,..,\theta_{2n-2}) }
T_{R,R'}^{A,A'} 
o^{(2n) R}_{A}( {\bf k}_1,{\bf k}_2,.., {\bf k}_{2n})
\Dtwon^{R'}_{A'}({\bf q}). \nn
\eqa
The nature of the resulting non-Fermi liquid depends on whether $e^{(2n)}$ is relevant or irrelevant at the decoupled fixed point with $e^{(2n)}=0$.
While the coupling is clearly irrelevant near the upper critical dimension, the fate of the coupling is unclear near $d=2$. 
Here, we leave this as an open question and remark on the nature of the critical non-Fermi liquid in the two scenarios.
If $e^{(2n)}$ remains irrelevant in $d=2$, the critical non-Fermi liquid is described by a theory composed of two decoupled sectors: one for the stable non-Fermi liquid phase and the other for the critical boson $\Dtwon$.
In this case, the universal low-energy physics is essentially identical to that of the parent non-Fermi liquid state. 
If $e^{(2n)}$ is relevant, on the other hand, one expects that the low-energy physics is described by yet another type of non-Fermi liquid.
In particular, the emergent symmetry of the resulting metallic state crucially depends on $n$.
For $n=2$, the low-energy charge $4$ operator with zero center of mass momentum must include a fermion at angle $\theta$ and another at $\theta+\pi$
(see \fig{fig:4and6fermions} a).
Therefore, the critical fluctuations of the charge $4$ operator still leave the OLU(1) group intact.
For $n \geq 3$, however, the $2n$-fermion composite has enough internal degrees of freedom that one does not need to place a pair of fermions at anti-podal points to ensure net zero momentum 
(see \fig{fig:4and6fermions} b).
Therefore, the non-Fermi liquids that arise at the charge $6$ or higher superconducting phase transition are expected to possess only the global U(1) group
\cite{2025arXiv250406508G} in the low-energy limit 
if $e^{(2n)}$ is relevant.

\section{
Example 2: $C_{2n}$ Ising-nematic quantum critical metal}
\label{sec:ex2}

\begin{figure}[th]
\centering
\begin{subfigure}{.4\textwidth}
  \centering
  \includegraphics[width=1.0\linewidth]{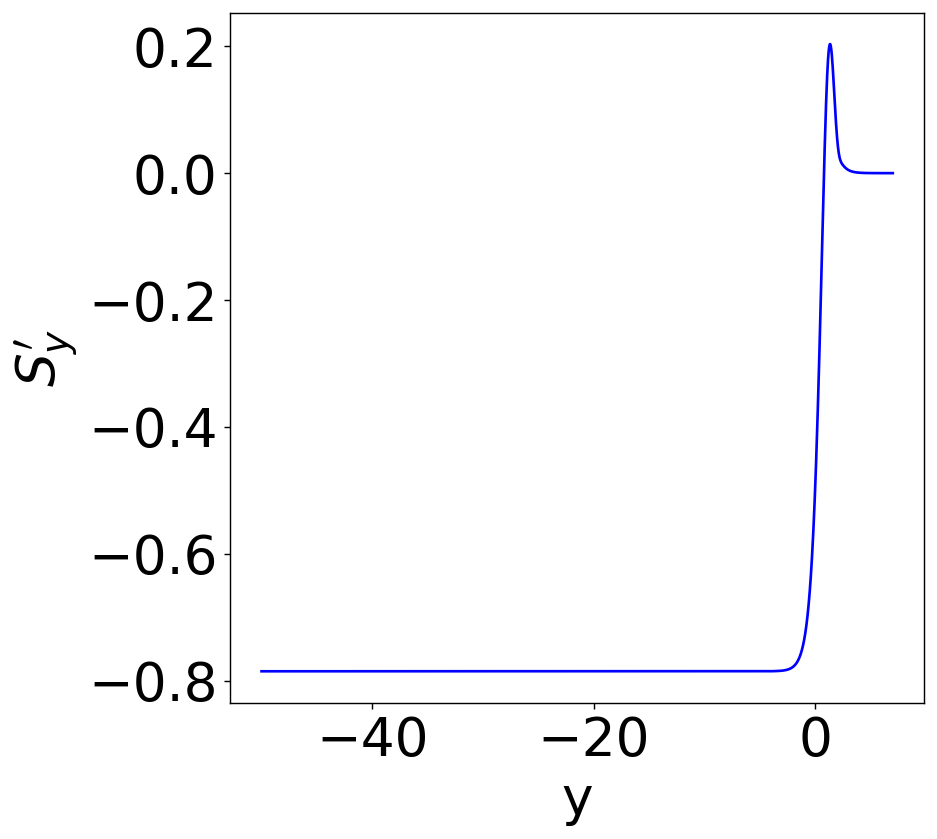}
  \caption{ 
}
  \label{fig:Source_d2_Ising}
\end{subfigure}%
\begin{subfigure}{.37\textwidth}
  \centering
  \includegraphics[width=1.0\linewidth]{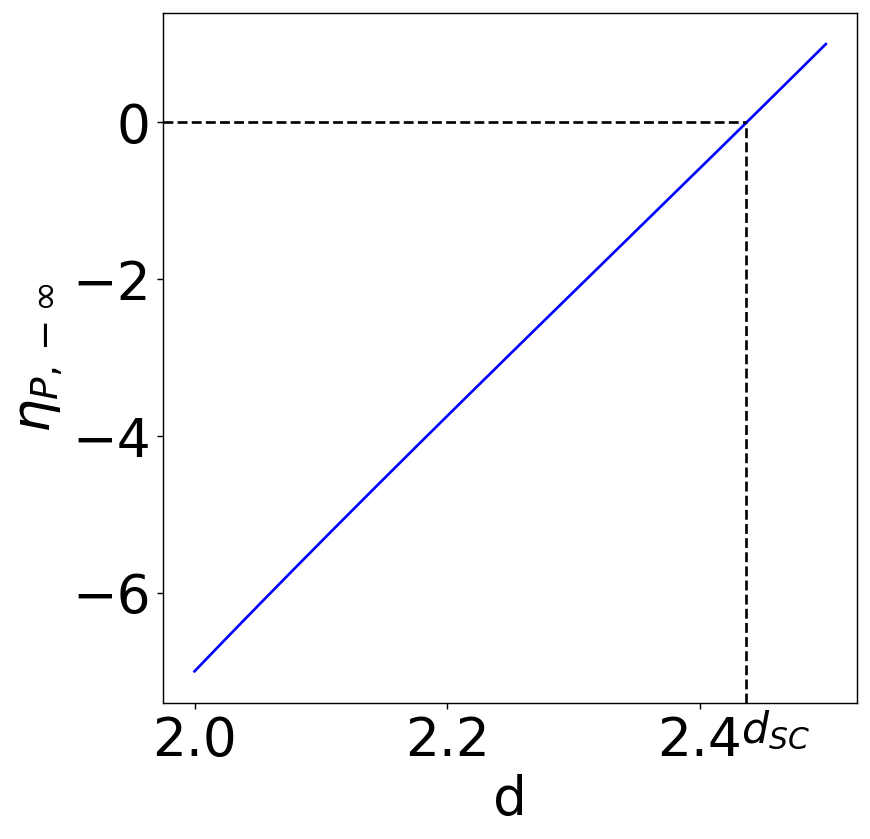}
  \caption{}
  \label{fig:disc_U(1)_Ising}
\end{subfigure}%
\caption{
(a) $\aS_y$ for the $C_4$-symmetric Ising-nematic quantum critical metal in $d=2$.
(b) $\discinf$ plotted as a function of $d$ for the Ising-nematic quantum critical metal.
}
\label{fig:SyetayforINQCM}
\end{figure}

In the Ising-nematic quantum critical metal,  \eq{eq:sttheta} is negative, and the pairing interaction generated from the critical boson is the opposite of that in the gauge theory, 
as shown in \fig{fig:Source_d2_Ising}.
Consequently, $\aS_y$ generates an attractive interaction in the s-wave limit and becomes repulsive at intermediate angular momenta before approaching zero in the large angular momentum limit.
Whether the system has the s-wave pairing instability or not
is determined by $\discinf$.
In \fig{fig:disc_U(1)_Ising}, we plot $\discinf$ as a function of $d$.
Near the upper critical dimension, the combination of the weak attractive interaction and the small density of states caused by the large co-dimension of the Fermi surface 
results in $\discinf>0$, excluding class C.
Near $d=2$, on the other hand, $\discinf$ becomes negative due to the strong attractive interaction and the large density of states,  despite the incoherence of fermions.\footnote{
Interestingly, $\HD$ remains $1$ in all $d$ to the leading order in $\epsilon$ as the $d$-dependence in the factor of $d-3/2$. 
As $d$ decreases, the decreasing co-dimension is compensated by the increasing anomalous dimensions, 
$3(d-1)(z-1) + 4 \eta_\psi$.}
The sign of $\discinf$  changes at a critical dimension
$d_{\text{SC}} \approx 2.44$\footnote{
Compared with $d_{\text{SC}}^{(0)}$ obtained in Ref.~\cite{PhysRevB.110.155142}, the current critical dimension is larger due to the contribution of the $\aV_y^2$ term in the beta function, which was not included in the previous study.
} at which the general PFP undergoes the topological phase transition. 
In the following, we discuss the three cases separately: 
(1) $d > d_{\text{SC}}$, (2) $d = d_{\text{SC}}$, and (3) $d < d_{\text{SC}}$.

\subsection{$d>d_{\text{SC}}$: stable non-Fermi liquid superuniversality class}


Since $\disc$ is positive at all $y$, the metallic PFP is regular.
From \eq{eq:ineq_1}, 
we conclude that the theory belongs to the stable NFL superuniversality class.
Most conclusions obtained for the U(1) gauge theory apply to this case with only quantitative changes.
At small $y$, the source term becomes
$\aS_y \approx -\frac{\eta_d}{4R_d} \left(1 - \frac{\alpha_d^{2/3}}{2} e^{2y}\right)$,
and the general PFP reads
\begin{equation}
    \begin{aligned}
\aV_{y} =-
 \frac{1}{4R_d}
\left(1+\tilde{\mathscr{C}}_y\left[\frac{c_2\Gamma\left(1-\frac{\sqrt{\discinf} }{2}\right)\left[I_{-1-\frac{1}{2} \sqrt{\discinf} }\left(\tilde{\mathscr{C}}_y\right)+I_{1-\frac{1}{2} \sqrt{\discinf} }\left(\tilde{\mathscr{C}}_y\right)\right]}{c_2\Gamma\left(1-\frac{\sqrt{\discinf} }{2}\right)I_{-\frac{1}{2} \sqrt{\discinf} }\left(\tilde{\mathscr{C}}_y\right)+i^{\sqrt{\discinf} }\Gamma\left(1+\frac{\sqrt{\discinf} }{2}\right)I_{\frac{1}{2} \sqrt{\discinf} }\left(\tilde{\mathscr{C}}_y\right)}\right.\right. \\
         \left.\left.+\frac{i^{\sqrt{\discinf} }\Gamma\left(1+\frac{\sqrt{\discinf} }{2}\right)\left[I_{-1+\frac{1}{2} \sqrt{\discinf} }\left(\tilde{\mathscr{C}}_y\right)+I_{1+\frac{1}{2} \sqrt{\discinf} }\left(\tilde{\mathscr{C}}_y\right)\right]}{c_2\Gamma\left(1-\frac{\sqrt{\discinf} }{2}\right)I_{-\frac{1}{2} \sqrt{\discinf} }\left(\tilde{\mathscr{C}}_y\right)+i^{\sqrt{\discinf} }\Gamma\left(1+\frac{\sqrt{\discinf} }{2}\right)I_{\frac{1}{2} \sqrt{\discinf} }\left(\tilde{\mathscr{C}}_y\right)}\right]\right),
         \label{eq:lbd_y_d>dsc}
    \end{aligned}
\end{equation}
where 
\bqa 
\tilde{\mathscr{C}}_y = \sqrt{\frac{\eta_d}{2}}\alpha^{\frac{1}{3}}_d e^y,
\label{eq:tildeC}
\eqa
and
$c_2$ is the constant of integration.
As in the U(1) gauge theory, $c_2=0$ for the separatrix PFP. 
\footnote{
In order for Eq. (\ref{eq:lbd_y_d>dsc}) to be real, 
$c_2 =i ^{\sqrt{\discinf}} \tilde c_2$  for a real $\tilde c_2$
with 
\bqa
\tilde c_{2} &=& -\frac{\Gamma \left(1+\frac{\sqrt{\discinf} ^{ }}{2}\right)}{\Gamma \left(1-\frac{\sqrt{\discinf} ^{ }}{2}\right)}\frac{\tilde{\mathscr{C}}_{0}^{-1}\left(1+4R_d\aVM_0\right) I_{\frac{\sqrt{\discinf} ^{ }}{2}}\left(\tilde{\mathscr{C}}_{0}\right)+ \ I_{-1+\frac{\sqrt{\discinf} ^{ }}{2}}\left(\tilde{\mathscr{C}}_{0}\right)+ I_{1+\frac{\sqrt{\discinf} ^{ }}{2}}\left(\tilde{\mathscr{C}}_{0}\right)}{\tilde{\mathscr{C}}_{0}^{-1}\left(1+4R_d\aVM_0\right)  I_{-\frac{\sqrt{\discinf} ^{ }}{2}}\left(\tilde{\mathscr{C}}_{0}\right)+  I_{-1-\frac{\sqrt{\discinf} ^{ }}{2}}\left(\tilde{\mathscr{C}}_{0}\right)+ I_{1-\frac{\sqrt{\discinf} ^{ }}{2}}\left(\tilde{\mathscr{C}}_{0}\right)}.
\label{eq:tildec_2}
\eqa
}
The separatrix PFP is given by Eqs. 
(\ref{eq:VySPXu1})
and
(\ref{eq:VthetaSPXu1})
with the replacement of
$\frac{d}{(4-d)^2(3-d)}\eta_d\rightarrow-\frac{\eta_d}{2}$ in $b^S$ and 
$\frac{d}{(4-d)^3(3-d)^2}\eta_d\rightarrow-\frac{\eta_d}{4}$ in $b^{S'}$.

\begin{figure}[th]
  \centering
  \includegraphics[width=0.4\linewidth]{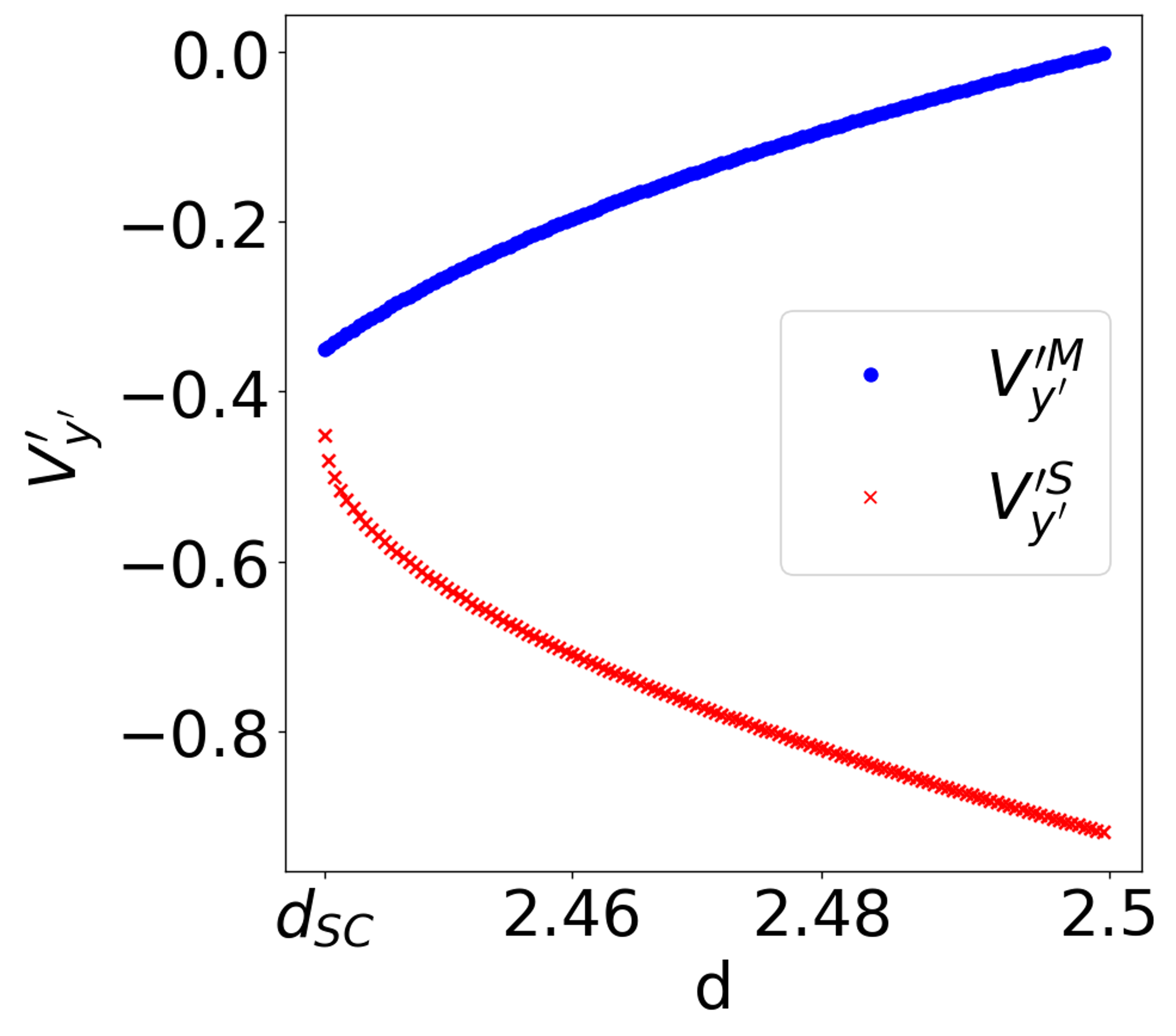}
 \caption{
 $\aVM_{y^\prime}$ and $\aVS_{y^\prime}$ 
 at $y^\prime \approx -7.2$ plotted 
 as functions of $d$ 
between $d_{SC} \approx 2.44$ and $d_C=5/2$.
Although the two PFPs approach as $d$ decreases, they remain separated all the way to $d=d_{SC}$.
}
  \label{fig:VMy_VSy_Ising}
  
\end{figure}
For $d \geq d_{SC}$, 
the metallic PFP is strictly above the separatrix PFP, as shown in Fig \ref{fig:VMy_VSy_Ising}.
One important quantitative difference from the U(1) gauge theory is that $\discinf <1$, 
as opposed to $\discinf >1$ in the U(1) gauge theory.
This gives rise to two observable differences from the U(1) gauge theory.
The first is that 
the metallic PFP, which is given by $\aVM_y = \aV^\bullet_{-\infty}+b^M e^{\sqrt{\discinf} y}$ with  
$b^M = -\frac{\sqrt{\discinf} }{
        2
        R_d
        c_2 i^{-\sqrt{\discinf}}}
        \left(
        \frac{\tilde{\mathscr{C}}_0}{2}
        \right)^{\sqrt{\discinf} }$,
decays as
$1/|\bar \theta_1 - \bar \theta_2|^\Delta$
with $\Delta < 2$.
\footnote{
We note that the Fermi liquid correction shifts the exponent from $2\Delta_d = 2-2\eta_d$ calculated in \cite{PhysRevB.110.155142} to $1+\sqrt{\discinf}$.}
The slower decay of the large-angle scattering, compared with the U(1) gauge theory, is caused by the attractive pairing interaction generated by the Ising-nematic fluctuations.
Despite this quantitative difference, 
the large-angle scatterings remain irrelevant above the critical dimension\cite{PhysRevB.110.155142} because
$\Delta > 1$ in 
$d>d_{SC}$.
The second difference is that $\etaPI/\etaPIII>1$ in the Ising-nematic critical metal.
This results in the increasing behavior of $T_c$ with increasing angular momentum $n$ when the pairing interaction in that angular momentum channel is tuned beyond the critical value, as shown in 
 \fig{eq:TcdelVA}.

\subsection{$d=d_{\text{SC}}$: NFL to s-wave SC critical superuniversality class}

\begin{figure}[h]
    \centering
        \includegraphics[height=6cm,width=7cm]{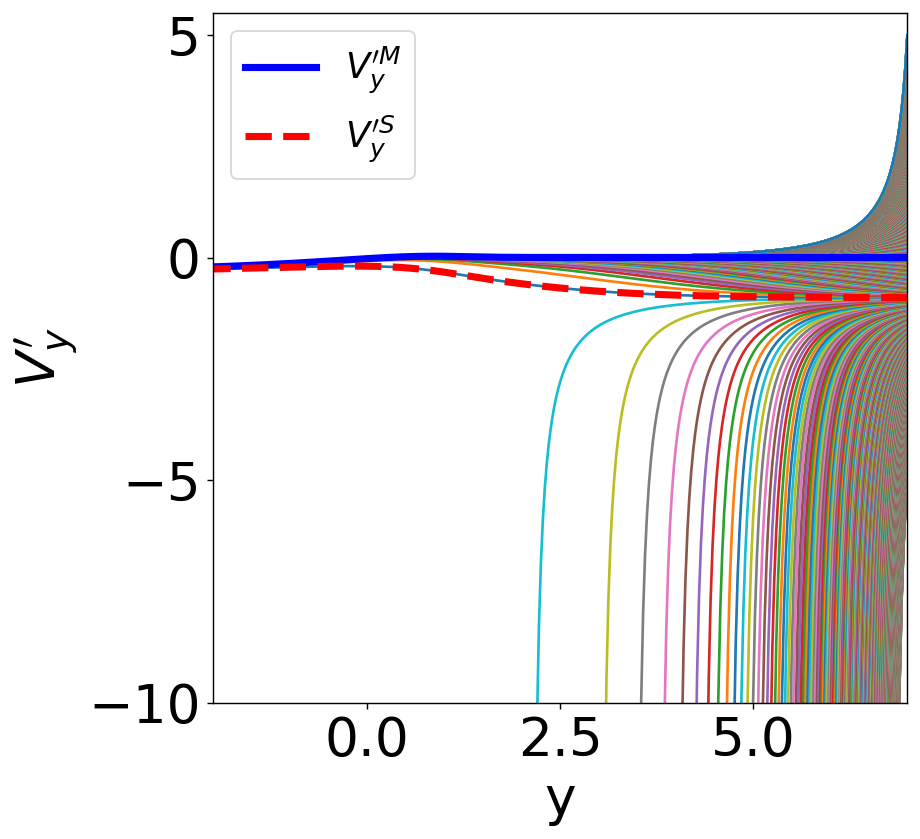}
        \caption{
The portrait of the general PFP for the $C_4$-symmetric Ising-nematic quantum critical metal at the critical dimension $d_{SC}$.
        }
        \label{Fig:Ising_phase_dsc}
\end{figure}

As $d$ approaches $d_{SC}$, two $-\infty$-asymptotic fixed points become closer.
At $d_{SC}$, they collide and merge into one marginal $-\infty$-asymptotic fixed point, 
$      \aV^{\halfominus}_{-\infty}=-\frac{1}{4R_d}$.
A small perturbation $\delta \aV_{-\infty}$ added to the asymptotic fixed point decays (grows) logarithmically in energy for positive (negative) $\delta \aV_{-\infty}$;
$\delta \acute V_{-\infty}(l) = \frac{\delta \acute V_{-\infty}(0)}{1+R_d\delta \acute V_{-\infty}(0) l}$.
In \fig{Fig:Ising_phase_dsc}, we show the general PFPs at $d_{SC}$.
The theory at $d_{SC}$ can be, in principle, in class
AC, BC, or ABC.
The deciding factor is whether the metallic PFP that emanates from $\aV^\bullet_\infty$ is in 
$
{\cal B}^{IR}_{\aV^\halfominus_{-\infty}}
-\partial {\cal B}^{IR}_{\aV^\halfominus_{-\infty}}$ (class AC),
$\partial  {\cal B}^{IR}_{\aV^\halfominus_{-\infty}}$ (class ABC),
or
${\cal B}^{IR}_{\aV^{SC}_{-\infty}}$ (class BC).
To determine this, we find the expression for the general PFP in the small $y$ limit,
\begin{equation}
    \begin{aligned}
\aV_y 
  = -\frac{1}{4R_d}
\left(1+\frac{\left(\alpha_d\right)^{\frac{1}{3}}e^y}{
 \sqrt{2}}\frac{
I_1\left(\frac{\left(\alpha_d\right)^{\frac{1}{3}}e^y}{2 \sqrt{2}}\right) -2 c_3 K_1\left(\frac{\left(\alpha_d\right)^{\frac{1}{3}}e^y}{2 \sqrt{2}}\right)}{
I_0\left(\frac{\left(\alpha_d\right)^{\frac{1}{3}}e^y}{2 \sqrt{2}}\right)  +2 c_3 K_0\left(\frac{\left(\alpha_d\right)^{\frac{1}{3}}e^y}{2
   \sqrt{2}}\right) }\right),
   \label{dsc_sol}
    \end{aligned}
\end{equation} 
where $I_{\alpha}(x)$ and $K_{\alpha}(x)$ are the modified Bessel functions of the first and second kind, respectively, and $c_{3}$ is the constant of integration.

\begin{figure}[h]
\centering
\includegraphics[width=0.4\linewidth]{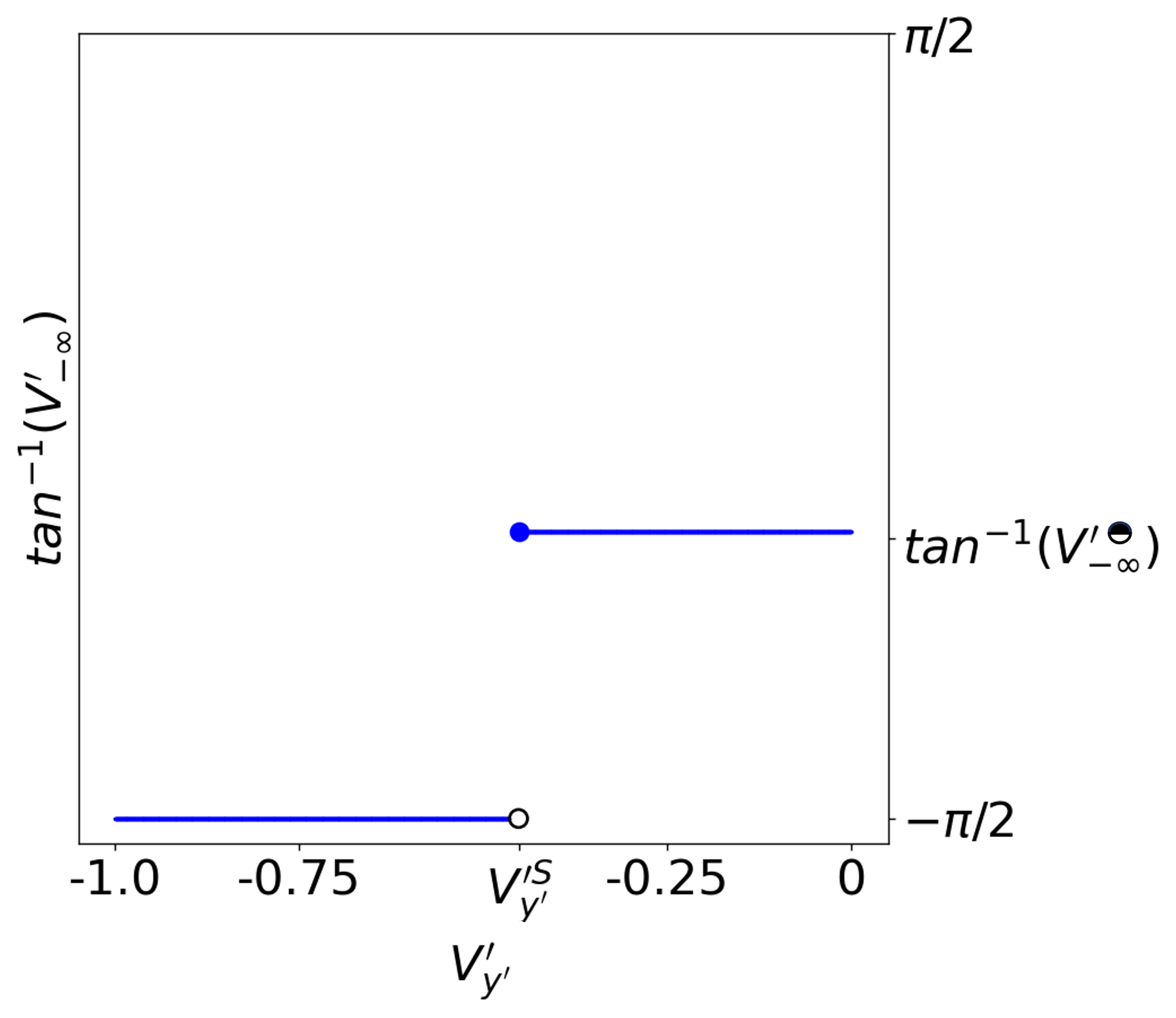}
\caption{
$\tan^{-1} \aV_{-\infty}$ in \eq{dsc_sol} plotted as a function of 
$\aV_{y^\prime}$ with $y^\prime\approx -7.2$.
As $\aV_{y^\prime}$ crosses the critical value,
which is close to $\aV^{S}_{y^\prime}$, 
$\aV_{-\infty}$ jumps from the marginal $-\infty$ asymptotic fixed point  to $-\infty$.
}
\label{Fig:V_IR_Ising_dsc} 
\end{figure}  

In Fig. \ref{Fig:V_IR_Ising_dsc}, we plot $\aV_{-\infty}$ as a function of  $\aV_{y^\prime}$ for a fixed $y'$.
As is shown in  \fig{fig:VMy_VSy_Ising},
$\aVM_{y'} > \aVS_{y'}$.
Since two distinct PFPs cannot cross, the metallic PFP is strictly above the separatrix PFP at all $y> -\infty$.
This can also be checked through the asymptotic analytic solutions.
The separatrix PFP, which arises at $c_3=0$, becomes
\begin{equation}
    \begin{aligned}
        \aVS_y = -\frac{1}{4R_d}-\frac{\alpha_d^{2/3}}{32R_d}e^{2y}-
        \frac{3\sqrt{3}}{512\pi}\frac{\alpha_d^{\frac{4}{3}}}{R_d}
        ye^{4y}
\label{eq:separatrixPFPdsc}
    \end{aligned}
\end{equation}
in the small $y$ limit.
%
The metallic PFP, which satisfies the asymptotic boundary condition of 
$\lim_{y \rightarrow \infty} \aVM_y = \aV^\bullet_{\infty}$\footnote{The constant of integration for the metallic PFP is $c_{3} \approx 0.838253$.}
becomes
\begin{equation}
    \begin{aligned}
        \aVM_y = -\frac{1}{
        4
        R_d
        }-\frac{1}{
        2
        R_d}\frac{1}{y}
        \label{eq:dsc_profile_y}
    \end{aligned}
\end{equation}
in the small $y$ limit.
Note that $\aVM_y$ is strictly above $\aVS_y$ for small $y$ where these solutions are valid.
The difference between $\aVM_y$ 
and $\aVS_y$ also shows up in its local stability.
The profile of a small perturbation added to $\aVM_y$ is given by
\begin{equation}
    \delta \aV_{y^\prime} =
    \left(\frac{y}{y^\prime}\right)^2 
    \delta \aV_{
    y
    }
    \label{eq:smally_perturbation_dsc}
\end{equation}
in the small $y$ limit.
Clearly, $\frac{\delta \aV_{y^\prime}}{\delta \aV_{y}}$ decays as $1/y^{\prime 2}$ in the small $y^\prime$ limit, irrespective of the sign of $\delta V_y$.
This implies that the metallic PFP is {\it in}
$
{\cal B}^{IR}_{
\aV^\halfominus_{-\infty}
}
-\partial {\cal B}^{IR}_{\aV^\halfominus_{-\infty}}$
and
the Ising-nematic quantum critical metal at $d=d_{SC}$ belongs to class AC (the NFL to s-wave SC superuniversality class).
%


If all bare couplings are in the basin of attraction of the metallic PFP, the normal state is stable down to zero temperature.
The universal pairing interaction of the resulting non-Fermi liquid is governed by the metallic PFP.
In the angular space, the metallic PFP takes the form of\cite{WONG1978173}
\begin{equation}
    \begin{aligned}
\aVM_{\bar\theta_1,\bar{\theta}_2}
\sim \frac{\sqrt{\mu}}{\left|\bar{\theta}_1-\bar{\theta}_2\right|\log^2\left(\frac{\left|\bar{\theta}_1-\bar{\theta}_2\right|}{\sqrt{\mu}}\right)}.
 \label{eq:stable_large_mom_dsc}
    \end{aligned}
\end{equation}
Unlike the stable non-Fermi liquids in class A,
 large-angle scattering is only marginally suppressed in class AC.

If the bare coupling in one (or more) angular momentum channel is tuned across the separatrix PFP, a quantum phase transition is induced from the stable non-Fermi liquid to a superconducting state.
In this superuniversality class AC, the critical non-Fermi liquid is characterized by a slower rise of $T_c$ with the deformation.
This is because the deformation only grows logarithmically in energy at small $y$ due to the marginal nature of the $-\infty$ asymptotic fixed point.
Suppose that the bare coupling in the angular momentum channel $n$ is deformed below the separatrix PFP. 
Let $y_0 = y^{(n)}(0)$ and $\deltaaVUV = \delta \aV_{y_0}(0)$ denote the logarithmic angular momentum $n$ and the deformation made in that angular momentum channel away from the separatrix.
At small $y$, the PFP 
connected to $\left( y_0,  \aV^S_{y_0}+  \deltaaVUV \right)$ can be written as
\begin{equation}
    \begin{aligned}
    \aV_y 
  = -\frac{1}{4R_d}
\left(1-
\frac{1}{
\frac{1}{4R_d \delta \aV_{y'}} 
- \frac{1}{2}(y-y')
}
\right),
        \label{appendix:ystar_ac}
    \end{aligned}
\end{equation}
where $y' \ll 0$ is a fixed reference scale below which the asymptotic expression of PFP in \eq{dsc_sol} is valid, and $\delta \aV_{y'}$ is the deformation away from the separatrix PFP at $y'$.
For $y_0$ that is not large,
$\delta \aV_{y'} = c \deltaaVUV$ 
for $O(1)$ coefficient $c$.
For $\deltaaVUV <0$, the PFP diverges to $-\infty$ at
$y_{SC} = y_0+ 
\frac{1}{2 c R_d \deltaaVUV}+O(1)$
and the superconducting transition temperature $T_c \sim \Lambda e^{-2z (y_0-y_{SC})}$ scales as
\bqa
T_{c} 
\sim
\Lambda
\exp\left\{
-\frac{z}{c R_d |\deltaaVUV|}
\right\} 
\eqa
for a constant $c$.
For large $y_0$, $c$ becomes large due to the growth of the deformation that takes place between $y_0$ and $y'$:
$c = c' e^{y_0}$, where $c'$ is a constant that depends only on $y'$ and the details of the separatrix PFP.
This leads to
\bqa
T_c \sim \Lambda
\left(
\frac{1}{n}
\sqrt{\frac{\mathbf{k}_F}{\Lambda}} \right)^{2z}
\exp\left\{-
\frac{2\gamma z}{c^\prime R_d n\pi}
\sqrt{\frac{\mathbf{k}_F}{\Lambda}}
\frac{1}{|\deltaaVUV| }\right\} 
\label{eq:TcdelaVclassAC}
\eqa
for a constant $c'$.
As $n$ increases for a fixed and small $|\deltaaVUV|$, $T_c$ increases.
This is because the faster growth of the coupling in the large $y$ region, compared with the growth in the small $y$ region, expedites superconducting instabilities with increasing $n$.
This trend is expected from  \eq{eq:TcdelVA2} because $\etaPI > \etaPIII=0$ in class AC.
The dependence of $T_c$ on $\deltaaVUV$ is similar to that of Fermi liquids;
in both cases, the perturbation grows logarithmically in energy.
However, the non-Fermi liquids in class AC exhibits the strong angular momentum dependence of $T_c$ unlike in Fermi liquids.
This is attributed to the universal pairing interaction that strongly depends on angular momentum.

\subsection{$d<d_{\text{SC}}$: s-wave superconducting superuniversality class}

\label{sec:d<dsc}

In $d<d_{SC}$, the $-\infty$-asymptotic fixed points become complex, and the s-wave pairing interaction exhibits a one-way flow to $-\infty$.
This can be also seen from the violation of \eq{eq:ineq_2_5} for small $d$ (see Appendix 
\ref{eq:absence_Hermitian_PFP}).
Therefore, it belongs to the s-wave SC superuniversality class (class C).
For non-s-wave pairing channels, instabilities can be suppressed  
at least above the s-wave superconducting transition temperature by tuning down the Fermi momentum, which shifts $y^{(m)}(0)$ to the region of larger $y$ with $\eta_{P,y}>0$,
and by making the bare four-fermion coupling more repulsive.
The flow of the s-wave pairing interaction can be obtained in closed form thanks to the lack of flow of $y$ at $y=-\infty$, 
\begin{equation}
    \aV_{-\infty}(l) = -\frac{1}{4R_d} + \frac{\sqrt{-\discinf}}{4R_d} \tan\left[-\frac{1}{4} \sqrt{-\discinf}\, l + \arctan\left(\frac{4R_d \aV^{\text{UV}}_{-\infty} + 1}{\sqrt{-\discinf}}\right)\right].
    \label{lambda0_l}
\end{equation}
Here, $\aV^{UV}_{-\infty}$ is the bare s-wave coupling defined at $l=0$.
The superconducting transition temperature becomes 
$T_c \sim \Lambda  e^{-z l_{\text{SC};-\infty} }$, 
where $z$ is the dynamical critical exponent and $l_{\text{SC}}$ is the logarithmic length scale at which the coupling diverges, 
\begin{equation}
    l_{\text{SC};-\infty} 
    = \frac{2\pi + 4 \arctan\left(\frac{4R_d \aV^{\text{UV}}_{-\infty} + 1}{\sqrt{-\discinf}}\right)}{\sqrt{-\discinf}}.
    \label{eq:lscswave}
\end{equation}
No matter how repulsive the bare coupling in the s-wave channel is, the superconducting scale and $T_c$ are bounded as
\bqa
l_{SC;-\infty} < l_{SC,max} = \frac{4\pi}{\sqrt{-\discinf}},
~~~~~
T_c > T_{c,min} = \Lambda  e^{- 
\frac{4\pi z}{\sqrt{-\discinf}}}.
\label{eq:Tcmin}
\eqa
At $d=2$, it is expected that both $z$ and $ -\discinf$ are $O(1)$, and the superconducting transition temperature may not be very different from $\Lambda$ below which the non-Fermi liquid sets in.
In that case, our low-energy effective theory does not have much to say about the normal state, which is largely governed by non-universal high-energy physics.
In $d$ close to $d_{SC}$, however, there is a large hierarchy between $\Lambda$ and $T_c$ because $\lim_{d \rightarrow d_{SC}^-} T_{c,min}/\Lambda = 0$.
In such cases, one expects universal behaviors to emerge in the normal state above the superconducting transition temperature but below the UV cutoff.
In the remainder of this section, we discuss the quasi-universal behaviors that emerge in the normal state of non-Fermi liquids in class C proximate to class A with $T_c \ll \Lambda$.
In Sec. \ref{sec:ex3}, we will discuss a more physical example that exhibits a large $\Lambda/T_c$. 
Everything we discuss here can be directly applied to that example.

\subsubsection{Complex asymptotic fixed points}

\begin{figure}[th]
\centering
\begin{subfigure}{.35\textwidth}
  \centering
\includegraphics[width=1.0\linewidth]{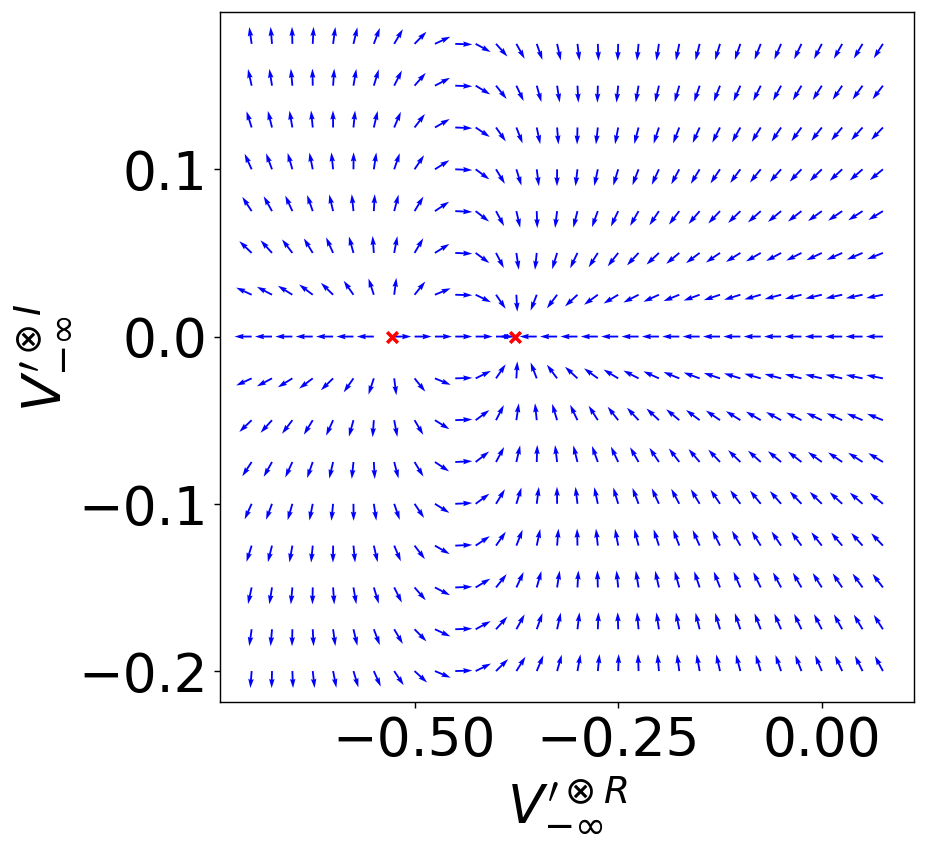}
  \caption{
}
 \label{fig:Complex_FP_d>dsc}
\end{subfigure}%
\begin{subfigure}{.35\textwidth}
  \centering
\includegraphics[width=1.0\linewidth]{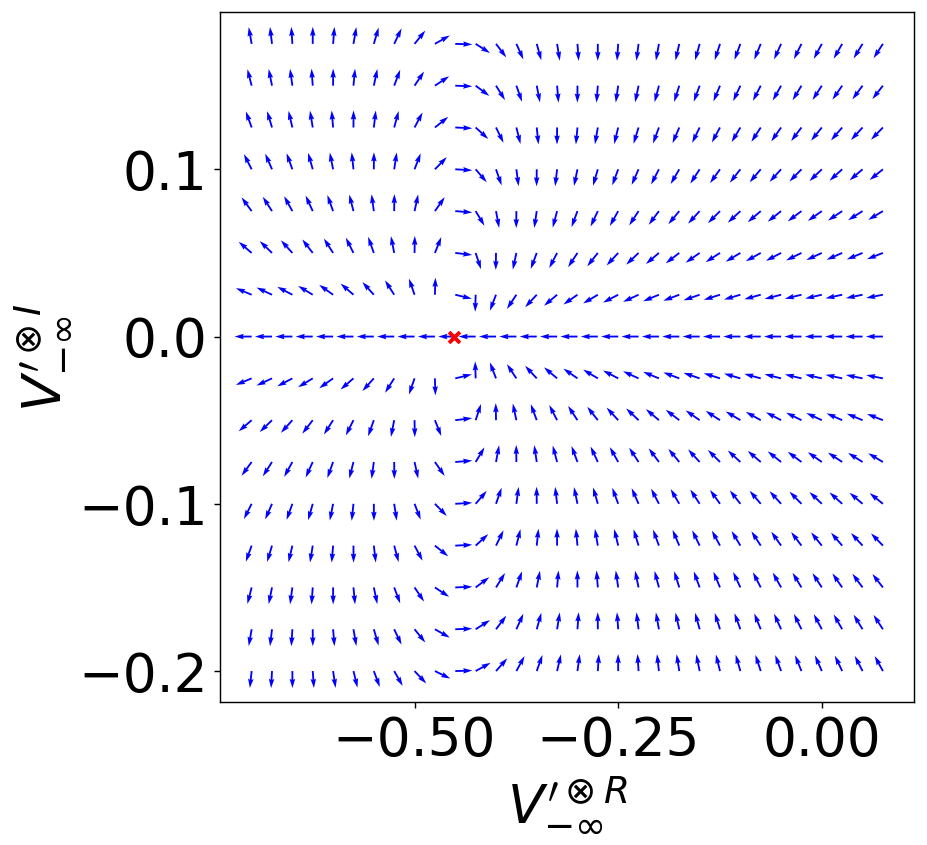}
  \caption{
}
\label{fig:Complex_FP_d=dsc}
\end{subfigure}%
\begin{subfigure}{.35\textwidth}
  \centering
\includegraphics[width=1.0\linewidth]{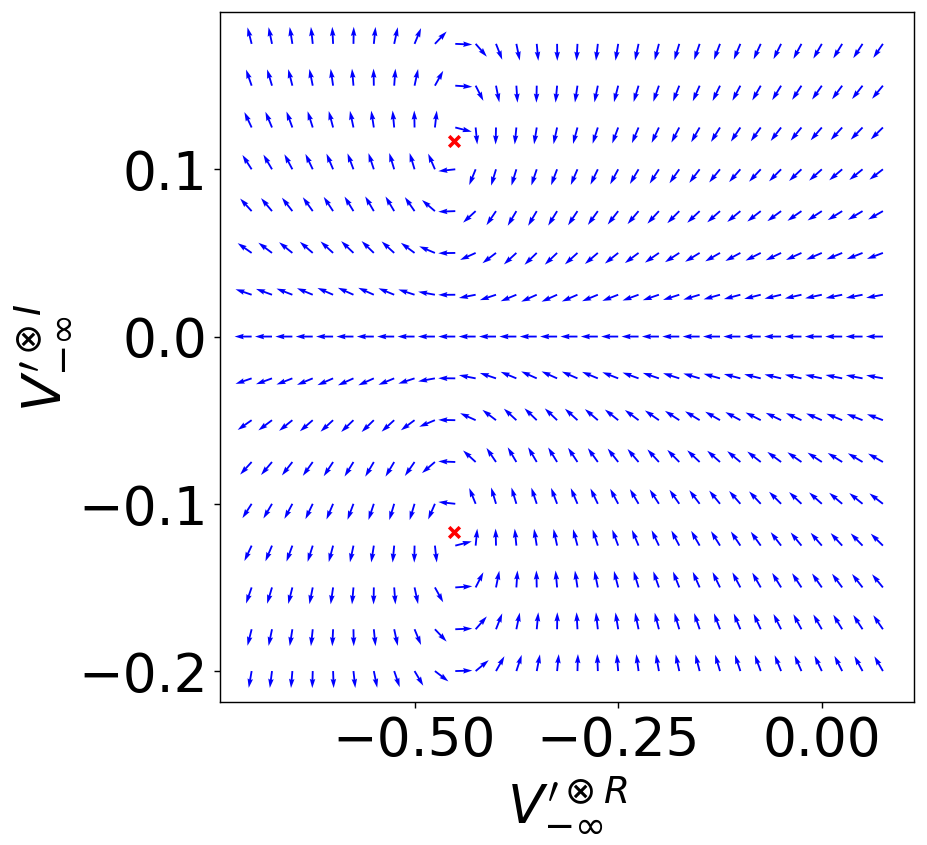}
  \caption{
}
\label{fig:Complex_FP_d<dsc}
\end{subfigure}
\caption{
As $d$ decreases from 
(a) above $d_{SC}$ 
to (c) below $d_{SC}$ 
through (b) the critical dimension $d_{SC}$, the stable and unstable $-\infty$ asymptotic fixed points denoted by the red dots in (a) collide,
becomes a marginal fixed point in (b), and become complex in (c).
Below $d_{SC}$, the physical coupling in the real axis flows to $-\infty$, leading to the superconducting instability in the s-wave channel.
Near the complex fixed points, the RG flow exhibits an oscillatory behavior with the `frequency' that is proportional to $\sqrt{-\discinf}$. 
}
    \label{fig:lambda0_fp}
\end{figure}

In $d<d_{SC}$, 
a pair of non-Hermitian (complex) fixed points arise at
(see \fig{fig:lambda0_fp})
\begin{equation}
    \aV^\otimes_{-\infty;\pm} = 
    \frac{1}{4R_d} \left(-1 \pm i \sqrt{-\discinf} \right).
    \label{eq:fixedpts_d<dsc}
\end{equation}
While the s-wave coupling runs toward $-\infty$ on the real axis, the coupling near the complex asymptotic fixed point exhibits the circular flow.
A small perturbation 
$\delta \aV_{-\infty}$
added to $\aV^\otimes_{-\infty;\pm}$ evolves as
\begin{equation}
\frac{d \delta \aV_{-\infty}}{dl} = \mp i 
\Omega^\otimes
\, \delta \aV_{-\infty},
\label{eq:delta_lbd0_RG_2}
\end{equation}
where $\Omega^\otimes
=\frac{\sqrt{-\discinf}}{2}$
is the frequency with which the perturbation oscillates around the complex fixed point with increasing $l$\cite{
VEYTSMAN1993315,
PhysRevLett.89.230401,
PhysRevB.69.020505,
PhysRevD.75.025005,
PhysRevLett.108.131601}.
This oscillatory behavior is shown in     \fig{fig:lambda0_fp}.

Near $d_{SC}$, the complex asymptotic fixed points are close to the real axis, and the RG flow becomes slow near 
\bqa
\aV_{b} =  -\frac{1}{4R_d}.
\label{eq:aVatbottleneck}
\eqa
We refer to this region of coupling as {\it bottleneck} region.
If the coupling enters this bottleneck region from the repulsive side, it takes a logarithmic scale of
\bqa
l_b = 
 \frac{2\pi}{\sqrt{-\discinf}}
 = \frac{l_{SC,max}}{2}
\label{eq:lthroughbottleneck}
\eqa
to pass through it.
This is the scale that it takes for the s-wave coupling to flow from the center of the bottleneck, $\aV_{-\infty}=-\frac{1}{4R_d}$
to $-\infty$, or alternatively from
$\infty$ to the center of the bottleneck.
For small $|\discinf|$, $l_b$ becomes large,
and the RG flow becomes severely stagnated in this region.

The complex fixed points themselves are not of direct interest to us because the physical theories are Hermitian with real $\aV_y$.
Nonetheless, the existence of the complex fixed points has non-trivial consequences for the behavior of the Hermitian theories due to the analyticity of the beta functional.
To see this, we consider the general PFP in the small $y$ limit,
\begin{equation}
    \begin{aligned}
         \aV_{y} =-\frac{1}{
         4R_d}\left(1+\tilde{\mathscr{C}}_y\left[\frac{c_4\Gamma\left(1-i\frac{\sqrt{-\discinf}}{2} \right)\left[I_{-1-i\frac{\sqrt{-\discinf}}{2} }\left(\tilde{\mathscr{C}}_y\right)+I_{1-i\frac{\sqrt{-\discinf}}{2} }\left(\tilde{\mathscr{C}}_y\right)\right]}{c_4\Gamma\left(1-i\frac{\sqrt{-\discinf}}{2} \right)I_{-i\frac{\sqrt{-\discinf}}{2} }\left(\tilde{\mathscr{C}}_y\right)+e^{-\frac{\pi}{2}\sqrt{-\discinf}}\Gamma\left(1+i\frac{\sqrt{-\discinf}}{2} \right)I_{i\frac{\sqrt{-\discinf}}{2} }\left(\tilde{\mathscr{C}}_y\right)}\right.\right.\\\left.\left.+\frac{e^{-\frac{\pi}{2}\sqrt{-\discinf}}\Gamma\left(1+i\frac{\sqrt{-\discinf}}{2} \right)\left[I_{-1+i\frac{\sqrt{-\discinf}}{2} }\left(\tilde{\mathscr{C}}_y\right)+I_{1+i\frac{\sqrt{-\discinf}}{2} }\left(\tilde{\mathscr{C}}_y\right)\right]}{c_4\Gamma\left(1-i\frac{\sqrt{-\discinf}}{2} \right)I_{-i\frac{\sqrt{-\discinf}}{2} }\left(\tilde{\mathscr{C}}_y\right)+e^{-\frac{\pi}{2}\sqrt{-\discinf}}\Gamma\left(1+i\frac{\sqrt{-\discinf}}{2} \right)I_{i\frac{\sqrt{-\discinf}}{2} }\left(\tilde{\mathscr{C}}_y\right)}\right]\right),
         \label{eq:qfp_profile_y}
    \end{aligned}
\end{equation}
where $c_4$ takes the form of  $c_4 = -e^{-\frac{\pi}{2}\sqrt{-\discinf} + i 2 \tau_d} $
with $\tau_d \in \mathbb{R}$ for real $\aV_y$.
For the metallic PFP, $\tau_d$ is fixed by the asymptotic boundary condition imposed at $y=\infty$, but its precise value is not important.
In the small $|\discinf|$ limit, $\tau_d$ for the metallic PFP scales as\footnote{
Alternatively, we can express $\tau_d$ in terms of $\aV^M_0$ as
\bqa
\tau_d^M = \arg\left\{
\Gamma \left(1+i\frac{\sqrt{-\discinf}}{2} \right)
\left( \tilde{\mathscr{C}}_0^{-1}\left(1+4R_d\aV^M_0\right) I_{i\frac{\sqrt{-\discinf}}{2} }\left(\tilde{\mathscr{C}}_0\right)+  I_{-1+i\frac{\sqrt{-\discinf}}{2} }\left(\tilde{\mathscr{C}}_0\right)+  I_{1+i\frac{\sqrt{-\discinf}}{2} }\left(\tilde{\mathscr{C}}_0\right)\right) \right\}.
\label{eq:tau_d}
\eqa
As $d\rightarrow d_{SC}^{-}$, $\tau_d \approx 
\arg\left\{
1.1332-0.363926 \discinf +i~0.0109144\sqrt{-\discinf}
\right\}
$
. In the limit of small $\discinf$, we obtain, $\tau_d \approx 0.01 \sqrt{-\discinf}$.
}
\bqa
\tau_d^M \sim \sqrt{-\discinf}.
\label{eq:taudscaling}
\eqa
Using $I_{\alpha}(x)$ $\approx$ $\frac{1}{\Gamma\left(1+\alpha\right)}\left(\frac{x}{2}\right)^{\alpha}$ for small $x$,
we can simplify the expression for the metallic PFP at small $y$ as
\begin{equation}
    \begin{aligned}
         \aVM_{y}  = 
         -\frac{1}{4R_d}\left\{1+\sqrt{-\discinf}\mathrm{cot}\left[
        \Omega^\otimes~\left(y+
         \mathrm{log}\left(\frac{\alpha_d^{\frac{1}{3}}\sqrt{\eta_d}}{2\sqrt{2}}\right)\right)-\tau_d^M 
         \right]\right\}.
         \label{eq:qfp_smallx}
    \end{aligned}
\end{equation}
This shows that the coupling function in the metallic PFP `oscillates' in the logarithmic angular momentum $y$ with a frequency $2 \Omega^\otimes$.
However, the oscillation of $\aVM_y$ in $y$ does not span more than one period because it ends when it diverges at 
\begin{equation}
    \begin{aligned}
       y^*_M = \frac{(-\pi + \tau_d^M)}{\Omega^\otimes}
         + \log\left( \frac{2\sqrt{2}}{\alpha_d^{1/3}\sqrt{\eta_d}} \right).
        \label{eq:ysc}
    \end{aligned}
\end{equation}

\subsubsection{Quasi-universality}
\label{sec:quasi_univ}

\begin{figure}[th]
\centering
\begin{subfigure}{.42\textwidth}
  \centering
\includegraphics[width=1.0\linewidth]{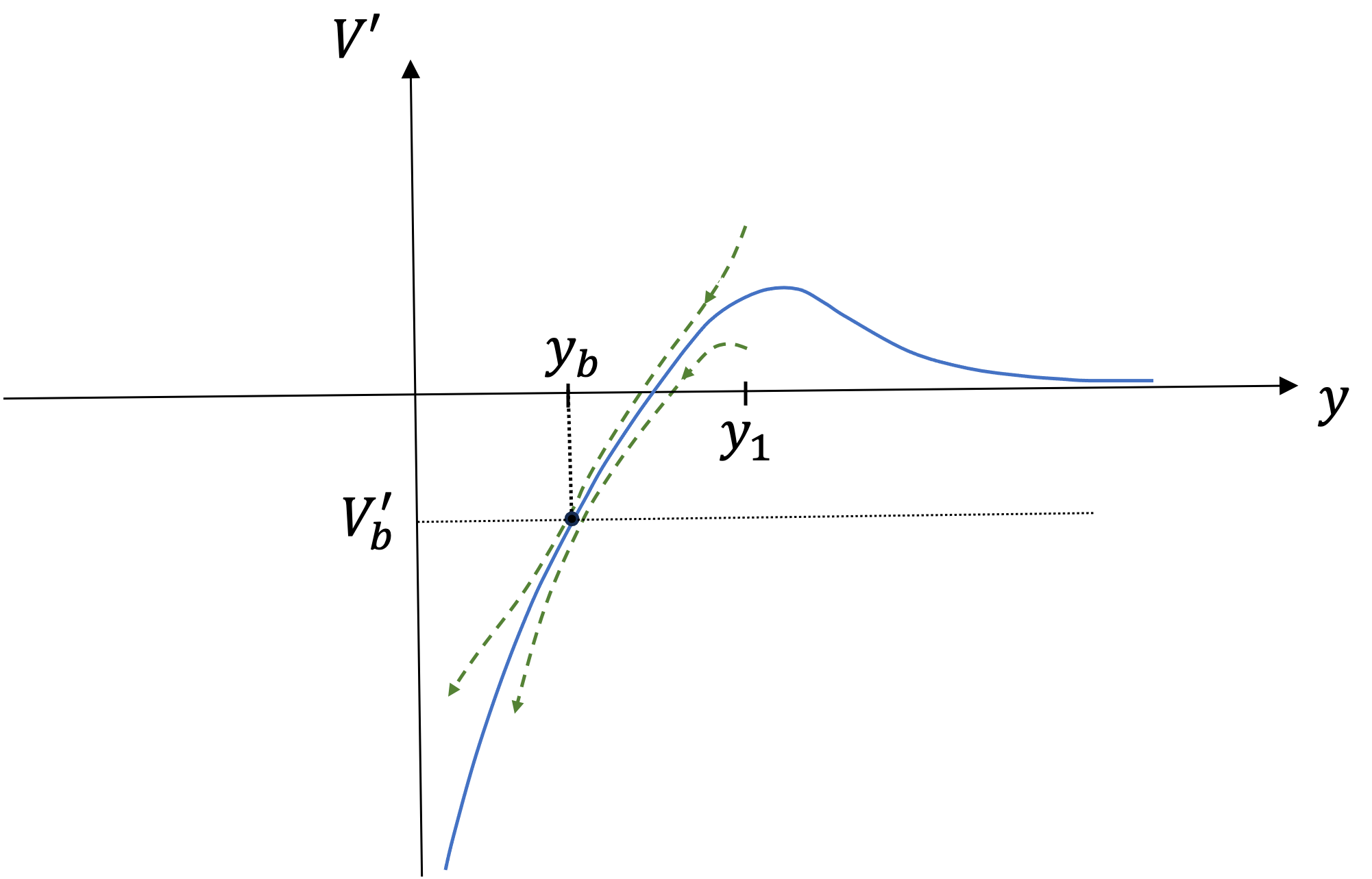}
  \caption{
}
\label{fig:changeofstability}
\end{subfigure}%
\begin{subfigure}{.29\textwidth}
  \centering
\includegraphics[width=1.0\linewidth]{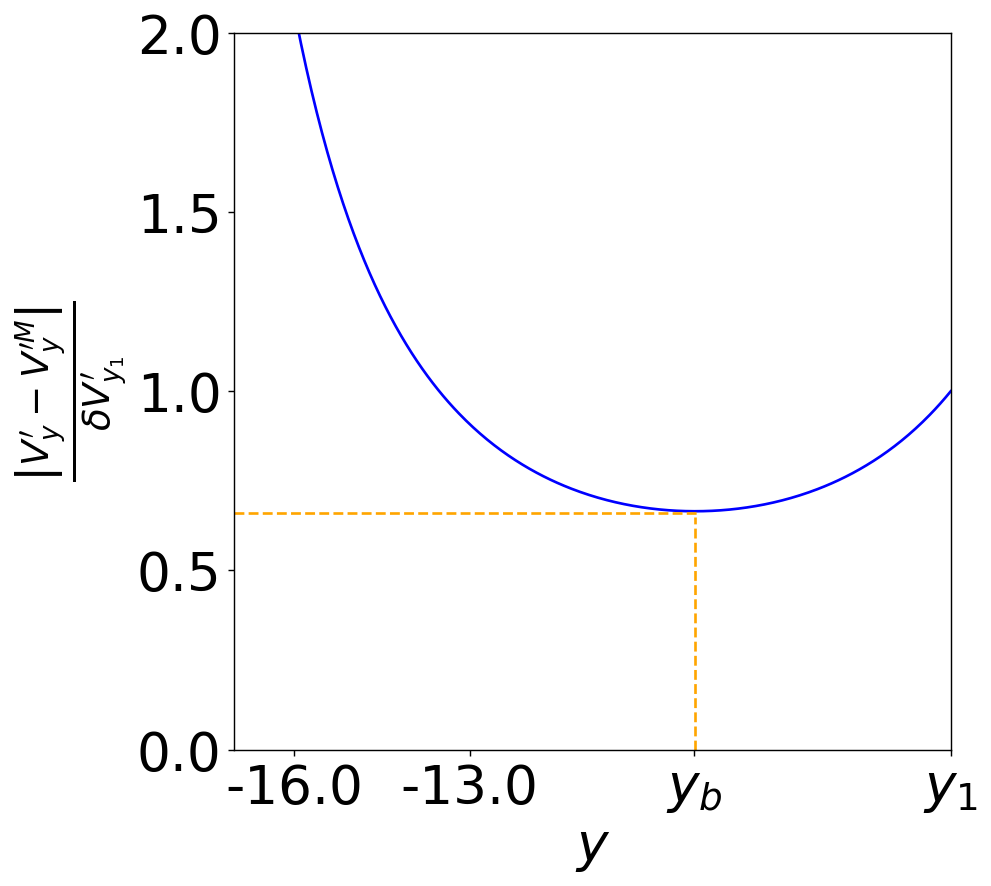}
  \caption{
}
\label{fig:bottlneck_focus1}
\end{subfigure}%
\begin{subfigure}{.29\textwidth}
  \centering
\includegraphics[width=1.0\linewidth]{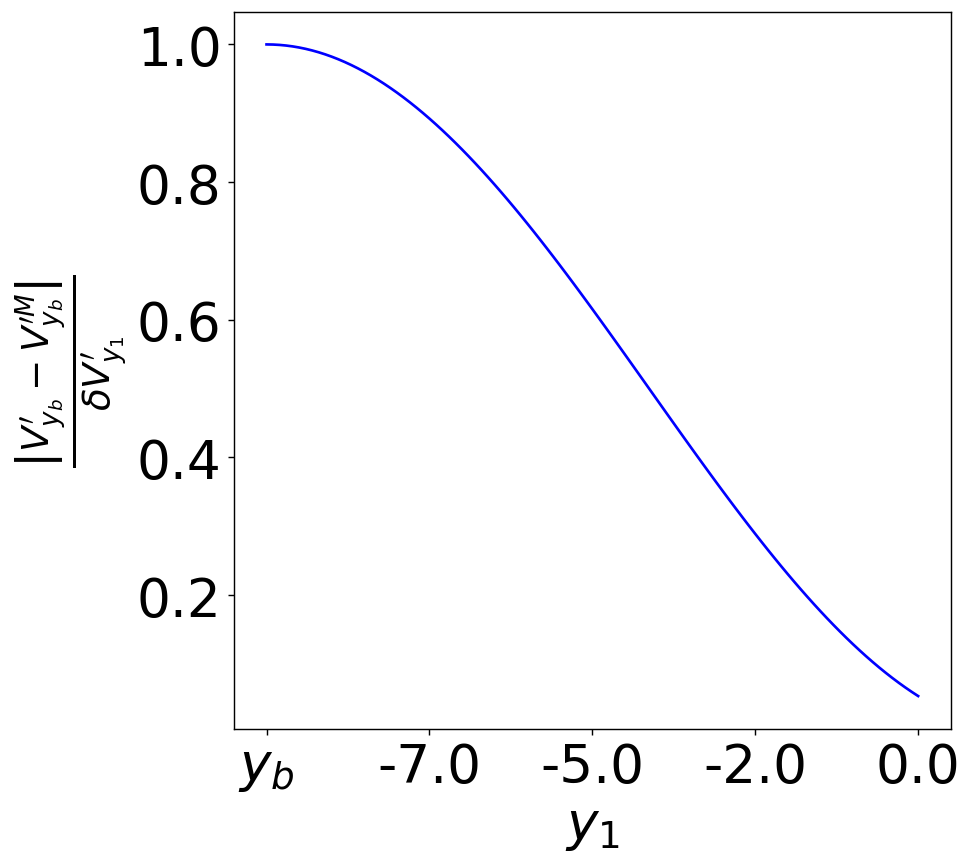}
  \caption{
}
\label{fig:bottlneck_focus2}
\end{subfigure}%
\caption{
(a) PFPs that are near the metallic PFP first approach the metallic PFP as $y$ decreases until they get closest at $y_b$ and deviate farther away from the metallic PFP as $y$ decreases below $y_b$.
(b) The vertical distance between the metallic PFP and a nearby PFP that passes through $(y_1, \aVM_{y_1} + \delta \aV_{y_1})$ plotted as a function of $y$.
(c) The minimum distance between the metallic PFP  and the nearby PFP considered in (b) plotted as a function of $y_1$ for a fixed $\delta \aV_{y_1}$.
For the plot, we use
$d \approx 2.432$.  
}
\label{fig:bottlneck_focus}
\end{figure}

Although a superconducting instability is inevitable in class C, one still expects to see some universal behaviors in theories with small $|\discinf|$ because the bottleneck generates a large window of energy scale in which the coupling function flows into a quasi-universal profile.
To understand this, we first examine the profile of individual PFPs that describe the RG flow of the couplings in individual angular momentum channels.
Suppose the metallic PFP enters the bottleneck region at $y_b$, 
where $y_b$ is determined by $\aV^M_{y_b}=\aV_b$.
Near $d_{SC}$, this scale is given by
\begin{equation}
    \begin{aligned}
        y_b = \frac{(- \frac{\pi}{2} + \tau_d^M)}{\Omega^\otimes}
         + \log\left( \frac{2\sqrt{2}}{\alpha_d^{1/3}\sqrt{\eta_d}} \right).
        \label{eq:yb}
    \end{aligned}
\end{equation}
For $|\etaPIII| \ll 1$, $y_b \sim y^*_M/2 \ll 0$.
The bottleneck also marks the region where the metallic PFP changes its local stability.
It is locally stable and unstable in $y>y_b$
and $y < y_b$, respectively\footnote{ The bottleneck $y_b$ also corresponds to the point at which the metallic PFP is flattest. This can be checked from the derivative of  \eq{eq:qfp_smallx} with respect to $y$, $ \partial_y \aVM_y = -\frac{\discinf}{8R_d} \csc^2 \left[ \Omega^\otimes~\left(y+\mathrm{log}\left(\frac{\alpha_d^{\frac{1}{3}}\sqrt{\eta_d}}{2\sqrt{2}}\right)\right)-\tau_d\right]$, which reaches a minimum at \( y = y_b \).}.
Consequently, PFPs that are near the metallic PFP at $y_1 \gg y_b$ first approach it as $y$ decreases, get closest to it at $y_b$, and then diverge away from it in $y<y_b$, as shown in \fig{fig:changeofstability}
and \fig{fig:bottlneck_focus1}. 
To be more quantitative, we consider a PFP that is $\delta \aV_{y_1}$ away from the metallic PFP at $y_1$. 
To the linear order in $\delta \aV_{y_1}$, the PFP is given by
$\aV_{y} =
\aVM_y + 
\left|
    \frac{
    \sin\left(
       \Omega^\otimes
\log\left(\frac{\mathscr{C}_d(y_1)}{2}\right) - \tau_d\right)
    }{
    \sin\left(
       \Omega^\otimes
\log\left(\frac{\mathscr{C}_d(y)}{2}\right) - \tau_d\right)
    }
    \right|^2
    \delta\aV_{y_1}$.
At $y=y_b$, the distance to the metallic PFP reaches the minimum,
$|\aV_{y_b} - \aVM_{y_b}|
=
\cos^2\left(
   \Omega^\otimes
(y_1 - y_b)\right)
\delta \aV_{y_1}$.
As the UV angular momentum $y_1$ increases, the minimum distance decreases because the deformation decays over a longer RG time,
as shown in  \fig{fig:bottlneck_focus2}.
At $y_1 \sim O(1)$, 
the distance becomes as small as
$|\etaPIII| \delta \aV_{y_1}$.
The PFPs that started at $y_1 \gg 1$ get even closer to the metallic PFP than this by an additional factor of $e^{-
\sqrt{\etaPI} y_1} \ll 1$ because the deformed PFP approaches the metallic PFP between $0 < y < y_1$.

Based on this information, we can now understand the quasi-universal profile of the renormalized coupling function that emerges around the bottleneck scale $l_b$ for $|\etaPIII| \ll 1$.
Those UV couplings of large angular momenta with $y^{(m)}(l_b) > y_b$ are still on the right side of the bottleneck and they are renormalized to the values that are very close to the metallic PFP.
Those UV couplings with $y^{(m)}(l_b) < y_b$ are first funneled to $\aV_b$ and stay at that coupling up to a long RG time $l_b$.
This is because the vertical speed 
of the RG flow is very slow 
at the bottleneck:
$\left. \beta_{\aV} \right|_{\aV=\aV_b} \sim \etaPIII$.
Consequently, the horizontal line $\aV=\aV_{b}$ serves as the quasi-attractor;
the couplings that started from the more repulsive UV couplings are attracted to $\aV_b$ in low angular momentum channels with $y^{(m)}(l_b) < y_b$.
In summary, the coupling function as a whole is attracted to a regular profile given by
\bqa
\aV^{RM}_y = 
    \begin{cases}
    \aVM_{y_b}
        &  y \leq y_b, 
        \\
    \aVM_y    
        &  y > y_b 
    \end{cases}
    \label{eq:quasi_universalV}
\eqa
at intermediate length scales around $l_b$.
We refer to this as the {\it regularized metallic PFP}.
This controls the quasi-universal behavior of the normal state that arises above $T_c$.
Since superconductivity cuts off the RG flow at low energies, the couplings that emerge at intermediate energies still depend on the bare couplings to some extent.
However, the theories with small $|\discinf|$ exhibit a strong universality due to the large hierarchy between $T_c$ and the UV cutoff.

\begin{figure}[th]
\centering
\begin{subfigure}{.35\textwidth}
  \centering
\includegraphics[width=1.0\linewidth]{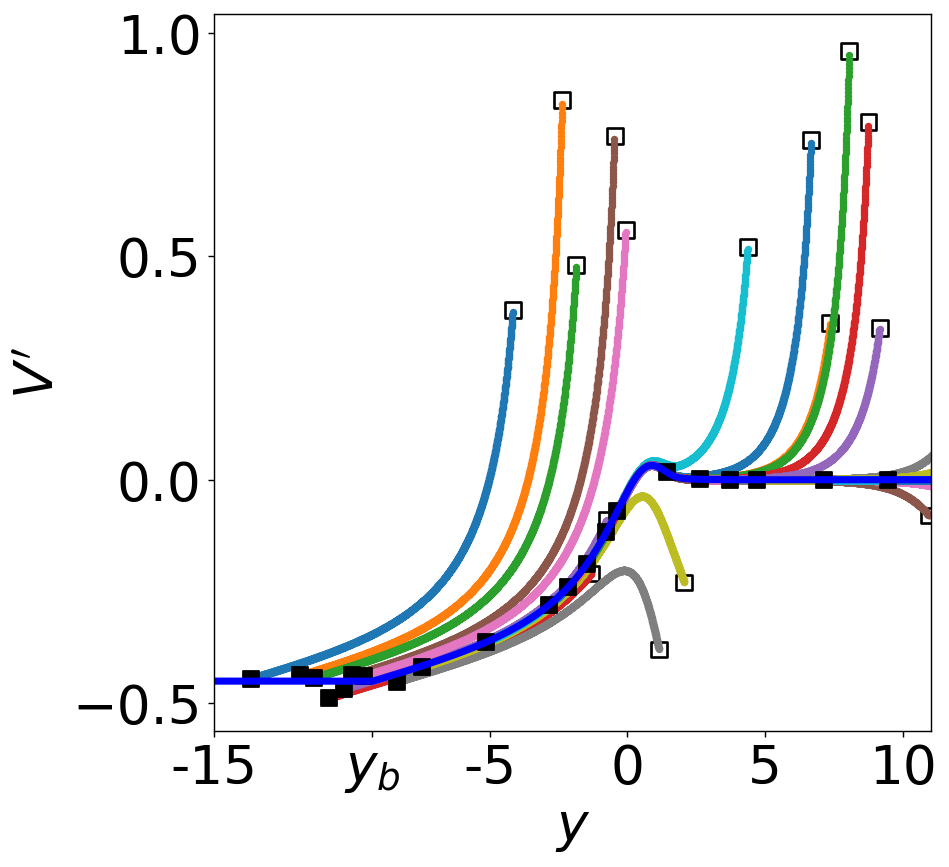}
  \caption{
}
 \label{fig:rmpfp1}
\end{subfigure}%
\begin{subfigure}{.35\textwidth}
  \centering
\includegraphics[width=1.0\linewidth]{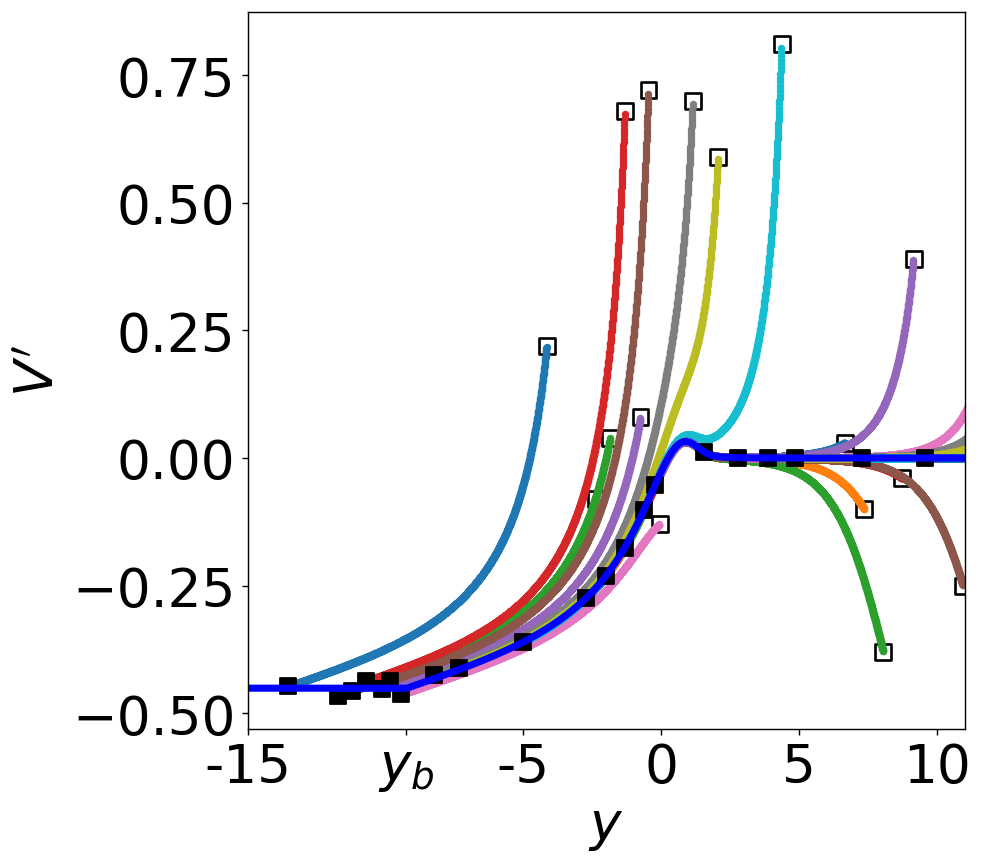}
  \caption{
}
\label{fig:rmfp2}
\end{subfigure}%
\begin{subfigure}{.35\textwidth}
  \centering
\includegraphics[width=1.0\linewidth]{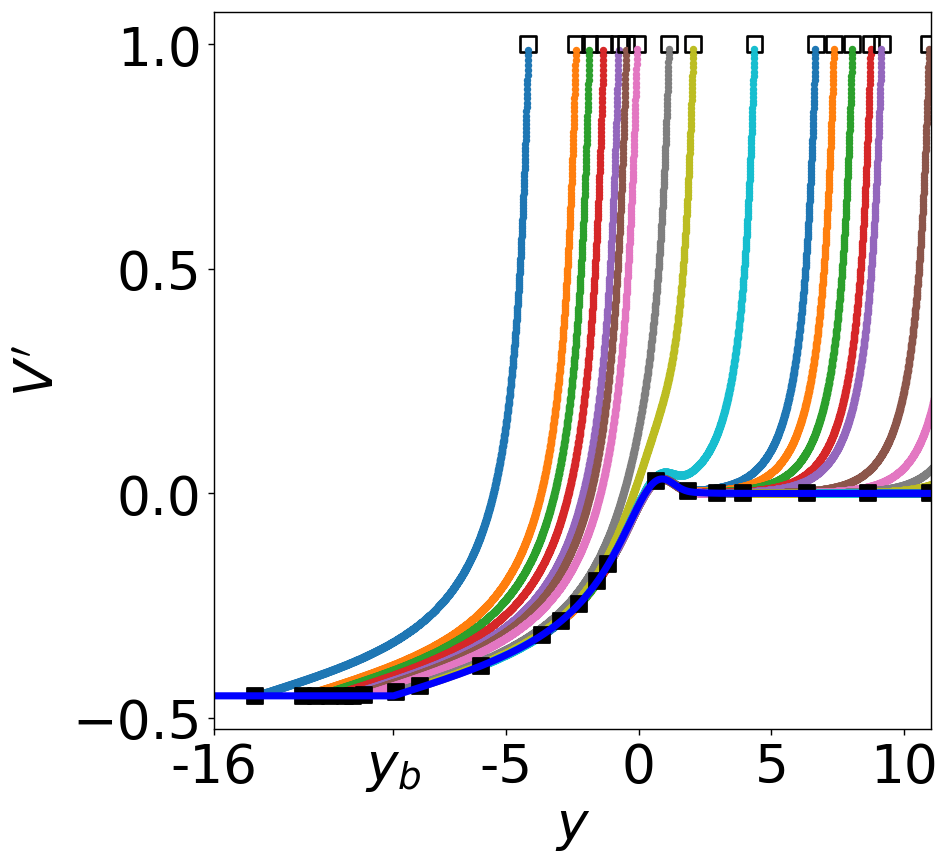}
  \caption{
}
\label{fig:rmfp3}
\end{subfigure}

\begin{subfigure}{.35\textwidth}
  \centering
\includegraphics[width=1.0\linewidth]{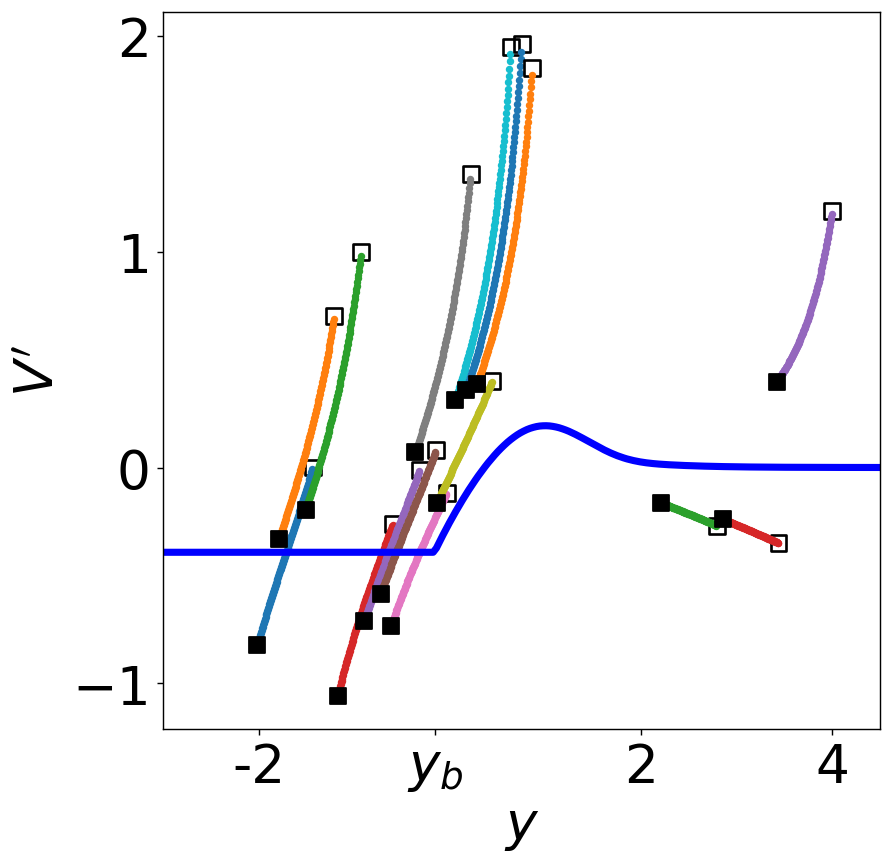}
  \caption{
}
 \label{fig:rmpfp1_d2}
\end{subfigure}%
\begin{subfigure}{.35\textwidth}
  \centering
\includegraphics[width=1.0\linewidth]{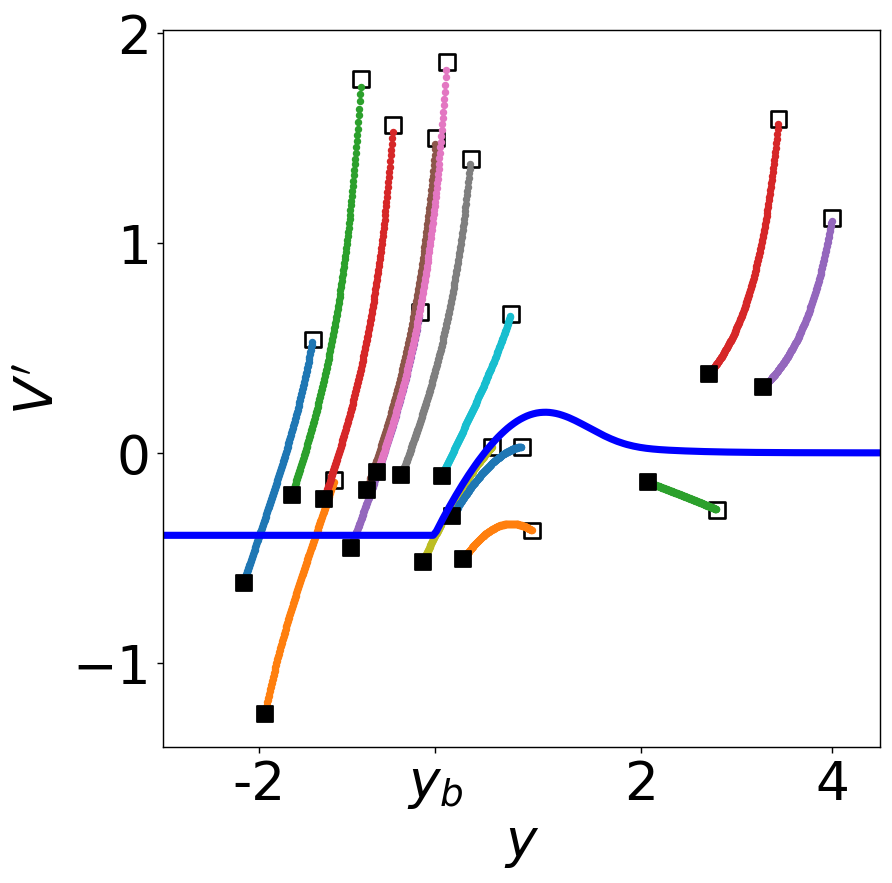}
  \caption{
}
\label{fig:rmfp2_d2}
\end{subfigure}%
\begin{subfigure}{.35\textwidth}
  \centering
\includegraphics[width=1.0\linewidth]{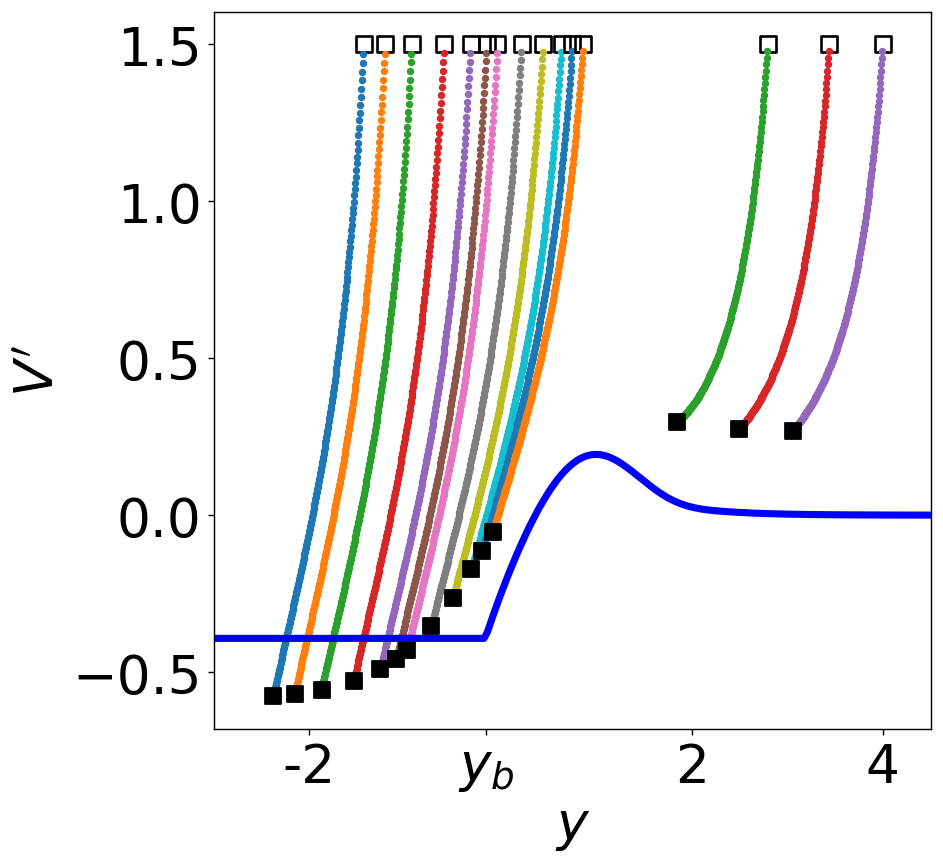}
  \caption{
}
\label{fig:rmfp3_d2}
\end{subfigure}
\caption{
Trajectories of PFPs that start from UV couplings specified by $(y^{(m)}(0), \av_m)$ with 
$m=1,6,10,17,30,40,60,200,500,5000,5\cross 10^4,1\cross 10^5,2\cross 10^5,4\cross 10^5,6\cross 10^5, 3.6\cross 10^6,1.2\cross 10^7, 3.6\cross 10^7, 9.6\cross 10^7, 1.08 \cross 10^9, 1.08 \cross 10^{10},1.08 \cross 10^{11}$
for (a)-(c), and $m=15,19,25,35,46,55,62,79,100,120,135,150,1050,1978,5111$ 
for (d)-(f). 
The UV couplings are denoted as the open squares.
Here, $\{ \av_m \}$ is chosen randomly above $\aV^M_{y}$ and
$\KFAVdim/\Lambda = 10000$ is used.
The RG flow is stopped at the scale $l_b$ that takes the renormalized couplings closest to the regularized metallic PFP. 
The renormalized couplings at $l_b$ are denoted as the filled squares.
(a), (b), (c) are for three choices of UV couplings in $d\approx 2.432$ slightly below $d_{SC}$.
(d), (e), (f) are for three choices of UV couplings in $d=2$.
}
\label{fig:rmfp}
\end{figure}

\begin{figure}[th]
\centering
\begin{subfigure}{.4\textwidth}
  \centering
  \includegraphics[width=1.0\linewidth]{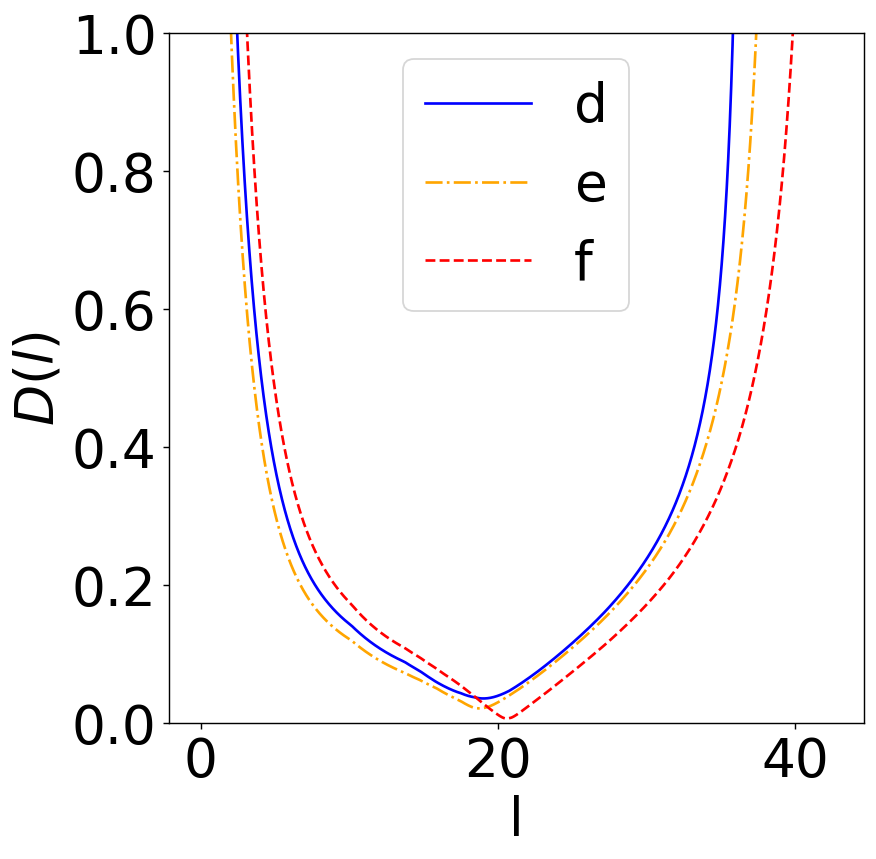}
  \caption{
}
  \label{fig:SEd24}
\end{subfigure}%
\begin{subfigure}{.365\textwidth}
  \centering
\includegraphics[width=1.0\linewidth]{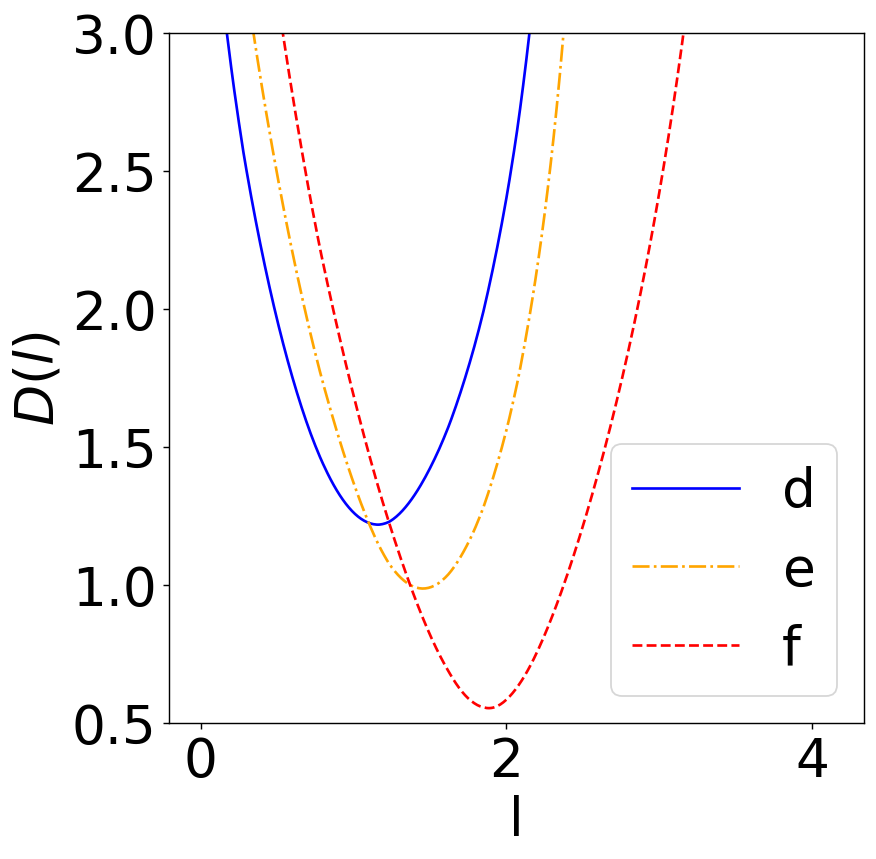}
  \caption{
}
  \label{fig:SE_d2}
\end{subfigure}
\caption{
The distance between the renormalized couplings and the regularized metallic PFP defined in \eq{eq:Dl2} for the UV couplings used in 
\fig{fig:rmfp} 
for (a) $d=2.432$
and (b) $d=2$.
The distance is minimized around the bottleneck scales, which are 
$l_b \approx 22.7$
and $2.4$ at $d=2.432$ and $2$, respectively.
}
\label{fig:SE}
\end{figure}

To confirm the attractive nature of the regularized metallic PFP,
we plot PFPs associated with a collection of UV couplings $(y^{(m)}(0), \av_m)$ in \fig{fig:rmfp}, where the bare couplings are denoted as open squares and $\av_m$'s are chosen randomly above the bottleneck coupling.
Along the RG flow, we monitor the `distance' of the renormalized couplings to the regularized metallic PFP using a measure of normalized distance defined as
\bqa
D(l)
=
\sqrt{
\frac{
\sum_m \left|
\aVP{y^{(m)}(0)}{\av_m}_{y^{(m)}(l)}
- \aV^{RM}_{y^{(m)}(l)} \right|^2
}{
\sum_m \left|
\aV^{RM}_{y^{(m)}(l)} \right|^2
}
},
\label{eq:Dl2}
\eqa
where
$\aVP{y^{(m)}(0)}{\av_m}_y$
denotes the PFP that goes through
$(y^{(m)}(0),\av_m)$.
The normalized distance reaches its minimum around $l_b$, as shown in \fig{fig:SE}.
The profiles of the renormalized couplings evaluated at $l_b$, denoted as the filled squares in \fig{fig:rmfp}, are indeed close to the regularized metallic PFP.
The emergent profile of the couplings indeed exhibits a stronger universality (less dependence on the bare couplings) near $d_{SC}$ than in $d=2$.

Having identified the quasi-attractor of the functional RG flow, we now describe the universal properties governed by it. 
In class C, the ground state is bound to be a superconductor.
Because the critical boson generates the strongest attractive interaction at $y=-\infty$, the superconducting instability is strongest in the s-wave channel for spinful fermions, and the p-wave channel for spinless fermions.
\footnote{Here, we assume $\KFthetadim \gg \Lambda$ that $y \ll 0$ for the $p$-wave channel.}
However, a superconducting instability may arise in a higher angular momentum channel if the bare coupling in that channel is significantly less repulsive than the rest.
Below, we discuss quasi-universal properties in various situations.\\

\subsubsection{Quasi-universal pairing interaction}
\label{sec:quasiuniversalpairing_classC}

Suppose the UV couplings are such that the coupling diverges to $-\infty$ in the lowest allowed angular momentum channel at scale $l_{SC}$. 
If $l_{SC} \gg 1$, 
which occurs near $d_{SC}$, there is a large window of length scale $0 \ll l \ll l_{SC}$ for couplings 
to converge to the regularized metallic PFP, and
the renormalized couplings exhibit quasi-universal behavior with a weak dependence on the UV data.
One can then define an approximate basin of attraction for the regularized metallic PFP.
Let the basin of attraction of tolerance $\mathcal{T}$ be the set of 
$(y_0, \aV^{\text{UV}}_{y_0})$ 
whose RG trajectory approaches the regularized metallic PFP within a tolerance \( \mathcal{T} \) 
before $l$ reaches $l_{SC}$.
Namely, 
($y_0, \aV^{\text{UV}}_{y_0}$) 
is in the basin of attraction of tolerance $\mathcal{T}$ if
there exists $l < l_{SC}$ at which
$|
\aV_{y_0-l/2}
-
\aV^{RM}_{y_0-l/2} 
| < \mathcal{T}$ 
where
$\aV_y$ is a PFP that satisfies the initial condition, 
$\aV_{y_0}=
\aV^{\text{UV}}_{y_0}$.

\begin{figure}[th]
\centering
\begin{subfigure}{.35\textwidth}
  \centering
\includegraphics[width=1.0\linewidth]{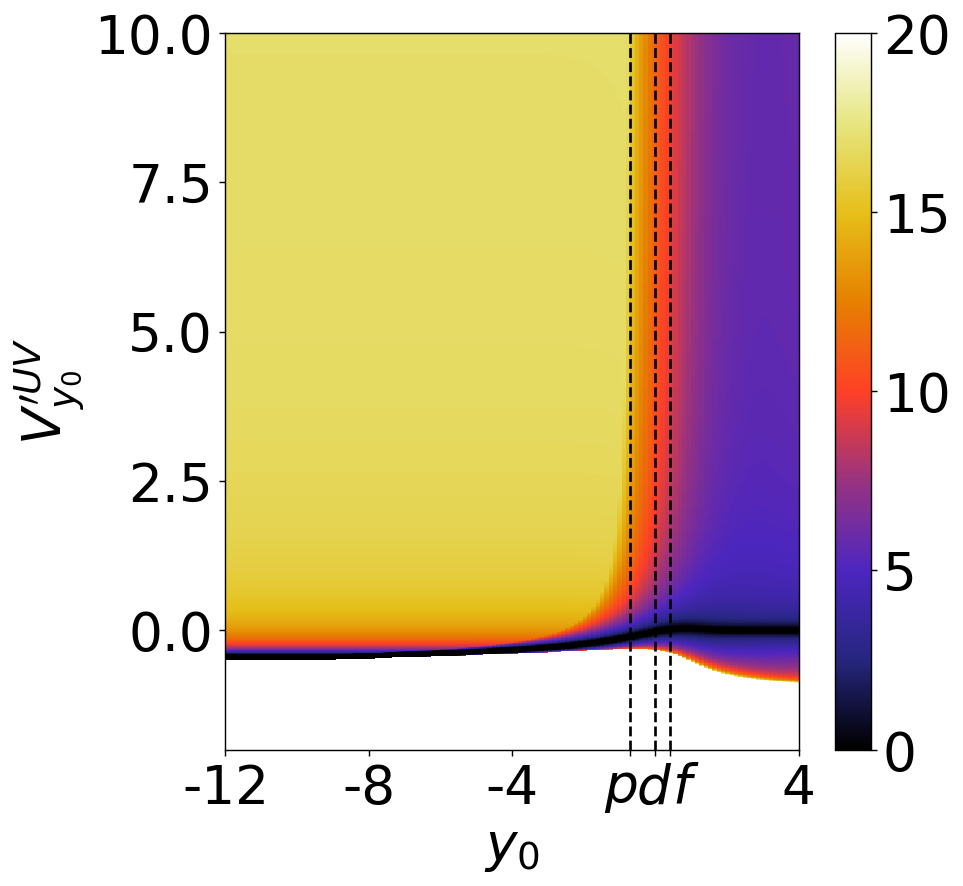}
  \caption{
}
 \label{fig:basin_attraction}
\end{subfigure}%
\begin{subfigure}{.35\textwidth}
  \centering
\includegraphics[width=1.0\linewidth]{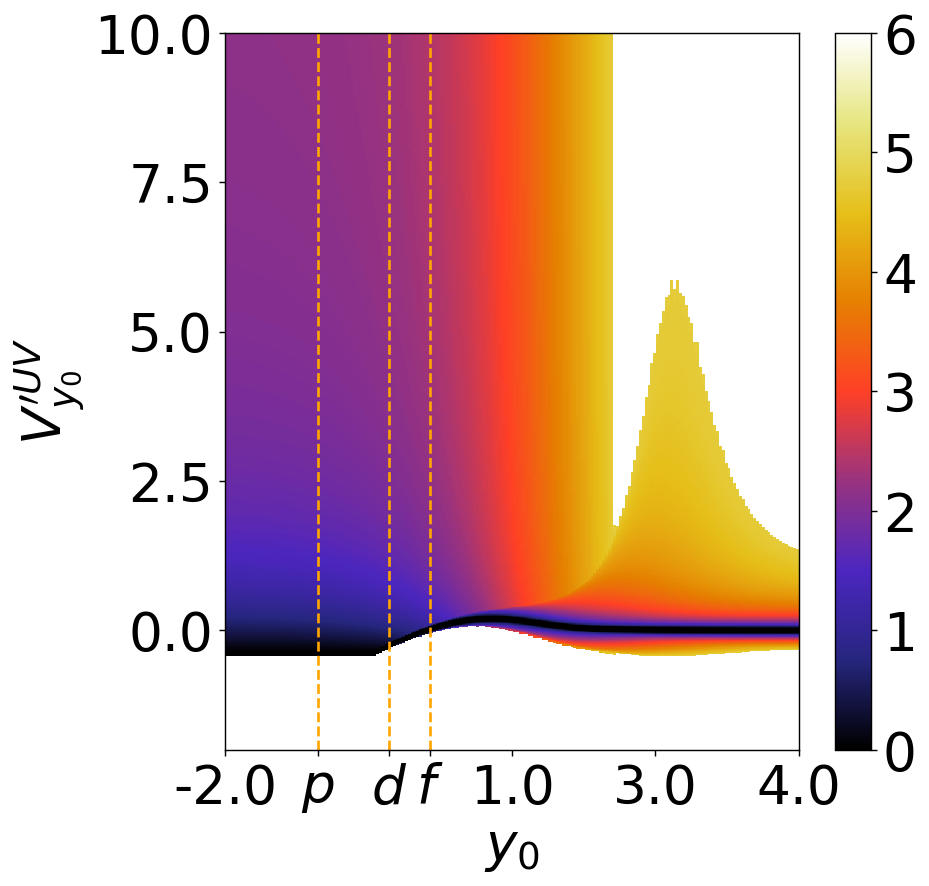}
  \caption{
}
\label{fig:basin_attraction_d2}
\end{subfigure}%
\begin{subfigure}{.35\textwidth}
  \centering
\includegraphics[width=1.0\linewidth]{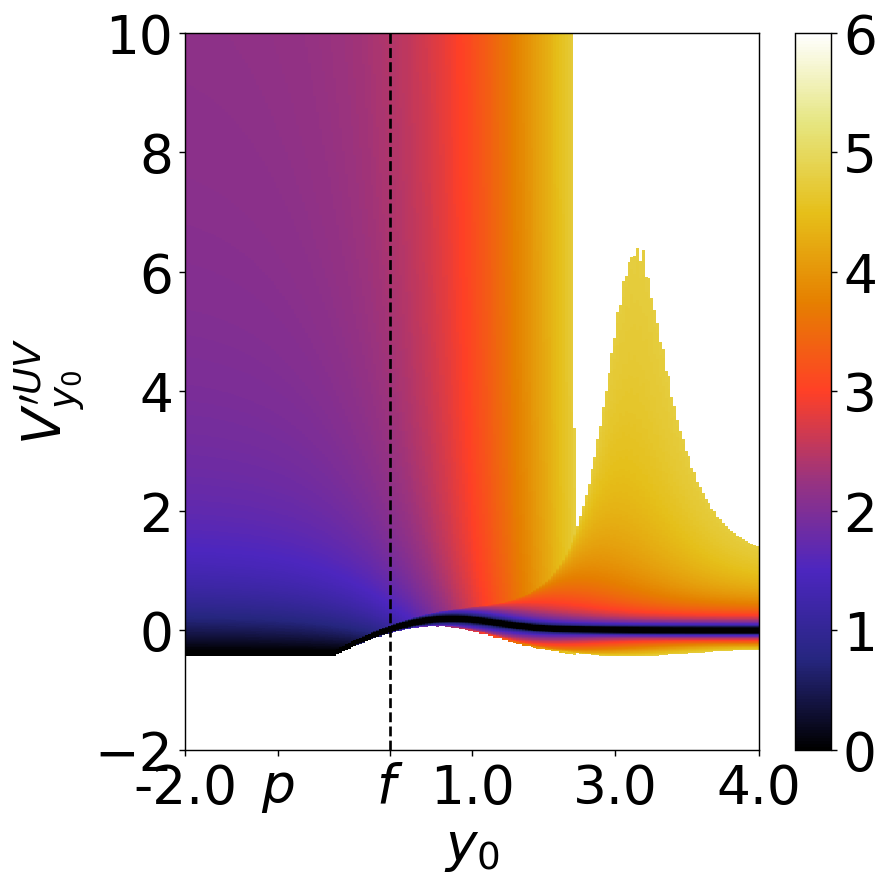}
  \caption{
}
\label{fig:Spinless_Fermion_basin_attraction_d2}
\end{subfigure}
\caption{
Basins of attraction with tolerance $\mathcal{T} = 0.045$ for
(a) the spinful fermions in $d=2.432$,
(b) the spinful fermions in $d=2$,
and 
(c) the spinless fermions in $d=2$
with 
$
\mathbf{K}_F =  10 \Lambda$. 
The color at each point in the basin of attraction represents the logarithmic length scale that it takes for the coupling to approach the regularized metallic PFP within tolerance $\mathcal{T}$.
The uncolored (white) region is outside the basin of attraction with tolerance $\mathcal{T}$ either because those points never approach the regularized PFP or because it takes RG time longer than that of the superconducting instability that arises in the lowest allowed angular momentum channel.
The vertical lines represent the $y$ values for a few low angular momenta next to the lowest angular momentum.
}
\label{fig:main_basin_attraction}
\end{figure}

In Figs.~\ref{fig:basin_attraction} and \ref{fig:basin_attraction_d2}, we show the basin of attraction
of tolerance $\mathcal{T}=0.045$ for the spinful case, assuming 
that $T_c$ in the s-wave channel is \eq{eq:Tcmin}
at \( d \approx 2.432 \) just below $d_{SC}$ and at \( d = 2 \), respectively.
The intersections between the vertical lines and the shaded region denote the basin of attraction in the $p$, $d$ and $f$-wave channels.
As expected, the basin of attraction is much larger in the dimension near \( d_{\text{SC}} \) than in $d=2$.
For the spinless case, the s-wave channel is not allowed. 
Before the p-wave superconductivity sets in, one can similarly identify the basin of attraction of the same tolerance as is shown in  Fig.~\ref{fig:Spinless_Fermion_basin_attraction_d2}.
We use $T_c$ for the $p$-wave pairing in the limit that the bare coupling is large in that channel.

\begin{figure}[th]
\centering
\begin{subfigure}{.33\textwidth}
  \centering
  \includegraphics[width=1.0\linewidth]{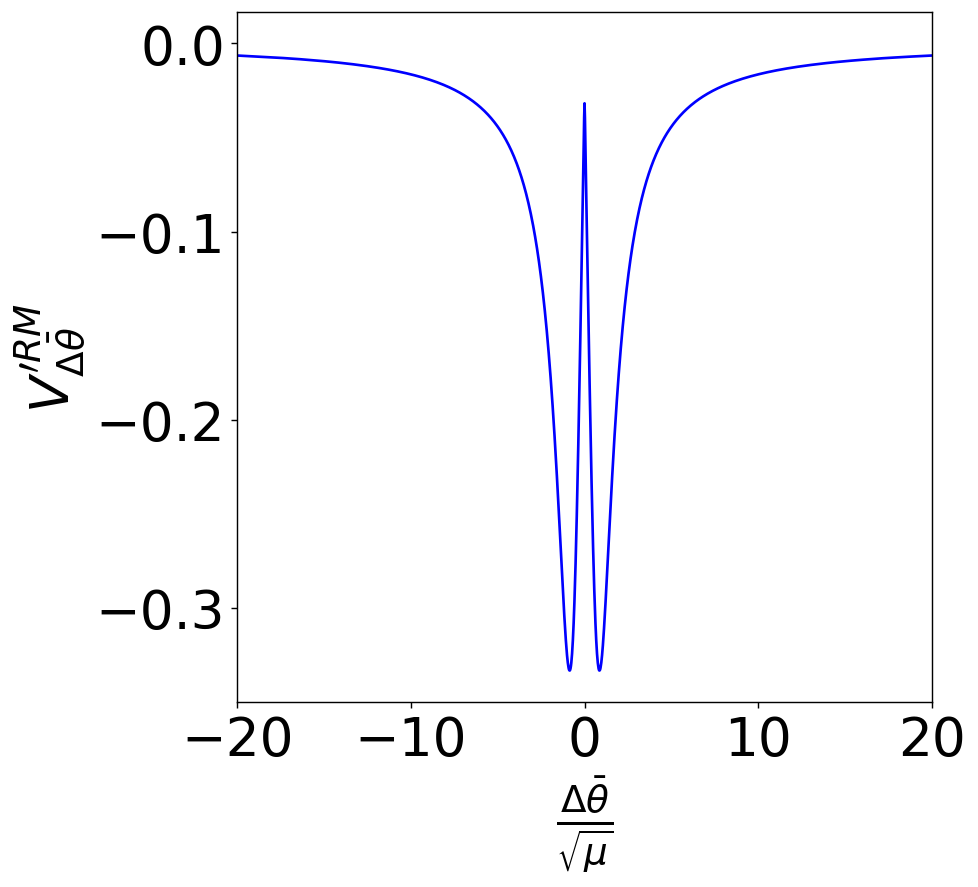}
  \caption{
}
  \label{fig:RMFP_theta}
\end{subfigure}%
\begin{subfigure}{.33\textwidth}
  \centering
\includegraphics[width=1.0\linewidth]{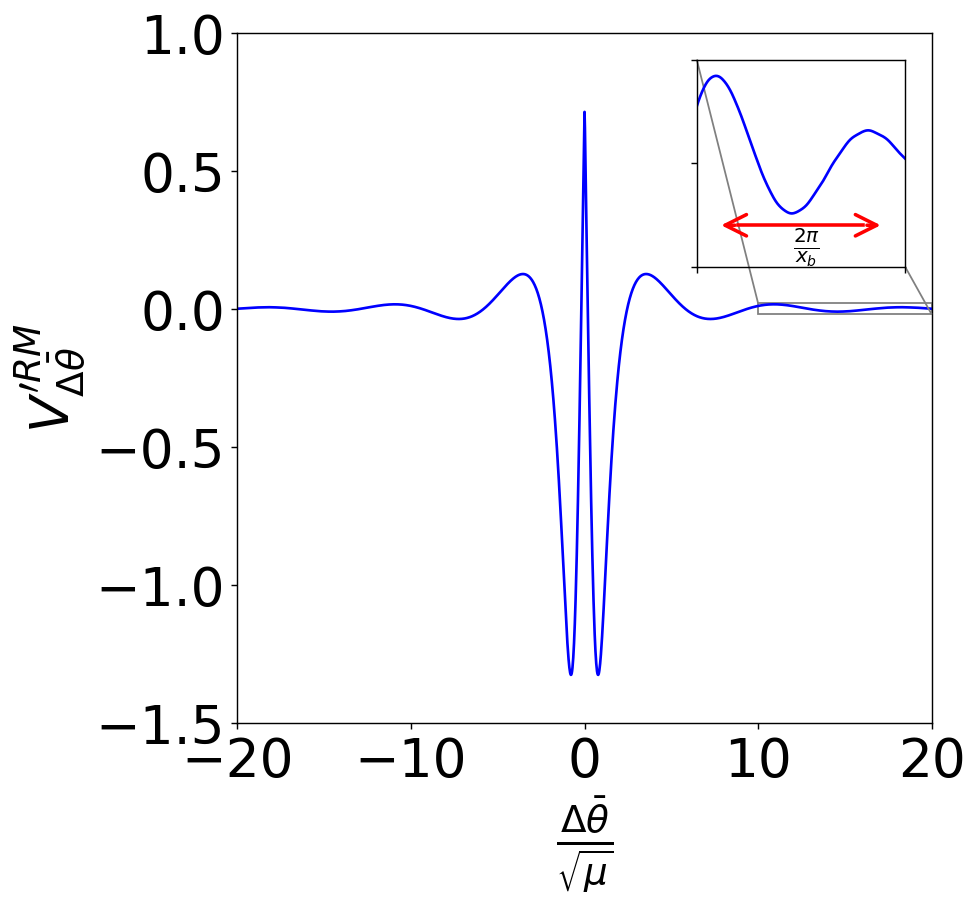}
  \caption{
}
  \label{fig:RMFP_theta_d2}
\end{subfigure}%
\begin{subfigure}{.33\textwidth}
  \centering
\includegraphics[width=1.0\linewidth]{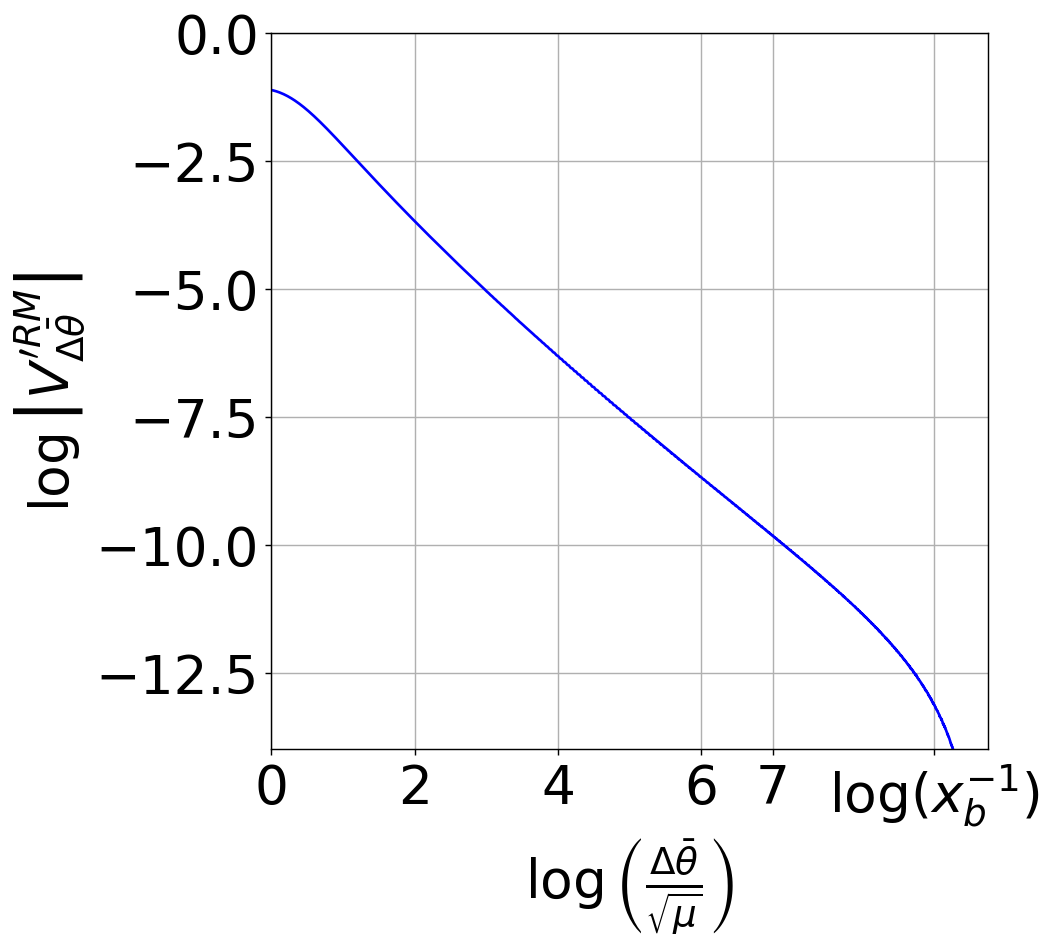}
  \caption{
}
  \label{fig:RMFP_decay}
\end{subfigure}
\caption{
The regularized metallic PFP plotted as a function of the rescaled angular coordinate 
for (a) $d=2.432$ and (b) $d=2$. 
As a function of 
$\bar \theta/\sqrt{\mu}$, 
the regularized metallic PFP exhibits oscillations with periodicity $2\pi/x_b$, where $x_b = e^{y_b}$. 
(c) 
The numerical integration of  \eq{eq:quasi_universalV}, shown in the logarithmic scale for $d = 2.432$. 
For a large range of  $\Delta \bar{\theta} \gg \sqmu$, it decays as a power law with exponent close to $-1$, but the oscillatory component obscures the power-law decay.
}
\label{fig:RMFP_theta_profiles}
\end{figure}

The renormalized couplings that become largely insensitive to the bare couplings will give rise to the quasi-universal pair susceptibility above $T_c$\footnote{
We will discuss the detailed expression of the universal pair susceptibility in the forthcoming paper.}. 
In the space of $\bar \theta$, the quasi-universal coupling 
that emerges at the bottleneck scale is given by the inverse Fourier transformation of the regularized metallic PFP 
(\eq{eq:InverseFourierV2}),
\bqa
\aV^{RM}_{ \bar{\theta}_1+ \bar{\theta}, \bar{\theta}_1} 
=  
2 \int dy ~ e^y ~ 
\aV^{RM}_{y}
e^{-i   \frac{\bar{\theta}}{\sqrt{\mu}} e^y }.
\eqa
At the bottleneck scale $\mu_b = \Lambda e^{-l_b}$,
this leads to
\footnote{
From an integration by part, we can write
$\frac{\hat \theta }{\sqrt{\Lambda}}
\aV^{RM}_{ \hat{\theta}_1+ \hat{\theta}, \hat{\theta}_1} 
=
2 i 
\int dx e^{i  
\frac{\hat{\theta}}{\sqrt{\Lambda}} 
x }
\partial_x \aV^{RM}_{\log |x|},
$
where $\hat \theta = \bar \theta e^{l/2}$.
The large $\hat \theta$ limit of this is governed by the kink of 
$\aV^{RM}_{\log |x|}$ at the bottleneck scale $|x|=e^{y_b}$. 
Writing 
$\aV^{RM}_{y} = -\frac{1}{4R_d} - \frac{\discinf}{8R_d}(y - y_b)  \Theta(y-y_b) $ 
near the kink, 
we have
$\partial_x \aV^{RM}_{\log x} = 
-  \frac{\discinf}{8R_d} 
\sum_{s=\pm} 
\left[
 \frac{1}{x}  \Theta(|x|-x_b)
+  \log \frac{|x|}{x_b} \delta(x-x_b)  
-  \log \frac{|x|}{x_b} \delta(x+x_b)  
\right]$, where $x_b=e^{y_b}$.
}
\bqa
\aV^{RM}_{ \bar{\theta}_1+ \bar{\theta}, \bar{\theta}_1} 
\sim
\frac{\sqrt{\mu_b}}{|\bar \theta|} 
\left[
\pi - 2 Si\left( x_b \frac{|\bar \theta|}{\sqrt{\mu_b}} \right)
\right],
\label{eq:VRMtheta}
\eqa
where 
\( 
x_b 
= e^{y_b} = 
\frac{
2\pi \sqrt{2}}{\pi \alpha_d^{1/3}\sqrt{\eta_d}}
e^{-\frac{(\frac{\pi}{2} - \tau_d)}{\Omega^\otimes}
}
\). 
The direct inverse Fourier transformation of the regularized metallic PFP is shown in Fig. \ref{fig:RMFP_theta_profiles}.
The profile of the quasi-universal coupling 
that emerges at the bottleneck scale exhibits two notable features. 
Firstly, the oscillatory modulation of the coupling function in the dimensionless angular coordinate $\bar \theta/\sqrt{\mu_b}$ has a pitch set by 
\( \frac{2\pi}{x_b} \), which depends on 
\( \Omega^\otimes \),
the `frequency' with which the deformation of the coupling rotates around the $-\infty$ asymptotic complex fixed point. 
Through the profile of the coupling function, we can indirectly probe this property of the complex asymptotic fixed points.
Secondly, the envelope of the coupling function decays as $\frac{\sqrt{\mu_b}}{|\bar \theta|}$ at large angles.
To see this, we note that the argument of the sine-integral function in \eq{eq:VRMtheta} is bounded by 
$x_b  \frac{ |\bar \theta_{max}|  }{\sqrt{\mu_{b}}} 
\sim 
\sqrt{ \frac{\KFAVdim}{\Lambda} }
e^{\frac{l_b}{2} + y_b}
\sim
\sqrt{ \frac{\KFAVdim}{\Lambda} }
$ due to 
Eqs. 
\eqref{eq:lthroughbottleneck},
\eqref{eq:taudscaling} 
and
\eqref{eq:yb}.
Since $\left[ \pi - 2 Si\left(  x_b  \frac{ |\bar \theta|  }{\sqrt{\mu_{b}}} \right) \right]$ in \eq{eq:VRMtheta}
approaches a non-zero constant at the largest possible $\bar \theta$, the quasi-universal coupling function scales as
$\aV_{\bar \theta_1+\bar \theta, \bar \theta_1}
(l_b) 
\sim 
\frac{\sqrt{\mu_{b}}}{|\bar \theta|}
$
at large $\bar \theta/\sqrt
{\mu_b}$.
Importantly, this non-zero constant does not decrease with increasing $l_b$.
The slow decay of the coupling function in the angular space originates from the kink of $\aV^{RM}_y$ at $y=y_b$  in \eq{eq:quasi_universalV}.
However, the kink that arises from actual RG flow is not infinitely sharp, as shown in \fig{fig:rmfp}.
Nonetheless, it is sharp enough to give rise to the $\sqmu/\bar \theta$ decay up to a range of angle that does not become small in the large $l_b$ limit.
This will be discussed in Sec. \ref{sec:ex4} and Appendix \ref{appendix:C}.
Due to the large-angle scattering, the quasi-universal non-Fermi liquid that arises in class C only respects OLU(1).

\subsubsection{Quasi-universal $T_c/\KFAVdim^z$ for non-s-wave superconductors}
\label{sec:quasiuniversalTckF}

\begin{figure}[th]
\centering
\begin{subfigure}{.33\textwidth}
  \centering
  \includegraphics[width=1.0\linewidth]{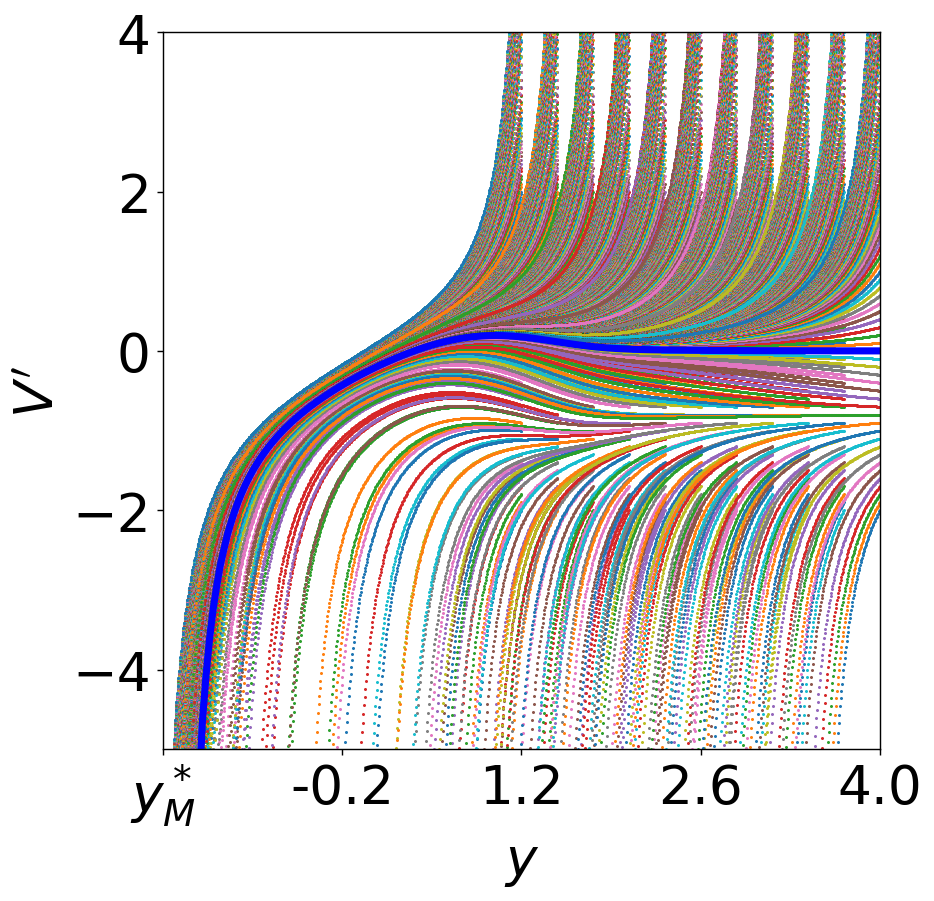}
  \caption{
}
 \label{fig:QFP_UVdistribution}
\end{subfigure}%
\begin{subfigure}{.33\textwidth}
  \centering
  \includegraphics[width=1.0\linewidth]{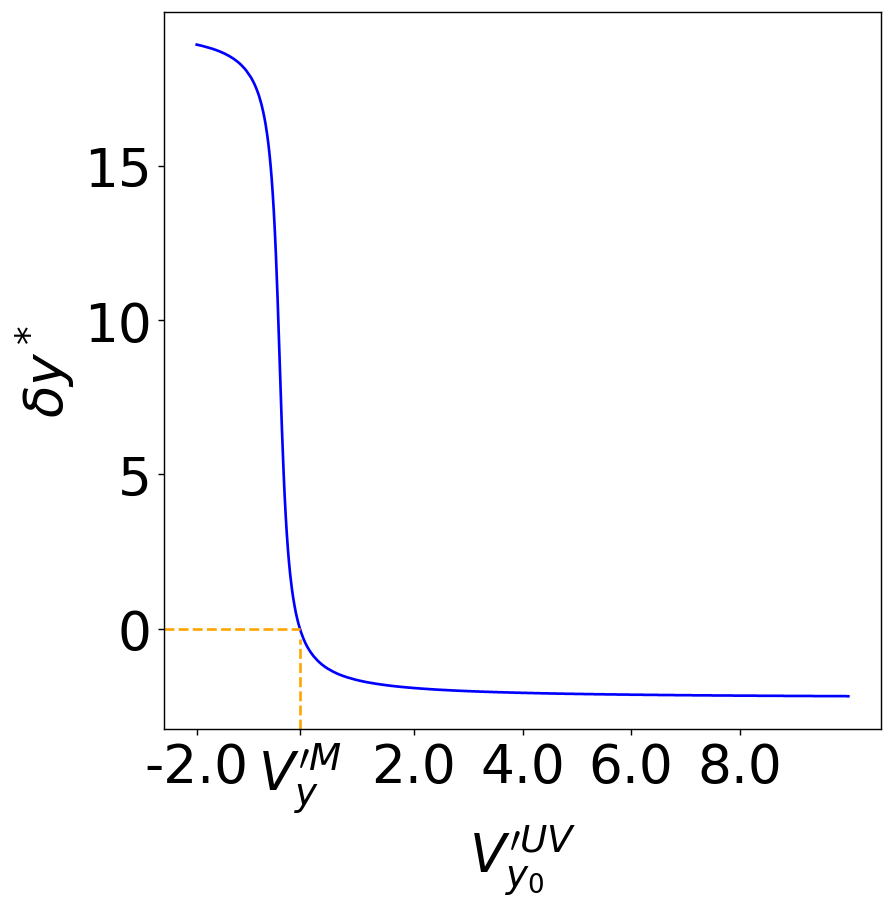}
  \caption{
}
  \label{fig:ysc_Vuv_yuv}
\end{subfigure}%
\begin{subfigure}{.33\textwidth}
  \centering
\includegraphics[width=1.0\linewidth]{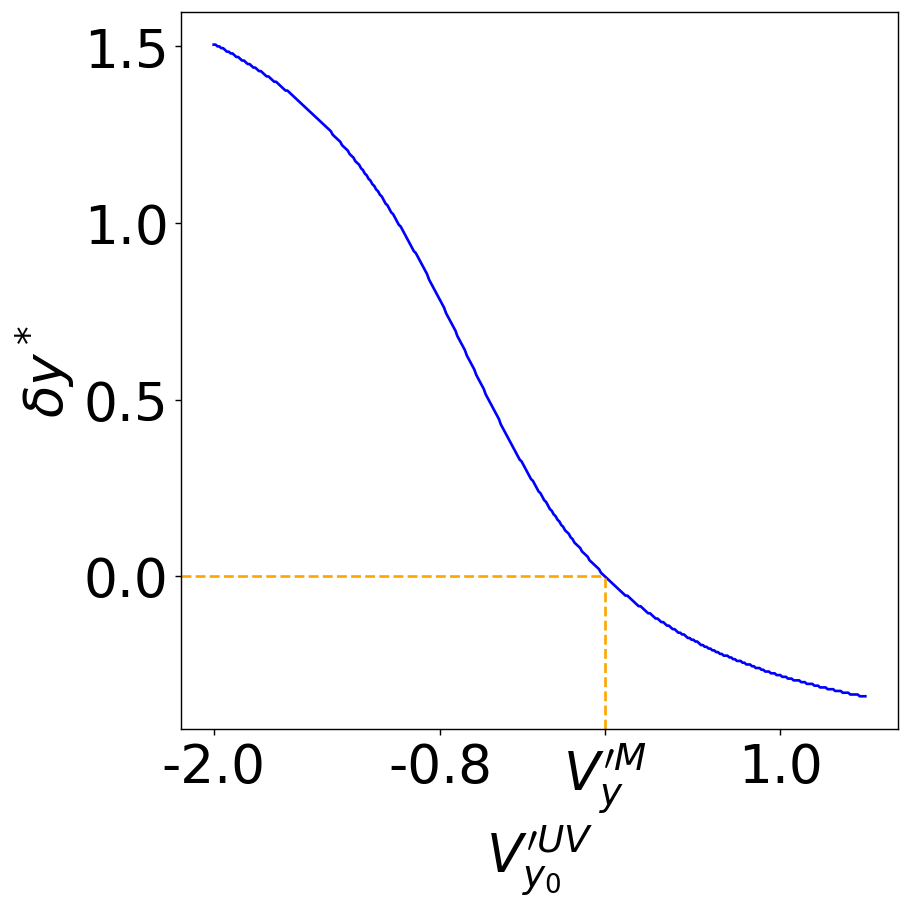}
  \caption{
}
  \label{fig:ysc_Vuv_yuv_d2}
\end{subfigure}
\caption{
(a) PFPs that emanate from $(y, \aV_y)$
with $y \gg y^b$ 
and $\aV_y > \aV^\circ_{\infty}$
first get attracted toward the metallic PFP, denoted as the thick blue line, and diverge to $-\infty$ close to $y^*_M$.
This leads $T_c/\KFAVdim^z$ that depends on the UV couplings only weakly.
(b) $\delta y^*$ in \eq{eq:deltaySC} for the p-wave channel plotted as a function of the UV coupling in $d=2.432$.
(c) $\delta y^*$ in \eq{eq:deltaySC} in the f-wave channel plotted as a function of the UV coupling in $d=2$.
}
\label{fig:main_ysc_Vuv_yuv}
\end{figure}

Suppose that superconductivity occurs in a non-zero angular momentum channel $n$ with 
$y_0 =  \log \left[ \frac{ \sqrt{\Lambda}}
     {2 \bar{\theta}_{\max}}  n \pi  \right]$ 
greater than $y_b$. 
This can happen if the bare couplings in other angular momentum channels are significantly more repulsive than the coupling in that channel.
If the bare coupling is on the metallic PFP, the superconducting transition temperature is given by
$T_c =\Lambda e^{-2z (y_0-y^*_M)}$, where $y^*_M \sim  2 y_b$
is the value of $y$ at which the metallic PFP diverges to $-\infty$.
If the bare coupling is deformed away from the metallic PFP as
$\aV_{y_0}^{UV}= \aVM_{y_0}+\deltaaVUV$,
$T_c$ is modified.
If $y_0 \gg y_b$, however,
the coupling is first attracted to the metallic PFP,
and the PFP is expected to diverge at $y_{SC}$ that is close to $y^*_M$.
This is shown in \fig{fig:QFP_UVdistribution}.
Let us examine how $y_{SC}$ depends on $\deltaaVUV$.
At a fixed intermediate scale $y'$ with $y_b \ll y' \ll 0$,
the deformation is reduced to 
$\delta \aV_{y'} = c' e^{-y_0} \deltaaVUV$ due to the attractive RG flow toward the metallic PFP,
where $c'$ is a constant that depends on $y'$ and the profile of the metallic PFP.
The PFP that goes through $(y', 
\aV_{y'}
=\aVM_{y'} + \delta \aV_{y'}
)$ takes the form of
\begin{equation}
    \begin{aligned}
 \aV_{y} = -\frac{1}{4R_d}\left\{
 1+
 \sqrt{-\discinf}
 \mathrm{cot}\left[
\Omega^\otimes\left(y+
         \mathrm{log}\left(\frac{\alpha_d^{\frac{1}{3}}\sqrt{\eta_d}}{2\sqrt{2}}\right)\right)-
\tau\left(y',\aV_{y'}\right)\right]
         \right\}
    \end{aligned}
\end{equation}
in $y < y'$, where
\begin{equation}
    \begin{aligned}
 \tau\left(y',\aV_{y'}\right) 
=
\arg\left\{\Gamma\left(1+i\frac{\sqrt{-\discinf}}{2}\right)\left[\tilde{\mathscr{C}}_{y'}^{-1}I_{
i\frac{\sqrt{-\discinf}}{2} ^{ }}\left(\tilde{\mathscr{C}}_{y'} \right)(1+4R_d\aV_{y'})+
I_{-1+
i\frac{\sqrt{-\discinf}}{2} ^{ }}\left(\tilde{\mathscr{C}}_{y'} \right)+I_{1+
i\frac{\sqrt{-\discinf}}{2} ^{ }}\left(\tilde{\mathscr{C}}_{y'} \right)\right]\right\}.
    \end{aligned}
\end{equation}
This PFP diverges to $-\infty$ at
\begin{equation}
    \begin{aligned}
      y_{SC}  \left(y',\aV_{y'}\right) =
\frac{-\pi+\tau\left(y',\aV_{y'}\right)}{\Omega^\otimes}+\log\left(\frac{2\sqrt{2}}{\alpha_d^{\frac{1}{3}}\sqrt{\eta_d}}\right).
        \label{eq:lsc_y_V_UV}
    \end{aligned}
\end{equation}
Using $\aV_{y'} = \aVM_{y'} + c' e^{-y_0} \deltaaVUV$, we obtain
$
y_{SC}(y_0,\aV_{y_0})
=
y_{SC}(y',\aV_{y'})
= y^*_M + \delta y^*$,
where\footnote{
Here, we use     
\begin{equation}
    \begin{aligned}
\tau\left(y,\aV_{y}\right) 
&= \arg \left\{ \left[1+4R_d \aV_{y}\right]\cos\left(\Omega^\otimes\log\left(\frac{\tilde{\mathscr{C}}_{y}}{2}\right)\right)-\sqrt{-\discinf}\sin\left(\Omega^\otimes\log\left(\frac{\tilde{\mathscr{C}}_{y}}{2}\right)\right)\right.\\
&\left.+i\left[\sqrt{-\discinf}\cos\left(\Omega^\otimes\log\left(\frac{\tilde{\mathscr{C}}_{y}}{2}\right)\right)+ \left(1+4R_d \aV_{y}\right)\sin\left(\Omega^\otimes\log\left(\frac{\tilde{\mathscr{C}}_{y}}{2}\right)\right)
 \right] \right\}.
    \end{aligned}
\end{equation} }\begin{equation}
    \delta y^* 
    = \left. \frac{d y_{SC}(y',\aV_{y'})}{d 
   \aV_{y'}  }
   \right|_{ \aV_{y'} = \aVM_{y'}}
   c' e^{-y_0} \deltaaVUV
   \sim  
\sqrt{\frac{\KFAVdim}{\Lambda}} 
\frac{\deltaaVUV}{n}.
\label{eq:deltaySC}
\end{equation}
$T_c = \Lambda e^{-2z ( y_0-y_{SC})}$
along with
Eqs. \eqref{eq:y} and \eqref{eq:thetamax}
gives a quasi-universal relation between the superconducting transition temperature and the Fermi momentum,
\begin{equation}
    \frac
    { \log \left[ \frac{\Lambda^{z - 1} T_{c,n}}{\KFAVdim^z}
   \left(  \frac{\pi n}{2 \gamma} \right)^{2z}
\right]_{\aV_{y_0}=\aVM_{y_0}+\deltaaVUV}}
    { \log \left[ \frac{\Lambda^{z - 1} T_{c,n}}{\KFAVdim^z}  
   \left(  \frac{\pi n}{2 \gamma} \right)^{2z}
\right]_{\aV_{y_0}=\aVM_{y_0}}}
    =
    1 + 
    \frac{\delta y^*}{y_M^*}.
  \label{eq:TC_KF_universal}
\end{equation}
Here, 
    $\left[ \frac{\Lambda^{z - 1} T_{c,n}}{\KFAVdim^z} \right]_{\aV_{y_0}=\aVM_{y_0}} =
    \left(\frac{8}{\alpha_d^{2/3} \eta_d}\right)^z
    \exp\left(\frac{ -2z(\pi - \tau_d^M)}{\Omega^\otimes} \right)
    \left( \frac{2\gamma}{n\pi} \right)^{2z}$
is the ratio between the superconducting transition temperature and the Fermi momentum for the bare coupling on the metallic PFP.
The correction scales as
\bqa
\frac{\delta y^*}{y_M^*} \sim 
\sqrt{\frac{\KFAVdim}{\Lambda}} 
\frac{\sqrt{-\etaPIII}}{n} \deltaaVUV.
\eqa
In class C proximate to class A,
$|\etaPIII| \ll 1$
and 
\eq{eq:TC_KF_universal}
becomes rather insensitive to $\deltaaVUV$.
This is confirmed through the direct integration of the PFP equation in Figs. \ref{fig:ysc_Vuv_yuv} and \ref{fig:ysc_Vuv_yuv_d2}.
The ratio becomes more weakly dependent on the bare coupling near $d_{SC}$ than in $d=2$ due to the proximity to class A.
For large $n$, $\delta y^*/y_M^*$ becomes further suppressed because there is a large window of RG time for the coupling to flow close to the metallic PFP before it diverges close to $y^*_M$.

\begin{figure}[th]
\centering

\begin{subfigure}{.35\textwidth}
  \centering
  \includegraphics[width=1\linewidth]{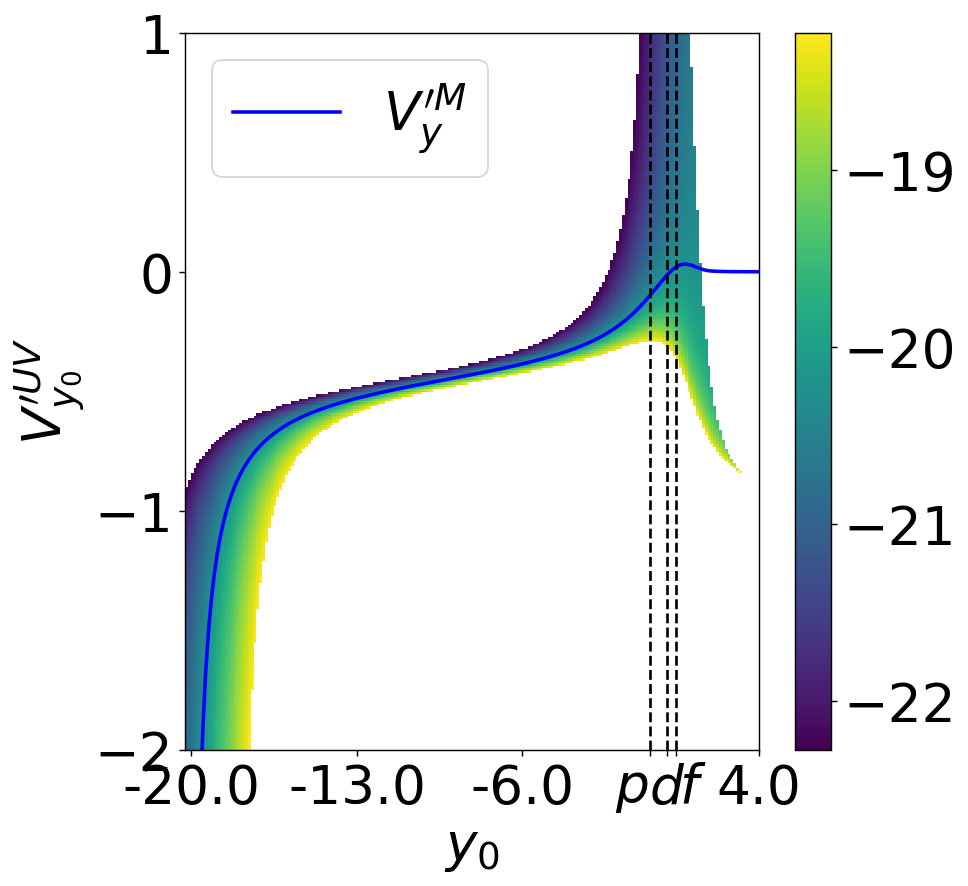}
  \caption{
}
  \label{fig:ysc}
\end{subfigure}%
\begin{subfigure}{.35\textwidth}
  \centering
  \includegraphics[width=1\linewidth]{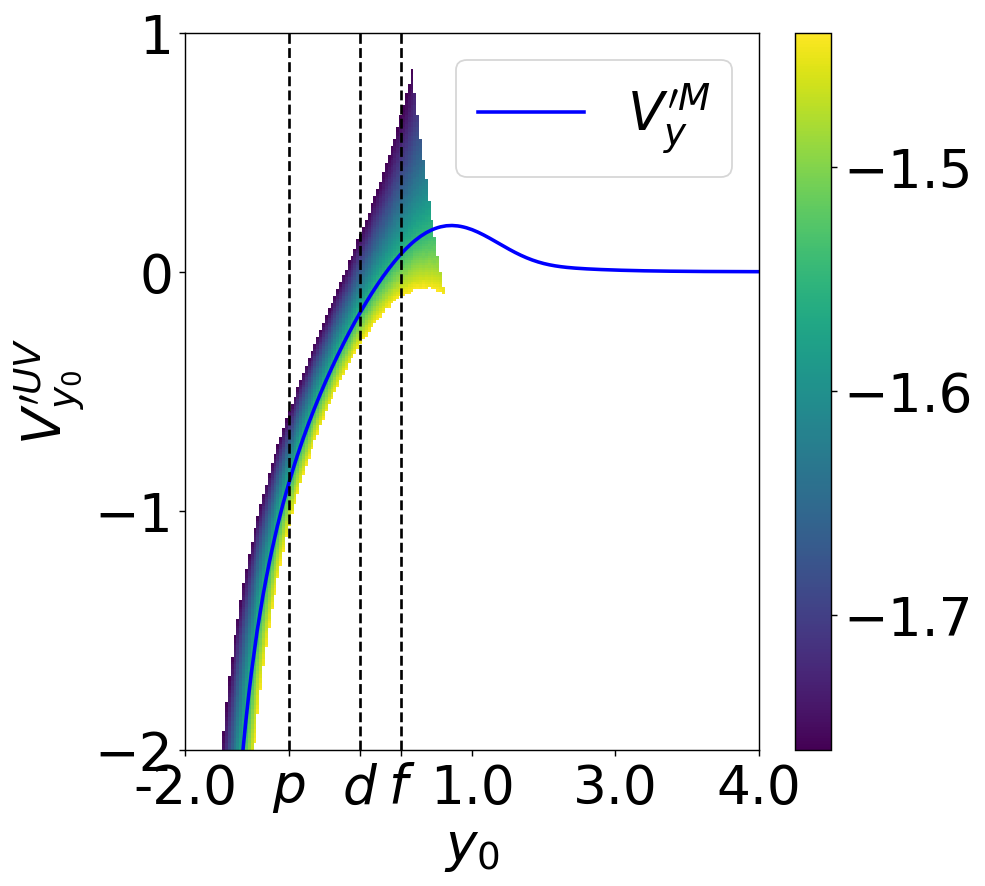}
  \caption{
}
  \label{fig:ysc_d2}
\end{subfigure}%
\begin{subfigure}{.35\textwidth}
  \centering
  \includegraphics[width=1\linewidth]{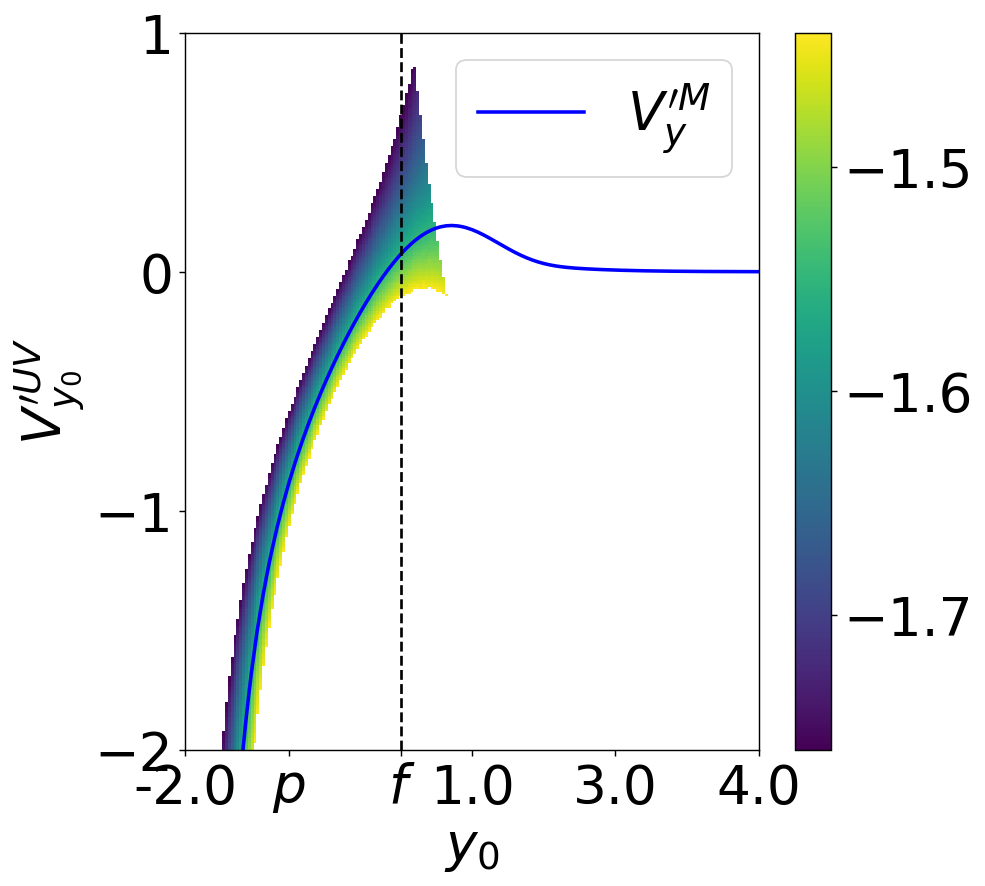}
  \caption{
}
  \label{fig:spinless_ysc_d2}
\end{subfigure}
\caption{
The shaded region shows the set of angular momenta and UV couplings $(y_0, \aV_{y_0}^{UV} )$ that exhibits superconducting instability at temperatures higher than $T_{c,min}$ in the lowest allowed angular momentum channel, while exhibiting the universal $T_{c,n}/\KFAVdim^z$ relation within tolerance
 \( \left| \frac{\delta y_{\text{SC}}}{y^*_M} \right| \leq 0.1 \) for
(a) spinful fermions in $d=2.2432$,
(b) spinful fermions in $d=2$,
and 
(c) spinless fermions in $d=2$.
The color at each point represents the value of $y_{SC}$ at which the associated PFP diverges to $-\infty$.
For the plot, we use
$\mathbf{K}_F/
\Lambda= 8$.
}
\label{fig:main_ysc}
\end{figure}


In \fig{fig:main_ysc}, we identify the set of angular momentum $y_0$ and its UV coupling 
$\aV^{UV}_{y_0}$ 
that becomes superconducting at a length scale shorter than $l_{SC;max}$ in \eq{eq:Tcmin},
and obeys the universal relation 
within a tolerance,
in a spatial dimension just below \( d_{\text{SC}} \) and in $d=2$, respectively. 
Near \( d_{\text{SC}} \), the $p$-wave channel exhibits a quasi-universal ratio for a large range of UV coupling.
In contrast, in two dimensions, the strong $s$-wave superconducting instability dominates, significantly reducing the region of quasi-universality.

As will be shown later,
the universal $T_c/\KFAVdim^z$ ratio becomes more robust for superconducting states that emerge in the superuniversality classes B and BC.
In those classes, there is no superconducting instability in low angular momentum channels, including the s-wave channel, if the bare couplings in those channels are chosen to be in the basin of attraction of the stable (marginal) $-\infty$-asymptotic fixed point.
As a result, there is a larger window of RG time for the couplings in large angular momentum channels to converge toward the metallic PFP and exhibit a stronger universality in $y_{SC}$.
Unfortunately, these superuniversality classes are not realized within the physical examples we consider here.
In Sec. \ref{sec:ex4}, we will use the toy model to discuss the physical properties of non-Fermi liquids realized in those superuniversality classes.

\subsubsection{Crossover from universal to non-universal superconductors}

\begin{figure}[h]
    \centering
        \begin{subfigure}{.36
        \textwidth}
            \centering
            \includegraphics[width = 1.0\linewidth]{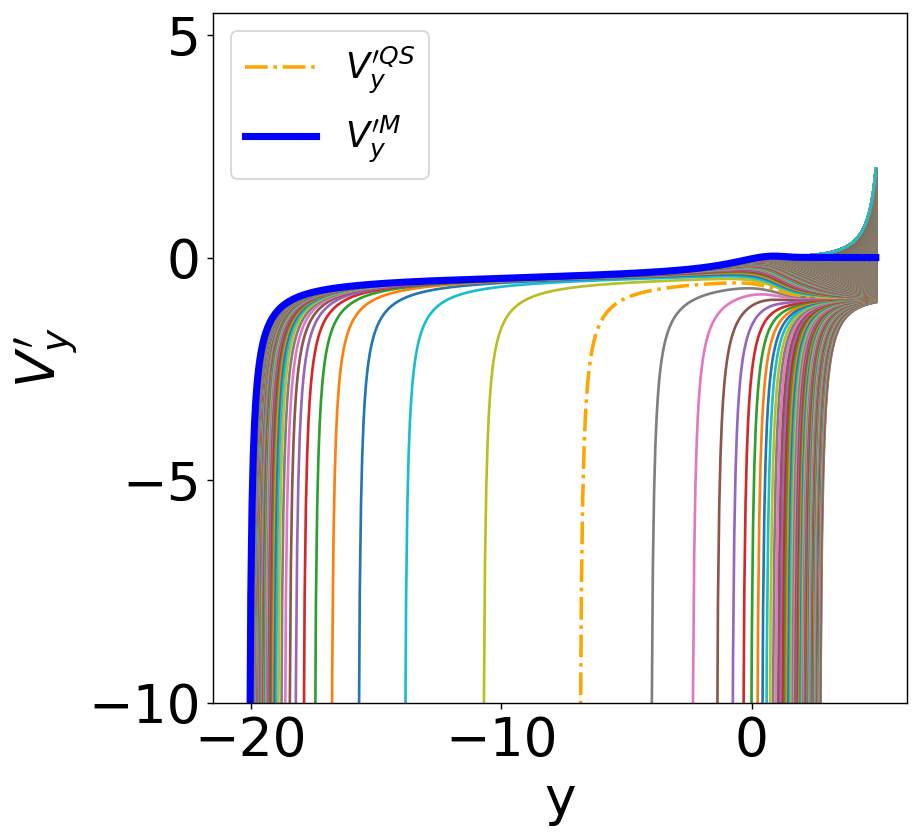}
        \caption{}
        \label{Fig:Ising_phase}
        \end{subfigure}%
         \begin{subfigure}{.36\textwidth}
            \centering
            \includegraphics[width=1.0\linewidth]{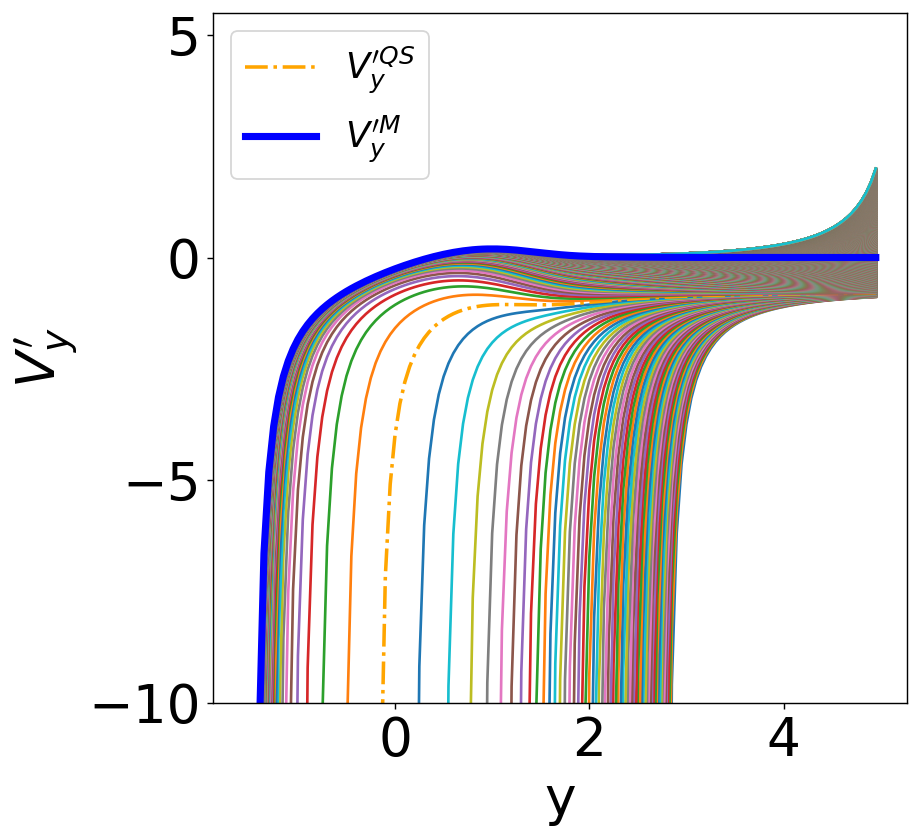}
        \caption{}
        \label{Fig:Ising_phase_d_2}
        \end{subfigure}

        \begin{subfigure}{.36
        \textwidth}
            \centering
            \includegraphics[width = 1.0\linewidth]{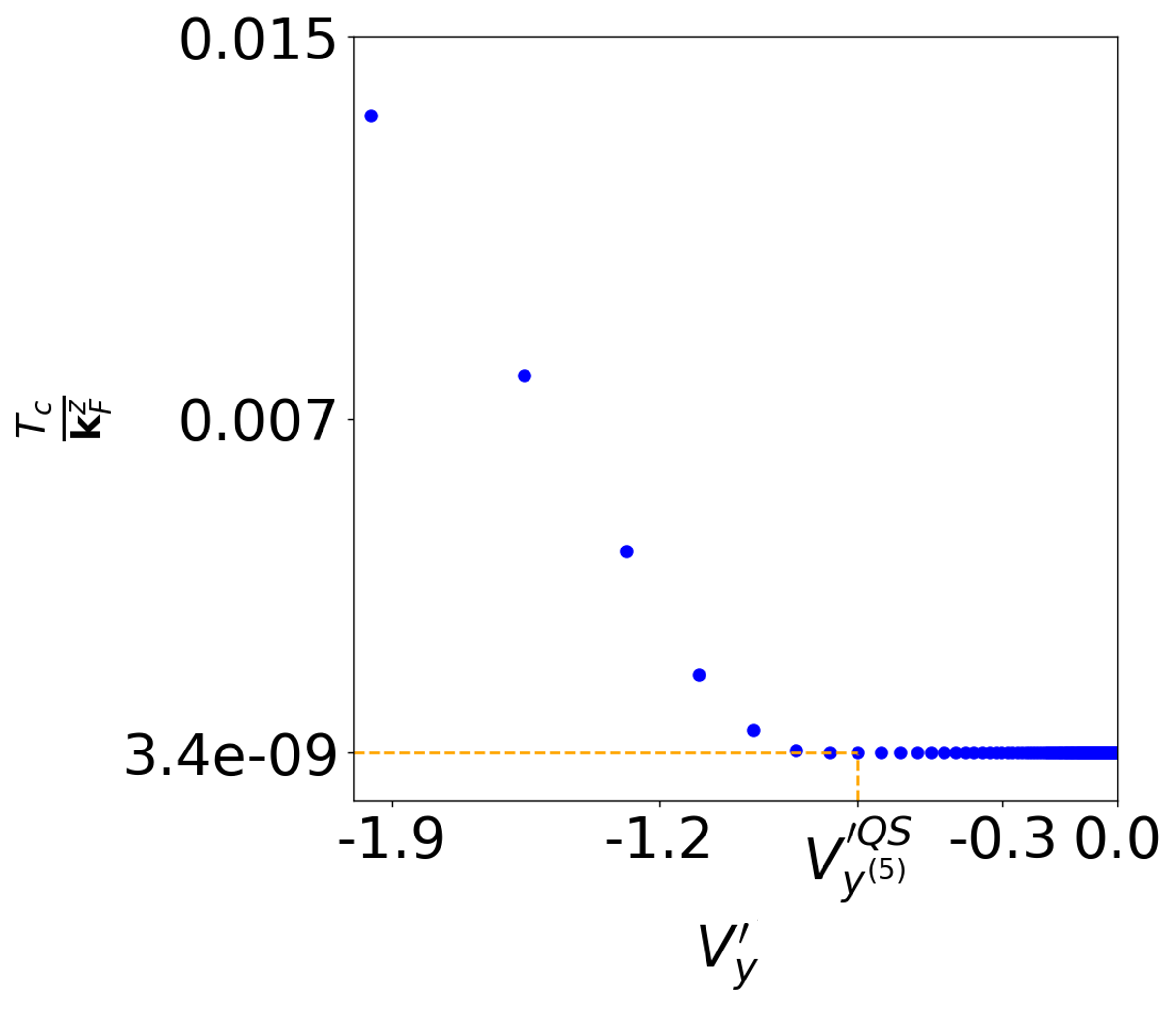}
        \caption{}
        \label{Fig:Tc_kf}
        \end{subfigure}%
         \begin{subfigure}{.36\textwidth}
            \centering
            \includegraphics[width=1.0\linewidth]{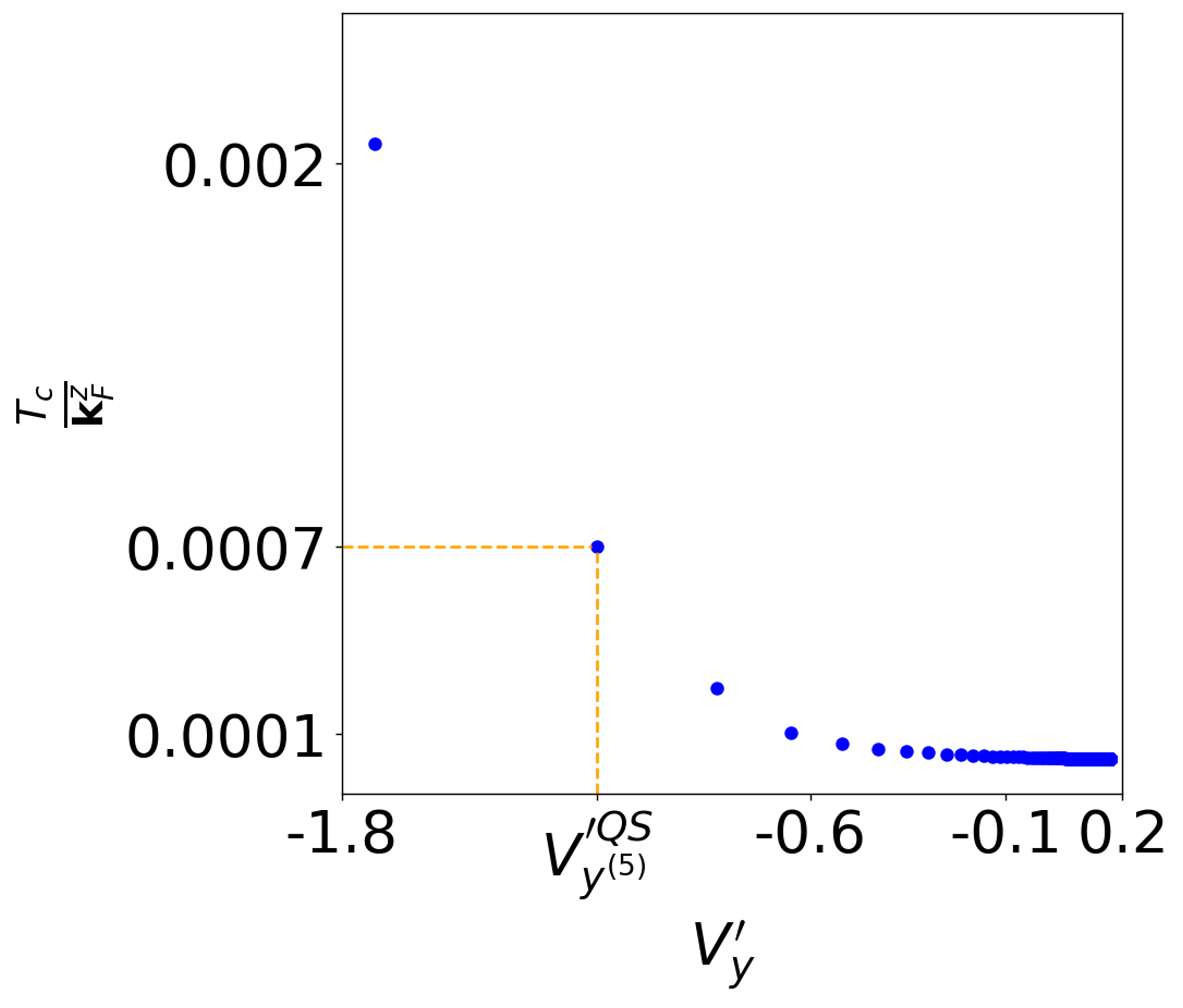}
        \caption{}
        \label{Fig:Tc_kf_d2}
        \end{subfigure}
\caption{
Families of PFPs with different UV couplings imposed at logarithmic angular momentum scale
$y=4.95$ for 
(a) $d=2.432$ and (b) $d=2$. 
The thick solid line is the metallic PFP ($\aVM_y$),
and the dashed dotted line represents the quasi-separatrix PFP ($\aV^{QS}_y$).
If the bare couplings lie above the quasi-separatrix PFP, the resulting superconducting state exhibits a quasi-universal behavior.
$\tfrac{T_c}{\KFAVdim^z}$ in angular mometum channel $n=5$ plotted for (c) $d=2.432$ and (d) $d=2$ as a function of the UV 
coupling in the corresponding angular momentum channel. 
The ratio becomes less sensitive to the bare coupling for the bare coupling chosen above the quasi-separatrix PFP.
Here, 
$\KFAVdim/\Lambda \approx 14$ is used. 
}
\label{Fig:}
\end{figure}

In class C, all PFPs that start below $\aV^\bullet_{\infty}$ in the large $y$ limit  
become singular at $y >y^*_M$ as their trajectories are below the metallic PFP which diverges to $-\infty$ at $y^*_M$.
Among those PFPs, there is a special one that emanates from $\aV^\circ_\infty$ in the large $y$ limit.
This singular PFP serves as a {\it quasi-separatrix} PFP:
PFPs that are above (below) this separatrix tend to get closer to (farther away from) the metallic PFP as $y$ decreases before they eventually diverge.
This is illustrated in   
Figs.
\ref{Fig:Ising_phase}
and
\ref{Fig:Ising_phase_d_2}.
This quasi-separatrix manifests itself as a sharp crossover in $T_c/\KFAVdim^z$.
Suppose one of the bare four-fermion couplings in $y>y^*_M$ is tuned across this quasi-separatrix PFP.
As discussed above, $y_{SC}$, at which the renormalized coupling diverges to $-\infty$, changes as a function of the bare coupling. 
Since the quasi-separatrix repels nearby PFPs, $y_{SC}$ changes most steeply when the bare coupling crosses the quasi-separatrix PFP, as is shown in 
Figs.
\ref{Fig:Tc_kf}
and
\ref{Fig:Tc_kf_d2}.
The weak (strong) dependence of $y_{SC}$ on the bare coupling is directly translated into the quasi-universality (and the lack thereof) in  $T_c/\KFAVdim^z$.

\begin{figure}[th]
    \centering
\includegraphics[width = 0.5\linewidth]{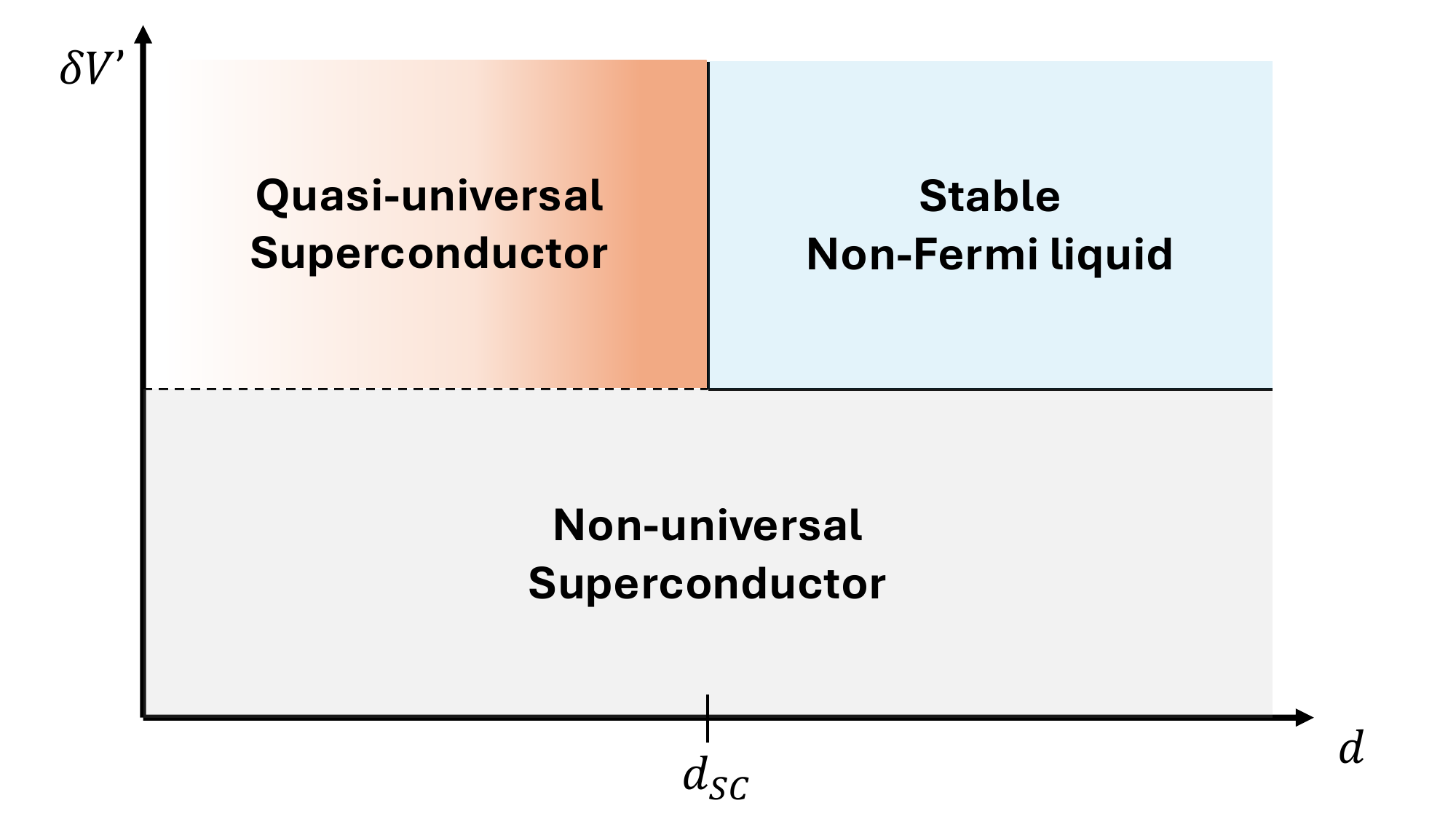}
\caption{
A phase diagram of the Ising-nematic quantum critical metal.
$d$ is the space dimension
and $\deltaaVUV$ is the deformation added to the pairing interaction at a non-zero angular momentum channel $n$ 
relative to the separatrix PFP 
in class A (for $d > d_{SC}$)
and class AC (for $d = d_{SC}$),
and
the quasi-separatrix PFP in class C (for $d< d_{SC}$).
In the non-universal superconducting state, 
$T_c/\KFAVdim^z$ is sensitive to 
$n$ and $\deltaaVUV$.
On the other hand, the ratio is only weakly dependent on 
$n$ and $\deltaaVUV$ in the quasi-universal superconducting state.
A stronger sense of universality arises closer to the boundary with the stable non-Fermi liquid.
Two superconducting states are separated by a crossover in $d < d_{SC}$.
In $d \geq d_{SC}$, the crossover becomes a sharp phase transition as $T_c$ becomes zero for $\deltaaVUV \geq 0$.
In the hybrid theory considered in the next section, the transition from classes A to C can be driven by tuning a marginal coupling at a fixed dimension.
}
\label{Fig:universal_to_nonuniversal_SC}
\end{figure} 

For the UV theories with the bare coupling above the quasi-separatrix, $T_c/\KFAVdim^z$ is largely insensitive to the bare coupling because the PFP trajectory is first attracted to the metallic PFP and diverges close to it.
For the UV theories whose bare couplings are below the quasi-separatrix,
$T_c/\KFAVdim^z$ is sensitive to the bare coupling because the PFP diverges far from the metallic PFP.
Since the nature of the ground state does not change, this is merely a crossover
from a quasi-universal superconductor 
to a non-universal superconductor.
However, the crossover becomes sharper near $d_{SC}$.
At $d_{SC}$, the quasi-separatrix becomes the true separatrix, and the crossover becomes the phase transition from the stable metal to a superconductor.
This is illustrated in \fig{Fig:universal_to_nonuniversal_SC}.


\section{Example 3: Hybrid theories}
\label{sec:ex3}

In this section, we consider an example with two types of critical bosons,
where one mediates an attractive interaction 
and the other a repulsive interaction
in the s-wave channel.
One such example is the $U(1)$ spin liquid with spinon Fermi surface that undergoes the Ising-nematic phase transition.
With the gauge boson being massless in the U(1) spin liquid phase, one only needs to tune the mass of the Ising-nematic critical boson to realize this state. 
Since the gauge boson is viewed as an anti-symmetric tensor in our dimensional regularization scheme,
in general $d$, there are  $\frac{(4-d)(3-d)}{2}$ repulsive (gauge) bosons and 
one attractive (Ising-nematic) boson with effective Yukawa couplings $g^{*}_1$ and $g^*_2$, respectively. 
The cases with  $(g^{1*},g^{2*})= \bar g^* (1,0)$ and  $(g^{1*},g^{2*})= \bar g^* (0,1)$ are the two limits considered in the previous sections.
In this hybrid theory, what controls the dynamical critical exponent and the anomalous dimension of the fermion is the sum of the Yukawa couplings in \eq{eq:sumofg},
which is fixed to be Eq.~(\ref{eq:gstar}) 
in the low-energy limit.
However, the relative coupling 
$g^{1*}/g^{2*}$ is marginal and freely tunable to leading order in the $\epsilon$-expansion.
While the relative coupling may not survive as a marginal parameter when higher-order effects are included, here we examine how the nature of the non-Fermi liquid evolves between the two limits.

\subsection{Candidate for the non-s-wave superconducting superuniversality class}

\begin{figure}[ht]
 \centering
 \begin{subfigure}{.45\textwidth}
 \centering
 \includegraphics[width=1.0\linewidth]{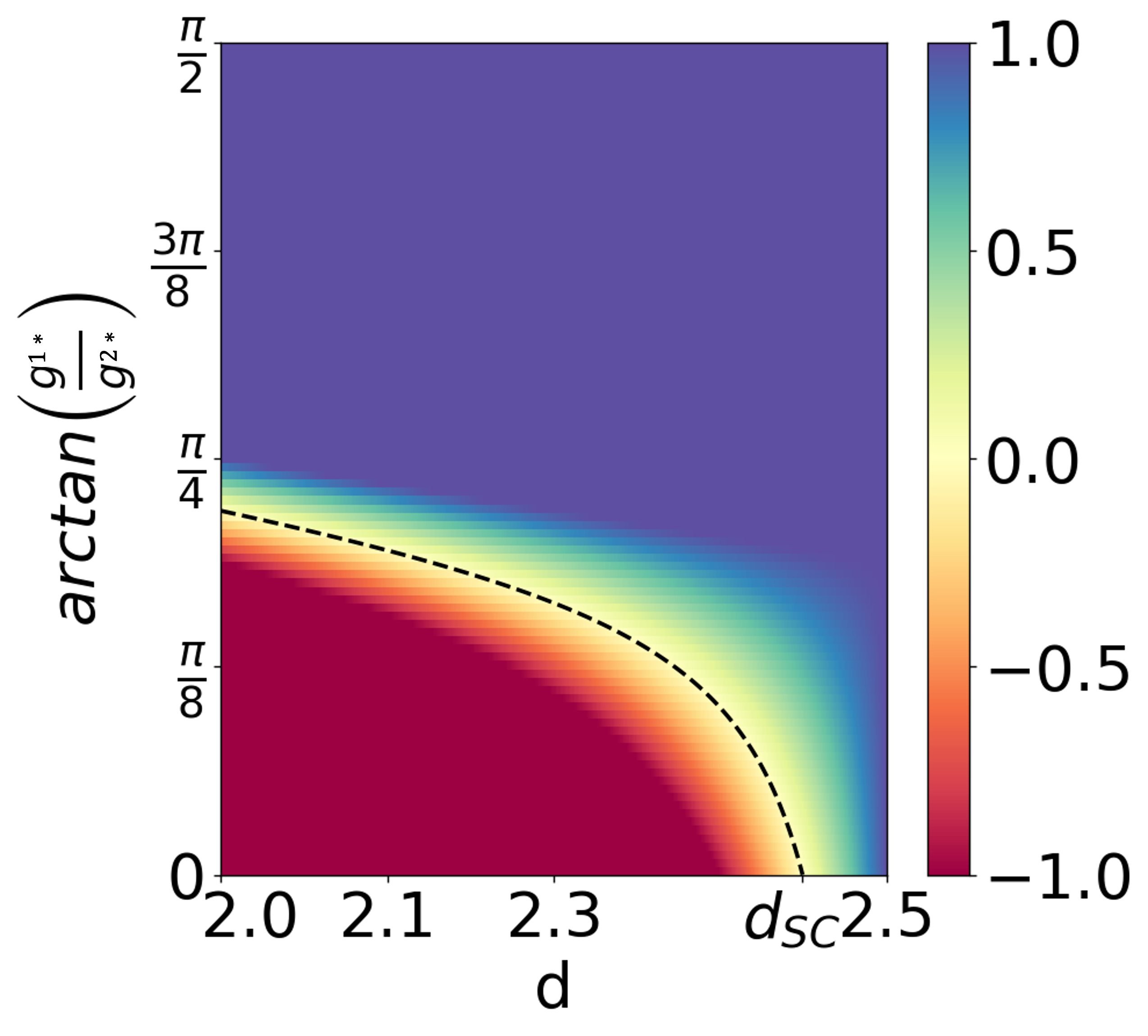}
 \caption{}
 \label{Fig:disc_swave_hybrid0}
 \end{subfigure}%
 \begin{subfigure}{.45\textwidth}
\centering
\includegraphics[width=1.0\linewidth]{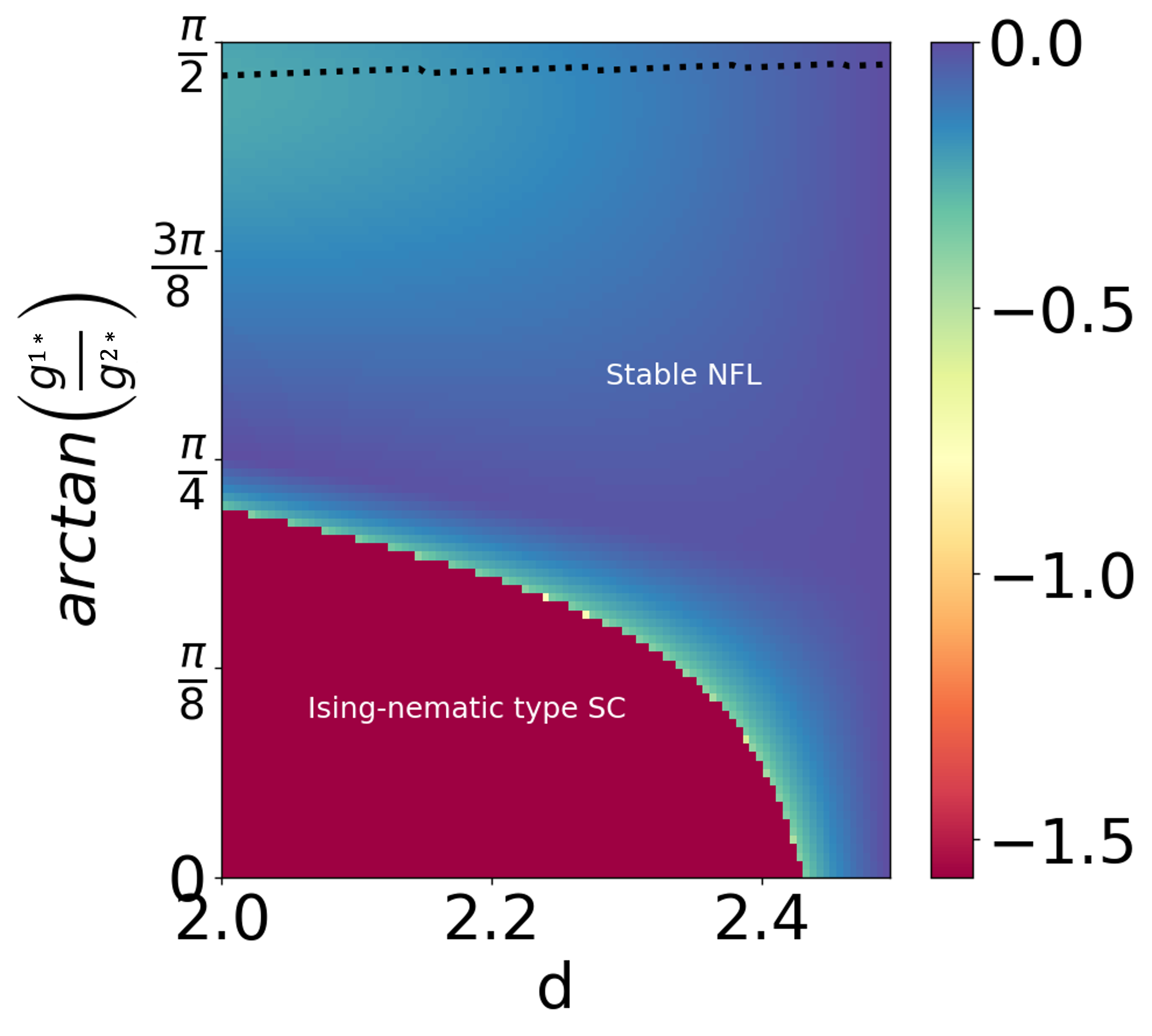}
\caption{}
\label{fig:Phase_diagram_hybrid0}
 \end{subfigure}%
\caption{
Phase diagram of a quantum critical metal coupled to attractive and repulsive bosons, characterized by effective Yukawa couplings $g^{1*}$ and $g^{2*}$, respectively. 
(a) Color plot of  $\discinf$ as a function of $\tan^{-1}\left(\frac{g^{1*}}{g^{2*}}\right)$ and spatial dimension.
The dashed line is the phase boundary with $\discinf=0$.
(b) The color map of $\arctan \aVM_{\text{min}}$. 
In the deep red region, $\aVM_{\text{min}}=-\infty$ and the system is unstable against superconductivity. 
The region of 
$\aVM_{\text{min}}=-\infty$ 
in (b) coincides with the region of $\discinf<0$ in (a).
Above (below) this line, superuniversality class A (C) is realized.
The dotted line in the stable non-Fermi liquid phase  denotes the location of the local minimum of
$\aVM_{\text{min}}$ as a function of $g^{1*}/g^{2*}$ at each $d$. 
}
\label{Fig:}
\end{figure} 
Since the fermions in this hybrid theory are as incoherent as they are in either the pure U(1) gauge theory or the Ising-nematic quantum critical metal, $\HD$
is independent of the relative coupling.
On the other hand, the pairing interaction 
generated from the critical bosons  
is sensitive to the relative coupling because
$\aS_y$ in \eq{eq:Sx} is the sum of the contributions with opposite signs.
The gauge boson generates a pairing interaction of overall strength $g^{1*}$, which is repulsive in the s-wave channel 
and becomes attractive beyond a crossover angular momentum 
$y'_1$.
On the other hand, the Ising-nematic boson generates an interaction with strength $g^{2*}$, which is attractive in the s-wave channel 
and becomes repulsive in angular momentum 
$y > y'_2$. 
The crossover angular momentum is determined from the effective Yukawa coupling for each boson,
\bqa
y'_t \approx 0.79 + \log\left(\alpha_d^{-\frac{1}{3}} \sqrt{\frac{\bar g^*}{g^{t*}}} \right).
\eqa
It is noted that $y'_t$ increases with decreasing $g^{t*}$.
This is because the patch size $\Delta \theta$ associated with a unit proper angular distance $\Delta \bth \sim 1$ is proportional to $\sqrt{g^{t*}}$ as can be seen in \eq{eq:theta_bartheta}.
$e^y$, which is the conjugate to the angle, should have a characteristic scale that is proportional to $1/\sqrt{g^{t*}}$.
As the relative coupling is tuned, the strengths and the crossover angular momenta change simultaneously.

The theory with $g^{2*}=0$ is the pure U(1) gauge theory.
In this theory, the repulsive pairing interaction in the s-wave channel prevents the s-wave superconducting instability.
Although the pairing interaction is attractive in large angular momentum channels, it does not lead to instability due to the flow of the rescaled angular momentum, as discussed in Sec. \ref{sec:ex1}. 
As $g^{2*}$ is increased from $0$,
$\discinf$ decreases monotonically as
\begin{equation}
    \discinf = 1 + 
    \frac{4\eta_d}{\bar g^*}
\left( \frac{(3-d)d}{2}g^{1*} - g^{2*} \right).
\label{eq:disc_swave_hybrid}
\end{equation}
This is expected because the Ising-nematic fluctuations promote the s-wave pairing instability.
Therefore, there is a phase boundary at which $\discinf=0$ 
in the plane of $d$ and $g^{1*}/g^{2*}$ 
as is shown in
\fig{Fig:disc_swave_hybrid0}.
Below this phase boundary, the system is unstable against s-wave pairing, and the theory belongs to the s-wave superconducting superuniversality class (class C), the same class to which the pure Ising-nematic critical metal belongs near $d=2$.
Above this boundary, there is no s-wave superconducting instability, and the theory is either in the stable NFL superuniversality class (class A) or 
the non-s-wave SC superuniversality class (class B), generically.
To determine which of the two superuniversality classes the theory of general $g^{1*}/g^{2*}$ belongs to, we plot the minimum of the metallic PFP in \fig{fig:Phase_diagram_hybrid0}. 
It turns out that $\aVM_{y}$ is regular everywhere in the region with $\discinf > 0$, and therefore, the theory outside the region of the s-wave superuniversality class belongs to the stable NFL superuniversality class. 

\begin{figure}[ht]
\centering
\begin{subfigure}{.4\textwidth}
\centering
\includegraphics[width=1.0\linewidth]{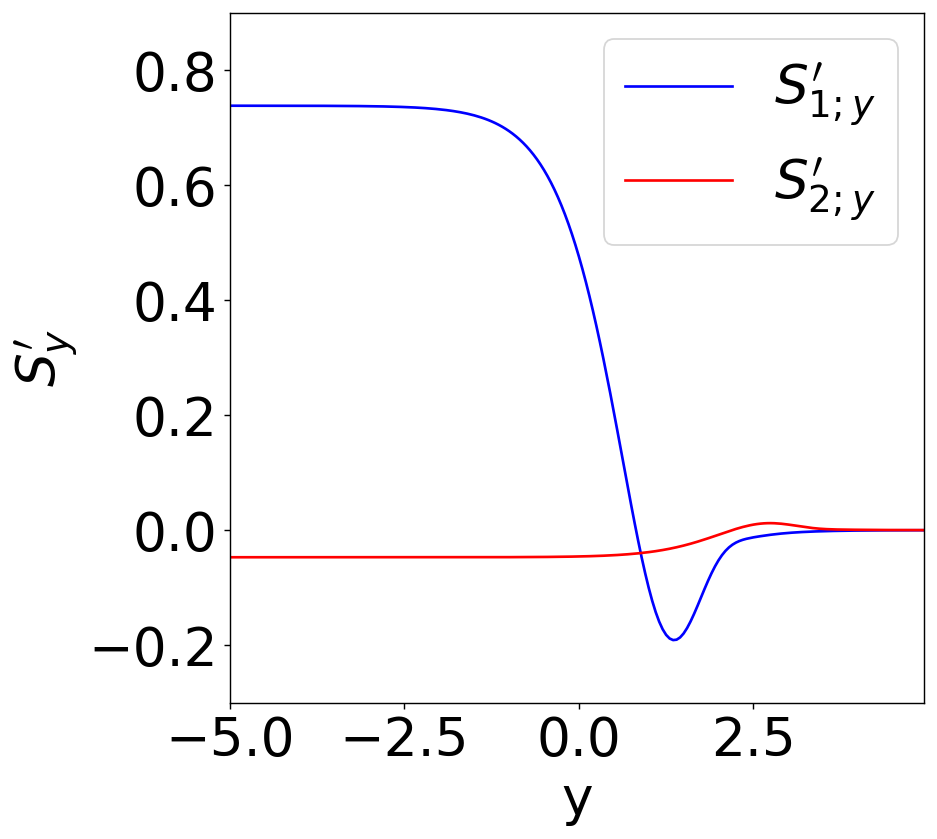}
\caption{}
\label{fig:S1S2}
 \end{subfigure}%
\begin{subfigure}{.4\textwidth}
\centering
\includegraphics[width=1.0\linewidth]{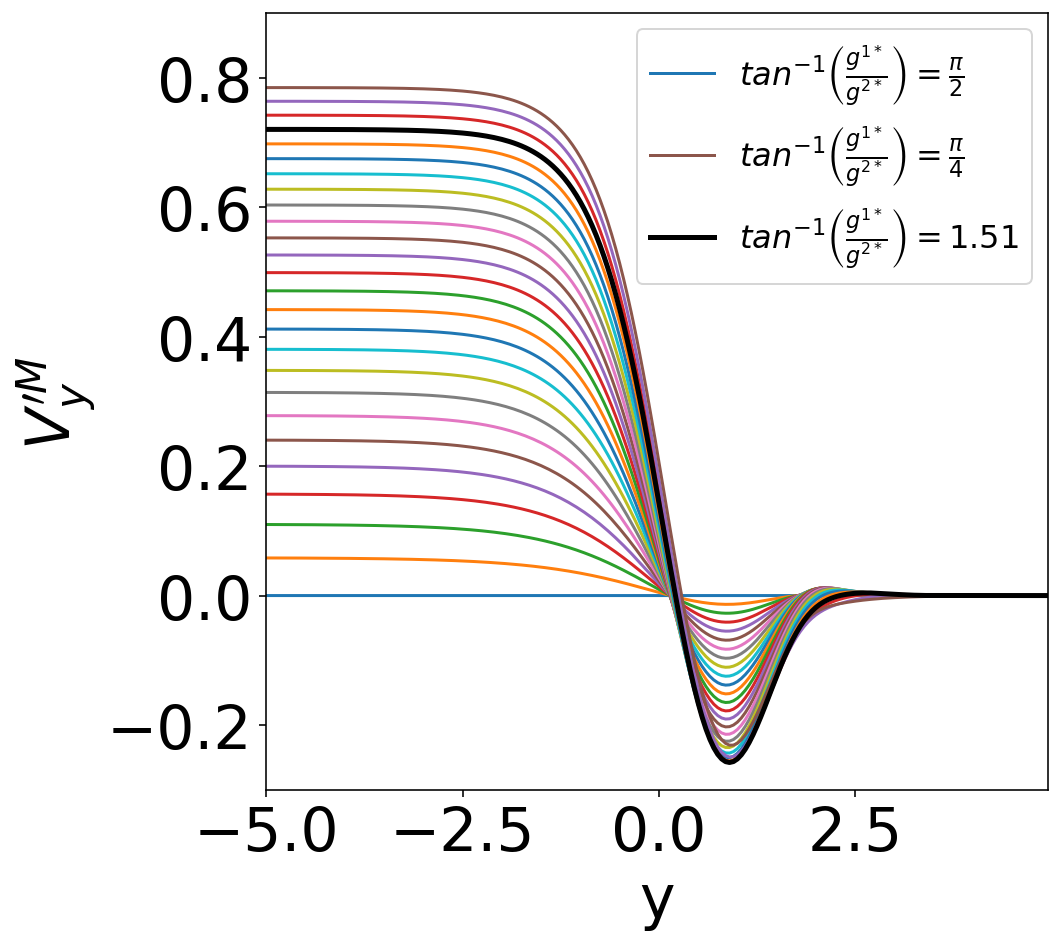}
\caption{}
\label{fig:Metallic_PFP_Hybrid_d2}
\end{subfigure}%
\caption{
(a) The angular momentum-dependent pairing interaction generated from the U(1) gauge field ($\aS_{1;y}$) and
the $C_4$ Ising-nematic boson ($\aS_{2;y}$) for a small but non-zero $g^{2*}$ with $\tan^{-1} g^{1*}/g^{2*} = 1.51$ in $d=2$.
For this choice of couplings, the universal pairing interactions generated from the gauge field and the Ising-nematic boson are both attractive within a range of angular momentum.
(b) 
The attractive interaction generated from both bosons causes the metallic PFP to dip to the lowest value for 
$\tan^{-1} g^{1*}/g^{2*} = 1.51$.
 }
\label{fig:Hybrid_profiles_source}
\end{figure}

Near $d=2$, however, higher-order effects can, in principle, alter the phase diagram.
Among various possibilities, let us consider the one that realizes the non-s-wave superuniversality class (class B).
Although the non-s-wave superuniversality class is not realized anywhere at the leading order,
the leading order result points to promising candidates for realizing class B upon including higher-order corrections.
In the pure U(1) gauge theory, the metallic PFP is regular, but it dips below zero in a range of $y$ due to the attractive pairing interaction at finite angular momenta. 
To realize the non-s-wave SC superuniversality class, one needs a stronger attractive interaction that makes the dip deeper to the extent that $\aVM_y$ diverge to $-\infty$.
As $g^{2*}$ is turned on, the overall strength of the interaction from the U(1) gauge field weakens, but the Ising-nematic boson kicks in and provides an attractive pairing interaction in $y<y_2'$.
At a small $g^{2*}$, the attractive interaction from the Ising-nematic fluctuations spans a wide range of $y$ from $y=-\infty$ to $y'_2 \gg 1$.
This creates a window of $y$ in which both the U(1) gauge field and the Ising-nematic fluctuations mediate an attractive pairing interaction for a small $g^{2*}$, as shown in \fig{fig:S1S2}.
As a result, $\aVM_y$ dips to the lowest value for a small but non-zero $g^{2*}$, as shown in  \fig{fig:Metallic_PFP_Hybrid_d2}. 
This suggests that the non-s-wave SC superuniversality class has the best chance of being realized close to the pure U(1) gauge theory with a small component of the Ising-nematic coupling in $d=2$.

The theory with $g^{1*}=g^{2*}$ is special in that the critical bosons generate no pairing interaction at all: $\aS_y=0$.
As a result, the metallic PFP has a vanishing universal pairing interaction, $\aVM_y=0$, as is shown in  \fig{fig:Metallic_PFP_Hybrid_d2}.
The vanishing pairing interaction results from the perfect cancelation between the pairing interaction generated by the gauge field and that from the Ising-nematic boson.
This corresponds to the most stable non-Fermi liquid in that
$\mbox{min}_y \eta_y$ is the largest.

\subsection{Class C proximate to class A}

In this section, we revisit the s-wave SC superuniversality class.
Although we already discussed this class extensively for the pure Ising-nematic critical metal, the hybrid theory provides an opportunity to realize a theory that is ultimately unstable against the s-wave superconductivity but is in a close proximity to the stable NFL superuniversality class with small $|\discinf|$.
In such cases, a stronger quasi-universal behavior emerges in the intermediate energy scales due to the large hierarchy between the superconducting transition temperature and the energy scale below which the non-Fermi liquid physics sets in.
While such quasi-universal behaviors also arise in the pure Ising-nematic critical metal in $d$ slightly below $d_{SC}$, the current example is more physical in that this can, in principle, arise in $d=2$.

\begin{figure}[th]
    \centering
        \begin{subfigure}{.4
        \textwidth}
            \centering
            \includegraphics[width = 1.0\linewidth]{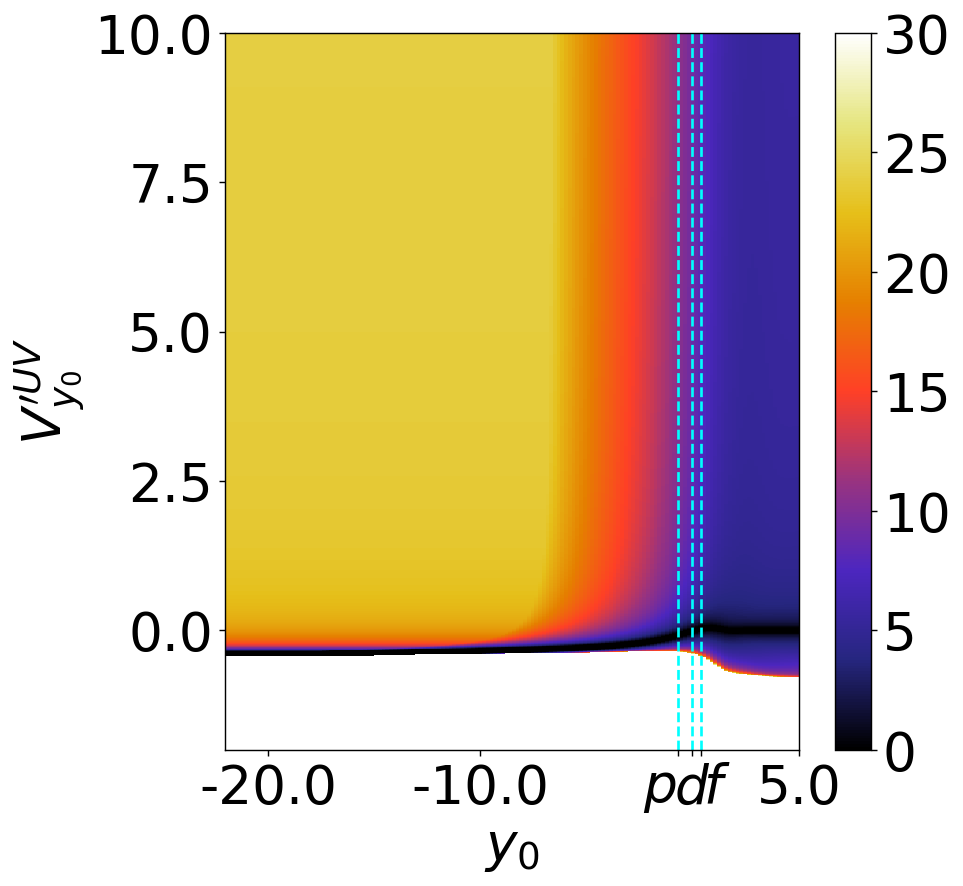}
        \caption{}
        \label{Fig:lb_contour_hybrid_d2}
        \end{subfigure}%
         \begin{subfigure}{.42
         \textwidth}
            \centering
            \includegraphics[width=1.0\linewidth]{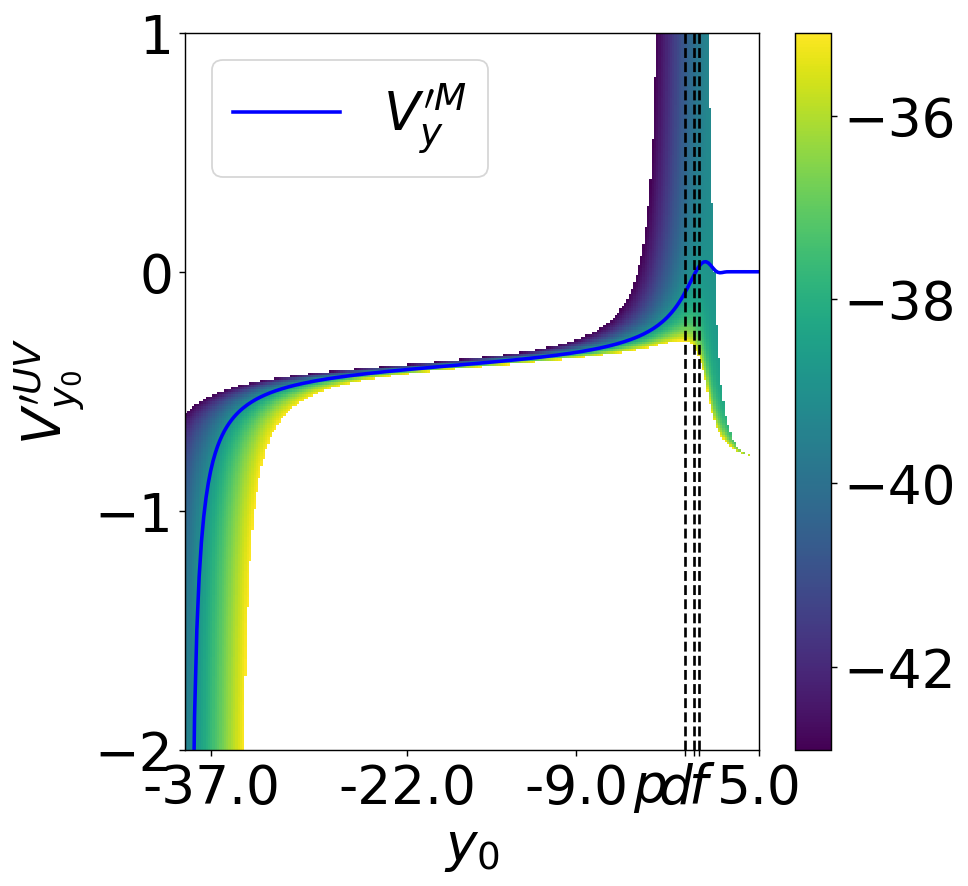}
        \caption{}
        \label{Fig:ysc_countour_hybrid_d2}
        \end{subfigure}%
    \caption{
(a) The color at each point represents the logarithmic scale $l$ that it takes for that coupling to approach the regularized metallic PFP within the tolerance $\mathcal{T}=0.045$ with $l$ less than $l_{SC;max}= 79.99$ in \eq{eq:Tcmin}
for the hybrid theory with $g^{1*} \approx 3.86$ and $g^{2*} \approx 4.99$ in $d=2$.
The white region is outside the basin of attraction.
(b) The set of UV couplings at non-zero angular momentum channels that exhibit the universal $T_c/\KFAVdim^z$ ratio within tolerance
$\left| \delta y^* / y^*_M \right| \leq 0.1$. 
For the plot, we use 
$\KFAVdim/\Lambda = 10$.
The color represents the value of $y_{SC}$ at which the associated PFP diverges to $-\infty$.
}
\label{Fig:}
\end{figure} 

To be concrete, let us consider the theory with  $g^{1*} \approx 3.86$ and $g^{2*} \approx 4.99$ in $d=2$.
In this theory, 
$\discinf \approx -0.02$
and the theory is barely in the s-wave SC superuniversality class.
The complex $-\infty$-asymptotic fixed points that are close to the real axis create a tight bottleneck. 
If a superconducting instability occurs in the s-wave channel, the transition temperature in \eq{eq:Tcmin} is low enough that the couplings in other pairing channels are attracted close to the metallic PFP as is shown in  \fig{Fig:lb_contour_hybrid_d2}.
In this case, the pairing interactions in high-angular momentum channels are controlled by the regularized metallic PFP.
If the superconducting instability occurs in a non-s-wave channel, either because the fermion is spinless or because the bare coupling is highly repulsive in the s-wave channel, the coupling in the non-s-wave channel first becomes attracted to the metallic PFP before it diverges near $y^*_M$.
Then, the ratio between $T_c$ and $\KFAVdim^z$ is expected to obey the universal relation in \eq{eq:TC_KF_universal} to a great degree of accuracy.
This is illustrated in         \fig{Fig:ysc_countour_hybrid_d2}.

\section{Example 4: Back to the toy model}
\label{sec:ex4}

The final example we consider is the toy model introduced in  Section \ref{sec:toymodel}, which has been demonstrated to be sufficiently flexible to realize all seven superuniversality classes. 
The primary purpose of returning to the toy model is to spell out the universal physics of the four remaining classes (B, AB, BC, and ABC) that are not discussed in our physical example. 
Having the universal properties of all superuniversality classes in one place also makes it easy to understand their differences.

\subsection{Superuniversal and universal observables}

The toy model is a useful tool for classifying the topology of the superuniversality classes.
However, can it also be used to predict more fine-grained observables, such as critical exponents?
We answer this question by discussing which universal properties the toy model can and cannot capture.
This is most conveniently done by comparing the predictions of the toy model against those obtained from the physical examples.
For this, we consider the U(1) gauge theory that is in class A. 
The universal superconducting fluctuations of the non-Fermi liquid are characterized by the metallic PFP.
For the U(1) gauge theory, we do not have a closed-form expression of the metallic PFP for general $y$, but its asymptotic form in the small $y$ limit is given by \eq{stable_sol}. 
In the toy model, the entire profile of the metallic PFP can be obtained as
\bqa
\aV_y^M = \frac{1}{4R_d}\begin{cases}
 \sqrt{\etaPI}-2\HD + 1 ~~~~& \text{for } y > 0\\
    -\sqrt{\etaPIII}\tanh\Big[\frac{1}{2}\sqrt{\etaPIII} y-\text{arctanh}\Big(\frac{\sqrt{\etaPI}}{\sqrt{\etaPIII}}\Big) \Big]-2\HD + 1 ~~~~ &\text{for } y \le  0  
\end{cases}.
\label{eq:toy_classa_vm_main}
\eqa
In the $y\to -\infty$ limit, it is reduced to
\bqa
\aV^M_y = \aV^\bullet_{-\infty} + b^M e^{\sqrt{\etaPIII}y} + \order{e^{2\sqrt{\etaPIII}y}},
\label{eq:aVM_classA_toy}
\eqa
where
$\aV^{\bullet}_{-\infty} = \frac{\sqrt{\etaPIII} -2\HD +1 }{4R_d}$
and
$b^M = \frac{\sqrt{\etaPIII}}{2R_d}\frac{\sqrt{\etaPI}-\sqrt{\etaPIII}}{\sqrt{\etaPIII}+\sqrt{\etaPI}}$.
This gives rise to the universal pairing interaction that decays as
\begin{equation}
    \begin{aligned}
\aV^{super}_{\bar{\theta}_1,\bar{\theta}_2} \sim
\left|\frac{\sqrt{\mu}}{\bar{\theta}_1-\bar{\theta}_2}\right|^{1+\sqrt{\discinf} }
\label{eq:VSU}
\end{aligned}
\end{equation}
at large angular separations with $\Delta \bth \gg \sqmu$.
This prediction from the toy model coincides with the  universal pairing interaction of the U(1) gauge theory in \eq{eq:stable_large_mom} to the leading order in $\sqmu/\Delta \bth$.
This contribution is entirely determined by the universal pairing interaction in the $-\infty$-asymptotic region.
It is common among all non-Fermi liquids within the superuniversality class A,
and is referred to as the 
{\it superuniversal pairing interaction}. 
This superuniversal pairing interaction is what is captured by the toy model.

However, the toy model has its limitations.
Since it models the scale-dependent pairing interaction through a piecewise-constant function, it does not encode the detailed angular momentum dependent crossover.
For example, the source for the universal pairing interaction in the U(1) gauge theory (\eq{eq:fp_eq_log}) approaches its asymptotic value as
\bqa
\aS_y = \aS_{-\infty} + s_1 e^{a y}
\label{eq:Syexpansion}
\eqa
in the small $y$ limit, where $a > 0$ is an exponent that depends on individual universality classes. 
This $y$-dependent source gives rise to an additional contribution to the universal pairing interaction, 
\bqa
\aV^{non-super}_{\bth,\bth+\Delta \bth} 
\sim 
\Big| \frac{\sqmu}{\Delta\bar{\theta}}\Big|^{1+a}.
\label{eq:SyContribution}
\eqa
This contribution is only universal but not superuniversal because it  generally varies among individual universality classes. 
Hence,  \eq{eq:SyContribution} is referred to as {\it non-superuniversal pairing interaction}. 
For the U(1) gauge theory, the non-superuniversal interaction gives rise to a subdominant contribution of
$\aV^{non-super}_{\bth,\bth+\Delta \bth} 
\sim 
\Big| \frac{\sqmu}{\Delta\bar{\theta}}\Big|^{5}$ to the leading order in the $\epsilon$-expansion.
However, we cannot exclude the possibility that the universal contribution dominates the superuniversal contribution.

With this limitation in mind, we extract superuniversal observables of all classes using the toy model in this section.
The derivations of the results discussed in this section can be found in Appendix
\ref{appendix:ex4}.

\subsection{Superuniversality  classes A, AB, AC, and ABC}

In this subsection, we collect superuniversal properties of the classes that contain stable non-Fermi liquids.

\subsubsection{Stable NFLs}

\begin{table}[h!]
\centering
\begin{tabular}{|c|c|c|c|c|}
\hline
& A & AB & AC & ~ABC~ \\
\hline
&&&&\\
~$
\aV^{stable}_{\bar{\theta}+\Delta \bar{\theta}, \bar{\theta} }
$~ 
& 
~$ \Big| \frac{\sqmu}{\Delta\bar{\theta}}\Big|^{1+\sqrt{\etaPIII}}$~ & 
~$\left| \frac{\sqmu}{\Delta \bar{\theta} } \right|
f\left(
\frac{\Delta \bth}{\bth_c(l)}
\right)
$
~&
~$\left| \frac{\sqmu}{|\Delta \bar{\theta} } \right|
\frac{1}{\log^2 \big(\frac{|\Delta \bar{\theta}|}{\sqmu}\big)}$
~& 
~$0$
~\\
&&&&\\
\hline
\end{tabular}
\caption{
The superuniversal pairing interactions that arise in the stable non-Fermi liquids 
for $\Delta \bar{\theta} \gg \sqmu$.
In class AB, $f(x)$ is a crossover function that decays to zero for $x \gg 1$,
where
$\bth_c(l)
\sim 
\sqrt{\Lambda}
\exp{-\frac{1}{2} \frac{\sqrt{\etaPIII}}{\sqrt{\etaPI}+\sqrt{\etaPIII}} l }
$ is the scale-dependent crossover angle in class AB. 
}
\label{tab:toy_stableclasses_universalvm}
\end{table}

In classes A, AB, AC, and ABC, the stable metallic states persist down to zero temperature as long as the bare couplings lie above the separatrix PFP.
In the low-energy limit, each non-Fermi liquid exhibits a unique universal pairing interaction.
The full pairing interaction consists of the superuniversal piece captured by the toy model
and
the non-superuniversal piece in \eq{eq:SyContribution}.
The superuniversal pairing interactions for the stable non-Fermi liquids are summarized in \tab{tab:toy_stableclasses_universalvm}.

In classes A, AC, and ABC, the entire region above the separatrix is the basin of attraction for the metallic PFP (see \fig{fig:PFP_nfl1} for class A, for example).
Therefore, all couplings that are placed above the separatrix at the UV scale flow to the metallic PFP at low energies, and the universal pairing interaction is given by the Fourier transform of $\aV^M_y$, as written in \eq{eq:InverseFourierV2}.
The expressions for the superuniversal pairing interaction in classes A and AC are shown in Eqs. \eqref{eq:stable_large_mom}
and \eqref{eq:stable_large_mom_dsc}, respectively.
They are precisely reproduced by the toy model.
In the small $y$ limit,
the metallic PFP in class ABC takes the same form as the separatrix PFP of class AC;
as one approaches ABC from AC, the metallic PFP merges with the separatrix PFP whose form does not change throughout that topological phase transition.
In the toy model, the separatrix PFP is $y$-independent in the small $y$ limit.
Accordingly, the superuniversal pairing interaction in class ABC is $0$, and the universal pairing interaction is solely given by the non-superuniversal contribution. 

From Sec. \ref{sec:symmetry}, we recall that large-angle scattering is marginal if the universal pairing interaction decays as 
$\sqrt{\mu}/\Delta \bth$ in the large $\Delta \bth$ limit.
In classes A and ABC, the superuniversal pairing interactions decay faster than the marginal case by a positive power of  $\sqmu/\Delta \bth$ at large angles.
Therefore, the large-angle scatterings are strictly irrelevant.
In class AC, the superuniversal pairing interaction is suppressed by $\log^2 \sqrt{\bth/\mu}$, and the large angle scatterings are marginally irrelevant.

The universal coupling in class AB is somewhat different from the three classes considered above.
Because the metallic PFP connects
$\aV^\bullet_{\infty}$ 
and
$\aV^\circ_{-\infty}$,
its local stability changes from stable to unstable as $y$ decreases, that is, $\chi_y$ changes from 
$-\frac{\sqrt{\etaPI}}{2} < 0$
to
$\frac{\sqrt{\etaPIII}}{2}>0$
as $y$ decreases from $\infty$ to $-\infty$\footnote{
Consequently, the basin of attraction of the metallic PFP is no more than the metallic PFP itself.}.
This implies that couplings at large angular momentum channels are first attracted toward the metallic PFP at intermediate energy scales before they flow to 
$\aV^\bullet_{-\infty}$ in the low-energy limit.
Therefore, the universal pairing interaction that emerges at low energies exhibits a crossover from the metallic PFP to $\aV^\bullet_{-\infty}$, as shown in 
\fig{fig:universalVstableNFLclassB}.
Since we will encounter such crossovers in classes B and C as well, here we discuss the angle-dependent pairing interactions that emerge in such cases in a general context.

\begin{figure}[th]
    \centering
\includegraphics[width=0.4\linewidth]{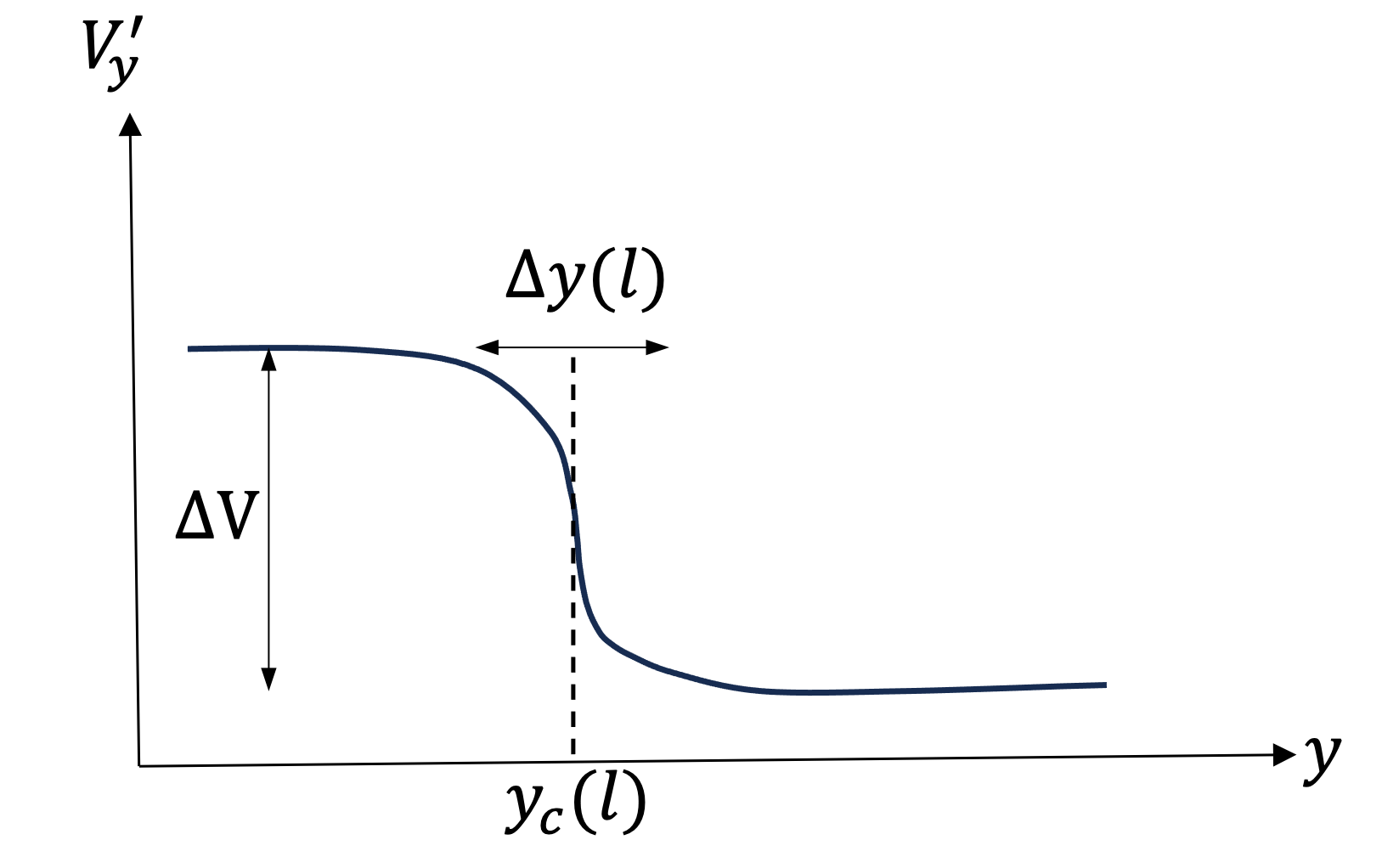}
\caption{
In class AB, the renormalized pairing interaction at scale $l$ exhibits a crossover from 
$\aV^\bullet_{-\infty}$ at low angular momentum 
to 
$\aV^M_{y}$ at large angular.
In other classes, similar crossovers occurs, but interpolating different limits.
The width $\Delta y(l)$ determines the range of the $\sqmu/|\Delta \bth|$ decay for the renormalized pairing interaction in the angular space through 
\eq{eq:theta_crossover}.
}
\label{fig:jump}
\end{figure} 

Suppose that the renormalized coupling function at scale $l$ exhibits a crossover at $y_c(l)$,
and let $\Delta V$ be the jump of the universal pairing interaction,
and $\Delta y(l)$, the range of $y$ over which the crossover occurs
(see \fig{fig:jump}).
If $\Delta y(l)$ was zero, the sharp step function would generate 
\bqa
\aV_{\bth_1, \bth_1+\Delta \bth} 
\sim
\Delta V
\sin \left(
e^{y_c(l)} 
 \frac{\Delta \bth}{\sqmu}
\right)
\left| \frac{\sqmu}{\Delta \bth} \right|.
\label{eq:aVforsharpjump}
\eqa
For a smooth profile with a non-zero width $\Delta y(l)$,
\eq{eq:aVforsharpjump} is valid 
only for $\Delta \bth < \bth_c(l)$.
Here,
\bqa
\bth_c(l) =
\frac{\sqrt{\Lambda}}{ \Delta \tilde x }
\label{eq:theta_crossover}
\eqa
is the crossover angle, where
$\Delta \tilde x = e^{l/2} \left( 
e^{y_c^{+}(l)} - 
e^{y_c^{-}(l)}
\right) $ 
with
$y_c^{\pm}(l) 
=y_c(l) \pm \frac{\Delta y(l) }{2}$
represents the width of the jump in the angular momentum space.
%
For $\theta \ll \bth_c$, the associated uncertainty in $y$ is larger than $\Delta y$ that the crossover can be regarded as a sharp jump
and 
\eq{eq:aVforsharpjump} is valid.
For $\theta \gg \bth_c$, 
however, the uncertainty in $y$ is small enough to resolve the smooth crossover,
and 
$\aV_{\bth_1, \bth_1+\Delta \bth}$ decays faster than $\sqmu/\Delta \bth$ 
for $\theta \gg \bth_c$. 
In class AB, the crossover occurs at
\bqa
y_c(l) \approx  
-\frac{1}{2} \frac{\sqrt{\etaPI}}{\sqrt{\etaPI}+\sqrt{\etaPIII}} l 
\eqa
at large $l$,
and
the width of the crossover is $\Delta y \sim O(1)$.
See Appendix \ref{appendix:ex4}-\ref{appendix:AB} for the derivations of $y_c(l)$ and $\Delta y$.
This causes the renormalized pairing interaction to have a sharp jump around
\bqa
n_c(l) 
\sim 
\sqrt{\frac{\KFAVdim}{\Lambda} }
\exp{\frac{1}{2} \frac{\sqrt{\etaPIII}}{\sqrt{\etaPI}+\sqrt{\etaPIII}} l }.
\label{eq:ncin AB}
\eqa
The $\sqmu/\bar \theta$ decay of the superuniversal pairing interaction is valid up to 
\eq{eq:theta_crossover}, which approaches zero in the low-energy limit. 
At larger angles, the pairing interaction decays faster due to the smooth profile in the $y$ space.

In all four stable non-Fermi liquid universality classes, the superuniversal large-angle scatterings decay faster than $\sqmu/\Delta \bth$ and the full LU(1) emerges in the low-energy limit.
However, there are subtle differences among them.
In class AB, the large-angle scattering is suppressed only for $\Delta \bth \gg  \bar \theta_c(l) \gg \sqmu$, whereas it is suppressed for $\Delta \bth \gg \sqmu$ in classes A and ABC.
Therefore, the number of conserved charges increases more slowly in class AB than in classes A and ABC.
In class AC, the large-angle scattering is only marginally suppressed.

\subsubsection{Critical NFLs} 

\begin{table}[h!]
\centering
\begin{tabular}{|c|c|c|c|c|}
\hline
& A & AB & AC & ABC \\
\hline
~$
\aV^{critical}_{\bar{\theta}, \bar{\theta} + \Delta \bar{\theta}} -
\aV^{stable}_{\bar{\theta}, \bar{\theta} + \Delta \bar{\theta}} 
$~ 
& 
~$\frac{\sqmu}{\bar\theta_{\text{max}}}
\cos\left(\frac{n \pi (\bar\theta_1-\bar\theta_2)}{
2 \bar\theta_{\text{max}}
}\right)$
~&  
~$ \begin{cases}
\frac{\sqmu}{\bar\theta_{\text{max}}}
\cos\left(\frac{n \pi (\bar\theta_1-\bar\theta_2)}{
2 \bar\theta_{\text{max}}
}\right) & 
\text{for $ n \ll n_c(l)$}
\\
0 & 
\text{for $ n \gg n_c(l)$}
\end{cases}
$
~& 
~$\frac{\sqmu}{\bar\theta_{\text{max}}
\log \left(\frac{\Lambda}{\mu}\right)
}
\cos\left(\frac{n \pi (\bar\theta_1-\bar\theta_2)}{
2 \bar\theta_{\text{max}}
}\right)$
~& ~0 ~\\
\hline
\end{tabular}
\caption{
The superuniversal pairing interaction for the critical non-Fermi liquid realized by tuning the pairing interaction in the angular momentum channel $n$ to its critical value.
$n_c(l)$ is the crossover angular momentum (see \eq{eq:ncin AB}) in class AB below (above) which the deviation from the parent stable non-Fermi liquid is significant (negligible).
}
\label{tab:toy_stableclasses_criticalprofiles}
\end{table}

In classes A, AB, AC and ABC, one can tune an irrelevant coupling to a critical strength to realize a critical non-Fermi liquid at the phase transition between the stable non-Fermi liquid and a superconductor.
Here, we examine the universal properties of the critical non-Fermi liquids realized at the charge-$2$ superconducting critical point.

\begin{figure}[th]
    \centering
\includegraphics[width=0.6\linewidth]{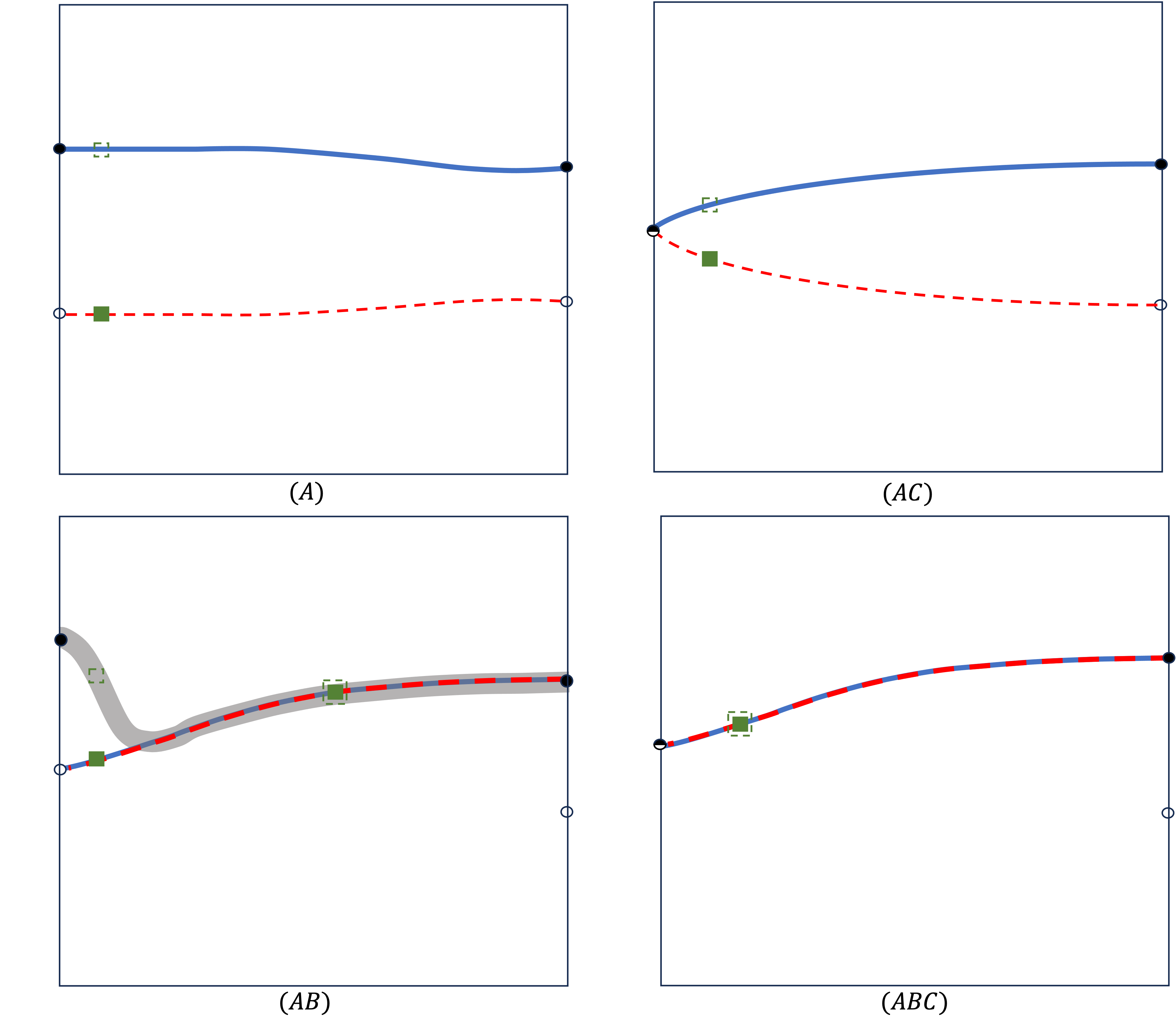}
\caption{
If the coupling at angular momentum channel $n$ is tuned to the critical value set by the separatrix PFP, denoted by the red dashed line, the emergent universal pairing interaction differs from that of the parent stable non-Fermi liquid by 
$\frac{2\pi\sqmu}{\bar\theta_{\text{max}}}
\Delta \aV
\cos\left(\frac{n \pi (\bar\theta_1-\bar\theta_2)}{
2 \bar\theta_{\text{max}}
}\right)$,
where $\Delta \aV$ denotes the difference between the coupling on the separatrix (solid square) and the coupling in the stable non-Fermi liquid (open square) in the angular momentum channel $n$.
Here, 
$\Delta \aV = 
\aV^\circ_{-\infty} -\aV^\bullet_{-\infty}$ in class A
and class AB with $y^{(m)}(l) \ll y_c(l)$,
and
$\Delta \aV = 0$ in classes AC (in the low-energy limit) and ABC as well as class AB with
$y^{(m)}(l) \gg y_c(l)$.
}
\label{fig:universalVcriticalNFL}
\end{figure} 

\tab{tab:toy_stableclasses_criticalprofiles} 
summarizes the universal pairing interaction that emerges at low energies when the four-fermion pairing interaction in the angular momentum channel $n$ is tuned to the critical value set by the separatrix PFP.
While the coupling in that angular momentum channel flows along the separatrix, all other couplings flow toward the universal profile associated with the parent stable non-Fermi liquid.
Therefore, the superuniversal pairing interaction in the critical non-Fermi liquid is given by the sum of  $\aV^{stable}_{\bar{\theta}+\Delta \bar{\theta}, \bar{\theta} }$  
given in \tab{tab:toy_stableclasses_universalvm}
and a correction that arises from the critically tuned coupling.
In class A, the correction is non-zero, as shown in \eq{eq:cNFLx},
because the separatrix PFP is distinct from the metallic PFP.
In class AB, the correction is negligible for $n \gg n_c(l)$, while it takes the same form as in class A for $n \ll n_c(l)$.
Since $n_c(l)$ increases with increasing $l$, the correction survives over a larger range of $n$ as the energy is lowered.
In class AC, the correction flows to zero logarithmically in energy because the difference between the separatrix PFP and the metallic PFP decays as $1/y$ in the $y \rightarrow -\infty$ limit
(see Eqs.
\eqref{eq:separatrixPFPdsc}
and
\eqref{eq:dsc_profile_y}).
In class ABC, 
the separatrix PFP is identical to the metallic PFP and the correction is absent.
This is illustrated in \fig{fig:universalVcriticalNFL}.

In the critical non-Fermi liquids, there exist components of superuniversal pairing interactions whose strengths do not decay as a function of $\bth$ with a $\mu$-dependent overall magnitude. 
As discussed in Sec. \ref{sec:symmetry}, such large angle scattering is exactly marginal if the overall magnitude scales as $\sqrt{\mu}$. 
This is the case for classes A and AB, and the symmetry of the critical non-Fermi liquids is lowered to OLU(1). 
In classes AC and ABC, the critical non-Fermi liquids have the full LU(1) in the low-energy limit as far as the superuniversal contribution is concerned.

\begin{table}[h!]
\centering
\begin{tabular}{|c|c|c|c|c|}
\hline
& A & AB & AC & ABC \\
\hline
$\deltaaVUV \to 0  $ & $|\deltaaVUV|^{ ^{\frac{2z}{\sqrt{\etaPIII}}}}$  & $|\deltaaVUV|^{ ^{\frac{2z}{\sqrt{\etaPIII}}}}$ & $ e^{-\frac{C}{|\deltaaVUV|} } $ & $ e^{-\frac{C}{|\deltaaVUV|}} $ \\
\hline
$n \to \infty$ & $ C'$ & $n^{-2z\Big(1 + \frac{\sqrt{\etaPI}}{\sqrt{\etaPIII}}\Big)}$ &  $C'$ & $n^{-2z} e^{-C' n^{\sqrt{\etaPI}}} $  \\
\hline
\end{tabular}
\caption{
The $T_c$ versus $\deltaaVUV$ scaling relation for the onset of superconductivity across the stable non-Fermi liquid to superconductor phase transition.
Here, $\deltaaVUV$ denotes the deviation of the bare pairing interaction from the critical value in angular momentum channel $n$.
The second row shows the $\deltaaVUV$ dependence of $T_c$ in the small $\deltaaVUV$ limit for a fixed $n$.
The third row shows the $n$ dependence of $T_c$ in the large $n$ limit for a fixed $\deltaaVUV$.
$C$ and $C'$ are constants that are independent of $\deltaaVUV$ and $n$, respectively.
The constants are different in different classes, in general.
}
\label{tab:toy_stableclasses_tcscaling}
\end{table}

Next, we consider the scaling relation that governs the growth of $T_c$ upon adding the bare four-fermion coupling beyond the critical strength.
We start with the critical non-Fermi liquid in which the pairing interaction in the angular momentum channel $n$ is tuned to the critical value.
If we further deform the bare coupling in that angular momentum channel by $\deltaaVUV < 0$, a superconductor is realized below the critical temperature $T_c$.
The universal relation between $T_c$ and $\deltaaVUV$ for small $|\deltaaVUV|$ 
is shown in \tab{tab:toy_stableclasses_tcscaling}
for each superuniversality class that supports the stable non-Fermi liquid to superconductor phase transition.

The second row of the table shows how $T_c$ scales with $\deltaaVUV$ in the small $|\deltaaVUV|$ limit for a fixed $n$.
In this case, the PFP associated with 
$\left(
y^{(n)}(0), 
\aV^S_{y^{(n)}(0)}
+ \deltaaVUV
\right)$ diverges to $-\infty$ in the region of small $y$.
In classes A and AB, the perturbation added to the separatrix PFP grows exponentially in $l$ (algebraically in energy) with the eigenvalue 
$\frac{\sqrt{\etaPIII}}{2}$
in the $y \rightarrow -\infty$  limit.
This results in a power-law scaling relation between $T_c$ and $\deltaaVUV$.
In classes AC and ABC, the separatrix is connected to the marginal $-\infty$ asymptotic fixed point,
and the perturbation added to the separatrix PFP grows only algebraically in $l$ (logarithmically in energy), leading to sub-algebraic growth of $T_c$ in $\deltaaVUV$ with an essential singularity.

\begin{figure}[th]
    \centering
\includegraphics[width=0.6\linewidth]{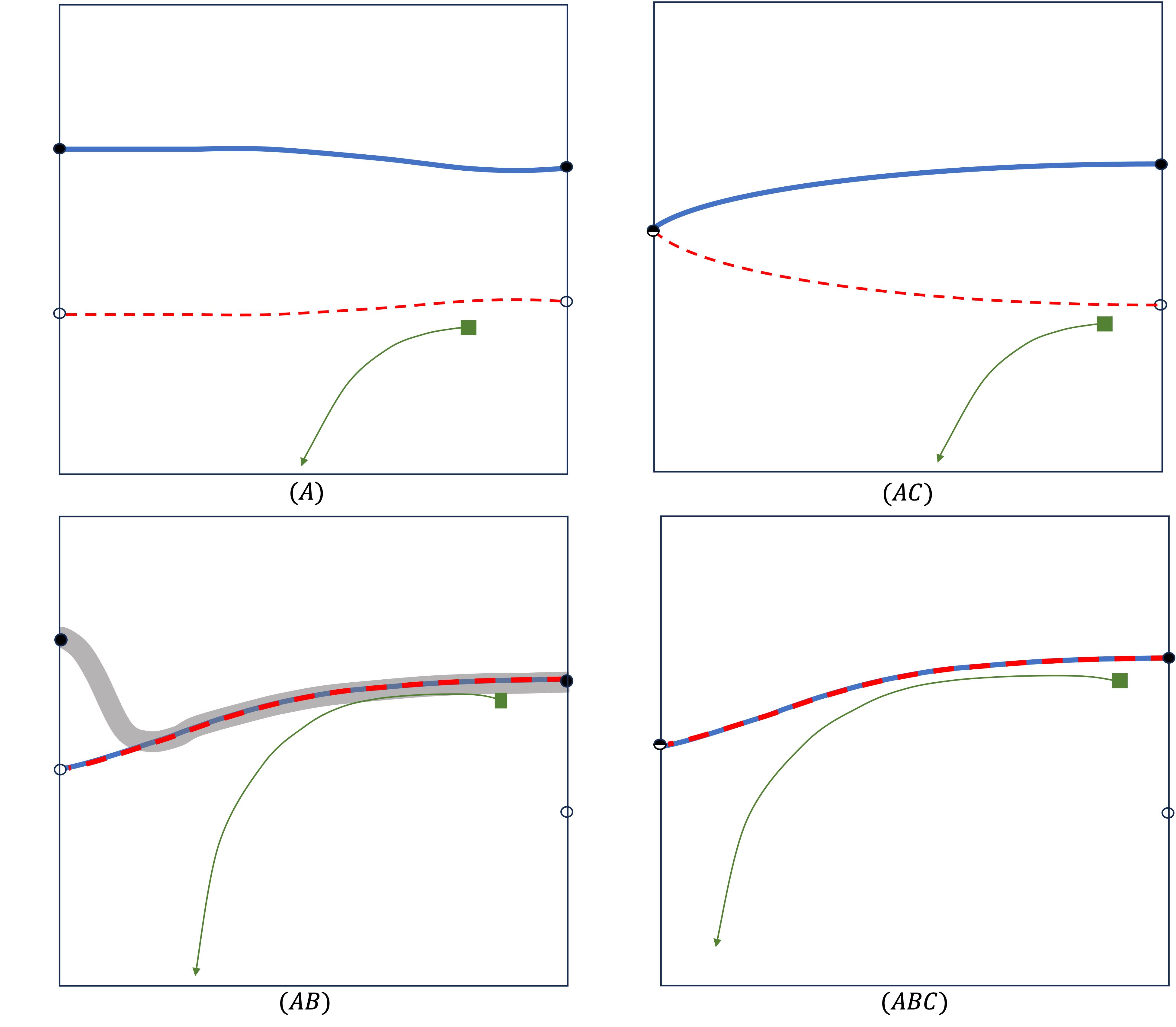}
\caption{
If the bare coupling in one angular momentum channel is tuned below the separatrix PFP, superconducting instability occurs in that channel.
In classes A and AC,
$T_c$ saturates to a finite value as the $n \rightarrow \infty$ limit is taken for a fixed $\deltaaVUV < 0$.
In those classes, the coupling placed below the separatrix decreases right from the high energy scale, and $T_c$ takes the same form as in the first row of \tab{tab:toy_stableclasses_tcscaling} with $\etaPIII$ replaced with $\etaPI$.
In classes AB and ABC, $T_c$ approaches zero in the large $n$ limit because the coupling placed below the separatrix is attracted to the separatrix PFP up to scale $l_1 \sim 2 \log n$. 
At scale $l_1$, the deviation away from the separatrix shrinks to 
$\delta \aV_{n,1} \sim
\deltaaVUV 
e^{-\frac{\sqrt{\etaPI}}{2} l_1}$.
Then, it takes another long RG scale $l_2$ for 
$\delta \aV_{n,1}$ to diverge $-\infty$.
Since the growth of the coupling occurs in the small $y$ limit, the relation between $l_2$ and 
$\delta \aV_{n,1}$ can be inferred from 
the first row of 
\tab{tab:toy_stableclasses_tcscaling}:
$l_2 \sim  \frac{2}{\sqrt{\etaPIII}} \log \frac{1}{\delta \aV_{n,1}} $ 
in class AB,
and 
$l_2 \sim  \frac{1}{\delta \aV_{n,1}}$ in class ABC.
This gives rise to $T_c \sim \Lambda e^{-z(l_1+l_2)}$ shown in 
the third row of \tab{tab:toy_stableclasses_tcscaling}. 
}
\label{fig:TcDeltaV}
\end{figure}

The third row shows how $T_c$ scales with $n$ when the large $n$ limit is taken at a fixed $\deltaaVUV$.
In classes A and AC, the separatrix is locally unstable in the large $y$ limit, and the PFP associated with the perturbed coupling diverges in the $y \rightarrow \infty$ asymptotic region.
Therefore, $T_c$ saturates to a constant in the large $n$ limit for a fixed $\deltaaVUV$.
In classes AB and ABC, however, the separatrix is locally stable ($\chi_y<0$) at large $y$ because it is connected to $\aV^\bullet_{\infty}$.
Therefore, a perturbation added at a large angular momentum channel first decays to a smaller value under the RG flow before it enters the small $y$ region with $\chi_y>0$.
The larger $n$ is, the smaller the perturbation becomes as it enters the region with $\chi_y<0$.
Consequently, $T_c$ decreases with increasing $n$.
$T_c$ decreases more steeply with increasing $n$ in class ABC compared with class AB 
because the perturbation grows more slowly in the region of small $y$ due to the marginality of the $-\infty$ asymptotic fixed point.
For more details, see
\fig{fig:TcDeltaV} and its caption.

In classes A and AC, the toy model reproduces the same critical exponents as the physical theories available for those classes.
This is expected because these exponents are solely controlled by the $y \rightarrow \pm \infty$ limits and are superuniversal.

\subsection{Superuniversality classes B and BC}

In this subsection, we discuss the superuniversal properties of the non-Fermi liquids that arise in classes B and BC, which are prone to non-s-wave superconducting instabilities.

\subsubsection{Quasi-universal superconductivity}

In superuniversality classes B and BC, the separatrix PFP diverges to $\infty$ at a finite $y$, denoted as $y^*_S$ (see \fig{fig:PFP}).
The location of divergence determines the critical angular momentum for non-s-wave superconductivity. 
For angular momentum channels with $y^{(m)}(0) > y^*_S$, superconducting instability is unavoidable, irrespective of the choice of bare coupling (see the discussion in Sec. \ref{sec:individual_classes}).
 In contrast, one can avoid superconducting instability in channels with $y^{(m)}(0) \leq y^*_S$ by choosing
sufficiently repulsive bare couplings.
The value of $y^*_S$ depends on the detailed profile of the separatrix PFP.
Since $y^*_S$ is infinite
in superuniversality classes (A, AB, AC, ABC) that support stable non-Fermi liquids, 
$y^*_S$ increases as classes B and BC approach classes AB and ABC, respectively. 
We can use $\delta w = \wII-\wII^c$ as the tuning parameter that drives the superuniversality class phase transitions from B to AB (or from BC to ABC), 
where $\wII$ is the logarithmic range of intermediate angular momenta in which the pairing interaction is strongly attractive, 
and $\wII^c$ is the critical width.
As $\delta w$ decreases, $y^*_S$ increases as 
$\frac{1}{\sqrt{\etaPI}} \ln \frac{1}{\delta w} $ in both classes B and BC.
This implies that the critical angular momentum for the superconducting instability is given by
\bqa
n_S = \frac{2 \gamma}{\pi} 
\sqrt{
\frac{\KFAVdim}{\Lambda}
}
\delta w^{
-1/\sqrt{\etaPI}
},
\label{eq:critical_n}
\eqa
where $\gamma$ is defined in \eq{eq:thetamax}.
The phase transition from class B to AB (or from class BC to ABC) occurs as the range of angular momentum that is `immune' to superconducting instability diverges as $\delta w$ approaches zero.
In this sense, $1/\sqrt{\etaPI}$ in \eq{eq:critical_n} corresponds to a critical exponent that governs the superuniversality phase transitions.

While $T_c$ depends on the bare couplings, the minimum $T_c$ is realized when the bare couplings are $\infty$ in all angular momentum channels.
If the bare coupling is infinitely repulsive, the PFP associated with angular momentum channel $m$ emanates from 
$( y^{(m)}(0), \infty )$.
In class B, the couplings in the channels with 
$y^{(m)}(0) < y^*_S$ 
($y^{(m)}(0) = y^*_S$) 
flow to 
$\aV^\bullet_{-\infty}$ 
($\aV^\circ_{-\infty}$)
in the low-energy limit.
In class BC, 
the couplings in channels with 
$y^{(m)}(0) \leq y^*_S$ 
flow to $\aV^\halfominus_{-\infty}$.
In both classes, the couplings in the channels with $y^{(m)}(0) > y^*_S$ diverge to $-\infty$ at sufficiently low energies.
The actual superconducting instability occurs in the channel with the smallest 
$y^{(m)}(0) - y^*_{SC}$, 
where 
$y^*_{SC}$ is the location at which the PFP that emanates from
$( y^{(m)}(0), \infty )$ diverges to $-\infty$.
As discussed in Sec. \ref{sec:individual_classes}, this occurs in the channel whose $y^{(m)}(0)$ is closest to the optimal value, $y^*_O$
(see \fig{fig:classBdetails} and the discussion below it).
In class B and BC, $y^*_O$ is given by
$y^*_{O,B} = -\frac{1}{\sqrt{\etaPI}}\ln \delta w  $
and $y^*_{O,BC}=-\frac{2}{\sqrt{\etaPI}} \ln(\delta w)  $, respectively.
This translates to the `optimal' angular momentum,
\bqa
n_O = 
\begin{cases}
\frac{2 \gamma}{\pi} 
\sqrt{
\frac{\KFAVdim}{\Lambda}
}
\delta w^{
-1/\sqrt{\etaPI}
} & \text{for class B}, \\
\frac{2 \gamma}{\pi} 
\sqrt{
\frac{\KFAVdim}{\Lambda}
}
\delta w^{
-2/\sqrt{\etaPI}
} & \text{for class BC}
\end{cases}
\label{eq:SCchannel_in_B_BC}
\eqa
to the leading order in $\delta w$.
In class $B$, $n_O \approx n_S$;
the angular momentum channel at which superconductivity arises for highly repulsive bare couplings is close to the critical angular momentum below which there is no instability.
In class $BC$, $n_O \gg n_S$.
At the optimal angular momentum, the superconducting transition temperature becomes
\bqa
T_{c,O} = 
\begin{cases}
 \delta w^{2z\Big(\frac{1}{\sqrt{\etaPI}} + \frac{1}{\sqrt{\etaPIII}}\Big)} 
& \text{for class B}, \\
  \delta w^\frac{4z}{\sqrt{\etaPI}}e^{\frac{8z}{\etaPII \delta w}} 
 & \text{for class BC}
\end{cases}
\label{eq:highestTc}
\eqa
to the leading order in $\delta w$.

\begin{figure}[H]
    \centering
    \includegraphics[width=0.8\linewidth]{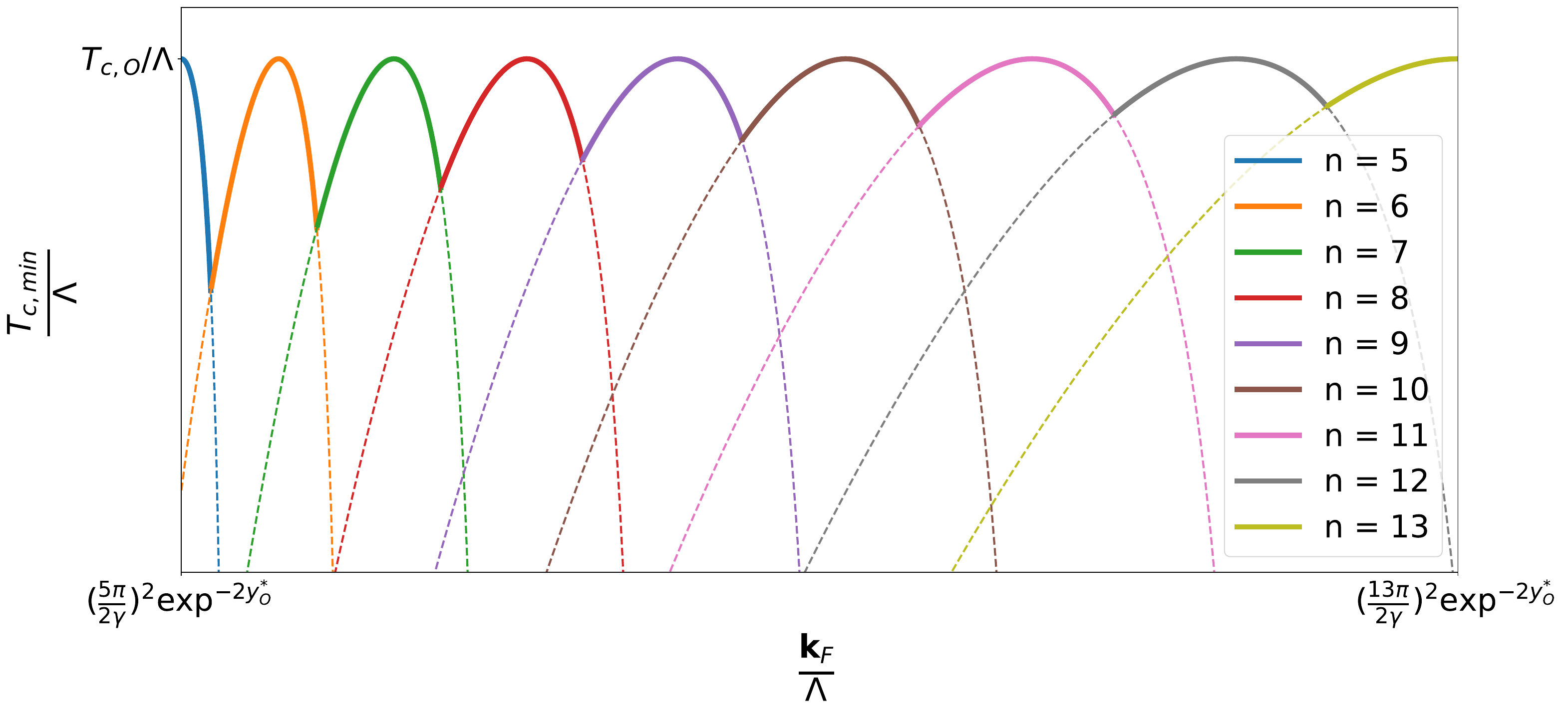}
    \caption{
The minimum superconducting transition temperature realized for the highly repulsive bare couplings plotted as a function of $\KFAVdim$.
The transition temperature reaches a local maximum as a function of $\KFAVdim$ whenever the pitch of the pairing wavefunction satisfies the resonance condition with the typical momentum carried by the critical boson in \eq{eq:resonantkF}.
In the toy model, we take $\etaPI = 1, \etaPII=-10$, and $ \etaPIII = 10$ for the discriminants and $\wII = \wII_c(1+0.1)\approx 0.2$. 
With $\gamma=\pi/2$, the resonance condition in 
\eq{eq:resonantkF} can be satisfied for $n \geq 5$ in the region of $\KFAVdim/\Lambda \geq 1$.
}
    \label{fig:toy_classb_oscillatorytcmin}
\end{figure}

For a generic $\KFAVdim$, 
there is no angular momentum channel that satisfies $y^{(m)}(0) = y^*_O$ because $m$ is discrete.
The superconducting instability occurs at angular momentum $m$ whose 
$y^{(m)}(0)$ is closest to $y^*_O$.
If the Fermi momentum is increased, 
$( y^{(m)}(0), \infty )$
shifts horizontally to the left (see \fig{fig:whyTc_oscillate_classB}).
Within a range of Fermi momentum, 
$y^{(m)}(0)$ remains closest to $y^*_O$ for one particular $m$, and superconductivity occurs in that channel.
Within that window, there exists a special choice of Fermi momentum 
\bqa
\KFAVdim^*{}_m = \left(
\frac{\pi m}{2 \gamma y_O^*}
\right)^2 \Lambda
\label{eq:resonantkF}
\eqa
at which $y^{(m)}(0)$ coincides with $y^*_O$.
At this Fermi momentum, $T_c$ reaches the maximum value shown in  
\eq{eq:highestTc}.
As the Fermi momentum is tuned away from the critical value, $T_c$ decreases as
\bqa
T_{c} = 
T_{c,O}
e^{-\frac{z \xi (\delta \KFAVdim)^2}{2 (\KFAVdim^*{}_m)^2} },
\label{eq:toy_classb_tcmin_main}
\eqa
where $\delta \KFAVdim = \KFAVdim - \KFAVdim^*{}_m$
and $\xi$ is constant.
See Appendix
\ref{appendix:ex4}-\ref{appendix:B}
and
\ref{appendix:ex4}-\ref{appendix:BC}
for the derivation and 
the expressions for $\xi$.
As $\KFAVdim$ increases beyond the window associated with angular momentum $m$, $y^*_O$ becomes closest to the next angular momentum channel, and superconductivity occurs in the angular momentum channel $m+1$.
Within the window of $\KFAVdim$ in which 
$y^{(m+1)}(0)$ is closest to $y^*_O$, the highest $T_c$ is realized at another Fermi momentum 
$\KFAVdim^*{}_{m+1}$.
The ratio between two consecutive optimal Fermi momenta is
\bqa
\frac{\KFAVdim^*{}_{m+1}}
{\KFAVdim^*{}_{m}}
= \frac{(m+1)^2}{m^2}.
\eqa
This causes $T_c$ to oscillate as a function of the Fermi momentum, as is shown  in \fig{fig:toy_classb_oscillatorytcmin} for class B.
Even if the bare couplings are not infinite, the oscillatory behavior of $T_c$ as a function of $\KFAVdim$ persists, but the local maximum of $T_c$ is no longer universal.
A similar oscillatory behavior arises in class BC as well.

\begin{figure}[ht]
    \centering
    \begin{subfigure}{0.4\textwidth}
        \centering
        \includegraphics[width=\textwidth]{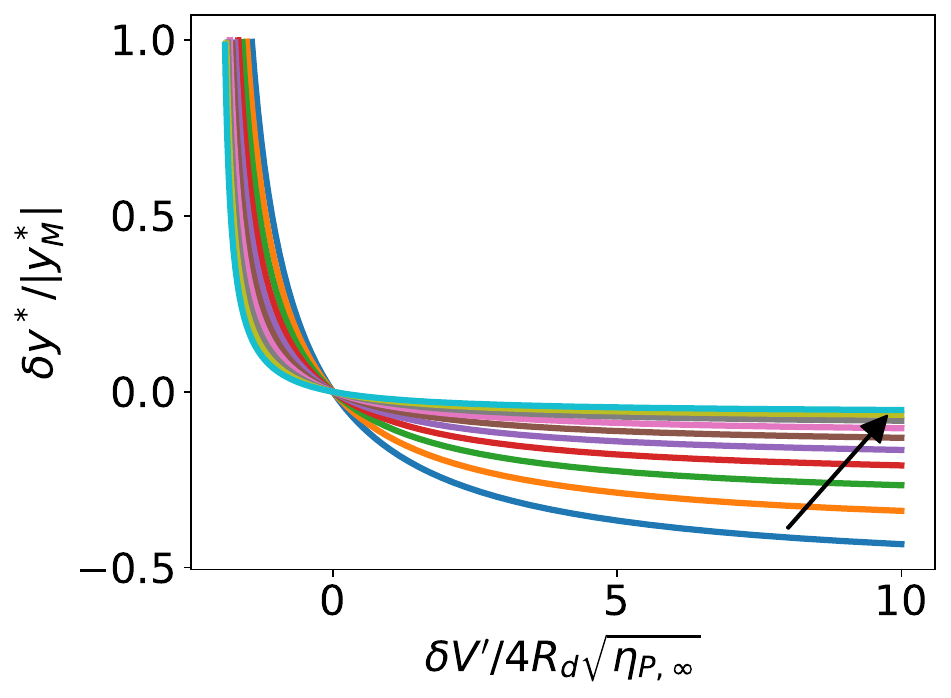}
        \caption{}
        \label{fig:toy_classb_ystariii}
    \end{subfigure}
        \begin{subfigure}{0.4 \textwidth}
        \centering
        \includegraphics[width=\textwidth]{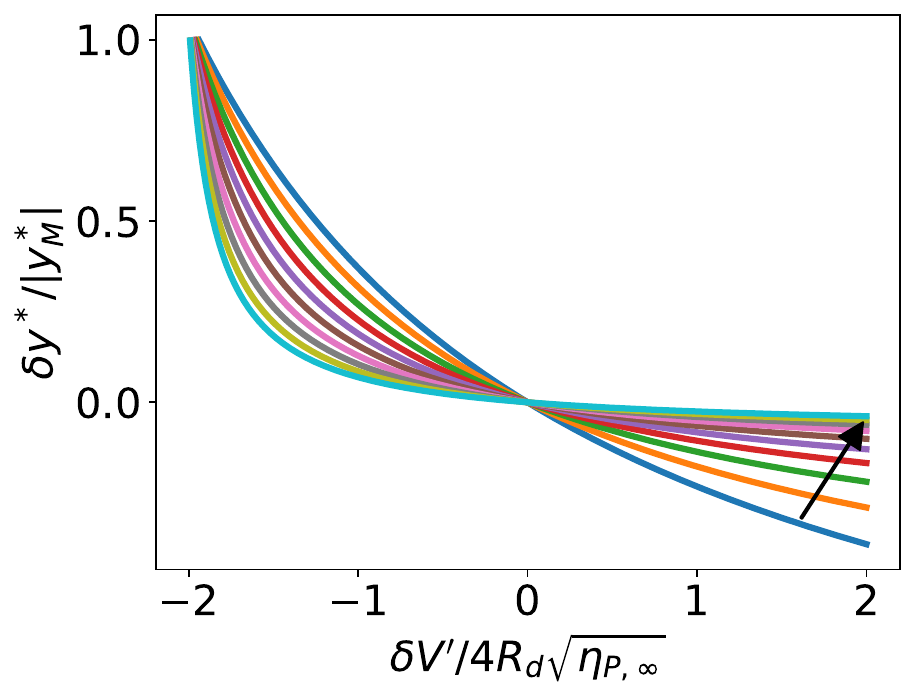}
        \caption{}
        \label{fig:toy_classbc_ystariii}
    \end{subfigure}
    \begin{subfigure}{0.4 \textwidth}
        \centering
        \includegraphics[width=\textwidth]{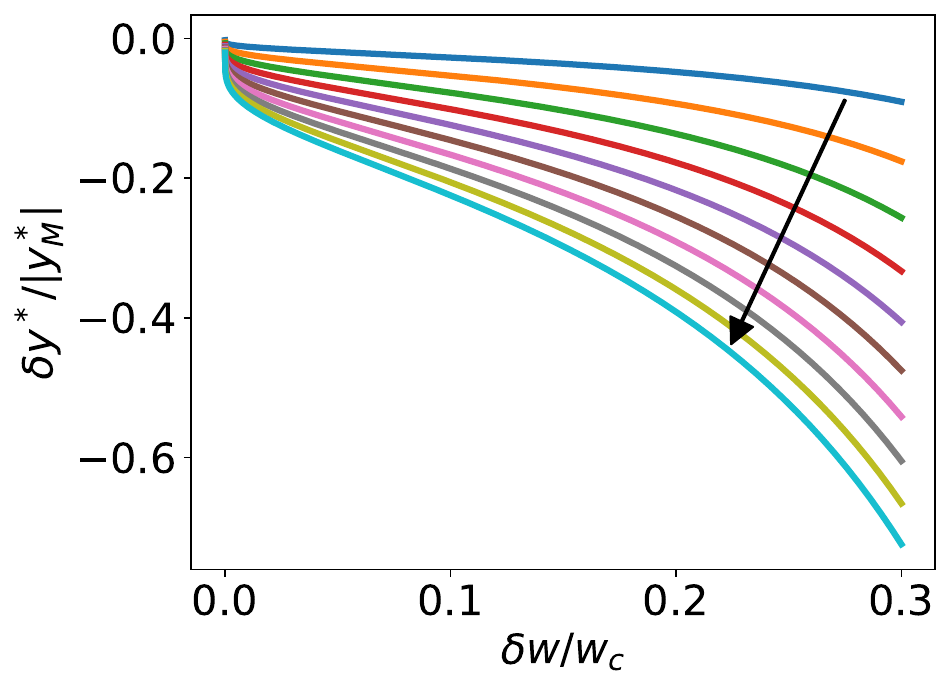}
        \caption{}
        \label{fig:toy_classb_ystariii2}
    \end{subfigure}
        \begin{subfigure}{0.4 \textwidth}
        \centering
        \includegraphics[width=\textwidth]{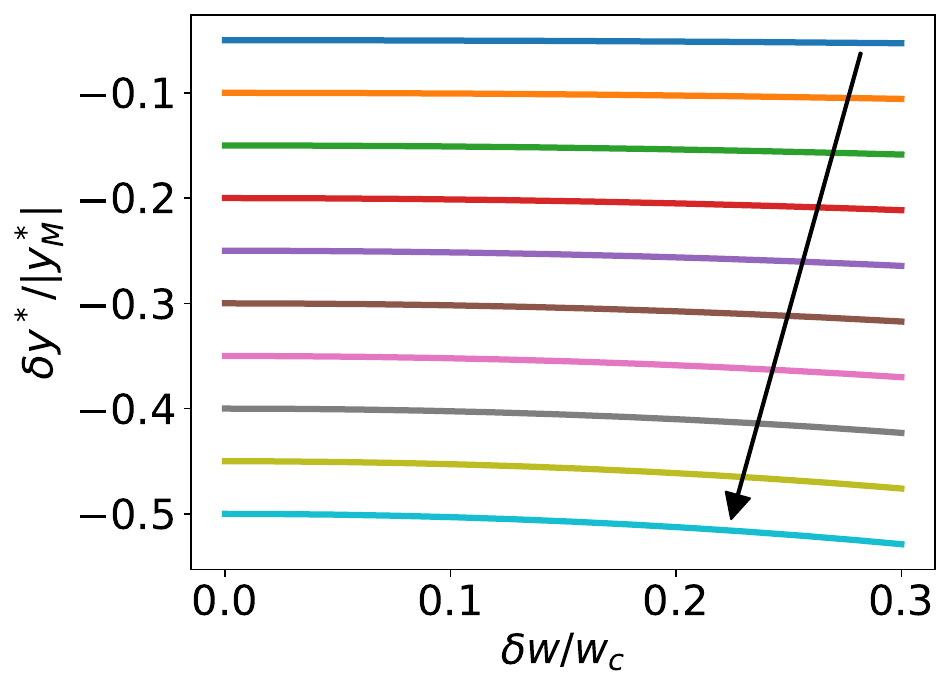}
        \caption{}
        \label{fig:toy_classbc_ystariii2}
    \end{subfigure}
    \caption{
    (a), (b)
$\delta y^*/|y^*_M|$ 
plotted as a function of deformation $\deltaaVUV$ added to the metallic PFP at logarithmic angular momentum $y_0$ for $\delta w = 0.3 w_c$ in 
(a) class B with $y_0$ increases from $1.5w $ to $2w$,
and (b) class BC with $y_0$ 
ranging from $2w$ to $3w $ along the directions of the arrows.
In both plots, $\delta w = 0.3w_c$ is used. 
As the angular momentum increases,
$\delta y^*/|y^*_M|$ becomes less sensitive to $\deltaaVUV$.
(c), (d)
$\delta y^*/|y^*_M|$ 
 plotted as a function of $\delta w / w_c$ for deformations $\deltaaVUV$ added to the metallic PFP at $y^*_S$ in 
 (c) class B 
 and (d) class BC 
 with $\deltaaVUV$
ranging from $0.1 \times 4R_d \sqrt{\etaPI}$ to $4R_d \sqrt{\etaPI}$ in both plots.
For all plots, we use
$\etaPI = \etaPIII = 1$, $\etaPII =-1$. 
Proximity to class A with smaller $\delta w$ reduces the dependence of the ratio on $\deltaaVUV$ in class B, but not in class BC.
}
 \label{fig:deltayinclassBandBC}
\end{figure}

The flip side of the existence of $n_c$ above which superconducting instability is unavoidable is that the metallic PFP diverges to $-\infty$ at a finite $y$ denoted as $y^*_M$.
Like $y^*_S$, the precise value of $y^*_M$ depends on the detailed profile of the metallic PFP.
However, $y^*_M$ diverges to $-\infty$ as one approaches the boundaries with class AB or ABC. 
The leading order expression of $y^*_M$ in the $\delta w$ limit is given by
\bqa
y^*_M \approx \begin{cases}
 -\frac{1}{\sqrt{\etaPIII}}\log
 \frac{1}{\delta w} & \text{for case B} \\
 -\frac{4}{\abs{\etaPII} \delta w}
& \text{for case BC}
\end{cases}.
\label{eq:yMinBandBC}
\eqa
Since the metallic PFP attracts nearby PFPs at large $y$, 
PFPs that start at large $y$ and large $\aV_y$ follow more or less along the metallic PFP at small $y$
and diverge close to $y^*_M$ 
(see the discussion in Sec. \ref{sec:individual_classes}).
To be quantitative,
let us suppose that superconducting instability occurs in the angular momentum channel $n$ and the bare coupling in that channel is 
$\aV^M_{y_0} + \deltaaVUV$, where 
$y_0= y^{(n)}(0)$
is the logarithmic angular momentum
and
$\deltaaVUV$ is the deviation from the metallic PFP.
Let the PFP that emanates from
$(y_0, \aV^M_{y_0}  + \deltaaVUV)$ diverge to $-\infty$ at 
$y^*_{SC} = y^*_M+\delta y^*$, 
where $\delta y^*$
is proportional to $\deltaaVUV$ for small $\deltaaVUV$.
As $n$ increases, the RG time during which the PFP is attracted to the metallic PFP increases as $l \sim 2 \log n$. 
This long RG time suppresses the deviation of the coupling from that of the metallic PFP to 
$e^{- \chi_\infty l} \deltaaVUV
\sim
\frac{ \deltaaVUV 
}{ n^{\sqrt{\etaPI}}
}$,
where
$\chi_\infty
=\frac{\sqrt{\etaPI}}{2}$ is the eigenvalue that controls the attraction of nearby PFPs to the metallic PFP at large $y$.
This causes $\delta y^*$ to depend on the bare coupling only through
$\frac{ \deltaaVUV }{ n^{\sqrt{\etaPI}} }$.
It gives rise to a quasi-universal ratio between $T_{c,n}$ and $\KFAVdim$ in
  \eq{eq:TC_KF_universal}
  with
\bqa
\frac{\delta y^*}{|y^*_M|} = 
- \frac{2 R_d}{\sqrt{\etaPI}} \times
\begin{cases}
\frac{
1
}{\left|\log\big( \frac{\etaPIII-\etaPII}{4\sqrt{\etaPIII}} \delta w\big) \right| } 
\left( \frac{n_s}{n} \right)^{\sqrt{\etaPI}}
\deltaaVUV, 
& \text{for class B}, \\
\left( \frac{n_s}{n} \right)^{\sqrt{\etaPI}}
\deltaaVUV
& \text{for class BC}
\end{cases}.
\label{eq:deltayyBC}
\eqa
In class B, 
$\delta y^*/|y^*_M|$ becomes less sensitive to the bare coupling as the angular momentum increases and $\delta w$ decreases. 
In class BC,  the ratio is suppressed only at large angular momentum.
This is shown in  \fig{fig:deltayinclassBandBC}.

Eqs. \eqref{eq:TC_KF_universal} 
and \eqref{eq:deltayyBC}
represent the insensitivity of $T_c/\KFAVdim^z$ to the bare coupling at large angular momentum channels.
A similar universality arises in the limit of highly repulsive bare couplings in classes B and BC proximate to class A. 
For infinitely positive bare couplings, the superconducting instability arises at the angular momentum channel closest to $n_O$ in \eq{eq:SCchannel_in_B_BC}.
In the small $\delta w$ limit, $n_O$ becomes large, and the spacing between 
$y^{(n)}(0)$ for discrete angular momenta becomes small:
$y^{(n_O+1)}(0) - y^{(n_O)}(0) \sim 1/n_O$.
Therefore, the discrete bare couplings can be approximately regarded as the line of bare coupling $(y, \aV_y(0))$ for continuously varying $y$ around $y \sim \log n_O$ to the leading order in $n_O$. 
At the infinite bare coupling,
superconducting instability arises in the angular momentum channel that is very close to $n_O$,
and the ratio between $T_c$ and $\KFAVdim^z$ is given by
\bqa
\frac{\Lambda^{z - 1} T_{c,n}}{\KFAVdim^z}
= 
   \left(  
   \frac{2 \gamma}{\pi n}
   \right)^{2z}
   e^{2 z y_{SC,O}^*},
\eqa
where $y_{SC,O}^*$ is the location at which the optimal PFP diverges to $-\infty$,
\bqa
y_{SC,O}^* = \begin{cases}
-\frac{1}{\sqrt{\etaPIII}}\ln
\frac{1}{\delta w} + \order{1} & \text{for class B}, \\
-\frac{4}{|\etaPII|}\frac{1}{\delta w} + \order{1}  & \text{for class BC}
\end{cases}.
\label{eq:ySOBandBC}
\eqa
$y^*_{SC,O}$ in \eq{eq:ySOBandBC} is the same as  $y^*_M$ in \eq{eq:yMinBandBC}
modulo $\order{1}$ correction; what is the same is the singular dependence on $\delta w$.
Because $y^*_O$ is large, the optimal PFP that emanates from $(y^*_O,\infty)$ has long RG time to be attracted to the metallic PFP.

For a finite bare coupling, $T_c/\KFAVdim$ is corrected but only slightly if the bare coupling is large.
Suppose the $y$-dependent bare coupling is given by
$\aV_{y} = \aV^{UV} + \Delta\aV(y)$,
where $\aV^{UV}$ is a large and positive constant, and 
$\Delta\aV(y)$ is an angular momentum-dependent component.
In this case, the ratio obeys
\begin{equation}
    \frac { \log \left[ \frac{\Lambda^{z - 1} T_{c,n}}{\KFAVdim^z}
   \left(  \frac{\pi n}{2 \gamma} \right)^{2z}
\right]_{
\aV_{y}=  \aV^{UV} + \Delta\aV(y)  }}
    { \log \left[ \frac{\Lambda^{z - 1} T_{c,n}}{\KFAVdim^z}  
   \left(  \frac{\pi n}{2 \gamma} \right)^{2z}
\right]_{\aV_{y}=\infty}}
    =
    1 + 
    \frac{\delta y^*}{y_{SC,O}^*}.
  \label{eq:TC_KF_universal_finiteV}
\end{equation}
where
$\delta y^* = - 
\frac{1}{4 R_d (\aV^{UV})^2}
\frac{1}{\xi}\frac{\text{d}\Delta \aV}{\text{d}y}\Big|_{y = y^*_O}$ is the correction that arises from finite couplings.
This expression holds for both classes B and BC, while the value of $\xi$ is different in the two classes.
It becomes small for large $\aV^{UV}$.

\begin{figure}[h]
\centering
\begin{subfigure}{.37\textwidth}
  \centering
  \includegraphics[width=1\linewidth]{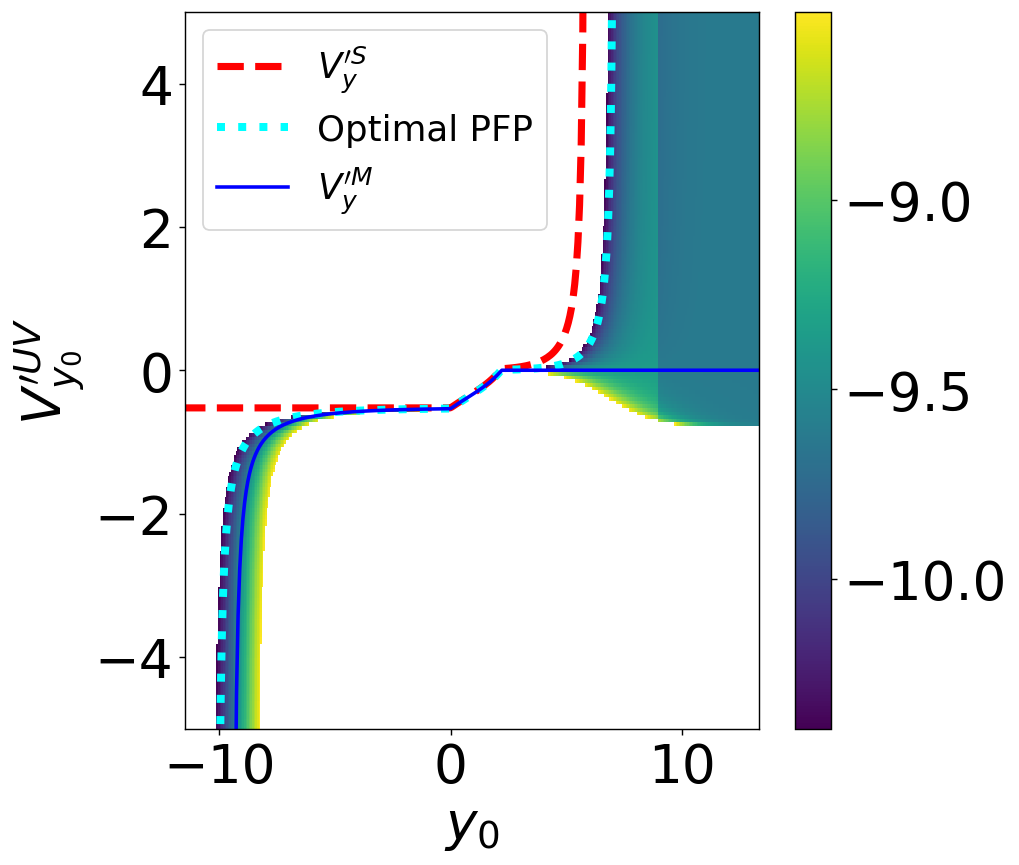}
  \caption{
}
  \label{fig:}
\end{subfigure}%
\begin{subfigure}{.35\textwidth}
  \centering
  \includegraphics[width=1\linewidth]{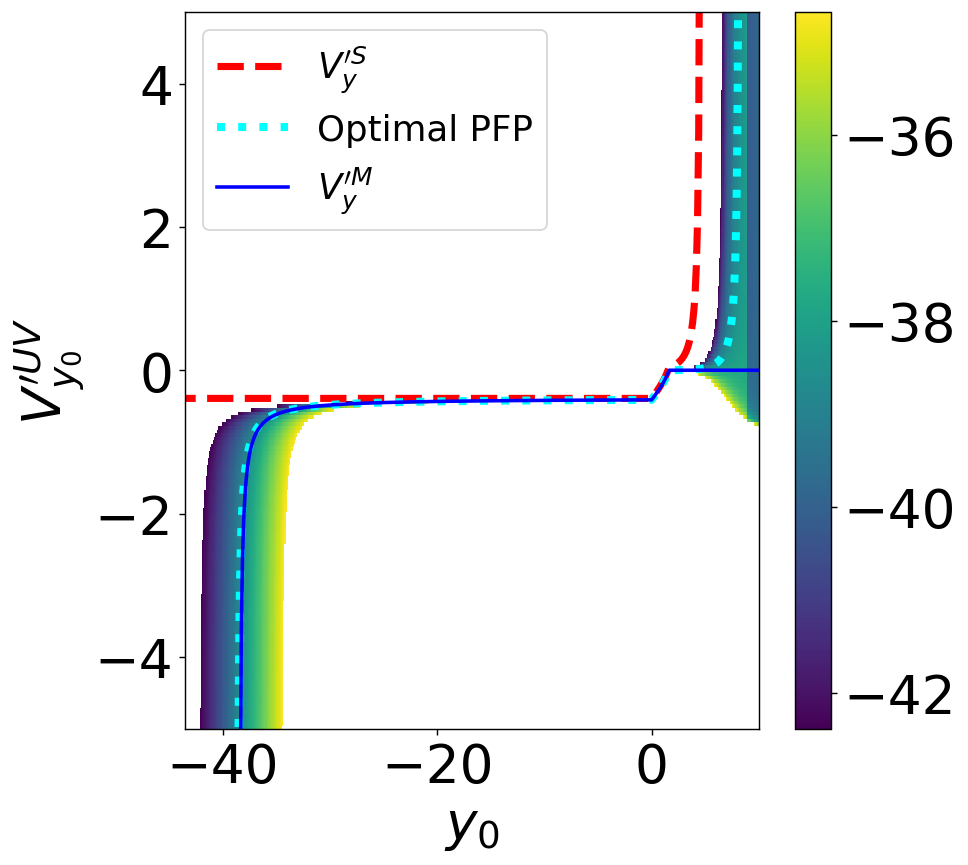}
  \caption{
}
  \label{fig:}
\end{subfigure}%
\caption{
The set of UV couplings  that exhibit quasi-universal $T_c/\KFAVdim^z$ 
within tolerance
$\left| \delta y^* / y^*_M \right| \leq 0.1$.
(a) Class  B 
with $\etaPI = 1$, $\etaPII \approx -1.04 $, $\etaPIII \approx 0.11$ and $\delta w = 0.05 $. 
(b) Class BC with
$\etaPI = 1$, $\etaPII \approx -1.04 $ and $\delta w = 0.1 $. 
The color represents the value of $y_{SC}$ at which the associated PFP diverges to $-\infty$.
}
\label{ysc_countour_B_BC} 
\end{figure}

\fig{ysc_countour_B_BC} shows the approximate basin of attraction in which the couplings diverge to $-\infty$ close to the metallic PFP within a finite tolerance.


\subsubsection{Quasi-universal NFLs}

\begin{figure}[ht]
    \centering
\includegraphics[width=0.7\linewidth]{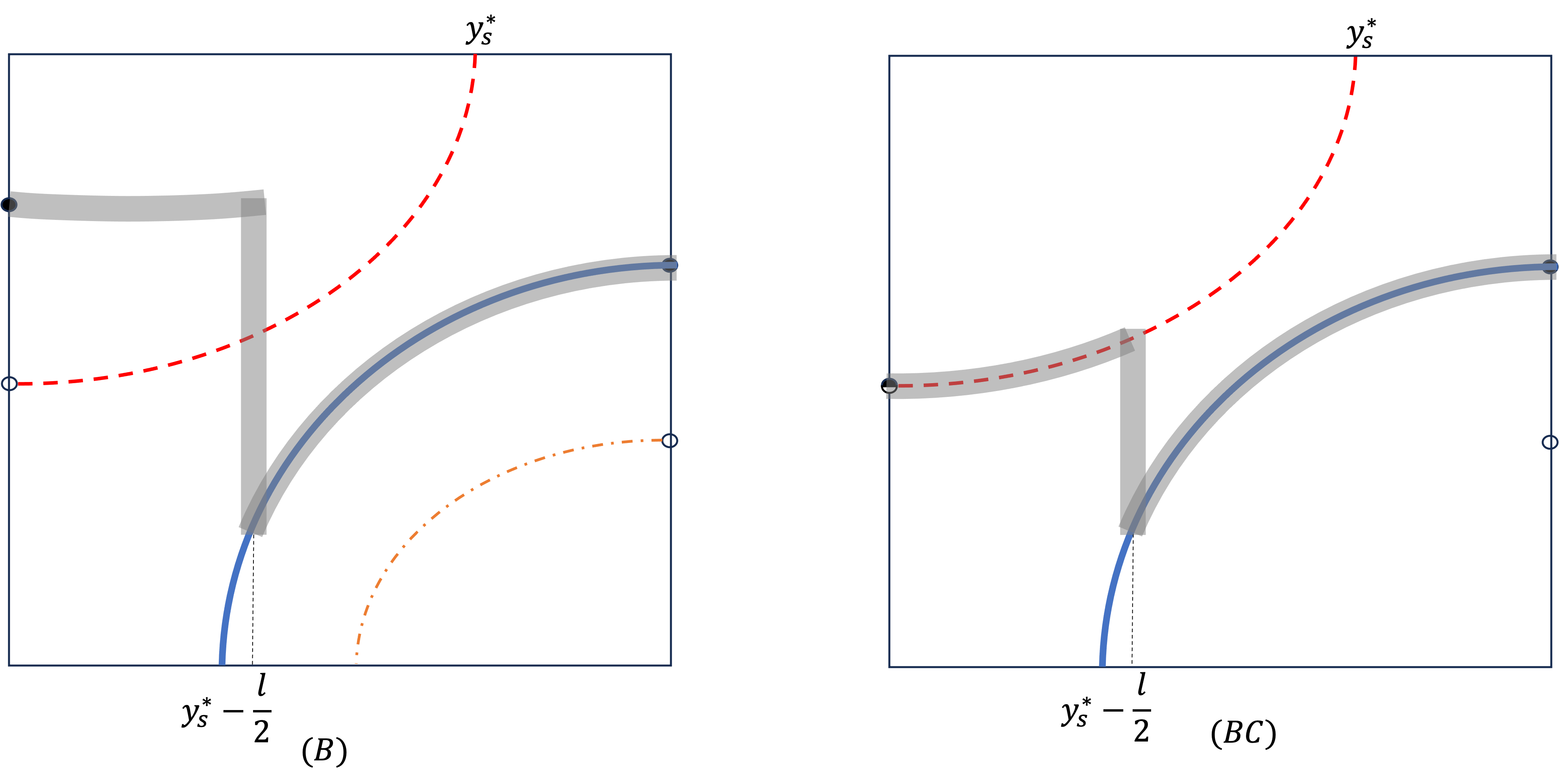}
\caption{
In classes B and BC, there is a sharp crossover in the regularized metallic PFP (denoted by the thick grey line) around $y_S^*-l/2$ in the renormalized pairing interaction if the bare couplings are large and repulsive.
This is because the couplings in $y^{(n)}(0) < y_S^*$ are attracted toward the stable or marginal $-\infty$ asymptotic fixed point,
whereas the couplings at larger angular momenta are attracted toward the metallic PFP.
The solid and dashed lines represent the metallic and separatrix PFPs, respectively.
}
\label{fig:RMPFP_classBandBC}
\end{figure}

If a non-Fermi liquid in the superuniversality class B or BC is proximate to class A, the hierarchy between $\Lambda$ and $T_c$ creates a large window of energy scales in which the normal state exhibits a quasi-universal behavior.
While the metallic PFP itself is singular, the universal pairing interaction that emerges at intermediate energy scales is controlled by the regularized metallic PFP, which is discussed in Sec. \ref{sec:quasiuniversalpairing_classC} for class C.
Here, we extend that discussion to classes B and BC.

In these classes, the regularized metallic PFP is determined by the fact that the couplings flow to different values depending on their locations in the $(y,\aV)$ plane relative to the separatrix PFP.
The couplings above the separatrix PFP are attracted to  $\aV^\bullet_{-\infty}$ in class B and the separatrix PFP in class BC\footnote{
In the toy model for class BC, the separatrix PFP is identical to $\aV^\halfominus_{-\infty}$ in the small $y$ limit.}.
On the other hand, the couplings below the separatrix are first attracted to the metallic PFP at intermediate energy scales and eventually diverge to $-\infty$ at sufficiently low energies.
If the bare couplings are highly repulsive, the relative location of the couplings with respect to the separatrix is determined by their angular momentum.
For angular momentum channel $m$ with $y^{(m)}(0) < y_S^*$, the couplings flow to 
$\aV^\bullet_{-\infty}$
or
$\aV^\halfominus_{-\infty}$
at large $l$.
On the other hand, 
the coupling at angular momentum channel $m$ with $y^{(m)}(0) > y_S^*$ 
becomes close to the metallic PFP. 
This inevitably creates a sharp crossover like  \fig{fig:jump} with $y_c(l) = y_s^* - l/2$.
Therefore, the profile of the renormalized couplings is attracted toward the regularized metallic PFP given by
\bqa
\aV^{RM}_y(l) 
= 
\begin{cases}
\begin{cases}
   \aV^\bullet_{-\infty} & \text{for $y < y^*_S - \frac{l}{2}$} \\
   \aVM_y & \text{for $y > y^*_S - \frac{l}{2}$} 
\end{cases} & \text{for class B} \\ & \\
\begin{cases}
   \aVS_y
   & \text{for $y < y^*_S - \frac{l}{2}$} \\
   \aVM_y & \text{for $y > y^*_S - \frac{l}{2}$} 
\end{cases} & \text{for class BC} 
\end{cases}.
\label{eq:RMPinBandBC}
\eqa
In contrast to class C,  \eq{eq:RMPinBandBC} depends on $l$ 
because the location of the crossover shifts as $y_c(l) \approx y^*_S-l/2$.
The scale-dependence of the regularized metallic PFP in classes B and BC originates from the fact that the Fermi momentum affects the non-s-wave pairing instability strongly, unlike in class C.
The schematic form of the regularized metallic PFP is depicted in \fig{fig:RMPFP_classBandBC}.
While the jump in the crossover is not infinitely sharp,
the width is small enough that
$\bar \theta_c(l)$ in \eq{eq:theta_crossover} is $O(1)$
in both classes B and BC.
This is shown in Appendices
\ref{appendix:ex4}-\ref{appendix:B}
and
\ref{appendix:ex4}-\ref{appendix:BC}. Consequently, the quasi-universal pairing interaction takes the form of
\bqa
\aV^{super}_{\bar{\theta},\bar{\theta}+\Delta \bar{\theta}} 
\sim
\frac{\sqmu}{|\Delta \bar \theta|} \sin\Big(
e^{y^*_S}
\frac{|\Delta\bar{\theta}|}{\sqrt{\Lambda}} \Big)
\eqa
up to $\Delta \bth / \sqrt{\Lambda} \sim e^{-y_S^*}$. 
The range of the large-angle scattering with the $\sqmu/\bth$ decay does not decrease with decreasing energy scale.
Therefore, LU(1) is lowered to OLU(1) in the quasi-universal non-Fermi liquids realized in classes B and BC.

\begin{figure}[h]
\centering
\begin{subfigure}{.35\textwidth}
  \centering
  \includegraphics[width=1\linewidth]{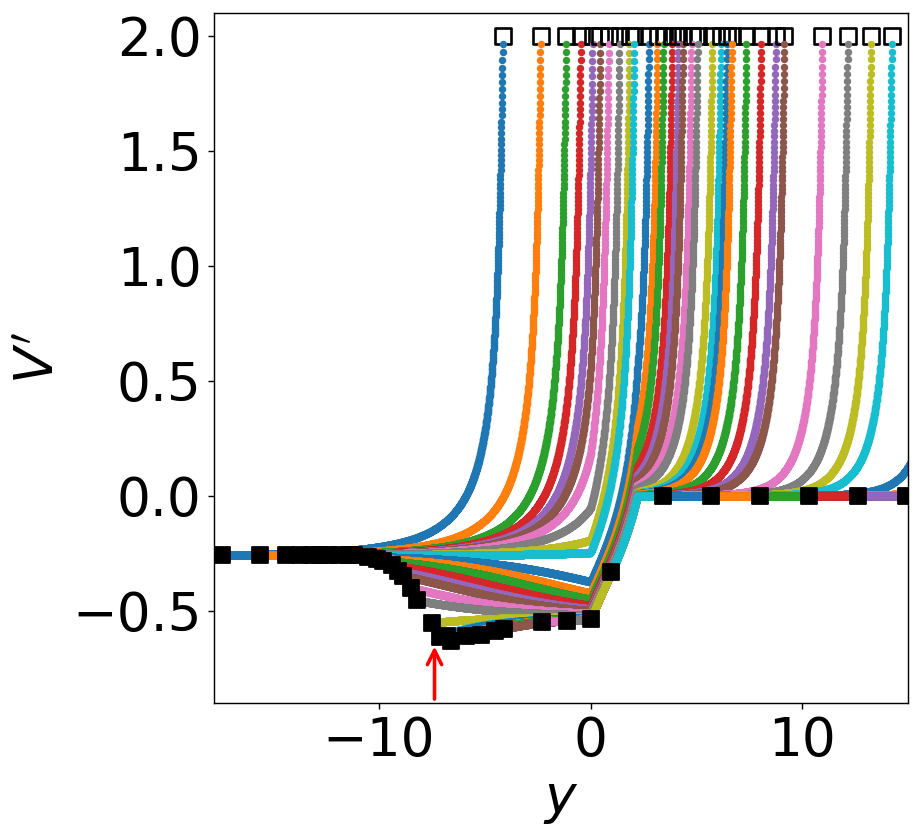}
  \caption{
}
  \label{fig:}
\end{subfigure}%
\begin{subfigure}{.35\textwidth}
  \centering
  \includegraphics[width=1\linewidth]{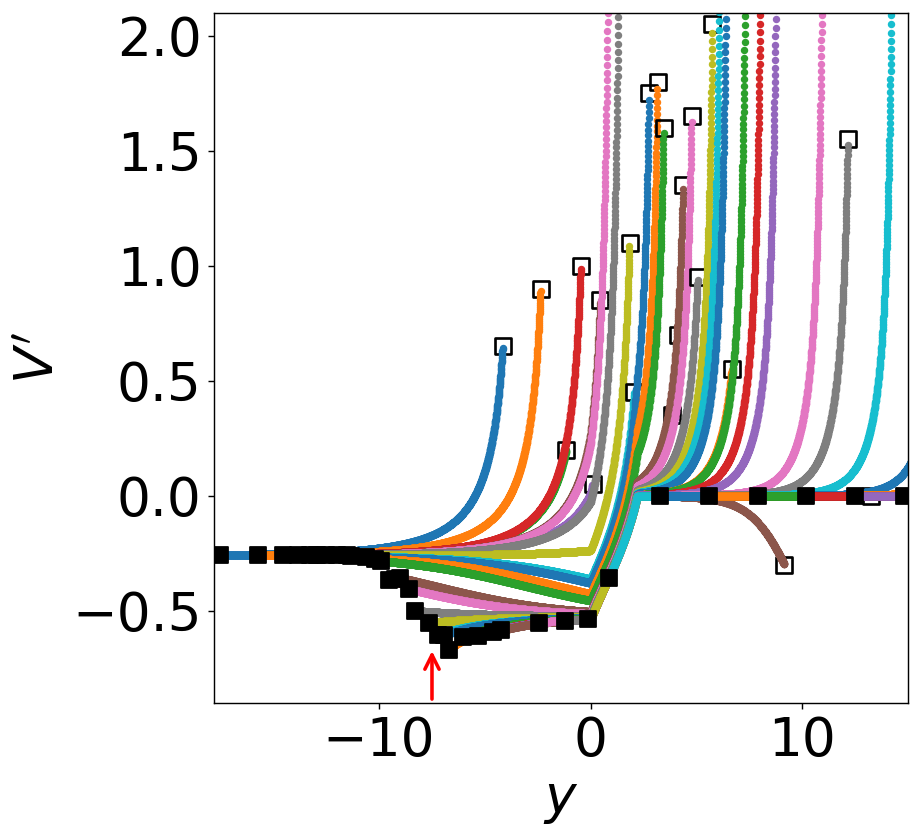}
  \caption{
}
  \label{fig:}
\end{subfigure}%
\begin{subfigure}{.35\textwidth}
  \centering
  \includegraphics[width=1\linewidth]{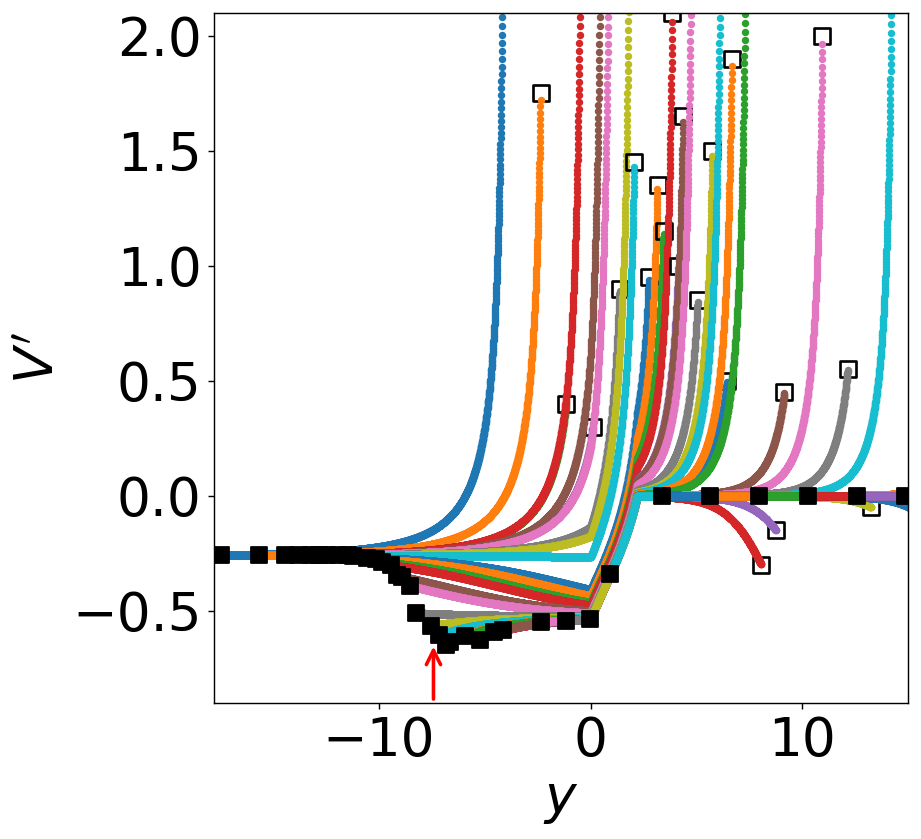}
  \caption{
}
  \label{fig:}
\end{subfigure}
\begin{subfigure}{.35\textwidth}
  \centering
  \includegraphics[width=1\linewidth]{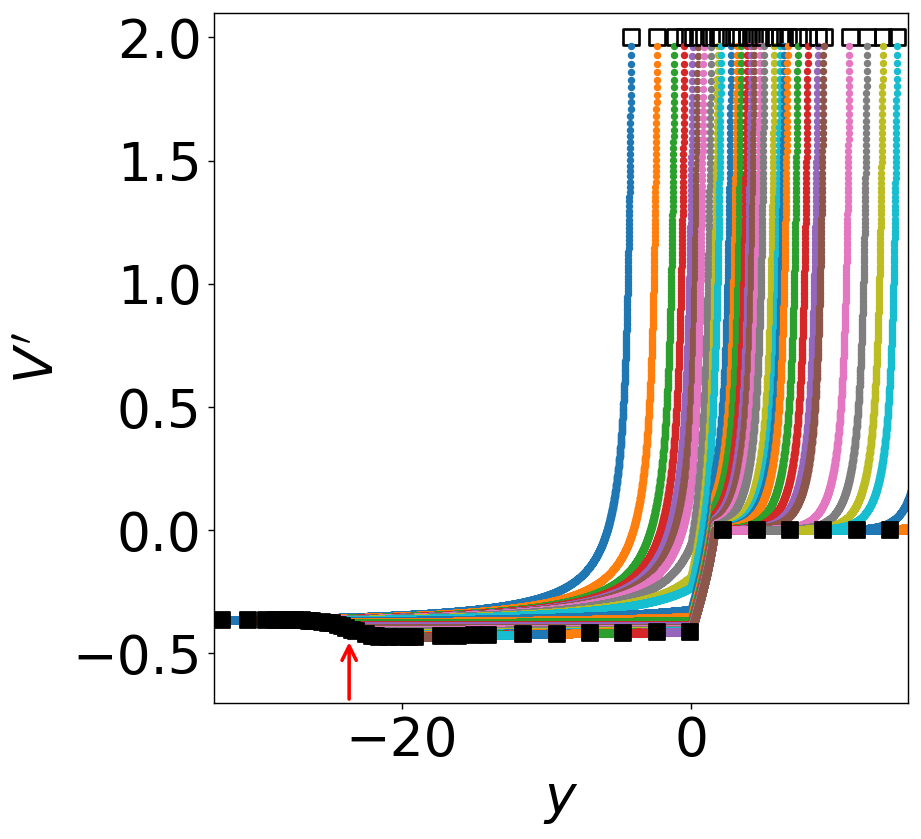}
  \caption{
}
  \label{fig:}
\end{subfigure}%
\begin{subfigure}{.35\textwidth}
  \centering
  \includegraphics[width=1\linewidth]{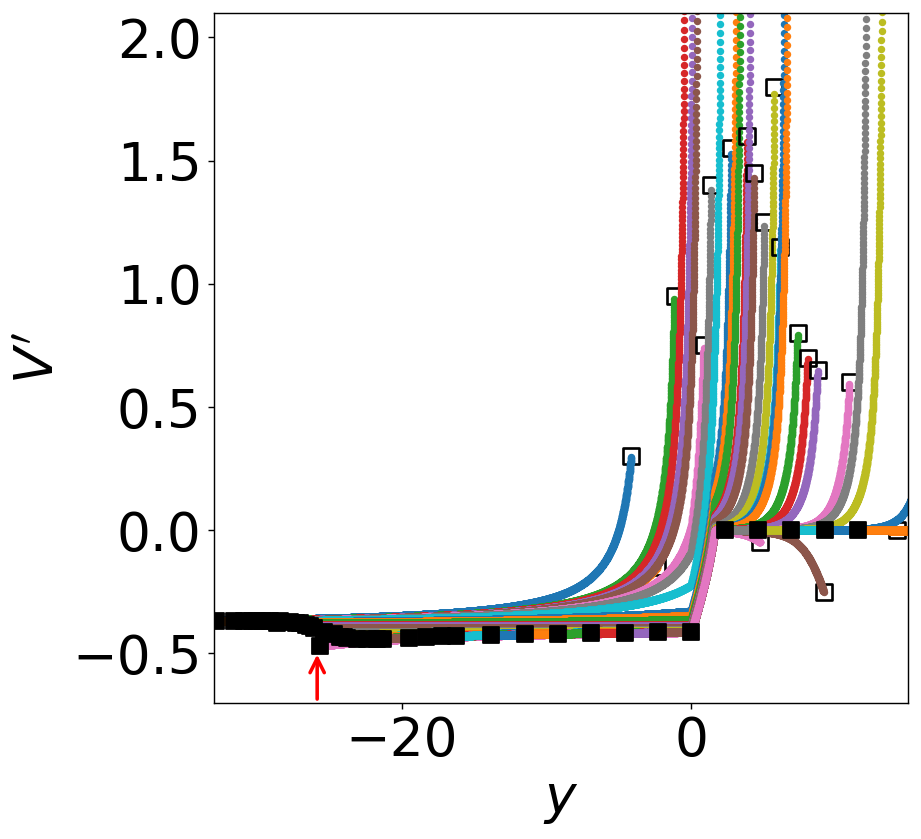}
  \caption{
}
  \label{fig:}
\end{subfigure}%
\begin{subfigure}{.35\textwidth}
  \centering
  \includegraphics[width=1\linewidth]{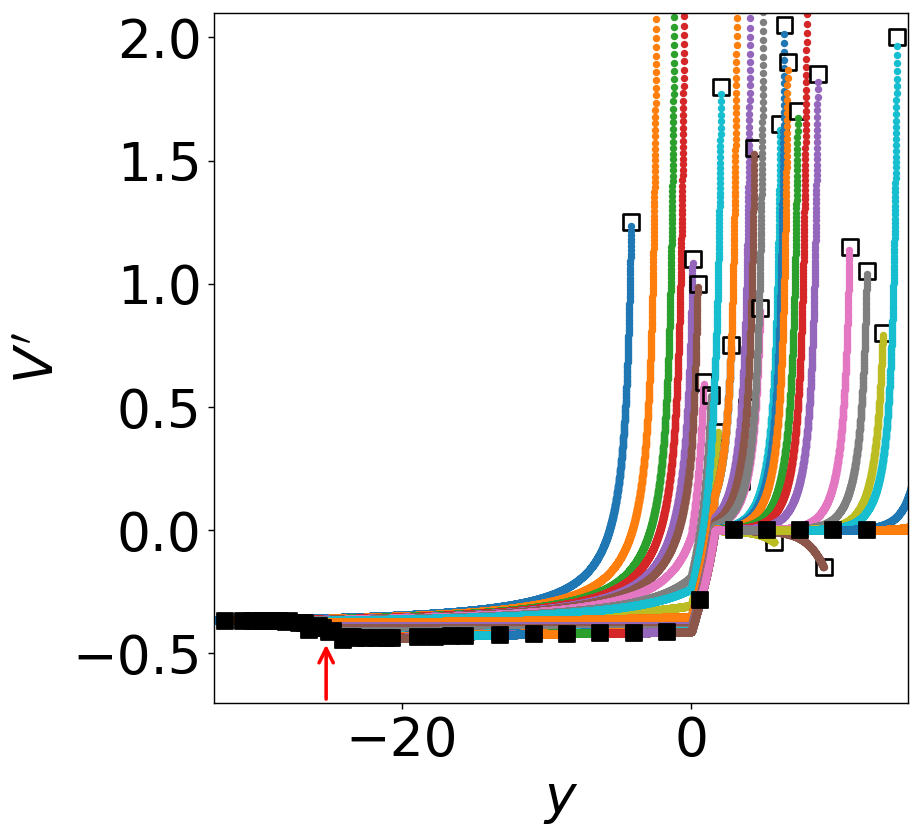}
  \caption{
}
  \label{fig:}
\end{subfigure}
\caption{
Trajectories of PFPs that start from UV couplings specified by $(y^{(m)}(0), \av_m)$ with 
$m=1,6,20,40,70,100,150,200,400,500,1000,1500,2000,3000,4000,5000,7500,10^4,2\cross 10^4,3\cross 10^4,4\cross 10^4,5\cross 10^4,1\cross 10^5,2\cross 10^5,4\cross 10^5,6\cross 10^5, 3.6\cross 10^6,1.2\cross 10^7, 3.6\cross 10^7, 9.6\cross 10^7, 1.08 \cross 10^9, 1.08 \cross 10^{10},1.08 \cross 10^{11},1.08 \cross 10^{12},1.08 \cross 10^{13},1.08 \cross 10^{14},1.08 \cross 10^{15},1.08 \cross 10^{16},1.08 \cross 10^{17},1.08 \cross 10^{18},1.08 \cross 10^{19},1.08 \cross 10^{20}$
with three sets of UV couplings: (a), (b), (c) in class B and (d), (e), (f) in class BC. 
The UV couplings are denoted as the open squares.
Here, $\av_m$ are chosen randomly in the range $-0.3\leq\av_m\leq 3$ and
$\KFAVdim/\Lambda = 10000$ is used.
The RG flow is stopped at the scale $l_b$ that takes the renormalized couplings closest to the regularized metallic PFP shown in Fig. \ref{fig:RMPFP_classBandBC}. 
The renormalized couplings at $l_b$ are denoted as the filled squares. 
For class B, we use $\etaPI = 1$, $\etaPII \approx -1.04 $, $\etaPIII \approx 0.11$ and $\delta w = 0.05 $.
For class BC, we use $\etaPI = 1$, $\etaPII \approx -1.04 $ and $\delta w = 0.1 $.
The jumps around $y=0$ are artifacts of the discontinuous source $\aS_y$ adopted in the toy model. 
In physical theories with continuous sources, those jumps will disappear.
However, the jumps that arise around $y_c(l_b)=y^*_S - l_b/2$ indicated by the arrows are genuine crossovers:
$y_c(l_b) \approx -7.34, -7.5, -7.43, -23.64, -25.86, -25.24$ for (a)-(f), respectively.
They are caused by the fact that the couplings with $y^{(m)}(0)<y^*_S$
are attracted to the $-\infty$ asymptotic fixed point, while those with
$y^{(m)}(0)>y^*_S$ are attracted toward  the metallic PFP.
}
\label{fig:rmfp_B_BC}
\end{figure}

The emergence of the regularized metallic PFP in intermediate energy scales can be directly confirmed through the solution of the PFP equation in the toy model.
In \fig{fig:rmfp_B_BC}, we show the RG flow of the couplings for various `random' choices of the bare couplings. 
With increasing $l$, the couplings at small angular momentum channels are attracted toward the $-\infty$ asymptotic fixed point.
On the other hand, the couplings at large angular momentum channels converge toward the metallic PFP, causing a crossover at $y_c(l) \sim y_S^*-l/2$.

\begin{figure}[H]
\centering
\begin{subfigure}{.35\textwidth}
  \centering
  \includegraphics[width=1\linewidth]{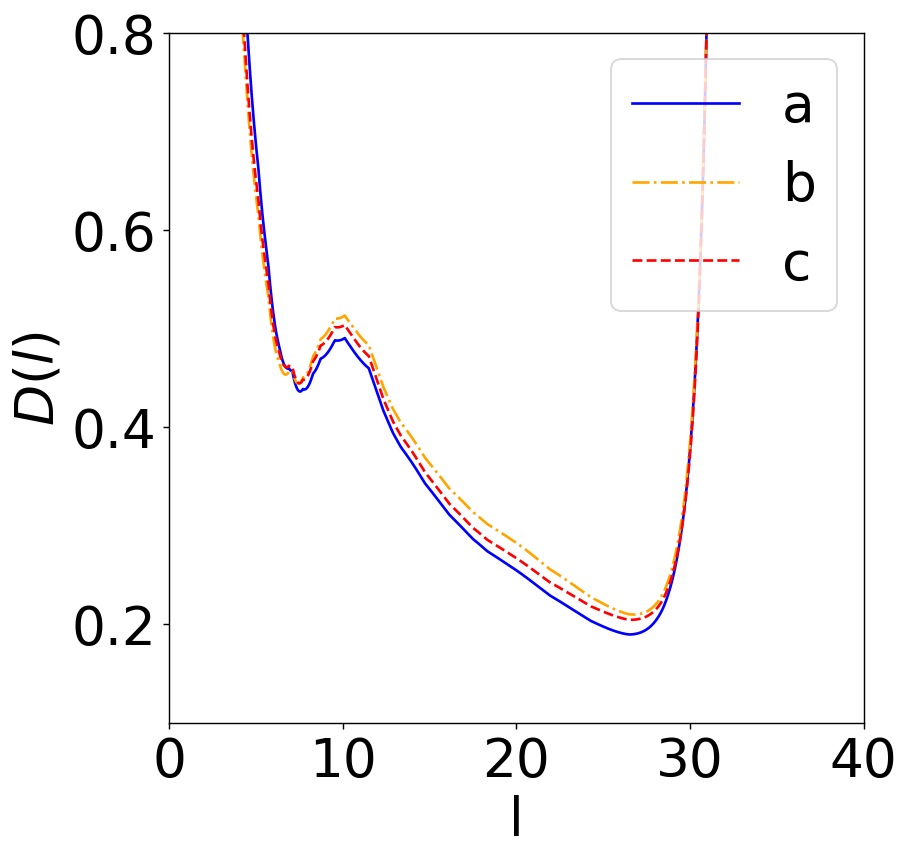}
  \caption{
}
  \label{fig:}
\end{subfigure}%
\begin{subfigure}{.35\textwidth}
  \centering
  \includegraphics[width=1\linewidth]{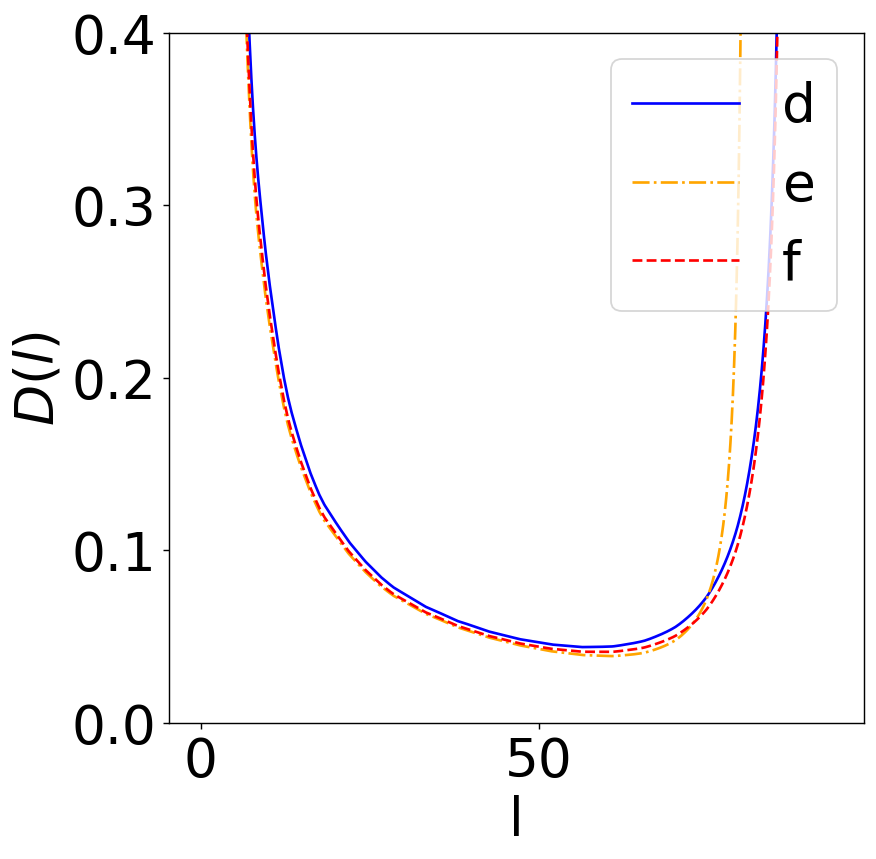}
  \caption{
}
  \label{fig:}
\end{subfigure}%
\caption{
The scale-dependent distance between the renormalized couplings and the regularized  metallic PFP shown in 
Fig. \ref{fig:RMPFP_classBandBC} 
for the UV couplings used in Fig. \ref{fig:rmfp_B_BC} 
for (a) class B
and (b) class BC.
The distance reaches the minimum around the bottleneck scale at which the renormalized couplings denoted by the filled squares in Fig. \ref{fig:RMPFP_classBandBC} approximately take the form of the regularized metallic PFP in \eq{eq:RMPinBandBC}.
The local minimum that arises at a short distance scale in (a) is of no great physical significance; it arises because the couplings temporarily move away from the regularized metallic PFP at high energy scales.
}
\label{fig:corr_B_BC} 
\end{figure}

In \fig{fig:corr_B_BC}, the distance  \eq{eq:Dl2}
between the renormalized coupling functions and the regularized metallic PFP 
is plotted as a function of RG length scale.
At small $l$, the profile of the couplings approach the regularized metallic PFP with increasing $l$ as the couplings are attracted to the $-\infty$ asymptotic fixed point and the metallic PFP, depending on their angular momenta. 
At the bottleneck scale $l_b$, the distance reaches the minimum.
As $l$ increases beyond $l_b$, the distance bounces up because the couplings with $y^{(m)}(0) > y^*_S$ head toward $-\infty$, reflecting the imminent superconducting instabilities.

\subsection{Superuniversality class C}

The physics of superuniversality class C is different from both the first group, which contains classes A, AB, AC, and ABC, 
and the second group, which contains classes B and BC.
As discussed in Sec.
\ref{sec:ex2}, the most prominent feature of class C is the presence of superconducting instability in the s-wave channel, with a minimum $T_c$ set by $\discinf$ (see 
\eq{eq:Tcmin}).
In non-Fermi liquids proximate to class A, the quasi-universal pairing interaction emerges in the intermediate energy scale.
The profile of the quasi-universal pairing interaction obtained from the toy model agrees with \eq{eq:VRMtheta} for $\Delta \bth \gg \sqmu$.
See Appendix \ref{appendix:ex4}-\ref{appendix:C} for details.
Even in class C, non-s-wave superconductors can be realized if the bare couplings in higher angular momentum channels are chosen such that $T_c$ in a non-zero angular momentum channel is higher than that of the s-wave channel.
For the reasons discussed in the previous section, as well as in Sec. \ref{sec:quasiuniversalTckF},
 the ratio between $T_c$ and $\KFAVdim^z$ for a non-s-wave superconductivity exhibits a weak dependence on the bare coupling.
If the superconducting instability occurs in the angular momentum channel $n$, 
\eq{eq:TC_KF_universal}
holds
with
\bqa
\frac{\delta y^*}{|y^*_M|} = 
- \frac{8R_d
\sqrt{|\etaPIII|}
}{\pi \big|\etaPI-\etaPIII\big|}\bigg(\frac{2\gamma \sqrt{\KFAVdim}}{\sqrt{\Lambda}\pi} \bigg)^{\sqrt{\etaPI}}  \frac{\deltaaVUV}{n^{\sqrt{\etaPI}}}. 
\eqa
Proximate to class A ($|\etaPIII| \ll 1$),
the dependence of $T_c/\KFAVdim^z$ on the UV coupling becomes weak.
All these features obtained from the toy model agree with the results obtained for the Ising-nematic critical metal in Sec. 
\ref{sec:ex2}
(see Appendix \ref{appendix:ex4} for details).

\section{Summary and Discussion}

\label{sec:conclusion}

In this paper, we classify non-Fermi liquids using their universal superconducting fluctuations as fingerprints.
Characterizing superconducting fluctuations in metals requires taking into account three ingredients.
The first is the incoherence of the fermions, which renders non-Fermi liquids less susceptible to superconducting instabilities.
The second is the universal pairing glue generated by quantum critical fluctuations.
The third is the lack of scale invariance caused by the 
Fermi momentum. 
Albeit a large momentum scale, the Fermi momentum does not decouple from low-energy physics, causing the balance between pair breaking incoherence and the universal pairing interaction to be intrinsically scale-dependent.

Since the RG flow is defined in the infinite dimensional space of couplings, the crossover energy scales can be arbitrarily small for some couplings. 
As a result, metals do not have scale invariant fixed points under the RG flow. 
Instead, metallic universality classes are identified with the universal RG trajectories  (projective fixed points) that emerge at low energies. 
Within each projective fixed point, which forms a one-dimensional manifold in the space of coupling functions, the dimensionless Fermi momentum continues to grow, creating explicit scale-dependence in the flow of other couplings.
A bundle of projective fixed points composed of RG trajectories with varying bare couplings defines a topological invariant associated with the connectivity of the short-distance part of the bundle to the long-distance part.
Based on the topology of the bundle of projective fixed points, universality classes of non-Fermi liquids are first grouped into seven topologically distinct superuniversality classes. 
Three of them are topologically stable, while the other four classes describe topological phase transitions from one class to another.
Within each superuniversality class, individual non-Fermi liquids are further classified by their universal superconducting fluctuations and emergent symmetries.

\begin{table}[h]
\centering
        \begin{tabular}{|c|c|c|}
            \hline
        ~Superuniversality class~ & Individual NFL & ~
        Low-energy symmetry~ \\
            \hline
            \hline
 \multirow{4}{*}{A, AB} & \multicolumn{1}{c|}{Stable NFLs} & \multicolumn{1}{c|}{LU(1)} \\
 \cline{2-3}
            & \multicolumn{1}{c|}{
2-critical NFLs 
            } & 
\multicolumn{1}{c|}{OLU(1)} \\
\cline{2-3} & \multicolumn{1}{c|}{ ~ $4$-critical NFLs ~  } & \multicolumn{1}{c|}{ ~ LU(1) or OLU(1)? ~ } \\ 
\cline{2-3} & \multicolumn{1}{c|}{ ~ $2n$-critical NFLs with $n \geq 3$~  } & \multicolumn{1}{c|}{ ~ LU(1) or U(1)? ~ } \\ \hline
           \hline
\multirow{4}{*}{AC, ABC} & \multicolumn{1}{c|}{Stable NFLs} & \multicolumn{1}{c|}{LU(1)} \\
            \cline{2-3}
            & \multicolumn{1}{c|}{
2-critical NFLs 
            } & \multicolumn{1}{c|}{LU(1)} \\
\cline{2-3} & \multicolumn{1}{c|}{ ~ $4$-critical NFLs ~  } & \multicolumn{1}{c|}{ ~ LU(1) or OLU(1)? ~ } \\ 
\cline{2-3} & \multicolumn{1}{c|}{ ~ $2n$-critical NFLs with $n \geq 3$~  } & \multicolumn{1}{c|}{ ~ LU(1) or U(1)? ~ } \\ \hline
\hline
 B, C, BC & ~Quasi-universal NFLs~ &  OLU(1) \\
           \hline
        \end{tabular}
    \caption{
Low-energy symmetries of various individual non-Fermi liquid universality classes.
Here, the $2n$-critical NFL refers to a critical non-Fermi liquid realized at the phase transition between the stable non-Fermi liquid and a charge $2n$ superconductor.
For $n \geq 2$, the low-energy symmetry is determined by whether the coupling between fermions and the charge-$2n$ critical boson is relevant or not at the decoupled fixed point.
}
\label{tab:emergent_symmetry}
\end{table}

Non-Fermi liquids that persist down to zero temperature can be categorized into stable non-Fermi liquids, which can be realized without fine tuning of the pairing interaction,
and critical non-Fermi liquids,  which require fine tuning.
In the superuniversality classes that are bound to become superconductors at low temperatures, the superconducting transition temperature and the pairing symmetry are largely determined by the quasi-universal non-Fermi liquids realized above $T_c$, provided that there is a hierarchy between the energy scale below which non-Fermi liquids set in and $T_c$.
For non-Fermi liquids that are prone to non-s-wave pairing instabilities and have a hierarchy of scales, we uncover robust relations between $T_c$ and the Fermi momentum, which are caused by the close interplay between the Fermi momentum and the pairing instabilities. 
They include a universal ratio between $T_c$ and $\KFAVdim^z$ with z being the dynamical critical exponent,
and an oscillatory behavior of $T_c$ as a function of $\KFAVdim$.

The emergent symmetries of stable, critical, and quasi-universal non-Fermi liquids are determined by the strengths of the inter-patch coupling generated by critical superconducting fluctuations.
In some non-Fermi liquids, the emergent symmetry remains the same as that of Fermi liquids\cite{PhysRevX.11.021005}, but in others, 
large-angle scatterings lower the symmetry to a proper subgroup.
The emergent symmetries of various individual non-Fermi liquids are summarized in \tab{tab:emergent_symmetry}.
Now, we conclude with several open questions and future directions.

\begin{itemize}

\item Stable metal and high-$T_c$ superconductor;

Although we use the $\epsilon$-expansion to determine the universality classes to which the physical theories belong near the upper critical dimension, we still do not know the definite answers in two spatial dimensions.
The final answer has to wait for non-perturbative solutions of the two-dimensional theories\cite{BORGES2023169221}.
However, our work reveals the physical mechanisms by which metals can evade superconducting instabilities down to zero temperature,
as well as the pathways through which metals can transition to superconducting states.
One way to evade superconducting instabilities is to keep fermions incoherent while the universal pairing interactions generated from multiple critical fluctuations are canceled, as discussed in Sec. \ref{sec:ex3}.
It will be interesting to establish the stability of such strongly interacting metallic states in two dimensions non-perturbatively.
On the other hand, one route to realizing higher $T_c$ is by increasing the energy scale $\Lambda$ below which non-Fermi liquid physics sets in for metals that are in one of the superconducting superuniversality classes.
In this regard, it will be interesting to consider metallic states realized by doping conformal field theories (CFTs).
Suppose that a CFT with global U(1) charge conservation and fermionic charged operators is realized below a UV cutoff scale $\Lambda_{CFT}$.
If the theory is doped with Fermi energy $E_F < \Lambda_{CFT}$, the crossover from the high-energy CFT to a low-energy metallic theory occurs around the energy scale $E \sim E_F$.
In this case, the superuniversality class and the `bare' coupling for the low-energy metal should be fixed by the CFT to the leading order in $E_F/\Lambda_{CFT}$
(see Ref. \cite{Delacretaz:2025aa} for constraints imposed on Landau parameters by such high-energy data).
Since $E_F$ plays the role of the UV cutoff for the non-Fermi liquid, it is expected that $T_c$ is of the order of $E_F$  for a generic strongly coupled CFT, 
if the resulting metallic state belongs to one of the superconducting superuniversality classes.
In this case, $T_c$ will increase linearly if both $\Lambda_{CFT}$ and $E_F$ are increased with a fixed ratio.

\item Toward the realization of the non-s-wave superconducting superuniversality class;

In the physical examples considered in this paper, the non-s-wave superconducting superuniversality class (classs B) is not realized at leading order in the $\epsilon$-expansion.
One possible route to realizing the class is to consider critical fluctuations that mediate more singular interactions.
For example, a Lifshitz critical boson with kinetic energy, $(q_0^2 + |\vec q|^4 ) |\phi(q)|^2$, is expected to promote pairing at higher angular momentum channels and may realize the non-s-wave superconducting superuniversality class.

\item Discrete scale invariance;

Due to Fermi momentum, metals do not possess scale invariance {\it as a whole}.
In other words, there is no scale transformation that maps the entire set of low-energy observables into itself.
However, this does not exclude the possibility that a subset of low-energy observables exhibits a scale invariance.
One such example is the universal pairing interaction that arises in the stable non-Fermi liquids.
The universal pairing interaction in  \eq{eq:stable_large_mom} is invariant under the scale transformation within the range of angles $\bth \sim \sqmu$.
Alternatively, one can have discrete scale invariance in Fourier space.
In stable non-Fermi liquids whose universal coupling is governed by a metallic projective fixed point, 
the coupling in angular momentum $m$ at scale $l$ is related to the coupling in angular momentum $m'$ at scale $l'$ provided that the two scales are related by
$l'-l = 2 \log m'/m$
(see \fig{fig:lack_of_scale_invariance}).
This has observable consequences for the angular momentum dependent pair susceptibility of stable non-Fermi liquids.

\item Lukewarm fermions;

In the present paper, we consider the superconducting fluctuations of the hot Fermi surface by ignoring the cold spots at which the coupling between the Fermi surface and critical fluctuations vanishes.
This is justified for the purpose of understanding the behavior of hot electrons at low energies because the size of the lukewarm region around the cold spots, if present, becomes vanishingly small in the low-energy limit. 
Although lukewarm fermions are minor players in the superconducting fluctuations of non-Fermi liquids, it will still be interesting to study the physics of lukewarm fermions for various reasons.
Firstly, the lukewarm regions act as buffers between the hot segments of the Fermi surface. 
The cold regions embedded in the Fermi surface play an important role in  regulating the entanglement formed across disjoint hot segments.
This is already manifest in \fig{Fig: C4FS_Rescaled_Single}, which shows that the parts of the Fermi surface that used to be lukewarm at high energy scales give rise to longer proper distances after they become hot at low energies, making it harder for Cooper pairs to traverse that region through scattering.
Secondly, it may be possible to understand the dynamics of the lukewarm fermions on their own in a controlled manner, even in two spatial dimensions.

\item Non-Fermi liquids with geometric criticality;

In globally convex Fermi surfaces, rotational invariance emerges in the proper angular space $\bth$.
This is why we could use the angular momentum, which is conjugate to the proper angle, as a good quantum number for Cooper pairs.
It will be interesting to consider non-Fermi liquids whose Fermi surfaces contain inflection points and to understand how geometric critical points associated with the change in the sign of local curvature affect the nature of non-Fermi liquids\cite{2023arXiv231007539S}.
This consideration can be further extended to the topological critical metals that exhibit van Hove singularities.

\item Non-universal superconductors;

The superconductors considered in this paper are the ones that emerge at temperatures well below the onset temperature of the non-Fermi liquid behavior.
Universal properties of such superconductors are determined by the projective fixed points of the parent non-Fermi liquids.
To understand non-universal superconductors whose transition temperatures are comparable to the onset temperature of non-Fermi liquid behavior, one must take into account the emergence of non-Fermi liquids along with superconductivity.
Technically, this requires including additional scale dependence in the beta functionals for the pairing interaction, which arises from
the running fermion-boson coupling function and the Fermi velocity. 
It will be interesting to see if there is any sense of universality in how such `non-universal' superconductors arise out of non-Fermi liquids in the making.

\item{Fermiological criticality;}

In this paper, we consider non-Fermi liquids that result from critical bosonic fluctuations.
If soft fluctuations originate from criticalities of  fermiology, 
such as
Kondo screening or
Fermi surface topology
\cite{
Senthil_Sachdev_Vojta_2003,
10.1143/PTP.32.37,1126-6708-2008-12-031,2005cond.mat..5529H,
Coleman_2001,
Georges_Kotliar_Hunds_popular,
Sunko:2019aa,
Shi:2017aa},
one should take into account more general form of critical fluctuations, which may alter the properties of metals differently.
It is of great interest to understand the universal physics of such critical metals within low-energy effective theories.

\end{itemize}


\appendix


\newpage

\section{Field-theoretic functional renormalization group scheme}
\label{app:RGscheme}

In this section, we briefly discuss the functional RG conditions. 
Let $\varGamma ^{(2m,n)}(\{\mathbf{k}_i\})$ denote the momentum-dependent vertex function for $2m$ fermions and $n$ bosons. 
The vertex functions evaluated as functions of angles around the Fermi surface at external frequencies $\sim \mu$ define the renormalized coupling functions at that energy scale.
The two-point and three-point functions define the 
Fermi momentum,
Fermi velocity,
the boson-fermion coupling
as well as the field renormalizations through\cite{PhysRevB.110.155142} 
\bqa
&&
\Re 
\left[\tr{ -i\varGamma^{(2,0)}( {\bf k})}_{{\bf k}=(\mu \hat 1_{d-1}, \delta=0, \theta)}\right] = 
0,
\label{eq:RG1}\\
&&\frac{i}{2(d-1)}
\tr{ {\tilde {\bf \Gamma}} \cdot \nabla_{{\bf K}}
  \varGamma^{(2,0)}({\bf k})}_{{\bf k}=(\mu \hat 1_{d-1},\delta=0, \theta )} = 1  + {\bf \mathcal{F}}_{1; \theta},\label{eq:RG2}\\
 && \frac{1}{2}
 \Re \left[ \tr{ -i 
 \frac{\partial }{\partial \delta}  
 \varGamma^{(2,0)}({\bf k}) }_{{\bf k}=(\mu \hat 1_{d-1},\delta=0, \theta )} 
 \right]
 = v_{F,\theta}  +\mathcal{F}_{2; \theta},\label{eq:RG3}\\
&&\frac{\partial}{\partial q^2}\varGamma^{(0,2)}\left(\mathbf{q}\right)\bigg|_{{\bf q}=
  \left(\mu \hat{1}_{d-1},\vec{q} = 0\right)} = 1+\mathcal{F}_{3},
 \label{eq:RG32} 
  \\
 &&-\frac{ \sqrt{N}}
 {2  \mu^{\frac{3-d}{2}}}
 \tr{\mathscr{T}_t \varGamma^{(2,1)}({\bf k +\bf q},{\bf k})}_{
 \scriptsize
   \begin{array}{l}
  {\bf q} =( \mu \hat 1_{d-1}, q(\theta',\theta), \varphi ) \\
  {\bf k}=( \mu \hat 1_{d-1}, \delta=0, \theta )  
 \end{array}
 } 
 = 
e^t_{\theta',\theta}(\varphi)  
+
\mathcal{F}^t_{4;\theta',\theta}(\varphi).
\label{eq:RG4}
\eqa
Here, the traces in the above expression are over the spinor indices.
The freedom to rescale frequency relative to momentum is used to set 
$v_{F,0} = 1$.
$q(\theta',\theta) = \sqrt{\mathbf{K}^2_{F,\theta'}+\mathbf{K}^2_{F,\theta}-2\mathbf{K}_{F,\theta'}\mathbf{K}_{F,\theta}\cos(\theta'-\theta)}
$
represents the momentum that connects two points on the Fermi surface at angles $\theta$ and $\theta'$.
$\hat{1}_{d-1}$ is a unit vector in the $(d-1)$-dimensional space, which can point in any direction due to
the $SO(d-1)$ symmetry. 
The four-fermion coupling functions is defined through the quartic vertex function,
\bqa
&&\frac{1}{
\mu^{1-d}}\int \left\{d\renf_i\right\}
\varGamma^{(4,0)}
_{\{j_i\};\{a_i\}}
 (\left\{{\bf k}_i\right\}
 )
 \bigg|_{{\bf k}_i=
  \renm_i}
 \dist(\tilde {\mathbf{K}}_1,\tilde {\mathbf{K}}_4
 )\dist(\renf_2,\renf_3)
  = 
T^{(\nu,s)}_{\left(\begin{smallmatrix}     j_1  & j_2      \\ j_4      & j_3       \end{smallmatrix}\right)} 
\left(\tilde I_m^{(\nu)}\right)_{a_1 a_2}\left(\tilde I_m^{(\nu)}\right)_{a_3 a_4}
\left[
\lambda^{(\nu,s)}_{ \theta_1,\theta_2}(\vec q)
+ \mathcal{F}^{(\nu,s)}_{5; \theta_1,\theta_2}(\vec q)
\right]. \nn
 \label{eq:RG5}
\eqa
The vertex function is evaluated at external momenta on the Fermi surface,
\begin{equation}
    \begin{aligned}
          \renm_1 &= \left(\renf_1, 0, \theta_1 \right), \quad 
        \renm_2 = \left(\renf_2, \Delta(0,\theta_2,\vec q), \Theta(\theta_2,\vec q)\right), \quad
        \renm_3 &= \left(\renf_3, 0, \theta_2 \right), \quad 
        \renm_4 = \left(\renf_4,\Delta(0,\theta_1,\vec q), \Theta(\theta_1,\vec q)\right).
    \end{aligned}
\end{equation}
Instead of choosing a fixed set of external frequencies, we average the vertex function over a distribution of external frequencies,
\begin{equation}
    \begin{aligned}
\dist(\mathbf{P},\mathbf{Q}) = 
\dist^+(\mathbf{P}+\mathbf{Q})
\dist^-(\mathbf{P}-\mathbf{Q}).
        \label{app:dist+_dist-}
    \end{aligned}
\end{equation}
Here, $\dist^-(\mathbf{P}-\mathbf{Q})$ determines the distribution of the net frequency that a particle-hole or particle-particle pair carries.
We choose
\bqa
\dist^-(\mathbf{P}-\mathbf{Q}) = 
\frac{1}{\Omega_{d-1}}
\int d\hat {\boldsymbol{\mu}} 
~\delta^{d-1}(\mathbf{P}-\mathbf{Q}-2\boldsymbol{\mu}),
\label{eq:dist_minus}
\eqa
where $\frac{1}{\Omega_{d-1}} \int d\hat {\boldsymbol{\mu}}$ averages over the direction of the $(d-1)$-dimensional unit vector.
$\dist^+(\mathbf{P}+\mathbf{Q})$ sets the distribution for the relative frequency.
The choice of 
$\dist^+(\mathbf{P}+\mathbf{Q})$ is part of the RG scheme. 
We will specify it in Appendix  \ref{app:lambda1} to simplify the beta functional for the pairing interaction while satisfying the normalization condition,
$\int d\mathbf{P}d\mathbf{Q}~\dist(\mathbf{P},\mathbf{Q}) = 1$. 
In Eqs. (\ref{eq:RG2})-(\ref{eq:RG5}), $\mathcal{F}_{i;\{\theta_i\}}$ are the scheme-dependent corrections that remain regular in the limit $\mu \to 0$. 
One must also ensure that the contribution of the corrections to physical observables are regular.
For this, we also require that 
$\int d\theta'\, K_{F,\theta'}\, \mathcal{F}_{5;\theta,\theta'}\, f_{\theta'}$
remains regular for any normalizable function $f_{\theta'}$\cite{BORGES2023169221}.

To enforce the renormalization conditions
in Eqs. \eqref{eq:RG1}-\eqref{eq:RG5},
we add a local counter-term action,
\begin{equation}\begin{aligned}
   \hspace{-60pt}
   S_{CT} & =   
  \int^{'}
d_{f}^{d+1} {\bf k}~
\tilde \Psi_j(\mathbf{K},\delta,\theta )
\Bigl[ 
iA_1(\theta ) \tilde{{\bf \Gamma}} \cdot{\bf K} + 
i
v_{F,\theta  }
\left\{
A_2(\theta ) 
\delta 
+
c(\theta ) 
\mu
\right\}
\Bigr]
\Psi_{j}(\mathbf{K},\delta,\theta )  \\ &
+ \frac{A_3}{2}
\int
d_{b}^{d+1} {\bf q}~
q^2
\left|\phi_t(\mathbf{q})\right|^2 
+ \frac{\mu^{\frac{3-d}{2}}}{\sqrt{N}}  
\int^{'}
d_{f}^{d+1} {\bf k}
\int
d_{b}^{d+1} {\bf q}~
A_4\left(
\thetasq ,
\theta,\varphi \right)~
e^t_{
\thetasq ,
\theta  }(\varphi)
\\ &~~~~\times
\phi_t(\mathbf{q}) 
\tilde \Psi_{j}\left(\mathbf{K}+\mathbf{Q},\Delta(\delta,\theta,\vec q),\thetasq \right)\mathscr{T}_t \Psi_{j}(\mathbf{K},\delta,\theta )
\\ &
+ \mu^{1-d}
\sum_{\nu}
\sum_{s=d,e}
\int^{'}
d_{f}^{d+1} {\bf k}
d_{f}^{d+1} {\bf k}'
\int
d_{b}^{d+1} {\bf q}~
A^{(\nu,s)}_{5}\left(\theta',\theta,\vec{q}\right)
\lambda^{(\nu,s)}_{ \theta',\theta}(\vec q) 
{\bf O}^{(\nu,s)}_{\Lc} ( \mathbf{K}^{\prime},  \delta^\prime, \theta'; \mathbf{K}, \delta, \theta; \mathbf{Q}, \vec q).
\label{eq:SCT}
\end{aligned}
\end{equation}
The bare quantities and the renormalized quantities are related through 
\begin{equation}\begin{aligned}
&	 
\mathbf{K}_B = Z_{\tau}\mathbf{K},
~~\delta_B = \delta+
\frac{c(\theta)}{Z_2(\theta)} \mu,
~~ \theta_B = \theta,
~~ \mathbf{Q}_B = Z_{\tau}\mathbf{Q},
~~q_{B}=q, 
~~\varphi_B = \varphi,\\ 
& \Psi_{jB}(\mathbf{K}_B, \delta_B,\theta_B)=\sqrt{Z_{\psi}(\theta )}\Psi_{j}(\mathbf{K}, \delta ,\theta  ),~~
\phi_B(\mathbf{Q}_B, q_{B},\varphi_B)=\sqrt{Z_{\phi}}\phi(\mathbf{Q},q,\varphi), \\
& v_{FB,\theta_B}=Z_{v_F}(\theta )v_{F,\theta  },~~
{\bf K}_{FB,\theta_B} = \mu
\left[
\KFtheta-
\frac{c(\theta)}{Z_2(\theta)}
\right],~~
\edim_{B,\theta_{1B},\theta_{2B}}(\varphi_B) = Z_{e}(\theta_1 ,\theta_2,\varphi)
\mu^{\frac{3-d}{2}}
e_{\theta_1 ,\theta_2 }(\varphi), \\
&\lambdadim^{(\nu,s)}_{B, \theta_B',\theta_B}(q_B,\varphi_B)=  Z^{(\nu,s)}_{\lambda}(\theta',\theta,\vec q) 
\mu^{1-d} \lambda^{(\nu,s)}_{\theta',\theta}(\vec q),
\label{eq:baretorenorm}
\end{aligned}\end{equation}
where 
\begin{equation}\begin{aligned}
  &  Z_{\tau} = \frac{Z_1(0)}{Z_2(0)}, ~~
    Z_{\psi}(\theta ) = Z_1(\theta )\left(\frac{Z_2(0)}{Z_1(0)}\right)^{d}, ~~
    Z_{\phi} = Z_3\left(\frac{Z_2(0)}{Z_1(0)}\right)^{d-1}, ~~
    Z_{v_F}(\theta ) = \frac{Z_2(\theta )}{Z_1(\theta )}\frac{Z_1(0)}{Z_2(0)},\\
    &
   Z_{e}(\theta_1 ,\theta_2) = \frac{Z_4(\theta_1 ,\theta_2)}{\sqrt{Z_1(\theta_1)Z_1(\theta_2  )Z_3}}\left(\frac{Z_1(0)}{Z_2(0)}\right)^{\frac{3-d}{2}}, ~~
Z^{(\nu,s)}_{\lambda} \left(\theta_1,\theta_2,\vec{q}\right)= Z^{(\nu,s)}_{5}\left(\theta_1,\theta_2,\vec{q}\right)\left(\frac{Z_1(0)}{Z_2(0)}\right)^{3-d} \prod_{i=1}^{4}\frac{1}{\sqrt{Z_1(\theta_i)}}
\label{eq:Z_multiplicative}
\end{aligned}\end{equation}
with $Z_i(\theta) = 1 + A_i(\theta)$.
Here, $Z_\tau$ is a factor by which the renormalized frequency is dilated relative to the bare frequency.
Because the frequency is dilated such that $v_{F,\theta=0}=1$, 
$Z_{v_F}(0)=1$
and $Z_\tau$ is determined from $Z_1(0)$ and $Z_2(0)$. The bare Fermi momentum is determined from ${\bf K}_{FB,\theta_B}+\delta_B = \KFthetadim+\delta$.
See Ref. \cite{PhysRevB.110.155142} for more details.

\section{Quantum corrections}
\label{app:qc}

\begin{figure}[h]
\begin{subfigure}{.225\textwidth}
  \includegraphics[width=0.9\linewidth]{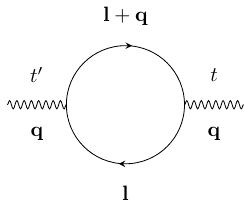}
  \caption{}
  \label{Fig:BSE}
\end{subfigure}
\begin{subfigure}{.225\textwidth}
  \includegraphics[width=0.9\linewidth]{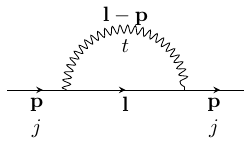}
  \caption{}
  \label{Fig:FSE}
\end{subfigure}
\begin{subfigure}{.225\textwidth}
  \includegraphics[width=0.9\linewidth]{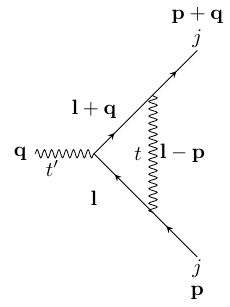}
  \caption{}
  \label{Fig:Yukawa_Vertex}
\end{subfigure}

\begin{subfigure}{.225\textwidth}
  \centering
  \includegraphics[width=0.65\linewidth]{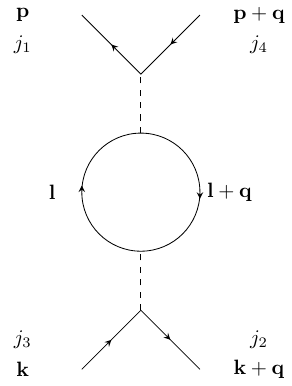}
  \caption{}
  \label{fish1}
\end{subfigure}
\begin{subfigure}{.225\textwidth}
  \centering
  \includegraphics[width=0.65\linewidth]{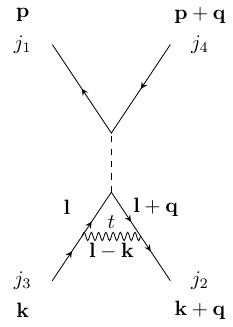}
  \caption{}
  \label{vertex3_1}
\end{subfigure}
\begin{subfigure}{.225\textwidth}
  \centering
  \includegraphics[width=0.65\linewidth]{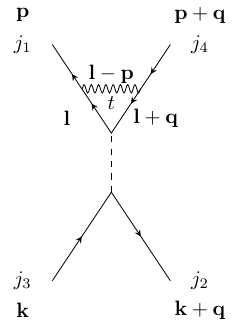}
  \caption{}
  \label{vertex3_2}
\end{subfigure}
\begin{subfigure}{.225\textwidth}
  \centering
  \includegraphics[width=1.0\linewidth]{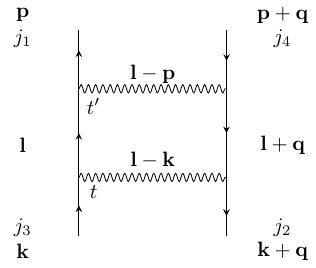}
  \caption{}
  \label{ladder1}
\end{subfigure}%
\caption{
The one-loop diagrams that contribute to the self-energies and the vertex corrections.
The solid line represents the fermion propagator.
The dashed line represents the short-range four-fermion interaction.
The wiggly line with label $t$ denotes the boson propagator of type $t$,
dressed with the self-energy in \fig{Fig:BSE}. 
For internal boson propagators, $t$ is summed over all types.
}
\label{fig:BSE_FSE_quarticfermion}
\end{figure}

In this section, we derive quantum corrections generated from the diagrams shown in Fig. \ref{fig:BSE_FSE_quarticfermion}.
We consider the theory that contains both types of critical bosons, namely, one repulsive and one attractive bosons. 
The results for the theory that has only one type of boson can be readily obtained by setting one of the fermion-boson couplings to zero.

\subsection{Boson self-energy}

The one loop boson quantum effective  action is given by
\begin{equation}\begin{aligned}
\delta S^{(0,2)} = \frac{1}{2}\int d_b^{d+1}~\mathbf{q}
\Pi_{t,t'}(\mathbf{q})
\phi_{t}(\mathbf{q})\phi_{t'}(-\mathbf{q}),
\label{Boson_SE_formal_action}
\end{aligned}\end{equation}
where the boson self-energy is written as 
\begin{equation}\begin{aligned}
\Pi_{t,t'}(\mathbf{q}) = -\left[-
\int^{'} 
d_f^{d+1}\mathbf{l}\left(-
\edim^{t}_{\thetasq,\theta}(\varphi)
\right)\left(-
\edim^{t'}_{\theta,\thetasq}(\varphi+\pi)
\right)
\mathrm{\mathrm{Tr}}\left\{\mathscr{T}_t\tilde G_{0}(\mathbf{l}+\mathbf{q})\mathscr{T}_{t^\prime} \tilde G_{0}(\mathbf{l})\right\}
\right].
\label{Boson_SE_formal}
\end{aligned}\end{equation}
Here $ \mathbf{l} = (\mathbf{L},\delta,\theta)$,
$\mathbf{q} =  (\mathbf{Q},\vec{q})$, and
$\mathbf{l+q} = (\mathbf{L}+\mathbf{Q},
\deltaq, \thetasq )$ with
$\deltaq$ and $\thetasq$ given in Eq. (\ref{eq:ThetaDelta}).
Here, \( \left(t,t'\right) \in \{1\} \cup
 \{ (\alpha< \beta) |  d-1 \leq \alpha, \beta \leq 2 \} \).
 The free fermion propagator $\tilde{G}_0$ is given as
\begin{equation}
    \begin{aligned}
        \tilde{G}_0(\mathbf{l}) = \frac{1}{i}\frac{1}{\tilde{\mathbf{\Gamma}}\cdot\mathbf{L}+v_{F,\theta}\delta} \equiv \frac{1}{i}\frac{-\tilde{\mathbf{\Gamma}}\cdot\mathbf{L}+v_{F,\theta}\delta}{|\mathbf{L}|^2+v_{F,\theta}^2\delta^2}.
\end{aligned}
\end{equation}
We write the trace inside 
Eq. (\ref{Boson_SE_formal}) as 
$\mathrm{\mathrm{Tr}}\left\{\mathscr{T}_t\tilde G_{0}(\mathbf{l}+\mathbf{q})\mathscr{T}_{t^\prime} \tilde G_{0}(\mathbf{l})\right\} = \frac{\mathcal{N}_{t,t'}}{\mathcal{D}}$
with
\begin{equation}\begin{aligned}
    \mathcal{N}_{t,t'} &= \mathrm{\mathrm{Tr}}\{\mathscr{T}_t\tilde{\Gamma}_k\mathscr{T}_{t^\prime}\tilde{\Gamma}_l(L+Q)_{k}L_l-\mathscr{T}_t\mathscr{T}_{t^\prime}\tilde{\Gamma}_l v_{F,\thetasq }\deltaq
    L_l 
    \\ &
    -\mathscr{T}_t\tilde{\Gamma}_k\mathscr{T}_{t^\prime} v_{F,\theta}\delta (L+Q)_k+\mathscr{T}_t\mathscr{T}_{t^\prime}
    v_{F,\thetasq }\deltaq v_{F,\theta }\delta \},\\
    \mathcal{D} &= i^2\left(|\mathbf{L}+\mathbf{Q}|^2+v^2_{F,\thetasq}\deltaq^{2}\right)\left(|\mathbf{L}|^2+v^2_{F,\theta }\delta^{2}\right).
    \label{appendix:num_denom_BSE}
\end{aligned}\end{equation}
Using 
\begin{equation}
    \begin{aligned}
    \left[\mathscr{T}_t,\tilde{\Gamma}_i\right]=0,
 \label{appendix:gamma_tilde_clifford}
 \end{aligned}
\end{equation}
the numerator in Eq. (\ref{appendix:num_denom_BSE}) can be simplified to 
\begin{equation}
    \begin{aligned}
        \mathcal{N}_{t,t'} = 2\delta_{t,t'}\left(\mathbf{L}\cdot(\mathbf{L+Q})-v_{F,\thetasq}v_{F,\theta}\delta\deltaq\right).
        \label{appendix:simplified_numerator_boson}
    \end{aligned}
\end{equation}
From Eqs. (\ref{Boson_SE_formal}) - (\ref{appendix:simplified_numerator_boson}), the boson self-energy can be derived in the same way as in Ref. \cite{PhysRevB.110.155142},
\begin{equation}
    \begin{aligned}
       \Pi_{t,t^\prime}(\mathbf{q})
       =v_d  \int^\prime
       \frac{d\theta}{2\pi}\frac{\KFthetadim }{v_{F,\theta }} \left|\edim^{t}_\theta(\varphi)\right|^2\frac{Q^2}{\left(Q^2+ \left(L_{\theta}\left(\vec{q}\right)\right)^2\right)^{\frac{4-d}{2}}}\delta_{t,t^\prime},
       \label{appendix:BSE_General}
    \end{aligned}
\end{equation}
where $v_d = \frac{1}{2}\frac{\Omega_{d-1}\Gamma\left(\frac{d-1}{2}\right)\Gamma\left(\frac{4-d}{2}\right)\Gamma^2\left(\frac{d}{2}\right)}{\Gamma\left(\frac{3}{2}\right)(2\pi)^{d-1}\Gamma\left(d\right)}$ and $\Omega_d = \frac{2\pi^{\frac{d}{2}}}{\Gamma\left(\frac{d}{2}\right)}$ is the solid angle subtended by a ($d-1$)-dimensional spherical surface. 
In the low-energy limit, the variation of the coupling functions over $\Delta \theta \sim q_\mu/\KFAVdim$ becomes negligible, where $q_\mu$ is the typical momentum that the critical boson carries at energy scale $\mu$. 
This allows us to use the adiabatic approximation to set
$\edim^{t}_{\thetasq,\theta}(\varphi)\approx \edim^{t}_{\theta,\theta}(\varphi)\equiv \edim^{t}_{\theta}(\varphi)$.
$L_{\theta}(\vec q)$ is the energy of the a particle-hole pair with momentum $\vec{q}$ created near the Fermi surface at angle $\theta$,
\begin{equation}
    \begin{aligned}
        L_{\theta}(\vec q) 
        = v_{F,\theta}\left(\mathscr{F}_{\varphi,\theta}q+
        \frac{\mathscr{G}_{\varphi,\theta}}{\KFthetadim}q^2\right).
        \label{app:Lqvarphi}
    \end{aligned}
\end{equation}
The most singular correction is given by
\begin{equation}\begin{aligned}
    \Pi_{t,t^\prime}(\mathbf{q})  = \boldsymbol{f}^t_{d,\varphi }\frac{|\mathbf{Q}|^{d-1}}{q}\delta_{t,t^\prime},
    \label{Boson_SE_Final_Expression}
\end{aligned}\end{equation}
where
$\boldsymbol{f}^t_{d,\varphi } = \beta_d\frac{\mathbf{K}_{F,\vartheta(\varphi ) } \left|\edim^{(t)}_{\vartheta(\varphi)}\right|^2}{|\chi_{\varphi}|v^2_{F,\vartheta(\varphi ) }}$ with 
$\beta_d = \frac{\Gamma^2\left(\frac{d}{2}\right)}{2^{d-1}\pi^{\frac{d-1}{2}}\Gamma(d)\Gamma\left(\frac{d-1}{2}\right)\left|\cos\left(\frac{\pi d}{2}\right)\right|}$
and
$\edim^{t}_{\vartheta(\varphi)} \equiv \edim^{t}_{\vartheta(\varphi)}(\varphi)$. In the $Q\gg q$ limit, Eq. (\ref{appendix:BSE_General})  becomes
\begin{equation}
    \begin{aligned}
         \Pi_{t,t^\prime}(\mathbf{q}) = v_d Q^{d-2}
         \int^\prime
         \frac{d\theta}{2\pi}\frac{\KFthetadim }{v_{F,\theta }}\left|\edim^{t}_\theta(\varphi)\right|^2\delta_{t,t^\prime}.
    \end{aligned}
\end{equation}
It is noted that there is no mixing between the attractive and repulsive bosons.
If there are multiple bosons in one sector, they can be mixed through the off-diagonal self-energies.

\subsection{Fermion self-energy}

The one-loop fermion self-energy 
reads 
\begin{equation}\begin{aligned}
\Sigma_{1}(\mathbf{k}) = -
\frac{2}{2!}\frac{1}
{N}\int\frac{d\mathbf{L}d\delta }{(2\pi)^{d}}
\int_0^{2\pi}\frac{\KFthetadim d\theta}{2\pi}(-\edim^{t}_{\theta,\theta_1}( \varphi))(-\edim^{t}_{\theta_1,\theta}(\varphi+\pi))(\mathscr{T}_t
\tilde G_{0}(\mathbf{l})\mathscr{T}_t
)D_1^t(\mathbf{l}-\mathbf{k}).
\label{fermion_SE_formal}
\end{aligned}\end{equation}
Here, $\varphi$ is the angular direction of the boson, which implicitly depends on loop momenta. 
The counter term to the self-energy then reads
\begin{equation}
    \begin{aligned}
        A_1\left(\theta_1\right) =
         \frac{
         1
         }{2N(d-1)}
        \tr\left\{\mathbf{\Gamma}\cdot\nabla_{\mathbf{K}}
        \int\frac{d\mathbf{L}d\delta \KFthetadim d\theta}{(2\pi)^{d+1}}
 D^t_{1;\mu}\left(\mathbf{L}-\mathbf{K},\theta,\theta_1\right)
 \frac{-\mathbf{\Gamma} \cdot
 \mathbf{L}
 +v_{F,\theta } \delta \gamma_{d-1}}
{
\left|\mathbf{L}\right|
^2 +v^2_{F,\theta }\delta^2} \left|\boldsymbol{e}^t_{\theta_1 ,\theta}\right| \left|\boldsymbol{e}^t_{\theta,\theta_1 }\right|\right\}\bigg|_{\mathbf{K} = \boldsymbol{\mu}
        },
        \label{appendix:Z1}
    \end{aligned}
\end{equation}
where
$D^t_{1;\mu}\left(\mathbf{L},\theta_1,\theta_2\right)= \frac{1}{q(\theta_1,\theta_2)^{2}+ \boldsymbol{f}^t_{d,\vartheta^{-1}\left( 
\mtheta
\right)}
\frac{|\mathbf{L}|^{d-1}}{\sqrt{q(\theta_1,\theta_2)^{2}+\mu^2}}}$ is the boson propagator regularized with an infrared energy cutoff $\mu$ with 
$q(\theta_1,\theta_2) =$
$ \sqrt{\mathbf{K}^2_{F,\theta_1}+\mathbf{K}^2_{F,\theta_2}-2\mathbf{K}_{F,\theta_1}\mathbf{K}_{F,\theta_2}\cos\left(\theta_1-\theta_2\right)}$ 
and 
$e_{\theta_!,\theta_2}\equiv e_{\theta_1,\theta_2}(\vartheta^{-1}\left(\mtheta\right))$ \footnote{$\varphi$ is evaluate at angle tangential to the average of the angles as it does not affect the singular part of the self-energy.}.
The log $\mu$ derivative of Eq. (\ref{appendix:Z1}) under the adiabatic approximation
can be calculated in the same way as was done in Ref. \cite{PhysRevB.110.155142},
\begin{equation}\begin{aligned}
   \tilde{\omega}_{\theta_1} \equiv \frac{\partial Z_1\left(\theta_1\right)}{\partial\log\mu} = \bar g_{\theta_1}
   u_1(d),
   \label{appendix:omega_bar_adiabatic}
\end{aligned}\end{equation}
to the leading order in $\epsilon = \frac{5}{2}-d$. Here, $\bar g_{\theta}= \sum_tg^t_{\theta_1}$ is the sum of all effective Yukawa couplings in the theory 
with $
g^{t}_{\theta} = \frac{1}{N}\frac{\left|e^t_{\theta }\right|^{4/3}\left|X_{\theta}\right||\chi_{\vartheta^{-1}(\theta)}|^{1/3}}{v^{1/3}_{F,\theta }\KFtheta^{1/3}}$.
$X_{\theta} = \sin\left(\vartheta^{-1}(\theta)-\theta\right)$ and $\chi_{\varphi} = \sin\left(\varphi-\vartheta(\varphi)\right)\left[1-\frac{\mathbf{K}^{\prime\prime}_{F,\vartheta(\varphi)}}{\mathbf{K}_{F,\vartheta(\varphi)}}+\left(\frac{\mathbf{K}^{\prime}_{F,\vartheta(\varphi)}}{\mathbf{K}_{F,\vartheta(\varphi)}}\right)^2\right]+\cos\left(\varphi-\vartheta(\varphi)\right)\frac{\mathbf{K}^{\prime}_{F,\vartheta(\varphi)}}{\mathbf{K}_{F,\vartheta(\varphi)}}$
.  
$ u_1(d) = \frac{1}{3\sqrt{3}} \frac{1}{\beta_d^{\frac{1}{3}}}\frac{\Omega_{d-1}}{(2\pi)^{d-1}}\frac{\Gamma(\frac{d}{2})\Gamma\left(\frac{d-1}{3}\right)\Gamma\left(\frac{d-1}{2}\right)\Gamma\left(\frac{11-2d}{6}\right)}{\Gamma\left(\frac{1}{2}\right)\Gamma\left(\frac{d-1}{6}\right)\Gamma\left(\frac{5d-2}{6}\right)}$ is an $O(1)$ dimension-dependent constant.

\subsection{Cubic vertex correction}
\label{sec:CVC}

The vertex correction 
can be written as
\begin{equation}\begin{aligned}
   \Gamma_{t'}(\mathbf{k},\mathbf{q}) = &-
  \frac{3\times2}{3!}
   \frac{1}{N^{\frac{3}{2}}}
\int\frac{d\mathbf{L}d\delta }{(2\pi)^{d}}
\int_0^{2\pi}\frac{\KFthetadim d\theta}{2\pi}
   (-\edim^t_{\Theta\left(\theta_1,\vec{q}\right),\thetasq}(\varphi_1+\pi))(-\edim^{t'}_{\thetasq ,\theta}(\varphi))(-\edim^t_{\theta  ,\theta_1}(\varphi_1))\\&
   \times \left(\mathscr{T}_t\tilde G_{0}(\mathbf{l}+\mathbf{q})\mathscr{T}_{t'}\tilde G_{0}(\mathbf{l})\mathscr{T}_t\right)D^t_1(\mathbf{l}-\mathbf{k}),
   \label{yukawa_vertex_formal}
\end{aligned}\end{equation}
where $t$ is summed over upto terms that are regular.
The rescaling $\delta\rightarrow \delta/v_{F,\theta }$ gives
\begin{equation}\begin{aligned}
   \Gamma_{t'}(\mathbf{k},0) = \frac{
   v_{t,t'}
   \mathscr{T}_{t'}
   }{N^{\frac{3}{2}}}\int\frac{d\mathbf{L}d\theta}{(2\pi)^{d}}\frac{\KFthetadim }{v_{F,\theta }} \left|\boldsymbol{e}^t_{\theta_1,\theta}\left(\varphi_1\right)\right|^2
    \boldsymbol{e}^{t'}_{\theta,\theta}\left(\varphi\right)
   D_1^t(\mathbf{l}-\mathbf{k})\int \frac{d\delta}{2\pi}\frac{\left(\delta^2-|\mathbf{L}|^2\right)
   }{\left(|\mathbf{L}|^2+\delta^2\right)^2}
   \label{appendix:cubic_vertex_zeroq}
\end{aligned}\end{equation}
at ${\bf q}=0$ with 
\begin{equation}
    \begin{aligned}
        v_{t,t'} = 
        \begin{cases}
            \frac{d^2+d-4}{(3-d)(4-d)},&~t=t' = \{(\alpha,\beta)|d-1\leq \alpha<\beta \leq 2\},\\
        1,&~\text{otherwise}
        \end{cases}
        .
    \end{aligned}
\end{equation}
For $t=t' = \{(\alpha,\beta)|d-1\leq \alpha<\beta \leq 2\}$, 
we use 
$\tilde \gamma_\alpha\tilde\gamma_\beta \tilde \gamma_\mu \tilde \gamma_\nu \tilde \gamma_\alpha \tilde \gamma_\beta = 4(\delta_{\nu,\alpha}\delta_{\beta,\mu}-\delta_{\mu,\alpha}\delta_{\nu,\beta})\tilde\gamma_\alpha\tilde\gamma_\beta+2\delta_{\mu,\alpha}\tilde\gamma_\alpha\tilde \gamma_\nu-2\delta_{\nu,\alpha}\tilde\gamma_\alpha\tilde \gamma_\mu+2\delta_{\mu,\beta}\tilde\gamma_\beta\tilde\gamma_\nu-2\delta_{\nu,\beta}\tilde\gamma_\beta\tilde\gamma_\mu-\tilde\gamma_\mu\tilde\gamma_\nu$ 
to obtain
\begin{equation}
    \begin{aligned}
        \frac{1}{(3-d)(4-d)}\sum_{\alpha\neq\beta} \tilde \gamma_\alpha\tilde\gamma_\beta \tilde \gamma_\mu \tilde \gamma_\nu \tilde \gamma_\alpha \tilde \gamma_\beta = -\frac{d^2+d-4}{(3-d)(4-d)}\tilde\gamma_\mu\tilde\gamma_\nu
    \end{aligned}
\end{equation}
For other cases, it is straightforward to derive $v_{t,t'}$ from Eq. (\ref{appendix:gamma_tilde_clifford}).
Eq. (\ref{appendix:cubic_vertex_zeroq}) vanishes upon integration over $\delta$.

\subsection{Four fermion coupling}
\label{app:fourfermion}

The quantum corrections to the four-fermion vertex can be written as
\begin{equation}
    \begin{aligned}
        \delta S_4 &= 
         \int^{'}
        \prod_{i=1}^{4}d_{f}^{d+1} {\bf k}_i ~
       \Gamma_{\{j_i\};abcd }\left(\mathbf{k}_1, \mathbf{k}_2, \mathbf{k}_3, \mathbf{k}_4\right)
       ~~
    \tilde{\Psi}_{j_1;a}
    (\mathbf{k}_1)
    \Psi_{j_4;b}
    \left(\mathbf{k}_4  \right)
    \tilde{\Psi}_{j_2;c}\left(\mathbf{k}_2
\right)
    \Psi_{j_3;d}(\mathbf{k}_3)
    \delta(\mathbf{k}_1+\mathbf{k}_2-\mathbf{k}_3-\mathbf{k}_4).
    \end{aligned}
\end{equation}
The vertex function evaluated over the distribution of the external frequencies 
in \eq{eq:RG5} is
\begin{equation}
    \begin{aligned}
        \int \{d\tilde{\boldsymbol{K}_i}\}
     \Gamma_{\{j_i\};abcd}
     \left(\renm_1, \renm_2, \renm_3, \renm_4\right)\dist(\renf_1,\renf_4)\dist(\renf_2,\renf_3)
=
\sum_{\nu,s}
\Gamma^{(\nu,s)}_{\mu}\left(\theta_1,\theta_2,\vec{q};\mu\right)
\left(\tilde{I}^{(\nu)}_m\right)_{ab} 
\left(\tilde{I}^{(\nu)}_m\right)_{cd} 
T^{(\nu,s)}_{\left(\begin{smallmatrix}     j_1  & j_2      \\ j_4      & j_3       \end{smallmatrix}\right)}.
\label{eq:generalQC4}
    \end{aligned}
\end{equation}
Here, $\tilde{I}^{(\nu)}_m$ and 
$T^{(\nu,s)}_{\left(\begin{smallmatrix} j_1 & j_2 \\ j_4 & j_3 \end{smallmatrix}\right)}$ 
are defined in Eqs.~(\ref{4f_channels}) and (\ref{Flavour_Tensor}), respectively. 
The distribution of the external frequency is to be chosen so that 
the counter term, 
$\Gamma^{CT;(\nu,s)}_{\mu,\theta_1,\theta_2}\left(\vec{q}\right)= -\Gamma^{(\nu,s)}_{\mu}\left(\theta_1,\theta_2,\vec{q};\mu\right)$
takes a simple form.

\subsubsection{$\lambda^2$}

We first consider the quantum correction that arises from Fig. \ref{fish1},
\begin{equation}
    \begin{aligned}
        \Gamma^{(2)}_{\{j_i\};abcd}
       \left(\renm_1, \renm_2, \renm_3, \renm_4\right) &= 
        -
        \frac{4}{2!}T^{(\nu_1,s_1)}_{\left(\begin{smallmatrix}     j_1  & j_2^{\prime}      \\ j_4      & j_1^{\prime}       \end{smallmatrix}\right)}T^{(\nu_2,s_2)}_{\left(\begin{smallmatrix}     j_1^{\prime}  & j_2      \\ j_2^{\prime}      & j_3       \end{smallmatrix}\right)}\left(\tilde{I}^{(\nu_1)}_m\right)_{ab}\left(\tilde{I}^{(\nu_2)}_{m^{\prime}} \right)_{cd}
        \times
        \\ &
        \left\{- \int^{'}d^{d+1}_f\mathbf{l} 
        \left(-\boldsymbol{\lambda}^{(\nu_1,s_1)}_{\theta_1,\theta}\left(\vec{q}\right)\right)\left(-\boldsymbol{\lambda}^{(\nu_2,s_2)}_{\theta,\theta_2}\left(\vec{q}\right)\right)
       \mathrm{Tr}\left\{\tilde{I}^{(\nu_1)}_m \tilde G_0(\mathbf{l})\tilde{I}^{(\nu_2)}_{m^{\prime}} \tilde G_0(\mathbf{l+q})\right\}
       \right\},
        \label{appendix:S4_(2)}
    \end{aligned}
\end{equation}
where
$\mathbf{l} = \left(\mathbf{L},\delta,\theta\right)$ is the loop momentum 
and $\mathbf{l+q} = (
\mathbf{L}
+\renf_2-\renf_3
,\deltaq,\thetasq)$.
Using Eq. (\ref{Flavour_Tensor}),
we can average Eq. (\ref{appendix:S4_(2)}) over external frequencies with the distribution function defined in Eqs. (\ref{app:dist+_dist-}) and (\ref{eq:dist_minus})
as
\begin{equation}
    \begin{aligned}
 &      \int \{d\tilde{\boldsymbol{K}_i}\}
     \Gamma_{\{j_i\};abcd}
     \left(\renm_1, \renm_2, \renm_3, \renm_4\right)\dist(\renf_1,\renf_4)\dist(\renf_2,\renf_3)
       = \sum_{\nu,s} \left(\tilde{I}^{(\nu)}_m\right)_{ab} 
\left(\tilde{I}^{(\nu)}_m\right)_{cd} 
T^{(\nu,s)}_{\left(\begin{smallmatrix}     j_1  & j_2      \\ j_4      & j_3       \end{smallmatrix}\right)}
\\&
\hspace{3cm}
\times
\int \{d\tilde{\boldsymbol{K}_i}\}\Gamma^{(\nu,s)}_{\mu}\left(\theta_1,\theta_2,\vec{q};\frac{\left|\renf_2-\renf_3\right|}{2}\right)
\dist(\renf_1,\renf_4)\dist(\renf_2,\renf_3)
   \label{appendix:diagonal_nu1_nu_2_lambda2}
    \end{aligned}
\end{equation}
with $\Gamma^{(\nu,s)}_{\mu}\left(\theta_1,\theta_2,\vec{q};\frac{\left|\renf_2-\renf_3\right|}{2}\right)$ being the same as in Eq. (D60) of Ref. \cite{PhysRevB.110.155142}, except for the replacement of 2$\mu$ by $|\renf_2-\renf_3|$, as this is the frequency flowing into the quantum correction,
\begin{equation}
    \begin{aligned}
    \Gamma^{(\nu,s)}_{\mu}\left(\theta_1,\theta_2,\vec{q};\frac{\left|\renf_2-\renf_3\right|}{2}\right) = &
        M^{(\nu,s)}_{s_1,s_2}
        \bigintss_{-\frac{\pi}{2}}^{\frac{\pi}{2}}\frac{d\theta}{2\pi}\frac{\KFthetadim }{v_{F,\theta }}
       \boldsymbol{\lambda}^{\left(\nu,s_1\right)}_{\theta_1,\theta}\left(\vec{q}\right)\boldsymbol{\lambda}^{\left(\nu,s_2\right)}_{\theta,\theta_2}\left(\vec{q}\right)
       \left[-\frac{1}
       {4(2-d)}
       \frac{A_+^{(\nu)}(d) T_+(d)}{\left( \left(L_{\theta}\left(\vec{q}\right)\right)^2+|\renf_2-\renf_3|^2\right)^{\frac{2-d}{2}}}
       \right.\\&\left.
       +\frac{A_-^{(\nu)}(d) T_-(d)\mu^2}
       {2\left( \left(L_{\theta}\left(\vec{q}\right)\right)^2+|\renf_2-\renf_3|^{2}\right)^{\frac{4-d}{2}}}
       \right].
       \label{app:lambda2_quantum_correction_genfreq}
    \end{aligned}
\end{equation}
Here, the matrix that determines the mixing between different flavour channels is given by
\begin{equation}
    \begin{aligned}
       M^{(F_{\pm},d)}_{s_1,s_2} = 
        \begin{cases}
            N,&~
            (s_1, s_2) = (d,d)\\
            1,&~
            ~(s_1,s_2)=(e,d) ~~\text{or}~~(d,e)\\
            0,&~\text{otherwise}
        \end{cases}, & ~~~~~~
       M^{(F_{\pm},e)}_{s_1,s_2} = 
        \begin{cases}
            1,&~
            (s_1,s_2) = (e,e)\\
            0,&~\text{otherwise}
        \end{cases},\\
        M^{(P,d)}_{s_1,s_2} = 
        \begin{cases}
            1,&~
            (s_1,s_2)=(d,d) ~\text{or}~~(e,e)\\
            0,&~\text{otherwise}
        \end{cases}, & ~~~~~~
        M^{(P,e)}_{s_1,s_2} = 
        \begin{cases}
            1,&~
            (s_1,s_2)=(e,d) ~~\text{or}~~(d,e) \\
            0&~\text{otherwise}
        \end{cases},
    \end{aligned}
\end{equation}
and $A_{\pm}^{(\nu)}(d) = A_1^{(\nu)}(d)\pm A_2^{(\nu)}(d)$ with
\begin{equation}
    \begin{aligned}
        A_1^{(F_+)}(d) = 1,~~ A_2^{(F_+)}(d)  = -1,~~
        A_1^{(F_-)}(d) = \frac{3-d}{d-1},~~ A_2^{(F_-)}(d)  = -1,~~
         A_1^{(P)}(d) =   1,~~ A_2^{(P)}(d) = 1
        .
        \label{A1A2}
    \end{aligned}
\end{equation}
$
        T_+(d)  =
        \frac{
        4
        \Omega_{d-1}\Gamma\left(\frac{d+1}{2}\right)\Gamma\left(\frac{4-d}{2}\right)\Gamma^2\left(\frac{d}{2}\right)}{\Gamma\left(\frac{3}{2}\right)(2\pi)^{d-1}\Gamma(d)},
        T_-(d)  =  \frac{
        4
        \Omega_{d-1}\Gamma\left(\frac{d-1}{2}\right)\Gamma\left(\frac{4-d}{2}\right)\Gamma^2\left(\frac{d}{2}\right)}{\Gamma\left(\frac{3}{2}\right)(2\pi)^{d-1}\Gamma(d)}$ are $O(1)$ dimension dependent constants
and $L_{\theta}\left(\vec{q}\right)$ is given in Eq. (\ref{app:Lqvarphi}). 
The integration over $\renf_1$ and $\renf_4$ can be done using the normalization condition of the distribution function 
in Eqs. (\ref{app:dist+_dist-}) and (\ref{eq:dist_minus}).
The integrations over $\renf_2$ and $\renf_3$ result in 
the contribution to the beta functional\cite{PhysRevB.110.155142},
\begin{equation}
    \begin{aligned}
        \mu^{d-1}\frac{d~\Gamma^{CT;(2);(\nu,s)}_{\mu,\theta_1,\theta_2}\left(\vec{q}\right)}{d~\mathrm{ln}~\mu} = - M^{(\nu,s)}_{s_1,s_2}
        \bigintss^\prime
        \frac{d\theta}{2\pi\mu}\frac{\KFthetadim }{v_{F,\theta }}
        \lambda^{\left(\nu,s_1\right)}_{ \theta_1,\theta}\left(\vec{q}\right)\lambda^{\left(\nu,s_2\right)}_{ \theta,\theta_2}\left(\vec{q}\right)\left[\frac{A_+^{(\nu)}(d) T_+(d)}{\left(\left(\mathscr{L}_{\mu,\theta}\left(\vec{q}\right)\right)^2+4\right)^{\frac{4-d}{2}}}
        \right.\\\left.
        +\frac{A_-^{(\nu)}(d) T_-(d)\left(\left(\mathscr{L}_{\mu,\theta}\left(\vec{q}\right)\right)^2+2(d-2)\right)}{\left(\left(\mathscr{L}_{\mu,\theta}\left(\vec{q}\right)\right)^2+4\right)^{\frac{6-d}{2}}}\right],
        \label{(2)_betafunctional_contribution}
    \end{aligned}
\end{equation}
where $        \mathscr{L}_{\mu,\theta}\left(\vec{q}\right) = \frac{L_{\theta}\left(\vec{q}\right)}{\mu}$.
It is noted that
\eq{(2)_betafunctional_contribution} does not depend on the form of the distribution function as long as it is normalized.

\subsubsection{$\lambda^1$}
\label{app:lambda1}

Next, we consider the quantum corrections that are linear in the quartic coupling in Figs. \ref{vertex3_1} and \ref{vertex3_2},
\begin{equation}\begin{aligned}
  \Gamma^{(1)}_{\{j_i\};abcd}
       \left(\renm_1,
       \renm_2,
       \renm_3,
       \renm_4
       \right)
 &=
  -
  \frac{3\times2}{3!}
  \frac{1}{N}
  T^{(\nu,s)}_{\left(\begin{smallmatrix}     j_1  & j_2      \\ j_4     & j_3       \end{smallmatrix}\right)}
  \int d_{f}^{d+1}\mathbf{l}
  ~ 
  \left\{\left(-\lambdadim^{(\nu,s)}_{\theta_1,\theta}\left(\vec{q}\right) \right)\left(-\edim^{t}_{\Theta\left(\theta_2,\vec{q}\right),\thetasq}(\varphi_1+\pi)\right)\left(-\edim^{t}_{\theta,\theta_2}(\varphi_1)\right)
  \right.\\&\left.
  \times D_1^t\left(\mathbf{l}-
  \renm_3
  \right)\left(\tilde{I}^{(\nu)}_m\right)_{ab}\left(\mathscr{T}_t \tilde G_0(\mathbf{l+q})\tilde{I}^{(\nu)}_m \tilde G_0(\mathbf{l})\mathscr{T}_t\right)_{cd}
  \right\}.
\label{quantum_correction_(1)_expression}
\end{aligned}\end{equation}
Here, $\mathbf{l}
= \left(\mathbf{L},\delta,\theta\right)$ and $\int d_{f}^{d+1}\mathbf{l} = \int\frac{d^{d-1}\mathbf{L}}{(2\pi)^{d-1}} \int\frac{d\delta}{2\pi} 
\int_0^{2\pi}\frac{\KFthetadim d\theta}{2\pi}$\footnote{The angular integration should be over the entire Fermi surface in Eq. (D65) in Ref. \cite{PhysRevB.110.155142}}.
$\varphi_1$ is  the angle of boson with momenta $\vec l - \vec{
\tilde {k}}_3
$.
Finally, $\mathbf{l+q} = (\mathbf{L}
+\renf_2-\renf_3
,\deltaq,\thetasq)$. 
Let us define a rank-4 tensor as
\begin{equation}
    \begin{aligned}
        \left(\tilde{I}^{(\nu)}_m\right)_{ab}\left(\mathscr{T}_t \tilde G_0(\mathbf{l+q})\tilde{I}^{(\nu)}_m \tilde G_0(\mathbf{l})\mathscr{T}_t\right)_{cd} = \frac{\mathcal{N}^{(1);(\nu)}_{abcd}}{\mathcal{D}^{(1)}}.
 \label{appendix:vertex_(1)}
    \end{aligned}
\end{equation}
Here, $\mathcal{D}^{(1)} = i^2\left(|\mathbf{L}
+
\renf_2-\renf_3|^2+v_{F,\thetasq}^2\deltaq^2\right)\left(|\mathbf{L}|^2+v_{F,\theta }^2\delta^2\right)$.
With the replacement, 
\begin{equation}\begin{aligned}
\left(\tilde \Gamma_{j}\right)_{ab}
\left(\tilde \Gamma_{k}\right)_{cd}
    L_{j}
    (L+\tilde K_2-\tilde K_3)_{k} 
    \rightarrow 
    \frac{\delta_{jk}}{d-1}
   ({\tilde{\bf \Gamma}}_{ab} 
   \cdot
   {\tilde{\bf \Gamma}}_{cd} )
    \left[
    \mathbf{L}\cdot(\mathbf{L}+\renf_2-\renf_3)
    \right],
\label{appendix:non_vanishing_freqprod}
\end{aligned}\end{equation}
the singular-contribution in the numerator is obtained to be
\begin{equation}
    \begin{aligned}
        \mathcal{N}^{(1);(F+)}_{abcd} 
 &= \frac{1}{d-1}
        \left(i\mathbbm{1}\right)_{ab}
        (\mathscr{T}_t \tilde{\Gamma}_{k}(i\mathbbm{1})\tilde{\Gamma}_{k}\mathscr{T}_t)_{cd}
        \mathbf{L}\cdot(\mathbf{L}+\renf_2-\renf_3)
        +\left(i\mathbbm{1}\right)_{ab}\left(\mathscr{T}_t(i \mathbbm{1})\mathscr{T}_t\right)_{cd}v_{F,\theta}v_{F,\thetasq}\delta\deltaq \\
        \mathcal{N}^{(1);(F-)}_{abcd} &=\frac{1}{d-1}
\left(\tilde{\Gamma}_m\right)_{ab}
     \left(\mathscr{T}_t\tilde{\Gamma}_k \tilde{\Gamma}_m\tilde{\Gamma}_{k}\mathscr{T}_t\right)_{cd}
     \mathbf{L}\cdot(\mathbf{L}+\renf_2-\renf_3) +\left(\tilde{\Gamma}_m\right)_{ab}\left(\mathscr{T}_t \tilde{\Gamma}_m\mathscr{T}_t\right)_{cd}v_{F,\theta}v_{F,\thetasq}\delta\deltaq
     \\
\mathcal{N}^{(1);(P)}_{abcd} 
 &= \frac{1}{d-1}
        \left(\tilde{I}^{(P)}_m\right)_{ab}
        (\mathscr{T}_t \tilde{\Gamma}_{k}\tilde{I}^{(P)}_m\tilde{\Gamma}_{k}\mathscr{T}_t)_{cd}
        \mathbf{L}\cdot(\mathbf{L}+\renf_2-\renf_3)
        +\left(\tilde{I}^{(P)}_m\right)_{ab}\left(\mathscr{T}_t\tilde{I}^{(P)}_m\mathscr{T}_t\right)_{cd}v_{F,\theta}v_{F,\thetasq}\delta\deltaq.
        \label{appendix:numerator_linear}
    \end{aligned}
\end{equation}
Eq. (\ref{appendix:vertex_(1)}) can be simplified as
\begin{equation}
\begin{aligned}
 \frac{\mathcal{N}^{(1);(\nu)}_{abcd}}{\mathcal{D}^{(1)}} = 
 \begin{cases}
 \begin{aligned}
     -\frac{\mathbf{L}\cdot(\mathbf{L}+\renf_2-\renf_3)-v_{F,\theta}v_{F,\thetasq}\delta\deltaq}{(|\mathbf{L}+\renf_2-\renf_3|^{2}+v_{F,\thetasq }^2\deltaq^2)(|\mathbf{L}|^{2}+v_{F,\theta }^2\delta^2)}\left(i\mathbbm{1}\right)_{ab}\left(i\mathbbm{1}\right)_{cd}, &~\nu = F_+~\text{and}~\forall t,\\
     -\frac{\frac{3-d}{d-1}\mathbf{L}\cdot(\mathbf{L}+\renf_2-\renf_3)-v_{F,\theta}v_{F,\thetasq}\delta\deltaq}{(|\mathbf{L}+\renf_2-\renf_3|^{2}+v_{F,\thetasq }^2\deltaq^2)(|\mathbf{L}|^{2}+v_{F,\theta }^2\delta^2)}
     \tilde{\mathbf{\Gamma}}_{ab}\cdot\tilde{\mathbf{\Gamma}}_{cd},&~\nu = F_-~\text{and}~\forall t,
 \end{aligned}
     \\
     \nu = P:
     \begin{cases}
         \frac{\mathbf{L}\cdot(\mathbf{L}+\renf_2-\renf_3)+v_{F,\theta}v_{F,\thetasq}\delta\deltaq}{(|\mathbf{L}+\renf_2-\renf_3|^{2}+v_{F,\thetasq }^2\deltaq^2)(|\mathbf{L}|^{2}+v_{F,\theta }^2\delta^2)}\left(\tilde{I}_m^{(P)}\right)_{ab}\left(\tilde{I}_m^{(P)}\right)_{cd}, &~t = 1,\\
         -\frac{d}{4-d}\frac{\mathbf{L}\cdot(\mathbf{L}+\renf_2-\renf_3)+v_{F,\theta}v_{F,\thetasq}\delta\deltaq}{(|\mathbf{L}+\renf_2-\renf_3|^{2}+v_{F,\thetasq }^2\deltaq^2)(|\mathbf{L}|^{2}+v_{F,\theta }^2\delta^2)}\left(\tilde{I}_m^{(P)}\right)_{ab}\left(\tilde{I}_m^{(P)}\right)_{cd}, &~t = \{(\alpha,\beta)|d-1\leq\alpha,\beta\leq 2\}.\\
     \end{cases}
     
 \end{cases}
\label{appendix:(2)_kernel_nu4}
\end{aligned}
\end{equation}
It is straightforward to verify that Eq.~(\ref{appendix:(2)_kernel_nu4}) follows from Eq.~(\ref{appendix:numerator_linear}) for $\nu = F_{\pm}$ channels.

It is noted that the quantum correction to the forward-scattering channels does not depend on whether the critical boson is attractive or repulsive in the pairing channel because the particle-hole pair is scattered within a small patch.
In the pairing channel, the sign of the vertex correction depends on the nature of the boson.
For the repulsive boson, we use
    $
    \sum_{\alpha \neq \beta}(\tilde{\gamma}_m)_{ab} \left( \tilde{\gamma}_\alpha \tilde{\gamma}_\beta \tilde{\gamma}_m \tilde{\gamma}_\alpha \tilde{\gamma}_\beta \right)_{cd} =  (\sum_{\alpha \neq \beta})
    \frac{d}{4-d} (\tilde{\gamma}_m)_{ab} (\tilde{\gamma}_m)_{cd}$.
With this replacement and a shift $\mathbf{L}\rightarrow \mathbf{L}+\renf_3$, the quantum correction becomes
\begin{equation}
    \begin{aligned}
         \Gamma^{(1);(\nu,s)}_{\mu
        }\left(\theta_1,\theta_2,\vec{q};\mu\right) & = 
         -\frac{\snut}{N}\int d\renf_2 d\renf_3 \int \frac{d\mathbf{L}d\delta \KFthetadim  d\theta}{(2\pi)^{d+1}}\boldsymbol{\lambda}^{(\nu,s)}_{
        \theta_1,\theta}\left(\vec{q}\right)
        \left|\edim^{t}_{\Theta\left(\theta_2,\vec{q}\right),\thetasq}\left(\varphi_1\right)\right|~\left|\edim^{t}_{\theta,\theta_2}\left(\varphi_1\right)\right|
        \\
        &\times
         D_1^t\left(\left(\mathbf{L},\vec{l}-\vec{\tilde k}_3\right)
         \right)K^{(\nu)}_d\left(\mathbf{L}+\renf_3,
         \frac{\renf_3-\renf_2}{2},\delta,\theta,\vec{q}\right) \dist(\renf_2,\renf_3),
        \label{quantum_correction_(1)_explicit}
    \end{aligned}
\end{equation}
where
\begin{equation}
    \begin{aligned}
 K^{(\nu)}_d\left(\mathbf{L},\boldsymbol{\mu},\delta,\theta,\vec{q}\right)= \frac{A^{(\nu)}_1(d)\mathbf{L}\cdot(\mathbf{L}-2\boldsymbol{\mu})+A^{(\nu)}_2(d)v_{F,\theta }v_{F,\thetasq}\delta\deltaq}{(|\mathbf{L}-2\boldsymbol{\mu}|^{2}+v_{F,\thetasq}^2\deltaq^2)(|\mathbf{L}|^{2}+v_{F,\theta }^2\delta^2)}
\label{appendix:general_4fkernel} \end{aligned} \end{equation}
and 
\begin{equation}
    \begin{aligned}
        \snut = 
        \begin{cases}
            1
            ~~&\nu = F_{\pm},\\
            N^t_p(d)
            &\nu = P
        \end{cases}
        \label{appendix:s_nu}
    \end{aligned}
\end{equation}
with
\begin{equation}
    \begin{aligned}
    N^t_p(d)= 
        \begin{cases}
            -1,~~&t \in \{1\}\\
            \frac{d}{4-d},~~&t\in
             \{ (\alpha, \beta) |  d-1 \leq \alpha, \beta \leq 2 \}\\
            \end{cases}.
        \label{appendix:Npt}
    \end{aligned}
\end{equation}
The derivative of the counter term with respect to $\log \mu$,
which determines the beta functional, becomes
\begin{equation}
    \begin{aligned}
        \mu^{d-1}\frac{d~\Gamma^{CT;(1);(\nu,s)}_{\mu,\theta_1,\theta_2}\left(\vec{q}\right)}{d~\log\mu} &= -\frac{\snut}
        {
        4}
        \int_0^{2\pi} \frac{d\theta}{2\pi\mu}\frac{\KFthetadim }{v_{F,\theta }}\lambda^{(\nu,s)}_{
        \theta_1,\theta}\left(\vec{q}\right)\left[\frac{A_+^{(\nu)}(d) T_+(d)}{\left(\left(\mathscr{L}_{\mu,\theta}\left(\vec{q}\right)\right)^2+4\right)^{\frac{4-d}{2}}}\left(h^{t}_{+;\theta,\theta_2}\left(\vec{q};\mu\right)\right)^{\dagger}
        \right.\\&\left.+\frac{A_-^{(\nu)}(d) T_-(d)\left(\left(\mathscr{L}_{\mu,\theta}\left(\vec{q}\right)\right)^2+2(d-2)\right)}{\left(\left(\mathscr{L}_{\mu,\theta}\left(\vec{q}\right)\right)^2+4\right)^{\frac{6-d}{2}}}\left(h^{t}_{-;\theta,\theta_2}\left(\vec{q};\mu\right)\right)^{\dagger}\right].
        \label{appenidx:(1)_general_beta_functional}
    \end{aligned}
\end{equation}
Similarly, the contribution of Fig. \ref{vertex3_2} can be written as
\begin{equation}
    \begin{aligned}
        \mu^{d-1}\frac{d~\Gamma^{CT;\left(1^{\prime}\right);(\nu,s)}_{\mu,\theta_1,\theta_2}\left(\vec{q}\right)}{d~\log\mu} &= -\frac{\snut}{4}\int_0^{2\pi} \frac{d\theta}{2\pi\mu}\frac{\KFthetadim }{v_{F,\theta }}\left[h^t_{+;\theta_1,\theta}\left(\vec{q};\mu\right)\frac{A_+^{(\nu)}(d) T_+(d)}{\left(\left(\mathscr{L}_{\mu,\theta}\left(\vec{q}\right)\right)^2+4\right)^{\frac{4-d}{2}}}
        \right.\\&\left.+h^t_{-;\theta_1,\theta}\left(\vec{q};\mu\right)\frac{A_-^{(\nu)}(d) T_-(d)\left(\left(\mathscr{L}_{\mu,\theta}\left(\vec{q}\right)\right)^2+2(d-2)\right)}{\left(\left(\mathscr{L}_{\mu,\theta}\left(\vec{q}\right)\right)^2+4\right)^{\frac{6-d}{2}}}\right]\lambda^{(\nu,s)}_{
        \theta,\theta_2}\left(\vec{q}\right),
        \label{appenidx:(1')_general_beta_functional}
    \end{aligned}
\end{equation}
where the mixing matrices are defined as\footnote{The Yukawa couplings have been redefined in the same manner as was done in the calculation of fermion self-energy.}
\begin{equation}
    \begin{aligned}
h^t_{+;\theta_1,\theta_2}\left(\vec{q};\mu\right)
&= -2\frac{\left|\edim^{t}_{\theta_1,\theta_2}\right|^2\mu}
{N}
\frac
{\left(\left(\mathscr{L}_{\mu,\theta_2}\left(\vec{q}\right)\right)^2+4\right)^{\frac{4-d}{2}}} 
{ T_+(d)}
\int d\renf_2 d\renf_3
\int\frac{d\mathbf{L}dE }{(2\pi)^{d}}
        D^t_{1;\mu}\left(\mathbf{L},\theta_1,\theta_2\right)
        \\&\times\tilde{K}_+\left(\mathbf{L}+\renf_3,\frac{\renf_3-\renf_2}{2},E,\theta_2,\vec{q}\right)\partial_{\log\mu}\dist(\renf_2,\renf_3)
        , 
        \label{appendix:h1_def} 
    \end{aligned}
\end{equation}
 \begin{equation}
    \begin{aligned}
h^t_{-;\theta_1,\theta_2}\left(\vec{q};\mu\right)
&= -2\frac{\left|\edim^{t}_{\theta_1,\theta_2}\right|^2\mu}
{N}
\frac{\left(\left(\mathscr{L}_{\mu,\theta_2}\left(\vec{q}\right)\right)^2+4\right)^{\frac{6-d}{2}} }{T_-(d)\left(\left(\mathscr{L}_{\mu,\theta_2}\left(\vec{q}\right)\right)^2+2(d-2)\right)}
\int d\renf_2 d\renf_3
\int\frac{d\mathbf{L}dE }{(2\pi)^{d}}
D^t_{1;\mu}\left(\mathbf{L},\theta_1,\theta_2\right)
\\&\times
\tilde{K}_-\left(\mathbf{L}+\renf_3,\frac{\renf_3-\renf_2}{2},E,\theta_2,\vec{q}\right)
\partial_{\log\mu}\dist\left(\renf_2,\renf_3\right)
       \label{main:h2_def}
    \end{aligned}
\end{equation}
with  $E = v_{F,\theta}\delta $ being the energy of the virtual fermions and 
\begin{equation}
    \begin{aligned}
        \tilde{K}_\pm\left(\mathbf{L},\boldsymbol{\mu},E,\theta,\vec{q}\right)
        =\frac{\mathbf{L}\cdot(\mathbf{L}-2\boldsymbol{\mu})\pm E\left(E+L_{\theta}(\vec{q})\right)}{\left(|\mathbf{L}-2\boldsymbol{\mu}|^{2}+\left(E+L_{\theta}(\vec{q})\right)^2\right)\left(|\mathbf{L}|^{2}+E^2\right)}.
        \label{appendix:tildeK}
    \end{aligned}
\end{equation}
$L_{\theta}(\vec{q})$ is defined in Eq. (\ref{app:Lqvarphi}).

\begin{figure}[H]
	\begin{center}
\includegraphics[height=6.0cm,width=6.0cm]{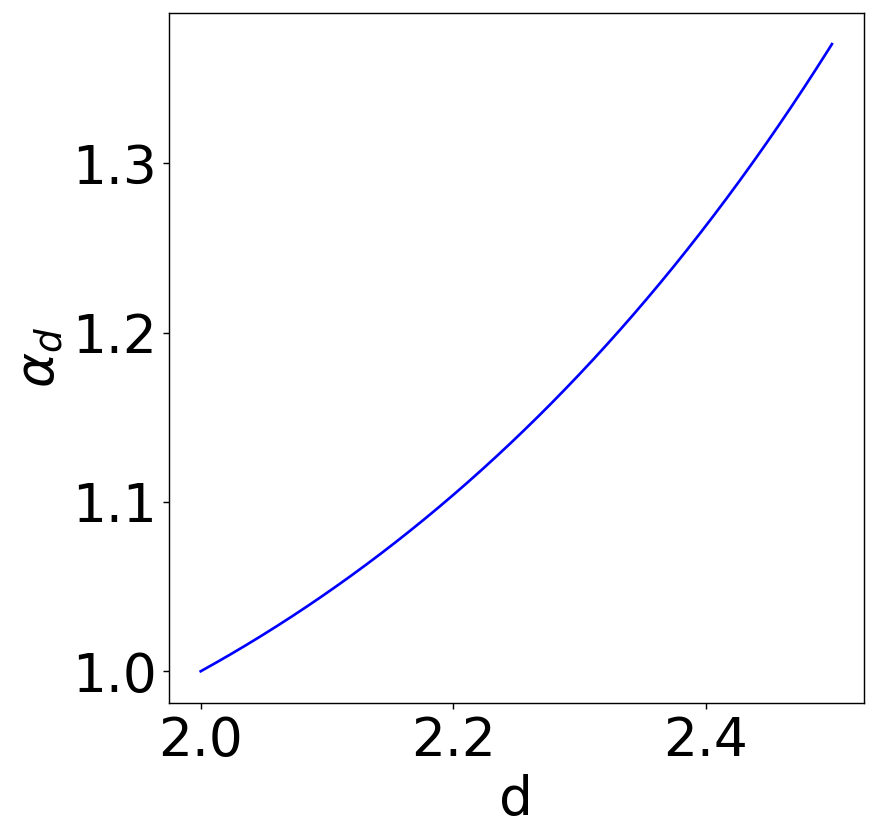}
	\end{center} 
	\caption{$\alpha_d$ as a function of dimension.
    }
    \label{fig:alphad}
\end{figure}

Now, we focus on the linear mixing matrices in the pairing channel at zero center of mass momentum, for which the contribution of $h^t_-$  vanishes due to Eq. (\ref{A1A2}). 
The mixing matrix that absorbs the angle-dependent phase space volume becomes
\begin{equation}
    \begin{aligned}
\tilde{h}_{\theta_1,\theta_2}\left(\mu\right) & = \sum_t\mathfrak{s}^{(P,t)}\sqrt{\frac{K_{F,\theta_1}K_{F,\theta_2}}{v_{F,\theta_1}v_{F,\theta_2}}}h^t_{+;\theta_1,\theta_2}\left(0;\mu\right) \\
&= -\frac{4}{R_d}\sum_t\mathfrak{s}^{(P,t)}\frac{\left|\tilde{\edim}^{t}_{\theta_1,\theta_2}\right|^2\mu}
{N}
\int\frac{d\mathbf{L}}{(2\pi)^{d-1}}
D^t_{1;\mu}\left(\mathbf{L},\theta_1,\theta_2\right)
 W(L,\mu),
        \label{appendix:htilde_def} 
    \end{aligned}
\end{equation}
where $h^t_+$ is defined in Eq. (\ref{appendix:h1_def}),
$\tilde{\edim}^{t}_{\theta_1,\theta_2} = \edim^{t}_{\theta_1,\theta_2}\left(\frac{K_{F,\theta_1}K_{F,\theta_2}}{v_{F,\theta_1}v_{F,\theta_2}}\right)^{\frac{1}{4}}$,
and
\begin{equation}
    \begin{aligned}
       W(L,\mu) = \partial_{\log\mu}\int d\renf_2 d\renf_3 \int\frac{dE }{2\pi}
       \frac{(\mathbf{L}+\renf_3)\cdot(\mathbf{L}+\renf_2)+E^2}{\left(|\mathbf{L}+\renf_2|^{2}+E^2\right)\left(|\mathbf{L}+\renf_3|^{2}+E^2\right)}\dist(\renf_2,\renf_3)
       \label{app:h_weight_dist}
    \end{aligned}
\end{equation}
with $L = | {\bf L}|$. 
Among many possible $\dist^+$, we choose the one in which the mixing matrix takes the same form as the boson propagator
regularized with an infrared cutoff that is proportional to $\mu$\cite{BORGES2023169221},
\bqa \tilde{h}_{\theta_1,\theta_2}\left(\mu\right) =\sum_t \mathfrak{s}^{(P,t)} \frac{\left|\tilde{e}^{t}_{\theta_1,\theta_2}\right|^2}
{N}\mu^2 D^t_{1;\mu}\left(\alpha_d^{\frac{1}{d-1}} \mu,\theta_1,\theta_2\right),
\label{eq:hD}
\eqa
where $\edim^t_{\theta_1,\theta_2} = \mu^{\frac{3-d}{2}}e^t_{\theta_1,\theta_2}$ and 
$\alpha_d$ is $O(1)$ number to be fixed later.
This can be achieved if we choose the distribution function that satisfies
\begin{equation}
 \begin{aligned}
 W(L,\mu) = -\frac{(2\pi)^{d-1}}{\Omega_{d-1}}\frac{R_d}{4}|\boldsymbol{\mu}|^{d-2}\int d\hat{\boldsymbol{\mu}}~\delta^{d-1}(\mathbf{L}-\alpha_d^{\frac{1}{d-1}}\boldsymbol{\mu}).
        \label{app:dist_delta}
    \end{aligned}
\end{equation}
To find the distribution
that satisfies \eq{app:dist_delta},
we first integrate over $E$ in Eq. (\ref{app:h_weight_dist}) to obtain
\begin{equation}
    \begin{aligned}
        W(L,\mu) =\frac{1}{2}\partial_{\log\mu}\int d\renf_2 d\renf_3 
       \fk
       \dist(\renf_2,\renf_3),
       \label{app:h_weight_dist_intE}
    \end{aligned}
\end{equation}
where
$\fk = \frac{1}{|\mathbf{L}+\renf_2|+|\mathbf{L}+\renf_3|}\left(\frac{(\mathbf{L}+\renf_3)\cdot(\mathbf{L}+\renf_2)}{|\mathbf{L}+\renf_3||\mathbf{L}+\renf_2|}+1\right)$.
We can rewrite \eq{app:h_weight_dist_intE} in Fourier space as 
\begin{equation}
    \begin{aligned}
        W(L,\mu) = \frac{1}{2}\partial_{\log\mu}\int d\renf_2 d\renf_3 \frac{d\renfc_2 d\renfc_3d\renfc^\prime_2 d\renfc^\prime_3}{(2\pi)^{4(d-1)}}
        e^{i\left(\renfc_2\cdot(\mathbf{L}+\renf_2)+\renfc_3\cdot(\mathbf{L}+\renf_3)\right)} \fkc^\prime\left(\renfc_2,\renfc_3\right)
        e^{i\left(\renfc^\prime_2\cdot\renf_2+\renfc^\prime_3\cdot\renf_3\right)} \dist^\prime(\renfc^\prime_2,\renfc^\prime_3).
    \end{aligned}
\end{equation}
Integrating over $\renf_2,\renf_3,\renfc^\prime_2$ and $\renfc^\prime_3$, we obtain 
\begin{equation}
    \begin{aligned}
        W(L,\mu) = \frac{(2\pi)^{-2d+2}}{2}\partial_{\log\mu}\int d\renfc_2 d\renfc_3
        e^{i\mathbf{L}\cdot \left(\renfc_2+\renfc_3\right)} \fkc^\prime\left(\renfc_2,\renfc_3\right)
        \dist^\prime(-\renfc_2,-\renfc_3).
    \end{aligned}
\end{equation}
A change of variables $\renfcs = \renfc_2+\renfc_3$, $\renfca = \renfc_2-\renfc_3$
leads to
\begin{equation}
    \begin{aligned}
        W(L,\mu) = \frac{2^{1-d}
        (2\pi)^{-2d+2}}{2}\partial_{\log\mu}\int d\renfca d\renfcs
        e^{i\mathbf{L}\cdot \renfcs} \tilde{\fkc}\left(\renfca,\renfcs\right)
        \tilde{\dist}^-(\renfca)\tilde{\dist}^+(\renfcs),
        \label{app:weight_simplified}
    \end{aligned}
\end{equation}
where $\tilde{\fkc}(\renfcs,\renfca) = \fkc^\prime(\frac{\renfcs+\renfca}{2},\frac{\renfcs-\renfca}{2})$, $\tilde{\dist}^-(\renfca) = \dist^{^\prime-}(-\renfca)$ and $\tilde{\dist}^+(\renfcs) = \dist^{^\prime+}(-\renfcs)$. $\dist^{^\prime+}(\renfcs)$ and $\dist^{-}(\renf_2-\renf_3)$ are defined as 
\begin{equation}
    \begin{aligned}
        \dist^{^\prime+}(\renfcs) = 2^{1-d}\int d(\renf_2+\renf_3) e^{-i\renfcs\cdot \frac{\renf_2+\renf_3}{2}}\dist^{+}(\renf_2+\renf_3),~~\dist^{^\prime-}(\renfca) = \int d(\renf_2-\renf_3) e^{-i\renfca\cdot \frac{\renf_2-\renf_3}{2}}\dist^{-}(\renf_2-\renf_3).
    \end{aligned}
\end{equation}
Rewriting the right hand side of Eq. (\ref{app:dist_delta}) as
\begin{equation}
    \begin{aligned}
        W(L,\mu)
        = -\frac{R_d}{4\Omega_{d-1}}|\boldsymbol{\mu}|^{d-2} \int d\hat{\boldsymbol{\mu}}~\int d\renfcs e^{i \renfcs\cdot(\mathbf{L}-\alpha_d^{\frac{1}{d-1}}\boldsymbol{\mu})},
        \label{app:delta_func_unpack}
    \end{aligned}
\end{equation}
and
equating Eqs. (\ref{app:weight_simplified}) and (\ref{app:delta_func_unpack}), we obtain
\begin{equation}
    \begin{aligned}
        \frac{
        2^{1-d}
        (2\pi)^{-2d+2}}{2}\partial_{\log \mu}\left(\tilde{\dist}^+(\renfcs)\int d\renfca
        \tilde{\fkc}\left(\renfca,\renfcs\right)
        \tilde{\dist}^-(\renfca)\right) =-\frac{R_d}{4\Omega_{d-1}}|\boldsymbol{\mu}|^{d-2}
        \int d\hat{\boldsymbol{\mu}}~
        e^{-i \renfcs\cdot\alpha_d^{\frac{1}{d-1}}\boldsymbol{\mu}}.
    \end{aligned}
\end{equation}
We use $\tilde{\dist}^-(\renfca) = \frac{1}{\Omega_{d-1}}\int d\hat{\boldsymbol{\mu}}~e^{i\renfca\cdot\boldsymbol{\mu}}$ and 
\begin{equation}
    \begin{aligned}
        \int d\renfca
        \tilde{\fkc}\left(\renfca,\renfcs\right)
        \tilde{\dist}^-(\renfca) = \frac{(2\pi)^{d-1}
        2^{d-1}}{\Omega_{d-1}}
        \int d\hat{\boldsymbol{\mu}}~
        \int d\renf^+ e^{-i\renfcs \cdot \renf^+} \frac{1}{|\renf^++\boldsymbol{\mu}|+|\renf^+-\boldsymbol{\mu}|}\left(\frac{(\renf^++\boldsymbol{\mu})\cdot(\renf^+-\boldsymbol{\mu})}{|\renf^++\boldsymbol{\mu}||\renf^+-\boldsymbol{\mu}|}+1\right)
    \end{aligned}
\end{equation}
to obtain 
\begin{equation}
    \begin{aligned}
        \tilde{\dist}^+(\renfcs) = \frac{R_d(2\pi)^{d-1}}{2}\frac{\int d\hat{\boldsymbol{\mu}^\prime}~\int^\infty_{\mu}d\mu^\prime\mu^{\prime(d-3)}e^{-i \renfcs\cdot\alpha_d^{\frac{1}{d-1}}\boldsymbol{\mu}^\prime}}{\int d\hat{\boldsymbol{\mu}}~\int d\renf^+ e^{-i\renfcs \cdot \renf^+} \frac{1}{|\renf^++\boldsymbol{\mu}|+|\renf^+-\boldsymbol{\mu}|}\left(\frac{(\renf^++\boldsymbol{\mu})\cdot(\renf^+-\boldsymbol{\mu})}{|\renf^++\boldsymbol{\mu}||\renf^+-\boldsymbol{\mu}|}+1\right)}.
        \label{app:symm_dist_explicit}
    \end{aligned}
\end{equation}
$\alpha_d$ in \eq{eq:generalsol} is fixed through the normalization condition,
$\tilde{\dist}^+(0) = 1$.
From
\begin{equation}
    \begin{aligned}
\lim_{{\bf T} \rightarrow 0 }        
\tilde{\dist}^+(\renfcs) = 
\frac{R_d(2\pi)^{d-1}}{2
}\alpha_d^{-\frac{d-2}{d-1}}
\lim_{{\bf T} \rightarrow 0 }        
\frac{\int d\hat{\boldsymbol{\mu}^\prime}~\int^\infty_{\alpha_d^{\frac{1}{d-1}}\mu}d\mu^\prime\mu^{\prime(d-3)}e^{-i \renfcs\cdot\boldsymbol{\mu}^\prime}}{\int d\hat{\boldsymbol{\mu}}~\int d\Omega_{d-1}\int_\mu^\infty d\tilde {K}^+
        \tilde {K}^{{}^+ d-3}e^{-i\renfcs \cdot \renf^+} }
=        
    \frac{R_d(2\pi)^{d-1}}{2\Omega_{d-1}}\alpha_d^{-\frac{d-2}{d-1}} =1,
        \label{app:G+_small_T}
    \end{aligned}
\end{equation}
we obtain\footnote{
As a consistency check, one can compute the weight of the linear mixing matrix as $\frac{R_d}{4}\int \frac{d\theta_2}{2\pi}\tilde{h}_{\theta_1,\theta_2}\left(\mu\right) =\bar g_{\theta_1}\rho_d $ with $\rho_d = \frac{1}{6\sqrt{3}}\frac{R_d}{(\alpha_d\beta_d)^{\frac{1}{3}}}\equiv\frac{1}{6\sqrt{3}}\left(\frac{2\Omega_{d-1}}{(2\pi)^{d-1}}\right)^{\frac{d-1}{3(d-2)}}\frac{R_d^{\frac{2d-5}{3(d-2)}}}{\beta_d^{\frac{1}{3}}} $. 
This is to be compared with $\rho_d$ used in Ref. \cite{PhysRevB.110.155142} in the pairing channel: $  \rho_d = \frac{1}{6\sqrt{3}}\frac{\Omega_{d-1}\Gamma\left(\frac{11-2d}{6}\right)\Gamma\left(\frac{d+1}{2}\right)}{(2\pi)^{d-1}\beta_d^{\frac{1}{3}}\Gamma\left(\frac{d-1}{6}\right)\Gamma\left(\frac{3}{2}\right)}
         \int_0^1\int_0^1 dx~dy~ y^{\frac{d-7}{6}}(1-y)^{\frac{1}{2}}
         \left(\{(2x-1)^{2}(-y^{2}+2y-1)-y+1\}^{\frac{2d-5}{6}}+\{-(2xy-2x+y+1)^2+3y+1\}^{\frac{2d-5}{6}}\right). 
$
These two expressions of $\rho_d$ agree at $d=5/2$ because both correspond to the coefficient of the logarithmically divergent term at the upper critical dimension: 
$\rho_{5/2} = \frac{1}{3\sqrt{3}}\frac{\Omega_{3/2}}{(2\pi)^{\frac{3}{2}}\beta_{5/2}^{\frac{1}{3}}}$.
} 
\begin{equation}
    \begin{aligned}
        \alpha_d = \left(\frac{R_d(2\pi)^{d-1}}{2\Omega_{d-1}}\right)^{\frac{d-1}{d-2}}.
        \label{app:alpha_d_new}
    \end{aligned}
\end{equation}
$\alpha_d$ is $O(1)$ number in $2 \leq d \leq 5/2$ as shown in Fig. \ref{fig:alphad}.

At $d=2$, we can find the distribution function in a closed form by noting that 
\begin{equation}
    \begin{aligned}
        \int d\renf^+ e^{-i\renfcs \cdot \renf^+} \frac{1}{|\renf^++\boldsymbol{\mu}|+|\renf^+-\boldsymbol{\mu}|}\left(\frac{(\renf^++\boldsymbol{\mu})\cdot(\renf^+-\boldsymbol{\mu})}{|\renf^++\boldsymbol{\mu}||\renf^+-\boldsymbol{\mu}|}+1\right) = -2\text{Ci}(T\mu),
        \label{app:fermion_integral_d2}
    \end{aligned}
\end{equation}
where Ci is cosine integral function. Using Eqs. (\ref{app:symm_dist_explicit}), (\ref{app:alpha_d_new}) and (\ref{app:fermion_integral_d2}), we obtain $\tilde{\dist}^+(\renfcs) = 1$, 
which in the frequency space becomes
\begin{equation}
    \begin{aligned}
         \dist^{+}(\renf_1+\renf_2) = \delta\left(\frac{\renf_1+\renf_2}{2}\right).
    \end{aligned}
\end{equation}
For $2<d<3$, we don't have the closed form expression for the distribution function.
However, it suffices to check that it is peaked at external frequencies that are $\order{\mu}$.
For $T\mu\ll 1$,
Eq. (\ref{app:G+_small_T}) gives
$\tilde{\dist}^+(\renfcs)\sim \frac{R_d(2\pi)^{d-1}}{2\Omega_{d-1}}\left(\alpha_d^{\frac{2-d}{d-1}}+ \frac{\alpha_d^{\frac{2-d}{d-1}}-1}{\Gamma(d-1)}|T\mu|^{d-2}\right)$.
For $|\renf_2+\renf_3|\gg \mu$, the distribution decays as
\begin{equation}
    \begin{aligned}
        \dist^{+}(\renf_1+\renf_2)\sim
        \mu^{1-d}\left|\frac{\mu}{\renf_1+\renf_2}\right|^{2d-3}
        .
    \end{aligned}
\end{equation}
As expected, the distribution decays faster than $\mu/|\renf|$ at large frequencies in $d \geq 2$, and the main contribution arises from $|\renf| \leq \mu$.

The contribution to the beta functional becomes
\begin{equation}
    \begin{aligned}
        \mu^{d-1}\sqrt{\frac{K_{F,\theta_1}K_{F,\theta_2}}{v_{F,\theta_1}v_{F,\theta_2}}}\frac{d~\left(\Gamma^{CT;(1);(P,d)}_{\mu,\theta_1,\theta_2}\left(0\right)\pm\Gamma^{CT;(1);(P,e)}_{\mu,\theta_1,\theta_2}\left(0\right)\right)}{d~\log\mu} &= -\frac{R_d}{4}\int_0^{2\pi} \frac{d\theta}{2\pi}\tilde V^{\pm}_{
        \theta_1,\theta}
        \tilde{h}_{\theta,\theta_2}\left(\mu\right)\equiv
         -\frac{R_d}{4}\int^\prime \frac{d\theta}{2\pi}\tilde V^{\pm}_{
        \theta_1,\theta}
        \tilde{h}^\pm_{\theta,\theta_2}\left(\mu\right),
        \label{appendix:}
    \end{aligned}
\end{equation}
where 
\begin{equation}
    \begin{aligned}
        \tilde{h}^\pm_{\theta_1,\theta_2}\left(\mu\right) =
        \tilde{h}_{\theta_1,\theta_2}\left(\mu\right)\mp \tilde{h}_{\theta_1+\pi,\theta_2}\left(\mu\right).
        \label{app:hpm}
    \end{aligned}
\end{equation}
Here, we have broken down the angular integration range into half and then used the fact that $\tilde V^\pm_{\theta_1,\theta_2+\pi} = \mp \tilde V^\pm_{\theta_1,\theta_2}$.

\subsubsection{$\lambda^0$}

The quantum correction from Fig. \ref{ladder1} reads\footnote{The angular integration associated with momenta $\mathbf{k}_1$ and $\mathbf{l}$ in Eqs. (D103) and (D107) in Ref. \cite{PhysRevB.110.155142} should be over the entire Fermi surface. }
\begin{equation}\begin{aligned}
   \delta S^{(0)}_4 
    = &-\frac{6\times2}{4!}
    \frac{1}{N^2}
\int
d_{f}^{d+1} {\bf k}_1
\int^\prime
d_{f}^{d+1} {\bf k}_3
\int
d_{b}^{d+1} {\bf q}
\int
d^{d+1}_{f}\mathbf{l}~
    \left(-\edim^{t}_{\theta_1,\theta}(\varphi_1)\right)\left(-\edim^{t'}_{\theta,\theta_2}(\varphi_2)\right)\left(-\edim^{t'}_{\Theta\left(\theta_2,\vec{q}\right),\thetasq}(\varphi_2+\pi)\right)
    \\&\times
    \left(-\edim^{t}_{\thetasq,\Theta\left(\theta_1,\vec{q}\right)}(\varphi_1+\pi)\right)
    D^{t'}_1\left(\mathbf{l}-\mathbf{k}_3\right)  D^t_1\left(\mathbf{l}-\mathbf{k}_1\right)\left(\mathscr{T}_t\tilde G_0(\mathbf{l})\mathscr{T}_{t^\prime}\right)_{ad}
    \left(\mathscr{T}_{t^\prime} \tilde G_0(\mathbf{l+q})\mathscr{T}_t\right)_{cb}
    \\&\times
    \tilde \Psi_{j_1;a}
    (\mathbf{k}_1)
     \Psi_{j_1;d}(\mathbf{k}_3)
     \tilde \Psi_{j_2;c}\left(\mathbf{k}_3
     + {\bf q}\right)
    \Psi_{j_2;b}
    \left(\mathbf{k}_1+{\bf q}  \right).
    \label{(0)_mu1_formal}
\end{aligned}\end{equation}
It is noted that the angles of the outgoing and virtual fermion can lie anywhere on the Fermi surface in the extended domain.
$\varphi_1$ and $\varphi_2$ are the the angles of bosons with momenta $\vec l - \vec k_1$ and $\vec l - \vec k_3$, respectively. Let us define $\left(\mathscr{T}_t\tilde G_0(\mathbf{l})\mathscr{T}_{t^\prime}\right)_{ad}
    \left(\mathscr{T}_{t^\prime} \tilde G_0(\mathbf{l+q})\mathscr{T}_t\right)_{cb} = \frac{\mathcal{N}^{(0)}_{adcb;(t,t')}}{\mathcal{D}^{(0)}}$ with $\mathcal{D}^{(0)} = i^2\left(|\mathbf{L}+\mathbf{Q}|^2+v_{F,\thetasq}^2\deltaq^2\right)
    \left(|\mathbf{L}|^2+v_{F,\theta }^2\delta^2\right)$ and the singular contribution in the numerator
\begin{equation}\begin{aligned}
    \mathcal{N}^{(0)}_{adcb;(t,t')} = \frac{1}{d-1}\mathbf{L} \cdot \left(\mathbf{L}+\mathbf{Q}\right)(\mathscr{T}_t\tilde\Gamma_n\mathscr{T}_{t^\prime})_{ad}(\mathscr{T}_{t^\prime}\tilde\Gamma_n\mathscr{T}_t)_{cb} + v_{F,\theta }v_{F,\thetasq}\delta\deltaq \left(\mathscr{T}_t\mathscr{T}_{t^\prime}\right)_{ad}\left(\mathscr{T}_{t^\prime}\mathscr{T}_t\right)_{cb}.
    \label{(0)_tensors}
\end{aligned}\end{equation}
Using the Fierz transformation \cite{PhysRevB.110.155142}, we obtain
\begin{equation}
    \begin{aligned}
       \mathcal{N}^{(0)}_{adcb;(t,t')} = 
        \left\{\frac{1}{d-1}\mathbf{L} \cdot \left(\mathbf{L}+\mathbf{Q}\right)C^{(1)}_{\nu;(t,t')}+v_{F,\theta }v_{F,\thetasq}\delta\deltaq C^{(2)}_{\nu;(t,t')} \right\}\left(\tilde I^{(\nu)}_m\right)_{ab}\left(\tilde I^{(\nu)}_m\right)_{cd},
        \label{app:Fierz_transformed_numerator}
    \end{aligned}
\end{equation}
where
\begin{equation}
    \begin{aligned}
        C^{(1)}_{\nu;(t,t')} &=  \mathscr{N}_{\nu}^{-2}\Tr\Bigg\{\mathscr{T}_t\tilde{\Gamma}_n\mathscr{T}_{t^\prime}\left(\tilde I^{\left(\nu\right)}_{m^{\prime}}\right)^{\dagger}\mathscr{T}_{t^\prime}\tilde\Gamma_n\mathscr{T}_t\left(\tilde I^{\left(\nu\right)}_{m^{\prime}}\right)^{\dagger}\Bigg\}, ~~C^{(2)}_{\nu;(t,t')} &=  \mathscr{N}_{\nu}^{-2}\Tr\Bigg\{\mathscr{T}_t\mathscr{T}_{t^\prime}\left(\tilde I^{\left(\nu\right)}_{m^{\prime}}\right)^{\dagger}\mathscr{T}_{t^\prime}\mathscr{T}_t\left(\tilde I^{\left(\nu\right)}_{m^{\prime}}\right)^{\dagger}\Bigg\}
        \label{appendix:Fierz_coeffs}
    \end{aligned}
\end{equation}
are the Fierz coefficients. To obtain Eq. (\ref{appendix:Fierz_coeffs}), we multiply $\left(\tilde I^{\left(\nu^\prime\right)}_{m^{\prime}}\right)^{\dagger}_{ba}\left(\tilde I^{\left(\nu^\prime\right)}_{m^{\prime}}\right)^{\dagger}_{dc}$, and sum over $m^\prime$ and spinor indices $a,b,c,d$ on the right-hand sides of Eqs. (\ref{(0)_tensors}) and (\ref{app:Fierz_transformed_numerator}). Then we can use the fact that $\sum_{m,m'}\left(\Tr\left\{\left(\tilde I^{\left(\nu\right)}_{m}\right)^{\dagger}\left(\tilde I^{\left(\nu^\prime\right)}_{m^{\prime}}\right)\right\}\right)^2 = \mathscr{N}_\nu^2\delta_{\nu\nu^\prime}$ with
$\mathscr{N}_\nu = \left(
 2,2\sqrt{d-1},
2\sqrt{4-d}\right)$ to simplify the right-hand side of Eq. (\ref{app:Fierz_transformed_numerator}). Solving for $C^{(1)}_{\nu;(t,t')}$ and $C^{(2)}_{\nu;(t,t')}$ yields Eq. (\ref{appendix:Fierz_coeffs}).

    The coefficients $C^{(1)}_{\nu;(t,t')}$ and $C^{(2)}_{\nu;(t,t')}$
    can be explicitly calculated as 
\begin{equation}
    \begin{aligned}
        C^{(1)}_{\nu;(t,t')} =&
          \frac{1}{2}\begin{cases}
          (d-1),~~&\nu = F+,~~\forall t,t'\\
           (3-d),~~&\nu = F-,~~\forall t,t'\\
           (d-1)N^t_p(d)N^{t'}_p(d),~~&\nu =P
        \end{cases},\\
        C^{(2)}_{\nu;(t,t')} =&
          \frac{1}{2}\begin{cases}
          -1,~~&\nu = F+,~~\forall t,t'\\
           -1,~~&\nu = F-,~~\forall t,t'\\
           N^t_p(d)N^{t'}_p(d),~~&\nu =P
        \end{cases},
        \label{appendix:Fierz_coeffs_explicit}
    \end{aligned}
\end{equation}
where $N^t_p(d)$ is given in Eq. (\ref{appendix:Npt}). 
To derive the Fierz coefficient, we first note from the cyclic property of the trace operation that $C^{(i)}_{\nu;(t,t')} = C^{(i)}_{\nu;(t',t)}$.  
These coefficients can be obtained straightforwardly in the $\nu = F_{\pm}$ channels 
$
\forall
t, t' 
$, as well as in the $\nu = P$ channel for $t, t' \in \{1\}$ and $t \neq t'$.  
To evaluate them in the pairing channel for $t, t' \in \{ (\alpha,\beta) \,|\, d-1 \leq \alpha,\beta \leq 2 \}$, we use the following steps.
The quantum correction has ten gamma matrices (two from the fermion propagators and eight from four fermion-boson vertices):
$\Tr\{
\tilde{\gamma}_\mu \tilde{\gamma}_\nu \tilde{\gamma}_\alpha \tilde{\gamma}_\beta \tilde{\gamma}_m
\tilde{\gamma}_\alpha \tilde{\gamma}_\beta \tilde{\gamma}_\mu \tilde{\gamma}_\nu \tilde{\gamma}_m
\}$,
where $m$ is summed over.  
Repeated use of the anticommutation relations for gamma matrices yields
\begin{equation}
\begin{aligned}
\Tr\{
\tilde{\gamma}_\mu \tilde{\gamma}_\nu \tilde{\gamma}_\alpha \tilde{\gamma}_\beta \tilde{\gamma}_m
\tilde{\gamma}_\alpha \tilde{\gamma}_\beta \tilde{\gamma}_\mu \tilde{\gamma}_\nu \tilde{\gamma}_m
\}
=&\, 2\big[
 -\Tr\{\tilde{\gamma}_\mu \tilde{\gamma}_\nu \tilde{\gamma}_\alpha \tilde{\gamma}_\beta \tilde{\gamma}_\beta \tilde{\gamma}_\mu \tilde{\gamma}_\nu \tilde{\gamma}_\alpha\} 
 +\Tr\{\tilde{\gamma}_\mu \tilde{\gamma}_\nu \tilde{\gamma}_\alpha \tilde{\gamma}_\beta \tilde{\gamma}_\alpha \tilde{\gamma}_\mu \tilde{\gamma}_\nu \tilde{\gamma}_\beta\} \\
&\quad
 -\Tr\{\tilde{\gamma}_\mu \tilde{\gamma}_\nu \tilde{\gamma}_\alpha \tilde{\gamma}_\beta \tilde{\gamma}_\alpha \tilde{\gamma}_\beta \tilde{\gamma}_\nu \tilde{\gamma}_\mu\} 
 +\Tr\{\tilde{\gamma}_\mu \tilde{\gamma}_\nu \tilde{\gamma}_\alpha \tilde{\gamma}_\beta \tilde{\gamma}_\alpha \tilde{\gamma}_\beta \tilde{\gamma}_\mu \tilde{\gamma}_\nu\}
\big] \\
&\quad -(4-d)\,\Tr\{\tilde{\gamma}_\mu \tilde{\gamma}_\nu \tilde{\gamma}_\alpha \tilde{\gamma}_\beta \tilde{\gamma}_\alpha \tilde{\gamma}_\beta \tilde{\gamma}_\mu \tilde{\gamma}_\nu\}.
\label{appendix:big_trace_source}
\end{aligned}
\end{equation}
The first trace reduces to $-\Tr\{\tilde{\gamma}_\mu \tilde{\gamma}_\nu \tilde{\gamma}_\alpha \tilde{\gamma}_\mu \tilde{\gamma}_\nu \tilde{\gamma}_\alpha\}$.  
To evaluate it, we make the replacement as an average over all allowed index permutations in the same sprit as was done for other quantum corrections:
\begin{equation}
\frac{1}{(3-d)^2(4-d)^2} \sum_{\mu \neq \nu, \, \alpha \neq \beta}
\Tr\{\tilde{\gamma}_\mu \tilde{\gamma}_\nu \tilde{\gamma}_\alpha \tilde{\gamma}_\mu \tilde{\gamma}_\nu \tilde{\gamma}_\alpha\}
= \frac{8}{4-d} - 2.
\label{appendix:averaged_trace_pairing_repulsive}
\end{equation}

The second trace in Eq.~\eqref{appendix:big_trace_source} gives the same magnitude as Eq.~\eqref{appendix:averaged_trace_pairing_repulsive} but with opposite sign.  
The fourth and fifth traces each evaluate to $2$, while the third trace is $-2$.  
Collecting all results, one can simplify Eq.~\eqref{appendix:big_trace_source} to
\begin{equation}
\frac{1}{(3-d)^2(4-d)^2} \sum_{\mu \neq \nu, \, \alpha \neq \beta}\Tr\{
\tilde{\gamma}_\mu \tilde{\gamma}_\nu \tilde{\gamma}_\alpha \tilde{\gamma}_\beta \tilde{\gamma}_m
\tilde{\gamma}_\alpha \tilde{\gamma}_\beta \tilde{\gamma}_\mu \tilde{\gamma}_\nu \tilde{\gamma}_m
\}
= -\frac{2d^2}{4-d}.
\end{equation}
Using Eqs. (\ref{(0)_tensors}) - (\ref{appendix:Fierz_coeffs_explicit}),
Eq. (\ref{(0)_mu1_formal}) now becomes
\begin{equation}\begin{aligned}
   \delta S^{(0)}_4 
    = &-
    \sum_{\nu}
    \frac{
    \snut\mathfrak{s}^{(\nu,t')}
    }{4N^2}
    \int
    d_{f}^{d+1} {\bf k}_1
    \int^\prime
    d_{f}^{d+1} {\bf k}_3
    \int
    d_{b}^{d+1} {\bf q}
    \int
    d_{f}^{d+1}
    \mathbf{l}~
    \\ &
    \left\{
    \left|\edim^{t}_{\theta_1,\theta}(\varphi_1)\right|\left|\edim^{t'}_{\theta,\theta_2}(\varphi_2)\right|\left|\edim^{t'}_{\Theta\left(\theta_2,\vec{q}\right),\thetasq}(\varphi_2)\right|\left|\edim^{t}_{\thetasq,\Theta\left(\theta_1,\vec{q}\right)}(\varphi_1)\right|\right.\\&\left.
    \times K^{(\nu)}_d\left(\mathbf{L},-\mathbf{Q}/2,\delta,\theta,\vec{q}\right) 
    D^{t'}_1\left(\mathbf{l}-\mathbf{k}_3\right) D^t_1\left(\mathbf{l}-\mathbf{k}_1\right)\left(\tilde{I}_m^{\left(\nu\right)}\right)_{ab}\left(\tilde{I}_m^{\left(\nu\right)}\right)_{cd}+\text{reg. terms}\right\}
    \\ &
    \times
    \tilde \Psi_{j_1;a}\left(\mathbf{k}_1\right)
    \Psi_{j_2;b}
    \left(\mathbf{k}_1+{\bf q}  \right)
    \tilde \Psi_{j_2;c}\left(\mathbf{k}_3+{\bf q}\right)
    \Psi_{j_1;d}\left(\mathbf{k}_3\right),
    \label{(0)_mu1_formal_postFierz}
\end{aligned}\end{equation}
where $K^{(\nu)}_d\left(\mathbf{L},\boldsymbol{\mu},\delta,\theta,\vec{q}\right)$ is defined in Eq. (\ref{appendix:general_4fkernel}).
The contribution to the beta function then reads 
\begin{equation}
    \begin{aligned}
&\mu^{d-1}\frac{d~\Gamma^{CT;\left(0\right);(\nu,s)}_{\mu,\theta_1,\theta_2}\left(\vec{q}\right)}{d~\log\mu}=
\snut\mathfrak{s}^{(\nu,t')}
        \frac{\delta_{s,s_{\nu}}}{
        8
        }
\mu^{6-d}   
\int_0^{2\pi} \frac{ d\theta}{2\pi\mu}\frac{\KFthetadim }{v_{F,\theta}}
   \frac{\left|e^t_{\theta_1,\theta}\right|^2
   \left|e^{t'}_{\theta,\theta_2}\right|^2
   }{N^2}
   \int \{d\renf_i\}
        \int\frac{d\mathbf{L}dE }{(2\pi)^{d}}
        \sum_{i=\pm}A_i^{(\nu)}(d)  
        \\&\times
D^t_{1;\mu}\left(\mathbf{L}-\renf_1,\theta_1,\theta\right)
\tilde{K}_i\left(\mathbf{L},\frac{\renf_1-\renf_4}{2},E,\theta,\vec{q}\right) 
D^{t^\prime}_{1;\mu}\left(\mathbf{L}-\renf_3,\theta,\theta_2\right)
\partial_{\log\mu}\left(\dist(\renf_2,\renf_3)\dist(\renf_1,\renf_4)\right)
        ,
        \label{appendix:(0)_general_beta_functional}
    \end{aligned}
\end{equation}
where 
$ s_{F_\pm}=e $,
$ s_{P}= d $ and $\tilde{K}_\pm\left(\mathbf{L},\boldsymbol{\mu},E,\theta,\vec{q}\right)$ is defined in Eq. (\ref{appendix:tildeK}).

In the pairing channel,
Eq. (\ref{appendix:(0)_general_beta_functional}) gives rise to the source,
\begin{equation}
    \begin{aligned}
&\tilde{S}_{\theta_1,\theta_2}\left(\mu\right)
= \mu^{d-1}\sqrt{\frac{K_{F,\theta_1}K_{F,\theta_2}}{v_{F,\theta_1}v_{F,\theta_2}}}  \frac{d~\Gamma^{CT;\left(0\right);(P,d)}_{\mu,\theta_1,\theta_2}\left(0\right)}{d~\log\mu}=
\sum_{t,t'}
\frac{\mathfrak{s}^{(P,t)}\mathfrak{s}^{(P,t')}}{4}
\mu^{6-d}
    \partial_{\log\mu}
    \left[
\int_0^{2\pi} \frac{ d\theta}
{2\pi}
   \frac{\left|\tilde{e}^t_{\theta_1,\theta}\right|^2
   \left|\tilde{e}^{t'}_{\theta,\theta_2}\right|^2
   }{N^2}
   \right. \\ & \left. \times
  \int\frac{d\mathbf{L}dE }{(2\pi)^{d}} \int \{d\renf_i\}
D^t_{1;\mu}\left(\mathbf{L}-\renf_1,\theta_1,\theta\right)
\tilde{K}_+\left(\mathbf{L},\frac{\renf_1-\renf_4}{2},E\right) 
D^{t^\prime}_{1;\mu}\left(\mathbf{L}-\renf_3,\theta,\theta_2\right)
\dist(\renf_2,\renf_3)\dist(\renf_1,\renf_4)\right].
 \end{aligned}
\label{appendix:source}
\end{equation} 
In the inter-patch limit, $q(\theta_1,\theta_2)\gg \sqrt{\mu\KFAVdim}$, 
and the loop variable $\theta$ can be close to either $\theta_1$ or $\theta_2$. 
For example, when $\theta$ is close to $\theta_1$,
one of the boson propagators in Eq. (\ref{appendix:source}) can be factored out of the frequency integrations and 
we can shift $\mathbf{L} \rightarrow\mathbf{L}+\renf_1$ followed by trivial integrations over $\renf_2$ and $\renf_3$. 
Then, one can perform the integration over $\renf_1, \renf_4$ and $\mathbf{L}$ using Eq. (\ref{app:dist_delta}) to obtain 
\begin{equation}
    \begin{aligned}
        \tilde{S}_{\theta_1,\theta_2}\left(\mu\right) =-\frac{R_d}{16}\sum_{t^\prime}\mathfrak{s}^{(P,t')}\frac{\left|\tilde{e}^{t'}_{\theta_1,\theta_2}\right|^2
   }{N}\frac{\mu^2}{q(\theta_1,\theta_2)^2}\int_0^{2\pi} \frac{d\theta}{2\pi}\tilde{h}_{\theta_1,\theta}(\mu) .
   \label{app:interpatch_1}
    \end{aligned}
\end{equation}
Similarly, when $\theta\sim \theta_2$, we can shift $\mathbf{L}\rightarrow \mathbf{L}+\renf_3$ to obtain
\begin{equation}
    \begin{aligned}
        \tilde{S}_{\theta_1,\theta_2}\left(\mu\right) =-\frac{R_d}{16}\sum_{t}\mathfrak{s}^{(P,t)}\frac{\left|\tilde{e}^{t}_{\theta_1,\theta_2}\right|^2
   }{N}\frac{\mu^2}{q(\theta_1,\theta_2)^2}\int_0^{2\pi} \frac{d\theta}{2\pi}\tilde{h}_{\theta,\theta_2}(\mu) .
   \label{app:interpatch_2}
    \end{aligned}
\end{equation}
In general, the source can be written as 
\begin{equation}
    \begin{aligned}
    \tilde{S}_{\theta_1,\theta_2}\left(\mu\right)
    &= -\frac{R_d}{16}\int_0^{2\pi} \frac{d\theta}{2\pi}\tilde{h}_{\theta_1,\theta}(\mu)\tilde{h}_{\theta,\theta_2}(\mu)
    +\delta \tilde S_{\theta_1,\theta_2}(\mu),
    \label{appendix:source_final}
    \end{aligned}
\end{equation}
where $\delta \tilde S_{\theta_1,\theta_2}(\mu)$ is the correction that arises for 
 $q(\theta_1,\theta_2) \sim \sqrt{\mu\KFAVdim}$.

To project Eq. (\ref{appendix:source_final}) into symmetric or antisymmetric channels, we first write down,
\begin{equation}\begin{aligned}
   \delta \Gamma_4^{(0)}
    = 
    \int \frac{d\boldsymbol{\mu}}{(2\pi)^{d-1}}
    \int^\prime \frac{d\theta_2}{2\pi}
    \int_{0}^{2\pi} \frac{d\theta_1}{2\pi}\left\{\left[-\frac{R_d}{16}\int_{0}^{2\pi} \frac{d\theta}{2\pi}
    \tilde{h}_{\theta_1,\theta}(\mu)\tilde{h}_{\theta,\theta_2}(\mu)\right]+
    \delta \tilde S_{\theta_1,\theta_2}(\mu)\right\}
    \left(
    o^{+}(\theta_1)o^{+}(\theta_2)+ o^{-}(\theta_1)o^{-}(\theta_2)\right),
    \label{app:source_external_sum}
\end{aligned}\end{equation}
where $o^\pm(\theta_1)o^\pm(\theta_2) =\tilde T^{(\pm)}_{\left(\begin{smallmatrix}     j_1  & j_2      \\ j_4      & j_3       \end{smallmatrix}\right)}
\Psi_{j_1}\left(\theta_1\right)
    \tilde{I}_m^{\left(P\right)}
    \Psi_{j_4}
    \left(\theta_1\right)\tilde \Psi_{j_2}\left(\theta_2\right)
    \tilde{I}_m^{\left(P\right)}\Psi_{j_3}\left(\theta_2\right)$ with $\tilde T^{\pm}$ being the tensors that are symmetric or antisymmetric for the flavors of outgoing or incoming fermions in the component form\footnote{The quantum correction independent of four-fermion coupling is only generated in the direct channel and corresponding flavor channel can be decomposed as: $T^{(P,d)}_{\left(\begin{smallmatrix}     j_1  & j_2      \\ j_4      & j_3       \end{smallmatrix}\right)} =\tilde T^{(+)}_{\left(\begin{smallmatrix}     j_1  & j_2      \\ j_4      & j_3       \end{smallmatrix}\right)}+\tilde T^{(-)}_{\left(\begin{smallmatrix}     j_1  & j_2      \\ j_4      & j_3       \end{smallmatrix}\right)}$ }.
We can rewrite Eq. (\ref{app:source_external_sum}) using Eq. (\ref{app:hpm}) as
\begin{equation}
    \begin{aligned}
        \delta \Gamma_4^{(0)}
    = &
    \int \frac{d\boldsymbol{\mu}}{(2\pi)^{d-1}}
    \int^\prime \frac{d\theta_2}{2\pi}
    \int_{0}^{2\pi} \frac{d\theta_1}{2\pi}
    \left\{\left[-\frac{R_d}{64}\int_{0}^{2\pi} \frac{d\theta}{2\pi}\left(\tilde{h}^+_{\theta_1,\theta}(\mu)+\tilde{h}^-_{\theta_1,\theta}(\mu)\right)\left(\tilde{h}^+_{\theta,\theta_2}(\mu)+\tilde{h}^-_{\theta,\theta_2}(\mu)\right)\right]+ \delta \tilde S_{\theta_1,\theta_2}(\mu)\right\}
    \\&\times\left(
    o^{+}(\theta_1)o^{+}(\theta_2)+ o^{-}(\theta_1)o^{-}(\theta_2)\right).
    \end{aligned}
\end{equation}
One can first break down the integration over $\theta_1$ into two terms spanning over the half angular range. Then, shifting  $\theta_1\rightarrow \theta_1+\pi$ and using the facts: $\tilde h^{\pm}_{\theta_1+\pi,\theta_2} = \mp\tilde h^{\pm}_{\theta_1,\theta_2} $ and $o^{\pm}(\theta+\pi) = \mp o^{\pm}(\theta)$, we obtain
\begin{equation}
    \begin{aligned}
&        \delta \Gamma_4^{(0)}
    = 
    \int \frac{d\boldsymbol{\mu}}{(2\pi)^{d-1}}
    \int^\prime \frac{d\theta_1d\theta_2}{(2\pi)^2}\left\{
    \left[-\frac{R_d}{32}\left(\int_{0}^{2\pi} \frac{d\theta}{2\pi}
    \left(\tilde{h}^+_{\theta_1,\theta}(\mu)\tilde{h}^+_{\theta,\theta_2}(\mu)+\tilde{h}^+_{\theta_1,\theta}(\mu)\tilde{h}^-_{\theta,\theta_2}(\mu)\right)\right)+\delta \tilde S_{\theta_1,\theta_2}(\mu)-\delta\tilde S_{\theta_1+\pi,\theta_2}\right]
    \right.\\&\left.
    \times o^{+}(\theta_1)o^{+}(\theta_2)
    +\left[-\frac{R_d}{32}\left(\int_{0}^{2\pi} \frac{d\theta}{2\pi}\left(\tilde{h}^-_{\theta_1,\theta}(\mu)\tilde{h}^+_{\theta,\theta_2}(\mu)+\tilde{h}^-_{\theta_1,\theta}(\mu)\tilde{h}^-_{\theta,\theta_2}(\mu)\right)\right)+\delta \tilde S_{\theta_1,\theta_2}(\mu)+\delta\tilde S_{\theta_1+\pi,\theta_2}(\mu)\right]
     o^{-}(\theta_1)o^{-}(\theta_2)\right\}.
    \end{aligned}
\end{equation}
From the same procedure for $\theta$, we obtain
\begin{equation}
    \begin{aligned}
        \delta \Gamma_4^{(0)}
    =& 
    \int \frac{d\boldsymbol{\mu}}{(2\pi)^{d-1}}
    \int^\prime \frac{d\theta_1d\theta_2}{(2\pi)^2}\left\{\left[
    \left(-\frac{R_d}{16}\int^\prime \frac{d\theta}{2\pi}\tilde{h}^+_{\theta_1,\theta}(\mu)\tilde{h}^+_{\theta,\theta_2}(\mu)\right)+\delta \tilde S^{+}_{\theta_1,\theta_2}(\mu)\right]o^{+}(\theta_1)o^{+}(\theta_2)
    \right.\\&\left.
    +\left[
    \left(-\frac{R_d}{16}\int^\prime \frac{d\theta}{2\pi}\tilde{h}^-_{\theta_1,\theta}(\mu)\tilde{h}^-_{\theta,\theta_2}(\mu)\right)+\delta \tilde S^{-}_{\theta_1,\theta_2}(\mu)\right]
     o^{-}(\theta_1)o^{-}(\theta_2)\right\},
    \end{aligned}
\end{equation}
where $\delta S^{\pm}_{\theta_1,\theta_2}(\mu) = \delta \tilde S_{\theta_1,\theta_2}(\mu)\mp\delta\tilde S_{\theta_1+\pi,\theta_2}(\mu)$.
Finally, Eq. (\ref{appendix:source_final}) projected onto symmetric and antisymmetric channels read
\begin{equation}
    \begin{aligned}
    \tilde{S}^{\pm}_{\theta_1,\theta_2}\left(\mu\right)
    &= -\frac{R_d}{16}\int^\prime \frac{d\theta}{2\pi}\tilde{h}^\pm_{\theta_1,\theta}(\mu)\tilde{h}^\pm_{\theta,\theta_2}(\mu)+\delta S^{\pm}_{\theta_1,\theta_2}(\mu).
    \label{appendix:source_final_pm}
    \end{aligned}
\end{equation}

\section{The absence of finite Hermitian fixed point
}

\label{eq:absence_Hermitian_PFP}

\begin{figure}[ht]
    \centering
    \begin{subfigure}{0.45\textwidth}
        \centering
        \includegraphics[width=\textwidth]{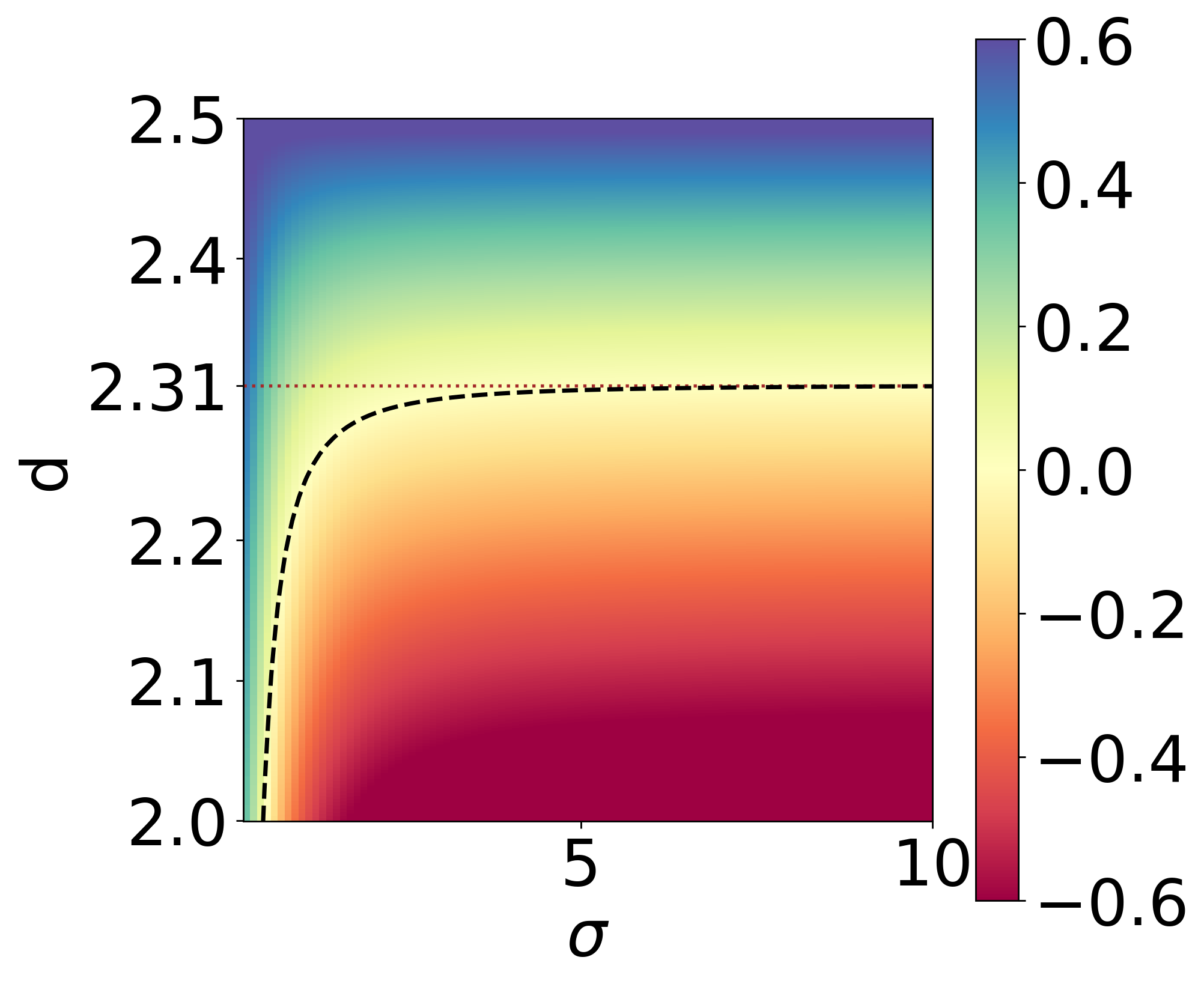}
        \caption{}
        \label{}
    \end{subfigure}
    \hfill
    \begin{subfigure}{0.35 \textwidth}
        \centering
        \includegraphics[width=\textwidth]{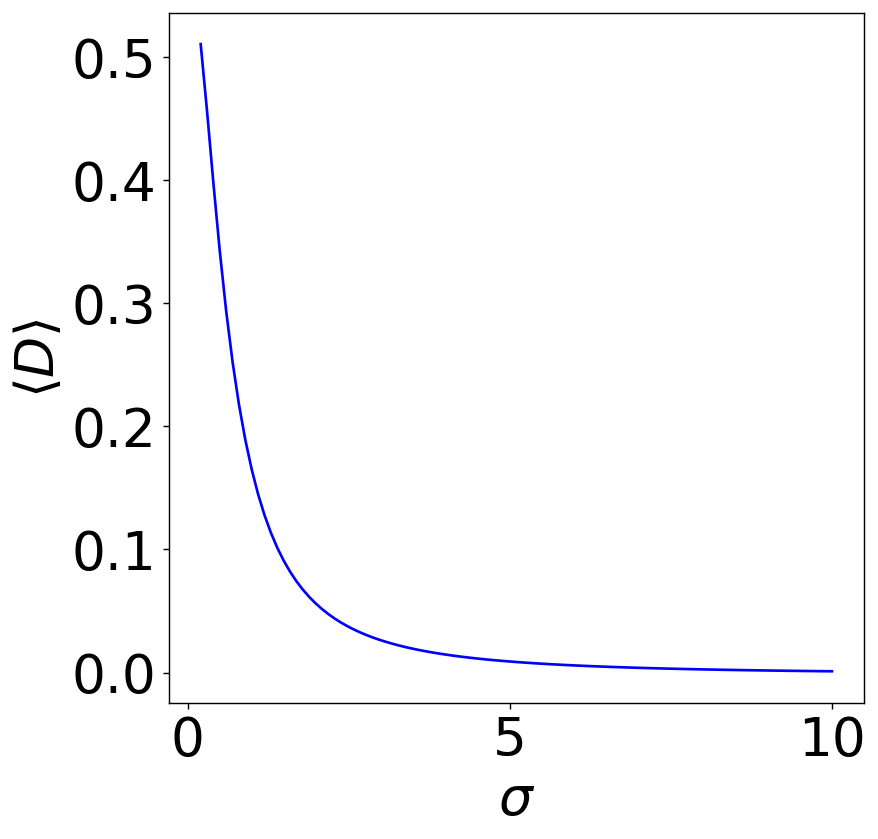}
        \caption{}
        \label{}
    \end{subfigure}
    \caption{
(a) A color plot of the expectation value of the discriminant as a function of the dimension and the width of the gaussian Ansatz in Eq. (\ref{eq:gaussian_ansatz}).
The dashed line denotes the points at which $\langle D\rangle = 0$.
It asymptotically approaches $d \approx 2.31$ in the large $\sigma$ limit. 
Being a variational calculation, this sets a lower bound for $d_{SC}$ below which superconducting instability is present.
The actual critical dimension is $d_{SC} \approx 2.44$, as discussed in Sec. \ref{sec:ex2}. 
(b) $\langle D \rangle$ plotted as a function of $\sigma$ at $d\approx 2.31$. 
    }
    \label{fig:disc_exp}
\end{figure}

In this Appendix, we use
\eq{eq:ineq_2_5} to show that the Ising-nematic quantum critical metal must be unstable against superconductivity below a critical dimension $d_{SC}$.
For this, we consider a square integrable Ansatz $F$ for the pairing wavefunction and evaluate
$\langle D \rangle \equiv \bra{F^\pm}D\ket{F^\pm} = \int \frac{d\hat{\theta}_1d\hat{\theta}_2}{(2\pi\sqrt{\Lambda})^{2}} F_{\hat{\theta}_1}^{*\pm}D_{\hat \theta_1,  \hat \theta_2}F^\pm_{\hat{\theta}_2}$,
where $F^\pm_{\hat \theta+2\bar\theta_{\text{max}}e^{l/2}} = \mp F^\pm_{\hat\theta}$.
The expectation value of the discriminant becomes
\begin{equation}
    \begin{aligned}
        \langle D \rangle &=
        \frac{1}{4R_d^2}
        \Bigg\{
        \int\frac{d\hat{\theta}_1}{2\pi\sqrt{\Lambda}}
        \left(\hat{F}^*_{\hat{\theta}_1}\hat{F}_{\hat{\theta}_1}(\HD-1)^2-(\HD-1)\hat{F}^*_{\hat{\theta}_1}
        \hat{\theta}_1\partial_{\hat\theta_1}\hat{F}_{\hat{\theta}_1}-(\HD-1)
        \hat{F}_{\hat{\theta}_1}\hat\theta_1\partial_{\hat\theta_1}
        \hat{F}^*_{\hat{\theta}_1}
        +
        \hat{\theta}_1^{2}\partial_{\hat{\theta}_1}\hat{F}_{\hat{\theta}_1}^*\partial_{\hat{\theta}_1}\hat{F}_{\hat{\theta}_1}\right)
        \\&
        - \frac{3\sqrt{3}}{2}\eta_d
        \int \frac{d\hat{\theta}_1d\hat{\theta}_2}{\left(2\pi\sqrt{\Lambda}\right)^2}\hat{F}_{\hat{\theta}_1}^*\frac{2(\HD-1)\left|\hat{\theta}_1-\hat{\theta}_2\right|^4\Lambda+(2\HD+1)\left|\hat{\theta}_1-\hat{\theta}_2\right|\Lambda^{\frac{5}{2}}}{\left(\left|\hat{\theta}_1-\hat{\theta}_2\right|^3+\Lambda^{\frac{3}{2}}\right)^2}\hat{F}_{\hat{\theta}_2}
        \Bigg\},
        \label{eq:expectation_disc_2}
    \end{aligned}
\end{equation}
where 
all dimension-dependent factors are pulled outside of the integration by rescaling $\hat \theta$ through
$\hat{F}_{\hat{\theta}} = F_{\alpha_d^{1/3}\hat{\theta}}$\footnote{
In the first term of Eq. (\ref{eq:expectation_disc_2}), the factors of $\alpha_d$ that arise from the rescaling are neutralized by the normalization of the eigenvector. 
The second term remains dependent on $\alpha_d$ through $\eta_d$.
}. 
The first term depends on $\HD$, which captures the degree of incoherence of fermions and the effect of the dilation of the rescaled angular variable that acquires a positive scaling dimension due to $\KFAVdim$.
The second term is negative for a positive definite Ansatz.
This reflects the attractive nature of the universal pairing interaction generated for the s-wave channel.
The prefactor is the anomalous dimension\cite{PhysRevB.110.155142} $\eta_d =  \frac{
      1
      }{3\sqrt{3}}\frac{R_d \bar g^{*}}{\left(\alpha_d\beta_d\right)^{\frac{1}{3}}}$.
As $d$ decreases, $\eta_d$ increases monotonically.
Therefore, the second term becomes increasingly more important relative to the first term as $d$ decreases.
For a gaussian Ansatz,
\begin{equation}
    \begin{aligned}
        \hat{F}_{\hat{\theta}_1} = 
        \frac{(8\pi)^{\frac{1}{4}}}{\sqrt{\sigma}}
        e^{-\left(\frac{\hat{\theta}_1}{\sqrt{\Lambda}\sigma}\right)^{2}},
        \label{eq:gaussian_ansatz}
    \end{aligned}
\end{equation}
the discriminant becomes
\begin{equation}
    \begin{aligned}
        \langle D \rangle =
        \frac{1}{4R_d^2}
        \left\{
        \frac{3}{4}
        -
        \frac{9 \eta _d }{4 \pi ^3}G_{2,5}^{5,2}\left(\frac{1}{216 \sigma ^6}|
\begin{array}{c}
 \frac{1}{6},\frac{2}{3} \\
 0,\frac{1}{3},\frac{2}{3},\frac{2}{3},\frac{7}{6} \\
\end{array}
\right)
        \right\}
        \label{eq:expectation_disc_ansatz_1}
    \end{aligned}
\end{equation}
for $\HD=1$,
where $G^{m,n}_{p,q}$ is Meiger G-function. 
As shown in \fig{fig:disc_exp}, the discriminant becomes negative for 
$d<2.31$.
This implies that $d_{SC}$ is bounded by $2.31$ from below.



\section{Derivations of superuniversal properties from the toy model}

\label{appendix:ex4}

The purpose of this appendix is to provide the details of the results quoted in Sec. \ref{sec:ex4}.
We first discuss the `A-family' composed of classes A, AB, AC, and ABC, which includes stable non-Fermi liquids.
Then, we discuss classes B, BC and C, which include non-Fermi liquids that inevitably become superconductors at low temperatures.

\subsection{Stable NFL superuniversality class (A)}
\label{appendix:A}


For class A, $\etaPIII>0$.
Since the intermediate region with negative discriminant does not play an important role, we set $\wII=0$ without loss of generality.  
Both the metallic and separatrix PFPs are regular,
\bqa
\aV_y^M = \frac{1}{4R_d}\begin{cases}
 \sqrt{\etaPI}-2\HD + 1 ~~~~& \text{for } y > 0\\
    -\sqrt{\etaPIII}\tanh\Big[\frac{1}{2}\sqrt{\etaPIII} y-\text{arctanh}\Big(\frac{\sqrt{\etaPI}}{\sqrt{\etaPIII}}\Big) \Big]-2\HD + 1 ~~~~ &\text{for } y \le  0  
\end{cases}\label{eq:toy_classa_vm}
\eqa
and 
\bqa
\aV_y^S = \frac{1}{4R_d}\begin{cases}
 -\sqrt{\etaPI}\tanh\Big[\frac{1}{2}\sqrt{\etaPI} y + \text{arctanh}\Big(\frac{\sqrt{\etaPIII}}{\sqrt{\etaPI}}\Big) \Big]-2\HD +1 ~~~~& \text{for } y > 0\\
    -\sqrt{\etaPIII}-2\HD +1  ~~~~ &\text{for } y \le  0 \label{eq:toy_classa_vs}
\end{cases}.
\eqa
For the U(1) gauge theory, the PFPs could be determined analytically only in the small $y$ limit in Eqs. \eqref{eq:VySPXu1} and \eqref{stable_sol}. 
In the toy model, the corresponding expressions in the $y\to -\infty$ limit are given by
\bqa
\aV^M_y = V^\bullet_{-\infty} + b^M e^{\sqrt{\etaPIII}y} + \order{e^{2\sqrt{\etaPIII}y}},
~~~~~
\aV^S_y = V^\circ_{-\infty},
\eqa
where
$\aV^{\bullet}_{-\infty} = \frac{\sqrt{\etaPIII} -2\HD +1 }{4R_d}$,
$b^M = \frac{\sqrt{\etaPIII}}{2R_d}\frac{\sqrt{\etaPI}-\sqrt{\etaPIII}}{\sqrt{\etaPIII}+\sqrt{\etaPI}}$ 
and
$\aV^{\circ}_{-\infty} = \frac{-\sqrt{\etaPIII} -2\HD +1 }{4R_d}$.
In the $y\to -\infty$ limit, these expressions agree with those obtained from the physical example to the leading order in $e^y$. 
At finite $y$, differences arise because the toy model ignores the sub-leading terms in $\aS_y$.
That is why the separatrix PFP has no $y$ dependence, and the metallic PFP retains only the $\order{e^{\sqrt{\eta_P,-\infty}y}}$ terms.
The same exercise can be done for the large-$y$ asymptotic region, where the solutions of the metallic and separatrix PFPs agree with those of the physical examples modulo vanishing terms in the asymptotic limit.
The universal coupling that emerges in the stable non-Fermi liquid is entirely determined by the metallic PFP.

\subsubsection{Non-Fermi liquid to superconductor phase transition}

When it comes to computing superuniversal observables that are controlled entirely by the asymptotic fixed points $\aV^\circ_{\pm\infty}$ and $\aV^\bullet_{\pm\infty}$, 
the toy model is expected to give the same answer as the physical theories.
To see this, let us extract the critical exponent $z \nu$ in 
\eq{eq:TcdeltaV}.
Recall that a critical non-Fermi liquid with one relevant direction for the four-fermion coupling is realized if the coupling in one angular momentum channel 
is on the separatrix.
Consider a PFP that is deformed away from the separatrix by $\deltaaVUV$ at the logarithmic angular momentum 
$y_0 \equiv y^{(n)}(0)$.
The PFP that goes through $(y_0, \aVS_{y_0}+\deltaaVUV)$ can be written as
\bqa
\aV_y = \frac{1}{4R_d}\begin{cases}
-\sqrt{\etaPI}\tanh\bigg\{\frac{1}{2}\sqrt{\etaPI} (y-y_0) - \text{arctanh}\bigg[\frac{4R_d\big(\aV^S_{y_0} + \deltaaVUV\big)+2\HD -1  }{\sqrt{\etaPI}}\bigg] \bigg\} -2\HD + 1 ~~~~& \text{for $y > 0$ with $y_0>0$}\\
     -\sqrt{\etaPIII} \tanh \bigg[ \frac{1}{2}\sqrt{\etaPIII} (y-y_0) - \text{arctanh}\bigg(\frac{4R_d \deltaaVUV}{\sqrt{\etaPIII}}-1\bigg)\bigg] -2\HD + 1 ~~~~ &\text{for $y \le 0$ with $y_0<0$}
\end{cases}.
\eqa
Thanks to the exact solvability of the toy model, this expression is valid for any $\deltaaVUV$.

Let us first consider the case with $y_0 \leq 0$.
For $\deltaaVUV < 0$, we use 
$\text{arctanh}(x) =  \text{arccoth}(x) \pm i \frac{\pi}{2} $ for $|x|>1$.
Using 
$\text{tanh}(x \pm i \pi/2) =  \text{coth}(x)$,
we find that the coupling diverges in region III at
\bqa
y_{SC}^* = y_0  + \frac{2}{\sqrt{\etaPIII}}
\text{arccoth}\bigg(\frac{4R_d\deltaaVUV}{\sqrt{\etaPIII}}-1\bigg)
\label{eq:toy_classa_y*2}.
\eqa
For small negative $\deltaaVUV$,
$\text{arccoth}\bigg(
\frac{4R_d\deltaaVUV}{\sqrt{\etaPIII}}
-1\bigg)
\approx
\frac{1}{2} \log \frac{2R_d |\deltaaVUV |}{\sqrt{\etaPIII}}
$
and
the superconducting transition temperature becomes
\bqa
T_c \approx  
\Lambda e^{-2z (y_0-y^*_{SC})}
=
\Lambda \left|
\frac{2R_d}{\sqrt{\etaPIII}}
\deltaaVUV \right|^{\frac{2z}{\sqrt{\etaPIII}}}.
\label{eq:toy_classa_tc20}
\eqa
For $y_0>0$, the coupling may diverge in either region.
If the coupling diverges in region I, the location of the divergence can similarly be written as
\bqa
y_{SC}^*  =  y_0 + 
 \frac{2}{\sqrt{\etaPI}} \text{arccoth}\bigg[\frac{4R_d\Big(\aV^S_{y_0} + \deltaaVUV\Big) +2\HD - 1}{\sqrt{\etaPI}}\bigg].
\label{eq:toy_classa_y*case1}
\eqa
If the coupling diverges instead in region III, $y_{SC}^*$ can be written as
\bqa
y_{SC}^*  =  
y_0^{III}+\frac{2}{\sqrt{\etaPIII}}\text{arccoth}\bigg(\frac{4R_d\delta\aV_{y^{III}_0}^*}{\sqrt{\etaPIII}}-1\bigg),
\label{eq:toy_classa_y*case2}
\eqa
where $\delta \aV_{y^{III}_0}^*$ is the the finite deformation measured at an intermediate $y=y_0^{III}$ in region III.

When we consider a small perturbation in a large angular momentum channel, the asymptotic form of $y^*_{SC}$ differs depending on whether we take the large $y_0$ limit first or the small $\deltaaVUV$ limit first. If we take the small $\deltaaVUV$ limit first for a fixed
$y_0 > 0$, 
the coupling diverges in region III 
and its divergence is controlled by \eq{eq:toy_classa_y*case2}.
$T_c$ is given by
\bqa
T_c \approx  \Lambda \Big|\kappa^{(A)}_{III} \deltaaVUV \Big|^{\frac{2z}{\sqrt{\etaPIII}}} 
\label{eq:toy_classa_tc2}
\eqa
with 
$\kappa^{(A)}_{III} = \frac{2R_d}{\sqrt{\etaPIII}}
    \frac{\etaPI-\etaPIII}{\etaPI} \cosh^2 \Big[\frac{1}{2}\sqrt{\etaPI} y_0 + \text{arctanh}\Big(\frac{\sqrt{\etaPIII}}{\sqrt{\etaPI}}\Big) \Big]e^{-\sqrt{\etaPIII}y_0}$.
This notation is not to be confused with $\kappa_{F,\theta}$ which denotes the shape of the Fermi surface.

Conversely,
in the large $y_0$ limit with a fixed $\deltaaVUV$,
the divergence of the PFP is controlled by \eq{eq:toy_classa_y*case1},
and $T_c$ becomes
\bqa
T_{c} 
\approx \Lambda  \Big|\kappa^{(A)}_{I} \deltaaVUV\Big|^{\frac{2z}{\sqrt{\etaPI}}}
\label{eq:toy_classa_tc1}
\eqa
to the leading order in $\deltaaVUV$ 
with $\kappa^{(A)}_I = 2R_d/\sqrt{\etaPI}$.
With $y_0  = y^{(n)}(0)$,
Eqs. 
\eqref{eq:toy_classa_tc2} 
and
\eqref{eq:toy_classa_tc1}
are to be directly compared with
Eqs. \eqref{eq:TcdelVA}-\eqref{eq:TcdelVA3}.
As expected, the critical exponent $\nu z$ precisely match.

\subsection{NFL to non-s-wave SC critical superuniversality class (AB)}
\label{appendix:AB}


In superuniversality class AB, the $-\infty$ asymptotic fixed points $\aV^\bullet_{-\infty}$ and $\aV^\circ_{-\infty}$ are present as in class A. 
However, the metallic and separatrix PFPs coincide.
In the toy model, this profile can be written 
\bqa
\aV_y^{M/S} =\frac{1}{4R_d} \begin{cases} 
    \sqrt{\etaPI} - 2\HD + 1 ~~~~ &\text{for } y > \wII_c\\
    \sqrt{-\etaPII} \tan \Big[ \frac{1}{2} \sqrt{-\etaPII}\, y - \arctan \Big(\frac{\sqrt{\etaPIII}}{\sqrt{-\etaPII}} \Big) \Big]- 2\HD + 1  &\text{for } 0 < y < \wII_c\\
    -\sqrt{\etaPIII} - 2\HD + 1 &\text{for } y \le 0
\end{cases} \label{eq:toy_classab_vms}
\eqa
with the critical width 
\bqa
\wII_c  = \frac{2}{\sqrt{-\etaPII}} \Bigg[ \arctan\Bigg( \frac{\sqrt{\etaPIII}}{\sqrt{-\etaPII}}\Bigg)  + \arctan\Bigg(\frac{\sqrt{\etaPI}}{\sqrt{-\etaPII}}\Bigg)\Bigg].
\eqa

\subsubsection{Universal pairing interaction}

\begin{figure}[H]
    \centering
    \includegraphics[width=0.7\linewidth]{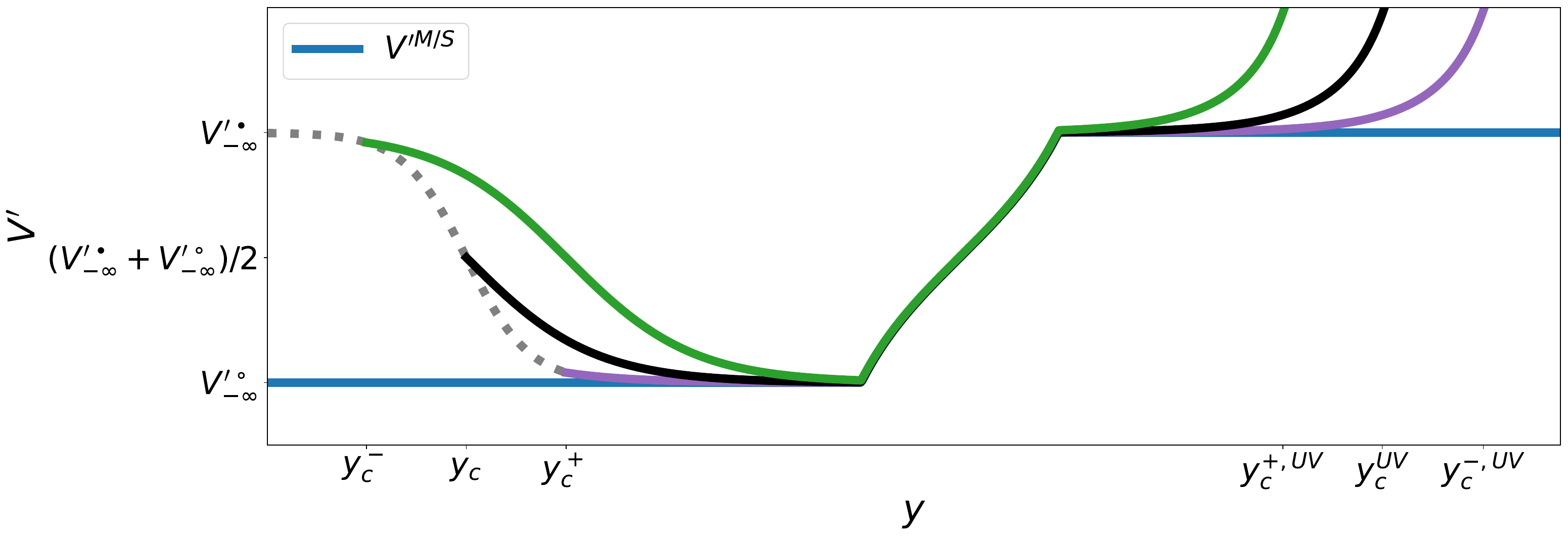}
\caption{
Schematic plot of the crossover that renormalized couplings exhibit in class AB.
The dotted line represents the profile of the renormalized coupling that emerges at large scale $l$ from the bare couplings that are highly repulsive.  
The solid blue curve connected to $\aV^\circ_{-\infty}$ represents $\aV^M_y$. 
The green, black and purple lines, which end on the dotted line from left to right, depict the PFPs that emanate from 
$(y^{-,UV}_c(l), \infty)$,
$(y^{UV}_c(l), \infty)$,
$(y^{+,UV}_c(l), \infty)$,
and 
end at
$\left(
y_c^-(l),
\aV^\bullet_{-\infty} 
+ \varepsilon 
( \aVM_{y_c^-(l)} -\aV^\bullet_{-\infty}   ) 
\right)$,
$\left( 
y_c(l), 
(\aV^\bullet_{-\infty}  +  \aVM_{y_c(l)})/2 
\right)$,
$\left( 
y_c^+(l),
\aVM_{y_c^+(l)}
+ \varepsilon 
(  \aV^\bullet_{-\infty}  - 
\aVM_{y_c^+(l)}) 
\right)$, respectively, 
where
$y_c^-(l) = y_c^{-,UV}-l/2 $,
$y_c(l)=y_c^{UV}-l/2$ 
and
$y_c^+(l) = y_c^{+,UV}-l/2$. 
It is understood that all crossover $y$ scales are generally $l$ dependent.
 }
\label{fig:toy_classab_jump}
\end{figure}

The main difference from class A is that here the metallic PFP changes stability as it passes from the large- to small-$y$ regions. 
This has a couple of important consequences. 
Firstly, the universal profile that emerges at low energies is not strictly controlled by the metallic PFP, since the metallic PFP becomes repulsive in the small $y$ asymptotic regime. 
In the strict $l \to \infty$ limit, all couplings saturate to $\aV^\bullet_{-\infty}$, instead of staying close to $\aV^M_y$. 
For a large but finite $l$,
the renormalized coupling interpolates 
the metallic PFP and $\aV^\bullet_{-\infty}$ as the angular momentum is lowered.
The precise form of the crossover that the renormalized couplings exhibit depends on the bare coupling.
However, the dependence of the renormalized coupling function on the bare coupling becomes weak as $l$ increases.
This is explicitly demonstrated in Sec. \ref{sec:quasiuniversalpairing_classC} in class C.
In the following, we consider the crossover that emerges when the bare couplings are highly repulsive.

We first find the center of the crossover $y_c^{UV}$. 
For this, and for later calculations, it is convenient to introduce the shifted and rescaled coupling 
\bqa
v'_y \equiv 4R_d \aV_y + 2\HD -1. 
\eqa
In this coordinate, 
$y_c^{UV}$ is defined by the condition that the PFP emanating from 
$(y_c^{UV}(l), \infty)$ reaches $v'_y=0$  at 
$y = y_c^{UV}(l) - l/2$. 
At large $l$, this results in the equation for  $y_c(l) \equiv y_c^{UV}(l) - l/2$, 
\bqa
-\sqrt{\etaPIII} \tanh\bigg[ \frac{1}{2}\sqrt{\etaPIII} y_c(l) + \text{arctanh}\bigg(1-2\frac{\sqrt{\etaPI}}{\sqrt{\etaPIII}} \frac{\etaPIII -\etaPII}{\etaPI - \etaPII} e^{\sqrt{\etaPI}(w_c -y_c(l)-l/2 )} \bigg) \bigg] = 0.
\eqa
Its solution is given by
\bqa
y_c(l) \approx 
-\frac{1}{2}\Bigg(\frac{\sqrt{\etaPI}}{\sqrt{\etaPI}+\sqrt{\etaPIII}} \Bigg)l 
+ \frac{1}{\sqrt{\etaPI}+\sqrt{\etaPIII}}\Bigg\{\sqrt{\etaPI} \wII_c + \ln \Bigg[\frac{\sqrt{\etaPI}(\etaPIII-\etaPII)}{\sqrt{\etaPIII}(\etaPI-\etaPII)} \Bigg] \Bigg\}.
\label{eq:toy_classab_ycl}
\eqa
Similarly, we determine $y_c^{+,UV}$  and $y_c^{-,UV}$ by requiring that the PFPs emanating from
$(y_c^{+,UV}(l), \infty)$ 
and
$(y_c^{-,UV}(l), \infty)$ 
to the right and left of $y_c^{UV}(l)$ reach close to the metallic PFP and 
$\aV^\bullet_{-\infty}$, respectively,
within a relative distance $\varepsilon$ (see \fig{fig:toy_classab_jump}).
In particular, they are obtained from setting 
$v^{\pm}_{y_c(l)} =\mp (1-2\varepsilon) \sqrt{\etaPIII}$,
where
$y_c^\pm(l)
=y_c^{\pm,UV}(l)-l/2$.
In the large $l$ limit, this becomes 
\bqa
\begin{aligned}
-\sqrt{\etaPIII} \tanh\bigg[ \frac{1}{2}\sqrt{\etaPIII} y^\pm_c(l) + \text{arctanh}\bigg(1-2\frac{\sqrt{\etaPI}}{\sqrt{\etaPIII}} \frac{\etaPIII -\etaPII}{\etaPI - \etaPII} e^{\sqrt{\etaPI}(w_c -y^\pm_c(l)-l/2 )} \bigg) \bigg]
= \mp ( 1-2 \varepsilon) \sqrt{\etaPIII} 
\end{aligned}, \nn
\eqa
which leads to 
\bqa
y_c^{\pm,UV}(l) 
\approx y_c^{UV}(l) \pm \frac{2}{\sqrt{\etaPI}+\sqrt{\etaPIII}} \text{arctanh}(1-2\varepsilon).
\eqa
It is noted that 
\bqa
\Delta y(l) = 
y_c^{+,UV}(l)
-
y_c^{-,UV}(l) \sim
 \frac{4}{\sqrt{\etaPI}+\sqrt{\etaPIII}} \text{arctanh}(1-2 \varepsilon)
\eqa
is independent of $l$. 
As a result, $\Delta \tilde x$ in 
\eq{eq:theta_crossover} increases as 
$\Delta \tilde x \sim 
\exp{
\frac{1}{2}\Bigg(\frac{\sqrt{\etaPIII}}{\sqrt{\etaPI}+\sqrt{\etaPIII}} \Bigg)l 
}
$ with increasing $l$.

\subsubsection{Non-Fermi liquid to superconductor phase transition}

When the bare interaction is deformed below the separatrix by $\deltaaVUV$ at angular momentum $n$ 
to induce the non-Fermi liquid to superconductor phase transition,
it does not matter whether the large $n$
or small $\deltaaVUV$ limit is taken first since the small deformation is reduced in region I and  diverges in region III.
Therefore, $T_c$ is determined by \eq{eq:toy_classa_tc2} with $\kappa^{(A)}_{III}$ replaced by $\kappa^{(AB)}_{III}$, 
\bqa
\kappa_{III}^{(AB)} = \frac{2R_d}{\sqrt{\etaPIII}}\begin{cases}
e^{\sqrt{\etaPI}\wII_c}\frac{\etaPIII-\etaPII}{\etaPI-\etaPII} e^{-\big(\sqrt{\etaPIII}+\sqrt{\etaPI}\big)y_0} ~~~~ & \text{for } y_0> \wII_c\\
1 ~~~~ & \text{for } y_0< 0
\end{cases},
\eqa
where $y_0 \equiv y^{(n)}(0)$.
Unlike class A, $T_c$ always decreases with increasing $n$, irrespective of $\etaPI$ and $\etaPIII$ in class AB.

\subsection{NFL to s-wave SC critical superuniversality class (AC) }
\label{appendix:AC}


Without loss of generality, we set $\wII = 0$. 
The metallic and separatrix PFPs can be obtained by taking the  $\etaPIII \to 0$ limit in \eq{eq:toy_classa_vm} and \eq{eq:toy_classa_vs}.
This yields
\bqa
\aV_y^M =  \frac{1}{4R_d}\begin{cases}
    \sqrt{\etaPI }-2\HD +1 ~~~~ &\text{for } y > 0 \\
    \frac{\sqrt{\etaPI}}{1 - \frac{1}{2}\sqrt{\etaPI} y } -2\HD + 1 &\text{for } y \le 0 
\end{cases},
\eqa
for the metallic PFP and 
\bqa
\aV_y^S =  \frac{1}{4R_d}\begin{cases}
    -\sqrt{\etaPI }\tanh\Big(\frac{1}{2}\sqrt{\etaPI} y\Big) - 2\HD +1 ~~~~ &\text{for } y > 0 \\
   -2\HD + 1 &\text{for } y \le 0 
\end{cases}
\eqa
for the separatrix. 
In the small-$y$ limit, these expressions become
\bqa
\aV_y^M = \aV^{\halfominus}_{-\infty} + c^M \frac{1}{y} + \order{\frac{1}{y^2}},
~~~~~ 
\aV^S_y = \aV^{\halfominus}_{-\infty},
\eqa
with $c^M = -\frac{1}{2R_d}$, which is consistent with the result obtained from the Ising-nematic theory at $d = d_{sc}$. 
Once again, discrepancies arise with the separatrix due to the fact that the toy model does not capture the leading $y$-dependence of the source.

\subsubsection{Non-Fermi liquid to superconductor phase transition}

We now consider how $T_c$ is turned on as the bare coupling in the angular momentum channel $n$ is deformed below the separatrix by $\deltaaVUV$. 
If we take 
$y_0 \equiv y^{(n)}(0)$
to be large first for a fixed $\deltaaVUV$, we again arrive at the expression in \eq{eq:toy_classa_tc1}. 
If we instead take
the small $\deltaaVUV$ limit for a fixed 
$y_0$, we obtain
\bqa
T_c \approx \Lambda \begin{cases}
    \exp \bigg\{-\left[2zy_0 + \frac{4z}{R_d}\, 
    e^{ -{\sqrt{\etaPI}} y_0 }
    \frac{1}{\big|\deltaaVUV\big|}
    \right] \bigg \} ~~~~&\text{for } 
    y_0 > 0\\
   \exp\bigg\{-\frac{z}{R_d}\frac{1}{\big|\deltaaVUV\big|}\bigg\} &\text{for } y_0 \le 0 
\end{cases}, 
\eqa
where  $y_0 \equiv y^{(n)}(0)
=
\log  
\frac{\sqrt{\Lambda}\pi n}{2\gamma \sqrt{\KFAVdim}}
$
and
we have combined the result obtained from \eq{eq:toy_classa_y*2}
for $y_0 \le 0$. 

\subsection{Double-critical superuniversality class (ABC)}
\label{appendix:ABC}


In the double-critical class, the metallic/separatrix PFP can be obtained from 
\eq{eq:toy_classab_vms} by setting $\etaPIII = 0$, giving
\bqa
\aV_y^{M/S} = \frac{1}{4R_d}\begin{cases}
    \sqrt{\etaPI} - 2\HD + 1~~~~ &\text{for } y > \wII_c\\
    \sqrt{-\etaPII} \tan \Big( \frac{1}{2} \sqrt{-\etaPII}\, y  \Big) -2\HD + 1 &\text{for } 0 < y < \wII_c\\
     -2\HD + 1&\text{for } y \le 0 
\end{cases} \label{eq:toy_class_abc_vms}
\eqa
with the critical width becoming 
\bqa
\wII_c = \frac{2}{\sqrt{-\etaPII}} \arctan\Bigg(\frac{\sqrt{\etaPI}}{\sqrt{-\etaPII}}\Bigg).
\eqa

\subsubsection{Non-Fermi liquid to superconductor phase transition}

For the non-Fermi liquid to superconductor phase transition, $T_c$ scales as
\bqa
T_c \approx \Lambda \begin{cases}
    \exp \bigg\{-\left[2zy_0 + \frac{z}{R_d}\, \Big(1 -\frac{\etaPI}{\etaPII}\Big)e^{-\sqrt{\etaPI} \wII_c}
    e^{\sqrt{\etaPI} y_0}
    \frac{1}{\big|\deltaaVUV\big|}\right] \bigg \} ~~~~&\text{for } y_0 > 0\\
   \exp\bigg\{-\frac{z}{R_d}\frac{1}{\big|\deltaaVUV\big|}\bigg\} &\text{for } y_0 \le 0 
\end{cases},
\eqa
where $y_0 \equiv y^{(n)}(0)$
and $\deltaaVUV$ is the deformation to the four-fermion coupling in angular momentum channel $n$.

The difference between classes A and AC and classes AB and ABC can be contrasted by considering how $\deltaaVUV$ should be tuned with increasing $n$ to keep $T_c e^{2zy_0}$ fixed.
Under the increase of $y_0 \to sy_0$ with $s>1$, the deformation scales as
\bqa
\deltaaVUV \to \begin{cases}
    s^{- \sqrt{\etaPI}} \deltaaVUV ~~~~ & \text{in classes A, AC} \\
    s^{\sqrt{\etaPI}} \deltaaVUV ~~~~ & \text{in classes AB, ABC}
\end{cases}.
\eqa
In classes AB and ABC, the separatrices are locally stable in the large-$y$ asymptotic region due to their coincidence with the respective metallic PFPs. 
Perturbations added below the separatrices are thus renormalized to smaller values in region I before the coupling enters region III. 
By increasing $y_0$, the RG time spent in region I is increased, and a larger perturbation is required in order to preserve the same $T_c e^{2zy_0}$.
This causes $\deltaaVUV$ to increase with increasing $s$.
Conversely, the separatrices in classes A and AC are unstable in both asymptotic regions, and the perturbation strictly grows. 
An increase in $y_0$ must then be balanced with a shrinking $\deltaaVUV$.

\subsection{Non-s-wave SC superuniversality class (B)}
\label{appendix:B}


In class B proximate to class A, the metallic and separatrix PFPs diverge in regions III and I respectively. 
The metallic PFP is given by 
\bqa
\aV_y^M = \frac{1}{4R_d}\begin{cases}
    \sqrt{\etaPI} -2\HD +1 ~~~~&\text{for } y \ge \wII \\
    \sqrt{-\etaPII} \tan \Big[\frac{1}{2}\sqrt{-\etaPII}\, (y-\wII)   + \arctan\Big(\frac{\sqrt{\etaPI}}{\sqrt{-\etaPII}} \Big)    \Big] -2\HD +1 &\text{for } 0 < y < \wII\\   
    -\sqrt{\etaPIII} \coth\Big[\frac{1}{2}\sqrt{\etaPIII}(y - y^*_M)\Big]  -2\HD +1&\text{for } y \le0
    \label{eq:toy_classb_vm}
    \end{cases}
\eqa
while the separatrix is given by
\bqa
\aV_y^S = \frac{1}{4R_d}
\begin{cases}
   - \sqrt{\etaPI} \coth\Big[ \frac{1}{2}\sqrt{\etaPI} \, (y-y^*_S)\Big]  -2\HD +1 & \text{for } y \ge \wII \\
 \sqrt{-\etaPII}  \tan \Big[ \frac{1}{2} \sqrt{-\etaPII}\, y   - \arctan \Big(\frac{\sqrt{\etaPIII}}{\sqrt{-\etaPII}} \Big)    \Big]  -2\HD + 1 & \text{for } 0 < y < \wII  \\  
  -\sqrt{\etaPIII} -2\HD +1 & \text{for } y \le  0 
\end{cases}.
\label{eq:toy_classb_vs}
\eqa
Here, $y^*_M$ and $y^*_S$ are the $y$ coordinates at which the metallic and separatrix PFPs diverge to $-\infty$ and $\infty$, respectively
(see Figs. \ref{fig:PFP} and \ref{fig:classB}),
\footnote{Since we commit to the regime proximate to class A, each profile has only one singularity.}
\bqa
y^*_M &=& -\frac{2}{\sqrt{\etaPIII}}\text{arccoth}\bigg\{\frac{\sqrt{-\etaPII}}{\sqrt{\etaPIII}} \tan\bigg[\frac{1}{2}\sqrt{-\etaPII}\, \wII   - \arctan\bigg(\frac{\sqrt{\etaPI}}{\sqrt{-\etaPII}} \bigg)    \bigg] \bigg\}, \nn
y^*_S &=& \wII + \frac{2}{\sqrt{\etaPI}}  \text{arccoth}\bigg\{ \frac{\sqrt{-\etaPII}}{\sqrt{\etaPI}} \tan\bigg[\frac{1}{2}\sqrt{-\etaPII}\, \wII   - \arctan\bigg(\frac{\sqrt{\etaPIII}}{\sqrt{-\etaPII}} \bigg)   \bigg]\bigg\}.
\eqa
At the leading order in the small $\delta w$ limit, these expressions become
\bqa
y^*_M \approx -\frac{1}{\sqrt{\etaPIII}}\log\bigg(\frac{4\sqrt{\etaPIII}}{\etaPIII - \etaPII}\frac{1}{\delta w} \bigg),
\eqa
\bqa
y^*_S \approx w_c + \frac{1}{\sqrt{\etaPI}}\log\bigg(\frac{4\sqrt{\etaPI}}{\etaPI - \etaPII}\frac{1}{\delta w} \bigg).
\eqa
$y^*_S$  and  $y^*_M$  are in regions I and III, respectively, for class B proximate to class A. 
$y^*_S$ acts as an angular momentum cutoff. In angular momentum channels with $y^{(m)}(0) \leq y^*_S$, superconductivity can be avoided as long as the bare couplings are not below the separatrix PFP.
For angular momentum channels with $y^{(m)}(0) > y^*_S$, however, superconducting instability is inevitable because the coupling is renormalized to $-\infty$ at low energies, irrespective of the bare coupling.
For large angular momentum channels, the couplings are first attracted toward the metallic PFP before they diverge close to $y^*_M$.
Therefore, $y^*_M$ controls the 
quasi-universal $T_c/\KFAVdim^z$, as discussed Sec. \ref{sec:quasiuniversalTckF}.

\subsubsection{Universal $T_c/\KFAVdim^z$}

Let us now discuss the quasi-universal superconductivity that arises in class B more quantitatively. 
Consider a channel at $y_0$ with the bare coupling $\aV^{UV}$.
If $(y_0, \aV^{UV})$ is below the separatrix, the PFP that emanates from   $(y_0, \aV^{UV})$ diverges to $-\infty$ at
\bqa
y^*_{SC} = \frac{2}{\sqrt{\etaPIII}} \text{arccoth} \bigg( \frac{4R_d \aV_{0}+2\HD - 1}{\sqrt{\etaPIII}} \bigg),
\label{eq:toy_classb_ystariii}
\eqa
where $\aV_0$
is the $y=0$ value of the PFP that passes through $(y_0, \aV^{UV})$. 
The argument of the inverse hyperbolic cotangent is less than $-1$.
Explicitly, the relation between $\aV_0$ and $(y_0, \aV^{UV})$ is given by 
\bqa
\begin{aligned}
\aV_0 = \frac{1}{4R_d}\bigg[\sqrt{-\etaPII} \tan\bigg(-\frac{1}{2}  \sqrt{-\etaPII} \wII -\text{arctan}\bigg\{&\frac{\sqrt{\etaPI}}{\sqrt{-\etaPII}}\text{coth}\bigg[\frac{1}{2}\sqrt{\etaPI}(\wII-y_0) \\&- \text{arccoth} \bigg(\frac{4R_d \aV^{UV} + 2\HD - 1}{\sqrt{\etaPI}}\bigg) \bigg] \bigg\}   \bigg) -2\HD +1 \bigg].
\end{aligned}
\eqa

Obviously, 
$y^*_{SC}=y^*_M$ if
$\aV^{UV}=\aV^M_{y_0}$.
As $\aV^{UV}$ increases (decreases) relative to $\aV^M_{y_0}$, superconducting instability occurs at larger (shorter) length scales, and 
$y^*_{SC}$ decreases (increases). 
However, the shift of  $y^*_{SC}$ becomes smaller as $y_0$ increases, as shown in  \fig{fig:deltayinclassBandBC}.
This is because, for large $y_0$, the PFP has a large window of length scale during which it is attracted to the metallic PFP and, as a consequence, diverges close to $y^*_M$.
Let us now consider a small $\deltaaVUV =
\aV^{UV}-\aV^M_{y_0}$,
for which we can use the linearized flow to write
$\aV_0 \approx \aV^M_0  + A_{0;y_0}\deltaaVUV$.
The shift of $y$ at which the coupling diverges is obtained to be
\bqa
\begin{aligned}
\delta y^* 
&
= -\frac{8R_d}{\sqrt{\etaPIII }(\etaPI - \etaPII)} \frac{e^{-\sqrt{\etaPI}(y_0-\wII_c)}}{\delta w} \deltaaVUV
\label{eq:toy_classb_deltay}
\end{aligned}
\eqa
at the leading order in $\delta w$. 
Relative to $|y^*_M|$, the shift becomes
\bqa
\frac{\delta y^*}{|y^*_M|} = - \frac{2 R_d}{\sqrt{\etaPI}} \frac{e^{-\sqrt{\etaPI} \Delta_S y_0}}{|\log\big( \frac{\etaPIII-\etaPII}{4\sqrt{\etaPIII}} \delta w\big) | } \deltaaVUV,
\eqa
where $\Delta_S y_0 \equiv y_0 - y^*_S$ is the deviation of the UV channel from $y^*_S$. 
For $y_0 > y^*_S$, the ratio depends weakly on $\deltaaVUV$ for small $\delta w$.

\subsubsection{Oscillation of $T_c$ with $\KFAVdim$} 

Let us now consider the behavior of the minimum superconducting temperature in class B. 
Here, the existence of the scale $y^*_S $ that divides the intrinsically unstable channels from the stable ones plays an important role as discussed in \fig{fig:classBdetails}. 
In order to see this explicitly, we first note that a PFP that diverges to $\infty$ at large $y$ can be labeled by the location of its divergence $y^*_\infty$ as
\bqa
\aV_{y,I} = \frac{1}{4R_d}\Big\{ -\sqrt{\etaPI} \coth \Big[ \frac{1}{2}\sqrt{\etaPI} \big(y-y^*_{\infty}  \big)\Big]  -2\HD + 1   \Big\} . \label{eq:toy_classb_region1div}
\eqa
In class B, the PFP with $y^*_{\infty} > y^*_S$ diverges to $-\infty$ at
\bqa
y^*_{SC}
(y^*_\infty)
= -\frac{2}{\sqrt{\etaPIII}}  \text{arcoth} \Bigg\{\frac{\sqrt{-\etaPII}}{\sqrt{\etaPIII}} \tan \Bigg[ \frac{1}{2}\sqrt{-\etaPII} \wII + \arctan \Bigg( \frac{\sqrt{\etaPI}}{\sqrt{-\etaPII}}\text{coth} \Bigg\{ \frac{1}{2} \sqrt{\etaPI}\Big[ w-y^*_{\infty}\Big] \Bigg\}\Bigg)\Bigg] \Bigg\} . \nn
\label{eq:toy_classb_region3div}
\eqa
We assume that $\delta w$ is small enough that $y^*_\infty$ and $y^*_{SC}$ are in region I and III, respectively.
The minimum $T_c$ arises when all bare couplings are $\infty$.
The superconducting transition temperature that arises in the angular momentum channel $n$ with $\infty$ bare coupling is 
given by 
$\Lambda e^{2z \Delta y_{SC}}$,
where $\Delta y_{SC} = y^*_{SC}(y_n(0)) -y_n(0)$. 
$T_c$ is set by the channel $n$ that exhibits the largest $ \Delta y_{SC}(n)$.

It is useful to first consider the optimal $y^*_O$ that maximizes $\Delta y_{SC}$ with respect to continuously varying $y^*_\infty$.
This is obtained from 
\bqa
\partial_{y^*_{\infty}} \Delta y_{SC} \big|_{y^*_{O}} = 0,
\eqa
which yields
\bqa
y_{O}^* =  \wII + \frac{2}{\sqrt{\etaPI}} \text{arcoth} \Bigg\{\frac{\sqrt{-\etaPII}}{\sqrt{\etaPI}} \csc \Big(\frac{1}{2}\sqrt{-\etaPII}\wII \Big) \Bigg[\frac{\sqrt{\etaPI-\etaPII}}{\sqrt{\etaPIII-\etaPII}}-\cos\Big( \frac{1}{2}\sqrt{-\etaPII}\wII\Big)\Bigg]    \Bigg\}.
\eqa
This sets the angular momentum channel with the highest $T_c$ when all bare couplings are highly repulsive.
For the profile that takes $y^*_{\infty} = y_{O}^*$, the coupling diverges to negative infinity at
\bqa
\begin{aligned}
y^*_{SC,O} = \frac{2}{\sqrt{\etaPIII}}  \text{arcoth} \Bigg[\frac{\sqrt{-\etaPII}}{\sqrt{\etaPIII}} &\tan \Bigg( -\frac{1}{2}\sqrt{-\etaPII}\wII \\&+ \arctan \Bigg\{  \csc \Big(\frac{1}{2}\sqrt{-\etaPII}\wII \Big) \Bigg[\frac{\sqrt{\etaPI-\etaPII}}{\sqrt{\etaPIII-\etaPII}}-\cos\Big( \frac{1}{2}\sqrt{-\etaPII}\wII\Big)\Bigg]   \Bigg\}\Bigg) \Bigg].
\end{aligned}
\eqa
Near the critical width, $\wII = \wII_c+\delta w$, the above expressions can be written to the leading order in $\delta w$ as 
\bqa
y^*_O = \wII_c 
+\frac{1}{\sqrt{\etaPI}}\ln \frac{1}{\delta w}  + \frac{1}{\sqrt{\etaPI}} \ln\Big[\frac{4(\etaPI + \sqrt{\etaPI}\sqrt{\etaPIII})}{\sqrt{\etaPIII}(\etaPI-\etaPII)} \Big] + \order{\delta w},
\eqa
\bqa
y^*_{SC, O} = 
-\frac{1}{\sqrt{\etaPIII}}\ln \frac{1}{\delta w} 
+ \frac{1}{\sqrt{\etaPI}}\ln\Big[\frac{\sqrt{\etaPI}(\etaPIII- \etaPII)}{4(\etaPIII+ \sqrt{\etaPI}\sqrt{\etaPIII})}\Big] + \order{\delta w},
\eqa
which result in
\bqa
\Delta y_{SC,O} \approx y^*_M - y^*_S - \ln\Bigg[ \Bigg(\frac{\sqrt{\etaPI}+\sqrt{\etaPIII}}{\sqrt{\etaPIII}}  \Bigg)^{\frac{1}{\sqrt{\etaPI}}} \Bigg( \frac{\sqrt{\etaPI}+\sqrt{\etaPIII}}{\sqrt{\etaPI}}\Bigg)^{\frac{1}{\sqrt{\etaPIII}}}  \Bigg].
\eqa
The optimal superconducting temperature associated with $y^*_O$ is
\bqa
T_{c,O} \approx  \varsigma_{c,O} \Lambda e^{2z( y^*_M - y^*_S)}, 
\eqa
where
\bqa
\varsigma_{c,O} = 
\Bigg[ \Bigg(\frac{\sqrt{\etaPIII}}{\sqrt{\etaPI}+\sqrt{\etaPIII}} \Bigg)^{\frac{1}{\sqrt{\etaPI}}} \Bigg( \frac{\sqrt{\etaPI}}{\sqrt{\etaPI}+\sqrt{\etaPIII}}\Bigg)^{\frac{1}{\sqrt{\etaPIII}}} \Bigg]^{2z}< 1 .
\eqa

For actual theories, there will generally not be an angular momentum channel $m$ that exactly satisfies $y_m(0) = y^*_{O}$ since the $m$ is discrete. 
Therefore, $T_{c}$ is set by the channel(s) whose $y_m(0)$ is closest to $y^*_{O}$. 
When the actual angular momentum deviates slightly from the optimal value as
$y_m(0) = y^*_{O} + \delta y$, we can write at the leading order in $\delta y$
\bqa
\Delta y_{SC} \approx \Delta y_{SC,O} - \xi \delta y^2 \label{eq:toy_classb_xi}
\eqa
where $\xi \equiv - \frac{1}{2} \partial^2_{y^*_\infty} \Delta y_{SC}|_{y_\infty^* = y^*_O}$ is a positive number given by
\bqa
\begin{aligned}
\xi = &\frac{1}{8\etaPII}
\Bigg\{
\Big(\frac{\etaPI-\etaPII}{\sqrt{\etaPI}}\Big)\Big(\etaPIII+\etaPII+(\etaPIII-\etaPII)\cos 
(\Phi)
\Big)\\ &\times\sinh
\Bigg[2\text{arccoth}\Big(\frac{\sqrt{-\etaPII}}{\sqrt{\etaPI}}\Big(\frac{\sqrt{\etaPI-\etaPII}}{\sqrt{\etaPIII-\etaPII}}-\cos\Big( \frac{1}{2}\sqrt{-\etaPII}w\Big) \Big) \csc\Big(\frac{1}{2}\sqrt{-\etaPII}w \Big) \Big) \Bigg]\\
&- 2\sqrt{\etaPII}\big(\etaPIII-\etaPII\big)
\sin
(\Phi)
\Bigg\},
\end{aligned}
\eqa
where
$
\Phi 
=
\sqrt{-\etaPII}w + 2\arctan\left[\cot\Big(\frac{1}{2}\sqrt{-\etaPII}w \Big) - \frac{\sqrt{\etaPI-\etaPII}}{\sqrt{\etaPIII-\etaPII}}\csc\Big(\frac{1}{2}\sqrt{-\etaPII}w\Big) \right] 
$.
This translates to the angular momentum-dependent superconducting transition temperature,
\bqa
T_{c} \approx T_{c,O} e^{-2z \xi \delta y^2 }
\label{eq:toy_classb_tcmin}
\eqa
for theories with highly repulsive bare couplings
in 
\eq{eq:toy_classb_tcmin_main}.
Since the values of $y_m(0)$ for fixed $m$ are shifted by tuning $\KFAVdim$, the channel most proximate to $y^*_O$ is modified as one varies the density of fermions. 
In addition, given that what matters is the deviation from $y^*_{O}$ through \eq{eq:toy_classb_tcmin}, the same values of $T_{c}$ are cycled as $\KFAVdim$ changes. 
This oscillatory behavior of $T_{c}$ as a function of $\KFAVdim$ is shown in \fig{fig:toy_classb_oscillatorytcmin}.

Now let us consider the behavior of the optimal $\Delta y_{SC}$ for a finite but large UV coupling $\aV_{y_0} = \aV^{UV} + \Delta\aV(y_0)$ in the large $\aV^{UV}$ limit. 
Once again, a generic profile in region I can be written as in \eq{eq:toy_classb_region1div} with $y^*_\infty$ given by
\bqa
y^*_\infty = y_0 + \frac{2}{\sqrt{\etaPI}} \text{arccoth} \Big\{\frac{4 R_d\big[\aV^{UV}+ \Delta\aV(y_0)\big] + 2\HD - 1}{\sqrt{\etaPI}}\Big\}.
\eqa
In the limit of large $\aV^{UV}$, this can be approximated 
as
\bqa
y^*_\infty \approx y_0 + 2 \Big(\frac{1}{v^{\prime \, UV}} \Big) - 8\Big(\frac{1}{v^{\prime\, UV}}\Big)^2 R_d \Delta\aV(y_0) 
\eqa
to the second order in $1/v^{\prime \,UV}$,
where $v^{\prime\, UV} \equiv  (4 R_d\aV^{UV}+2\HD - 1)$.
In order to solve for the optimal $y_O$ that gives rise to the highest $T_c$, we perform the same procedure as before; however, $y^*_\infty$ is no longer coincident with $y_0$. 
The maximization of $\Delta y_{SC}$ with respect to $y_0$ leads to
\bqa
\frac{\text{d}\Delta y_{SC}}{\text{d}y_0}  = \frac{\partial y^*_\infty}{\partial y_0} \frac{\partial y^*_{SC}}{\partial y^*_{\infty}} -1,
\eqa
where $y^*_{SC}(y^*_\infty)$ is the same as \eq{eq:toy_classb_region3div}. 
The optimal value of $y^*_\infty$ denoted as $\tilde{y}^*_O$ 
is computed from 
\bqa
0 = \bigg[1 -  8\Big(\frac{1}{v^{\prime\, UV}}\Big)^2 R_d \frac{\text{d}\Delta \aV}{\text{d}y_0} \Big|_{y_0 = \tilde{y}^*_O} \bigg]\frac{\partial y^*_{SC}}{\partial y^*_{\infty}} \Big|_{y^*_{\infty} = \tilde{y}^*_O} -1.
\eqa
Here, the derivative of $\Delta V(y_0)$ is evaluated at $\tilde{y}^*_O$,
which gives rise to corrections starting at the third order in $1/v^{\prime\, UV}$. The expression can be evaluated perturbatively in $1/v^{\prime\, UV}$ up to corrections that begin at $\order{\Delta \aV^2/(v^{\prime\, UV })^3}$ or $\order{1/(v^{\prime\, UV})^3}$.
This yields
\bqa
y_O \approx y^*_O - 2 \Big(\frac{1}{v^{\prime\, UV}}\Big) + 4 \Big(\frac{1}{v^{\prime\, UV}}\Big)^2 R_d \bigg[2\Delta \aV(y^*_O)- \frac{1}{\xi} \frac{\text{d}\Delta \aV}{\text{d}y_0} \Big|_{y_0 = y^*_O} \bigg], 
\eqa
where $\xi$ is in \eq{eq:toy_classb_xi}. 
The PFP diverges to $-\infty$ at 
\bqa
y^*_{SC,y_O} \approx y^*_{SC,O} - 4\Big(\frac{1}{v^{\prime\, UV}}\Big)^2 \frac{R_d}{\xi}\frac{\text{d}\Delta \aV}{\text{d}y_0}\Big|_{y_0 = y^*_O} .
\eqa
This results in  \eq{eq:TC_KF_universal_finiteV}.

\subsubsection{Universal pairing interaction}

In class B, over some finite window of energy scales, couplings are generically attracted towards the metallic PFP or towards the stable asymptotic fixed point $\aV^\bullet_{-\infty}$ in the small $y$ regime depending on whether they begin to the left or right of $y^*_S$. 
This creates a sharp crossover in the profile of the renormalized couplings at intermediate energy scales.
More precisely, such a profile emerges over the range of scales $l_b$ satisfying 
\bqa
y^*_{SC, O} \ll y^*_O - \frac{l_b}{2} \ll y^*_O.
\label{eq:toy_classb_lbrange}
\eqa
Here, the left inequality imposes the condition that superconductivity does not set in at scale $l_b$ yet.
The right inequality implies that $l_b$ is large so that couplings have had a long RG time to be attracted towards the metallic PFP or stable asymptotic fixed point. 
In the following, we focus on subsets of $l_b$ for which the crossovers occur in the $-\infty$ asymptotic region, and their forms become superuniversal. 
As will be discussed below, there are two distinct crossovers in class B and they become superuniversal at different energy scales.

To understand the crossover that arises in theories with highly repulsive UV couplings, we consider PFPs that emanate from a set of bare couplings given by $(y^{UV},\infty)$ with $-\infty < y^{UV} < \infty$.
We can use the separatrix PFP, which emanates from $y_S^*$, as the reference of crossover. 
For $y^{UV}<y_S^*$, a crossover interpolates between the separatrix PFP and $\aV^\bullet_{-\infty}$.
For $y^{UV}>y_S^*$, a crossover interpolates between the separatrix PFP and the metallic PFP.

\begin{figure}[H]
         \centering
         \includegraphics[width=0.7\linewidth]{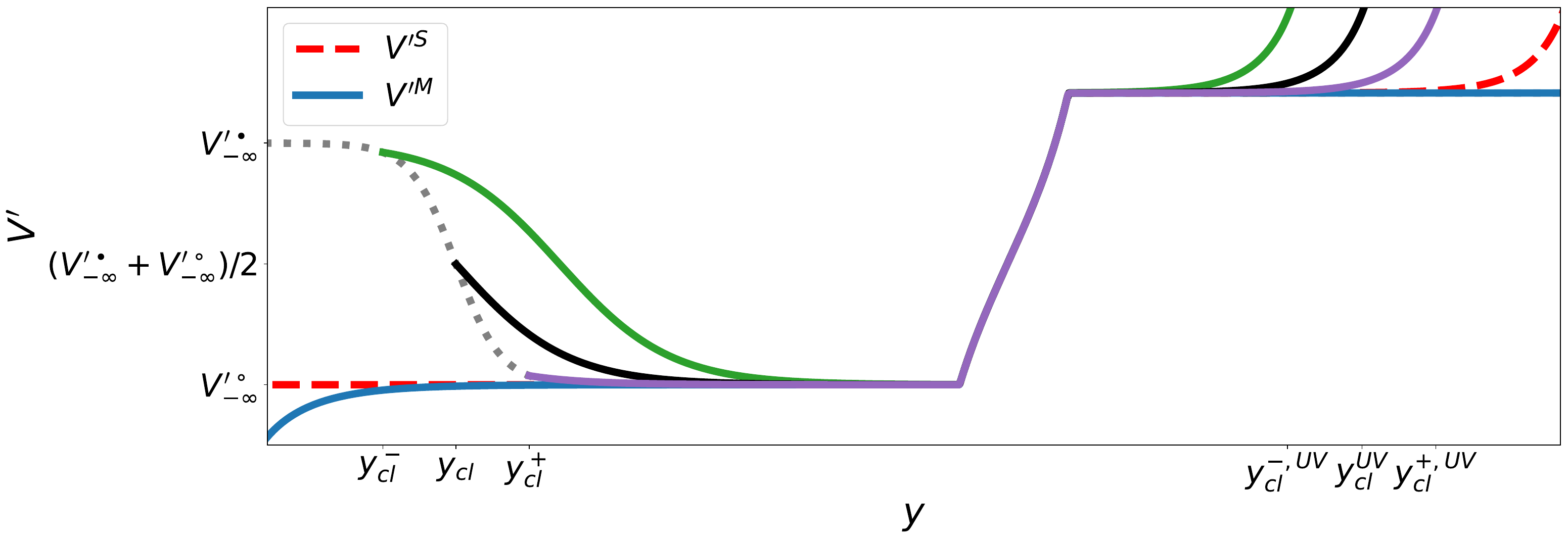}
         \caption{
Schematic plot of the crossover that renormalized couplings with $y^{(m)}(0)<y^*_S$ exhibit in class B.
The dotted line represents the profile of the renormalized coupling that emerges at large scale $l$ from the bare couplings that are highly repulsive.  
The curves have the same meaning as in \fig{fig:toy_classab_jump}, except that  $\aV^\circ_{-\infty}$ plays the role of $\aVM_y$.
The $y$ coordinates of the beginnings and ends of the PFPs are now labeled with a modified subscript to emphasize that they correspond to the jump that arises from couplings on the left-hand side of $y^*_S$.
 }
\label{fig:toy_classb_left_jump}
\end{figure}

Let us first consider the crossover between the separatrix and $\aV^\bullet_{-\infty}$.
To the left of $y^*_S$, the values of $y^{UV}$ whose renormalized couplings become 
$
\aV^\bullet_{-\infty} 
+
\varepsilon( \aV^\circ_{-\infty}  - \aV^\bullet_{-\infty}  )
$,
$\frac{
\aV^\bullet_{-\infty}
+ 
\aV^\circ_{-\infty}}{2}
$
and
$\aV^\circ_{-\infty} + 
\varepsilon( 
\aV^\bullet_{-\infty}  
-
\aV^\circ_{-\infty}  
)
$
at scale $l$ are denoted as
$y^{-,UV}_{cl}(l)$,
$y^{UV}_{cl}(l)$
and
$y^{+,UV}_{cl}(l)$, respectively
(see \fig{fig:toy_classb_left_jump}).
The associated `IR' logarithmic angular momenta shifted by $-l/2$ are denoted as
$y^{\pm}_{cl}(l)=
y^{\pm,UV}_{cl}(l)-l/2$
and
$y_{cl}(l)=
y^{UV}_{cl}(l) - l/2$.
For small $\delta w$,
the crossover angular momentum 
obeys the transcendental equation given by
\bqa
-\sqrt{\etaPIII}\Delta_S y^{UV}_{cl}(l) + \ln \Big(e^{{-\sqrt{\etaPI}}\Delta_S y^{ UV}_{cl}(l)} - 1 \Big) \approx -\frac{1}{2} \sqrt{\etaPIII} (l -l_c),
\label{eq:toy_classb_dsyc}
\eqa
where 
$\Delta_S y^{UV}_{cl}(l) \equiv y^{UV}_{cl}(l) - y_S^* $
and
\bqa
l_c \equiv 2( y^*_S - y^*_M )
\eqa
is a positive number.
For $l \ll l_c$,
the right-hand side of the expression is large and positive, 
and 
$\Delta_Sy_{cl}^{UV}$ must be large and negative.
For $|\Delta_S y_{cl}^{UV}| \gg 1/\sqrt{\etaPI}$ this leads to  
\bqa
\Delta_S y^{UV}_{cl}(l) \approx -\frac{1}{2} \frac{\sqrt{\etaPIII}}{\sqrt{\etaPI}+\sqrt{\etaPIII}}(l_{c}-l). \label{eq:toy_classb_yccase1}
\eqa 
In the other limit of $l \gg l_c$, 
the negative logarithm on the LHS of \eq{eq:toy_classb_dsyc} dominates and one obtains
\bqa
\Delta_S y^{UV}_{cl}(l) \approx -\frac{1}{\sqrt{\etaPI}}e^{-\frac{1}{2}\sqrt{\etaPIII}(l-l_c)}. \label{eq:toy_classb_yccase2}
\eqa
In class B, the metallic phase that exhibits universal behavior is confined to a finite window of energy scales since the system is unavoidably superconducting in the large $l$ limit. 
This limits the bottleneck scale $l_b$ in which the quasi-universal behavior arises to $l_b \ll l_{SC} \sim 2(y_O^*-y_{SC,O}^*)$. 
Within this range, if $l_b$ is larger than $l_{b,min}^{cl} = 2 y^*_S$,
$y_{cl}^{\pm}(l_b)$ is in the $-\infty$ asymptotic region, and the form of the crossover becomes superuniversal.
This leads us to focus on the length scales in 
\bqa
2(y^*_S - y^*_M ) +  2\ln\Bigg[ \Bigg(\frac{\sqrt{\etaPI}+\sqrt{\etaPIII}}{\sqrt{\etaPIII}}  \Bigg)^{\frac{1}{\sqrt{\etaPI}}} \Bigg( \frac{\sqrt{\etaPI}+\sqrt{\etaPIII}}{\sqrt{\etaPI}}\Bigg)^{\frac{1}{\sqrt{\etaPIII}}}  \Bigg] \gg l_b \gg 2 y^*_S,
\label{eq:toy_classb_lbrange1}
\eqa
which is a subset of  \eq{eq:toy_classb_lbrange}.
The difference $l_b - l_c$ over the range in \eq{eq:toy_classb_lbrange1} satisfies 
\bqa
C  \gg l_b - l_c \gg 2y^*_M , \label{eq:toy_classb_lb-lc}
\eqa
where $C$ is a positive constant 
given by the logarithmic term in \eq{eq:toy_classb_lbrange1}. 
Since $l_b - l_c$ is large and negative for essentially the entire range of permissible bottleneck $l_b$, the form of $y_{cl}$ that is appropriate is the one given by \eq{eq:toy_classb_yccase1}. 

The crossover momenta $y_{cl}^\pm$ are similarly obtained from equation \eq{eq:toy_classb_dsyc} under the replacements $\Delta_S y_{cl}^{UV} \to \Delta_S y_{cl}^{\pm,UV} $ and $l_c \to l_c^\pm$, with 
\bqa
l_c^\pm \equiv l_c  \mp \frac{4}{\sqrt{\etaPIII}} \text{arctanh}(1-2\varepsilon).
\eqa
This leads to $\Delta_S y^{\pm,UV}_{cl}(l)$ given by the same approximate expressions as in \eq{eq:toy_classb_yccase1} and \eq{eq:toy_classb_yccase2} in the appropriate regimes, but with $l_c^\pm$ now playing the role of $l_c$. 
\eq{eq:toy_classb_lb-lc} implies that $l_b - l_c^\pm$ is bounded as
$C \pm \frac{4}{\sqrt{\etaPIII}} \text{arctanh}(1-2\varepsilon)   \gg l_b - l_c^\pm \gg 2y^*_M \pm \frac{4}{\sqrt{\etaPIII}} \text{arctanh}(1-2\varepsilon)$,
which becomes
\bqa
C \pm \frac{2}{\sqrt{\etaPIII}} \ln\bigg(\frac{1}{\epsilon} \bigg) \gg l_b - l_c^\pm \gg 2y^*_M \pm \frac{2}{\sqrt{\etaPIII}} \ln\bigg(\frac{1}{\epsilon} \bigg)
\eqa
for $\varepsilon \ll 1$.
For small $\varepsilon$,
$y_{cl}^{\pm,UV}$ can behave differently over a large range of energy scales depending on $l_b$ relative to their respective crossovers. For this reason, we consider separately the jumps that exist on either side of $y_{cl}$. 


For $y_{cl}^-$, the appropriate form to use is the same as for $y_{cl}$ since the effect of $\varepsilon \neq 1$ is merely to shift the range of $l_b - l_c^- $ into more negative values. 
The width associated with jump between $y_{cl}$ and $y_{cl}^-$ is
\bqa
|\Delta \tilde{x}_{cl}^-| 
\equiv 
\left| 
e^{y_{cl}^{-,UV}}-e^{y_{cl}^{UV}} \right| 
\approx e^{y_S^* - \frac{1}{2} \frac{\sqrt{\etaPIII}}{\sqrt{\etaPI}+\sqrt{\etaPIII}}(l_c -l_b)}\Big|\varepsilon^{\frac{1}{\sqrt{\etaPI}+\sqrt{\etaPIII}}} - 1\Big|.
\eqa
Since $l_b$ is less than $l_c$,  the width of this jump is bounded from above by $e^{y^*_S} \approx e^{w_c} \Big[\frac{4\sqrt{\etaPI}}{(\etaPI - \etaPII)\delta w}\Big]^\frac{1}{\sqrt{\etaPI}}$.
%

For $y_{cl}^+$, $l_b-l_c^+$ is positive within a window of energy scales for small $\varepsilon$\footnote{
We still choose $\varepsilon >
\varepsilon_{min}(\delta l_b)$
so that $y_{cl}^{\pm,UV}$ are deep in region I, where
$\varepsilon_{min}(\delta l_b) \sim  \delta w ^{\frac{\sqrt{\etaPIII}}{\sqrt{\etaPI}}}e^{-\frac{1}{2}\sqrt{\etaPIII}\delta l_b}$
with
$\delta l_b \equiv l_b -l^{cl}_{b,min}$.}.
The width of the second-half of the jump can similarly be computed, giving
\bqa
|\Delta \tilde{x}_{cl}^+| 
\equiv  \left| e^{y_{cl}^{UV}}-e^{y_{cl}^{+,UV}} \right|
\approx e^{y^*_S} \begin{cases}
       e^{- \frac{1}{2} \frac{\sqrt{\etaPIII}}{\sqrt{\etaPI}+\sqrt{\etaPIII}}(l_c-l_b)}\varepsilon^{-\frac{1}{\sqrt{\etaPI}+\sqrt{\etaPIII}}} ~~~~~ &\text{for } l_b \ll l_c^+\\
     1~~~~~ &\text{for } l_b \gg l_{c}^+
\end{cases}.
\eqa
$|\Delta \tilde{x}_{cl}^+|$ is also bounded by $e^{y^*_S}$.

\begin{figure}[H]
         \centering
         \includegraphics[width=0.7\linewidth]{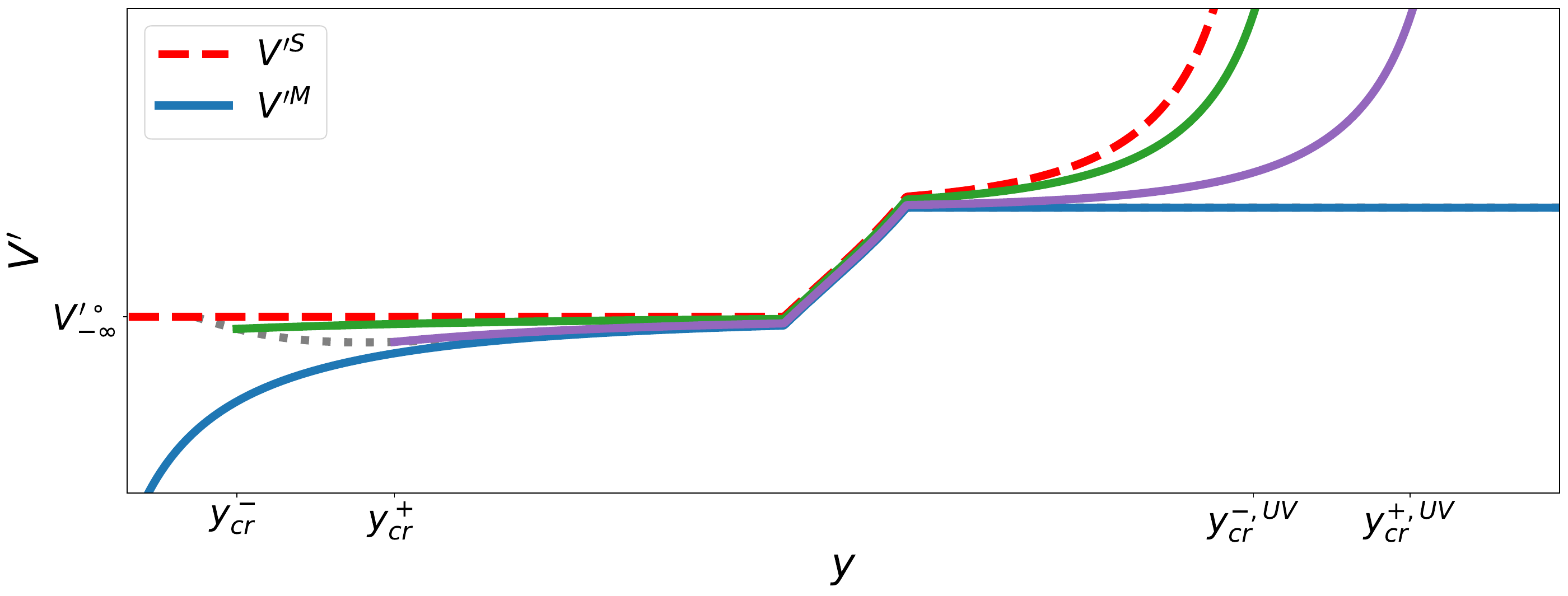}
         \caption{
Schematic plot of the crossover that renormalized couplings with 
$y^{(m)}(0)>y^*_S$ exhibit in class B.
The dotted line represents the profile of the renormalized coupling that emerge at scale $l_b$ from the highly-repulsive UV couplings. 
The green curve emanates from $(y^{-,UV}_{cr}(l_b), \infty)$ and terminates at $\Big(y^{-}_{cr}(l_b),\aV^S_{y^-_{cr}(l_b)}+ \varepsilon (\aV^M_{y^-_{cr}(l_b)} - \aV^S_{y^-_{cr}(l_b)} )\Big) $, while the purple curve emanates from $(y^{+,UV}_{cr}(l_b), \infty)$ and terminates at $\Big(y^{+}_{cr}(l_b),\aV^M_{y^+_{cr}(l_b)}+ \varepsilon (\aV^S_{y^+_{cr}(l_b)} - \aV^M_{y^+_{cr}(l_b)} )\Big) $.   
The shifted logarithmic angular momenta are related to their bare values through $y_{cr}^\pm(l_b) = y_{cr}^{\pm,UV}(l_b) - l_b/2$.}
\label{fig:toy_classb_right_jump}
\end{figure}

Now, let us consider the crossover of the couplings with angular momenta larger than the critical angular momentum, which begin to the right of $y^*_S$.
To the right of $y^*_S$, the values of $y^{UV}$ whose renormalized couplings become 
$
\aVS_{ y^{-}_{cr}(l)} 
+\varepsilon
(
\aVM_{ y^{-}_{cr}(l)} 
-
\aVS_{ y^{-}_{cr}(l)}  
)
$,
$\frac{
\aVS_{ y_{cr}(l)}  
+\aVM_{ y_{cr}(l)} 
}{2}
$
and
$
\aVM_{ y^{+}_{cr}(l)} 
+\varepsilon
(\aVS_{ y^{+}_{cr}(l)}  - \aVM_{ y^{+}_{cr}(l)} )
$
at scale $l$ are denoted as
$y^{-,UV}_{cr}(l)$,
$y^{UV}_{cr}(l)$
and
$y^{+,UV}_{cr}(l)$, respectively,
where
$y^{\pm}_{cr}(l) = y^{\pm,UV}_{cr}(l)-l/2$
and
$y_{cr}(l) = y^{UV}_{cr}(l)-l/2$.
In the small $\delta w$ limit, all trajectories are within a narrow gap between the metallic and separatrix PFPs. 
The width of the crossovers can be found by equating 
$\aV^\pm_0 - \aV^S_0  = (\frac{1}{2}\pm \varepsilon') (\aV^M_0 - \aV^S_0)$ 
with $\varepsilon' \equiv \frac{1}{2} -\varepsilon$,
where $\aV^\pm_y$ is the PFP that emanates from $y^{\pm,UV}_{cr}$.
This leads to
\bqa
\Delta_S y_{cr}^{\pm,UV} = -\frac{1}{\sqrt{\etaPI}} \ln \big(\frac{1}{2} \mp \varepsilon' \big),\label{eq:toy_classb_deltasyright}
\eqa
where 
$\Delta_Sy^{\pm,UV}_{cr} = 
y_{cr}^{\pm,UV}- y^*_S$.
Therefore, the width of the crossover to the right of $y^*_S$ is given by 
\bqa
\Delta \tilde{x}_{cr} \approx e^{y^*_S}\varepsilon^{-\frac{1}{\sqrt{\etaPI}}}. \label{eq:toy_classb_rightjumpwidth}
\eqa
The minimum $l_b$ required for this crossover to occur deep in region III and thus for \eq{eq:toy_classb_rightjumpwidth} to hold is given by $l_{b,min}^{cr} \equiv 2y^{+,UV}_{cr} \approx 2y^*_S + \frac{2}{\sqrt{\etaPI}}\ln\frac{1}{\varepsilon}$.
In the small $\delta w$ limit with a small but fixed $\varepsilon$, there is a hierarchy of length scales:
$l_{b,min}^{cl} \ll  l_{b,min}^{cr} \ll  l_{SC}$.
In  $l_{b,min}^{cl} \ll l \ll  l_{b,min}^{cr}$, the crossover to the left of $y^*_S$ takes the superuniversal form.
In $l_{b,min}^{cr} \ll l \ll  l_{SC}$, the crossovers on both sides of $y_S^*$ become superuniversal.
As long as $l \gg l_{b,min}^{cl}$, there exists a crossover with $\Delta \tilde x \sim \order{1}$ 
and the universal pairing interaction in \eq{eq:aVforsharpjump} exhibits the $\sqmu/\bth$ decay upto $\bth_c \sim \order{1}$.

\subsection{S-wave to non-s-wave SC critical superuniversality class (BC)}
\label{appendix:BC}

The critical class BC is similar to the class B except that the stable and unstable asymptotic fixed points in small $y$ regime have merged into a single marginal one. 
Similarly to what was done in the analysis of class B, here we focus only on the class BC proximate to class A. 
The solutions for the metallic and separatrix PFPs may be obtained from taking the limit as $\etaPIII \to 0$ in \eq{eq:toy_classb_vm} and \eq{eq:toy_classb_vs}, giving
\bqa
\aV_y^M = \frac{1}{4R_d}\begin{cases}
    \sqrt{\etaPI} -2\HD + 1  ~~~~&\text{for } y \ge \wII \\
    \sqrt{-\etaPII} \tan \Big[\frac{1}{2}\sqrt{-\etaPII}\, (y-\wII)   + \arctan\Big(\frac{\sqrt{\etaPI}}{\sqrt{-\etaPII}} \Big)    \Big] -2\HD + 1 &\text{for } 0 < y < \wII \\   
  -\frac{2}{y-y^*_M}-2\HD +1 &\text{for } y \le 0\label{eq:toy_classb_vm_bc}
    \end{cases}
\eqa
\bqa
\aV_y^S = \frac{1}{4R_d}\begin{cases}
   - \sqrt{\etaPI} \coth\Big[\frac{1}{2}\sqrt{\etaPI} \, (y-y^*_S)\Big]   -2\HD + 1 ~~~~&\text{for } y \ge \wII \\
    \sqrt{-\etaPII} \tan \Big(\frac{1}{2}\sqrt{-\etaPII}\, y     \Big) -2\HD + 1 &\text{for } 0 < y < \wII \\   
    -2\HD + 1 &\text{for } y \le  0\label{eq:toy_classb_vs_bc}
    \end{cases}.
\eqa
Here, $y_M^*$ and $y_S^*$ are the locations at which the metallic and separatrix PFPs diverge to $-\infty$
and $\infty$, respectively,
\bqa
y_M^* &=& - \frac{2}{\sqrt{-\etaPII}} \cot \bigg[\frac{1}{2}\sqrt{-\etaPII} \wII - \arctan \bigg( \frac{\sqrt{\etaPI}}{\sqrt{-\etaPII}}\bigg)  \bigg],  \nn
y_S^* &=& \wII + \frac{2}{\sqrt{\etaPI}}  \text{arccoth}\bigg[ \frac{\sqrt{-\etaPII}}{\sqrt{\etaPI}} \tan\bigg(\frac{1}{2}\sqrt{-\etaPII}\, \wII    \bigg)\bigg].
\eqa
In the small $\delta w$ limit, they become
\bqa
y^*_M \approx -\frac{4}{\abs{\etaPII} \delta w},
~~~~~
y^*_S \approx w_c + \frac{1}{\sqrt{\etaPI}}\log\bigg(\frac{4\sqrt{\etaPI}}{\etaPI - \etaPII}\frac{1}{\delta w} \bigg).
\eqa

\subsubsection{
Universal $T_c/\KFAVdim^z$
and oscillation of $T_c$ with $\KFAVdim$
}

As in the classes B and C, class BC exhibits universal $T_c$. 
The discussion from class B is largely applicable here, 
except that the starting point \eq{eq:toy_classb_ystariii} is replaced with its $\etaPIII\to 0$ limit.
That PFP that emanates from $(y_0, \aV_{y_0})$ diverges to $-\infty$ at
\bqa
y^*_{SC} = \frac{2}{4R_d \aV_0 + 2\HD -1},
\eqa
where $\aV_0$ is the PFP evaluated at $y=0$.
For  $\aV_{y_0} = \aVM_{y_0} + \deltaaVUV$, the PFP diverges at
$y^*_{SC}=y^*_M + \delta y^*$  with
\bqa
\delta y^* \approx - \frac{32 R_d}{\big|\etaPII (\etaPI-\etaPII)   \big|}\frac{e^{-\sqrt{\etaPI}(y_0-w_c)}}{\delta w^2 } \deltaaVUV
\eqa 
to the leading order in $\delta w$.
The reduced deviation becomes
\bqa
\frac{\delta y^*}{|y^*_M|}\approx -\frac{2R_d}{\sqrt{\etaPI}} e^{-\sqrt{\etaPI}\Delta_S y_0} \deltaaVUV,
\eqa
where  $\Delta_S y_0 = y_0 - y^*_S$.

For highly repulsive bare coupling, the optimal value of angular momentum $y^*_O$ for the highest $T_c$
and the location at which the optimal PFP diverges to $-\infty$ are given by
\bqa
y^*_O & \approx & \frac{2}{\sqrt{\etaPI}} \ln \frac{1}{\delta w} + \frac{1}{\sqrt{\etaPI}}\ln\Big[\frac{16 \etaPI}{|\etaPII|(\etaPI-\etaPII)} \Big] + \wII_c  + \order{\delta w}, \nn
y^*_{O,SC} & \approx & -\frac{4}{|\etaPII|}\frac{1}{\delta w} - \frac{1}{\sqrt{\etaPI}} + \order{\delta w}.
\eqa
Although $y^*_{O,SC} \approx y^*_M + \order{1}$ as is the case in class B, $y^*_O$ is generally far to the right of $y^*_S$ by  $ \sim - \frac{1}{\sqrt{\etaPI}}\ln{\delta w}$. 
The discrepancy arises from the fact that while $y^*_S$ is oblivious to the marginality of $\aV^\halfominus_{\infty}$, $y^*_O$ is the result of an optimization that is sensitive to the nature of the flow proximate to $\aV^\halfominus_{\infty}$ in the small $y$ regime.   
In class BC, $T_c$ also exhibits the same oscillatory behavior as a function of $\KFAVdim$ as observed in class B, with the precise form of the oscillations being modified by the marginality of the small-$y$ asymptotic fixed point. 
Near one of the resonance Fermi momenta that give rise to the maximum $T_c$, $T_c$ still obeys
\eq{eq:toy_classb_tcmin_main}
with
\bqa
\begin{aligned}
\xi = - &\Bigg[  \etaPI \csc^2 \bigg(\frac{1}{2}\sqrt{-\etaPII} w+ \arctan\bigg[\cot\bigg( \frac{1}{2}\sqrt{-\etaPII} w \bigg)- \frac{\sqrt{\etaPI - \etaPII}}{\sqrt{-\etaPII}}\csc\bigg( \frac{1}{2} \sqrt{-\etaPII}w\bigg) \bigg]  \bigg)\\ &\times \Bigg(\frac{2\etaPI \sqrt{-\etaPII}\sqrt{\etaPI -\etaPII} \sin \bigg( \frac{1}{2}\sqrt{-\etaPII }w \bigg) }{\sqrt{\etaPI - \etaPII}\cos \bigg( \frac{1}{2}\sqrt{-\etaPII }w \bigg)-\sqrt{-\etaPII}} - \sqrt{\etaPI}(\etaPI - \etaPII)\\&\times \sinh \bigg\{2 \text{arccoth}\bigg[ \frac{\sqrt{\etaPI - \etaPII} -\sqrt{-\etaPII} \cos\bigg(\frac{1}{2} \sqrt{-\etaPII} w  \bigg) }{\sqrt{\etaPI}} \csc\bigg( \frac{1}{2} \sqrt{-\etaPII} w\bigg) \bigg]  \bigg\} \Bigg)  \Bigg]/
\\& \bigg[\etaPI + \etaPII + (\etaPI - \etaPII) \cosh \bigg\{2 \text{arccoth}\bigg[ \frac{\sqrt{\etaPI - \etaPII} -\sqrt{-\etaPII} \cos\bigg(\frac{1}{2} \sqrt{-\etaPII} w  \bigg) }{\sqrt{\etaPI}} \csc\bigg( \frac{1}{2} \sqrt{-\etaPII} w\bigg) \bigg]  \bigg\}  \bigg]^2.
\end{aligned}
\eqa

\subsubsection{Universal pairing interaction}


\begin{figure}
    \centering
    \includegraphics[width=0.7\linewidth]{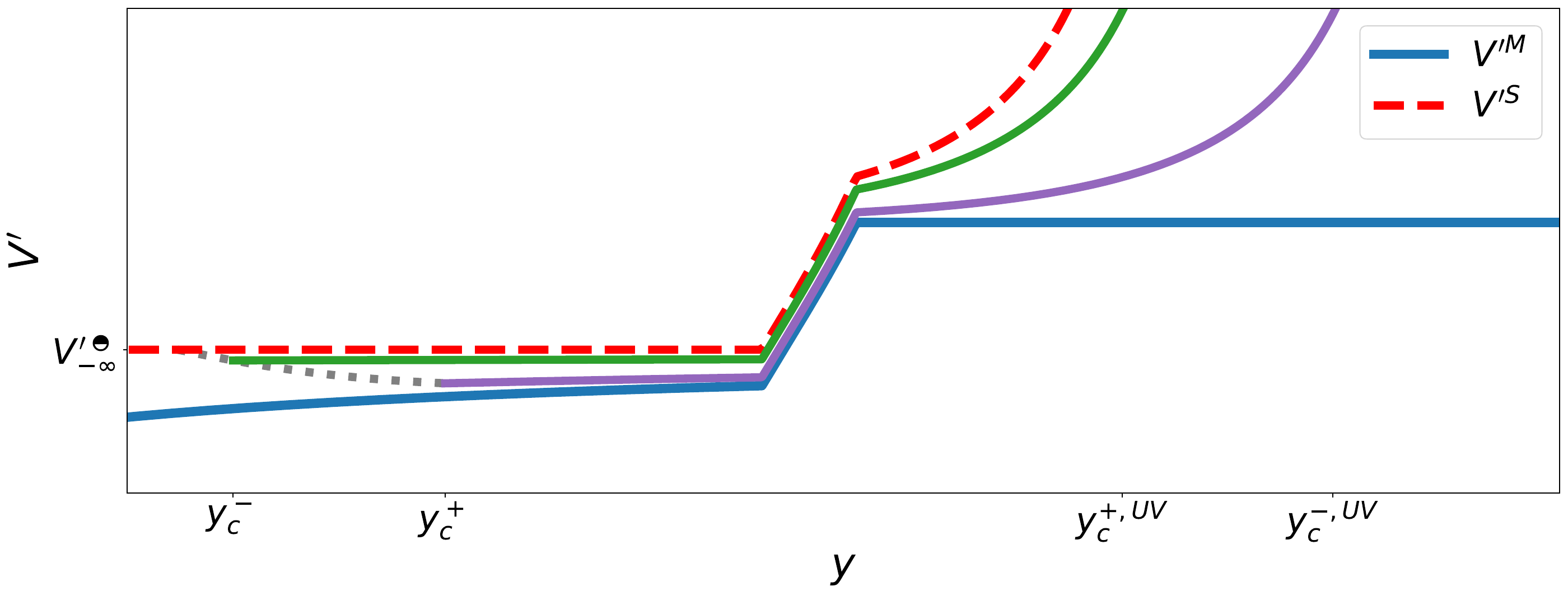}
    \caption{
Schematic plot of the crossover that renormalized couplings
exhibit in class BC.
The curves have the same meaning as in \fig{fig:toy_classb_right_jump} 
once 
$\aV^\circ_{-\infty}$
is replaced  with
$\aV^\halfominus_{-\infty}$.
}
\label{fig:toy_classbc_jump}
\end{figure}

As in class B, the renormalized couplings that emerge at intermediate energies exhibit a crossover.
In class BC, we only need to consider the crossover of couplings that originate to the right of $y^*_S$ at UV, since there is only a single marginal fixed point in region III.
We once again consider bare couplings that are highly repulsive. 
The values of the logarithmic angular momentum $y^{UV}$ whose renormalized couplings become  
$  \aV^\halfominus_{-\infty}  
+  
\varepsilon  
(\aVM_{y_c^-(l)}- \aV^\halfominus_{-\infty}) $ 
and  
$\aVM_{y_c^+(l)}
+\varepsilon  
( \aV^\halfominus_{-\infty} - \aVM_{y_c^+(l)} )$ 
at scale $l$ are denoted as 
$y^{-,UV}_{c}(l)$ and $y^{+,UV}_{c}(l)$, respectively,  where $y^{\pm}_{c}(l)  =  y^{\pm,UV}_{c}(l)-l/2$. 
This is illustrated in \fig{fig:toy_classbc_jump}.
With $\varepsilon' \equiv 1/2 - \varepsilon$, $y_c^{\pm,UV}$ obeys 
\bqa
\frac{1}{v'^{-1}_0(y_c^{\pm,UV}) - \frac{1}{2}y_c^\pm}=
\left(\frac{1}{2} \pm \varepsilon'\right) \frac{2}{y^*_M - y_c^\pm },
\label{eq:toy_classbc_yceq}
\eqa
with
\bqa
v'_0(y_c^{\pm,UV}) = \sqrt{-\etaPII}\tan\Bigg(-\frac{1}{2}\sqrt{-\etaPII}\wII - \arctan\Bigg\{ \frac{\sqrt{\etaPI}}{\sqrt{-\etaPII}} \coth\Big[ \frac{1}{2} \sqrt{\etaPI}(\wII - y_c^{\pm, UV})\Big] \Bigg\} \Bigg).
\eqa
To the leading order in $\delta w$, this is simplified as 
\bqa
v'_0(y_c^{\pm,UV}) \approx -\frac{1}{2}\etaPII \delta w\Big(e^{-\sqrt{\etaPI}\Delta_S y_c^{\pm,UV}} -1 \Big),
\eqa
where
$\Delta_S y_c^{\pm,UV} 
= y^{\pm,UV}_c(l) - y^*_S$.
For $l < l_{b,max} \sim 1/\delta w$, $y_c^\pm $ 
are sub-leading in the denominators of equation \eq{eq:toy_classbc_yceq}. 
To the leading-order in $y_c^{\pm}\delta w \ll 1$, we obtain
\bqa
\Delta_S y_c^{\pm, UV} \approx - \frac{1}{\sqrt{\etaPI}}\ln\Big(\frac{1}{2} \mp \varepsilon' \Big),
\eqa
which follows from expanding both sides of \eq{eq:toy_classbc_yceq}.
This gives rise to the width of the crossover,
\bqa
|\Delta \tilde{x}| 
\equiv 
|e^{y_{c}^{+,UV}}-e^{y_{c}^{-,UV}}| 
= 
e^{y^*_S}\Big| \varepsilon^{-\frac{1}{\sqrt{\etaPI}}} - (1-\varepsilon)^{-\frac{1}{\sqrt{\etaPI}}} \Big|.
\eqa
For small $\varepsilon$, the width scales as $\Delta\tilde{x} \sim \big(\varepsilon \delta w\big)^{- \frac{1}{\sqrt{\etaPI}}}$, which is independently of $l$. 
Therefore, the crossover angle in
\eq{eq:aVforsharpjump} is independent of $l$.

\subsection{S-wave SC superuniversality class (C)}
\label{appendix:C}


In class C, it suffices to set $\wII = 0$ since the intermediate region does not play any essential role in realizing the class. 
The metallic PFP, which emenates from $\aV^\bullet_{\infty}$ is given by
\bqa
\aV^M_y = \frac{1}{4R_d}\begin{cases}
    \sqrt{\etaPI}-2\HD + 1 ~~~~ &\text{for } y > 0\\
    \sqrt{-\etaPIII} \tan \Big[\frac{1}{2}\sqrt{-\etaPIII} \,(y-y^*_M)  - \frac{\pi}{2}]-2\HD + 1 &\text{for } y\le 0
\end{cases},
\eqa
where $y_M^*$ is the shifted logarithmic angular momentum at which the metallic PFP diverges to $-\infty$, 
\bqa
y^*_M = - \frac{2}{\sqrt{-\etaPIII}} \Bigg[ \frac{\pi}{2}  +  \arctan \bigg(\frac{\sqrt{\etaPI}}{\sqrt{-\etaPIII}} \bigg)\Bigg].
\eqa
Since there is no real asymptotic fixed-point in the small-$y$ region, there is no separatrix PFP.

\subsubsection{Universal $T_c/\KFAVdim^z$}

The PFP that goes through
$(y_0, \aV^M_{y_0}+\deltaaVUV)$ 
with $y_0>0$ diverges at
\bqa
y^*_{SC} = - \frac{2}{\sqrt{-\etaPIII}} \Bigg[ \frac{\pi}{2}  +  \arctan \bigg(\frac{4R_d \aV_0 + 2\HD -1}{\sqrt{-\etaPIII}} \bigg)\Bigg],
\eqa
where 
$\aV_0$ is the coupling of the PFP evaluated at $y=0$.
For small 
$\deltaaVUV = 
\aV_{y_0}-\aV^M_{y_0}$,
the perturbation 
$\delta \aV_0 = \aV_0 - \aV^M_0$ at $y=0$ is reduced to
$\delta\aV_0  \approx A_{0;y_0} \deltaaVUV$ 
with
$A_{0;y_0} = 
e^{-\sqrt{\etaPI}y_0}$.
Since $\delta \aV_0$ becomes smaller with increasing $y_0$, the deformed PFP diverges  
$y^*_{SC} = y^*_M + \delta y^* $ closer to the metallic PFP,
\bqa
\delta y^* = - \frac{8R_d e^{-\sqrt{\etaPI}y_0}}{\big|\etaPI-\etaPIII\big|} \deltaaVUV.
\eqa
This leads to the weak dependence of $T_c/\KFAVdim^z$ on the bare coupling when the superconducting instability occurs at a large angular momentum.
This is consistent with the result obtained in the Ising-nematic theory  (see \eq{eq:deltaySC}).

\subsubsection{Universal pairing interaction}

Now, we calculate the profile of the regularized metallic PFP that arises at
the bottleneck scale
$l_b = \frac{2\pi}{\sqrt{-\discinf}}$.
For the bare couplings that are highly repulsive, the regularized metallic PFP that emerges at the bottleneck scale $l_b$ becomes
\begin{equation}
    \begin{aligned}
        \aV^{RM}_y = \frac{1}{4R_d}\begin{cases}
    \sqrt{\etaPI}\coth\left[
    \frac{\sqrt{\etaPI}}{4} l_b
    \right]-2\HD + 1 ~~~~ &\text{for } y > 0,\\
    \sqrt{-\etaPIII} \tan \Big\{\frac{\sqrt{-\etaPIII}}{2} \,y  + \arctan \Big[\frac{\sqrt{\etaPI}}{\sqrt{-\etaPIII}} \coth\left(
        \frac{\sqrt{\etaPI}}{2}
\left(y + \frac{l_b}{2} \right)
\right)\Big]\Big\}-2\HD + 1 &\text{for } -\frac{l_b}{2} <y\leq 0,\\
    -2\HD + 1 &\text{for } y\leq -\frac{l_b}{2},
\end{cases}.
\label{eq:toy_c_rmfp}
    \end{aligned}
\end{equation}

\begin{figure}[ht]
    \centering

        \includegraphics[width=0.45 \textwidth]{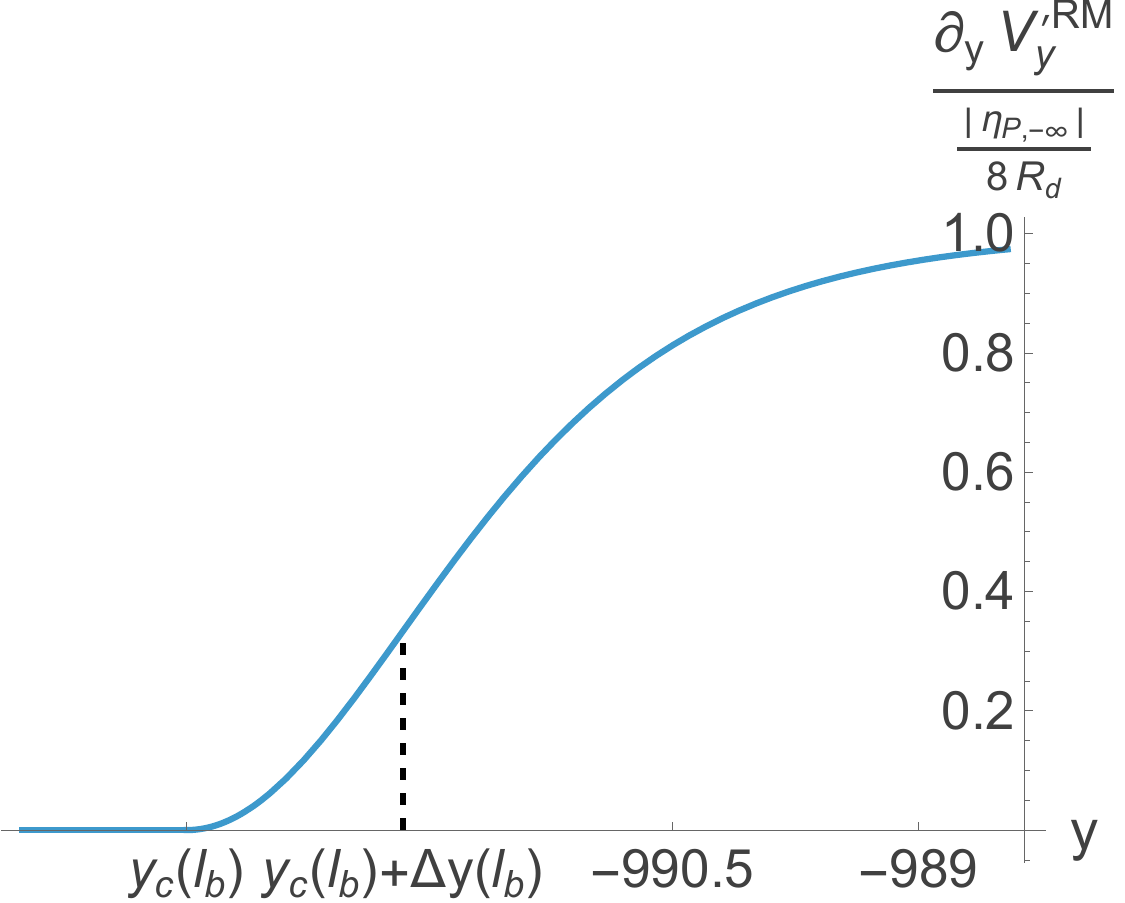}
    \caption{
The derivative of \eq{eq:toy_c_rmfp} with respect to $y$ plotted as a function of $y$ near the kink at the bottleneck scale $l_b$.
For the ideal kink in Eq. (\ref{eq:quasi_universalV}), the derivative is the step function.
For the actual profile that emerges at $l_b$, however, the `jump' occurs over a finite width, starting at $y_c(l_b)=-l_b/2$.
At the center of the width indicated by the dashed line,
the second derivative is maximum.
For the plot, we use
$\etaPIII = -0.00001$, $\etaPI = 1$ and $R_d = 1$.
}
\label{fig:first_deriv_sharp_jump}
\end{figure}

While the regularized metallic PFP is continuous,
it exhibits a crossover at  $y_b^{RM} = -\frac{l_b}{2}$.
We note that 
$\partial_y \aV^{RM}_y = 0$ 
for  
$y< y_b^{RM}$
and 
$\partial_y \aV^{RM}_y \approx \frac{|\etaPIII|}{8 R_d}$ 
for $y \gg y_b^{RM}$.
These two limits are interpolated by a smooth crossover,
as shown in \fig{fig:first_deriv_sharp_jump}. 
The center of the crossover is identified as the point at which 
$\partial_y \aV^{RM}_y $ changes most rapidly and 
the second derivative of $\aV^{RM}_y$ reaches the maximum.
At this point, the third derivative should vanish.
Proximate to class A, $\Delta y = y-y_b^{RM}$ satisfies
\begin{equation}
    \begin{aligned}
        \partial_y^3 \aV^{RM}_y \approx \frac{\etaPI\etaPIII}{16R_d}\tanh^4\left(\frac{\sqrt{\eta _{\infty }}}{2}
  \Delta y \right)
\csch^2\left(\frac{\sqrt{\eta _{\infty }}}{2}
  \Delta y \right)\left[2-\mathrm{csch}^2\left(\frac{\sqrt{\eta _{\infty }}}{2}
   \Delta y  \right)\right] = 0
    \end{aligned}
\end{equation}
to the leading order in $|\etaPIII|$. 
The solution is given by
\begin{equation}
    \begin{aligned}
        \Delta y = \frac{2}{\sqrt{\etaPI}}~\mathrm{arcsinh}\left(\frac{1}{\sqrt{2}}\right),
    \end{aligned}
\end{equation}
and the width of the crossover  at scale $l_b$ becomes
\bqa
|\Delta \tilde{x}| 
\equiv 
e^{l_b/2} 
\left|
e^{y_{c}(l_b)+2 \Delta y}  
-
e^{y_{c}(l_b)} 
\right|
\sim \order{1}. 
\eqa

\section*{Acknowledgement}

This research was supported by the Natural Sciences 
and Engineering Research Council of
Canada. Research at the Perimeter Institute is supported in part by the
Government of Canada through Industry Canada, and by the Province of
Ontario through the Ministry of Research and Information.

\bibliographystyle{unsrtnat}

\bibliography{references-2}

\end{document}